 \documentclass[twocolumn]{aastex62}

\shorttitle{Light Curves of Classical Novae}
\shortauthors{Hachisu \& Kato}

\begin{document}

\title{A Light Curve Analysis of 32 Recent Galactic Novae ---
Distances and White Dwarf Masses}


\author{Izumi Hachisu}
\affil{Department of Earth Science and Astronomy, 
College of Arts and Sciences, The University of Tokyo,
3-8-1 Komaba, Meguro-ku, Tokyo 153-8902, Japan} 
\email{hachisu@ea.c.u-tokyo.ac.jp}


\author{Mariko Kato}
\affil{Department of Astronomy, Keio University, 
Hiyoshi, Kouhoku-ku, Yokohama 223-8521, Japan} 

%


%
%



\begin{abstract}
We obtained the absolute magnitudes, distances, and white dwarf (WD) masses
of 32 recent galactic novae based on the time-stretching method for nova
light curves.  A large part of the light/color curves of two classical novae
often overlap each other if we properly squeeze/stretch their timescales.
Then, a target nova brightness is related to the other template
nova brightness by $(M_V[t])_{\rm template} = (M_V[t/f_{\rm s}] -
2.5 \log f_{\rm s})_{\rm target}$, where $t$ is the time, $M_V[t]$
is the absolute $V$ magnitude, and $f_{\rm s}$ is their timescaling ratio.  
Moreover, when these two
time-stretched light curves, $(t/f_{\rm s})$-$(M_V-2.5 \log f_{\rm s})$,
overlap each other, $(t/f_{\rm s})$-$(B-V)_0$ do too,
where $(B-V)_0$ is the intrinsic $B-V$ color.  Thus, the two nova
tracks overlap each other in the $(B-V)_0$-$(M_V-2.5 \log f_{\rm s})$ diagram.
Inversely using these properties,
we obtain/confirm the distance and reddening by comparing each nova
light/color curves with the well calibrated template novae.   
We classify the 32 novae into two types, LV~Vul and
V1500~Cyg types, in the time-stretched $(B-V)_0$-$(M_V-2.5 \log f_{\rm s})$
color-magnitude diagram.
The WD mass is obtained by direct comparison of the model $V$ light curves
with the observation.  Thus, we obtain a uniform set of 32 galactic
classical novae that provides the distances and WD masses from 
a single method.  Many novae broadly follow the 
universal decline law and the present method can be applied to them,
while some novae largely deviate from the universal decline law and so 
the method cannot be directly applied to them.  We discuss such examples.
\end{abstract}




\section{Introduction}
\label{introduction}
A nova is a thermonuclear explosion event on a mass-accreting 
white dwarf (WD) in a close binary.  A hydrogen-rich envelope 
accumulates on the WD.  When the envelope mass increases and
reaches a critical value, hydrogen ignites to trigger a nova explosion.
The envelope expands to giant size and blows strong winds.
The nova peak brightness depends on the radius of 
the photosphere and the wind mass-loss rate \citep[e.g.,][]{hac15k}.
After the maximum expansion of the photosphere or the maximum wind
mass-loss rate is attained, the nova brightness begins to decline
\citep[e.g.,][]{hac17k}.  
A large part of the hydrogen-rich envelope is blown in the wind.
After the winds stop, the envelope mass decreases further
by nuclear burning.  When it decreases to below a critical mass
(minimum mass for steady hydrogen-shell burning), 
nuclear burning extinguishes \citep[e.g.,][]{kat14shn}.
The WD cools down, and the nova ends.

After the optical peak, optical/IR fluxes are dominated by
free-free emission.  \citet{hac06kb} calculated many free-free emission
model light curves based on the optically thick wind model calculated
by \citet{kat94h}, in which the wind is accelerated deep inside 
the photosphere (owing to the iron peak of opacity
at $\log T {\rm ~(K)} \sim 5.2$).  Their theoretical light curves show
a homologous shape.  If we properly normalize the timescales, 
all the light curves overlap each other, independently of the WD mass
and chemical composition \citep[see, e.g., Figure 8 of][]{hac06kb}.  
They called this property of nova model light curves the universal decline law.

\citet{hac14k} clarified that many classical novae also show a similarity
even in their color evolutions.  Physically, these properties mean
that two novae with different speed classes experience a similar
temporal change relative to the total duration.
The timescaling law is very useful for analyzing a (target) nova
light curve because the distance modulus, extinction, and other various
quantities can be derived from a well calibrated (template) nova
by comparing/overlapping with the light/color curves.
\citet{hac10k, hac15k, hac16k} formulated the method and called
it the time-stretching method.

\begin{figure*}
\epsscale{1.1}
\plottwo{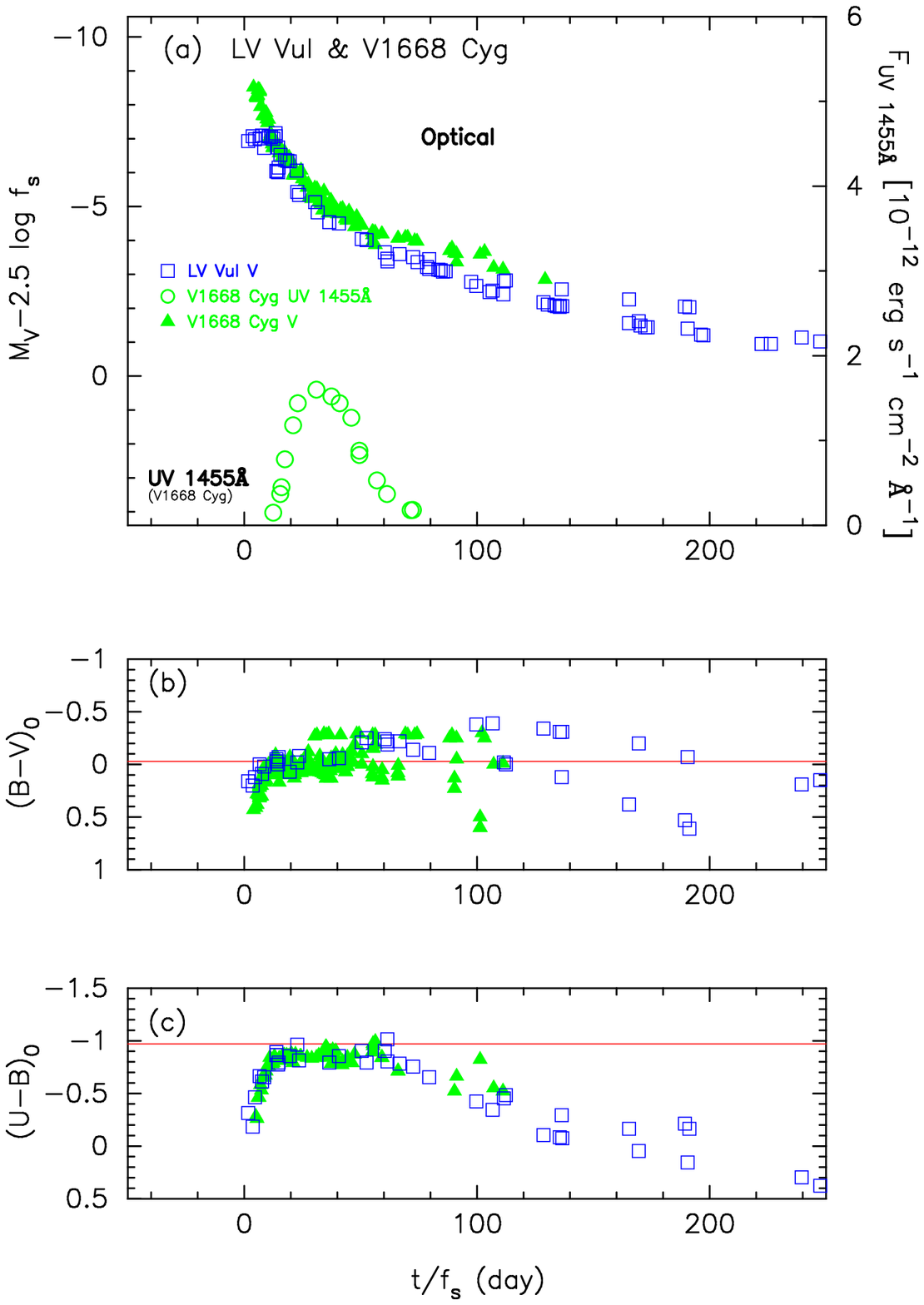}{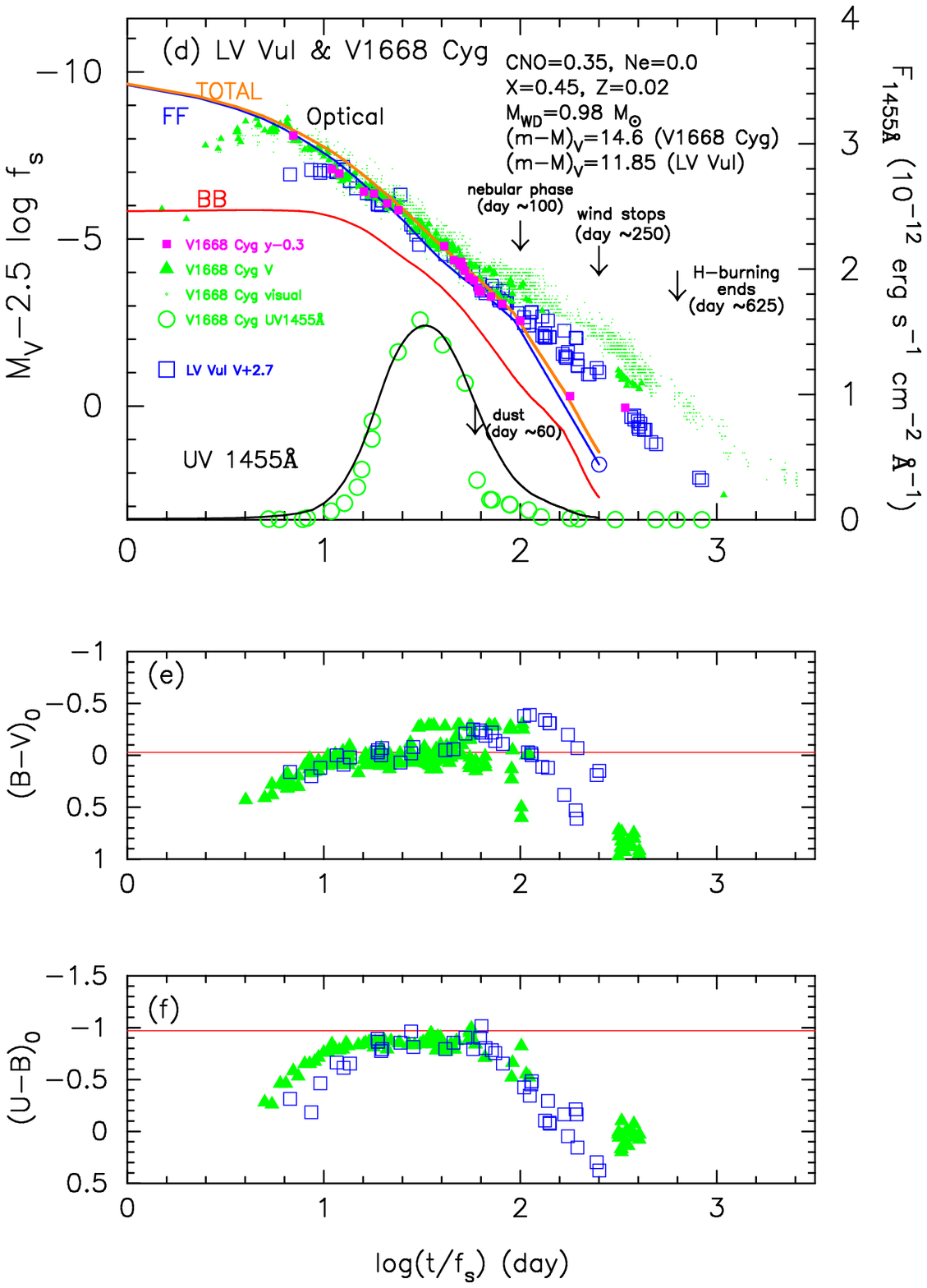}
\caption{
The light/color curves of LV~Vul and V1668~Cyg on a linear (left column)
and logarithmic (right column) timescale.  Here, $t$ is the time since
the outburst, $M_V$ is the absolute $V$ magnitude, and $f_{\rm s}$ is
the stretching factor of a target nova.  Here, we adopt 
$f_{\rm s}= 1.0$ for both LV~Vul and V1668~Cyg.  
(a) and (d) The $V$ light curves of LV~Vul and V1668~Cyg.  We added the
UV~1455\AA\  light curve of V1668~Cyg.  The data of V1668~Cyg and
LV~Vul are the same as those in Figures 1 and 4 of \citet{hac16kb}.
We adopt $(m-M)_V=11.85$ for LV~Vul and $(m-M)_V=14.6$ for V1668~Cyg
\citep{hac19k}.  In panel (d), we plot the model light curves of 
a $0.98~M_\sun$ WD \citep[CO3,][]{hac16k}.  The 
orange line labeled ``TOTAL'' denotes the total $V$ flux 
of photospheric (red line labeled ``BB'') 
plus free-free (blue line labeled ``FF'') emissions.
The blue unfilled circle at the right end of the blue line corresponds
to the end of the optically thick wind phase.
The black solid line represents the UV~1455\AA\  flux.
(b) and (e) The $(B-V)_0$ color curves of V1668~Cyg and LV~Vul are 
dereddened with $E(B-V)=0.30$ and $E(B-V)=0.60$, respectively. 
The horizontal red lines show $(B-V)_0= -0.03$,
which are the intrinsic colors of
optically thick free-free emission \citep{hac14k}.  
(c) and (f) The $(U-B)_0$ color curve of LV~Vul and V1668~Cyg.
The horizontal red lines show $(U-B)_0= -0.97$, the intrinsic color of
optically thick free-free emission \citep{hac14k}.  
\label{all_mass_v1668_cyg_lv_vul_v_bv_ub_x45z02c15o20_no2}}
\end{figure*}

\citet{hac10k, hac15k, hac16k, hac18kb} showed that if a substantial part
of two light curves overlap each other by stretching the timescale of the
template nova, the absolute magnitude $(M_V[t])_{\rm target}$ of the target
nova is related to the $(M_V[t])_{\rm template}$ of the template nova by 
$(M_V[t] )_{\rm target} = (M_V[f_{\rm s}\times t] 
+ 2.5 \log f_{\rm s})_{\rm template}$ or $(M_V[t])_{\rm template} = 
(M_V[t/f_{\rm s}] - 2.5 \log f_{\rm s})_{\rm target}$, where $t$ is 
the time since the outburst and $f_{\rm s}$ is the timescaling factor
of the target against the template nova.
Note that they required the overlap of a large (substantial)
part of the light curves, including the nebular phase (and supersoft X-ray
source phase, if available), to accurately determine the timescaling factor.
This means that these two time-stretched absolute light curves overlap
each other in the $(t/f_{\rm s})$-$(M_V-2.5 \log f_{\rm s})$ plane, and
the two time-stretched color curves do too in the 
$(t/f_{\rm s})$-$(B-V)_0$ plane.  
Here, $M_V$ is the absolute $V$ magnitude and $(B-V)_0$ is 
the intrinsic $B-V$ color.  (See, e.g., Figure 
\ref{all_mass_v1668_cyg_lv_vul_v_bv_ub_x45z02c15o20_no2}
for such an example.)  Then, a large part of the two nova tracks overlap
each other in the time-stretched $(B-V)_0$-$(M_V-2.5 \log f_{\rm s})$ 
color-magnitude diagram.  
We applied these properties to 32 recent novae and analyzed 
their distances and reddenings
by comparing each nova light/color curve and time-stretched
color-magnitude track with those of well-calibrated
(known distance and reddening) template novae 
(see Appendix \ref{light_curve_appendix} for the 32 recent novae and
\citet{hac19k} for the other 20 novae).

The color-magnitude diagram of stars plays an important role in the
study of stellar evolution and is also used to estimate the absolute
magnitude of a star.  The color-magnitude diagram of classical novae
in outburst was extensively studied by \citet{hac16kb}.
They proposed several ``template tracks'' of novae in outburst
\citep[see Figure 12 of][]{hac16kb}, where the template tracks are
taken from novae that have a well-defined distance modulus in the $V$ band,
$(m-M)_V$; color excess, $E(B-V)$; and shape of the track.
Using these template tracks, they obtained the $(m-M)_V$ and $E(B-V)$
of a target nova by directly comparing the target nova track with one
of these template tracks.  
However, they were not fully successful because the absolute $V$
magnitudes of each track have very different brightnesses 
and never overlap each other.  They found six different types
of tracks in the $(B-V)_0$-$M_V$ diagram, but they could not dtermine
the reason for the differences or similarities.

\citet{hac19k} revised Hachisu \& Kato's (2016b) method.
\citet{hac19k} used the $(B-V)_0$-$(M_V-2.5\log f_{\rm s})$
diagram instead of the $(B-V)_0$-$M_V$ diagram.  This is because
the $(t/f_{\rm s})$-$(M_V-2.5\log f_{\rm s})$ light curves overlap each other
and therefore the time-stretched absolute magnitude, $M_V-2.5\log f_{\rm s}$,
guarantees the same brightness in the time-stretched 
$(B-V)_0$-$(M_V-2.5\log f_{\rm s})$ diagram.

\citet{hac06kb} showed that nova light curves follow the universal 
decline law when free-free emission dominates the nova spectra.
\citet{hac10k} further showed that the time-stretched absolute $V$
brightnesses, $M_V-2.5\log f_{\rm s}$,
overlap each other in the $(t/f_{\rm s})$-$(M_V-2.5\log f_{\rm s})$
light curve \citep[see, e.g., Figures 48 and 49 of][]{hac18kb}.  
We show this property in more detail in 
Appendix \ref{timescaling_law_free-free_emission}.
If we are able to overlap the light curve of the template nova
to the target nova by stretching/squeezing its timescale
as $t'=t/f_{\rm s}$, we have the relation
\begin{eqnarray}
\left( M_V[t] \right)_{\rm template} 
&=& \left( M'_V[t'] \right)_{\rm target} \cr
&=& \left( M_V[t/f_{\rm s}]-2.5\log f_{\rm s} \right)_{\rm target}
\label{time-stretching_general}
\end{eqnarray}
between the $(t', M'_V)$ and $(t, M_V)$ coordinate systems
\citep[see Appendix A of][]{hac18kb}, where $M_V[t]$ is the original
brightness and $M'_V[t']$ is the time-stretched brightness 
after time stretching by $t'=t/f_{\rm s}$.
  
The color $(B-V)_0$ is not changed by this time stretch
because the both $M_B$ and $M_V$ are shifted by $-2.5\log f_{\rm s}$ 
after time stretching \citep[e.g.,][]{hac18k}.
\citet{hac19k} called this $(B-V)_0$-$(M_V-2.5\log f_{\rm s})$ diagram 
the time-stretched color-magnitude diagram.
They applied this method to 20 well-observed novae and
confirmed that each nova track follows a template track in 
the time-stretched $(B-V)_0$-$(M_V-2.5 \log f_{\rm s})$ diagram.  
If a target nova track overlaps to the template
nova track in the $(B-V)_0$-$(M_V-2.5 \log f_{\rm s})$ diagram,
we can confirm that (1) our method gives the correct value of the absolute
$V$ magnitude, $M_V-2.5\log f_{\rm s}$, from the vertical fit, and, 
at the same time, (2) the correct value of the color excess from the 
horizontal fit.  Thus, we can check the distance to the target nova
because the $M_V-2.5\log f_{\rm s}$ and $E(B-V)$ of the template
nova are known.  In the present paper, we apply this method to 
32 recent novae and obtain their distances and reddenings.
 
We also obtain the $(m-M)_V$ and the WD mass, $M_{\rm WD}$, from 
our model light-curve fitting.  
\citet{hac15k} calculated many $M_V[t]$ model light curves for various
WD masses and envelope chemical compositions based on 
Kato \& Hachisu's (1994) results.  Their model $M_V$ light curve
is composed of photospheric emission and free-free emission.  
The photospheric emission is approximated by blackbody emission 
at the photosphere while the free-free emission originates from
optically thin plasma outside the photosphere.
Fitting an $M_V$ model light curve with the observed apparent $m_V$,
we specify the $(m-M)_V$ and $M_{\rm WD}$ for a nova.

Our paper is organized as follows.  First we describe our method
based on the time-stretched $(B-V)_0$-$(M_V-2.5 \log f_{\rm s})$ diagram 
in Section \ref{method_example}.  
We apply this method to 32 recent novae, and determine their 
distances and WD masses in Section \ref{gcmd_recent_novae}.
A discussion and conclusions follow in Sections
\ref{discussion} and \ref{conclusions}, respectively.
The model light-curve fitting and time-stretching method for each of the
32 novae are summarized in Appendix \ref{light_curve_appendix}.
In Appendix
\ref{timescaling_law_free-free_emission}, we explain a timescaling law
of novae based on optically thick wind solutions.
In Appendix \ref{m31n200812a_v1500_cyg}, we discuss exceptional novae
that deviate largely from the universal decline law.  We show that
the present method applies to them if we adopt a different template
nova depending on the reason for deviation.

\begin{figure*}
\epsscale{1.1}
\plottwo{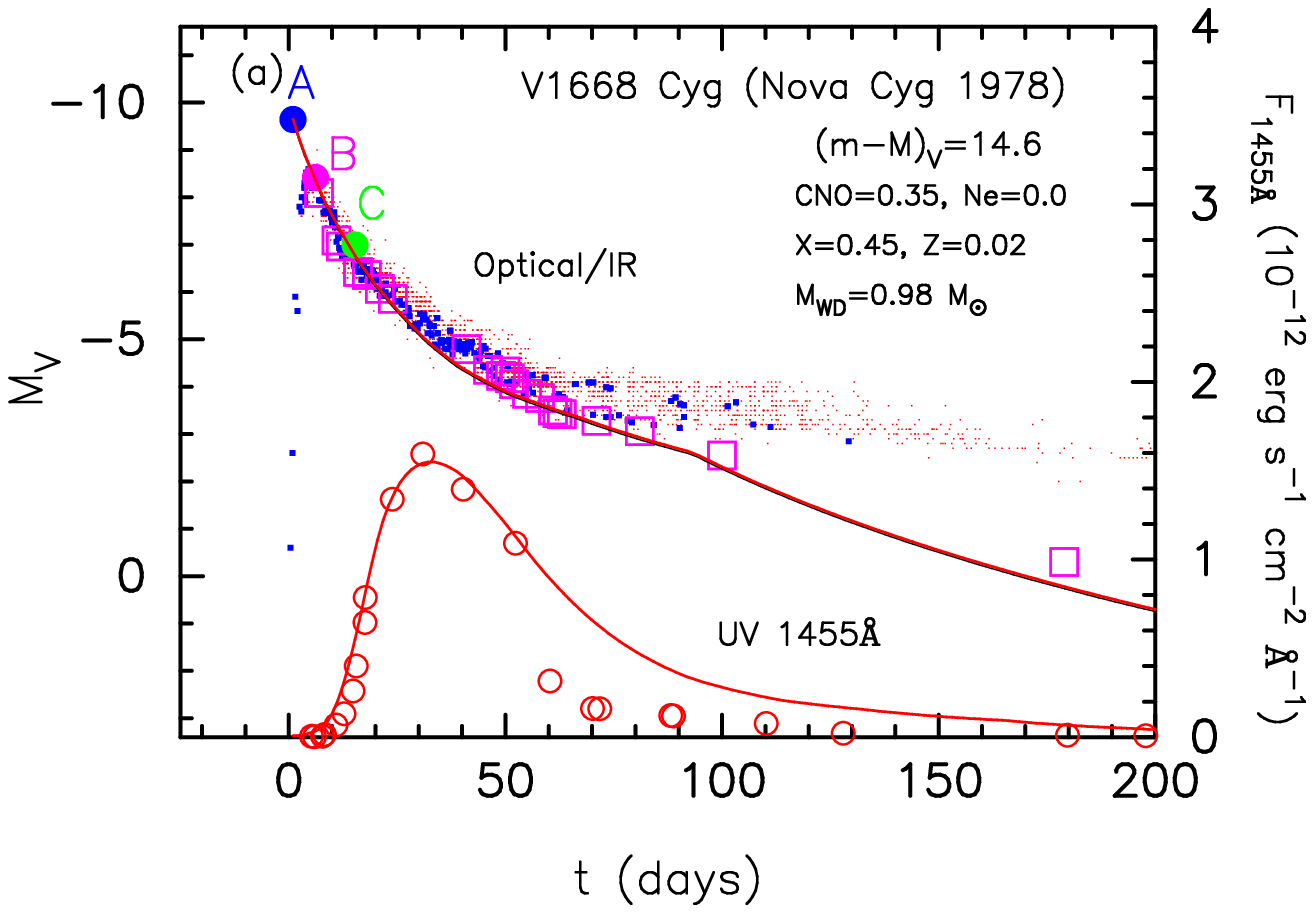}{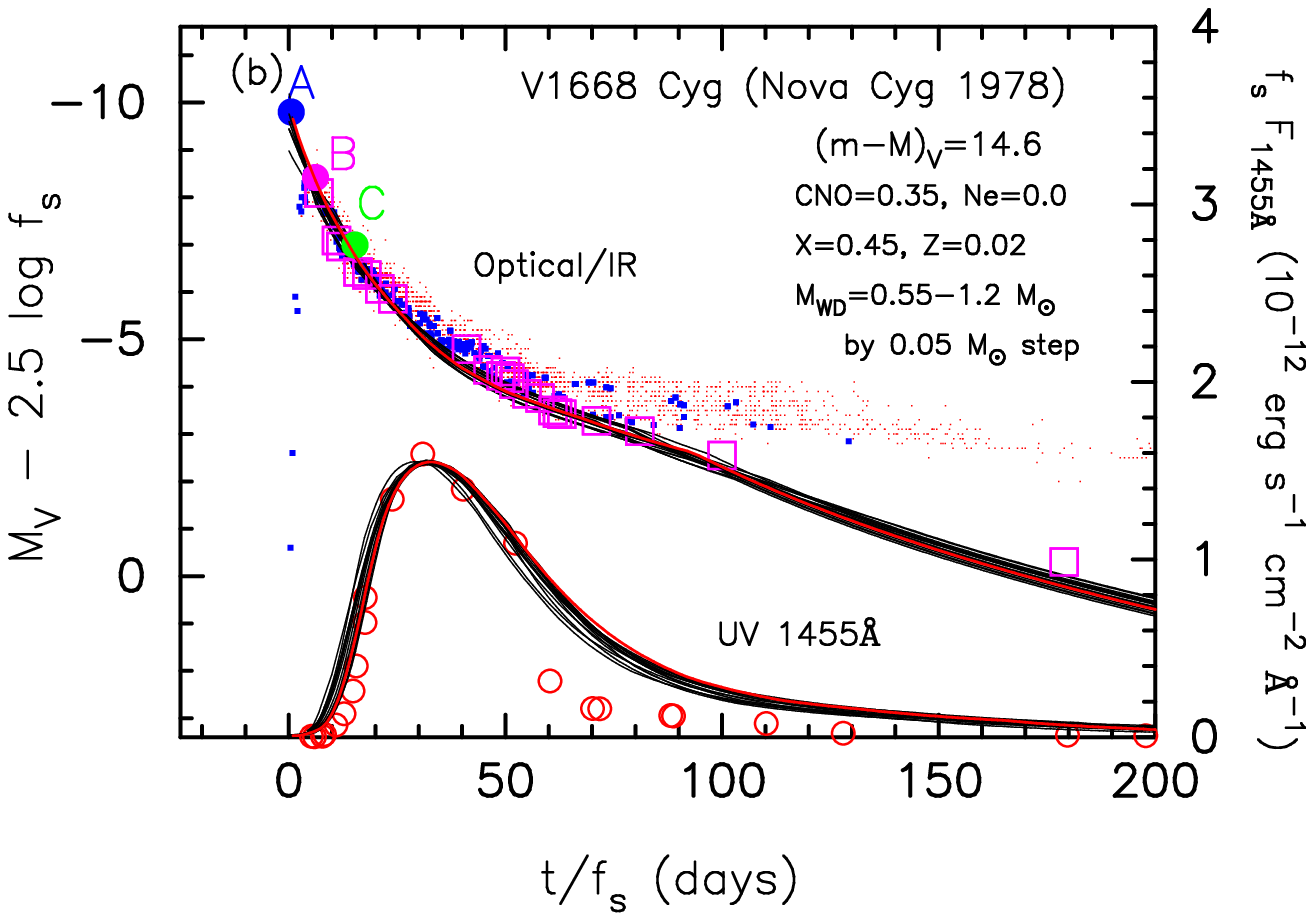}
\caption{
(a) A slight close-up view of the V1668~Cyg light curves in
Figure \ref{all_mass_v1668_cyg_lv_vul_v_bv_ub_x45z02c15o20_no2}(a).
The $0.98 ~M_\sun$ WD (CO3) model reasonably reproduces the V1668~Cyg
optical $y$ (unfilled magenta squares), $V$ (filled blue squares),
and UV 1455 \AA\  (large unfilled red circles) light curves.
We add the visual magnitudes (red dots).
These data are all the same as Figure 46 of \citet{hac18kb}.
We place three points, A, B, and C, on the model $V$ light curve, 
corresponding to three different initial envelope masses,
$M_{\rm env,0} = 2.0 \times 10^{-5}~M_\sun$, $1.4 \times 10^{-5}~M_\sun$,
and $0.93 \times 10^{-5}~M_\sun$, respectively. 
Point B is the optical peak of V1668~Cyg, $m_V=6.2~(M_V=6.2 - 14.6=-8.4)$.
(b) Free-free emission model $V$ and UV 1455~\AA\  model light curves  
for 0.55, 0.60, 0.65, 0.70, 0.75, 0.80,
0.85, 0.90, 0.95, 1.0, 1.05, 1.1, 1.15, and $1.20~M_\sun$ WDs
(solid black lines), but in the $(t/f_{\rm s})$-$(M_V-2.5\log f_{\rm s})$
coordinates for the $V$ and the $(t/f_{\rm s})$-$(f_{\rm s} 
F_{\rm 1455\AA})$ coordinates for the UV~1455\AA.
The timescaling factor $f_{\rm s}$ of each model, tabulated
in Table 3 of \citet{hac16k}, is measured against that
of the V1668~Cyg light curves.
\label{all_mass_v1668_cyg_x45z02c15o20_calib_linear_m098}}
\end{figure*}


\begin{figure*}
\epsscale{1.1}
\plottwo{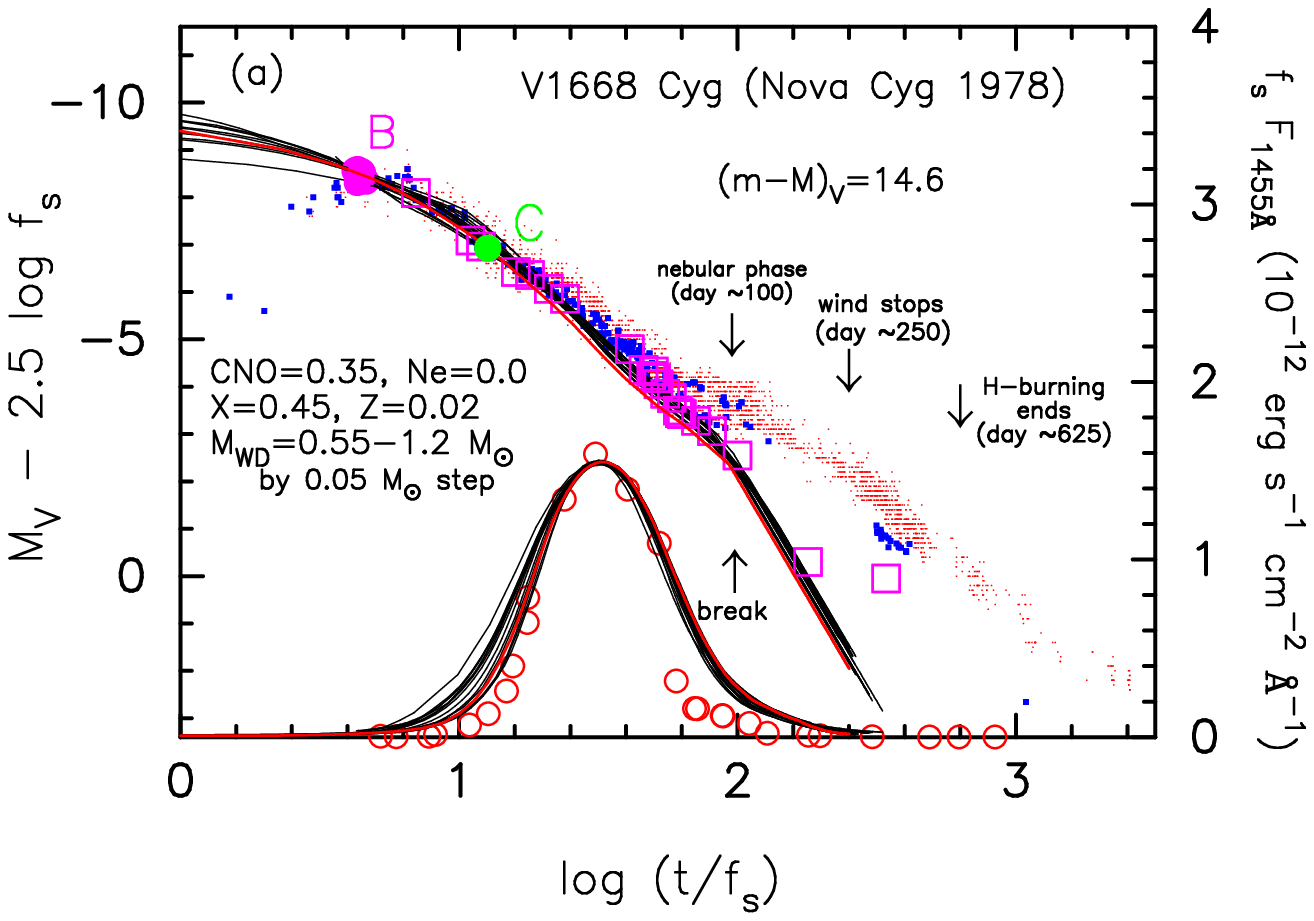}{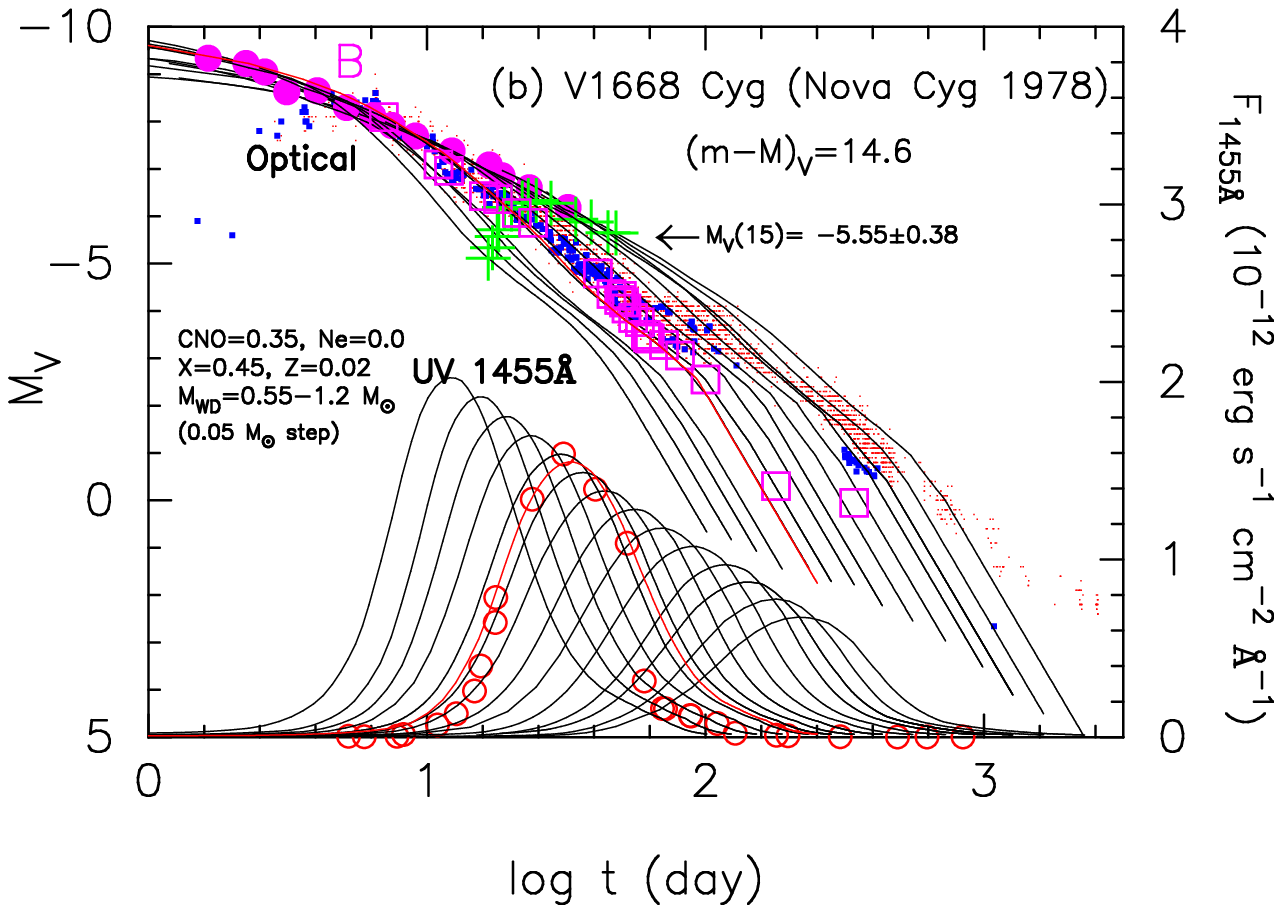}
\caption{
(a) Same as Figure
\ref{all_mass_v1668_cyg_x45z02c15o20_calib_linear_m098}(b), but on 
logarithmic timescales. 
(b) Same model light curves as those in panel (a), but for real
timescale and absolute $V$ magnitude.  The magenta filled circles
are point B on each $V$ light curve in panel (a).
The green pluses represent the $M_V(15)$, the absolute $V$ magnitude
15 days after the $V$ maximum at each point B.  
We also recover the real flux of the UV~1455\AA\  band.
\label{all_mass_v1668_cyg_x45z02c15o20_real_scale_universal_no2}}
\end{figure*}


\begin{figure}
\epsscale{1.1}
\plotone{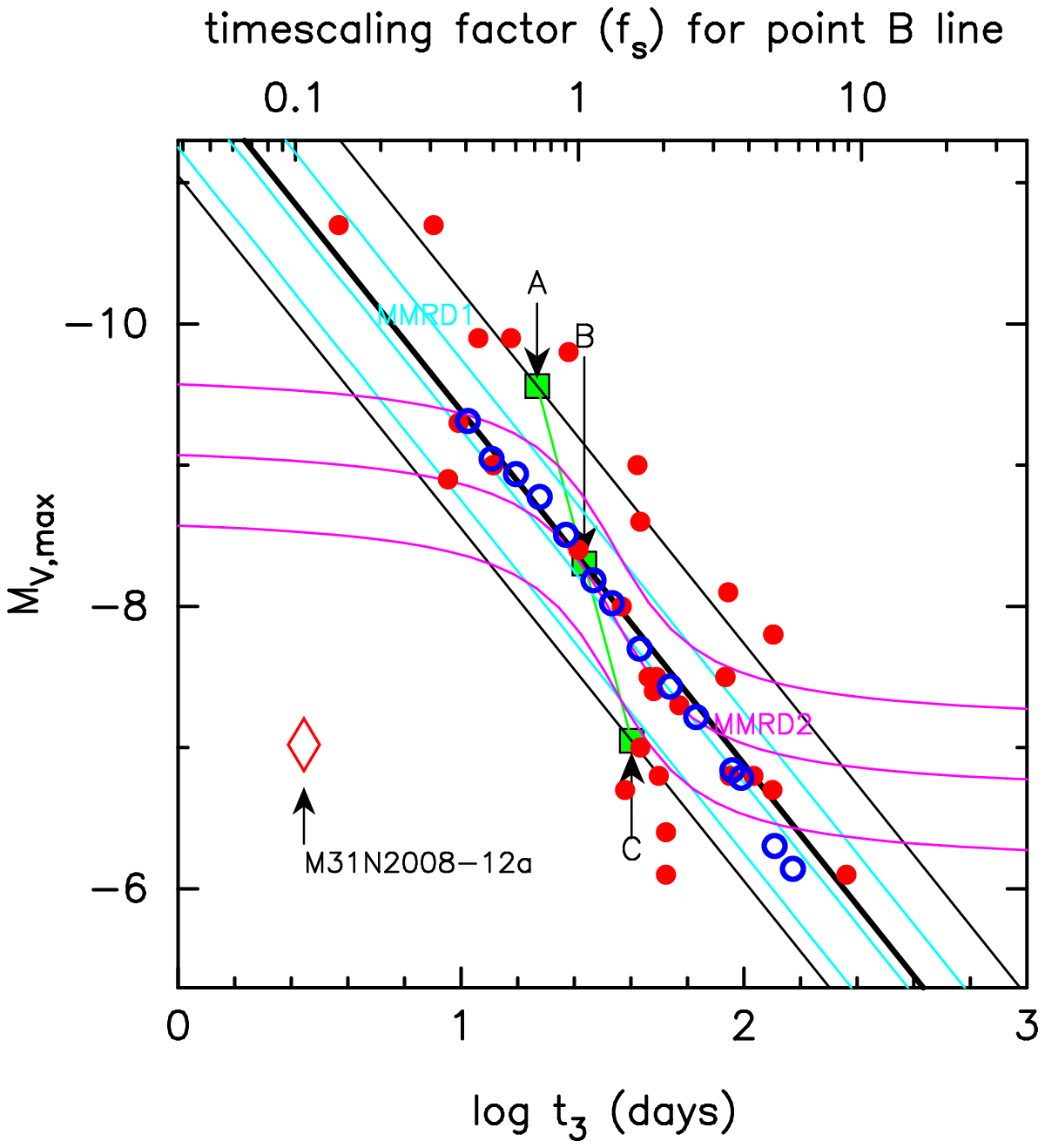}
\caption{
Maximum magnitude versus rate of decline (MMRD) relation for classical
novae.  The ordinate is the $V$ peak $M_{V, \rm max}$ and the bottom
abscissa is the $t_3$ time, during which the nova brightness decays
by 3 mag from the peak.  Point B (filled green square with black outline)
is the MMRD point of V1668~Cyg on the $0.98 M_\sun$ WD (CO3) model light
curve.  We assume that point B on each WD mass model in Figure
\ref{all_mass_v1668_cyg_x45z02c15o20_real_scale_universal_no2}
is a typical initial envelope mass for each WD mass.
Therefore, the thick black line
passing through point B is an MMRD relation for a typical nova with
different WD masses.  The top abscissa is the timescaling factor of a
different WD mass model on this line, in which $f_{\rm s}$ is measured
against the $0.98 M_\sun$ WD model.
The upper solid black line passing through point A 
is an MMRD relation for a much brighter nova
than that for point B, which corresponds to a much lower mass
accretion rate and, as a result, a much larger initial envelope mass,
than that for point B.  The lower solid black line passing through
point C is an MMRD relation
for a much fainter nova than that for point B, which corresponds to a higher
mass accretion rate and, as a result, a much smaller initial envelope
mass, than that for point B.
The unfilled blue circles are the MMRD points for different WD masses
calculated from each point B in Figure
\ref{all_mass_v1668_cyg_x45z02c15o20_real_scale_universal_no2}(b).
We plot MMRD points (filled red circles) for individual galactic novae,
which are taken from Table 5 of \citet{dow00}.
We also add the MMRD point for M31N~2008-12a \citep[unfilled red 
diamond;][]{dar16}.  We further add the two well-known empirical
MMRD relations: one is Kaler-Schmidt's law 
\citep[solid cyan lines labeled ``MMRD1'';][]{sch57},
and the other is della Valle \& Livio's law
\citep[solid magenta lines labeled ``MMRD2'';][]{del95}.
We plot $\pm 0.5$ mag upper/lower bounds for each of these two relations.
}
\label{max_t3_scale_no8}
\end{figure}

\section{Method and Examples}
\label{method_example}
We briefly summarize the time-stretching method of nova light curves
and how to determine the distance and WD mass.

\subsection{Model Light Curves of Novae}
\label{model_light_curve}
We determine the WD mass of a target nova by directly fitting 
the model light curve with the observation.
Figure \ref{all_mass_v1668_cyg_lv_vul_v_bv_ub_x45z02c15o20_no2} shows
the light/color curves of V1668~Cyg and LV~Vul on linear 
(left side column) and logarithmic (right side column) timescales.
We explain our method based on these figures.
  
\citet{kat94h} calculated evolutions of nova outbursts based on the optically
thick wind theory.  \citet{hac15k} obtained the $M_V$ light curves
based on Kato \& Hachisu's (1994) envelope models. 
Their $M_V$ flux is composed of photospheric emission
and free-free emission. The photospheric emission is approximately
calculated from blackbody emission at the photosphere (red line labeled ``BB''
in Figure \ref{all_mass_v1668_cyg_lv_vul_v_bv_ub_x45z02c15o20_no2}(d))
while the free-free emission is estimated from plasma outside 
the photosphere (blue line labeled ``FF'').
The observed flux (orange line labeled ``TOTAL'') is the sum of the
free-free and photospheric emissions.

The flux of free-free emission can easily exceed the Eddington limit
(which is as close as the flat part of the red line (BB) in Figure 
\ref{all_mass_v1668_cyg_lv_vul_v_bv_ub_x45z02c15o20_no2}(d)), if the 
wind mass-loss rate is large enough \citep[see also, e.g.,][]{hac17k}.
Fitting an $M_V$ model light curve with the observed $m_V$,
we specify the $(m-M)_V$ and $M_{\rm WD}$ for a nova \citep[e.g.,][for
such examples]{hac15k, hac16k, hac17k, hac18k, hac18kb, hac19k}.  

We can determine the WD mass more accurately if the UV~1455\AA\  band
and/or supersoft X-ray fluxes are available.  The UV~1455\AA\  band
is a narrowband (1445--1465~\AA) flux that represents well the UV
continuum fluxes of novae \citep{cas02}.
The UV~1455\AA\   flux is calculated from the photospheric
emission assumed to be blackbody.
The soft X-ray band (0.3--1.5~keV) flux is also calculated from
the photospheric emission (blackbody assumption).

Figure \ref{all_mass_v1668_cyg_lv_vul_v_bv_ub_x45z02c15o20_no2}(d)
shows such an example of model light-curve fitting.  In the case of V1668~Cyg,
we adopt $(m-M)_V= 14.6$ and $M_{\rm WD}= 0.98~M_\sun$
with the chemical composition of CO nova 3 \citep[CO3;][]{hac16k}.
We estimate the fitting error to be $\epsilon((m-M)_V)= \pm0.2$ mag and
$\epsilon(M_{\rm WD})= \pm0.02~M_\sun$.  The simultaneous fitting of the
$V$ and UV~1455\AA\  light curves gives a relatively
accurate mass of white dwarf rather than fitting only of a $V$ light curve.

\subsection{Deviation from a Model $V$ Light Curve}
\label{deviation_model}
Our model $V$ light curve is composed of free-free emission and
blackbody emission.  Therefore, the model light curves deviate
from observation when the effect of strong emission lines or
dust absorption is important.

Nova ejecta become optically thin when the nebular phase starts.
In Figure \ref{all_mass_v1668_cyg_lv_vul_v_bv_ub_x45z02c15o20_no2}(d),
we denote the start of the nebular phase of V1668~Cyg by the arrow
labeled ``nebular phase'' about 100 days after the outburst. 
Our $M_V$ light curve (orange line) begins to deviate from 
the $V$ observation (filled green triangles) about 80--100 days after
the outburst.

In the nebular phase, strong [\ion{O}{3}] lines
start to contribute to the $V$ band.  This makes $B-V$ color redder.
The $(B-V)_0$ color turns to the red in Figure
\ref{all_mass_v1668_cyg_lv_vul_v_bv_ub_x45z02c15o20_no2}(b) and (e).  
The response function of each $V$ filter sometimes shows a slight
difference in its blue edge.  Such a small difference in the $V$ filter
makes a large difference in the $V$ magnitude because [\ion{O}{3}]
contribute to the blue edge of the $V$ filter 
\citep[see, e.g., Figure 1 of][]{mun13b}.
We observe a large scatter or multiple branches
of $(B-V)_0$ color evolution, which start at the nebular phase (Figure
\ref{all_mass_v1668_cyg_lv_vul_v_bv_ub_x45z02c15o20_no2}(b) and (e)).  
Therefore, we do not expect a unique path of nova evolution in the
$(B-V)_0$ color curve especially in the nebular phase.

The intermediate-width band, $y$, is designed to avoid strong [\ion{O}{3}] 
lines \citep[see, e.g., Figure 1 of][]{mun13b}.
So, the $y$ magnitude can follow the continuum well.
We add the $y$ observations (filled magenta square) 
of V1668~Cyg to Figure
\ref{all_mass_v1668_cyg_lv_vul_v_bv_ub_x45z02c15o20_no2}(d).  
Our model $V$ light curve shows a good agreement with the $y$ observation
even after the nebular phase started, except for the last data point.
This confirms that our model $V$ light curve follows well the continuum
flux of a nova. 

Dust formation usually occurs far outside the photosphere whereas
continuum free-free emission comes mainly from near outside the photosphere.
Therefore, dust absorbs optical and UV photons and re-emit IR photons.
For example,
V1668~Cyg showed an optically thin dust blackout on day 60.
\citet{hac06kb} discussed such optical and near infrared (NIR)
light curves of V1668~Cyg. 
The UV~1455\AA\  light curve shows a small drop around day 60 as plotted
in Figure \ref{all_mass_v1668_cyg_lv_vul_v_bv_ub_x45z02c15o20_no2}(d).  
A much thicker dust formation is shown, e.g., in Figure
\ref{v5579_sgr_lv_vul_v1668_cyg_v1535_sco_v_bv_ub_color_logscale}(a),
in which the optical flux suddenly drops due to dust shell formation. 
Our model $V$ light curve does not include the effect of dust shell
formation far outside the photosphere.  Therefore, we do not follow
well the dust blackout phase.

To summarize, we basically exclude the nebular phase and dust blackout
phase, but we include the nebular phase when their light/color
curves overlap in the nebular phase (e.g., V1663 Aql, V2575 Oph, 
V5117 Sgr, V2576 Oph, V390 Nor, V459 Vul, V2670 Oph, QY Mus, NR TrA,
V5585 Sgr, PR Lup, V834 Car, V2677 Oph, V5592 Sgr, V962 Cep, 
V2659 Cyg, V5667 Sgr, V5668 Sgr, V2944 Oph).

\subsection{Peak Brightness versus Timescaling Factor}
\label{peak_timescale}
Figure \ref{all_mass_v1668_cyg_x45z02c15o20_calib_linear_m098}(a)
shows that the peak $V$ brightness of our model light curve
depends on the initial envelope mass.  In general, the initial envelope
mass ($M_{\rm env,0}$) is close to the ignition mass ($M_{\rm ig}$),
which is the hydrogen-rich envelope mass at the start of hydrogen burning.
The ignition mass depends mainly on the mass accretion rate 
($\dot M_{\rm acc}$) and WD mass \citep[see, e.g.,][for a recent
calculation]{kat14shn}.  The larger the mass-accretion rate is,
the smaller the ignition mass.  (The smaller the WD mass is,
the larger the ignition mass.)

Points A, B, and C in 
Figure \ref{all_mass_v1668_cyg_x45z02c15o20_calib_linear_m098}(a)
correspond to the different initial envelope mass, i.e., 
$M_{\rm env,0} = 2.0 \times 10^{-5}~M_\sun$,
$1.4 \times 10^{-5}~M_\sun$, and
$0.93 \times 10^{-5}~M_\sun$, respectively. 
When the ignition mass is $M_{\rm env,0} = 2.0 \times 10^{-5}~M_\sun$,
the peak brightness reaches $M_{V, \rm max}= -9.7$ (point A), which is
very bright compared with the ignition mass of $1.4 \times 10^{-5}~M_\sun$,
for which the peak brightness reaches $M_{V, \rm max}= -8.4$ (point B), 
which is moderately bright.  

The peak brightness of LV~Vul is about $M_{V, \rm max}= -7.15$ (point C).
This means that, if both WD masses are the same, the mass-accretion
rate to the WD is smaller in V1668~Cyg than in LV~Vul.  This also shows
that, even if the timescaling factors are the same between V1668~Cyg
and LV~Vul, the peak $V$ brightnesses are very different from each other.

\subsection{MMRD Relations}
\label{mmrd_relatons}
We explain the theory behind the maximum magnitude versus rate of decline
(MMRD) relations based on the universal decline law.  
The MMRD methods require the decline rates 
in the very early phase of a nova outburst,
usually $t_2$ or $t_3$, which are the times during which
the $V$ brightness decays by 2 or 3 mag from the peak, respectively.  
Many empirical MMRD relations between the peak $V$ brightness, 
$M_{V, {\rm max}}$, and the decline timescale, $t_2$ or $t_3$, have been
proposed \citep[e.g.,][]{sch57, dev78, coh88, del95, dow00}.
\citet{hac10k, hac15k, hac16k, hac18kb} formulated a theory 
of MMRD relations.  Here, we summarize their main points.

It should be noted that the timescales $t_2$ and $t_3$ are 
local timescales just after the $V$ peak but not global timescales like
$f_{\rm s}$.  For example, LV~Vul and V1668~Cyg have the same
global timescale of $f_{\rm s}=1$ (Figure 
\ref{all_mass_v1668_cyg_x45z02c15o20_calib_linear_m098}), 
but their $t_3$ times are different from each other, that is,
$t_3= 43$ days (point C in Figure \ref{max_t3_scale_no8})
for LV~Vul \citep{dow00} and $t_3= 26$ days (point B in 
Figure \ref{max_t3_scale_no8}) for V1668~Cyg \citep{hac10k}.
Thus, the MMRD methods are theoretically not the same as 
the time-stretching method.

First, we explain the main trend of the MMRD relations,
which is derived from the universal decline law.
The brightness of the free-free emission declines along
the universal decline law, as shown in Figures
\ref{all_mass_v1668_cyg_x45z02c15o20_calib_linear_m098}(b)
and \ref{all_mass_v1668_cyg_x45z02c15o20_real_scale_universal_no2}(a).
This is the sequence of decreasing envelope mass, $M_{\rm env}$,
and thus, the peak $V$ brightness is closely related to 
the initial envelope mass, $M_{\rm env,0}$, which is almost
equal to the envelope mass at maximum.  We can determine the
peak brightness on the $V$ light curve if we specify
the WD mass and initial envelope mass.

   Point B is the peak $V$
brightness of V1668~Cyg as shown in Figure 
\ref{all_mass_v1668_cyg_x45z02c15o20_calib_linear_m098}(a).
We assume that the brightness at point B is typical of the $0.98~M_\sun$ WD. 
We further assume that point B 
is also a typical brightness for each WD mass
in the $(t/f_{\rm s})$-$(M_V-2.5\log f_{\rm s})$ plane in Figure 
\ref{all_mass_v1668_cyg_x45z02c15o20_calib_linear_m098}(b).
For the other speed classes (other WD masses, $M_{\rm WD}$) of novae,
we have the typical peak $V$ brightness at point B, that is,
$M_V({\rm B}, M_{\rm WD}) -2.5\log f_{\rm s}= M_V({\rm B}, 0.98~M_\sun)$,
in Figure \ref{all_mass_v1668_cyg_x45z02c15o20_calib_linear_m098}(b).
A more massive WD ($M_{\rm WD} > 0.98~M_\sun$ ) 
with a smaller $f_{\rm s}<1$ than
that of V1668~Cyg has a brighter maximum magnitude of 
$M_{V, {\rm max}}({\rm B}, M_{\rm WD})= 
M_V({\rm B}, 0.98~M_\sun) + 2.5 \log f_{\rm s}$ over that of V1668~Cyg
and vice versa (see Figure
\ref{all_mass_v1668_cyg_x45z02c15o20_real_scale_universal_no2}(b)).

Figure
\ref{all_mass_v1668_cyg_x45z02c15o20_real_scale_universal_no2}(a)
shows the same model light curves of Figure
\ref{all_mass_v1668_cyg_x45z02c15o20_calib_linear_m098}(b),
but on a logarithmic timescale.
Figure
\ref{all_mass_v1668_cyg_x45z02c15o20_real_scale_universal_no2}(b)
shows the real $M_V$ magnitudes and real timescales of each model
light curves.
We plot point B on each nova model light curve by the filled magenta
circles. 
We read the value of $M_{V, \rm max}$ at point B (filled magenta circle)
on each model light curve (different WD mass) and obtain its $t_3$ time
on the $\log t$-$M_V$ light curve.
Thus, we plot the MMRD relations for point B on each WD mass.
Such MMRD points are shown in Figure \ref{max_t3_scale_no8}, that is,
the unfilled blue circles on the $\log t_3$ versus $M_{V, \rm max}$ plot,  
which show the main (central) trend of the MMRD distribution.
Thus, we can explain the main trend of MMRD relations
by typical brightness (point B) for different WD masses.
We approximate this trend as $M_{V, \rm max}= 2.5 \log t_3 -11.9$, 
the line of which (thick solid black line) has the slope
of 2.5 and is passing through point B (filled green square with black
outline) in Figure \ref{max_t3_scale_no8}.  

Second, we explain the large scatter of individual novae from the main
trend of the MMRD relation.  The ignition mass depends on the mass
accretion rate if we fix the WD mass (for example, $M_{\rm WD}= 
0.98~M_\sun$ in Figure
\ref{all_mass_v1668_cyg_x45z02c15o20_calib_linear_m098}(a)).
When the mass accretion rate is smaller than that of point B,
the initial envelope mass, $M_{\rm env,0}$, is larger than 
that at point B.  Then, the peak $V$ brightness reaches, for example,
point A, which is much brighter than that for point B.
This clearly shows a different peak $V$ brightness (and initial envelope
mass) for a different mass-accretion rate even if the WD mass is the same as
$0.98~M_\sun$ and the chemical composition is also the same as CO3.
Plotting this brighter MMRD relation of point A for the other speed classes
of novae (different WD masses), we have a brighter end for the
$(t_3,~M_{V, \rm max})$ distribution, that is, 
$M_{V, {\rm max}}({\rm A}, M_{\rm WD})= 
M_V({\rm A}, 0.98~M_\sun) + 2.5 \log f_{\rm s}$,
which is depicted by the upper thin solid black line passing through point A
in Figure \ref{max_t3_scale_no8}.  
If we plot a fainter MMRD relation for point C for the other speed
classes of novae (different WD masses), we have a fainter end for the
$(t_3,~M_{V, \rm max})$ distribution, that is, $M_{V, {\rm max}}({\rm C},
M_{\rm WD})= M_V({\rm C}, 0.98~M_\sun) + 2.5 \log f_{\rm s}$,
which is represented by the lower solid black line
passing through point C in Figure \ref{max_t3_scale_no8}.  

The three black lines passing through points A, B, and C indicate a degree
of scatter, originating from the difference in the mass-accretion rate,
that is, the difference in the initial envelope mass at ignition.
These upper and lower black lines broadly envelop the MMRD points
of galactic novae (filled red circles in Figure \ref{max_t3_scale_no8}),
which were studied by \citet{dow00}.
Very fast and very faint novae detected in M31 \citep{kas11, dar16}
can be also explained in the same context.  
We plot the MMRD point of the one-year recurrence period nova 
M31N~2008-12a (unfilled red diamond in Figure \ref{max_t3_scale_no8})
for such an example.
\citet{hac18kb} examined such very fast and faint novae in our Galaxy
and nearby galaxies, i.e., YY Dor, LMC N2009a, LMC N 2012a, LMC N 2013,
in LMC, SMC N 2016 in SMC, and M31N 2008-12a in M31.  They applied 
the time-stretching method to these novae and concluded that
their time-stretched light curves show good agreement with the known
distance to each nearby galaxy except for SMC~2016 \citep[e.g.,][]{ori19}.
These very fast and faint novae should have extremely
large mass-accretion rates onto very massive WDs
\citep[e.g.,][]{kat15sh, kat17sha, kat17shb}.

To summarize the theory of MMRD relations, the main trend of MMRD relations
is closely related to the WD mass, which is represented by the timescaling
factor $f_{\rm s}$.  The large scatter from the main trend originates
from the different mass accretion rates and, as a result, different initial
envelope masses at ignition.  This second parameter of the mass accretion
rate can reasonably explain the degree of scatter of individual novae
from the proposed empirical MMRD relations (see Figure 
\ref{max_t3_scale_no8}).

\subsection{Properties of Universal Decline Law}
\label{universal_decline_law}
\citet{hac06kb} found that, when free-free emission dominates the spectrum,
there is a universal decline law of novae.  Figures 
\ref{all_mass_v1668_cyg_x45z02c15o20_calib_linear_m098}(b) and
\ref{all_mass_v1668_cyg_x45z02c15o20_real_scale_universal_no2}(a)
demonstrate that 14 free-free emission model light curves (solid black
lines) for different WD masses overlap each other
if they are properly stretched in the time
direction, $t/f_{\rm s}$, and shifted by $-2.5 \log f_{\rm s}$
(see Appendix \ref{timescaling_law_free-free_emission} for the derivation).
\citet{hac06kb} called this property of the model light curves
the universal decline law.  They determined the $f_{\rm s}$
of the 14 theoretical light curves against the optical/UV
light curves of V1668~Cyg.  The $0.98~M_\sun$ WD (CO3) model reasonably
reproduces the observed optical/UV light curves of V1668~Cyg, that is,
$f_{\rm s}=1.0$.  The other $f_{\rm s}$ of each WD mass is tabulated
in Table 3 of \citet{hac16k}.

Figures \ref{all_mass_v1668_cyg_x45z02c15o20_calib_linear_m098}(b) and
\ref{all_mass_v1668_cyg_x45z02c15o20_real_scale_universal_no2}(a)
also show UV~1455~\AA\ model light curves.
Assuming blackbody emission at the photosphere, \citet{hac06kb}
calculated the model UV~1455~\AA\ light curves.  We stretch them
with the same factor $f_{\rm s}$ as those of the optical/NIR light curves
(see Appendix \ref{timescaling_law_free-free_emission} for details).
Strictly speaking, we obtain each UV~1455~\AA\  flux with the reddening
of $E(B-V)=0.30$ and the distance of $d= 5.4$~kpc.  These two parameters
are the same as those of V1668~Cyg.

Because we already know the absolute $V$ magnitude of the template nova,
we obtain the absolute $V$ magnitude of the target nova from Equation
(\ref{time-stretching_general}).  If we convert the $(t,~M_V)$
of the target nova to ($t', ~M'_V$) by a time stretch of
$t'=t/f_{\rm s}$, we obtain $M'_V[t']=M_V[t/f_{\rm s}] - 2.5 \log f_{\rm s}$.
This relation is derived in Appendix \ref{timescaling_law_free-free_emission}.
Equation (\ref{time-stretching_general}) is equivalent to
\begin{eqnarray}
\left( M_V[t\times f_{\rm s}] \right)_{\rm template}
&=& \left(M'_V[t]\right)_{\rm target} \cr
&=& \left( M_V[t]\right)_{\rm target} - 2.5 \log f_{\rm s}.
\label{time-stretching_absolute_equivalent}
\end{eqnarray}
Using the time stretch of $t'=t\times f_{\rm s}$, we obtain
$M'_V[t']=M_V[t\times f_{\rm s}] + 2.5 \log f_{\rm s}$ for the template,
and we can rewrite Equation (\ref{time-stretching_general}) as
\begin{eqnarray}
\left( M_V[t] \right)_{\rm target}
&=& \left(M'_V[t']\right)_{\rm template} \cr
&=& \left( M_V[t\times f_{\rm s}]\right)_{\rm template} + 2.5 \log f_{\rm s}.
\label{time-stretching_absolute_inverse}
\end{eqnarray}

The observed $V$ light curves of the novae are given in the apparent $V$
magnitude of $m_V$.  If we can overlap the two novae 
after horizontally shifting by $\log f_{\rm s}$ and vertically shifting 
by $\Delta V$ the template nova light curve in the
$(\log t, ~m_V)$ plane, we have the relation 
\begin{equation}
(m_V[t])_{\rm target} = (m_V[t\times f_{\rm s}]
+ \Delta V)_{\rm template}.
\label{differnce-apparent_magnitude}
\end{equation}
Subtracting Equation (\ref{time-stretching_absolute_inverse}) from
Equation (\ref{differnce-apparent_magnitude}),
we obtain 
\begin{eqnarray}
 \left( m[t] - M[t] \right)_{V,\rm target} &=& \cr
 ( (m[t\times f_{\rm s}]-M[t\times f_{\rm s}] )_V
&+& \Delta V  )_{\rm template} \cr
  &-& 2.5 \log f_{\rm s}.
\label{distance_modulus_general_apparent}
\end{eqnarray}
It should be noted that the absolute $V$ light curve of $M_V[t]$
has the same shape as the apparent $V$ light curve of $m_V[t]$,
both for the target and template novae.  Therefore, we can suppose
that $(m[t]-M[t])_V =$ constant in time.  Finally, we obtain 
\begin{equation}
(m-M)_{V,\rm target} = \left( (m-M)_V
+ \Delta V - 2.5 \log f_{\rm s}\right)_{\rm template},
\label{distance_modulus_general_temp}
\end{equation}
where $(m-M)_{V, \rm template}$ is the distance modulus of the template
nova and is already known.  Equation (\ref{distance_modulus_general_apparent})
is a function of time but Equation (\ref{distance_modulus_general_temp})
is just a constant value.  Thus, we can derive the distance modulus
of a target nova from the template nova.
\citet{hac10k} called this method (Equation 
(\ref{distance_modulus_general_temp})) the time-stretching method.

It should be noted that individual novae more or less deviate
from the universal decline law.  We suppose that the overall trends
of these novae broadly follow the universal decline law and 
Equation (\ref{distance_modulus_general_temp}) approximately applies
to such novae.  We will really see it in the rest of this paper.

\subsection{Distance from the Time-Stretched Color-Magnitude Diagram Method}
\label{time_stretched_cmd}
We analyze the light/color curves of novae in the following procedure.\\
{\bf 1.} We plot the $V$ light curve and dereddened 
of $(B-V)_0$ and $(U-B)_0$ color curves of a target nova 
on a logarithmic timescale (for example, Figure 
\ref{all_mass_v1668_cyg_lv_vul_v_bv_ub_x45z02c15o20_no2}).  
The color excess $E(B-V)$ of the target nova is taken from the literature
or our time-stretching method mentioned below.
We obtain the dereddened colors of $(B-V)_0$ and $(U-B)_0$ via
\begin{equation}
(B-V)_0 = (B-V) - E(B-V),
\label{dereddening_eq_bv}
\end{equation}
and
\begin{equation}
(U-B)_0 = (U-B) - 0.64 E(B-V),
\label{dereddening_eq_ub}
\end{equation}
where the factor of $0.64$ is taken from \citet{rie85}. \\
{\bf 2.} We use a well-calibrated nova as a template nova, for which the
distance modulus, $\mu_V\equiv (m-M)_V$, and color excess, $E(B-V)$,
are well defined.  We add the light/color curves of the template novae and
overlap them with the target nova as much as possible by shifting them
back and forth ($\log f_{\rm s}$) and up and down ($\Delta V$).  Then, we
estimate the distance modulus of the target nova, $(m-M)_{V, \rm target}$,
from Equation (\ref{distance_modulus_general_temp}).
We shift the light/color curves in steps of $\delta \log f_{\rm s} = 0.01$
and $\delta (\Delta V)= 0.1$ mag and find the best match by eye.
In our previous paper \citep{hac18k}, we estimated the error
for V959~Mon.  Its typical allowance is about 
$\log f_{\rm s} = 0.14 \pm 0.05$ and $\Delta V= +1.6\pm0.2$
from a least-square fit.
Therefore, we usually obtain typical errors of $\epsilon( \log f_{\rm s})
= 0.05$ and $\epsilon (\Delta V)= 0.2$ even when fitting by eye unless 
otherwise mentioned.  We adopt LV~Vul (or V1500~Cyg, V1668~Cyg, and
V1974~Cyg) as a template nova, unless otherwise specified.  
Theoretically, the spectrum of free-free emission is almost
independent of frequency, i.e., $F_\nu \propto \nu^0$.
This means that the other broadband light curves follow
the same universal decline law as the $V$ band.  
We apply the same relation to the $B$ and $I_{\rm C}$ bands, that is,
\begin{equation}
(m-M)_{B,\rm target} = ((m-M)_B + \Delta B
- 2.5 \log f_{\rm s})_{\rm template},
\label{distance_modulus_general_temp_b}
\end{equation}
\begin{equation}
(m-M)_{I,\rm target} = ((m-M)_I + \Delta I_{\rm C}
- 2.5 \log f_{\rm s})_{\rm template},
\label{distance_modulus_general_temp_i}
\end{equation}
where the timescaling factor $f_{\rm s}$ is the same as that in
Equation (\ref{distance_modulus_general_temp}).  Thus, we obtain
the three distance moduli in the $B$, $V$, and $I_{\rm C}$ bands
(for example, Figure \ref{v1663_aql_yy_dor_lmcn_2009a_b_v_i_logscale_3fig}).
\\
{\bf 3.} 
We adopt the relations between the distance $d$ and color excess
$E(B-V)$ to the target nova, i.e.,
\begin{eqnarray}
(m-M)_V = 3.1 E(B-V) + 5 \log (d/10~{\rm pc}),
\label{distance_modulus_rv}
\end{eqnarray}
where the factor $R_V=A_V/E(B-V)=3.1$ is the ratio of total to selective
extinction \citep[e.g.,][]{rie85},
\begin{eqnarray}
(m-M)_B = 4.1 E(B-V) + 5 \log (d/10~{\rm pc}),
\label{distance_modulus_rb}
\end{eqnarray}
where $A_B/E(B-V)=4.1$ \citep{rie85},
and 
\begin{eqnarray}
(m-M)_I = 1.5 E(B-V) + 5 \log (d/10~{\rm pc}),
\label{distance_modulus_ri}
\end{eqnarray}
where $A_I/E(B-V)= 1.5$ from \citet{rie85}.\\
{\bf 4.} Using the $(m-M)_V$, $(m-M)_B$, and $(m-M)_I$ obtained 
in step {\bf 2}, we plot the three relations of Equations
(\ref{distance_modulus_rv}), 
(\ref{distance_modulus_rb}),  and 
(\ref{distance_modulus_ri}) in the reddening-distance plane,
respectively.  
If the three lines cross at the same point, this point gives the
correct values of reddening $E(B-V)$ and distance $d$ (for example, 
Figure \ref{distance_reddening_v1663_aql_v5116_sgr_v2575_oph_v5117_sgr}).\\
{\bf 5.} Using $E(B-V)$, $(m-M)_V$, and $f_{\rm s}$ obtained above 
in steps {\bf 1} - {\bf 4}, we plot the 
$(B-V)_0$-$(M_V - 2.5\log f_{\rm s})$ diagram of the target nova
(for example, Figure
\ref{hr_diagram_v1663_aql_v5116_sgr_v2575_oph_v5117_sgr_outburst}).\\
{\bf 6.} If the track of the target nova overlaps one of the
template novae (LV~Vul or V1500~Cyg) 
in the $(B-V)_0$-$(M_V - 2.5\log f_{\rm s})$ diagram,
we regard our adopted values of $f_{\rm s}$, $E(B-V)$,
$(m-M)_V$, and $d$ to be reasonable.  When the target nova overlaps 
the track of LV~Vul, we classify it as the LV~Vul type.

Many novae, including the present 32 novae, approximately follow
the universal decline law, and we apply our method to them.
However, it should be noted that some exceptional novae deviate
greatly from the universal decline law, and the present method cannot be
directly applied to them.  We discuss such examples in more detail
in Appendix \ref{m31n200812a_v1500_cyg}.

\subsection{Comparison with Galactic Reddening-Distance Relations}
\label{reddening_distance}
We compare the obtained $E(B-V)$ and $d$ of a nova with several
galactic reddening-distance relations toward $(l, b)$, where
$(l, b)$ is the galactic coordinates of the nova.  We use six results:\\
(1) \citet{mar06} published a three-dimensional (3D) extinction
map of the Galaxy in the direction of $-100\fdg0 \le l \le 100\fdg0$
and $-10\fdg0 \le b \le +10\fdg0$ with grids of $\Delta l=0\fdg25$
and $\Delta b=0\fdg25$.  We convert their $A_{K_{\rm s}}$
to our $E(B-V)$ with the relation of $A_{K_{\rm s}} = A_K/0.95 = 
0.112 A_V /0.95 = (0.112 \times 3.1/0.95) E(B-V) = 0.365 E(B-V)$.\\   
(2) \citet{sal14} calculated a 3D reddening map
for a region toward $30\arcdeg \le l < 215\arcdeg$ and $|b|<5\arcdeg$
based on the INT Photometric H-Alpha Survey photometry.
We convert their $A_0$ to our $E(B-V)$ using the relation 
$A_0 = A_V = 3.1 E(B-V)$.\\
(3) \citet{schu14} presented a 3D dust extinction map
toward the galactic bulge covering $-10\fdg0 < l < 10\fdg0$ and 
$-10\fdg0 <b< 5\fdg0$, using data of the VISTA Variables 
in the Via Lactea (VVV) survey together with the Besan\c{c}on stellar
population synthesis model of the Galaxy.  The resolution is 
$0\fdg1\times0\fdg1$ and the distance is extended up to 10 kpc
in 0.5~kpc steps.  We convert their $E(J-K_s)$ to our $E(B-V)$
using the relations $A_{K_s}=0.364~E(B-V)$ 
\citep{sai13} and $A_{K_s}=0.528~E(J-K_s)$ \citep{nis09shogo}.\\
(4) \citet{gre15} presented a 3D extinction map,
which covers three quarters of the sky with grids $(\Delta l, \Delta b)$
of 3\farcm4 to 13\farcm7.
The data were recently revised \citep{gre18}.\\
(5) \citet{ozd16} provided distance-reddening relations toward 46 novae.
They used the unique position of the red clump giants
in the color-magnitude diagram.  The data have been revised
and extended \citep{ozd18}.  We convert their $E(J-K_{\rm s})$ to 
our $E(B-V)$ using the relation $A_{K_{\rm s}} = 0.346 E(B-V)$
and $A_{K_{\rm s}} = 0.657 E(J-K_{\rm s})$ \citep{rie85}.\\
(6) \citet{chen18} obtained 3D reddening maps
in the three colors of $E(G - K_{\rm s})$,
$E(G_{\rm BP}-G_{\rm RP})$,  and $E(H - K_{\rm s})$.
We convert their $E(G_{\rm BP}-G_{\rm RP})$ to our $E(B-V)$ using
the relation $E(B-V)= 0.75 E(G_{\rm BP}-G_{\rm RP})$
\citep{chen18}.  The maps have a spatial angular resolution of 6$\arcmin$ 
and covers $0\arcdeg < l < 360\arcdeg$ and $|b| < 10\arcdeg$.  The maps
are based on the distances from the {\it Gaia} Data Release 2
({\it Gaia} DR2) and the photometry of 2MASS, {\it WISE}, 
and {\it Gaia} DR2 surveys.

We regard our set of reddening and distance to be reasonable when
the result $(E(B-V), d)$ for the target nova is consistent with one of
the above six distance-reddening relations.


\startlongtable
\begin{deluxetable*}{lllllrll}
\tabletypesize{\scriptsize}
\tablecaption{Extinctions, distance moduli, and distances for selected novae
\label{extinction_various_novae}}
\tablewidth{0pt}
\tablehead{
\colhead{Object} & \colhead{Outburst} & \colhead{$E(B-V)$} 
& \colhead{$(m-M)_V$} & 
\colhead{$d$} & \colhead{$\log f_{\rm s}$\tablenotemark{a}} &
\colhead{$(m-M')_V$} & \colhead{References\tablenotemark{b}} \\
  & (year) &  &  &  (kpc) &  &  & 
} 
\startdata
OS~And & 1986 & 0.15 & 14.8 & 7.3 & $-0.15$ & 14.4 & 2 \\
CI~Aql & 2000 & 1.0 & 15.7 & 3.3 & $-0.22$ & 15.15 & 4 \\
V1370~Aql & 1982 & 0.35 & 16.3 & 11 & 0.05 & 16.4 & 6 \\
V1419~Aql & 1993 & 0.52 & 15.0 & 4.7 & 0.15 & 15.4 & 5 \\
V1663~Aql & 2005 & 1.88 & 18.15 & 2.9 & $-0.08$ & 17.95 & 6 \\
V679~Car & 2008 & 0.69 & 16.1 & 6.2 & $0.0$ & 16.1 & 5 \\
V834~Car & 2012 & 0.50 & 17.25 & 14 & $-0.19$ & 16.75 & 6 \\
V705~Cas & 1993 & 0.45 & 13.45 & 2.6 & 0.45 & 14.55 & 5 \\
V1065~Cen & 2007 & 0.45 & 15.0 & 5.3 & 0.0 & 15.0 & 4 \\
V1213~Cen & 2009 & 0.78 & 16.95 & 8.1 & 0.05 & 17.05 & 6 \\
V1368~Cen & 2012 & 0.93 & 17.6 & 8.8 & 0.10 & 17.85 & 6 \\
V1369~Cen & 2013 & 0.11 & 10.25 & 0.96 & 0.17 & 10.65 & 5 \\
IV~Cep & 1971 & 0.65 & 14.5 & 3.1 & 0.0 & 14.5 & 3 \\
V962~Cep & 2014 & 1.10 & 18.45 & 10.2 & 0.12 & 18.75 & 6 \\
T~CrB & 1946 & 0.056 & 10.1 & 0.96 & $-1.32$ & 13.4 & 4 \\
V407~Cyg & 2010 & 1.0 & 16.1 & 3.9 & $-0.37$ & 15.2 & 4 \\
V1500~Cyg & 1975 & 0.45 & 12.3 & 1.5 & $-0.22$ & 11.75 & 2 \\
V1668~Cyg & 1978 & 0.30 & 14.6 & 5.4 & 0.0 & 14.6 & 2 \\
V1974~Cyg & 1992 & 0.30 & 12.2 & 1.8 & 0.03 & 12.3 & 2 \\
V2362~Cyg & 2006 & 0.60 & 15.4 & 5.1 & 0.25 & 16.0 & 5 \\
V2468~Cyg & 2008 & 0.65 & 16.2 & 6.9 & 0.38 & 17.15 & 5 \\
V2491~Cyg & 2008 & 0.45 & 17.4 & 15.9 & $-0.34$ & 16.55 & 5 \\
V2659~Cyg & 2014 & 0.80 & 15.7 & 4.4 & 0.52 & 17.0 & 6 \\
YY~Dor & 2010 & 0.12 & 18.9 & 50 & $-0.72$ & 17.1 & 4 \\
V446~Her & 1960 & 0.40 & 11.95 & 1.38 & 0.0 & 11.95 & 5 \\
V533~Her & 1963 & 0.038 & 10.65 & 1.28 & 0.08 & 10.85 & 5 \\
V838~Her & 1991 & 0.53 & 13.7 & 2.6 & $-1.22$ & 10.65 & 4 \\
PR~Lup & 2011 & 0.74 & 16.1 & 5.8 & 0.23 & 16.65 & 6 \\
V959~Mon & 2012 & 0.38 & 13.15 & 2.5 & 0.14 & 13.5 & 3 \\
QY~Mus & 2008 & 0.58 & 14.65 & 3.7 & 0.35 & 15.55 & 6 \\
V390~Nor & 2007 & 0.89 & 16.6 & 5.8 & 0.45 & 17.75 & 6 \\
RS~Oph  & 2006 & 0.65 & 12.8 & 1.4 & $-1.02$ & 10.25 & 4 \\
V2575~Oph  & 2006\#1 & 1.43 & 17.85 & 4.9 & 0.11 & 18.15 & 6 \\
V2576~Oph  & 2006\#2 & 0.62 & 16.65 & 8.8 & $-0.15$ & 16.25 & 6 \\
V2615~Oph  & 2007 & 0.90 & 15.95 & 4.3 & 0.20 & 16.45 & 5 \\
V2670~Oph  & 2008\#1 & 1.05 & 17.6 & 7.4 & 0.33 & 18.45 & 6 \\
V2676~Oph  & 2012\#1 & 0.90 & 17.3 & 8.0 & 0.53 & 18.65 & 6 \\
V2677~Oph  & 2012\#2 & 1.30 & 19.2 & 10.7 & $-0.17$ & 18.75 & 6 \\
V2944~Oph  & 2015 & 0.62 & 16.5 & 8.2 & 0.25 & 17.15 & 6 \\
V574~Pup  & 2004 & 0.45 & 15.0 & 5.3 & 0.10 & 15.25 & 5 \\
V597~Pup  & 2007 & 0.24 & 16.4 & 13.5 & $-0.18$ & 15.95 & 6 \\
U~Sco & 2010 & 0.26 & 16.3 & 12.6 & $-1.32$ & 13.0 & 4 \\
V745~Sco & 2014 & 0.70 & 16.6 & 7.8 & $-1.32$ & 13.3 & 4 \\
V1281~Sco & 2007\#2 & 0.82 & 17.4 & 9.4 & $-0.07$ & 17.25 & 6 \\
V1313~Sco & 2011\#2 & 1.30 & 19.0 & 9.9 & $-0.22$ & 18.45 & 6 \\
V1324~Sco & 2012 & 1.32 & 16.95 & 3.7 & 0.28 & 17.75 & 6 \\
V1534~Sco & 2014 & 0.93 & 17.6 & 8.8 & $-1.22$ & 14.55 & 4 \\
V1535~Sco & 2015 & 0.78 & 18.3 & 15 & 0.38 & 19.25 & 6 \\
V475~Sct & 2003 & 0.55 & 15.4 & 5.5 & 0.36 & 16.3 & 2 \\
V496~Sct & 2009 & 0.45 & 13.7 & 2.9 & 0.30 & 14.45 & 5 \\
V5114~Sgr & 2004 & 0.47 & 16.65 & 10.9 & $-0.12$ & 16.35 & 5 \\
V5116~Sgr & 2005\#2 & 0.23 & 16.05 & 12 & 0.20 & 16.55 & 6 \\
V5117~Sgr & 2006 & 0.53 & 16.0 & 7.5 & 0.05 & 16.15 & 6 \\
V5579~Sgr & 2008 & 0.82 & 15.95 & 4.8 & 0.28 & 16.65 & 6 \\
V5583~Sgr & 2009\#3 & 0.30 & 16.3 & 12 & $-0.29$ & 15.55 & 6 \\
V5584~Sgr & 2009\#4 & 0.70 & 16.7 & 8.0 & 0.13 & 17.05 & 6 \\
V5585~Sgr & 2010 & 0.47 & 16.7 & 11 & 0.10 & 16.95 & 6 \\
V5589~Sgr & 2012\#1 & 0.84 & 17.6 & 10 & $-0.67$ & 15.95 & 6 \\
V5592~Sgr & 2012\#4 & 0.33 & 16.05 & 10 & 0.13 & 16.35 & 6 \\
V5666~Sgr & 2014 & 0.50 & 15.4 & 5.8 & 0.25 & 16.0 & 5 \\
V5667~Sgr & 2015\#1 & 0.63 & 15.4 & 4.9 & 0.57 & 16.85 & 6 \\
V5668~Sgr & 2015\#2 & 0.20 & 11.0 & 1.2 & 0.27 & 11.65 & 6 \\
NR~TrA & 2008 & 0.24 & 15.35 & 8.3 & 0.43 & 16.45 & 6 \\
V382~Vel & 1999 & 0.25 & 11.5 & 1.4 & $-0.29$ & 10.75 & 5 \\
LV~Vul & 1968\#1 & 0.60 & 11.85 & 1.0 & 0.0 & 11.85 & 2,3,5 \\
PW~Vul & 1984\#1 & 0.57 & 13.0 & 1.8 & 0.35 & 13.85 & 5 \\
QU~Vul & 1984\#2 & 0.55 & 13.6 & 2.4 & 0.33 & 14.45 & 1 \\
V459~Vul & 2007\#2 & 0.90 & 15.45 & 3.4 & $-0.15$ & 15.05 & 6 \\  
LMC~N~2009a & 2014 & 0.12 & 18.9 & 50 & $-0.52$ & 17.6 & 4 \\
LMC~N~2012a & 2012 & 0.12 & 18.9 & 50 & $-1.22$ & 15.85 & 4 \\
LMC~N~2013 & 2013 & 0.12 & 18.9 & 50 & $-0.42$ & 17.85 & 4 \\
SMC~N~2016 & 2016 & 0.08 & 16.8 & 20.4 & $-0.72$ & 15.0 & 4 \\
M31~N~2008-12a & 2015 & 0.30 & 24.8 & 780 & $-1.32$ & 21.5 & 4
\enddata
\tablenotetext{a}{
$f_{\rm s}$ is the timescale against that of LV~Vul.}
\tablenotetext{b}{
(1) \citet{hac16k},
(2) \citet{hac16kb},
(3) \citet{hac18k},
(4) \citet{hac18kb},
(5) \citet{hac19k},
(6) present paper.
} 
\end{deluxetable*}


\startlongtable
\begin{deluxetable*}{lllllll}
\tabletypesize{\scriptsize}
\tablecaption{White dwarf masses of selected novae
\label{wd_mass_novae}}
\tablewidth{0pt}
\tablehead{
\colhead{Object} & \colhead{$\log f_{\rm s}$}
& \colhead{$M_{\rm WD}$}
& \colhead{$M_{\rm WD}$}
& \colhead{$M_{\rm WD}$}
& \colhead{$M_{\rm WD}$}
& \colhead{Chem. Comp.}
\\
\colhead{} & \colhead{}
& \colhead{$f_{\rm s}$\tablenotemark{a}}
& \colhead{UV~1455~\AA\tablenotemark{b}}
& \colhead{$t_{\rm SSS-on}$\tablenotemark{c}}
& \colhead{$t_{\rm SSS-off}$\tablenotemark{d}}
&
\\
\colhead{} & \colhead{} & \colhead{($M_\sun$)} & \colhead{($M_\sun$)}
& \colhead{($M_\sun$)} & \colhead{($M_\sun$)} &
}
\startdata
OS~And   & $-0.15$ & 1.05  & --- & --- & --- & CO3 \\
CI~Aql   & $-0.22$ & 1.18  & --- & --- & --- & interp.\tablenotemark{e}\\
V1419~Aql & $+0.15$ & 0.90 & --- & --- & --- & CO3 \\
V1663~Aql & $-0.08$ & 0.95 & --- & --- & --- & CO2 \\
V1663~Aql & $-0.08$ & 1.0 & --- & --- & --- & CO3 \\
V679~Car & $+0.0$ & 0.98 & --- & --- & --- & CO3 \\
V834~Car & $-0.19$ & 1.20 & --- & --- & --- & Ne3 \\
V705~Cas & $+0.45$ & 0.78 & 0.78 & --- & --- & CO4 \\
V1065~Cen & $+0.0$ & 0.98 & --- & --- & --- & CO3 \\
V1368~Cen & $+0.10$ & 0.95 & --- & --- & --- & CO3 \\
V1369~Cen & $+0.17$ & 0.90 & --- & --- & --- & CO3 \\
IV~Cep & $+0.0$ & 0.98 & --- & --- & --- & CO3 \\
V962~Cep & $+0.12$ & 0.95 & --- & --- & --- & CO3 \\
T~CrB   & $-1.32$ & 1.38  & --- & --- & --- & interp.\\
V407~Cyg   & $-0.37$ & 1.22  & --- & --- & --- & interp.\\
V1500~Cyg & $-0.22$ & 1.20 & --- & --- & --- & Ne2 \\
V1668~Cyg & $+0.0$ & 0.98 & 0.98 & --- & --- & CO3 \\
V1974~Cyg & $+0.03$ & 0.98 & 0.98 & 0.98 & 0.98 & CO3 \\
V2362~Cyg & $+0.25$ & 0.85 & --- & --- & --- & interp. \\
V2468~Cyg & $+0.38$ & 0.85 & --- & --- & --- & CO4 \\
V2491~Cyg & $-0.34$ & 1.35 & --- & 1.35 & 1.35 & Ne2 \\
V2659~Cyg & $+0.52$ & 0.75 & --- & --- & --- & CO4 \\
YY~Dor  & $-0.72$ & 1.29  & ---  & --- & --- & interp. \\
V446~Her & $+0.0$ & 0.98 & --- & --- & --- & CO3 \\
V533~Her & $+0.08$ & 1.03 & --- & --- & --- & Ne2 \\
V838~Her & $-1.22$ & 1.35 & 1.35 & --- & --- & Ne2 \\
V838~Her & $-1.22$ & 1.37 & 1.37 & --- & --- & Ne3 \\
PR~Lup & $+0.23$ & 0.90 & --- & --- & --- & CO3 \\
V959~Mon & $+0.14$ & 0.95 & --- & 0.95 & --- & CO3 \\
V959~Mon & $+0.14$ & 1.05 & --- & 1.05 & --- & Ne2 \\
V959~Mon & $+0.14$ & 1.1 & --- & 1.10 & --- & Ne3 \\
QY~Mus & $+0.35$ & 0.75 & --- & --- & --- & CO2 \\
QY~Mus & $+0.35$ & 0.80 & --- & --- & --- & CO3 \\
QY~Mus & $+0.35$ & 0.85 & --- & --- & --- & CO4 \\
V390~Nor & $+0.45$ & 0.75 & --- & --- & --- & CO3 \\
RS~Oph   & $-1.02$ & 1.35 & 1.35 & 1.35 & 1.35 & evol.\tablenotemark{f} \\
V2575~Oph  & $+0.11$ & 0.90 & --- & --- & --- & CO3 \\
V2576~Oph  & $-0.15$ & 1.15 & --- & --- & --- & Ne2 \\
V2615~Oph  & $+0.20$ & 0.90 & --- & --- & --- & CO3 \\
V2670~Oph  & $+0.33$ & 0.80 & --- & --- & --- & CO3 \\
V2676~Oph  & $+0.53$ & 0.70 & --- & --- & --- & CO2 \\
V2677~Oph  & $-0.17$ & 1.15 & --- & --- & --- & Ne2 \\
V2944~Oph  & $+0.25$ & 0.85 & --- & --- & --- & CO3 \\
V574~Pup  & $+0.10$ & 1.05 & --- & 1.05 & 1.05 & Ne2 \\
V597~Pup  & $-0.18$ & 1.2 & --- & 1.2 & 1.2 & Ne3 \\
U~Sco    & $-1.32$ & 1.37 & --- & 1.37 & 1.37 & evol. \\
V745~Sco & $-1.32$ & 1.38 & --- & 1.385 & 1.385 & evol. \\
V1281~Sco & $-0.07$ & 1.13 & --- & --- & --- & Ne2 \\
V1313~Sco & $-0.22$ & 1.20 & --- & --- & --- & Ne2 \\
V1324~Sco & $+0.28$ & 0.80 & --- & --- & --- & CO2 \\
V1534~Sco & $-1.22$ & 1.37 & --- & --- & --- & interp.\\
V1535~Sco & $+0.38$ & 0.85 & --- & --- & --- & CO4 \\
V475~Sct & $+0.36$ & 0.80 & --- & --- & --- & CO3 \\
V496~Sct & $+0.30$ & 0.85 & --- & --- & --- & CO3 \\
V5114~Sgr & $-0.12$ & 1.15 & --- & --- & --- & Ne2 \\
V5116~Sgr & $+0.20$ & 0.90 & --- & --- & --- & CO2 \\
V5116~Sgr & $+0.20$ & 1.07 & --- & --- & --- & Ne3 \\
V5117~Sgr & $+0.05$ & 0.95 & --- & --- & --- & CO3 \\
V5579~Sgr & $+0.28$ & 0.85 & --- & --- & --- & CO3 \\
V5583~Sgr & $-0.29$ & 1.23 & --- & 1.23 & 1.23 & Ne2 \\
V5584~Sgr & $+0.13$ & 0.90 & --- & --- & --- & CO3 \\
V5585~Sgr & $+0.10$ & 0.95 & --- & --- & --- & CO3 \\
V5589~Sgr & $-0.67$ & 1.33 & --- & --- & --- & Ne2 \\
V5592~Sgr & $+0.13$ & 0.93 & --- & --- & --- & CO4 \\
V5666~Sgr & $+0.25$ & 0.85 & --- & --- & --- & CO3 \\
V5667~Sgr & $+0.57$ & 0.78 & --- & --- & --- & CO4 \\
V5668~Sgr & $+0.27$ & 0.85 & --- & --- & --- & CO3 \\
NR~TrA & $+0.43$ & 0.75 & --- & --- & --- & CO3 \\
V382~Vel & $-0.29$ & 1.23 & --- & --- & 1.23 & Ne2 \\
LV~Vul   &  $+0.0$  & 0.98 & --- & --- & --- & CO3 \\
PW~Vul & $+0.35$ & 0.83 & 0.83 & --- & --- & CO4 \\
QU~Vul & $+0.33$ & 0.96 & --- & --- & --- & Ne3 \\
QU~Vul & $+0.33$ & 0.90 & --- & --- & --- & Ne2 \\
QU~Vul & $+0.33$ & 0.86 & --- & --- & --- & CO4 \\
QU~Vul & $+0.33$ & 0.82 & --- & --- & --- & CO2 \\
V459~Vul & $-0.15$ & 1.15 & --- & --- & --- & Ne2 \\
LMC~N~2009a & $-0.52$ & 1.25 & --- & 1.25 & 1.25 & Ne3 \\
LMC~N~2012a & $-1.22$ & 1.37 & --- & --- & --- & interp.\\
LMC~N~2013 & $-0.42$ & 1.23 & --- & --- & --- & interp. \\
SMC~N~2016 & $-0.72$ & 1.29 & --- & --- & 1.3 & Ne3 \\
SMC~N~2016 & $-0.72$ & 1.29 & --- & --- & 1.25 & Ne2 \\
M31~N~2008-12a & $-1.32$ & 1.38 & --- & 1.38 & 1.38 & evol. \\
\enddata
\tablenotetext{a}{WD mass estimated from the $f_{\rm s}$ timescale.}
\tablenotetext{b}{WD mass estimated from the UV~1455~\AA\  fit.}
\tablenotetext{c}{WD mass estimated from the $t_{\rm SSS-on}$ fit.}
\tablenotetext{d}{WD mass estimated from the $t_{\rm SSS-off}$ fit.}
\tablenotetext{e}{WD mass estimated from the linear interpolation
of the $\log f_{\rm s}$ versus WD mass relation \citep[see][]{hac18kb}.}
\tablenotetext{f}{WD mass estimated from the time-evolution
calculation with a Henyey type code \citep[see, e.g.,][]{hac18kb}.}
\end{deluxetable*}

\section{Time-Stretched Color-Magnitude Diagram of 32 Novae}
\label{gcmd_recent_novae}
We select 32 novae that have enough $BVI_{\rm C}$ data and 
analyze their light and color curves in the order of discovery date.
To avoid repetitive descriptions,
we put all routines of the time-stretching method into Appendix
\ref{light_curve_appendix}.  The results of our light/color
curve fittings are summarized in Table \ref{extinction_various_novae}
for distance moduli and in Table \ref{wd_mass_novae} for WD masses.
The WD mass is estimated from
the model (free-free plus blackbody) light-curve fitting.
The total of 73 novae including previous data
\citep{hac18kb, hac19k} provide a uniform quality data set
obtained with a single method, that is,
the time-stretching method of nova light curves. 


\begin{figure*}
\plotone{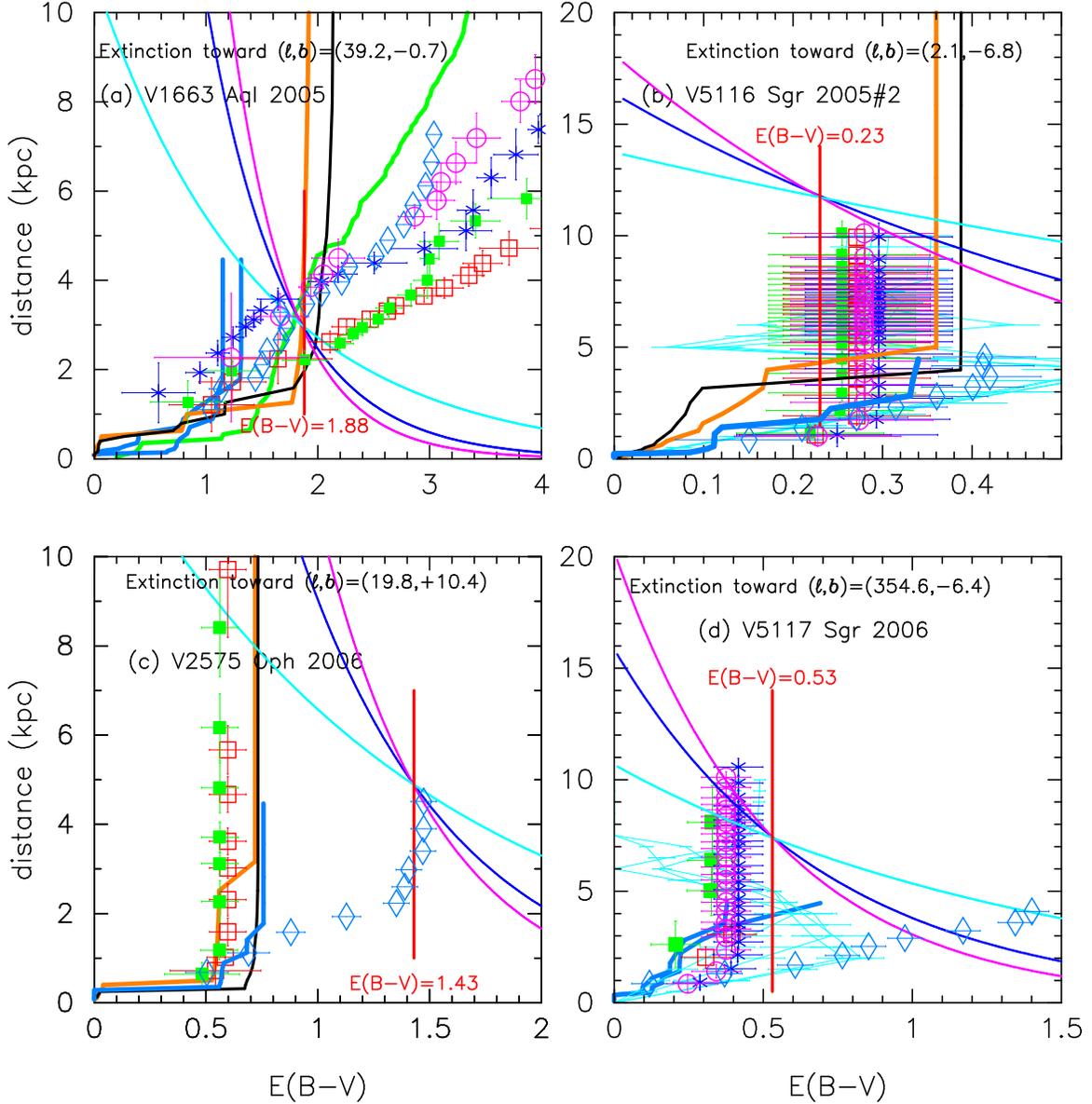}
\caption{
Distance-reddening relations toward
(a) V1663~Aql, (b) V5116~Sgr, (c) V2575~Oph, and (d) V5117~Sgr.
In panel (a), we plot $(m-M)_B= 19.98$ and Equation 
(\ref{distance_modulus_rb}), $(m-M)_V= 18.16$ and Equation 
(\ref{distance_modulus_rv}), and $(m-M)_I= 15.17$ and
Equation (\ref{distance_modulus_ri}) with the thin magenta, blue,
and cyan lines, respectively.
The three lines cross at $d=2.9$~kpc and $E(B-V)=1.88$.
The vertical solid red line indicates the reddening of $E(B-V)=1.88$.
The four relations toward $(l, b)=(39\fdg0, -0\fdg50)$, 
$(39\fdg25, -0\fdg50)$, $(39\fdg0, -0\fdg75)$, and $(39\fdg25, -0\fdg75)$
are taken from \citet{mar06} and depicted by unfilled red squares,
filled green squares, blue asterisks, and unfilled magenta circles,
respectively, each with error bars.  Among the four nearby directions,  
the direction close to V1663~Aql is that of the unfilled magenta circles.
The thick solid green line denotes the relation of \citet{sal14}.
The thick solid black line indicates the relation of
\citet{gre15} and the orange one is their revised version \citep{gre18}.
The unfilled cyan-blue diamonds represent the relation of \citet{ozd18}.
We also plot the two thick solid cyan-blue lines of
\citet{chen18} for $(l, b)=(39\fdg15, -0\fdg65)$ and
$(l, b)=(39\fdg15, -0\fdg75)$.
In panels (b) and (d), we add the relations of \citet{schu14}
with the very thin cyan lines.
\label{distance_reddening_v1663_aql_v5116_sgr_v2575_oph_v5117_sgr}}
\end{figure*}


\begin{figure*}
\plotone{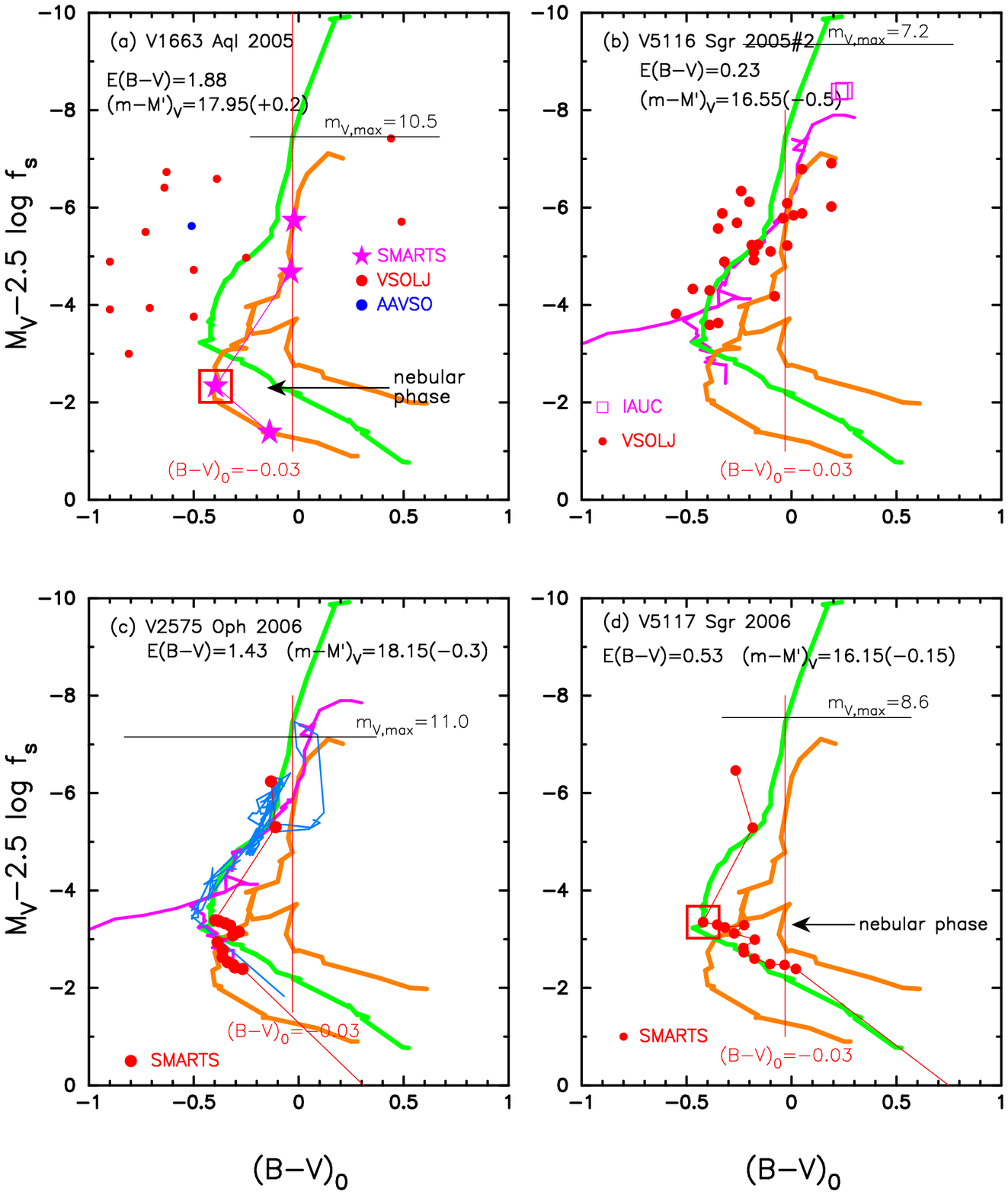}
\caption{
Time-stretched color-magnitude diagram for (a) V1663~Aql, (b) V5116~Sgr,
(c) V2575~Oph, and (d) V5117~Sgr.  The timescaling factor of $f_{\rm s}$
is measured against LV~Vul.
The thick solid orange lines represent the template track of LV~Vul while
the thick solid green lines denote that of V1500~Cyg.
In panels (b) and (c), we add the track of V1974~Cyg (solid magenta lines).
In panel (c), we add the track of PW~Vul (solid cyan-blue lines).
\label{hr_diagram_v1663_aql_v5116_sgr_v2575_oph_v5117_sgr_outburst}}
\end{figure*}

\subsection{V1663~Aql 2005}
\label{v1663_aql_cmd}
V1663~Aql reached $m_V=10.5$ at $V$ maximum on UT 2005 July 10.226 
\citep{poj05o}.
Based on the \ion{O}{1} lines of their $0.47$--$2.5~\mu$m spectroscopy,
\citet{pue05} suggested a large reddening of $E(B-V)\sim2$.
\citet{boy06} reported that the nova reached its maximum light on
HJD 2,453,531.2$\pm0.2$, when the apparent $V$ magnitude was $10.7\pm0.1$.
They estimated the decline times to be $t_{2,V}= 17$~days and
$t_{3,V}= 32$ days, and derived a maximum absolute $V$ magnitude of
$M_{V,\rm max}=-7.8\pm0.2$ (from the MMRD and $M_{V,15}$ relations),
color excess of $E(B-V)= 2$, and distance of $d=2.9\pm0.4$~kpc.
Here, $t_{2,V}$ and $t_{3,V}$ are the times during which the $V$ 
magnitude decays by 2 and 3 mag from maximum, respectively.  
\citet{pog06} obtained $t_2=10.9\pm2.0$ days and $t_3=22.0\pm3.5$ days,
and derived a maximum $V$ magnitude of $M_{V,\rm max}=-8.4$ (based on
the MMRD and $M_{V, 15}$ relations), color excess of $E(B-V)=1.22$,
and distance of $d=7.3$--$11.3$~kpc.
\citet{lan07} resolved sizes of the nova 5--18 days after the outburst,
using NIR interferometry, and derived the distance of
$8.9\pm3.6$~kpc.  \citet{hac07k} analyzed the light curve of
V1663~Aql based on their free-free emission model light curves
and concluded that the $V$ and $I_{\rm C}$ light curves are consistent
with those of a $0.95~M_\sun$ WD (CO2).

We obtain three distance moduli, $(m-M)_B= 19.98$, $(m-M)_V= 18.16$,
and $(m-M)_I= 15.17$, in Appendix \ref{v1663_aql} and plot them in Figure
\ref{distance_reddening_v1663_aql_v5116_sgr_v2575_oph_v5117_sgr}(a)
with the magenta, blue, and cyan lines, 
respectively, which cross at $d=2.9$~kpc and $E(B-V)=1.88$.
Taking into account our fitting accuracy, we obtain $E(B-V)=1.88\pm0.05$
and $d=2.9\pm0.3$~kpc.
The distance modulus in the $V$ band is $(m-M)_V=18.15\pm0.2$,
and the timescaling factor is $f_{\rm s}= 0.83$ against that of LV~Vul.
These values are listed in Table \ref{extinction_various_novae}.

Our distance estimate of $d=2.9\pm0.3$~kpc is largely different from that
of \citet{lan07}, i.e., $8.9\pm3.6$~kpc, 
based on the expansion parallax method.  
In general, the result of the expansion parallax method depends
on the assumed expansion velocity and asphericity of ejecta.
\citet{lan07} adopted $v_{\rm exp}=1375\pm500$~km~s$^{-1}$.
If we adopt the $v_{\rm exp}\sim700$~km~s$^{-1}$ given by \citet{den05}
from the P-Cygni profiles at H$\alpha$ and \ion{O}{1} (777 nm),
the distance becomes $4.5\pm1.8$~kpc, which is reasonably consistent
with our results.

For the reddening toward V1663~Aql, \citet{pog06} obtained $E(B-V)=1.22$
from the intrinsic color at $t_2$, i.e., $(B-V)_{0,t2}=-0.02\pm0.04$
\citep{van87}.  However, the $B-V$ data of 
the Variable Star Observers League of Japan (VSOLJ) 
and the American Association of Variable Star Observers (AAVSO) are rather
scattered and about 0.7 mag bluer than those of the Small and Medium 
Aperture Telescope System (SMARTS) \citep{wal12} as shown in Figures 
\ref{hr_diagram_v1663_aql_v5116_sgr_v2575_oph_v5117_sgr_outburst}(a)
and \ref{v1663_aql_v_bv_ub_color_curve}(b).
If we use only the SMARTS data, we obtain $E(B-V)=1.88\pm0.1$.
Then, the $(B-V)_0$ colors of LV~Vul and V1668~Cyg are similar to
that of V1663~Aql during $\log t {\rm ~(day)~}= 1 - 2$ as shown in Figure
\ref{v1663_aql_lv_vul_v1668_cyg_v_bv_logscale}(b).
Our reddening value of V1663~Aql is consistent with the estimates
of $E(B-V)\sim2$ given by \citet{pue05} and \citet{boy06}.  The $(B-V)_0$
color curve is located on the line of $B-V=-0.03$ (denoted 
by the thin red line in Figures 
\ref{hr_diagram_v1663_aql_v5116_sgr_v2575_oph_v5117_sgr_outburst}(a)
and \ref{v1663_aql_v_bv_ub_color_curve}(b)), which is the color of
optically thick free-free emission \citep{hac14k}.

We further check our results by comparing with the galactic 
distance-reddening relations
toward V1663~Aql, $(l,b)=(39\fdg1610, -0\fdg6648)$,
in Figure 
\ref{distance_reddening_v1663_aql_v5116_sgr_v2575_oph_v5117_sgr}(a).
The four relations toward
$(l, b)=(39\fdg0, -0\fdg50)$, $(39\fdg25, -0\fdg50)$,
$(39\fdg0, -0\fdg75)$, and $(39\fdg25, -0\fdg75)$
are taken from \citet{mar06} and depicted 
by unfilled red squares, filled green squares, blue asterisks,
and unfilled magenta circles, respectively, each with error bars.
The closest direction toward V1663~Aql in the galactic coordinates
is that of the unfilled magenta circles.
The thick solid green line denotes the relation of \citet{sal14}.
The thick solid black line indicates the relation of \citet{gre15}, 
and the orange one is their revised version \citep{gre18}.
The unfilled cyan-blue diamonds represent the relation of \citet{ozd18}.
We plot two thick solid cyan-blue lines of 
\citet{chen18} for $(l, b)=(39\fdg15, -0\fdg65)$ and
$(l, b)=(39\fdg15, -0\fdg75)$.
Our crossing point of $d=2.9$~kpc and $E(B-V)=1.88$ is consistent with
Green et al.'s relation (orange line). 

Using $E(B-V)=1.88$ and $(m-M')_V=17.95$ from Equation
(\ref{absolute_mag_v1663_aql}), we plot the time-stretched
color-magnitude diagram of V1663~Aql in Figure
\ref{hr_diagram_v1663_aql_v5116_sgr_v2575_oph_v5117_sgr_outburst}(a).
The text ``$(m-M')_V=17.95(+0.2)$'' in the figure means 
$(m-M')_V=17.95$ and $(m-M)_V=17.95 + 0.2 = 18.15$.
The data from SMARTS \citep[filled magenta stars;][]{wal12}
are just on the track of LV~Vul.  
Therefore, we regard V1663~Aql to belong to the LV~Vul type
in the $(B-V)_0$-$(M_V-2.5 \log f_{\rm s})$ diagram.
\citet{pue05} showed that the nova had already entered the nebular
phase on UT 2005 November 14.16 (JD 2,453,688.66), 164 days
after the outburst \citep[the outburst day, $t_{\rm OB}=
2,453,525.0$, is taken from][]{hac07k}.
The SMARTS spectra also showed that the nebular phase had already
started at least on UT 2005 September 2 (JD 2,453,615.5).
We regard the nova to have entered the nebular phase
about $80-100$ days after the outburst.
The start of the nebular phase is close to the third data point of SMARTS,
so we plot a large unfilled red square in Figure
\ref{hr_diagram_v1663_aql_v5116_sgr_v2575_oph_v5117_sgr_outburst}(a).
This overlap of the SMARTS data with the track of LV~Vul suggests that
our adopted values of $E(B-V)=1.88$ and $(m-M')_V=17.95$ are reasonable,
that is, $E(B-V)=1.88\pm0.05$, $f_{\rm s}=0.83$, $(m-M)_V=18.15\pm0.2$,
and $d=2.9\pm0.3$~kpc. 

We check the value of $(m-M)_V=18.15$ by comparing
our model $V$ light curve with the observation.
Figure \ref{v1663_aql_lv_vul_v1668_cyg_v_bv_logscale}(a) shows the
model $V$ and UV~1455\AA\  light curves (thin solid black lines)
of a $1.0~M_\sun$ WD (CO3).
We add another model $V$ light curve (green lines) of a $0.98~M_\sun$ WD 
(CO3).  This is a model for V1668~Cyg,
which is time-stretched by $\Delta \log t= \log f_{\rm s}= \log 0.83 = -0.08$.
The $V$ light curve of V1668~Cyg follows the model $V$ light curve
until day $\sim 100$.  V1668~Cyg entered the nebular phase on day $\sim 100$.
Strong [\ion{O}{3}] lines make a significant contribution
to the $V$ band and, as a result, the observed $V$ light curve keeps
a similar decay trend of $F_\nu\propto t^{-1.75}$ in the nebular phase.
Because our model light curve does not include the effect of emission lines,
it begins to deviate from the observed $V$ light curve
(see Figure \ref{v1663_aql_lv_vul_v1668_cyg_v_bv_logscale}(a)).

The optically thick winds eventually stop.
The end of the optically thick winds is denoted by the unfilled circle
at the right end of the model $V$ light curve (day $\sim 230$) 
in Figure \ref{v1663_aql_lv_vul_v1668_cyg_v_bv_logscale}(a).
After the optically thick winds stop, the $V$ light curve follows
the trend of $F_\nu\propto t^{-3}$ (solid blue line)
in Figure \ref{v1663_aql_lv_vul_v1668_cyg_v_bv_logscale}(a).
This line indicates the decay of homologously expanding ejecta, i.e.,
the total mass of the ejecta is kept constant in time 
\citep[see, e.g.,][]{woo97, hac06kb}.
The $V$ light curve of V1663~Aql decays along $t^{-3}$ after day $\sim 230$.
Thus, the end of the model $V$ light curve corresponds reasonably
well to the change of slope at the epoch of ``wind stops'' (day $\sim 230$).
Thus, we confirm that the values of $(m-M)_V=18.15$ 
and $M_{\rm WD}= 1.0~M_\sun$ (CO3) are reasonable.

As already discussed in \citet{hac16k}, if we fix the chemical composition
of the hydrogen-rich envelope, the mass determination has an accuracy 
of $\pm0.01~M_\sun$ for the case of V1668~Cyg because we simultaneously
fit the $V$ and UV~1455\AA\  light curves with the observation.  
On the other hand, \citet{hac10k} obtained a $0.95~M_\sun$ WD as a best fit
for V1663~Aql assuming the other chemical composition of CO2.
To summarize, the WD mass determination depends on the chemical composition
of the hydrogen-rich envelope, especially on the hydrogen content
of $X$ and the carbon-nitrogen-oxygen abundance ($X_{\rm CNO}$) by weight,
as discussed by \citet{hac07k}.
We again discuss the accuracy of WD mass determination for more detail
in Section  \ref{wd_mass_determination}.

\subsection{V5116~Sgr 2005\#2}
\label{v5116_sgr_cmd}
The nova reached 7.2 mag at maximum on UT 2005 July 5.085 \citep{lil05}.
\citet{hac07k} analyzed the light curve of V5116~Sgr based on
their free-free emission model light curves and concluded that
the $V$ light curve is consistent with that of a $0.90\pm0.1~M_\sun$ WD (CO2).
The nova became a supersoft X-ray source \citep{sal08} at least
674 days after the outburst \citep[the outburst day of
$t_{\rm OB}=$JD 2,453,552.0, i.e., UT 2005 July 1.5, taken from][]{hac07k}.
\citet{sal08} reported that the X-ray light curve shows abrupt decreases
and increases of the flux by a factor of $\sim 8$. This periodicity
is consistent with a 2.97 hr orbital period suggested by \citet{dob08}. 
These authors speculated that the X-ray light curve may result
from a partial coverage by an asymmetric accretion disk in a
high-inclination system \citep[see also][]{sal17}.
\citet{nes07c} also reported the supersoft X-ray detection 
with {\it Swift} 768 days after the outburst (UT 2007 August 7.742).

\citet{sal08} estimated the absorption of $A_V= 3.1 E(B-V)= 0.8\pm0.2$
from the color at maximum, $(B-V)_{0,\rm max}=0.23\pm0.06$ \citep{van87},
and the distance of $d=11\pm3$~kpc from $t_2=6.5\pm1.0$~days \citep{dob08}
together with the MMRD relation proposed by \citet{del95}.
\citet{sal08} concluded that O/Ne-rich WD atmosphere models
provide a better fit than C/O-rich WD atmosphere models.
\citet{hac10k} adopted the chemical composition of Ne nova 3 (Ne3)
and obtained a new best-fitting model of a $1.07 ~M_\sun$ WD
\citep[see Figure 24 of][]{hac10k} based on their
free-free emission model light curves.  They also obtained
the distance modulus of $(m-M)_V=16.2\pm0.2$ and the distance of
$d=12\pm1$~kpc based on their light-curve fitting with their model
light curve.  \citet{ozd16} obtained the very different distance of
$d=1.55\pm0.70$~kpc from their distance-reddening relation
together with $E(B-V)=0.23\pm0.06$.

We obtain $(m-M)_B= 16.28$, $(m-M)_V= 16.07$, and $(m-M)_I= 15.68$,
which cross at $d=11.7$~kpc and $E(B-V)=0.23$,
in Appendix \ref{v5116_sgr} and plot them in Figure
\ref{distance_reddening_v1663_aql_v5116_sgr_v2575_oph_v5117_sgr}(b).
Taking into account our fitting accuracy, we obtain $E(B-V)=0.23\pm0.05$
and $d=12\pm2$~kpc.
The distance modulus in the $V$ band is $(m-M)_V=16.05\pm0.1$
and the timescaling factor is $f_{\rm s}= 1.58$ against LV~Vul.

For the reddening toward V5116~Sgr, $(l,b)=(2\fdg1363, -6\fdg8326)$, the VVV
survey catalog \citep{sai13} gives $E(B-V)=A_{K_s}/0.36=0.084/0.36=0.23$.
Also \citet{sal08} adopted $E(B-V)=A_V/3.1=0.8/3.1=0.25$.
The NASA/IPAC Infrared Science
Archive,\footnote{http://irsa.ipac.caltech.edu/applications/DUST/}
which is calculated using the data from \citet{schl11},
gives $E(B-V)=0.224\pm0.003$ toward V5116~Sgr.
All of these reddening values are consistent with our crossing point. 

We further examine our result
in Figure
\ref{distance_reddening_v1663_aql_v5116_sgr_v2575_oph_v5117_sgr}(b).
The vertical solid red line represents $E(B-V)=0.23$ and 
the thick solid blue line denotes the relation of Equation
(\ref{distance_modulus_rv}) together with $(m-M)_V=16.05$.  These two
lines cross each other at $d=12$~kpc.  The four relations toward 
$(l, b)=(2\fdg00, -6\fdg75)$, $(2\fdg25, -6\fdg75)$, $(2\fdg00, -7\fdg00)$,
and $(2\fdg25, -7\fdg00)$ are taken from \citet{mar06}. 
The closest direction is that of filled green squares.
We denote four of Schultheis et al.'s (2014)
distance-reddening relations toward $(l,b)=(2\fdg1, -6\fdg8)$, 
$(2\fdg1, -6\fdg9)$, $(2\fdg2, -6\fdg8)$, and $(2\fdg2, -6\fdg9)$
with very thin solid cyan lines in Figure 
\ref{distance_reddening_v1663_aql_v5116_sgr_v2575_oph_v5117_sgr}(b).
These four lines show zigzag patterns, although the reddening should
increase monotonically with the distance.  In this sense,
Schultheis et al.'s relation may not be appropriate
in the middle distance.
The other symbols/lines have the same meanings as those in Figure
\ref{distance_reddening_v1663_aql_v5116_sgr_v2575_oph_v5117_sgr}(a).

The reddening toward V5116 Sgr seems to saturate at $d > 5$~kpc
as shown in Figure
\ref{distance_reddening_v1663_aql_v5116_sgr_v2575_oph_v5117_sgr}(b).
Our crossing point, $d=12$~kpc and $E(B-V)=0.23$, is
consistent with Marshall et al.'s relations considering their error bars,
although it ends at $d = 10$~kpc.  On the other hand, Green et al.'s
black/orange lines and Chen et al.'s cyan-blue line
deviate slightly from Marshall et al.'s relation.
\citet{ozd16} obtained $d=1.55\pm0.70$~kpc as mentioned earlier,
assuming $E(B-V)=0.23\pm0.06$ and using their distance-reddening relation.
We think that this distance is too small.  
Their results come from their distance-reddening relation,
which is similar to the distance-reddening relations of \citet{schu14}.
This distance-reddening relation is rather different from the 
distance-reddening relations of \citet{mar06}. 
We confirmed that the adopted set of $E(B-V)=0.23$ and
$d=12$~kpc is consistent with Marshall et al.'s relation.

We plot the track of V5116~Sgr in Figure
\ref{hr_diagram_v1663_aql_v5116_sgr_v2575_oph_v5117_sgr_outburst}(b),
using $E(B-V)=0.23$ and $(m-M')_V=16.55$ ($f_{\rm s}=1.58$) in
Equation (\ref{absolute_mag_v5116_sgr}).
The track of V5116~Sgr almost follows that of V1974~Cyg, although
the data points of VSOLJ are rather scattered around the track of
V1974~Cyg (solid magenta lines) and V1500~Cyg (solid green line).
In general, the VSOLJ data were obtained by many amateur astronomers, and
their typical errors were not reported.  Therefore, we do not examine
the typical deviation of their data.  However, the mean trend
of the $V$ light and $(B-V)_0$ color curves in Figure 
\ref{v5116_sgr_v1974_cyg_x65z02o03ne03_v_bv_ub_color_logscale}(a)
and (b) broadly follow the trends of V1974 Cyg.  
This suggests that the mean trend of the track in Figure 
\ref{hr_diagram_v1663_aql_v5116_sgr_v2575_oph_v5117_sgr_outburst}(b)
broadly overlaps the track of V1974~Cyg.

Basically, we overlap the track of a target nova to that of the template
nova, LV Vul, and measure the timescaling factor against LV Vul.
The V1500~Cyg and V1974~Cyg tracks have been calibrated in our previous
paper \citep{hac19k} and their timescaling factors were measured against
LV~Vul.  Therefore, their time-stretched color-magnitude tracks are
also well calibrated.  This means that we can use either V1500~Cyg or
V1974~Cyg as another template nova.  However, it should be noted that
the timescaling factor of $f_{\rm s}$ is measured against LV Vul even if
we use the V1500~Cyg track as a template nova, because we have to use
common timescales for various novae that belong to one of the LV Vul
and V1500~Cyg types.

We regard V5116~Sgr to belong to the V1500~Cyg type.
This rough overlapping with the V1974~Cyg track may support
our adopted values of $E(B-V)=0.23$ and $(m-M')_V=16.55$,
that is, $E(B-V)=0.23\pm0.05$, $(m-M)_V=16.05\pm0.1$, 
$d=12\pm2$~kpc, $f_{\rm s}=1.58$.

We further check our value of $(m-M)_V=16.05$ by fitting our model
$V$ light curve with the observation as shown in Figure
\ref{v5116_sgr_v1974_cyg_x65z02o03ne03_v_bv_ub_color_logscale}(a).
Assuming $(m-M)_V=16.05$, our model $V$ light curve (black line) and the
soft X-ray light curve (red line) of a $1.07~M_\sun$ WD 
\citep[Ne3,][]{hac16k} fit reasonably well with the observed $V$ 
and X-ray light curves.
This confirms that our result of $(m-M)_V=16.05$ is reasonable.
We also add a $0.98~M_\sun$ WD (CO3) model for V1974~Cyg
\citep[see also][]{hac16k}.
If we adopt \"Ozd\"ormez et al.'s results of $d=1.55$~kpc and
$E(B-V)=0.23$, then the distance modulus in the $V$ band should be
$(m-M)_V= 11.7$, being much smaller than $(m-M)_V=16.05$ and
not consistent with our model light curve.   This means that,
if we plot our model light curve of $1.07~M_\sun$ WD
assuming $(m-M)_V = 11.7$, its $V$ model light curve is located 
$-4.35$ mag far above the observed $V$ light curve.

\subsection{V2575~Oph 2006}
\label{v2575_oph_cmd}
The nova reached $m_{V, \rm max}=10.96$ on
JD 2,453,779.0 (taken from the AAVSO data).
\citet{rus06b} reported that V2575~Oph is an \ion{Fe}{2} type nova and
the reddening is $E(B-V)=1.42$ from their $0.8$--$5.4~\mu$m spectroscopy.
\citet{rud06} obtained $E(B-V)=1.5$ from the \ion{O}{1} lines.

We obtain $(m-M)_B= 19.3$, $(m-M)_V= 17.88$, and $(m-M)_I= 15.59$,
which cross at $d=4.9$~kpc and $E(B-V)=1.43$,
in Appendix \ref{v2575_oph} and plot them in Figure
\ref{distance_reddening_v1663_aql_v5116_sgr_v2575_oph_v5117_sgr}(c).
Thus, we obtain $E(B-V)=1.43\pm0.05$ and $d=4.9\pm0.5$~kpc.
Our value of $E(B-V)=1.43\pm0.05$ is consistent with and close to
the values obtained by \citet{rud06} and \citet{rus06b}.

Figure \ref{hr_diagram_v1663_aql_v5116_sgr_v2575_oph_v5117_sgr_outburst}(c)
shows the $(B-V)_0$-$(M_V-2.5 \log f_{\rm s})$ diagram of V2575~Oph 
for $E(B-V)=1.43$ and $(m-M')_V=18.15$
in Equation (\ref{absolute_mag_v2575_oph}).
The data taken from SMARTS (filled red circles) follow well
the track of V1974~Cyg (solid magenta lines) and PW~Vul (solid cyan-blue
lines).  We regard V2575~Oph to belong to the V1500~Cyg type,
because both V1974~Cyg and PW~Vul belong to
the V1500~Cyg type \citep[see][]{hac19k}.
This overlapping with the PW~Vul track may support the fact that our adopted
values of $E(B-V)=1.43$ and $(m-M')_V=18.15$ are reasonable,
that is, $E(B-V)=1.43\pm0.05$, $(m-M)_V=17.85\pm0.1$, $f_{\rm s}=1.29$, 
and $d=4.9\pm0.5$~kpc.

We examine the distance and reddening toward V2575~Oph,
$(l,b)=(19\fdg7995, +10\fdg3721)$,
in Figure
\ref{distance_reddening_v1663_aql_v5116_sgr_v2575_oph_v5117_sgr}(c).
Two relations toward $(l, b)=(19\fdg75, +10\fdg00)$ and 
$(20\fdg00, +10\fdg00)$ are taken from \citet{mar06} and
depicted by the unfilled red squares and filled green squares, respectively,
each with error bars.
The closer direction in the galactic coordinates
is that of the unfilled red squares.  
The solid black/orange lines are taken from the relations given
by \citet{gre15, gre18}, respectively, which deviate 
slightly from Marshall et al.'s relations.
The unfilled cyan-blue diamonds are the relation of \citet{ozd16}.
The solid cyan-blue line corresponds to the relation of \citet{chen18}.
Our crossing point, $d=4.9$~kpc and $E(B-V)=1.43$, is rather 
far from Marshall et al.'s, Green et al.'s, and Chen et al.'s relations.
The NASA/IPAC absorption map gives $E(B-V)=0.64\pm0.03$
toward V2575~Oph, which is consistent with Marshall et al.'s,
Green et al.'s, and Chen et al.'s relations.  On the other hand, the VVV
catalog \citep{sai13} gives $E(B-V)=A_{K_s}/0.36=0.488/0.36=1.35$,
which is close to our value of $E(B-V)=1.43$.
\"Ozd\"ormez et al.'s relation is also consistent with our crossing point. 

We further check our value of $(m-M)_V=17.85$ by fitting our model
$V$ light curve with the observation as shown in Figure
\ref{v2575_oph_v1668_cyg_lv_vul_v_bv_ub_logscale}(a).  
Assuming $(m-M)_V=17.85$, our model light curve of a $0.90~M_\sun$ WD
\citep[CO3;][]{hac16k} fits reasonably well with the observed $V$ light
curve.  This confirms that our result of $(m-M)_V=17.85$ is reasonable.

\subsection{V5117~Sgr 2006\#2}
\label{v5117_sgr_cmd}
  The nova reached 
$m_{V, \rm max}=8.6$ on JD~$2,453,785.0\pm1.0$ (UT 2006 February 
$18.5\pm1.0$, IAU Circular 8673, 1).
\citet{lyn06} reported that V5117~Sgr is an \ion{Fe}{2} type nova and
the reddening is $E(B-V)=0.50\pm0.15$ from their
$0.8$--$5.5~\mu$m spectroscopy on UT 2006 May 1.

We obtain $(m-M)_B= 16.53$, $(m-M)_V= 16.01$, and $(m-M)_I= 15.12$,
which cross at $d=7.5$~kpc and $E(B-V)=0.53$,
in Appendix \ref{v5117_sgr} and plot them in Figure
\ref{distance_reddening_v1663_aql_v5116_sgr_v2575_oph_v5117_sgr}(d).
Thus, we obtain $d=7.5\pm0.8$~kpc and $E(B-V)=0.53\pm0.05$.

For the reddening toward V5117~Sgr, 
$(l,b)=(354\fdg6241, -6\fdg3774)$,
our obtained value of $E(B-V)=0.53$ is consistent with
Lynch et al.'s result as mentioned above.
On the other hand, the NASA/IPAC absorption map gives $E(B-V)=0.37\pm0.02$.
The VVV catalog \citep{sai13} gives
$E(B-V)=A_{K_{\rm s}}/0.36=0.098/0.36=0.27$.
We further check our result toward V5117~Sgr
in Figure
\ref{distance_reddening_v1663_aql_v5116_sgr_v2575_oph_v5117_sgr}(d).
We plot Marshall et al.'s (2006) relations toward
$(l, b)=(354\fdg50, -6\fdg50)$, $(354\fdg75,-6\fdg50)$,
$(354\fdg50, -6\fdg25)$, and $(354\fdg75, -6\fdg25)$. 
Green et al.'s (2015, 2018) relations are not available for this direction.
We also add the 3D reddening map obtained by \citet{schu14}
with the very thin solid cyan lines (four close directions toward V5117~Sgr).
The two thick solid cyan-blue lines correspond to
the relations of \citet{chen18}, toward
$(l, b)=(354\fdg65, -6\fdg35)$ and $(l, b)=(354\fdg65, -6\fdg45)$.  
The other symbols/lines have the same meanings as those in Figure
\ref{distance_reddening_v1663_aql_v5116_sgr_v2575_oph_v5117_sgr}(a).
These 3D maps are largely different from each other, but 
Marshall et al.'s relations with error bars are roughly close to
our crossing point.

Using $E(B-V)=0.53$ and $(m-M')_V=16.15$ in Equation
(\ref{absolute_mag_v5117_sgr}), we plot the $(B-V)_0$-$(M_V-2.5 
\log f_{\rm s})$ diagram of V5117~Sgr in Figure
\ref{hr_diagram_v1663_aql_v5116_sgr_v2575_oph_v5117_sgr_outburst}(d).
The track of V5117~Sgr almost follows that of V1500~Cyg, so we regard 
V5117~Sgr to belong to the V1500~Cyg type.
The nova had already entered the nebular phase on UT 2006 September 6
from the SMARTS spectra, corresponding to $m_V=13.6$ ($M_V=-2.4$).  We
consider that V5117~Sgr entered the nebular phase at the turning point
from blue toward red \citep{hac16kb}
as shown by the large unfilled red square in Figure
\ref{hr_diagram_v1663_aql_v5116_sgr_v2575_oph_v5117_sgr_outburst}(d).
This overlapping with the V1500~Cyg track supports 
$E(B-V)=0.53$ and $(m-M')_V=16.15$, that is,
$E(B-V)=0.53\pm0.05$, $(m-M)_V=16.0\pm0.1$, $f_{\rm s}=1.12$,
and $d=7.5\pm0.8$~kpc.  If the distance of 7.5~kpc is correct, 
V5117~Sgr belongs to the galactic bulge.

We further check the distance modulus of $(m-M)_V=16.0$ by comparing 
our model $V$ light curve with the observation in Figure 
\ref{v5117_sgr_v1668_cyg_lv_vul_v_bv_ub_logscale}(a).
Assuming that $(m-M)_V=16.0$, we plot a model $V$ light curve 
(solid black lines) of
a $0.95~M_\sun$ WD \citep[CO3;][]{hac16k}.
The model light curve reproduces well the observation, supporting
our value of $(m-M)_V=16.0$.
We also plot a model $V$ light curve (solid green lines) of
a $0.98~M_\sun$ WD (CO3),
assuming that $(m-M)_V=14.6$ for V1668~Cyg.


\begin{figure*}
\plotone{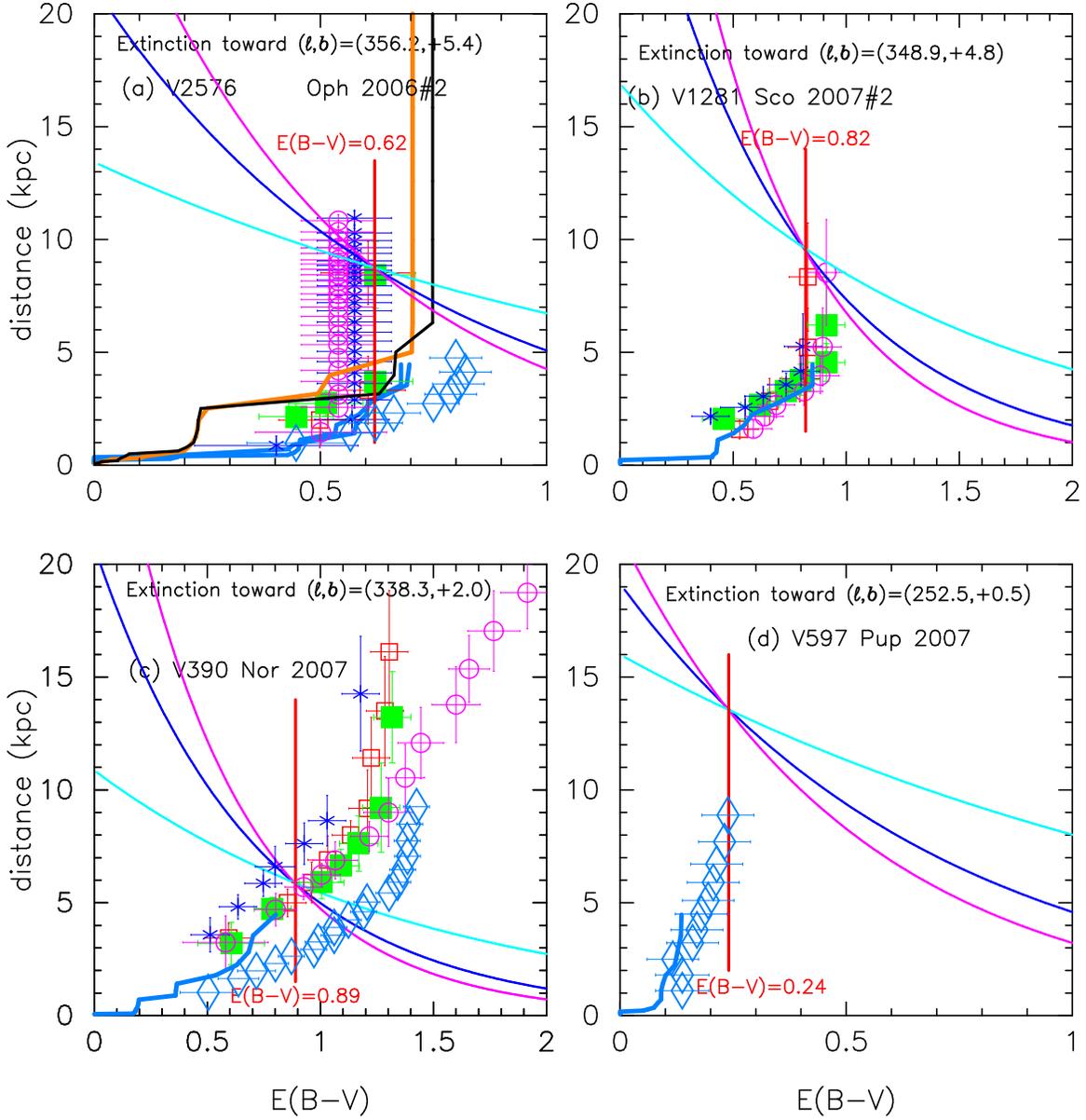}
\caption{
Same as Figure 
\ref{distance_reddening_v1663_aql_v5116_sgr_v2575_oph_v5117_sgr},
but for (a) V2576~Oph, (b) V1281~Sco, (c) V390~Nor, and (d) V597~Pup.
\label{distance_reddening_v2576_oph_v1281_sco_v390_nor_v597_pup}}
\end{figure*}


\begin{figure*}
\plotone{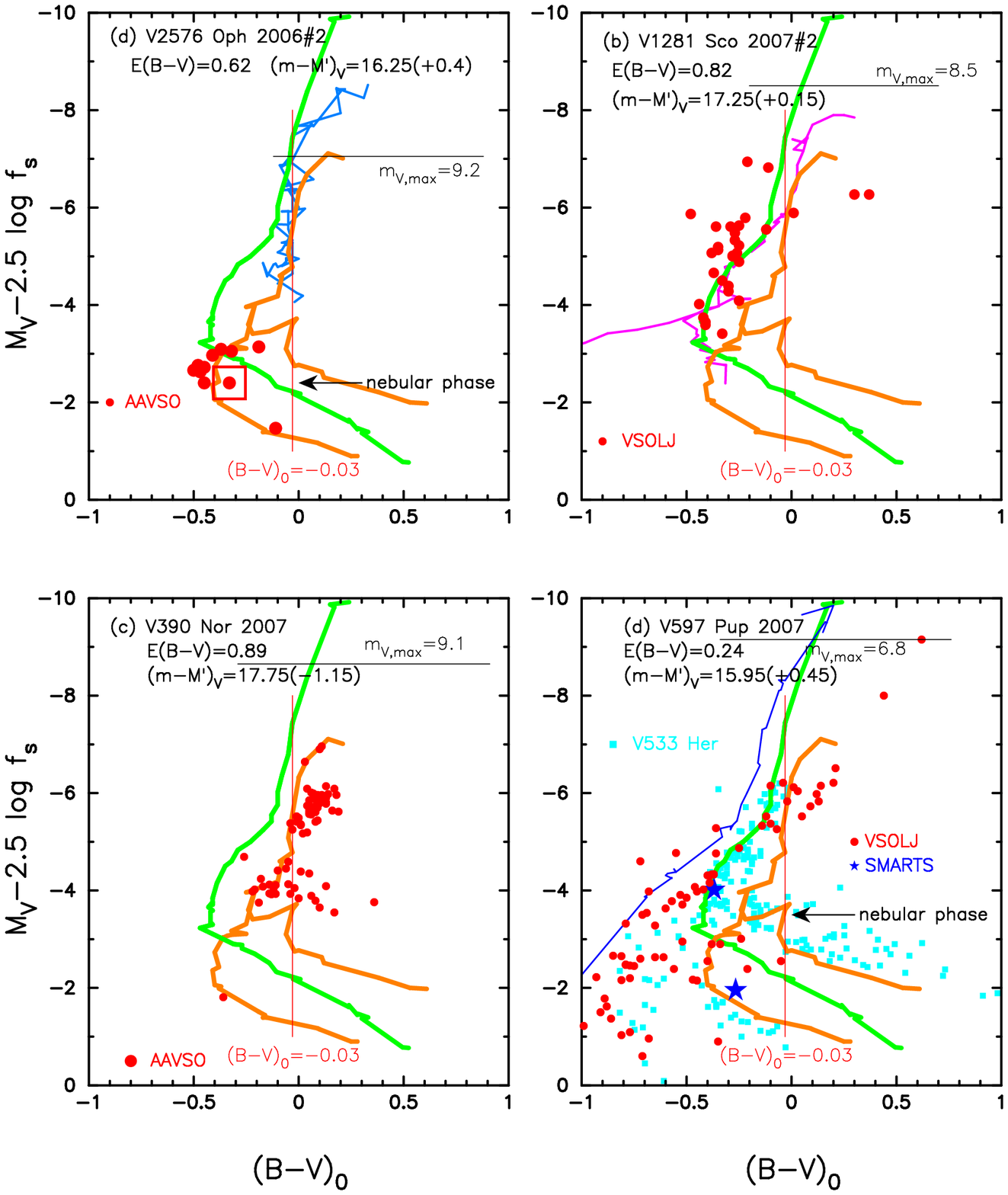}
\caption{
Same as Figure 
\ref{hr_diagram_v1663_aql_v5116_sgr_v2575_oph_v5117_sgr_outburst},
but for (a) V2576~Oph, (b) V1281~Sco, (c) V390~Nor,
and (d) V597~Pup.  The solid green lines show the template track
of V1500~Cyg.  The solid orange lines represent that of LV~Vul.
In panel (a), we add the track of V1668~Cyg (thin solid cyan-blue lines).
In panel (b), we add the track of V1974~Cyg (thin solid magenta lines).
In panel (d), we add the track of V1500~Cyg (thin solid blue lines)
obtained by \citet{pfa76} and the track of V533~Her (filled cyan 
squares), which are taken from Figure 25(b) of \citet{hac19k}.
\label{hr_diagram_v2576_oph_v1281_sco_v390_nor_v597_pup_outburst}}
\end{figure*}

\subsection{V2576~Oph 2006\#2}
\label{v2576_oph_cmd}
\citet{rus06b} reported that the nova showed an of \ion{Fe}{2}
type spectrum from their $0.8$--$5.4~\mu$m spectroscopy
on UT 2006 April 30.  They obtained $E(B-V)=0.62$.
\citet{lyn06b} reported a much smaller value of $E(B-V)=0.25$ and
no evidence of dust in the nova ejecta based on their
$0.47$--$2.5~\mu$m spectroscopy on UT 2006 June 14.4.

We obtain $(m-M)_B= 17.25$, $(m-M)_V= 16.63$, and $(m-M)_I= 15.64$
(see Appendix \ref{v2576_oph}), 
which cross at $d=8.8$~kpc and $E(B-V)=0.62$, and plot them in Figure
\ref{distance_reddening_v2576_oph_v1281_sco_v390_nor_v597_pup}(a).
Thus, we have $E(B-V)=0.62\pm0.05$ and $d=8.8\pm1$~kpc.

For the reddening toward V2576~Oph, $(l,b)=(356\fdg2004, +5\fdg3694)$,
the NASA/IPAC absorption map gives $E(B-V)=0.62\pm0.01$,
which is consistent with our result and the value obtained by \citet{rus06b}. 
We further check our result
in Figure
\ref{distance_reddening_v2576_oph_v1281_sco_v390_nor_v597_pup}(a).
We plot four relations of \citet{mar06} toward
$(l, b)=(356\fdg00, +5\fdg25)$, $(356\fdg25, +5\fdg25)$, 
$(356\fdg00, +5\fdg50)$, and $(356\fdg25, +5\fdg50)$.
The closest direction is that of the filled green squares.
We add the 3D reddening maps calculated by \citet[][unfilled cyan-blue
diamonds with error bars]{ozd16} and by \citet[][thick solid cyan-blue
lines toward $(l, b)=(356\fdg15, +5\fdg35)$ and 
$(l, b)=(356\fdg25, +5\fdg35)$]{chen18}.
The other symbols/lines have the same meanings as those in Figure
\ref{distance_reddening_v1663_aql_v5116_sgr_v2575_oph_v5117_sgr}(a).
These 3D maps are largely different from each other.
Only Marshall et al.'s relation, which is shown by the filled green squares
with error bars, is consistent with our crossing point.

Figure
\ref{hr_diagram_v2576_oph_v1281_sco_v390_nor_v597_pup_outburst}(a) 
shows the $(B-V)_0$-$(M_V-2.5 \log f_{\rm s})$ diagram of V2576~Oph
for $E(B-V)=0.62$ and $(m-M')_V=16.25$ 
in Equation (\ref{absolute_mag_v2576_oph}).
The track of V2576~Oph almost follows that of LV~Vul in the middle
phase of nova evolution, so we regard V2576~Oph to belong to the
LV~Vul type.
From the SMARTS spectra, the nova had already entered the nebular phase
on UT 2006 August 9 (JD~2,453,956.5), corresponding to $m_V=15.0$ 
($M_V=-1.65$).  We consider that V2576~Oph entered the nebular phase
at the point denoted by the large unfilled red square in Figure
\ref{hr_diagram_v2576_oph_v1281_sco_v390_nor_v597_pup_outburst}(a),
i.e., $M_V=-2.5$ and $(B-V)_0=-0.31$.
This overlapping with the LV~Vul track suggests that
$E(B-V)=0.62$ and $(m-M')_V=16.25$ are reasonable.  
Thus, we confirm $E(B-V)=0.62\pm0.05$, $(m-M)_V=16.65\pm0.1$,
$f_{\rm s}=0.71$, and $d=8.8\pm1$~kpc.
If the distance is correct, V2576~Oph belongs to the galactic bulge.

We finally check the distance modulus of $(m-M)_V=16.65$ by comparing
our model $V$ light curve with the observation.
Assuming that $(m-M)_V=16.65$, we plot a model light curve of a
$1.15~M_\sun$ WD \citep[Ne2; solid black line,][]{hac10k}
for V2576~Oph as shown in Figure 
\ref{v2576_oph_v1668_cyg_lv_vul_v_bv_ub_logscale}(a).
Our model light curve reasonably follows the $V$ light curve of V2576~Oph.
This again confirms that the distance modulus of $(m-M)_V=16.65$ 
is reasonable.  For comparison, we add the model light curve of 
a $0.98~M_\sun$ WD \citep[CO3; solid green lines,][]{hac16k} 
and show that it follows well the light curve of V1668~Cyg,
assuming that $(m-M)_V=14.6$ for V1668~Cyg.

\subsection{V1281~Sco 2007\#2}
\label{v1281_sco_cmd}
The nova was as bright as 9.3 mag  when Y. Nakamura discovered the nova
\citep{yam07c}.  \citet{hen07} estimated a maximum around UT 2007 February
20.5 at $m_{V,\rm max}=8.5$ from their extrapolation of the available data.
The nova had declined to 8.8 mag by UT 2007 February 20.85 \citep{yam07c}.
\citet{rus07c} reported  the reddening of $E(B-V) = 0.7$ derived from 
the \ion{O}{1} lines based on their $0.8$--$5.5~\mu$m spectroscopy
on UT 2007 May 6.  \citet{hac10k} obtained a best-fitting model of
$1.13 ~M_\sun$ WD (Ne2) based on their free-free emission model light
curves, $(m-M)_V=17.8\pm0.2$, and $d=13\pm1$~kpc for $E(B-V)=0.7$.
The nova was detected with {\it Swift} as a supersoft X-ray source on
UT 2008 January 24.18 \citep{schw11}.

We obtain $(m-M)_B= 18.25$, $(m-M)_V= 17.43$, and $(m-M)_I= 16.14$,
which cross at $d=9.4$~kpc and $E(B-V)=0.82$,
in Appendix \ref{v1281_sco} and plot them in Figure
\ref{distance_reddening_v2576_oph_v1281_sco_v390_nor_v597_pup}(b).
Thus, we have $E(B-V)=0.82\pm0.05$ and $d=9.4\pm1$~kpc.

For the reddening toward V1281~Sco, 
$(l,b)=(348\fdg8564, +4\fdg7984)$,
the NASA/IPAC absorption map gives $E(B-V)=0.77\pm0.02$,
consistent with our value of $E(B-V)=0.82\pm0.05$.
Our result is slightly larger than the $E(B-V)=0.7$ obtained
by \citet{rus07c}.  We further check our result toward V1281~Sco
in Figure 
\ref{distance_reddening_v2576_oph_v1281_sco_v390_nor_v597_pup}(b).
We plot Marshall et al.'s (2006) relations toward
$(l, b)=(348\fdg75, +5\fdg00)$, $(349\fdg00, +5\fdg00)$,
$(348\fdg75, +4\fdg75)$, and $(349\fdg00, +4\fdg75)$.
The closest direction is that of the blue asterisks.
The solid cyan-blue line corresponds to the relation of \citet{chen18}.
Our crossing point of $d=9.4$~kpc and $E(B-V)=0.82$
is consistent with Marshall et al.'s relation (blue asterisks).
This again supports our result that $E(B-V)=0.82$ is reasonable.

We plot the $(B-V)_0$-$(M_V-2.5 \log f_{\rm s})$ diagram of V1281~Sco
in Figure \ref{hr_diagram_v2576_oph_v1281_sco_v390_nor_v597_pup_outburst}(b)
for $E(B-V)=0.82$ and $(m-M')_V=17.25$ in Equation 
(\ref{absolute_mag_v1281_sco}).
Unfortunately, we only have the VSOLJ data.
The typical errors of the VSOLJ data were not reported, so
we do not have a way to evaluate the typical deviation of their data.
The $V$ light and $(B-V)_0$ color evolutions of V1281~Sco in Figure 
\ref{v1281_sco_v1500_cyg_m1130_x55z02o10ne03_logscale}(a) and (b)
are also scattered, but they broadly follow the trend of V1500 Cyg.
We may conclude that the mean trend in the time-stretched
color-magnitude diagram in Figure 
\ref{hr_diagram_v2576_oph_v1281_sco_v390_nor_v597_pup_outburst}(b)
broadly follows the track of V1500~Cyg,
so we regard V1281~Sco to belong to the V1500~Cyg type.
This rough overlapping supports our adopted values of $E(B-V)=0.82$ 
and $(m-M')_V=17.25$.  Thus, we obtain $E(B-V)=0.82\pm0.05$,
$(m-M)_V=17.4\pm0.1$, $f_{\rm s}= 0.80$, and $d=9.4\pm1$~kpc.
If this distance is correct, V1281~Sco belongs to the galactic bulge.

We check the distance modulus of $(m-M)_V=17.4$ by comparing
our model $V$ light curve with the observation.
Assuming that $(m-M)_V=17.4$, we plot a model light curve of a
$1.13~M_\sun$ WD \citep[Ne2, solid black line;][]{hac10k} as shown
in Figure \ref{v1281_sco_v1500_cyg_m1130_x55z02o10ne03_logscale}(a).
Our model light curve reasonably follows the $V$ light curve of V1281~Sco.
This again confirms that the distance modulus of $(m-M)_V=17.4$ 
is reasonable.  For comparison,
we add the model light curve of a $1.20~M_\sun$ WD (Ne2, solid blue lines)
and show that it follows well the light curve of V1500~Cyg,
assuming that $(m-M)_V=12.3$ for V1500~Cyg \citep{hac19k}.

\citet{hac10k} obtained $(m-M)_V= 17.8$ using the $V$ model light
curve fitting of a $1.13~M_\sun$ WD (Ne2).  The VSOLJ data of $V$ magnitude
are quite scattered.  They fitted the $V$ model light curve
with the lower bound of the broad distribution of data points, which
resulted in the distance modulus of $(m-M)_V= 17.8$.  If we adopt
the central value of the broad distribution of data points,
we obtain $(m-M)_V= 17.4$ as shown in Figure 
\ref{v1281_sco_v1500_cyg_m1130_x55z02o10ne03_logscale}(a).
This is because the scatter of $V$ data points is about $\pm0.5$ mag.
\citet{hac10k} also obtained $(m-M)_V=17.5$ from the light curve
fitting between V1281~Sco and V1500~Cyg, which is consistent with
the present result.

\subsection{V390~Nor 2007}
\label{v390_nor_cmd}
The nova reached 9.1 mag at maximum on UT 2007 June 7.2 (JD~2,454,258.7)
\citep{lil07}.
\citet{lyn07a} reported the reddening of $E(B-V)\sim1.0$ derived
from the strong \ion{O}{1} lines based on their $0.8$--$5.5~\mu$m
spectroscopy on UT 2007 June 21.

We obtain $(m-M)_B=17.48$, $(m-M)_V=16.59$, and $(m-M)_I=15.19$,
which cross at $d=5.8$~kpc and $E(B-V)=0.89$,
in Appendix \ref{v390_nor}, and plot them in Figure
\ref{distance_reddening_v2576_oph_v1281_sco_v390_nor_v597_pup}(c).
Thus, we obtained $E(B-V)=0.89\pm0.05$ and $d=5.8\pm0.6$~kpc
toward V390~Nor, whose galactic coordinates are 
$(l,b)=(338\fdg3371, +2\fdg0061)$. 

We check our result 
in Figure
\ref{distance_reddening_v2576_oph_v1281_sco_v390_nor_v597_pup}(c).
We plot four relations of \citet{mar06} toward
$(l, b)=(338\fdg25, +2\fdg25)$, $(338\fdg50, +2\fdg25)$,
$(338\fdg25, +2\fdg00)$, and $(338\fdg50, +2\fdg00)$.
The closest direction is that of the blue asterisks.
The other symbols/lines have the same meanings as those in Figure
\ref{distance_reddening_v1663_aql_v5116_sgr_v2575_oph_v5117_sgr}(a).
Our crossing point of $d=5.8$~kpc and $E(B-V)=0.89$
is consistent with Marshall et al.'s relation (unfilled red squares,
filled green squares, and unfilled magenta circles)
and a linear extension of Chen et al.'s relation (solid cyan-blue line).

We plot the $(B-V)_0$-$(M_V-2.5 \log f_{\rm s})$ diagram of V390~Nor in Figure
\ref{hr_diagram_v2576_oph_v1281_sco_v390_nor_v597_pup_outburst}(c)
for $E(B-V)= 0.89$ and $(m-M')_V= 17.75$ in Equation
(\ref{absolute_mag_v390_nor}).
The track of V390~Nor almost follows that of LV~Vul, although 
the AAVSO data are rather scattered.
Therefore, we regard V390~Nor to belong to the LV~Vul type. 
Thus, we confirmed that $E(B-V)=0.89$ and $(m-M')_V=17.75$ are reasonable,
that is,  $E(B-V)=0.89\pm0.05$, $(m-M)_V=16.6\pm0.2$, 
$f_{\rm s}=2.82$, and $d=5.8\pm0.6$~kpc.

We check the distance modulus of $(m-M)_V=16.6$ by comparing
our model $V$ light curve with the observation.
Assuming that $(m-M)_V=16.6$, we plot a model light curve of a
$0.75~M_\sun$ WD \citep[CO3, solid red line;][]{hac16k} as shown in Figure 
\ref{v390_nor_v5666_sgr_lv_vul_v1668_cyg_x55z02c10o10_logscale}(a).
Our model light curve reasonably follows
the $V$ light curve of V390~Nor.  This again confirms that 
the distance modulus of $(m-M)_V=16.6$ is reasonable.  For comparison,
we add the model light curve of a $0.98~M_\sun$ WD (CO3, solid green lines)
and show that it follows the light curve of V1668~Cyg,
assuming that $(m-M)_V=14.6$ for V1668~Cyg.

\subsection{V597~Pup 2007}
\label{v597_pup_cmd}
The nova reached $m_{V, \rm max}=6.8$ on JD~2,454,419.2 from 
the VSOLJ data.  \citet{nai07} reported that
the nova belongs to the \ion{Fe}{2} type from a spectrum obtained
on UT 2007 November 14.77.  Rudy et al. obtained a small
reddening of $E(B-V)\sim0.3$ from the \ion{O}{1} lines
based on their $0.8$--$2.42~\mu$m spectroscopy obtained on
UT 2008 January 7.48 \citep{nes08c}.  The nova was observed as a
supersoft X-ray source on UT 2008 January 8.02 and 17.18
\citep{nes08c, schw11}.
\citet{war09} reported that V597~Pup is an intermediate polar with
the orbital period of $P_{\rm orb}=2.6687$~hr and the spin period of
$P_{\rm spin}=261.9$~s.
\citet{nai09} obtained NIR spectra in the decline phase
and concluded that the nova belongs to the He/N type.
\citet{hac10k} analyzed the optical and supersoft X-ray light curves
based on their free-free emission model light curves, 
and obtained the WD mass of $M_{\rm WD}=1.2~M_\sun$ (Ne3)
and the distance modulus of $(m-M)_V=17.0\pm0.2$.
\citet{hou16} revealed the rising phase of the nova
from the data obtained with the {\it Solar Mass Ejection Imager (SMEI)}.

We obtain $(m-M)_B= 16.63$, $(m-M)_V= 16.41$, and $(m-M)_I= 16.02$,
which broadly cross at $d=13.5$~kpc and $E(B-V)=0.24$,
in Appendix \ref{v597_pup} and plot them in Figure
\ref{distance_reddening_v2576_oph_v1281_sco_v390_nor_v597_pup}(d).
Thus, we have $E(B-V)=0.24\pm0.05$ and $d=13.5\pm2$~kpc.

For the reddening toward V597~Pup, $(l,b)=(252\fdg5295, +0\fdg5453)$,
the NASA/IPAC absorption map gives $E(B-V)=0.381\pm0.003$.
This value is slightly larger than our value of $E(B-V)= 0.24$ 
and $E(B-V)\sim 0.3$ obtained by Rudy et al. \citep{nes08c}.
We plot two distance-reddening relations in Figure
\ref{distance_reddening_v2576_oph_v1281_sco_v390_nor_v597_pup}(d).
The unfilled cyan-blue diamonds are the relation of \citet{ozd16}.  
The solid cyan-blue line corresponds to the relation of \citet{chen18}.
No data of \citet{mar06} and \citet{gre15, gre18} are available.
It seems that our crossing point of $d= 13.5$~kpc and $E(B-V)= 0.24$ is
consistent with the relation of \citet{ozd16} if the reddening saturates
at the distance of $d > 10$~kpc.

We plot the $(B-V)_0$-$(M_V-2.5 \log f_{\rm s})$ diagram of V597~Pup in Figure
\ref{hr_diagram_v2576_oph_v1281_sco_v390_nor_v597_pup_outburst}(d)
for $E(B-V)=0.24$ and $(m-M')_V=15.95$ in Equation
(\ref{absolute_mag_v597_pup}).
In this case, the scatter of the $(B-V)_0$ colors is too large to
immediately conclude overlapping.
As already mentioned, the VSOLJ data have large errors and
their degrees of errors were not reported.  
The scatter of the $B-V$ color may come from the different responses of
each $V$ filter among various observers.  
We plot another track of V1500~Cyg obtained by \citet{pfa76} with
the thin solid blue line, which deviates from the solid green line when
strong emission lines contribute to the $V$ filter.
This is because strong [\ion{O}{3}] lines
contribute to the blue edge of the $V$ filter and a small difference
in the $V$ filter response makes a large difference in the $B-V$ color.
To demonstrate such effects, we add the track of V533~Her, 
which belongs to the V1500~Cyg type \citep{hac19k}.
The V533~Her data bifurcate into three major branches
because of different $V$ filter systems after the nebular phase started
\citep[see, e.g., Figure 11(b) of][]{hac19k}.  Similarly, the data of
V597~Pup seem to bifurcate into two (or three) branches.  There are two
data points of SMARTS that have very small errors (0.003 mag).
The first data point is located on the V1500~Cyg (and V533~Her) track
and the second one is near the second (from left to right) branch
of V533~Her.  Considering such bifurcation in the nebular phase, we may
conclude a broad overlapping of the V597~Pup track with that of V1500~Cyg
or V533~Her.
This may confirm that $E(B-V)=0.24$ and $(m-M')_V=15.95$ are reasonable,
that is, $E(B-V)=0.24\pm0.05$, $(m-M)_V=16.4\pm0.1$, $f_{\rm s}=0.66$,
and $d=13.5\pm2$~kpc.

We check the distance modulus of $(m-M)_V=16.4$ by comparing
our model $V$ light curve with the observation
in Figure \ref{v597_pup_v1500_cyg_lv_vul_v_bv_ub_x65z02o03ne03_logscale}.
Assuming the distance modulus of $(m-M)_V=16.4$,
we plot the $V$ light curve of our $1.2~M_\sun$ WD (Ne3).
Our model light curves simultaneously follows the $V$ and supersoft 
X-ray fluxes of V597~Pup,
confirming that our adopted value of $(m-M)_V=16.4$ is reasonable.

Our WD model is essentially the same as that obtained by 
\citet{hac10k}, but the distance modulus in the $V$ band
is improved from their $(m-M)_V=17.0\pm0.2$ to our $(m-M)_V=16.4\pm0.1$.
This is partly because \citet{hac10k} assumed 
$(m-M)_{V, \rm V1500~Cyg}= 12.5$ while we adopt 
$(m-M)_{V, \rm V1500~Cyg}= 12.3$ \citep{hac14k} 
and partly because we improve the vertical fit of $\Delta V$
including the effect of photospheric emission \citep{hac15k}.


\begin{figure*}
\plotone{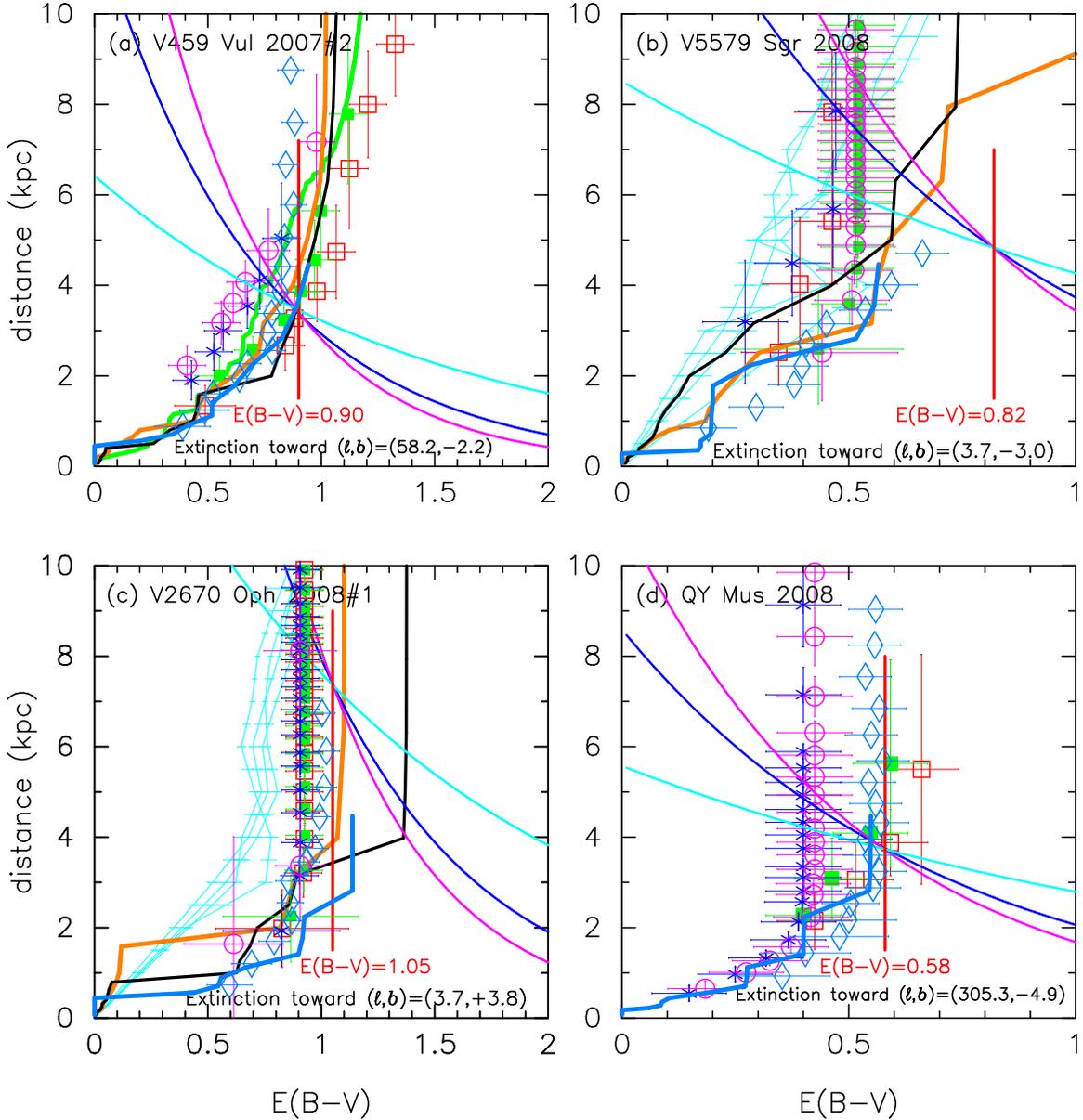}
\caption{
Same as Figure 
\ref{distance_reddening_v1663_aql_v5116_sgr_v2575_oph_v5117_sgr},
but for (a) V459~Vul, (b) V5579~Sgr, (c) V2670~Oph, and (d) QY~Mus.
\label{distance_reddening_v459_vul_v5579_sgr_v2670_oph_qy_mus}}
\end{figure*}


\begin{figure*}
\plotone{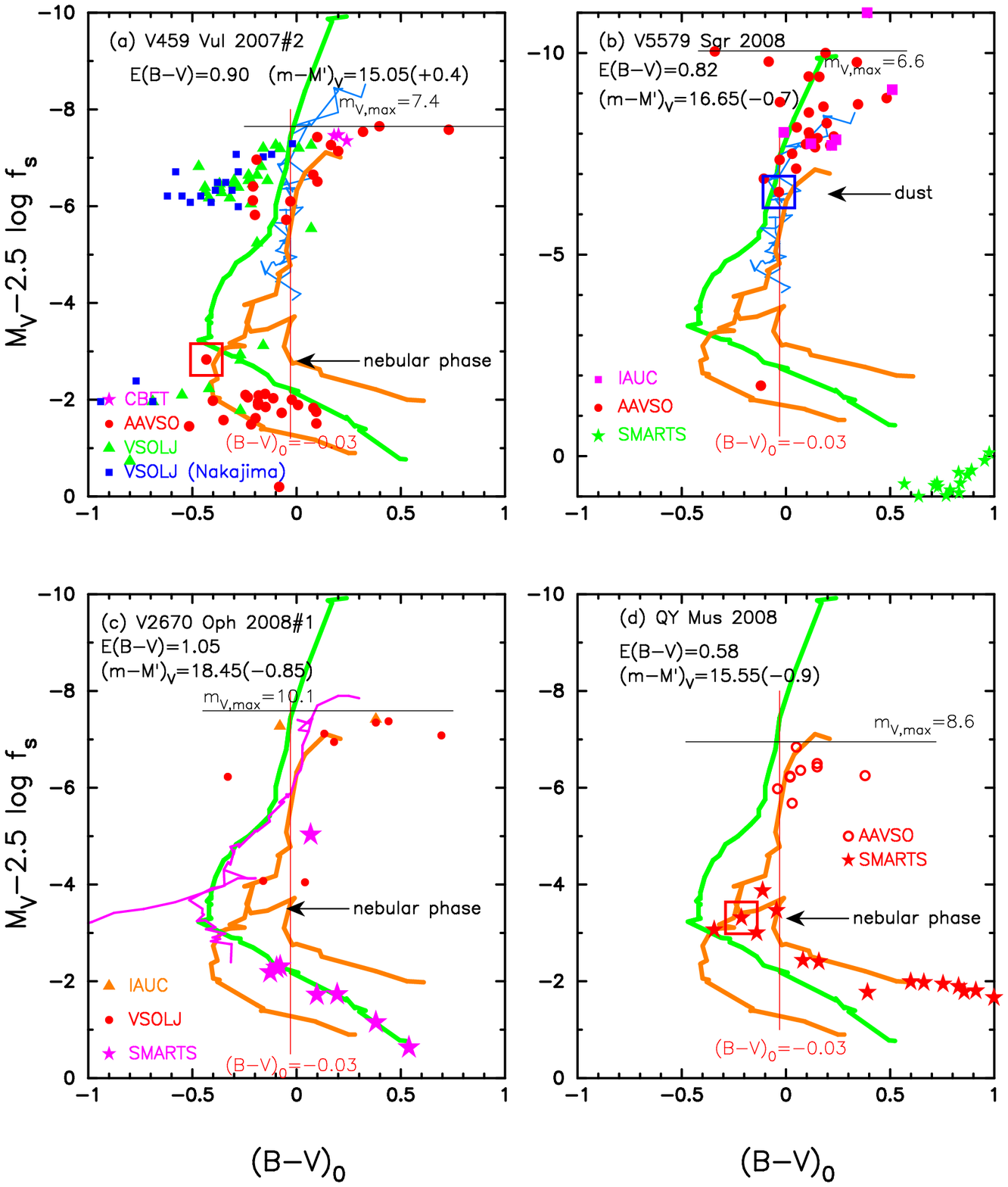}
\caption{
Same as Figure 
\ref{hr_diagram_v1663_aql_v5116_sgr_v2575_oph_v5117_sgr_outburst},
but for (a) V459~Vul, (b) V5579~Sgr, (c) V2670~Oph,
and (d) QY~Mus.  The solid green lines show the template track
of V1500~Cyg.  The solid orange lines represent that of LV~Vul.
We add the track of V1668~Cyg (thin solid cyan-blue lines) and
V1974~Cyg (thin solid magenta lines).
\label{hr_diagram_v459_vul_v5579_sgr_v2670_oph_qy_mus_outburst}}
\end{figure*}

\subsection{V459~Vul 2007\#2}
\label{v459_vul_cmd}
The nova reached $m_{V, \rm max}=7.4$ on JD 2,454,462.47 from the AAVSO data.
The {\it SMEI} light curve showed several oscillations in the decline with
amplitudes on the order of a few tenths of a magnitude 
\citep[see Figure 8 of][]{hou16}.
The spectra show prominent Balmer lines and \ion{Fe}{2} lines
with P-Cygni profile \citep{yam07}, indicating that the nova is
an \ion{Fe}{2} type \citep{hou16}.
\citet{mun07c} reported that strong diffuse interstellar bands and
\ion{Na}{1} interstellar lines suggest a large reddening,
possibly $E(B-V) > 0.7$.
\citet{rus08a} reported, based on their IR spectroscopy, that
the \ion{O}{1} lines indicate a reddening of $E(B-V)\sim1.0$,
a part of which is local to the nova because the spectrum
shows dust thermal emission beyond $1.5~\mu$m.
\citet{pog10} obtained the reddening of
$E(B-V)=(2.75\pm0.38)/3.1=0.89\pm0.12$, $d=2.3$--$5.0$~kpc,
The spectra at day 114 (UT 2008 April 19) show strong [\ion{O}{3}] lines
\citep{pog10}, so the nova had already entered the nebular phase.

We obtain $(m-M)_B= 16.35$, $(m-M)_V= 15.43$, and $(m-M)_I= 14.04$,
which broadly cross at $d=3.4$~kpc and $E(B-V)=0.90$,
in Appendix \ref{v459_vul} and plot them in Figure 
\ref{distance_reddening_v459_vul_v5579_sgr_v2670_oph_qy_mus}(a).
Thus, we have $E(B-V)=0.90\pm0.05$ and $d=3.4\pm0.4$~kpc.

For the reddening toward V459~Vul, \\$(l,b)=(58\fdg2138, -2\fdg1673)$,
\citet{mun07c} reported $E(B-V) > 0.7$.  \citet{rus08a} also reported
$E(B-V)\sim1.0$.  \citet{pog10} obtained the reddening of
$E(B-V)= 0.89\pm0.12$.  All of the reported
reddenings are consistent with our value of $E(B-V)=0.90\pm0.05$.
The distance of $d=2.3$--$5.0$~kpc obtained by \citet{pog10} 
is also consistent with our value of $d=3.4\pm0.4$~kpc.
We further check our result
in Figure
\ref{distance_reddening_v459_vul_v5579_sgr_v2670_oph_qy_mus}(a).
We plot four relations of \citet{mar06} toward
$(l, b)=(58\fdg00, -2\fdg00)$, $(58\fdg25, -2\fdg00)$,
$(58\fdg00, -2\fdg25)$, and $(58\fdg25, -2\fdg25)$.
The nova direction is almost between that of the unfiflled magenta circles 
and filled green squares in the galactic coordinates.
The other symbols/lines have the same meanings as those in Figure
\ref{distance_reddening_v1663_aql_v5116_sgr_v2575_oph_v5117_sgr}(a).
Our crossing point at $d=3.4$~kpc and $E(B-V)=0.90$
is consistent with Green et al.'s, Marshall et al.'s 
(filled green squares), and Chen et al.'s relations.
Thus, we again confirm that our adopted values of $d=3.4$~kpc and 
$E(B-V)=0.90$ are reasonable.

We plot the $(B-V)_0$-$(M_V-2.5 \log f_{\rm s})$ diagram of V459~Vul in Figure
\ref{hr_diagram_v459_vul_v5579_sgr_v2670_oph_qy_mus_outburst}(a)
for $E(B-V)=0.90$ and $(m-M')_V=15.05$ in Equation
(\ref{absolute_mag_v459_vul}).
Basically, the error of the $B-V$ data was not reported for the VSOLJ
and AAVSO data.  So, we cannot select $(B-V)_0$ data with high accuracy
from them.   The CBET data are on the LV Vul track (orange line).
The AAVSO data are consistent with the LV Vul track.  The redder side
of the $(B-V)_0$ data distribution including CBET data almost follows
the track of LV~Vul.  This suggests a rough overlapping
of the V459~Vul track with the LV~Vul track
in the $(B-V)_0$-$(M_V-2.5 \log f_{\rm s})$ diagram.
This rough agreement of the redder side of the $(B-V)_0$ distribution
with the track of LV~Vul may support that
$E(B-V)=0.90$ and $(m-M')_V=15.05$ are reasonable, i.e., 
$E(B-V)=0.90\pm0.05$, $(m-M)_V=15.45\pm0.1$, $f_{\rm s}=0.71$,
and $d=3.4\pm0.5$~kpc.

We check the distance modulus of $(m-M)_V=15.45$ by comparing
our model $V$ light curve with the observation.
Assuming that $(m-M)_V=15.45$,
we plot the $V$ light curve of our $1.15~M_\sun$ WD 
\citep[Ne2;][]{hac10k} in Figure
\ref{v459_vul_v533_her_v1668_cyg_lv_vul_v_bv_ub_logscale}(a).
The model light curve fits well the $V$ magnitude of V459~Vul,
suggesting that our adopted value of $(m-M)_V=15.45$ is reasonable.

\subsection{V5579~Sgr 2008}
\label{v5579_sgr_cmd}
The nova reached $m_{V, \rm max}=6.6$ at maximum on JD~2,454,579.9 
from the AAVSO data.   \citet{rus08b} identified the nova as being of
\ion{Fe}{2} class.  They also estimated the reddening of the nova
to be $E(B-V)\sim1.2$ based on the \ion{O}{1} lines of their infrared
spectrum obtained on UT 2008 May 9 (JD~2,454,595.5), and
suggested that a part of the reddening may be local to the nova
because the nova already formed dust (denoted by the arrow labeled
``dust'' in Figure 
\ref{hr_diagram_v459_vul_v5579_sgr_v2670_oph_qy_mus_outburst}(b)).
\citet{raj11} reported their NIR spectroscopy and photometry.
They estimated the reddening to be $E(B-V)=0.72\pm0.06$ from
$(B-V)_{0,\rm max}=0.23\pm0.06$ \citep{van87}, the maximum brightness to
be $M_{V,\rm max}=-8.8$ from the MMRD relation of \citet{del95}
together with $t_2=8\pm0.5$~days, and the distance to be $d=4.4\pm0.2$~kpc.
The VVV catalog \citep{sai13} gives $E(B-V)=A_{K_s}/0.36=0.237/0.36=0.65$
toward V5579~Sgr.

We obtain 
$(m-M)_B= 16.78$, $(m-M)_V= 15.96$, and $(m-M)_I= 14.67$,
which cross at $d=4.8$~kpc and $E(B-V)=0.82$,
in Appendix \ref{v5579_sgr} and plot them in Figure
\ref{distance_reddening_v459_vul_v5579_sgr_v2670_oph_qy_mus}(b).
Thus, we have $E(B-V)=0.82\pm0.05$ and $d=4.8\pm0.5$~kpc.

For the reddening toward V5579~Sgr, $(l,b)=(3\fdg7342, -3\fdg0225)$,
the NASA/IPAC absorption 2D map gives $E(B-V)=0.86\pm0.04$.
This value is roughly consistent with our result of $E(B-V)=0.82\pm0.05$.
\citet{rus08b} obtained $E(B-V)\sim1.2$ from the \ion{O}{1} lines.
But, they suggested that a part of the reddening is local to the nova.
\citet{raj11} estimated the reddening to be $E(B-V)=0.72\pm0.06$ from
$(B-V)_{0,\rm max}=0.23\pm0.06$ \citep{van87}, which is
slightly smaller than our result.
However, if we adopt $B-V= 1.21$ on UT 2008 April 23.12 \citep{mun08c},
we obtain the reddening of $E(B-V)=0.98\pm0.06$.  The arithmetic average
of these two estimates, $E(B-V)=(0.72 + 0.98)/2= 0.85$, is roughly   
consistent with our value of $E(B-V)=0.82\pm0.05$.
The VVV catalog \citep{sai13} gives $E(B-V)=A_{K_s}/0.36=0.237/0.36=0.65$
toward V5579~Sgr, which is slightly smaller than our value.

We further check our result
in Figure \ref{distance_reddening_v459_vul_v5579_sgr_v2670_oph_qy_mus}(b).
We plot four relations of \citet{mar06} toward
$(l, b)=(3\fdg50, -3\fdg00)$, $(3\fdg75,-3\fdg00)$,
$(3\fdg50,-3\fdg25)$, and $(3\fdg75,-3\fdg25)$.
The closest direction is that of the filled green squares.
The other symbols/lines have the same meanings as those in Figure
\ref{distance_reddening_v1663_aql_v5116_sgr_v2575_oph_v5117_sgr}(a).
Our crossing point of $d=4.8$~kpc and $E(B-V)=0.82$ slightly deviates
from Green et al.'s results but is roughly consistent with
a linear extension of \"Ozd\"ormez et al.'s relation.
These 3D dust maps give the average value of a relatively broad region.
The pinpoint reddening could be different from it because
their resolutions are considerably larger than those of the molecular
cloud structures.

We plot the $(B-V)_0$-$(M_V-2.5 \log f_{\rm s})$ diagram of V5579~Sgr in Figure
\ref{hr_diagram_v459_vul_v5579_sgr_v2670_oph_qy_mus_outburst}(b)
for $E(B-V)=0.82$ and $(m-M')_V=16.65$ in Equation
(\ref{absolute_mag_v5579_sgr}).
The data are scattered but the redder side distribution almost
follows the track of V1668 Cyg before it abruptly goes down due to
dust blackout.   This overlapping of the distribution of the redder side 
of V5579~Sgr with the track of V1668~Cyg may support
$(B-V)=0.82$ and $(m-M')_V=16.65$,
that is, $E(B-V)=0.82\pm0.05$, $(m-M)_V=15.95\pm0.1$,
$f_{\rm s}=1.91$, and $d=4.8\pm0.5$~kpc.

We check our distance modulus of $(m-M)_V=15.95$ by comparing our model
light curve with the observation.  In Figure 
\ref{v5579_sgr_lv_vul_v1668_cyg_v1535_sco_v_bv_ub_color_logscale}(a),
we plot a $0.85~M_\sun$ WD \citep[CO3, solid black lines;][]{hac16k},
assuming that $(m-M)_V=15.95$.
The model $V$ light curve is calculated based on the free-free emission
plus blackbody emission as explained in Sections \ref{introduction}
and \ref{method_example}.  Because the model does not
include the effects of dust absorption and strong emission lines,
it cannot correctly follow the dust blackout and nebular phases in Figure 
\ref{v5579_sgr_lv_vul_v1668_cyg_v1535_sco_v_bv_ub_color_logscale}.
The model light curves fit well the $V$ magnitude of V5579~Sgr in the
early phase before the dust blackout started,
suggesting that $(m-M)_V=15.95$ is reasonable.
For comparison, we also add $0.85~M_\sun$ WD (CO4, solid red line)
and $0.98~M_\sun$ WD (CO3, solid green lines),
assuming that $(m-M)_V=18.3$ and $(m-M)_V=14.6$, respectively, 
for V1535~Sco and V1668~Cyg.

\subsection{V2670~Oph 2008\#1}
\label{v2670_oph_cmd}
The nova reached $m_{V,\rm max}=10.1$ on UT 2008 May 26 \citep{pog08b}.
\citet{rus08c} identified the nova as probably an \ion{Fe}{2} type,
which was later confirmed by \citet{hel08} and \citet{pog08b}.
\citet{rus08c} and \citet{sit08} reported that there was no evidence
for dust formation.  \citet{rus08c} obtained the reddening of
$E(B-V)\sim1.3$ from \ion{O}{1} lines based on their $0.8$--$5.5~\mu$m
spectroscopy on UT 2008 June 14.
\citet{pog08b} estimated the absolute magnitude at maximum to be
$M_{V,\rm max}=-7.9$ to $-7.4$ from the MMRD relation \citep{del95}
together with $t_3=42\pm2$~days and the distance to be $d=4.7$--$5.8$~kpc
together with $E(B-V)=1.3$.
The VVV catalog gives $E(B-V)=A_{K_s}/0.36=0.356/0.36=0.98$ toward
V2670~Oph \citep{sai13}.

We obtain $(m-M)_B= 18.65$, $(m-M)_V= 17.6$, and $(m-M)_I= 15.92$,
which cross at $d=7.4$~kpc and $E(B-V)=1.05$,
in Appendix \ref{v2670_oph} and plot them in Figure
\ref{distance_reddening_v459_vul_v5579_sgr_v2670_oph_qy_mus}(c).
Thus, we have $E(B-V)=1.05\pm0.1$ and $d=7.4\pm0.8$~kpc.

For the reddening toward V2670~Oph, $(l,b)=(3\fdg6656, +3\fdg7799)$,
the NASA/IPAC galactic absorption 2D map gives $E(B-V)=0.90\pm0.01$ and
the VVV catalog gives $E(B-V)= 0.98$ \citep{sai13}.
On the other hand, \citet{rus08c} obtained $E(B-V)\sim1.3$
from the \ion{O}{1} lines as mentioned above.
The arithmetic average of these three values is 
$E(B-V) = (0.90 + 0.98 + 1.3)/3= 1.06$, which is
consistent with our value of $E(B-V) = 1.05$ at the crossing point.  
Figure \ref{distance_reddening_v459_vul_v5579_sgr_v2670_oph_qy_mus}(c)
shows various distance-reddening relations.
We plot four relations of \citet{mar06} toward
$(l, b)=(3\fdg50, 3\fdg75)$, $(3\fdg75, 3\fdg75)$,
$(3\fdg50, 4\fdg00)$, and $(3\fdg75, 4\fdg00)$.
The closest direction is that of the filled green squares.
The other symbols/lines have the same meanings as those in Figure
\ref{distance_reddening_v1663_aql_v5116_sgr_v2575_oph_v5117_sgr}(a).
Our crossing point of $d=7.4\pm0.8$~kpc and $E(B-V)=1.05\pm0.1$
is consistent with Green et al.'s (2018) orange line and
\"Ozd\"ormez et al.'s result.

We plot the $(B-V)_0$-$(M_V-2.5 \log f_{\rm s})$ diagram of V2670~Oph in Figure
\ref{hr_diagram_v459_vul_v5579_sgr_v2670_oph_qy_mus_outburst}(c)
for $E(B-V)=1.05$ and $(m-M')_V=18.45$ in Equation
(\ref{absolute_mag_v2670_oph}).
Although the number of data points are few, the track follows
that of V1500~Cyg (solid green line) in the later phase.
Therefore, we regard V2670~Oph to belong to the V1500~Cyg type 
because its light curve is similar to that of QU Vul and NR TrA
(see Figures \ref{v2670_oph_qu_vul_x55z02c10o10_logscale}
and \ref{v2670_ooph_qu_vul_nr_tra_b_v_i_logscale_3fig})
and these two novae also belong to the V1500 Cyg type.
The SMARTS spectra \citep{wal12} showed that the nova had already entered
the nebular phase on UT 2008 June 20 (JD~2,454,637.5) at $m_V=17.4$.
Therefore, we suppose that the nebular phase began at the turning 
point denoted by the arrow labeled ``nebular phase'' in Figure
\ref{hr_diagram_v459_vul_v5579_sgr_v2670_oph_qy_mus_outburst}(c).

We check our distance modulus of $(m-M)_V=17.6$ by comparing our model
light curve with the observation.  
Assuming that $(m-M)_V=17.6$, we plot the $V$ (free-free plus blackbody
emission) light curves of a $0.80~M_\sun$ WD model 
(CO3, solid green/red lines) in Figures
\ref{v2670_oph_qu_vul_x55z02c10o10_logscale}(a) and
\ref{v2670_oph_v1668_cyg_lv_vul_v_bv_ub_color_logscale}(a),
respectively.  The model light curve follows well the observed $V$
light curve of V2670~Oph, confirming that $(m-M)_V=17.6$ is reasonable.
In Figure \ref{v2670_oph_v1668_cyg_lv_vul_v_bv_ub_color_logscale}(a),
for comparison, we plot the $V$ and UV 1455\AA\  light curves of
a $0.98~M_\sun$ WD (CO3, solid green lines),
assuming that $(m-M)_V=14.6$ for V1668~Cyg.

\subsection{QY~Mus 2008}
\label{qy_mus_cmd}
The nova reached $m_{V, \rm max}=8.6$ at maximum on JD~2,454,748.9
from the VSOLJ data.  The SMARTS spectra showed that QY~Mus
is a typical \ion{Fe}{2}-type nova \citep{lil09}.

We obtain $(m-M)_B=15.23$, $(m-M)_V=14.67$, and $(m-M)_I=13.74$,
which cross at $d=3.7$~kpc and $E(B-V)=0.58$,
in Appendix \ref{qy_mus} and plot them in Figure 
\ref{distance_reddening_v459_vul_v5579_sgr_v2670_oph_qy_mus}(d). 
Thus, we obtain $d=3.7\pm0.4$~kpc, $E(B-V)=0.58\pm0.05$,
$(m-M)_V=14.65\pm0.1$, and $f_{\rm s}=2.2$ against LV~Vul.

For the reddening toward QY~Mus, \\$(l,b)=(305\fdg3330, -4\fdg8590)$,
the 2D NASA/IPAC galactic dust absorption map gives $E(B-V)=0.58\pm0.03$,
which is consistent with our result.  We further check our result
in Figure
\ref{distance_reddening_v459_vul_v5579_sgr_v2670_oph_qy_mus}(d).
We plot four relations of \citet{mar06} toward
$(l, b)=(305\fdg25, -4\fdg75)$, $(305\fdg50, -4\fdg75)$,
$(305\fdg25, -5\fdg00)$, and $(305\fdg50, -5\fdg00)$.
The closest direction is that of the unfilled red squares.
The other symbols/lines have the same meanings as those in Figure
\ref{distance_reddening_v1663_aql_v5116_sgr_v2575_oph_v5117_sgr}(a).
Our crossing point of $d=3.7$~kpc and $E(B-V)=0.58$
is consistent with Marshall et al.'s,
\"Ozd\"ormez et al.'s, and Chen et al.'s relations.

We plot the $(B-V)_0$-$(M_V-2.5 \log f_{\rm s})$ diagram of QY~Mus in Figure
\ref{hr_diagram_v459_vul_v5579_sgr_v2670_oph_qy_mus_outburst}(d)
for $E(B-V)=0.58$ and $(m-M')_V=15.55$ in Equation
(\ref{absolute_mag_qy_mus}).
The track of QY~Mus almost follows that of LV~Vul.
Therefore, we identify QY~Mus as an LV~Vul type. 
This matching with the LV~Vul track also supports our results
of $E(B-V)=0.58$ and $(m-M')_V=15.55$, that is, 
$E(B-V)=0.58\pm0.05$, $(m-M)_V=14.65\pm0.1$, $f_{\rm s}=2.2$,
and $d=3.7\pm0.4$~kpc.
The SMARTS spectra of QY~Mus showed that the nova had entered the
nebular phase on UT 2009 May 12 (JD~2,454,963.5) at $m_V=12.5$.
We plot the start of the nebular phase by a large unfilled red square
in Figure
\ref{hr_diagram_v459_vul_v5579_sgr_v2670_oph_qy_mus_outburst}(d).

We check the distance modulus of $(m-M)_V=14.65$ by comparing
our $V$ model light curve with the observation.  We also discuss
the accuracy of the WD mass determination from the light curve modeling
in Section \ref{wd_mass_determination}.
Our model light-curve fitting gives the WD mass of
$M_{\rm WD}=0.75$, $0.80$, and $0.85~M_\sun$ for the hydrogen content of
$X=0.35$, $0.45$, and $0.55$ by weight, respectively.  
It should also be noted that our model light
curves reasonably reproduce the $V$ light curve of QY~Mus.
This confirms that the distance modulus of $(m-M)_V=14.65$ is
reasonable.


\begin{figure*}
\plotone{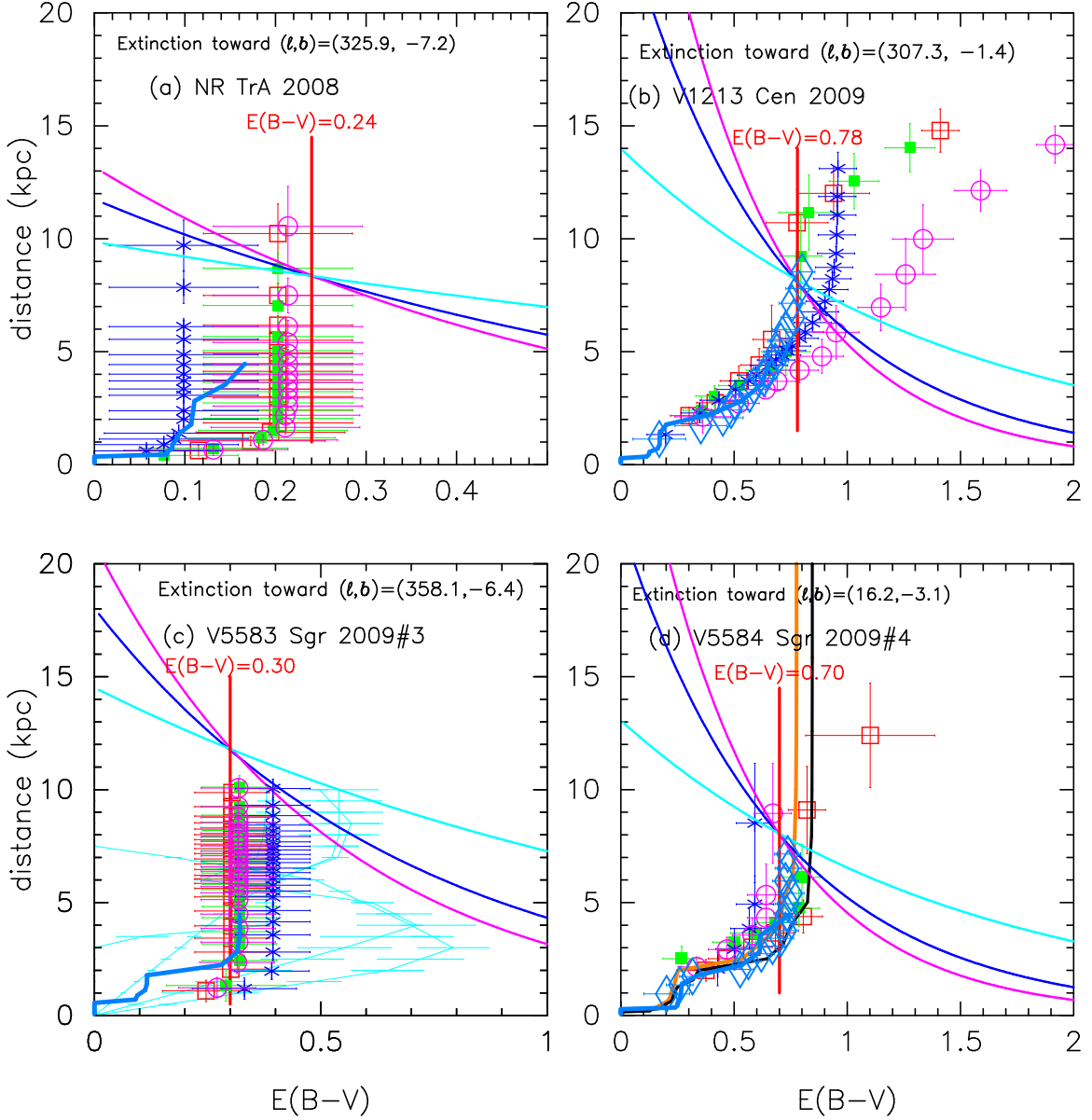}
\caption{
Same as Figure 
\ref{distance_reddening_v1663_aql_v5116_sgr_v2575_oph_v5117_sgr},
but for (a) NR~TrA, (b) V1213~Cen, (c) V5583~Sgr, and (d) V5584~Sgr.
\label{distance_reddening_nr_tra_v1213_cen_v5583_sgr_v5584_sgr}}
\end{figure*}


\begin{figure*}
\plotone{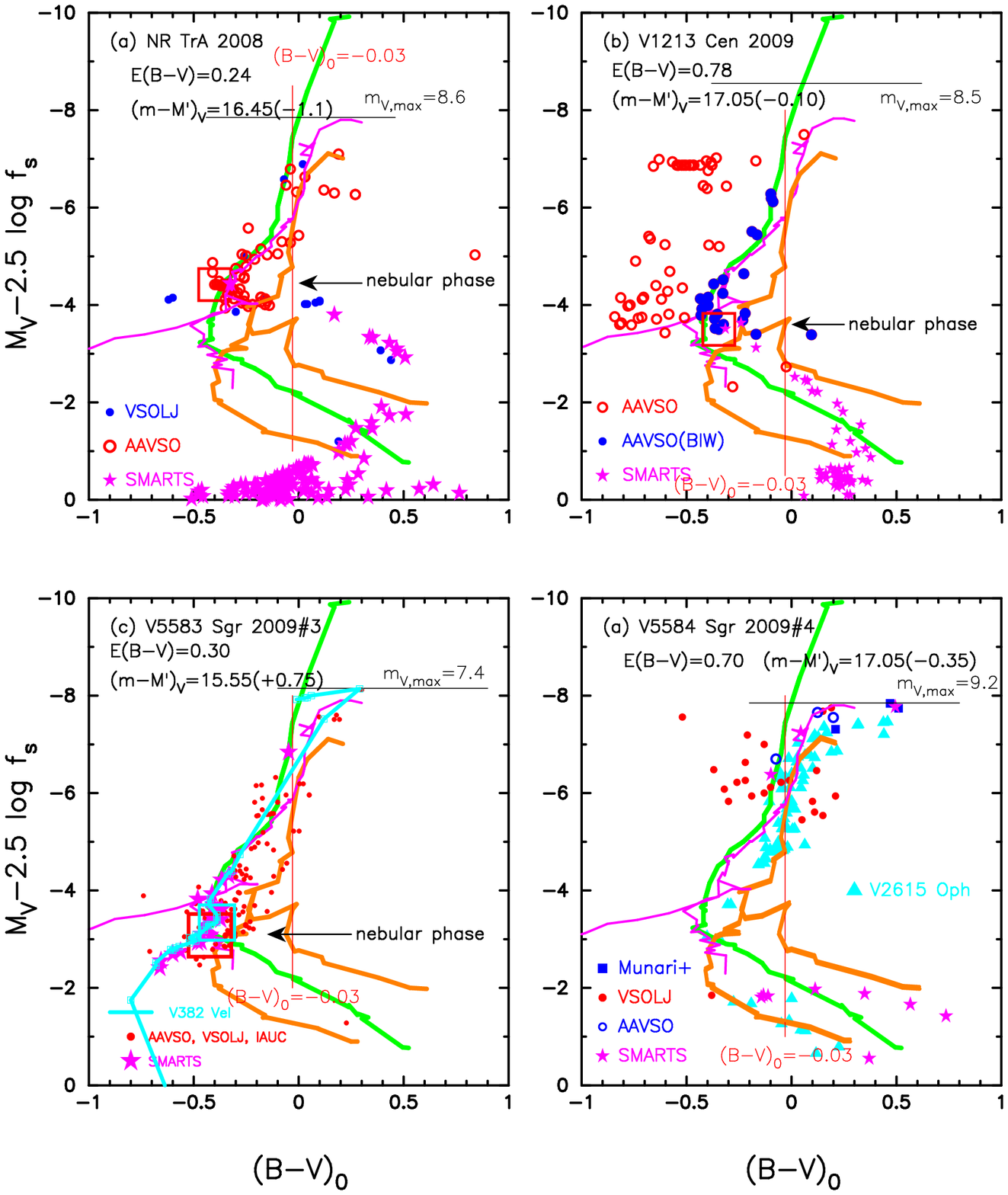}
\caption{
Same as Figure 
\ref{hr_diagram_v1663_aql_v5116_sgr_v2575_oph_v5117_sgr_outburst},
but for (a) NR~TrA, (b) V1213~Cen, (c) V5583~Sgr,
and (d) V5584~Sgr.  The solid green lines show the template track
of V1500~Cyg.  The solid orange lines represent that of LV~Vul.
We add the track of V1974~Cyg (solid magenta lines).
In panel (c), we add the tracks of V382~Vel (cyan line).
In panel (d), we add the track of V2615~Oph (filled cyan triangles).
\label{hr_diagram_nr_tra_v1213_cen_v5583_sgr_v5584_sgr_outburst}}
\end{figure*}

\subsection{NR~TrA 2008}
\label{nr_tra_cmd}
NR~TrA was discovered by N. J. Brown at mag 9.2 
on UT 2008 April 1.73 \citep[$=$JD~2,454,558.23;][]{bro08}.
It reached $m_{V, \rm max}=8.6$ at maximum on JD~2,454,570.15
and showed a 2.5 mag largest secondary peak and then a few
smaller secondary peaks with half a magnitude.  This feature of
multiple secondary peaks is similar to that of V2670~Oph in
Section \ref{v2670_oph_cmd}.  The novae is in the nebular phase
at least until Day 1237 \citep{wal15}.  \citet{wal15} also reported
that the nova light curve showed an eclipse of 5.25~hr binary period
and obtained the mass function of $M_2^3/(M_1+M_2)^2= 0.024~M_\sun$.

We obtain $(m-M)_B= 15.6$, $(m-M)_V= 15.35$, and $(m-M)_I=14.97$, which
cross at the distance of $d=8.3$~kpc and the reddening of $E(B-V)=0.24$,
in Appendix \ref{nr_tra} and plot them in Figure
\ref{distance_reddening_nr_tra_v1213_cen_v5583_sgr_v5584_sgr}(a).
Thus, we obtain $d=8.3\pm1.0$~kpc, $E(B-V)=0.24\pm0.05$,
$(m-M)_V=15.35\pm0.2$, and $f_{\rm s}=2.7$ against LV~Vul.

For the reddening toward NR~TrA, \\$(l,b) = (325\fdg9316,  -7\fdg2184)$,
the 2D NASA/IPAC galactic dust absorption map gives $E(B-V)=0.19\pm0.01$,
which is consistent with our value of $E(B-V)=0.24\pm0.05$ 
at the crossing point.  We check our result
in Figure
\ref{distance_reddening_nr_tra_v1213_cen_v5583_sgr_v5584_sgr}(a).
We plot four relations of \citet{mar06} toward
$(l, b)=(325\fdg75, -7\fdg00)$, $(326\fdg00, -7\fdg00)$,
$(325\fdg75, -7\fdg25)$, and $(326\fdg00, -7\fdg25)$.
The closest direction is that of the unfilled magenta circles.
The other symbols/lines have the same meanings as those in Figure
\ref{distance_reddening_v1663_aql_v5116_sgr_v2575_oph_v5117_sgr}(a).
Our crossing point of $d=8.3$~kpc and $E(B-V)=0.24$
is consistent with Marshall et al.'s relation (unfilled magenta circles with
error bars).
\citet{wal15} discussed his spectra and concluded that $E(B-V) < 0.3$
if the blue continuum arose from the accretion disk.  Our value 
of $E(B-V)=0.24$ is also consistent with his estimate.
 
We plot the $(B-V)_0$-$(M_V-2.5 \log f_{\rm s})$ diagram of NR~TrA in Figure
\ref{hr_diagram_nr_tra_v1213_cen_v5583_sgr_v5584_sgr_outburst}(a)
for $E(B-V)=0.24$ and $(m-M')_V=16.45$ in Equation
(\ref{absolute_mag_nr_tra}).
The track is close to that of V1974~Cyg (and V1500~Cyg) until
the nebular phase starts.
Therefore, we regard NR~TrA to belong to the V1500~Cyg type.
The broadband filter light curves are contaminated by strong emission
lines in the nebular phase.  This effect results in a large difference
especially in the $(B-V)_0$ color, because the color is the difference
between two magnitudes and a strong line usually contributes to one of
the bands.  Similarly, a slight difference in the filter response makes
a large difference in the $(B-V)_0$ color.  We sometimes find large
scatters in the $(B-V)_0$ color among various observers.
The effect of strong [\ion{O}{3}] lines could be
slightly different from nova to nova, even if they are globally similar
in their light curves.  This means that, even if their early 
color-magnitude tracks overlap the template tracks, 
the later tracks in the nebular phase separate from the template tracks.
The early-phase data of NR~TrA follows well the V1974~Cyg track.
The SMARTS spectra of NR~TrA showed that the nova had entered the
nebular phase on UT 2008 August 26 (JD~2,454,704.5) at $m_V=11.9$.
We denote the start of the nebular phase with a large unfilled red square
at $M_V= m_V - (m-M)_V= 11.9 - 15.35 = -3.45$ and $(B-V)_0= (B-V) -
E(B-V)=-0.16 -0.24=-0.40$. 
We may conclude that the track of NR~TrA overlaps
the V1974~Cyg track except for the nebular phase.
The similarity of NR~TrA to V1974~Cyg in the $(B-V)_0$-$(M_V-2.5 
\log f_{\rm s})$ diagram may support our obtained values,
$E(B-V)=0.24$ and $(m-M')_V=16.45$, that is, 
$E(B-V)=0.24\pm0.05$, $(m-M)_V=15.35\pm0.2$, $f_{\rm s}=2.7$,
$d=8.3\pm1.0$~kpc.

We check our distance modulus of $(m-M)_V=15.35$ by comparing
our model light curve with the observation.  
Assuming that $(m-M)_V=15.35$, we plot the $V$ light curves of 
a $0.75~M_\sun$ WD model (CO3, solid green/red line) in Figures
\ref{nr_tra_v2670_oph_qu_vul_x55z02c10o10_logscale}(a) and
\ref{nr_tra_v1668_cyg_lv_vul_v_bv_ub_color_logscale}(a), respectively.
The distance modulus of $(m-M)_V=15.35$ shows 
good agreement with the observed $V$ light curve of NR~TrA,
confirming that $(m-M)_V=15.35$ is reasonable.
For comparison, assuming that $(m-M)_V=14.6$ for V1668~Cyg,
we plot the $V$ and UV 1455\AA\  light curves of a $0.98~M_\sun$ WD model
(CO3, solid green lines) 
in Figure \ref{nr_tra_v1668_cyg_lv_vul_v_bv_ub_color_logscale}(a).
If we adopt the WD mass of $0.75~M_\sun$, then the companion mass is
about $M_2=0.30~M_\sun$ from the mass function of $M_2^3/(M_1+M_2)^2=
0.024~M_\sun$ obtained by \citet{wal15}.  The Roche lobe radius of
the donor star is $R_2= 0.47~R_\sun$, slightly evolving from the main
sequence.

\subsection{V1213~Cen 2009}
\label{v1213_cen_cmd}
V1213~Cen was discovered on UT
2009 May 8.235 at $m_V=8.53$ by ASAS-3 \citep{poj09}.
Then, the nova declined with $t_2=11\pm2$~days and
became a supersoft X-ray source on UT 2010 March 29.6 \citep{schw10}.
\citet{schw10} obtained a blackbody fit with $kT = 31\pm4$~eV and
$N_{\rm H} = (6.4\pm1.8)\times10^{21}$~cm$^{-2}$.
We assume that $m_{V, \rm max}=8.5$.
We estimate the reddening to be $E(B-V)= N_{\rm H}/8.3\times 10^{21}
=(6.4\pm1.8)\times10^{21}/ 8.3\times 10^{21}= 0.77\pm0.22$,
based on the relation of \citet{lis14}.  

We obtain $(m-M)_B= 17.74$, $(m-M)_V= 16.95$, and $(m-M)_I= 15.75$, 
which cross at $d=8.1$~kpc and $E(B-V)=0.78$,
in Appendix \ref{v1213_cen} and plot them in Figure 
\ref{distance_reddening_nr_tra_v1213_cen_v5583_sgr_v5584_sgr}(b). 
Thus, we obtain $d=8.1\pm1$~kpc, $E(B-V)=0.78\pm0.05$,
$(m-M)_V=15.95\pm0.1$, and $f_{\rm s}=1.12$ against LV~Vul.

For the reddening toward V1213~Cen, $(l,b)=(307\fdg2862, -1\fdg4263)$,
the hydrogen column density obtained by \citet{schw10} suggests
$E(B-V)= 0.77\pm0.22$, consistent with our crossing point, but
the 2D NASA/IPAC galactic dust absorption map gives $E(B-V)=1.77\pm0.14$,
which is much larger than our results.  This implies that the reddening
does not saturate yet at the distance of $d = 8.1$~kpc as shown in
Figure \ref{distance_reddening_nr_tra_v1213_cen_v5583_sgr_v5584_sgr}(b).
We further check our result
in Figure
\ref{distance_reddening_nr_tra_v1213_cen_v5583_sgr_v5584_sgr}(b).
We plot four relations of \citet{mar06} toward
$(l, b)=(307\fdg25, -1\fdg50)$, $(307\fdg50, -1\fdg50)$,
$(307\fdg25, -1\fdg25)$, and $(307\fdg50, -1\fdg25)$.
The closest direction is that of the unfilled red squares.
The other symbols/lines have the same meanings as those in Figure
\ref{distance_reddening_v1663_aql_v5116_sgr_v2575_oph_v5117_sgr}(a).
Our crossing point of $d=8.1$~kpc and $E(B-V)=0.78$
is consistent with Marshall et al.'s (unfilled red squares with error bars),
\"Ozd\"ormez et al.'s, and Chen et al.'s relations.

We plot the $(B-V)_0$-$(M_V-2.5 \log f_{\rm s})$ diagram of V1213~Cen in Figure
\ref{hr_diagram_nr_tra_v1213_cen_v5583_sgr_v5584_sgr_outburst}(b)
for $E(B-V)=0.78$ and $(m-M')_V=17.05$ in Equation
(\ref{absolute_mag_v1213_cen}).
We plot the data points of AAVSO and SMARTS, but separately depict
BIW's data (filled blue circles) from the other AAVSO data (unfilled
red circles), because BIW's data are systematically redder than
the others and consistent with the SMARTS data.  If we use BIW's data,
the track of V1213~Cen is close to that of V1500~Cyg and V1974~Cyg,
so we regard V1213~Cen to belong to the V1500~Cyg type.
The SMARTS spectra \citep{wal12} of V1213~Cen showed that the nova
had already entered the nebular phase on UT 2009 December 20 
(JD~2,455,185.5) at $m_V=14.5$.
We regard the nova to have entered the nebular phase soon after
UT 2009 September 13 (JD~2,455,087.5) at $m_V=13.6$.
We denote the start of the nebular phase by a large unfilled red square
at $M_V= m_V - (m-M)_V= 13.6 - 16.95 = -3.35$ and $(B-V)_0= (B-V) -
E(B-V)=0.43 - 0.78=-0.35$.
The overlapping of V1213~Cen with the V1500~Cyg track may support
that our adopted values of $E(B-V)=0.78$ and $(m-M')_V=17.05$ are reasonable,
that is, $E(B-V)=0.78\pm0.05$,  $(m-M)_V=16.95\pm0.2$, 
$f_{\rm s}=1.12$, and $d=8.1\pm1$~kpc.

We check the distance modulus of $(m-M)_V=16.95$ by comparing
our model $V$ light curve with the observation.
Assuming the distance modulus of $(m-M)_V=16.95$,
we plot the $V$ light curve of our $1.0~M_\sun$ WD 
\citep[Ne2, solid blue lines;][]{hac10k} in Figure 
\ref{v1213_cen_lv_vul_v4743_sgr_v_bv_ub_color_curve_logscale}(a).
The model light curve fits the $V$ magnitude of V1213~Cen
as well as the supersoft X-ray light curve,
suggesting that our adopted value of $(m-M)_V=16.95$ and
the WD mass of $1.0~M_\sun$ are reasonable.
For comparison, we also plot the $V$ and supersoft X-ray light curves
of V4743~Sgr and compare our model light curves of a $1.15~M_\sun$ WD
(Ne2, solid red lines).

\subsection{V5583~Sgr 2009\#3}
\label{v5583_sgr_cmd}
The light curve obtained with
NASA's {\it Solar TErrestrial RElations Observatory}/Heliospheric Imager
({\it STEREO}/HI) instruments constrained the optical peak
at JD~2,455,050.54$\pm0.17$ \citep[UT 2009 August 7.04;][]{hol14}.
This time is almost the same as the peak time of JD~2,455,050.6$\pm0.04$
(UT 2009 August 7.08) determined from the {\it SMEI} light curve \citep{hou16}.
Therefore, we regard the $V$ maximum to be $m_{V, \rm max}=7.43$
on JD~2,455,050.635 from the VSOLJ data (observed by H. Maehara).
\citet{hol14} suggested that the nova is most likely a neon nova,
based on their optical spectroscopy and abundance analysis.
\citet{hac09kkm} analyzed the early-phase $BVyR_{\rm C}I_{\rm C}$
light curves and estimated the WD mass and supersoft X-ray source phase
based on their model light curves.  They concluded that the light curves
of V5583~Sgr is almost identical with those of V382~Vel.

We obtain $(m-M)_B= 16.6$, $(m-M)_V= 16.28$, and $(m-M)_I= 15.79$,
which cross at $d=12$~kpc and $E(B-V)=0.30$,
in Appendix \ref{v5583_sgr} and plot them in Figure
\ref{distance_reddening_nr_tra_v1213_cen_v5583_sgr_v5584_sgr}(c).
Thus, we have $E(B-V)=0.30\pm0.05$ and $d=12\pm2$~kpc.

For the reddening toward V5583~Sgr, 
$(l,b)=(358\fdg1004, -6\fdg3941)$,
the 2D NASA/IPAC galactic dust absorption map gives $E(B-V)=0.337\pm0.005$.
The VVV survey catalog \citep{sai13} gives $E(B-V)=A_{K_s}/0.36
=0.107/0.36=0.30$ toward V5583~Sgr.  Both values of reddening are
consistent with our crossing point of $E(B-V)=0.30\pm0.05$ 
and $d=12\pm2$~kpc.
We further check our result
in Figure \ref{distance_reddening_nr_tra_v1213_cen_v5583_sgr_v5584_sgr}(c).
We plot four relations of \citet{mar06} toward 
$(l, b)=(358\fdg00, -6\fdg50)$, $(358\fdg25, -6\fdg50)$,
$(358\fdg00, -6\fdg25)$, and $(358\fdg25, -6\fdg25)$.
The closest direction is that of the unfilled red squares.
Unfortunately, Green et al.'s (2015, 2018) data are not available for this
direction.
The other symbols/lines have the same meanings as those in Figure
\ref{distance_reddening_v1663_aql_v5116_sgr_v2575_oph_v5117_sgr}(a).
Our crossing point of $d=12$~kpc and $E(B-V)=0.30$
is consistent with Marshall et al.'s relation (unfilled red squares)
and Chen et al.'s relation.

We plot the $(B-V)_0$-$(M_V-2.5 \log f_{\rm s})$ diagram of V5583~Sgr in Figure
\ref{hr_diagram_nr_tra_v1213_cen_v5583_sgr_v5584_sgr_outburst}(c)
for $E(B-V)=0.30$ and $(m-M')_V=15.55$ in Equation
(\ref{absolute_mag_v5583_sgr}).
We add the track of V382~Vel (solid cyan line),
the data of which are the same as those in \citet{hac19k}.
The data of V5583~Sgr are scattered in the early phase but its track
is very similar to that of V382~Vel. 
Therefore, we can conclude that V5583~Sgr belongs to the V1500~Cyg type
in the $(B-V)_0$-$(M_V-2.5 \log f_{\rm s})$ diagram
because V382~Vel belongs to the V1500~Cyg type \citep{hac19k}.
This overlapping with the V382~Vel, V1974~Cyg, and V1500~Cyg tracks
supports $E(B-V)=0.30$ and $(m-M')_V=15.55$,
that is, $E(B-V)=0.30\pm0.05$, $(m-M)_V=16.3\pm0.1$,
$f_{\rm s}=0.51$, and $d=12\pm2$~kpc.

The SMARTS spectra of V5583~Sgr showed that the nova had already entered
the nebular phase on UT 2009 October 2 (JD~2,455,106.5) at $m_V=12.5$.
We regard the nova to have entered the nebular phase soon after
UT 2009 September 13 (JD~2,455,087.5) at $m_V=12.5$.
We denote the start of the nebular phase by a large unfilled red square
at $M_V= m_V - (m-M)_V= 12.5 - 16.3 = -3.8$ and $(B-V)_0= (B-V) -
E(B-V)=-0.123 - 0.30=-0.42$.   This phase is also similar to that of
V382~Vel (a large unfilled cyan square).  

We check the distance modulus of $(m-M)_V=16.3$ by comparing
our model $V$ light curve with the observation.
Assuming $(m-M)_V=16.3$, we plot a model $V$ light curve of a $1.23~M_\sun$
WD \citep[Ne2, solid black line;][]{hac10k, hac16k} for V5583~Sgr
and V382~Vel in Figure
\ref{v5583_sgr_v382_vel_v_bv_ub_x55z02o10ne03_logscale}(a).  
We also plot the supersoft X-ray flux (solid black line: 0.1--2.4~keV)
of the $1.23~M_\sun$ WD model.  Both the $V$ and supersoft X-ray
light curve models reasonably reproduce the observation.
We show the X-ray flux for V382~Vel in units of erg~s$^{-1}$~cm$^{-2}$
but only the count rate in units of cts~s$^{-1}$ for V5583~Sgr
(scaled from 100 at the upper bound to 10$^{-4}$ at the lower bound).
The X-ray flux of V382~Vel is taken from \citet{ori02} and
\citet{bur02} while the X-ray data of V5583~Sgr are from
the {\it Swift} website \citep{eva09}.  
This confirms that the distance modulus of $(m-M)_V=16.3$ is reasonable.
The WD mass of $1.23~M_\sun$ is consistent with the neon nova
suggestion by \citet{hol14}. 
For comparison, we add a light curve of a $0.98~M_\sun$ WD 
\citep[CO3, solid green line;][]{hac16k} for LV~Vul.

\subsection{V5584~Sgr 2009\#4}
\label{v5584_sgr_cmd}
V5584~Sgr reached $m_{V, \rm max}= 9.2$ at $V$ maximum on JD~2,455,134.7
\citep[UT 2009 October 29;][]{pog11}.
\citet{pog11} estimated $t_2=27\pm2$ days and the distance of
5.1--7.8~kpc based on various MMRD relations.
\citet{mun09} suggested that V5584~Sgr is an \ion{Fe}{2}-type nova based
on their spectra obtained on UT October 28 and 29.
\citet{rus10} reported that V5584~Sgr had already formed dust 
on UT 2010 February 10. 

We obtain $(m-M)_B= 17.39$, $(m-M)_V= 16.69$, and $(m-M)_I= 15.58$,
which cross at $d=8.0$~kpc and $E(B-V)=0.70$,
in Appendix \ref{v5584_sgr} and plot them in Figure
\ref{distance_reddening_nr_tra_v1213_cen_v5583_sgr_v5584_sgr}(d).
Thus, we have $E(B-V)=0.70\pm0.05$ and $d=8.0\pm1$~kpc.
The distance modulus in the $V$ band is $(m-M)_V=16.7\pm0.1$.

For the reddening toward V5584~Sgr, $(l,b)=(16\fdg1682, -3\fdg1003)$,
the 2D NASA/IPAC galactic dust absorption map gives $E(B-V)=0.91\pm0.08$.
We further check our result
 in Figure
\ref{distance_reddening_nr_tra_v1213_cen_v5583_sgr_v5584_sgr}(d).
We plot four relations of \citet{mar06} toward 
$(l, b)=(16\fdg00, -3\fdg00)$, $(16\fdg25, -3\fdg00)$,
$(16\fdg00, -3\fdg25)$, and $(16\fdg25, -3\fdg25)$.
The closest direction is that of the filled green squares.
The other symbols/lines have the same meanings as those in Figure
\ref{distance_reddening_v1663_aql_v5116_sgr_v2575_oph_v5117_sgr}(a).
Our crossing point of $d=8.0$~kpc and $E(B-V)=0.70$ is consistent with
these relations.

We plot the $(B-V)_0$-$(M_V-2.5 \log f_{\rm s})$ diagram of V5584~Sgr in Figure
\ref{hr_diagram_nr_tra_v1213_cen_v5583_sgr_v5584_sgr_outburst}(d)
for $E(B-V)=0.70$ and $(m-M')_V=17.05$ in Equation
(\ref{absolute_mag_v5584_sgr}).
The track of V5584~Sgr almost follows that of V2615~Oph, so we regard 
V5584~Sgr to belong to the LV~Vul type 
because V2615~Oph belongs to the LV~Vul type
\citep{hac19k}.  The broad overlapping of the V5584~Sgr track
to the V2615~Oph track may support that our adopted values of $E(B-V)=0.70$
and $(m-M')_V=17.05$, that is, $E(B-V)=0.70\pm0.05$, 
$(m-M)_V=16.7\pm0.1$, $f_{\rm s}=1.35$, and $d=8.0\pm1$~kpc.

We check the distance modulus of $(m-M)_V=16.7$ by comparing
our model $V$ light curve with the observation.
Assuming $(m-M)_V=16.7$, we plot a model $V$ light curve of 
a $0.90~M_\sun$ WD \citep[CO3, solid red line;][]{hac16k} in Figure
\ref{v5584_sgr_v2615_oph_v496_sct_v_bv_ub_logscale}(a).
The $V$ light-curve model reproduces the observation well.
This confirms that distance modulus of $(m-M)_V=16.7$ is reasonable.


\begin{figure*}
\plotone{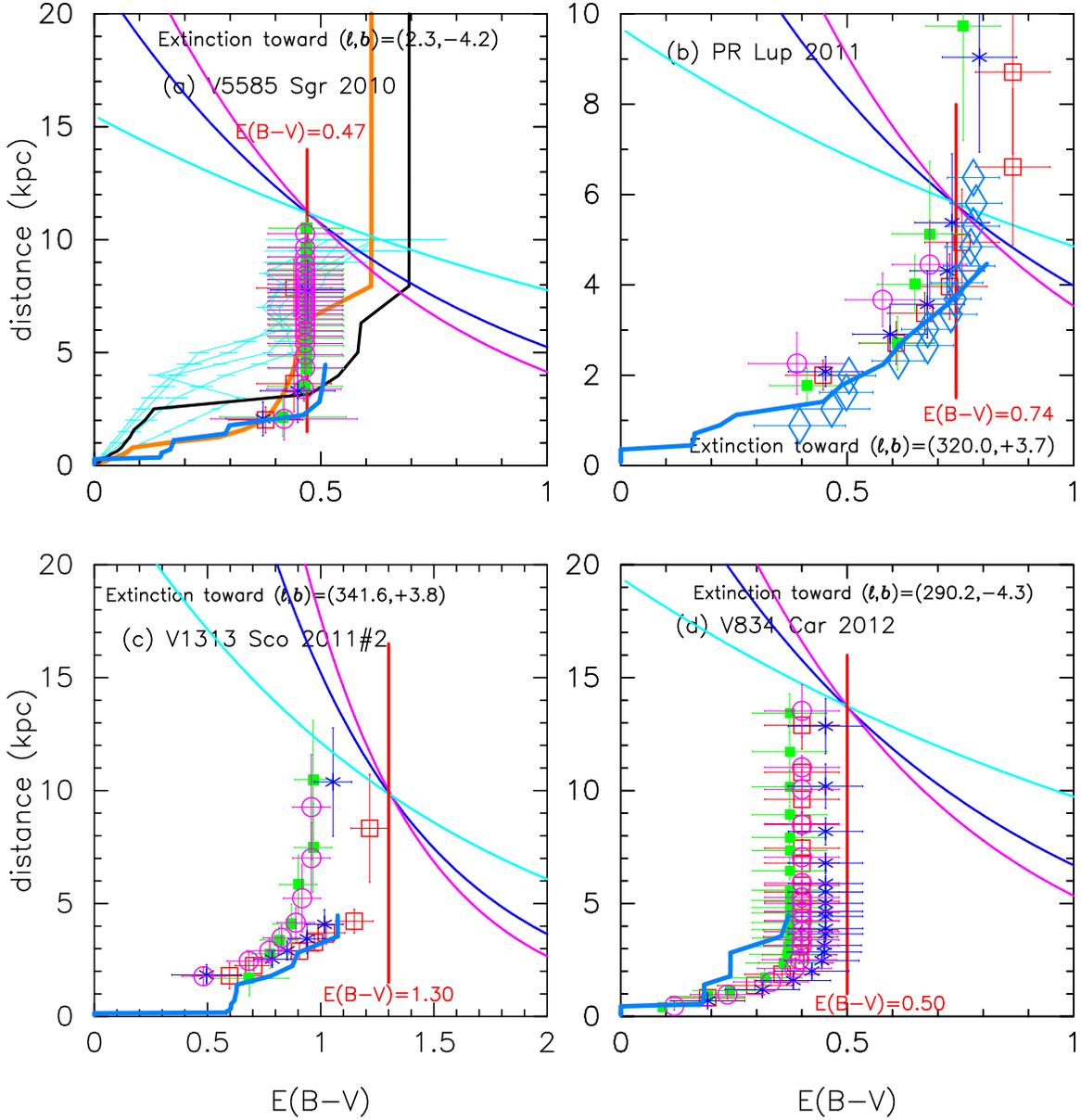}
\caption{
Same as Figure 
\ref{distance_reddening_v1663_aql_v5116_sgr_v2575_oph_v5117_sgr},
but for (a) V5585~Sgr, (b) PR~Lup, (c) V1313~Sco, and (d) V834~Car.
\label{distance_reddening_v5585_sgr_pr_lup_v1313_sco_v834_car}}
\end{figure*}


\begin{figure*}
\plotone{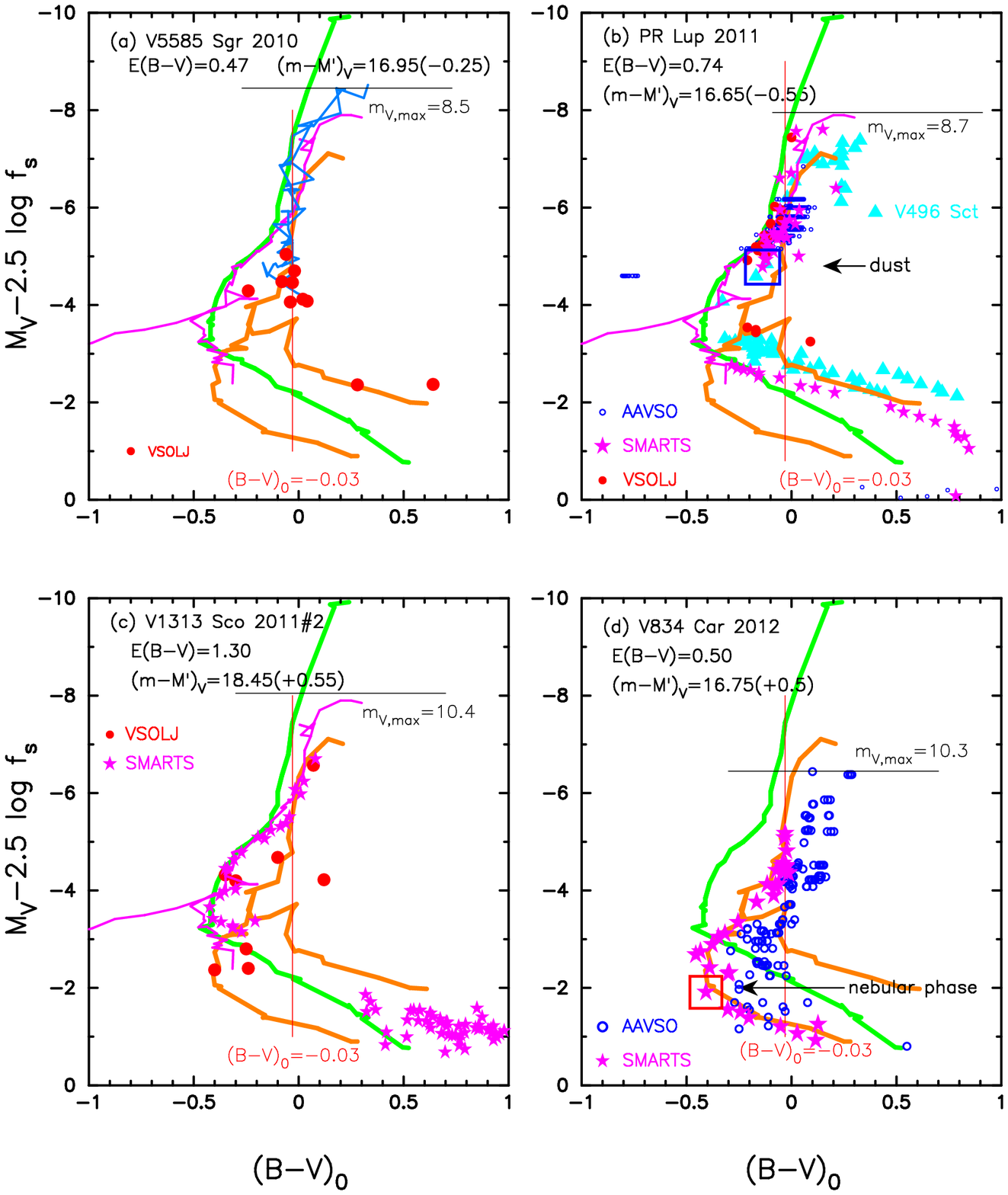}
\caption{
Same as Figure 
\ref{hr_diagram_v1663_aql_v5116_sgr_v2575_oph_v5117_sgr_outburst},
but for (a) V5585~Sgr, (b) PR~Lup, (c) V1313~Sco,
and (d) V834~Car.  The solid green lines show the template track
of V1500~Cyg.  The solid orange lines represent that of LV~Vul.
In panels (a), (b), and (c), we add the track of V1974~Cyg
(solid magenta lines).  In panel (a), we add the track of V1668~Cyg
(solid cyan-blue lines).  In panel (b), we add the track of V496~Sct
(filled cyan triangles).  
\label{hr_diagram_v5585_sgr_pr_lup_v1313_sco_v834_car_outburst}}
\end{figure*}

\subsection{V5585~Sgr 2010}
\label{v5585_sgr_cmd}
The nova was discovered by \citet{sea10} on UT 2010 January 20.72
at mag $\sim 8.5$.
V5585~Sgr reached its optical maximum sometime between UT 2009 November
15.89 and UT 2010 January 20.72, but the optical peak was missed because of
its proximity in the sky to the Sun \citep[e.g.,][]{mro15}.
The spectrum obtained by \citet{mae09b} showed that the nova is of
\ion{Fe}{2} class.  This nova showed sharp-peaked brightenings (flares)
in the $I$-band more than 15 times as shown in Figure 3 of \citet{mro15}.
\citet{mro15} found that the nova shows eclipses with an orbital period
of 3.3~hr.  They discussed the resemblance of V5585~Sgr to
V4745~Sgr and V5588~Sgr, both of which show
similar sharp-peaked brightenings  \citep[see also][]{tan11, mun15}.

We obtain $(m-M)_B= 17.23$, $(m-M)_V= 16.72$, and $(m-M)_I= 15.93$,
which cross at $d=11$~kpc and $E(B-V)=0.47$,
in Appendix \ref{v5585_sgr} and plot them in Figure
\ref{distance_reddening_v5585_sgr_pr_lup_v1313_sco_v834_car}(a).
Thus, we have $E(B-V)=0.47\pm0.05$ and $d=11\pm2$~kpc.
The distance modulus in the $V$ band is $(m-M)_V=16.7\pm0.2$.

For the reddening toward V5585~Sgr, $(l,b)=(2\fdg3311,-4\fdg1679)$,
the VVV catalog \citep{sai13} gives $E(B-V)=A_{K_s}/0.36=0.170/0.36=0.47$
toward V5585~Sgr.  The 2D NASA/IPAC galactic dust absorption map gives 
$E(B-V)=0.49\pm0.04$ for V5585~Sgr. The both reddening values are
consistent with our obtained value of $E(B-V)=0.47\pm0.05$.
We further check our result
in Figure
\ref{distance_reddening_v5585_sgr_pr_lup_v1313_sco_v834_car}(a).
We plot four relations of \citet{mar06} toward
$(l, b)=(2\fdg25, -4\fdg00)$, $(2\fdg50, -4\fdg00)$,
$(2\fdg25, -4\fdg25)$, and $(2\fdg50, -4\fdg25)$.
The closest direction is that of the blue asterisks.
The other symbols/lines have the same meanings as those in Figure
\ref{distance_reddening_v1663_aql_v5116_sgr_v2575_oph_v5117_sgr}(a).
Our crossing point of $d=11$~kpc and $E(B-V)=0.47$
is consistent with Marshall et al.'s relations.

We plot the $(B-V)_0$-$(M_V-2.5 \log f_{\rm s})$ diagram of V5585~Sgr in Figure
\ref{hr_diagram_v5585_sgr_pr_lup_v1313_sco_v834_car_outburst}(a)
for $E(B-V)=0.47$ and $(m-M')_V=16.95$ in Equation
(\ref{absolute_mag_v5585_sgr}).
The data are taken from VSOLJ.
The track of V5585~Sgr is close to that of LV~Vul and V1668~Cyg, 
so we regard V5585~Sgr to belong to the LV~Vul type. 
This overlapping of the V5585~Sgr track with the LV~Vul track 
confirms $E(B-V)=0.47$ and $(m-M')_V=16.95$,
that is, $E(B-V)=0.47\pm0.05$, $(m-M)_V=16.7\pm0.2$,
$f_{\rm s}=1.26$, and $d=11\pm2$~kpc.

We check the distance modulus of $(m-M)_V=16.7$ by comparing
our model $V$ light curve with the observation.
Assuming $(m-M)_V=16.7$, we plot a model $V$ light curve of 
a $0.95~M_\sun$ WD \citep[CO3, solid red line;][]{hac16k} in Figure
\ref{v5585_sgr_v2576_oph_v1668_cyg_lv_vul_v_bv_ub_logscale}(a).
The $V$ light-curve model reasonably reproduces the observation.
This confirms again that the distance modulus of $(m-M)_V=16.7$ is reasonable.
For comparison, we add light curves of a $0.98~M_\sun$ WD 
(CO3, solid green lines), assuming that $(m-M)_V = 14.6$ for V1668~Cyg.

\subsection{PR~Lup 2011}
\label{pr_lup_cmd}
PR~Lup reached $m_{V, \rm max}=8.7$ on UT 2,455,787.867 
(UT 2011 August 14.367).
The early spectra showed that PR~Lup belongs to the \ion{Fe}{2} type
\citep{mal11, wal11}.  The $K_{\rm s}$ magnitudes of SMARTS showed a rise 
at JD~2,455,833.5 (UT 2011 September 29.0) about 50 days after the optical
maximum.  This clearly indicates the formation of a dust shell.  
The $V$ light curve correspondingly drops at this epoch
(see Figure \ref{pr_lup_v_bv_ub_color_curve}).

We obtain $(m-M)_B=16.84$, $(m-M)_V=16.1$, and $(m-M)_I=14.94$,
which cross at $d=5.8$~kpc and $E(B-V)=0.74$,
in Appendix \ref{pr_lup} and plot them in Figure
\ref{distance_reddening_v5585_sgr_pr_lup_v1313_sco_v834_car}(b).
Thus, we obtain $d=5.8\pm0.6$~kpc, $E(B-V)=0.74\pm0.05$,
$(m-M)_V=16.1\pm0.1$, and $f_{\rm s}=1.7$ against LV~Vul.

For the reddening toward PR~Lup, \\$(l,b)=(319\fdg9767, +3\fdg6641)$,
the 2D NASA/IPAC galactic dust absorption map gives $E(B-V)=0.78\pm0.02$,
being roughly consistent with our obtained value of $E(B-V)=0.74\pm0.05$.
Figure \ref{distance_reddening_v5585_sgr_pr_lup_v1313_sco_v834_car}(b)
shows several distance-reddening relations toward PR~Lup.
We plot Marshall et al.'s (2006) relations toward
$(l, b)=(319\fdg75, +3\fdg75)$, $(320\fdg00, +3\fdg75)$,
$(319\fdg75, +3\fdg50)$, and $(320\fdg00. +3\fdg50)$.
The closest direction is that of filled green squares.
The other symbols/lines have the same meanings as those in Figure
\ref{distance_reddening_v1663_aql_v5116_sgr_v2575_oph_v5117_sgr}(a).
Our crossing point of $d=5.8\pm0.6$~kpc and $E(B-V)=0.74\pm0.05$
is consistent with these relations.

We plot the $(B-V)_0$-$(M_V-2.5 \log f_{\rm s})$ diagram of PR~Lup in Figure
\ref{hr_diagram_v5585_sgr_pr_lup_v1313_sco_v834_car_outburst}(b)
for $E(B-V)=0.74$ and $(m-M')_V=16.65$ in Equation
(\ref{absolute_mag_pr_lup}).
We also add the data points of V496~Sct (filled cyan triangles)
for comparison.  The PR~Lup track is close to that of V496~Sct
and LV~Vul.  Therefore, we regard PR~Lup to belong to the LV~Vul type.
This similarity of PR~Lup to V496~Sct and LV~Vul supports the fact that
our adopted values of $E(B-V)=0.74$ and $(m-M')_V=16.65$,
that is, $E(B-V)=0.74\pm0.05$, $(m-M)_V=16.1\pm0.1$, $f_{\rm s}=1.70$,
and $d=5.8\pm0.6$~kpc are reasonable.

We check the distance modulus of $(m-M)_V=16.1$ by comparing
our model $V$ light curve with the observation.
Assuming $(m-M)_V=16.1$, we plot a model $V$ light curve of 
a $0.80~M_\sun$ WD \citep[CO2, solid red line;][]{hac10k}  in Figure
\ref{pr_lup_v1369_cen_v496_sct_v_bv_ub_color_logscale}(a).
The $V$ light-curve models reproduces the observation well.
This confirms that distance modulus of $(m-M)_V=16.1$ is reasonable.

\subsection{V1313~Sco 2011\#2}
\label{v1313_sco_cmd}
V1313~Sco reached $m_{V,\rm max}=10.4$ on JD~2,455,812.0 
\citep[UT 2011 September 7.5;][]{waa11}.
The nova was identified as an \ion{Fe}{2} type \citep{ara11}, but
later \citet{wal12} classified V1313~Sco as a hybrid of the
\ion{Fe}{2} and He/N types.

We obtain $(m-M)_B= 20.33$, $(m-M)_V= 19.01$, and $(m-M)_I= 16.92$,
which cross at $d=9.9$~kpc and $E(B-V)=1.30$,
in Appendix \ref{v1313_sco} and plot them in Figure
\ref{distance_reddening_v5585_sgr_pr_lup_v1313_sco_v834_car}(c).
Thus, we have $E(B-V)=1.30\pm0.1$ and $d=9.9\pm2$~kpc.

For the reddening toward V1313~Sco, $(l,b)=(341\fdg5552, +3\fdg8397)$,
the 2D NASA/IPAC galactic dust absorption map gives $E(B-V)=1.20\pm0.10$,
which is consistent with our crossing point of $E(B-V)=1.30\pm0.1$.
We further check our result
in Figure \ref{distance_reddening_v5585_sgr_pr_lup_v1313_sco_v834_car}(c).
We plot four of Marshall et al.'s (2006) relations toward
$(l, b)=(341\fdg50,+3\fdg75)$, $(341\fdg75, +3\fdg75)$,
$(341\fdg50, +4\fdg00)$, and $(341\fdg75, +4\fdg00)$.
The closest direction is that of the unfilled red squares.
The other symbols/lines have the same meanings as those in Figure
\ref{distance_reddening_v1663_aql_v5116_sgr_v2575_oph_v5117_sgr}(a).
Our crossing point is consistent with Marshall et al.'s
distance-reddening relation (unfilled red squares with error bars).

We plot the $(B-V)_0$-$(M_V-2.5 \log f_{\rm s})$ diagram of V1313~Sco
in Figure \ref{hr_diagram_v5585_sgr_pr_lup_v1313_sco_v834_car_outburst}(c)
for $E(B-V)=1.30$ and $(m-M')_V=18.45$ in Equation
(\ref{absolute_mag_v1313_sco}).
The track of V1313~Sco closely follows that of V1974~Cyg and V1500~Cyg.
Thus, we regard V1313~Sco to belong to the V1500~Cyg type. 
This similarity of the V1313~Sco track to the V1974~Cyg and V1500~Cyg tracks
supports our adopted values of $E(B-V)=1.30$ and $(m-M')_V=18.45$,
that is, $E(B-V)=1.30\pm0.1$, $(m-M)_V=19.0\pm0.1$, $f_{\rm s}=0.60$,
and $d=9.9\pm2$~kpc.

We check the distance modulus of $(m-M)_V=19.0$ by comparing
our model $V$ light curve with the observation.  
Assuming that $(m-M)_V=19.0$, we plot a model $V$ light curve of 
a $1.20~M_\sun$ WD \citep[Ne2, solid red line;][]{hac10k} in Figure
\ref{v1313_sco_v533_her_v2576_oph_v1668_cyg_lv_vul_v_bv_ub_logscale}(a).
The $V$ light-curve model reproduces the observation.
This confirms that the distance modulus of $(m-M)_V=19.0$ is reasonable.
For comparison, we add light curves of a $0.98~M_\sun$ WD 
(CO3, solid green lines), assuming that $(m-M)_V=14.6$ for V1668~Cyg.

\subsection{V834~Car 2012}
\label{v834_car_cmd}
V834~Car reached $m_{V, \rm max}=10.3$ on JD~2,455,987.8 
(UT 2012 March 1.3) from the AAVSO data.
\citet{wal12} identified the nova as an \ion{Fe}{2} type
and estimated the decline rates of $t_2=20\pm3$~days and $t_3=38\pm1$~days.

We obtain $(m-M)_B= 17.75$, $(m-M)_V= 17.23$, and $(m-M)_I= 16.44$
in Appendix \ref{v834_car} and plot them in Figure
\ref{distance_reddening_v5585_sgr_pr_lup_v1313_sco_v834_car}(d),
which broadly cross at $d=14$~kpc and $E(B-V)=0.50$.
Thus, we have $E(B-V)=0.50\pm0.05$ and $d=14\pm2$~kpc.
The distance modulus in the $V$ band is $(m-M)_V=17.25\pm0.1$.

For the reddening toward V834~Car, \\$(l,b) = (290\fdg1799, -4\fdg2820)$,
the 2D NASA/IPAC galactic dust absorption map gives $E(B-V)=0.52\pm0.01$,
which is consistent with our value of $E(B-V)=0.50\pm0.05$. 
We further check our result
in Figure \ref{distance_reddening_v5585_sgr_pr_lup_v1313_sco_v834_car}(d).
Marshall et al.'s (2006) relations are plotted toward
$(l, b)=(290\fdg00, -4\fdg25)$, $(290\fdg25, -4\fdg25)$,
$(290\fdg00, -4\fdg50)$, and $(290\fdg25, -4\fdg50)$.
The closest direction is that of the filled green squares.
The other symbols/lines have the same meanings as those in Figure
\ref{distance_reddening_v1663_aql_v5116_sgr_v2575_oph_v5117_sgr}(a).
Our crossing point is roughly consistent with Marshall et al.'s
relation (blue asterisks with error bars).

We plot the $(B-V)_0$-$(M_V-2.5 \log f_{\rm s})$ diagram of V834~Car in Figure
\ref{hr_diagram_v5585_sgr_pr_lup_v1313_sco_v834_car_outburst}(d)
for $E(B-V)=0.50$ and $(m-M')_V=16.75$ in Equation
(\ref{absolute_mag_v834_car}).
The track of V834~Car (SMARTS data; filled magenta stars)
follows that of LV~Vul.
Thus, we regard V834~Car to belong to the LV~Vul type. 
This overlapping with the LV~Vul track supports the fact that
$E(B-V)=0.50$ and $(m-M')_V=16.75$, that is,
$E(B-V)=0.50\pm0.05$, $(m-M)_V=17.25\pm0.1$, $f_{\rm s}=0.65$,
and $d=14\pm2$~kpc.

The nova had already entered the nebular phase on UT 2012 May 27
(JD~2,456,074.5), because [\ion{O}{3}] lines developed as shown in the
SMARTS spectra \citep{wal12}.
We identify the start of the nebular phase as
$M_V=14.82 - 17.25= -2.43$ and $(B-V)_0=0.105 - 0.50= -0.395$
denoted by the large unfilled red square in Figure
\ref{hr_diagram_v5585_sgr_pr_lup_v1313_sco_v834_car_outburst}(d).

We check the distance modulus of $(m-M)_V=17.25$ by comparing
our model $V$ light curve with the observation.  
Assuming that $(m-M)_V=17.25$, we plot a $V$ model light curve 
(solid red line) of a $1.20~M_\sun$ WD \citep[Ne2, ][]{hac10k} in 
Figure \ref{v834_car_v2576_oph_v1668_cyg_lv_vul_v_bv_ub_logscale}(a).
The model $V$ light curve reasonably reproduces the observation.
This confirms that the distance modulus of $(m-M)_V=17.25$ is
reasonable.  In the same figure, we also plot the model light curves of
a $0.98~M_\sun$ WD (CO3, solid green lines),
assuming $(m-M)_V=14.6$ for V1668~Cyg.


\begin{figure*}
\plotone{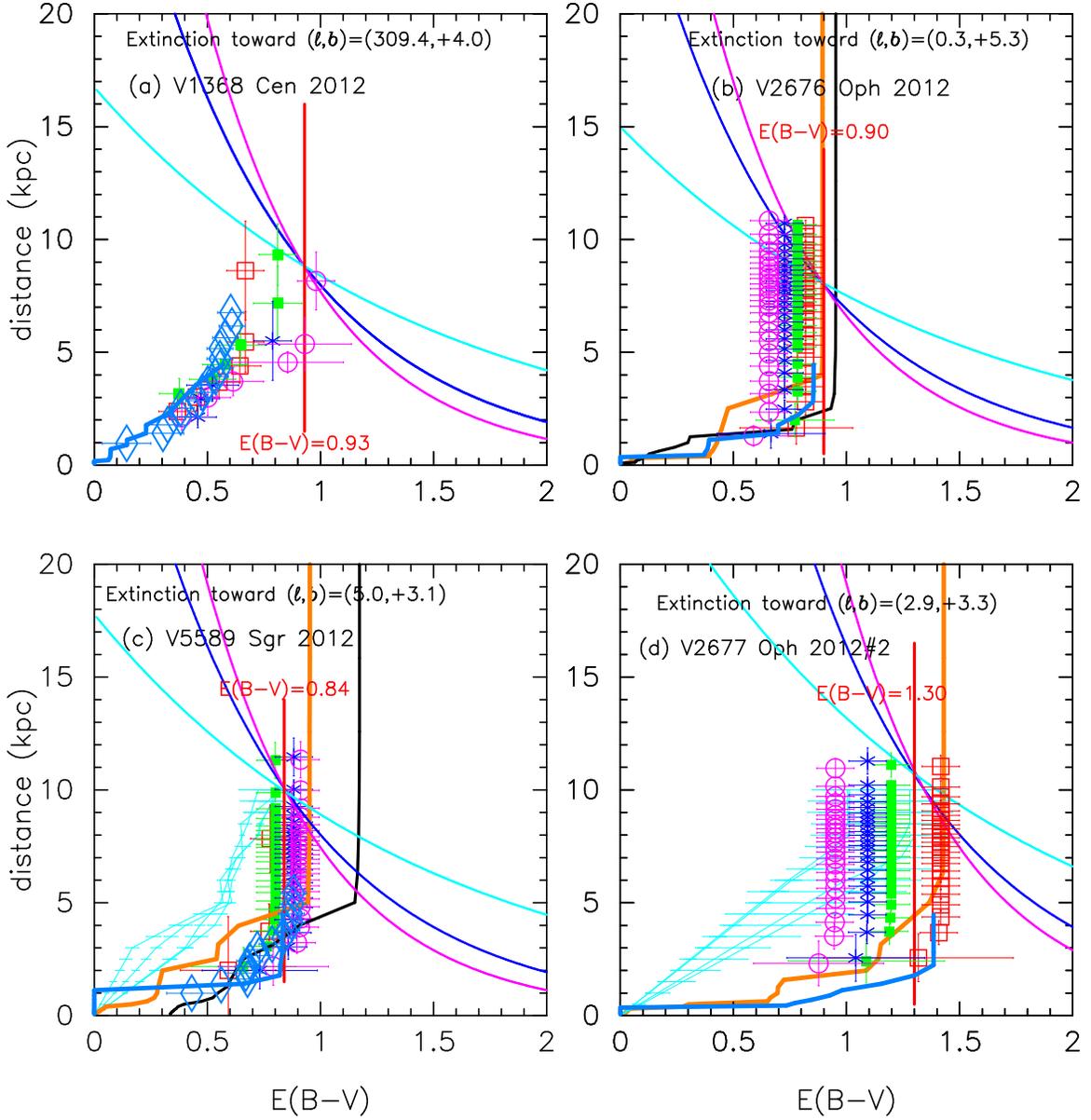}
\caption{
Same as Figure 
\ref{distance_reddening_v1663_aql_v5116_sgr_v2575_oph_v5117_sgr},
but for (a) V1368~Cen, (b) V2676~Oph, (c) V5589~Sgr, and (d) V2677~Oph.
\label{distance_reddening_v1368_cen_v2676_oph_v5589_sgr_v2677_oph}}
\end{figure*}


\begin{figure*}
\plotone{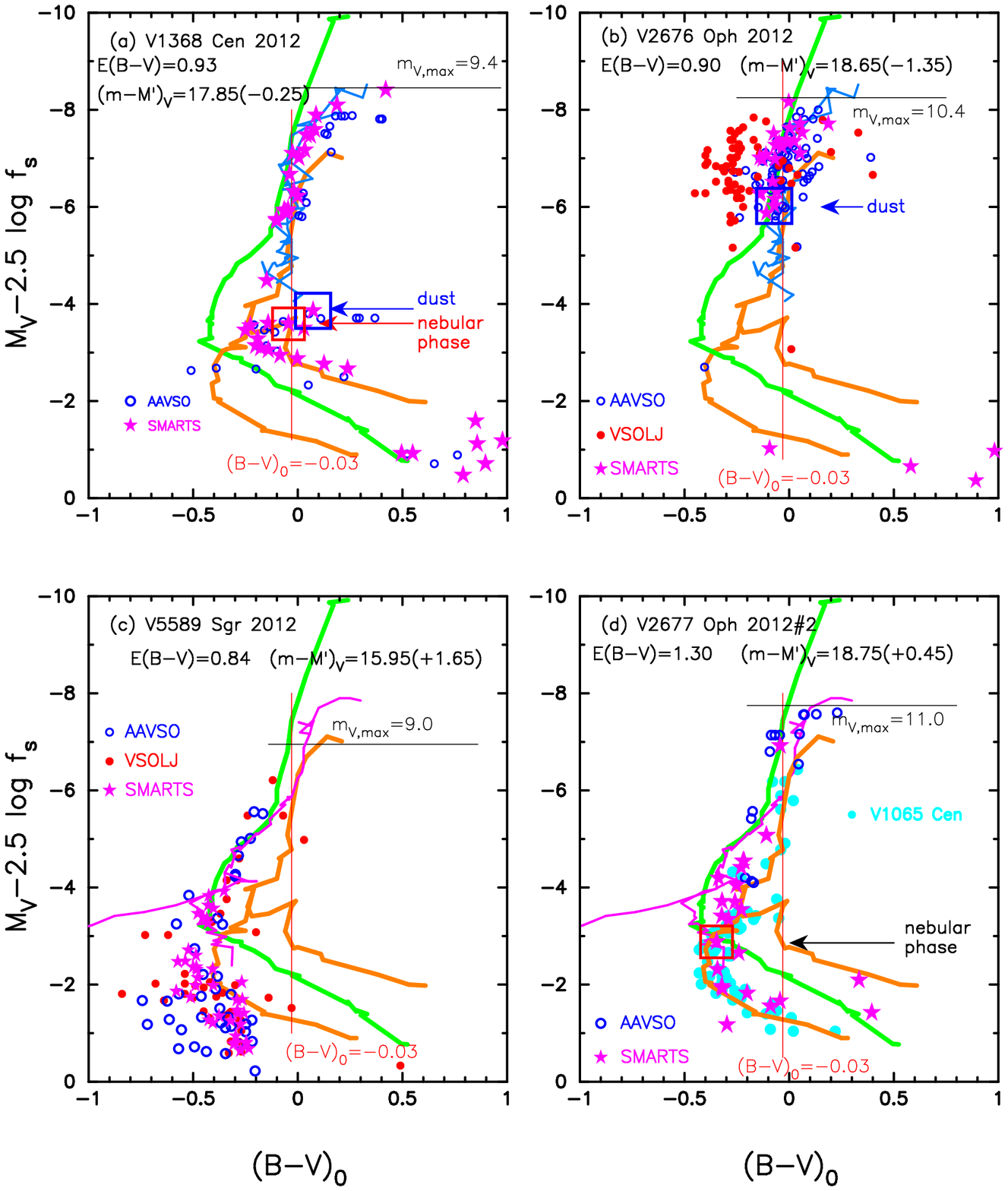}
\caption{
Same as Figure 
\ref{hr_diagram_v1663_aql_v5116_sgr_v2575_oph_v5117_sgr_outburst},
but for (a) V1368~Cen, (b) V2676~Oph, (c) V5589~Sgr, and (d) V2677~Oph.
The solid green lines show the template track of V1500~Cyg.  
The solid orange lines show the template tracks of LV~Vul.
In panels (a) and (b), we add the track of V1668~Cyg (solid cyan-blue lines).
In panels (c) and (d), we add the tracks of V1974~Cyg
(solid magenta lines).    In panel (d), we add the track of V1065~Cen
(filled cyan circles).
\label{hr_diagram_v1368_cen_v2676_oph_v5589_sgr_v2677_oph_outburst}}
\end{figure*}

\subsection{V1368~Cen 2012}
\label{v1368_cen_cmd}
V1368~Cen reached $m_{V, \rm max}=9.4$ on JD~2,456,010.6.
The $K_{\rm s}$ magnitudes of SMARTS showed a rise on JD~2,456,042.7
(UT 2012 April 25.2) about 32 days after the optical maximum.
This clearly indicates the formation of a dust shell \citep{wal12}.
The $V$ light curve correspondingly drops at this epoch as shown by
the arrow in Figure \ref{v1368_cen_v_bv_ub_color_curve}.
\citet{wal12} identified this nova to be of \ion{Fe}{2} type and
estimated the decline rates as $t_2=16\pm1$~days and $t_3\sim32$~days.

We obtain $(m-M)_B= 18.53$, $(m-M)_V= 17.6$, and $(m-M)_I= 16.12$,
which broadly cross at $d=8.8$~kpc and $E(B-V)=0.93$,
in Appendix \ref{v1368_cen} and plot them in Figure
\ref{distance_reddening_v1368_cen_v2676_oph_v5589_sgr_v2677_oph}(a).
Thus, we have $E(B-V)=0.93\pm0.05$ and $d=8.8\pm1$~kpc.
The distance modulus in the $V$ band is $(m-M)_V=17.6\pm0.1$.

For the reddening toward V1368~Cen, $(l,b)=(309\fdg4457, +3\fdg9788)$,
the 2D NASA/IPAC galactic dust absorption map gives $E(B-V)=0.97\pm0.06$,
which is consistent with our value of $E(B-V)=0.93\pm0.05$.
We further check our result
in Figure
\ref{distance_reddening_v1368_cen_v2676_oph_v5589_sgr_v2677_oph}(a).
Marshall et al.'s (2006) relations are plotted toward
$(l, b)=(309\fdg25,+3\fdg75 )$, $(309\fdg50, +3\fdg75)$,
$(309\fdg25, +4\fdg00)$, and $(309\fdg50, +4\fdg00)$.
The closest direction is that of the unfilled magenta circles.
The other symbols/lines have the same meanings as those in Figure
\ref{distance_reddening_v1663_aql_v5116_sgr_v2575_oph_v5117_sgr}(a).
Our crossing point is consistent with Marshall et al.'s 
distance-reddening relation (unfilled magenta circles with error bars).

We plot the $(B-V)_0$-$(M_V-2.5 \log f_{\rm s})$ diagram of V1368~Cen in Figure
\ref{hr_diagram_v1368_cen_v2676_oph_v5589_sgr_v2677_oph_outburst}(a)
for $E(B-V)=0.93$ and $(m-M')_V=17.85$ from Equation
(\ref{absolute_mag_v1368_cen}).
The track of V1368~Cen almost follows V1668~Cyg in the early phase.
The V1368~Cen track follows the LV~Vul track after the dust blackout ended.
We plot the start of the dust blackout with the arrow labeled ``dust.''
Therefore, we regard V1368~Cen to belong to the LV~Vul type.
This overlapping of V1368~Cen with the V1668~Cyg and LV~Vul tracks
supports $E(B-V)=0.93$ and $(m-M')_V=17.85$, that is,
$E(B-V)=0.93\pm0.05$, $(m-M)_V=17.6\pm0.1$, $f_{\rm s}=1.26$,
and $d=8.8\pm1$~kpc.
The nova had already entered the nebular phase on UT 2012 May 27
(JD~2,456,074.5), because [\ion{O}{3}] lines developed as shown
in SMARTS spectra \citep{wal12}.
We identify the start of the nebular phase as
$M_V=14.23 - 17.6 = -3.37$ and $(B-V)_0=0.799 - 0.93= -0.13$
denoted by a large unfilled red square.

We check the distance modulus of $(m-M)_V=17.6$ by comparing
our model $V$ light curve with the observation.  
Assuming $(m-M)_V=17.6$, we plot a $V$ model light curve of 
a $0.95~M_\sun$ WD \citep[CO3, solid green line;][]{hac16k} in Figure
\ref{v1368_cen_lv_vul_v1668_cyg_os_and_v_bv_ub_logscale}(a).
The model $V$ light curve reasonably reproduces the observation.
This confirms that the distance modulus of $(m-M)_V=17.6$ is
reasonable.  In the same figure, we also plot model light curves of
a $0.98~M_\sun$ WD (CO3, solid blue lines) for V1668~Cyg
and a $1.05~M_\sun$ WD (CO3, solid red lines) for OS~And,
assuming that $(m-M)_V = 14.6$ for V1668~Cyg
and $(m-M)_V = 14.8$ for OS~And.

\subsection{V2676~Oph 2012}
\label{v2676_oph_cmd}
V2676~Oph reached $m_{V, \rm max}=10.4$ on JD~2,456,038.97 
(UT 2012 April 21.47) from the VSOLJ data.  
\citet{ara12} identified the nova as an \ion{Fe}{2} type.
The $K_{\rm s}$ magnitudes of SMARTS showed a rise on JD~2,456,103.7
(UT 2012 June 25.2), about 95 days after outburst
(see Figure \ref{v2676_oph_v_bv_ub_color_curve}).
This clearly indicates the formation of a dust shell \citep{wal12}.
The $V$ light curve correspondingly drops at this epoch as shown by
the arrow in Figure \ref{v2676_oph_v_bv_ub_color_curve}.
\citet{rud12} found carbon-monoxide emission from the fundamental and the
first and second overtones on UT 2012 May 1 and 2 before the dust shell
formation. \citet{nag14} and \citet{kaw15} also found strong emission from
C$_2$ and CN molecules.  The mid-infrared spectroscopic and photometric
observations in 2013 and 2014 (452 and 782 days after its discovery,
respectively) showed no significant \ion{Ne}{2} emission at $12.8~\mu$m,
suggesting evidence for a CO-rich WD \citep{kaw17}. 

We obtain $(m-M)_B= 18.2$, $(m-M)_V= 17.3$, and $(m-M)_I= 15.88$,
which cross at $d=8.0$~kpc and $E(B-V)=0.90$,
in Appendix \ref{v2676_oph} and plot them in Figure
\ref{distance_reddening_v1368_cen_v2676_oph_v5589_sgr_v2677_oph}(b).
Thus, we have $E(B-V)=0.90\pm0.05$ and $d=8.0\pm1$~kpc.
The distance modulus in the $V$ band is $(m-M)_V=17.3\pm0.1$.

For the reddening toward V2676~Oph, $(l,b)=(0\fdg2631, +5\fdg3013)$,
the 2D NASA/IPAC galactic dust absorption map gives $E(B-V)=0.91\pm0.05$,
consistent with our value of $E(B-V)=0.90\pm0.05$.
\citet{nag15} obtained $E(B-V)=0.71\pm0.02$ from the Balmer decrement.
\citet{kaw16} obtained $E(B-V)= A_V/3.1= (2.65\pm0.15)/3.1= 0.85\pm0.05$
from the $T_{\rm eff}$ versus $E(V-I)$ relation.  
\citet{raj17} obtained $E(B-V)= A_V/3.1= (2.9\pm0.1)/3.1= 0.94\pm0.03$
from the Balmer decrement.
However, hydrogen recombination lines are usually not described by 
Case B approximation, and the extinction estimated from this method
likely has a large error.  Our value of $E(B-V)= 0.90\pm0.05$ at the
crossing point is roughly consistent with $E(B-V)= 0.85\pm0.05$
\citep{kaw16} and $E(B-V)= 0.94\pm0.03$ \citep{raj17}.

We further check our result
 in Figure
\ref{distance_reddening_v1368_cen_v2676_oph_v5589_sgr_v2677_oph}(b).
Marshall et al.'s (2006) relations are plotted toward
$(l, b)=(0\fdg25,+5\fdg25)$, $(0\fdg50,+5\fdg25)$,
$(0\fdg25,+5\fdg50)$, and $(0\fdg50,+5\fdg50)$.
The closest direction is that of the unfilled red squares.
The other symbols/lines have the same meanings as those in Figure
\ref{distance_reddening_v1663_aql_v5116_sgr_v2575_oph_v5117_sgr}(a).
Our crossing point is consistent with Marshall et al.'s (unfilled red 
squares with error bars),  Green et al.'s (orange line), and
Chen et al.'s (cyan-blue line) distance-reddening relations.

We plot the $(B-V)_0$-$(M_V-2.5 \log f_{\rm s})$ diagram of V2676~Oph in Figure
\ref{hr_diagram_v1368_cen_v2676_oph_v5589_sgr_v2677_oph_outburst}(b)
for $E(B-V)=0.90$ and $(m-M')_V=18.65$ in Equation
(\ref{absolute_mag_v2676_oph}).
The track of V2676~Oph (SMARTS data) is close to that of V1668~Cyg
and LV~Vul until the dust blackout started.
Thus, we regard V2676~Oph to belong to the LV~Vul type in the 
$(B-V)_0$-$(M_V-2.5 \log f_{\rm s})$ diagram like V1065~Cen and V1368~Cen.
This overlapping with the LV~Vul track is consistent with our values of
$E(B-V)=0.90$ and $(m-M')_V=18.65$, that is, 
$E(B-V)=0.90\pm0.05$, $(m-M)_V=17.3\pm0.2$, $f_{\rm s}=3.4$,
and $d=8.0\pm1.0$~kpc.

We check the distance modulus of $(m-M)_V=17.3$ by comparing
our model $V$ light curve with the observation in Figure 
\ref{v2676_oph_v496_sct_v475_sct_lv_vul_v_bv_ub_color_logscale}(a).  
Assuming $(m-M)_V=17.3$, we plot a $V$ model light curve of 
a $0.70~M_\sun$ WD \citep[CO2, solid red line;][]{hac10k}.
The model $V$ light curve reasonably reproduces the observation,
although the light curve has a wavy structure before the dust blackout.
This may confirm that the distance modulus of $(m-M)_V=17.3$ is
reasonable.  In the same figure, we also plot model light curves of
a $0.80~M_\sun$ WD (CO3, solid black line),
assuming that $(m-M)_V=14.65$ for QY~Mus.

\subsection{V5589~Sgr 2012\#1}
\label{v5589_sgr_cmd}
V5589~Sgr reached $m_{V, \rm max}\sim9.0$ on JD~2,456,039.56 
\citep[UT 2012 April 22.06; e.g.,][]{mro15}.
\citet{wal12} estimated the decline rates of $t_2=4.5\pm1.5$~days
and $t_3=7$~days, while \citet{wes16} obtained $t_2=6.8\pm0.8$~days
and $t_3=12.8\pm1.5$~days.  
\citet{thomp17} analyzed {\it Solar TErrestrial RElations Observatory
(STEREO)} data and obtained optical maximum at mag $8.23\pm0.03$ in 
the Heliospheric Imager (HI)-1 bandpass on JD~2,456,039.3224.
They also obtained $t_2 = 5.0\pm0.6$ day and $t_3 = 10.9\pm0.7$ day.
So the nova belongs to the very fast nova
speed class defined by \citet{pay57}.  

V5589~Sgr was identified
as a hybrid nova from \ion{Fe}{2} to He/N type \citep{wal12}.
The nova had entered the coronal phase of [\ion{Fe}{10}],
[\ion{Fe}{11}],  and [\ion{Fe}{14}] by day 65 \citep{wal12}.
The detection of the coronal phase sometimes
indicates that the nova had already entered the supersoft X-ray
source (SSS) phase.  The {\it Swift}/XRT observation showed that
the nova increased its soft X-ray flux on day 64.5 \citep{wes16}.
We may regard the nova to have entered the SSS phase by at least
day 64.5.   \citet{mro15} obtained the orbital period
of this nova as $P_{\rm orb}=1.5923$~days ($P_{\rm orb}=38.215$~hr).
They suggested that the companion has already evolved off
the main sequence and is a subgiant like the recurrent nova U~Sco.

We obtain $(m-M)_B= 18.45$, $(m-M)_V= 17.62$, and $(m-M)_I= 16.25$,
which cross at $d=10$~kpc and $E(B-V)=0.84$,
in Appendix \ref{v5589_sgr} and plot them in Figure
\ref{distance_reddening_v1368_cen_v2676_oph_v5589_sgr_v2677_oph}(c).
Thus, we have $E(B-V)=0.84\pm0.05$ and $d=10\pm1$~kpc.
The distance modulus in the $V$ band is $(m-M)_V=17.6\pm0.1$.

For the reddening toward V5589~Sgr, $(l,b)=(4\fdg9766, +3\fdg0724)$,
the 2D NASA/IPAC galactic dust absorption map gives $E(B-V)=0.84\pm0.04$.
The VVV survey catalog gives $E(B-V)= A_{K_s}/0.36= 0.274/0.36= 0.76$
\citep{sai13}.  \citet{wes16} estimated the reddening of 
$E(B-V)=0.8\pm0.19$ from four diffuse interstellar band (DIB) features.
These reddening values are roughly consistent with our crossing point
of $E(B-V)=0.84\pm0.05$.

We further check our result
in Figure
\ref{distance_reddening_v1368_cen_v2676_oph_v5589_sgr_v2677_oph}(c).
Marshall et al.'s (2006) relations are plotted toward
$(l, b)=(4\fdg75,  +3\fdg00)$, $(5\fdg00,+3\fdg00)$,
$(4\fdg75,+3\fdg25)$, and $(5\fdg00,+3\fdg25)$.
The closest direction is that of the filled green squares.
The other symbols/lines have the same meanings as those in Figure
\ref{distance_reddening_v1663_aql_v5116_sgr_v2575_oph_v5117_sgr}(a).
Our crossing point is consistent with Marshall et al.'s 
(filled green squares with error bars), \"Ozd\"ormez et al.'s
(unfilled cyan-blue diamonds), and Chen et al.'s (cyan-blue line)
distance-reddening relations.

We plot the $(B-V)_0$-$(M_V-2.5 \log f_{\rm s})$ diagram of V5589~Sgr in Figure
\ref{hr_diagram_v1368_cen_v2676_oph_v5589_sgr_v2677_oph_outburst}(c)
for $E(B-V)=0.84$ and $(m-M')_V=15.95$ in Equation
(\ref{absolute_mag_v5589_sgr}).
The track of V5589~Sgr follows that of V1500~Cyg and V1974~Cyg
in the early phase.
Thus, we regard V5589~Sgr to belong to the V1500~Cyg type.
This overlapping of the V5589~Sgr track with the V1500~Cyg and V1974~Cyg
tracks supports $E(B-V)=0.84$ and $(m-M')_V=15.95$, that is,
$E(B-V)=0.84\pm0.05$, $(m-M)_V=17.6\pm0.1$, $d=10\pm1$~kpc, and
$f_{\rm s}=0.21$ against LV~Vul.

We check the distance modulus of $(m-M)_V=17.6$ by comparing
our model $V$ light curve with the observation.  
Assuming that $(m-M)_V=17.6$, we plot a $V$ model light curve 
and supersoft X-ray flux (0.1--2.4~keV) of a $1.33~M_\sun$ WD 
\citep[Ne2, solid black lines;][]{hac10k} in Figure
\ref{v5589_sgr_v5583_sgr_v382_vel_v_bv_ub_logscale}(a).
The $V$ light curves reasonably reproduce the observation.  
This again confirm that the distance modulus of $(m-M)_V=17.6$ is reasonable.
The blue lines are the model light curves of $1.23~M_\sun$ WD (Ne2),
assuming that $(m-M)_V=11.5$ for V382~Vel.

\citet{wes16} obtained $E(B-V)=0.80\pm0.19$ as mentioned above and
derived the distance of $d=3.6^{+1.0}_{-2.4}$~kpc in 
conjunction with the 3D reddening map of \citet{gre15}.
This distance estimate is not reasonable, because Green et al.'s 
(2015, black line) 3D reddening map deviates largely from 
Marshall et al.'s (2006, filled green squares) and revised
Green et al.'s (2018, orange line) 3D reddening maps as shown in Figure
\ref{distance_reddening_v1368_cen_v2676_oph_v5589_sgr_v2677_oph}(c)
and the NASA/IPAC absorption map of $E(B-V)=0.84\pm0.04$.
The reason why \citet{wes16} obtained such a short distance is
that they used the distance-reddening relation of \citet{gre15},
which is not consistent with other 2D and 3D dust maps.

Assuming a Hubble flow of the ejecta,
\citet{wes16} estimated the ejecta mass as
$M_{\rm ej}=2.6\times 10^{-5}~M_\sun$ together with the other
parameters of $d=4$~kpc, $T=1.2\times 10^4$~K,
$v_{\rm HWZI}=4000$~km~s$^{-1}$, and $\zeta=v_{\rm min}/v_{\rm HWZI}=0.84$.
The ejected mass estimated from our model of a $1.33~M_\sun$ WD (Figure
\ref{v5589_sgr_v5583_sgr_v382_vel_v_bv_ub_logscale})
is $M_{\rm ej}=1.7\times10^{-6}~M_\sun$.
Weston et al.'s value is about 10 times larger than our value.
Theoretically, the ignition mass, i.e., the hydrogen-rich envelope mass
at the nova ignition, is much smaller in a high-mass WD
than in a low-mass WD, that is, for the $1.33~M_\sun$ WD,
$M_{\rm ig}\sim 1\times10^{-6}$ -- $3\times10^{-6}~M_\sun$,
depending on the mass accretion rate onto the WD 
\citep[see, e.g., Figure 3 of][for a recent
estimate of ignition masses]{kat14shn}.

\citet{wes16} reported the {\it Swift} X-ray observation of V5589~Sgr.
The X-ray spectrum showed an increase of soft X-ray flux
(lower than 1~keV) on day 64.5 compared with the previous
X-ray observation on day 53.  They regarded this rise of soft
X-ray flux as the appearance of the SSS phase.
They further obtained the blackbody temperature of $50^{+36}_{-18}$~eV
from the spectrum on day 68.  The X-ray count rate had declined
on day 78.  Figure \ref{v5589_sgr_v5583_sgr_v382_vel_v_bv_ub_logscale}(a)
shows our supersoft X-ray light curve (solid black lines)
of a $1.33~M_\sun$ WD, which shows a rise on day $\sim 55$
(optically thick wind stops) and fall on day $\sim 75$
(hydrogen-shell burning ends).  These features are
consistent with the {\it Swift}/XRT observation.

\subsection{V2677~Oph 2012\#2}
\label{v2677_oph_cmd}
V2677~Oph reached $m_{V,\rm max}=11.0$ on JD~2,456,068.84 
(UT 2012 May 21.34) from AAVSO data. 
\citet{wal12} identified the nova as an \ion{Fe}{2} type.
\citet{mro15} presented their OGLE photometric observation of V2677~Oph
and obtained a possible orbital period of $P_{\rm orb}=8.607$~hr.

We obtain $(m-M)_B= 20.5$, $(m-M)_V= 19.18$, and $(m-M)_I= 17.1$,
which broadly cross at $d=10.7$~kpc and $E(B-V)=1.30$,
in Appendix \ref{v2677_oph} and plot them in Figure
\ref{distance_reddening_v1368_cen_v2676_oph_v5589_sgr_v2677_oph}(d).
Thus, we have $E(B-V)=1.30\pm0.10$ and $d=10.7\pm2$~kpc.
The distance modulus in the $V$ band is $(m-M)_V=19.2\pm0.1$.

For the reddening toward V2677~Oph, $(l,b)=(2\fdg8584, +3\fdg2550)$,
the 2D NASA/IPAC galactic dust absorption map gives $E(B-V)=1.30\pm0.05$,
which is consistent with our value of $E(B-V)=1.30\pm0.10$.
We further check our result
in Figure
\ref{distance_reddening_v1368_cen_v2676_oph_v5589_sgr_v2677_oph}(d).
Marshall et al.'s (2006) relations are plotted toward
$(l, b)=(2\fdg75,+3\fdg25)$, $(3\fdg00,+3\fdg25)$, 
$(2\fdg75,+3\fdg50)$, and $(3\fdg00,+3\fdg50)$.
The direction toward V2677~Oph is between the directions of the
filled green squares and unfilled red squares.
The other symbols/lines have the same meanings as those in Figure
\ref{distance_reddening_v1663_aql_v5116_sgr_v2575_oph_v5117_sgr}(a).
Our crossing point is in reasonable agreement with Marshall et al.'s
distance-reddening relations.

We plot the $(B-V)_0$-$(M_V-2.5 \log f_{\rm s})$ diagram of V2677~Oph in Figure
\ref{hr_diagram_v1368_cen_v2676_oph_v5589_sgr_v2677_oph_outburst}(d)
for $E(B-V)=1.30$ and $(m-M')_V=18.75$ in Equation
(\ref{absolute_mag_v2677_oph}).
The track of V2677~Oph is very close to that of LV~Vul and
V1065~Cen (filled cyan circles).
Thus, we regard V2677~Oph to belong to the LV~Vul type.
This similarity of the V2677~Oph track to the LV~Vul track
supports our values of $E(B-V)=1.30$ and $(m-M')_V=18.75$,
that is, $E(B-V)=1.30\pm0.10$, $(m-M)_V=19.2\pm0.1$, $f_{\rm s}=0.68$,
and $d=10.7\pm2$~kpc.

The nova had already entered the nebular phase on JD~2,456,118.5
(UT 2012 July 10) in the spectra of SMARTS \citep{wal12}.
We specify this point in the $(B-V)_0$-$(M_V-2.5 \log f_{\rm s})$ diagram
by a large unfilled red square, i.e., $(B-V)_0=0.974-1.30= -0.33$ and 
$M_V=15.84 - 19.2= -3.36$ (Figure 
\ref{hr_diagram_v1368_cen_v2676_oph_v5589_sgr_v2677_oph_outburst}(d)).

We check the distance modulus of $(m-M)_V=19.2$ by comparing
our model $V$ light curve with the observation (filled green circles) 
in Figure \ref{v2677_oph_v1065_cen_lv_vul_v_bv_ub_logscale}(a).  
Assuming $(m-M)_V=19.2$, we plot a $V$ model light curve of 
a $1.15~M_\sun$ WD \citep[Ne2, solid green line;][]{hac10k}.
For comparison, we add the LV~Vul light/color curves (unfilled blue squares).
We also add the $V$ light/color curves of V1065~Cen 
(unfilled magenta diamonds).
The model $V$ light curve reasonably reproduces the observation.
However, the observed $V$ magnitudes are slightly ($\sim 0.8$ mag)
fainter than the model $V$ light curve in the middle phase.
The $V$ light curve of V2677~Oph is very similar to that of V1065~Cen
in the middle phase.  Therefore, we suppose that a similar effect
that we do not include in our model occurs for both novae.
We regard our model $V$ light curve to reproduce the $V$ light curve
of V2677~Oph except for the middle phase.
This again confirms that the distance modulus of $(m-M)_V=19.2$ is
reasonable.


\begin{figure*}
\plotone{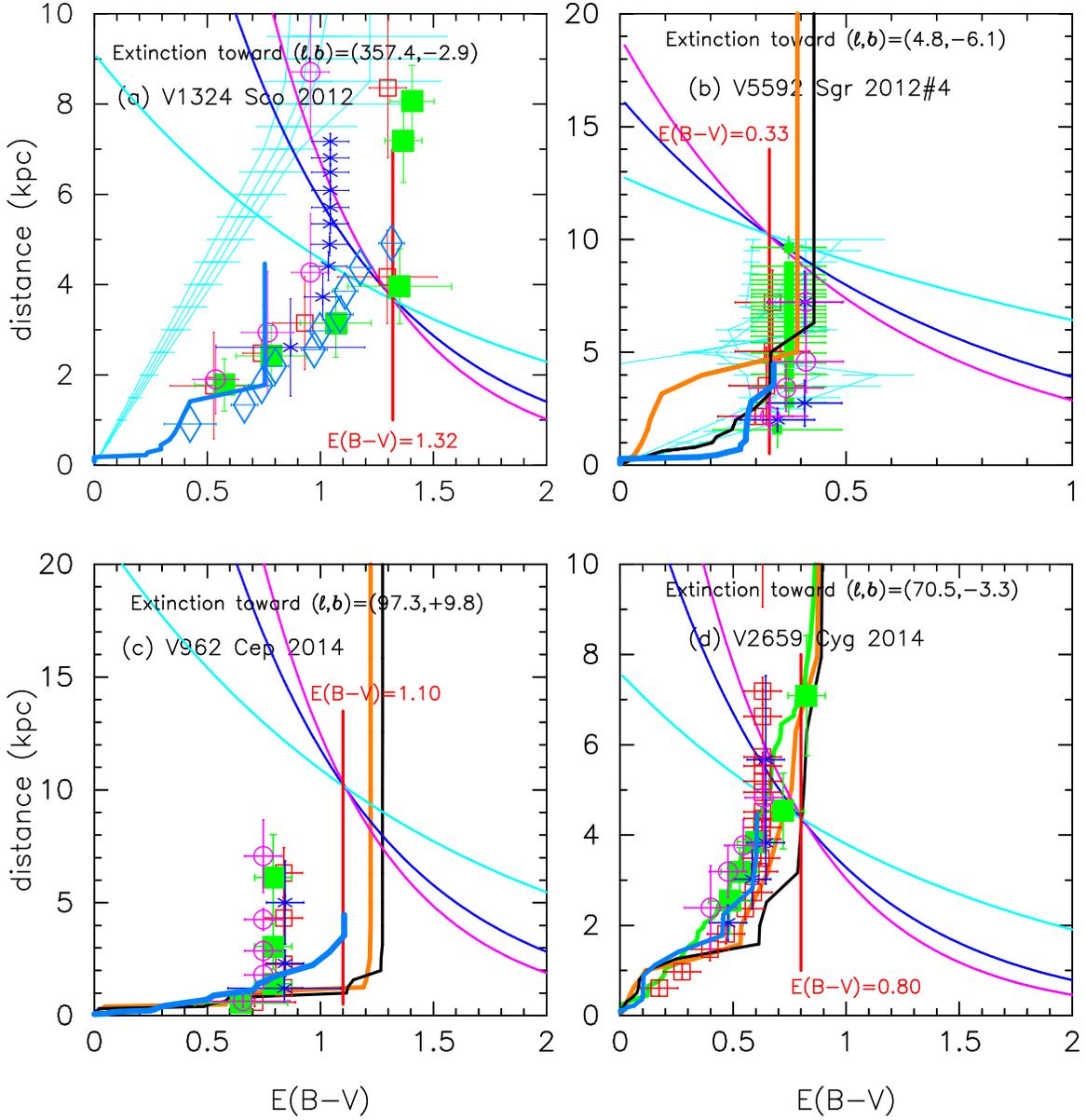}
\caption{
Same as Figure 
\ref{distance_reddening_v1663_aql_v5116_sgr_v2575_oph_v5117_sgr},
but for (a) V1324~Sco, (b) V5592~Sgr, (c) V962~Cep, and (d) V2659~Cyg.
\label{distance_reddening_v1324_sco_v5592_sgr_v962_cep_v2659_cyg}}
\end{figure*}


\begin{figure*}
\plotone{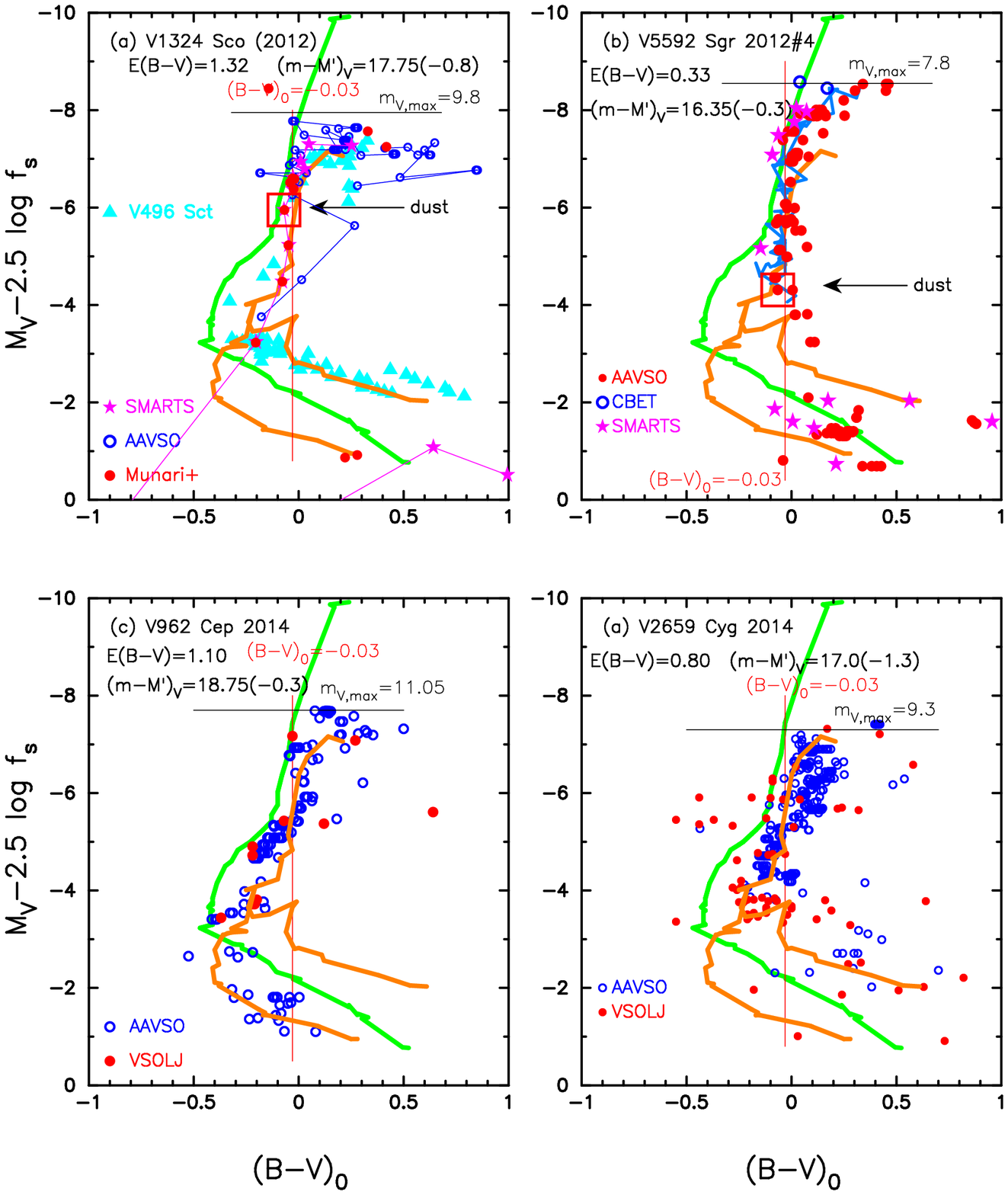}
\caption{
Same as Figure
\ref{hr_diagram_v1663_aql_v5116_sgr_v2575_oph_v5117_sgr_outburst},
but for (a) V1324~Sco, (b) V5592~Sgr, (c) V962~Cep, and (d) V2659~Cyg.
The solid orange/green lines show the template tracks of LV~Vul/V1500~Cyg,
respectively.  In panel (a), we add the track of V496~Sct with the 
filled cyan triangles.  In panel (b), we add the tracks of V1668~Cyg 
with the solid cyan-blue lines.    
\label{hr_diagram_v1324_sco_v5592_sgr_v962_cep_v2659_cyg_outburst}}
\end{figure*}

\subsection{V1324~Sco 2012}
\label{v1324_sco_cmd}
V1324~Sco reached $m_{V, \rm max}=9.8$ on JD~2,456,098.45 
(UT 2012 June 19.95) from the AAVSO data.
\citet{wag12} detected a periodicity of 1.6~hr with an amplitude of 0.1 mag
on UT 2012 May 28 and 31.  \citet{fin15} suggested a main-sequence
companion if this periodicity is the orbital period
($P_{\rm orb}\sim1.6$~hr or $P_{\rm orb}\sim3.2$~hr
for the ellipsoidal modulation).  \citet{ack14} detected gamma-rays
from this nova with the {\it Fermi}/Large Area Telescope (LAT).
Assuming the distance of $d=4.5$~kpc to V1324~Sco, they concluded that
V1324~Sco is the strongest among the detected gamma-ray sources 
V407~Cyg (2.7~kpc), V1324~Sco (4.5~kpc), V959~Mon (3.6~kpc),
and V339~Del (4.2~kpc).  However, the distances of novae are always debated.
For example, \citet{hac18k} have already redetermined the distance of 
V959~~Mon to be $d=2.5\pm0.5$~kpc based on the time-stretching method
and the time-stretched color-magnitude diagram.  This value is much smaller
than Ackerman et al.'s assumed value of 3.6~kpc for V959~Mon.
\citet{hac18kb} also redetermined the distance of V407~Cyg to be
$d= 3.9\pm0.5$~kpc, which is much larger than Ackerman et al.'s assumed
value of 2.7~kpc for V407~Cyg.
Ackerman et al. estimated the distance of V1324~Sco, $d= 4.5$~kpc,
from various MMRD relations, but the MMRD relations are
statistical relations and not that accurate for individual novae
\citep[see, e.g.,][]{schaefer18}.
\citet{mun15wh} obtained the reddening of $E(B-V)=1.23\pm0.12$ from
the colors of novae at maximum and $t_2$ time.
\citet{fin15} estimated the reddening of $E(B-V)=1.16\pm0.12$ and
the distance of $d > 6.5$~kpc.  \citet{fin18} presented the comprehensive
behavior of V1324~Sco.

We obtain $(m-M)_B=18.24$, $(m-M)_V=16.95$, and $(m-M)_I=14.78$,
which cross at $d=3.7$~kpc and $E(B-V)=1.32$,
in Appendix \ref{v1324_sco} and plot them in Figure 
\ref{distance_reddening_v1324_sco_v5592_sgr_v962_cep_v2659_cyg}(a).
Thus, we obtain $d=3.7\pm0.6$~kpc, $E(B-V)=1.32\pm0.1$,
$(m-M)_V=16.95\pm0.2$, and $f_{\rm s}=2.1$ against LV~Vul.

For the reddening toward V1324~Sco, 
i.e., $(l,b)=(357\fdg4255, -2\fdg8723)$,
the 2D NASA/IPAC galactic dust absorption map gives $E(B-V)=1.53\pm0.14$.
We suppose that the reddening does not saturate yet
at the distance of V1324~Sco as shown in Figure
\ref{distance_reddening_v1324_sco_v5592_sgr_v962_cep_v2659_cyg}(a).
Marshall et al.'s (2006) relations are plotted toward
$(l, b)=(357\fdg25,-2\fdg75)$, $(357\fdg50,-2\fdg75)$,
$(357\fdg25,-3\fdg00)$, and $(357\fdg50,-3\fdg00)$.
The closest direction is that of the filled green squares.
The other symbols/lines have the same meanings as those in Figure
\ref{distance_reddening_v1663_aql_v5116_sgr_v2575_oph_v5117_sgr}(a).
Our crossing point of $d=3.7$~kpc and $E(B-V)=1.32$ is
consistent with Marshall et al.'s
distance-reddening relations.

\citet{fin15} obtained the reddening toward V1324~Sco to be
$E(B-V)=1.16\pm0.12$ from five diffuse interstellar bands (DIBs)
and \ion{Na}{1}~D and \ion{K}{1} absorption features, and
obtained the distance of $d > 6.5$~kpc
in the conjunction with the 3D reddening map of \citet{schu14}.
We plot Schultheis et al.'s data by very thin cyan lines in Figure
\ref{distance_reddening_v1324_sco_v5592_sgr_v962_cep_v2659_cyg}(a).
Schultheis et al.'s distance-reddening relations
show different trends from Marshall et al.'s and \"Ozd\"ormez et al.'s 
relations.  We may not use Schultheis et al.'s
distance-reddening relations because their trends are sometimes largely
different from the trends of Marshall et al.'s, Green et al.'s, and
\"Ozd\"ormez et al.'s relations.

We plot the $(B-V)_0$-$(M_V-2.5 \log f_{\rm s})$ diagram of V1324~Sco in Figure
\ref{hr_diagram_v1324_sco_v5592_sgr_v962_cep_v2659_cyg_outburst}(a)
for $E(B-V)=1.32$ and $(m-M')_V=17.75$ from Equation 
(\ref{absolute_mag_v1324_sco}).
The track of V1324~Sco is very close to those of LV~Vul (orange line)
and V496~Sct (filled cyan triangles) until the dust blackout started.
Thus, we regard V1324~Sco to belong to the LV~Vul type.
This overlapping of V1324~Sco with LV~Vul and V496~Sct 
supports $E(B-V)=1.32$ and $(m-M')_V=17.75$,
that is, $E(B-V)=1.32\pm0.1$, $(m-M)_V=16.95\pm0.2$, $f_{\rm s}=2.1$,
and $d=3.7\pm0.6$~kpc.   

We check the distance modulus of $(m-M)_V=16.95$ by comparing
our model $V$ light curve with the observation in Figure
\ref{v1324_sco_lv_vul_v496_sct_v1369_cen_v_bv_ub_color_logscale}(a).  
We plot a $V$ model light curve of a $0.80~M_\sun$ WD 
\citep[CO2, solid red line;][]{hac10k}, 
assuming $(m-M)_V=16.95$ for V1324~Sco.
The $V$ light curves reasonably reproduce the observation.
This again confirms that the distance modulus of $(m-M)_V=16.95$ is
reasonable.

\subsection{V5592~Sgr 2012\#4}
\label{v5592_sgr_cmd}
V5592~Sgr reached $m_{V, \rm max}=7.8$ on JD~2,456,116.12
(UT 2012 July 7.62) from CBET No. 3166.  The nova was identified as
an \ion{Fe}{2} type by M. Fujii \citep{nis12}.

We obtain $(m-M)_B=16.38$, $(m-M)_V=16.06$, and $(m-M)_I=15.59$,
which cross at $d=10$~kpc and $E(B-V)=0.33$,
in Appendix \ref{v5592_sgr} and plot them in Figure 
\ref{distance_reddening_v1324_sco_v5592_sgr_v962_cep_v2659_cyg}(b).
Thus, we obtain $d=10\pm1$~kpc, $E(B-V)=0.33\pm0.05$,
$(m-M)_V=16.05\pm0.1$, and $f_{\rm s}=1.35$ against LV~Vul.

For the reddening toward V5592~Sgr, $(l,b)=(4\fdg8122, -6\fdg0895)$,
the 2D NASA/IPAC galactic dust absorption map gives $E(B-V)=0.33\pm0.01$,
which is consistent with our crossing point.
We examine our results
in Figure \ref{distance_reddening_v1324_sco_v5592_sgr_v962_cep_v2659_cyg}(b).
Marshall et al.'s (2006) relations are plotted 
toward $(l, b)=(4\fdg75,-6\fdg00)$, $(5\fdg00,-6\fdg00)$,
$(4\fdg75,-6\fdg25)$, and $(5\fdg00,-6\fdg25)$.
The closest direction is that of the unfilled red squares.
The other symbols/lines have the same meanings as those in Figure
\ref{distance_reddening_v1663_aql_v5116_sgr_v2575_oph_v5117_sgr}(a).
Our crossing point of $E(B-V)=0.33$ and $d=10$~kpc is consistent 
with Marshall et al.'s (unfilled red squares) distance-reddening relation.
The reddening toward V5592~Sgr already saturates at the distance of $10$~kpc.

We plot the $(B-V)_0$-$(M_V-2.5 \log f_{\rm s})$ diagram of V5592~Sgr in Figure
\ref{hr_diagram_v1324_sco_v5592_sgr_v962_cep_v2659_cyg_outburst}(b)
for $E(B-V)=0.33$ and $(m-M')_V=16.35$ in Equation
(\ref{absolute_mag_v5592_sgr}).
The track of V5592~Sgr almost follows the V1668~Cyg track until
the dust blackout starts.
Thus, we regard V5592~Sgr to belong to the LV~Vul type,
because V1668~Cyg belongs to the LV~Vul type \citep{hac19k}.
This overlapping of V5592~Sgr with V1668~Cyg supports $E(B-V)=0.33$
and $(m-M')_V=16.35$, that is, $E(B-V)=0.33\pm0.05$, 
$(m-M)_V=16.05\pm0.1$, $f_{\rm s}=1.35$, and $d=10\pm1$~kpc.

We check the distance modulus of $(m-M)_V=16.05$ by comparing
our model $V$ light curve with the observation in Figure
\ref{v5592_sgr_v705_cas_lv_vul_v_bv_ub_color_logscale}(a).  
We plot a $V$ model light curve of a $0.93~M_\sun$ WD 
\citep[CO4, solid red line;][]{hac15k}, 
assuming $(m-M)_V=16.05$ for V5592~Sgr.
The $V$ light curves reasonably reproduce the observation.
This again confirms that the distance modulus of $(m-M)_V=16.05$ is
reasonable.

\subsection{V962~Cep 2014}
\label{v962_cep_cmd}
V962~Cep reached $m_{V,\rm max}=11.05$ on UT 2014 March 13.92 
(JD~2,456,730.42) from the AAVSO data.  Then, the nova
declined with $t_2=22\pm2$ days and $t_3=42\pm1$~days \citep{sri15}.
\citet{sri15} obtained the reddening of $E(B-V)=0.935$ from
the empirical relation of $(B-V)_0= 0.23\pm0.06$ at optical maximum
and $(B-V)_0=-0.02\pm0.04$ at $t_2$ time \citep{van87}.
Using the MMRD relation proposed by \citet{del95}, \citet{sri15} obtained
the distance of $d=15.8\pm4$~kpc.  They identified V962~Cep as an 
\ion{Fe}{2} type.

We obtain $(m-M)_B= 19.58$, $(m-M)_V= 18.46$, and $(m-M)_I= 16.67$,
which cross at $d=10.2$~kpc and $E(B-V)=1.10$,
in Appendix \ref{v962_cep} and plot them in Figure
\ref{distance_reddening_v1324_sco_v5592_sgr_v962_cep_v2659_cyg}(c).
Thus, we have $E(B-V)=1.10\pm0.1$ and $d=10.2\pm2$~kpc.
The distance modulus in the $V$ band is $(m-M)_V=18.45\pm0.2$.

For the reddening toward V962~Cep, \\$(l,b)=(97\fdg3128, +9\fdg8202)$,
the 2D NASA/IPAC galactic dust absorption map gives $E(B-V)=0.96\pm0.02$,
which is slightly smaller than our value at the crossing point.
We examine our result
in 
Figure \ref{distance_reddening_v1324_sco_v5592_sgr_v962_cep_v2659_cyg}(c).
Marshall et al.'s (2006) relations are plotted
toward $(l, b)=(97\fdg25, +9\fdg75)$, $(97\fdg50, +9\fdg75)$,
$(97\fdg25, +10\fdg00)$, and $(97\fdg50, +10\fdg00)$.
The closest direction is that of the unfilled red squares.
The other symbols/lines have the same meanings as those in Figure
\ref{distance_reddening_v1663_aql_v5116_sgr_v2575_oph_v5117_sgr}(a).
Our crossing point at $E(B-V)=1.10$ and $d=10.2$~kpc is consistent with
Chen et al.'s (cyan-blue line) but located between Marshall et al.'s 
(unfilled red squares with error bars) and Green et al.'s 
(black/orange lines) relations.

We plot the $(B-V)_0$-$(M_V-2.5 \log f_{\rm s})$ diagram of V962~Cep in Figure
\ref{hr_diagram_v1324_sco_v5592_sgr_v962_cep_v2659_cyg_outburst}(c)
for $E(B-V)=1.10$ and $(m-M')_V=18.75$ in Equation
(\ref{absolute_mag_v962_cep}).
The track of V962~Cep almost follows the LV~Vul track, although the data
points are quite scattered.  Thus, we regard V962~Cep to belong to
the LV~Vul type.
This rough overlapping of V962~Cep to LV~Vul supports our values of 
$E(B-V)=1.10$ and $(m-M')_V=18.75$, that is, $E(B-V)=1.10\pm0.1$, 
$(m-M)_V=18.45\pm0.2$, $d=10.2\pm2$~kpc, and $f_{\rm s}=1.32$ against LV~Vul.

We examine the distance modulus of $(m-M)_V=18.45$ for V962~Cep in Figure
\ref{v962_cep_v1668_cyg_lv_vul_v_bv_ub_logscale}(a). 
We plot an absolute $V$ model light curve (solid red line) 
of a $0.95~M_\sun$ WD \citep[CO3;][]{hac16k},
assuming that $(m-M)_V=18.45$ for V962~Cep.
The $V$ light curve reasonably reproduces the observation.
This confirms that the distance modulus of $(m-M)_V=18.45$ is
reasonable.  For comparison,
we also plot a $0.98~M_\sun$ WD model light curve (solid green lines),
assuming that $(m-M)_V=14.6$ for V1668~Cyg.

\subsection{V2659~Cyg 2014}
\label{v2659_cyg_cmd}
V2659~Cyg reached $m_{V,\rm max}=9.3$ on JD~2,456,757.27 
(UT 2014 April 9.77) from the VSOLJ data.
The nova was identified as an \ion{Fe}{2} type by A. Arai, K. Ayani,
and M. Fujii \citep{nis14}.  \citet{raj14} obtained the reddening
of $E(\bv)=0.63$ from the optically thin \ion{K}{1} 7699 line.
\citet{cho14} obtained the reddening of $E(\bv)=0.77\pm0.06$ from
the arithmetic average of the values obtained with various methods including
the intrinsic colors at maximum and $t_2$ time \citep{van87}, 
the equivalent widths of \ion{Na}{1} and \ion{K}{1}, and Raj et al.'s value.
\citet{cho14} also derived the distance of $d=5.5\pm0.3$~kpc from the MMRD
relation \citep{dow00}.

We obtain $(m-M)_B=16.5$, $(m-M)_V=15.68$, and $(m-M)_I=14.41$,
which cross at $d=4.4$~kpc and $E(B-V)=0.80$,
in Appendix \ref{v2659_cyg} and plot them in Figure 
\ref{distance_reddening_v1324_sco_v5592_sgr_v962_cep_v2659_cyg}(d).
Thus, we obtain $d=4.4\pm0.5$~kpc, $E(B-V)=0.80\pm0.05$,
$(m-M)_V=15.7\pm0.1$, and $f_{\rm s}=3.3$ against LV~Vul.

For the reddening toward V2659~Cyg, $(l,b)=(70\fdg5250,  -3\fdg2860)$,
the 2D NASA/IPAC galactic dust absorption map gives $E(B-V)=0.85\pm0.02$.
This value is slightly larger than our value at the crossing point.
This is because the reddening does not saturate yet at the distance
of V2659~Cyg as shown in Figure
\ref{distance_reddening_v1324_sco_v5592_sgr_v962_cep_v2659_cyg}(d).
We examine our result in this plot.
Marshall et al.'s (2006) relations are plotted
toward $(l, b)=(70\fdg50, -3\fdg25)$, $(70\fdg75, -3\fdg25)$,
$(70\fdg50, -3\fdg50)$, and $(70\fdg75, -3\fdg50)$.
The closest direction is that of the unfilled red squares.
The other symbols/lines have the same meanings as those in Figure
\ref{distance_reddening_v1663_aql_v5116_sgr_v2575_oph_v5117_sgr}(a).
Our crossing point at $E(B-V)=0.80$ and $d=4.4$~kpc is roughly
consistent with Marshall et al.'s (filled green squares) 
and Green et al.'s (solid black line) relations.

We plot the $(B-V)_0$-$(M_V-2.5 \log f_{\rm s})$ diagram of V2659~Cyg in Figure
\ref{hr_diagram_v1324_sco_v5592_sgr_v962_cep_v2659_cyg_outburst}(d)
for $E(B-V)=0.80$ and $(m-M')_V=17.0$ in Equation
(\ref{absolute_mag_v2659_cyg}).
The track of V2659~Cyg almost follows the LV~Vul track.
Thus, we regard V2659~Cyg to belong to the LV~Vul type in the 
$(B-V)_0$-$(M_V-2.5 \log f_{\rm s})$ diagram.  This overlapping supports 
$E(B-V)=0.80$ and $(m-M')_V=17.0$, that is,
$E(B-V)=0.80\pm0.05$, $(m-M)_V=15.7\pm0.1$, $f_{\rm s}=3.3$,
and $d=4.4\pm0.5$~kpc.

We examine the distance modulus of $(m-M)_V=15.7$ for V2659~Cyg in Figure
\ref{v2659_cyg_v5666_sgr_v1369_cen_v496_sct_lv_vul_v_bv_ub_color_logscale}(a). 
We plot a $V$ model light curve of a $0.75~M_\sun$ WD 
\citep[CO4, solid red line;][]{hac15k}, 
assuming that $(m-M)_V=15.7$ for V2659~Cyg. 
The $V$ light curve reasonably reproduces the observation.
This confirms that the distance modulus of $(m-M)_V=15.7\pm0.1$ is
reasonable.


\begin{figure*}
\plotone{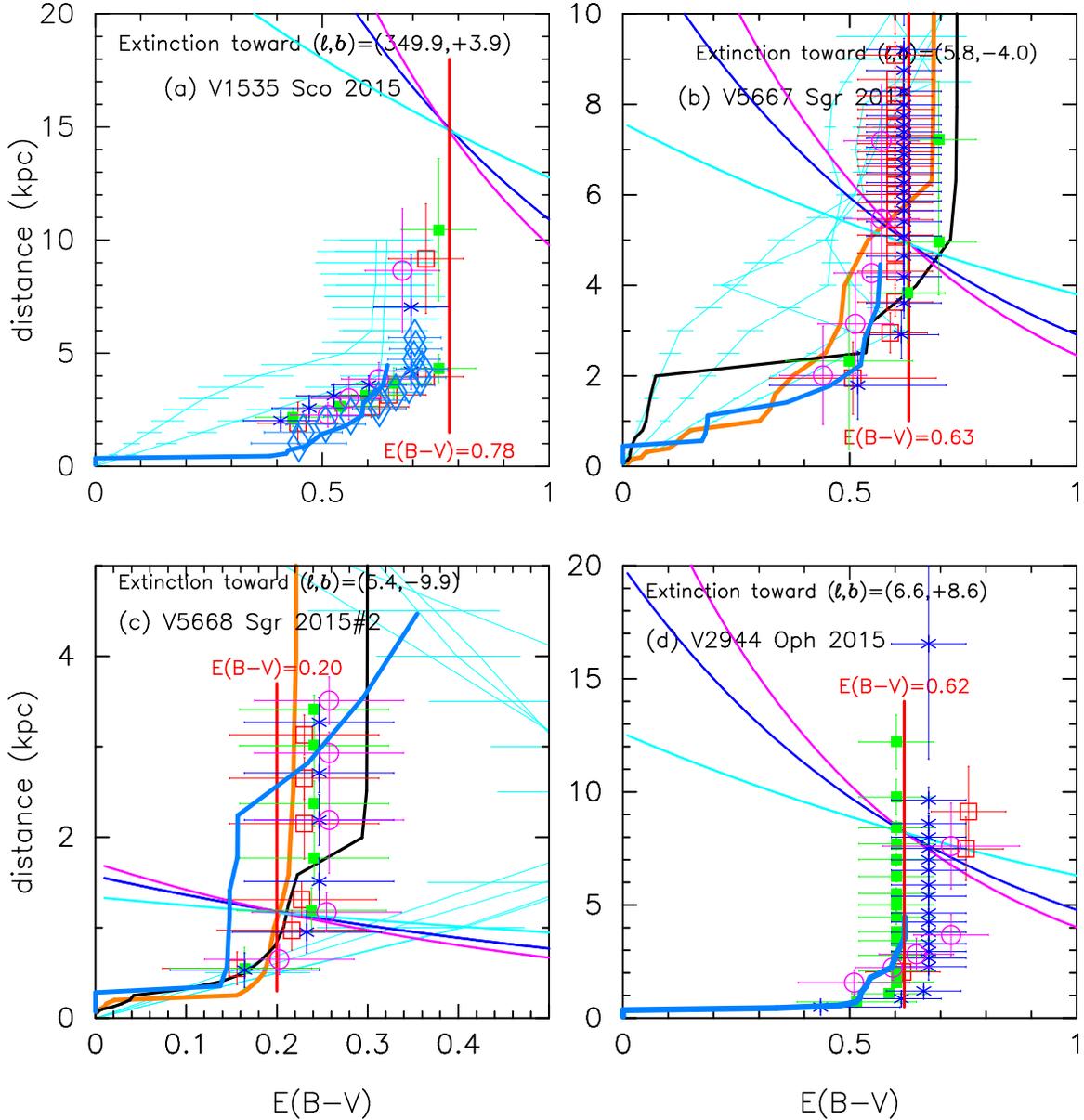}
\caption{
Same as Figure 
\ref{distance_reddening_v1663_aql_v5116_sgr_v2575_oph_v5117_sgr},
but for (a) V1535~Sco, (b) V5667~Sgr, (c) V5668~Sgr, and (d) V2944~Oph.
\label{distance_reddening_v1535_sco_v5667_sgr_v5668_sgr_v2944_oph}}
\end{figure*}


\begin{figure*}
\plotone{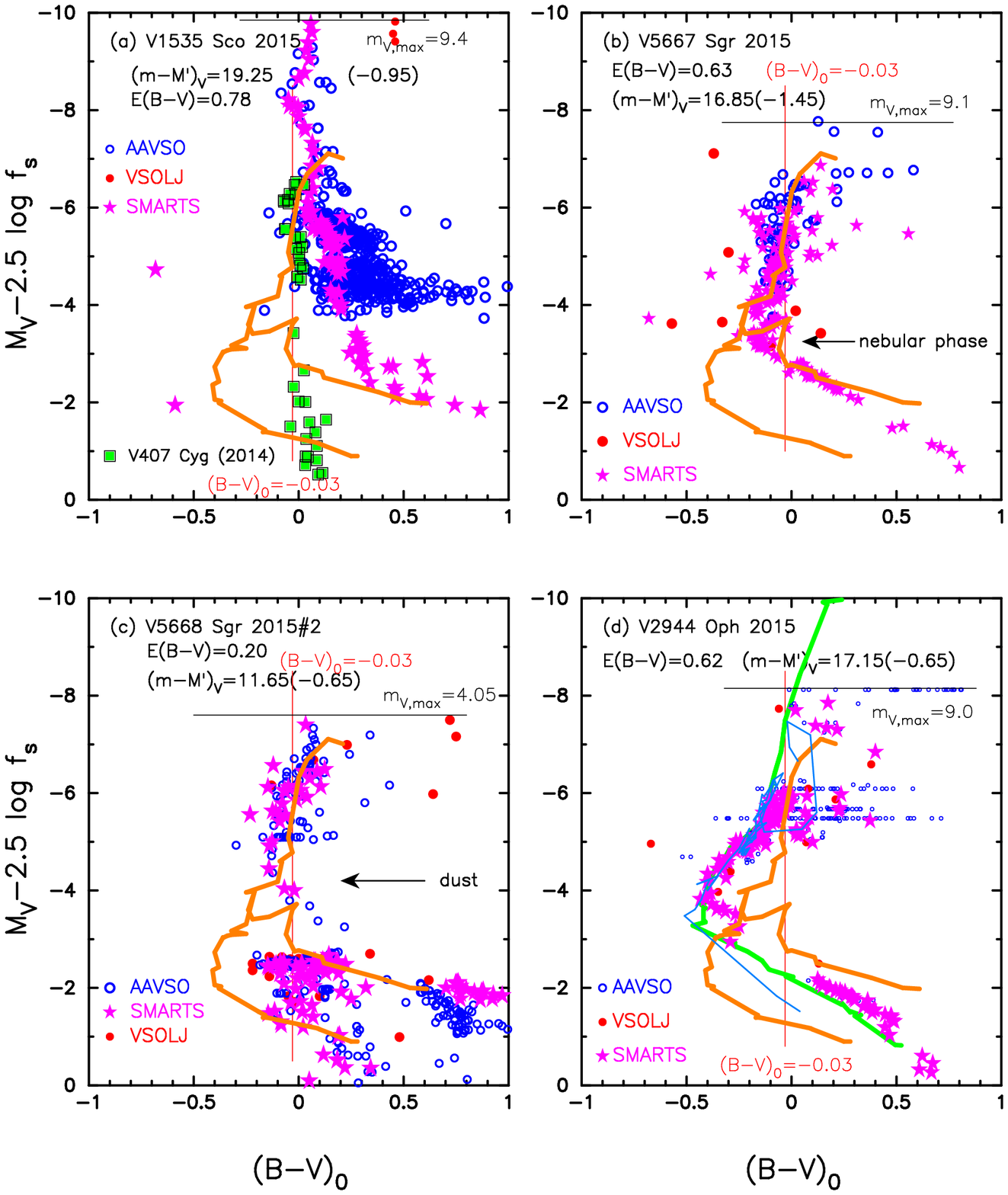}
\caption{
Same as Figure 
\ref{hr_diagram_v1663_aql_v5116_sgr_v2575_oph_v5117_sgr_outburst},
but for (a) V1535~Sco, (b) V5667~Sgr, (c) V5668~Sgr, and (d) V2944~Oph.
The thick solid orange lines show the template track of LV~Vul.
The thick solid green line shows the template track of V1500~Cyg.
In panel (a), we add the track of V407~Cyg (filled green squares with
black outline).
In panel (d), we add the track of PW~Vul (thin solid cyan-blue line),
the data of which are the same as those in Figure 7(b) of \citet{hac16kb}.
\label{hr_diagram_v1535_sco_v5667_sgr_v5668_sgr_v2944_oph_outburst}}
\end{figure*}

\subsection{V1535~Sco 2015}
\label{v1535_sco_cmd}
V1535~Sco reached $m_{V, \rm max}=9.4$ on JD~2,457,065.34 
(UT 2015 February 11.84) based on the VSOLJ data.
\citet{wal15a} suggested that the companion to the exploding WD
is a red giant with the brightness of $J=13.4$, $H=12.5$, and $K=12.2$
from the 2MASS archival database \citep[see also][]{waa15}.
\citet{sri15} estimated the position in the color-color diagram
($(J-H)_0$ versus $(H-K)_0$) in quiescence, and it is consistent with 
a giant of type K3III/K4III, although their spectral energy distribution
(SED) analysis gave a different type of giant, M4III/M5III.
\citet{nel15} detected hard X-rays from hot plasma, showing a strong
deceleration of the ejecta by circumstellar matter (CSM).
\citet{sri15} obtained $t_2=14\pm2$~days and $t_3=19\pm1$~days
from the $V$ light curve of AAVSO data and
estimated the reddening of $E(B-V)=0.72$ from the empirical
relations of \citet{van87}. They also estimated the distance
of $d=13.7\pm0.4$~kpc from the MMRD relation of \citet{del95}
and $d=14.7\pm3.8$~kpc from the MMRD relation of \citet{dow00}.
\citet{linf17} reported a comprehensive observation of V1535~Sco and
concluded that a strong shock was present two weeks earlier, and
the companion star is likely a K giant from the optical, NIR, and 
X-ray neutral hydrogen column density measurements.
\citet{mun17} also proposed that the companion is likely a K3--4III giant
from the broadband spectra of the progenitor.

We obtain $(m-M)_B= 19.04$, $(m-M)_V= 18.3$, and $(m-M)_I= 17.03$,
which cross at $d=15$~kpc and $E(B-V)=0.78$,
in Appendix \ref{v1535_sco} and plot them in Figure
\ref{distance_reddening_v1535_sco_v5667_sgr_v5668_sgr_v2944_oph}(a).
Thus, we have $E(B-V)=0.78\pm0.05$ and $d=15\pm2$~kpc.
The distance modulus in the $V$ band is $(m-M)_V=18.3\pm0.1$.

For the reddening toward V1535~Sco, $(l,b)=(349\fdg8984, +3\fdg9382)$,
the 2D NASA/IPAC galactic dust absorption map gives $E(B-V)=0.65\pm0.04$,
which is slightly smaller than our value at the crossing point.
We further examine our result
in Figure 
\ref{distance_reddening_v1535_sco_v5667_sgr_v5668_sgr_v2944_oph}(a). 
Marshall et al.'s (2006) relations are plotted
toward $(l, b)=(349\fdg75, +3\fdg75)$, $(350\fdg00, +3\fdg75)$,
$(349\fdg75, +4\fdg00)$, and $(350\fdg00, +4\fdg00)$.
The direction toward V1535~Sco is in the middle of these four directions.
The other symbols/lines have the same meanings as those in Figure
\ref{distance_reddening_v1663_aql_v5116_sgr_v2575_oph_v5117_sgr}(a).
Our crossing point at $E(B-V)=0.78$ and $d=15$~kpc is roughly
consistent with Marshall et al.'s relation (filled green squares with
error bars).

We plot the $(B-V)_0$-$(M_V-2.5 \log f_{\rm s})$ diagram of V1535~Sco
in Figure \ref{hr_diagram_v1535_sco_v5667_sgr_v5668_sgr_v2944_oph_outburst}(a)
for $E(B-V)=0.78$ and $(m-M')_V= 19.25$ in Equation
(\ref{absolute_mag_v1535_sco}).
The track of V1535~Sco goes almost straight down, and this kind of behavior
is common among the symbiotic classical novae that have a red-giant
companion \citep[see, e.g.,][]{hac18kb}.
The track of V1535~Sco follows the LV~Vul and V407~Cyg tracks
in the early phase, although the peak of V1535~Sco is quite bright
compared with that of V407~Cyg and LV~Vul.  In the later phase,
the track of V1535~Sco is close to the upper branch of LV~Vul.
These similarities and overlapping support
our value of $E(B-V)=0.78$ and $(m-M')_V=19.25$, that is,
$E(B-V)=0.78\pm0.05$, $d= 15\pm2$~kpc, $(m-M)_V=18.3\pm0.1$, and
$f_{\rm s}= 2.4$ against LV~Vul.  We regard V1535~Sco to belong to
the LV~Vul type.

We check the distance modulus of $(m-M)_V=18.3$ by comparing our model
light curve with the observation in Figure 
\ref{v1535_sco_lv_vul_v1668_cyg_v2468_cyg_v_bv_logscale}(a).
We plot a model $V$ light curve of a $0.85~M_\sun$ WD 
\citep[CO4, solid red line;][]{hac15k},
assuming $(m-M)_V=18.3$ for V1535~Sco.
The model absolute $V$ light curve reasonably fits with the observed 
apparent $V$ light curve of V1535~Sco.
This again confirms that $(m-M)_V=18.3$ is reasonable for V1535~Sco.

\subsection{V5667~Sgr 2015\#1}
\label{v5667_sgr_cmd}
Walter identified the nova to be of \ion{Fe}{2} type \citep{nis15}.
\citet{rud15} reported the $0.38-2.5~\mu$m spectroscopy of V5667~Sgr 
and V5668~Sgr.  They derived the extinction of $E(B-V)= 0.2\pm0.15$ 
for V5667~Sgr and $E(B-V)= 0.5\pm0.2$ for V5668 Sgr, both 
from the \ion{O}{1} line.  However, we suppose that the reported
extinctions are typographical errors 
and should be $E(B-V)= 0.5\pm0.2$ for V5667~Sgr and
$E(B-V)= 0.2\pm0.15$ for V5668~Sgr as shown in Figure
\ref{distance_reddening_v1535_sco_v5667_sgr_v5668_sgr_v2944_oph}(b) and (c).   

We obtain $(m-M)_B=16.08$, $(m-M)_V=15.42$, and $(m-M)_I=14.39$,
which cross at $d=4.9$~kpc and $E(B-V)=0.63$,
in Appendix \ref{v5667_sgr} and plot them in Figure
\ref{distance_reddening_v1535_sco_v5667_sgr_v5668_sgr_v2944_oph}(b).
Thus, we obtain $d=4.9\pm0.5$~kpc, $E(B-V)=0.63\pm0.05$,
$(m-M)_V=15.4\pm0.2$, and $f_{\rm s}=3.7$ against LV~Vul.

For the reddening toward V5667~Sgr, $(l,b)=(5\fdg8040, -4\fdg0427)$,
the 2D NASA/IPAC galactic dust absorption map gives $E(B-V)=0.63\pm0.02$,
which is consistent with our value of $E(B-V)=0.63\pm0.05$.
Our value of $E(B-V)=0.63$ is also consistent with Rudy et al.'s (2015)
result of $E(B-V)= 0.5\pm0.2$ from the \ion{O}{1} line.
We examine our results based on distance-reddening relations in Figure 
\ref{distance_reddening_v1535_sco_v5667_sgr_v5668_sgr_v2944_oph}(b). 
Marshall et al.'s (2006) relations are plotted
toward $(l, b)=(5\fdg75,-4\fdg00)$, $(6\fdg00,-4\fdg00)$,
$(5\fdg75,-4\fdg25)$, and $(6\fdg00,-4\fdg25)$.
The closest direction is that of the unfilled red squares.
The other symbols/lines have the same meanings as those in Figure
\ref{distance_reddening_v1663_aql_v5116_sgr_v2575_oph_v5117_sgr}(a).
Our crossing point at $E(B-V)=0.63$ and $d=4.9$~kpc is consistent with
both Marshall et al.'s (unfilled red squares and blue asterisks) and
Green et al.'s (orange line) relations.

We plot the $(B-V)_0$-$(M_V-2.5 \log f_{\rm s})$ diagram of V5667~Sgr in Figure
\ref{hr_diagram_v1535_sco_v5667_sgr_v5668_sgr_v2944_oph_outburst}(b)
for $E(B-V)=0.63$ and $(m-M')_V=16.85$ in Equation
(\ref{absolute_mag_v5667_sgr}).
The track of V5667~Sgr almost follows the LV~Vul track,
so we regard V5667~Sgr to belong to the LV~Vul type in the
$(B-V)_0$-$(M_V-2.5 \log f_{\rm s})$ diagram.
This overlapping supports $E(B-V)=0.63$ and $(m-M')_V=16.85$,
that is, $E(B-V)=0.63\pm0.05$, $(m-M)_V=15.4\pm0.2$, 
$f_{\rm s}=3.7$, and $d=4.9\pm0.5$~kpc.  

We check the distance modulus of $(m-M)_V=15.4$ by comparing our model
light curve with the observation in Figure 
\ref{v5667_sgr_lv_vul_v5666_sgr_v1369_cen_v496_sct_v_bv_ub_color_logscale}(a).
We plot a model $V$ light curve of a $0.78~M_\sun$ WD \citep[CO4, 
solid red line;][]{hac15k}, assuming that $(m-M)_V=15.4$ for V5667~Sgr.
The model absolute $V$ light curve reasonably follows the observed 
apparent $V$ light curve, although the $V$ light curve has a wavy 
structure in the early phase.
This again confirms that $(m-M)_V=15.4$ is reasonable.

\subsection{V5668~Sgr 2015\#2}
\label{v5668_sgr_cmd}
V5668~Sgr reached
$m_{V, \rm max}=4.05$ on JD~2,457,102.9 from the SMARTS data.
\citet{rud15} reported $0.38-2.5~\mu$m spectroscopy of V5667~Sgr 
and V5668~Sgr and derived the extinction of $E(B-V)= 0.2\pm0.15$ 
for V5667~Sgr and $E(B-V)= 0.5\pm0.2$ for V5668 Sgr, both 
from the \ion{O}{1} line.  As mentioned in the previous subsection
(Section \ref{v5667_sgr_cmd}), we assume that the reported
extinctions are typographical errors
and should be $E(B-V)= 0.5\pm0.2$ for V5667~Sgr and
$E(B-V)= 0.2\pm0.15$ for V5668~Sgr as shown in Figure
\ref{distance_reddening_v1535_sco_v5667_sgr_v5668_sgr_v2944_oph}(b) and (c).
Gamma-ray emission was detected  with the {\it Fermi}/LAT 
on UT 2015 March 15.6 \citep[JD~2,457,097.1;][]{che16}.
\citet{taj16} reported the detection of $^7$Be in their Subaru HDS spectrum
obtained on UT 2015 May 29.67 (JD 2,457,103.17). 
\citet{ban16} reported their NIR observation and classified the nova as an
\ion{Fe}{2}.  They also estimated the distance of $d=1.54$~kpc
from the blackbody angular diameter of a dust shell, that is, 42 mas.

We obtain $(m-M)_B=11.16$, $(m-M)_V=10.98$, and $(m-M)_I=10.64$,
which cross at $d=1.2$~kpc and $E(B-V)=0.20$,
in Appendix \ref{v5668_sgr} and plot them in Figure 
\ref{distance_reddening_v1535_sco_v5667_sgr_v5668_sgr_v2944_oph}(c).
Thus, we obtain $d=1.2\pm0.2$~kpc, $E(B-V)=0.20\pm0.03$,
$(m-M)_V=11.0\pm0.2$, and $f_{\rm s}=1.86$ against LV~Vul.

For the reddening toward V5668~Sgr, 
$(l,b)=(5\fdg3799, -9\fdg8668)$,
the 2D NASA/IPAC galactic dust absorption map gives $E(B-V)=0.204\pm0.002$,
which is consistent with our value of $E(B-V)=0.20\pm0.05$.
We examine our results based on the distance-reddening relations in Figure 
\ref{distance_reddening_v1535_sco_v5667_sgr_v5668_sgr_v2944_oph}(c). 
Marshall et al.'s (2006) relations are plotted
toward $(l, b)=(5\fdg25, -9\fdg75)$, $(5\fdg50, -9\fdg75)$,
$(5\fdg25, -10\fdg00)$, and $(5\fdg50, -10\fdg00)$.
The direction toward V5668~Sgr is in the middle of these four directions.
The other symbols/lines have the same meanings as those in Figure
\ref{distance_reddening_v1663_aql_v5116_sgr_v2575_oph_v5117_sgr}(a).
Our crossing point at $E(B-V)=0.20$ and $d=1.2$~kpc is 
consistent with Marshall et al.'s and Green et al.'s relations.

We plot the $(B-V)_0$-$(M_V-2.5 \log f_{\rm s})$ diagram of V5668~Sgr in Figure
\ref{hr_diagram_v1535_sco_v5667_sgr_v5668_sgr_v2944_oph_outburst}(c)
for $E(B-V)=0.20$ and $(m-M')_V=11.65$ in Equation
(\ref{absolute_mag_v5668_sgr}).
The track of V5668~Sgr is just on the LV~Vul track (thick orange lines)
until the start of the dust blackout phase.
Therefore, V5668~Sgr belongs to the LV~Vul type.
This similarity supports
our adopted values of $E(B-V)=0.20$ and $(m-M')_V=11.65$, that is,
$E(B-V)=0.20\pm0.03$ and $(m-M)_V=11.0\pm0.2$, $f_{\rm s}=1.86$,
and $d=1.2\pm0.2$~kpc.

We check the distance modulus of $(m-M)_V=11.0$ by comparing our model
light curve with the observation in Figure 
\ref{v5668_sgr_lv_vul_v496_sct_v1369_cen_v_bv_ub_color_logscale}(a).
We plot a model $V$ light curve of a $0.85~M_\sun$ WD \citep[CO3, 
solid red line;][]{hac16k}, assuming that $(m-M)_V=11.0$ for V5668~Sgr.
The model absolute $V$ light curve reasonably follows the observed 
apparent $V$ light curve, although the $V$ light curve has a wavy 
structure in the early phase.
This again confirms that $(m-M)_V=11.0$ is reasonable.

\subsection{V2944~Oph 2015}
\label{v2944_oph_cmd}
V2944~Oph reached
$m_{V, \rm max}=9.0$ on JD~2,457,127.245 from the AAVSO data.
The nova was identified to be of He/N type in the early pre-maximum
phase \citep[$\sim 3$ mag below the maximum brightness;][]{dani15h},
but later as an \ion{Fe}{2} type at maximum \citep[e.g.,][]{mun15w}.
\citet{mun15w} obtained $E(B-V)=0.52$ from the relation of \citet{mun14}
or $E(B-V)=0.56$ from the relation of \citet{kos13} between the reddening
and the equivalent width of the diffuse interstellar band (DIB) at 6614\AA.
\citet{taj16} identified the line of unstable $^7$Be in the spectrum
of V2944~Oph obtained on UT 2015 July 3 (JD~2,457,206.5) about 80 days
after the outburst.  This clearly suggests that thermonuclear
runaway occurred deep in the hydrogen-rich envelope,
and its product was blown in the outburst wind.
\citet{sri16} obtained NIR spectra of V2944~Oph.  Their power-law fit
to the spectra showed a fairly constant slope, which differs from the 
usual trend expected during a nova spectral evolution.  They obtained
$E(B-V)=0.59\pm0.04$ from the strengths of the $0.8446~\mu$m
and $1.1287~\mu$m lines. 

We obtain $(m-M)_B=17.11$, $(m-M)_V=16.5$, and $(m-M)_I=15.5$, 
which cross at $d=8.2$~kpc and $E(B-V)=0.62$,
in Appendix \ref{v2944_oph} and plot them in Figure 
\ref{distance_reddening_v1535_sco_v5667_sgr_v5668_sgr_v2944_oph}(d).
Thus, we obtain $d=8.2\pm1$~kpc, $E(B-V)=0.62\pm0.05$,
$(m-M)_V=16.5\pm0.2$, and $f_{\rm s}=1.78$ against LV~Vul.

For the reddening toward V2944~Oph, 
$(l,b)=(6\fdg6421, +8\fdg5773)$,
the 2D NASA/IPAC galactic dust absorption map gives $E(B-V)=0.55\pm0.02$.
\citet{mun15w} obtained $E(B-V)=0.52$ and $E(B-V)=0.56$ from the
6614 \AA\  diffuse interstellar band while \citet{sri16} reported
$E(B-V)=0.59\pm0.04$ from the strengths of the $0.8446~\mu$m and $1.1287~\mu$m
lines as mentioned above.  These values are roughly consistent with
our obtained value of $E(B-V)=0.62\pm0.05$.
We examine our results 
in Figure \ref{distance_reddening_v1535_sco_v5667_sgr_v5668_sgr_v2944_oph}(d).
Marshall et al.'s (2006) relations are plotted
toward $(l, b)=(6\fdg50, +8\fdg75)$, $(6\fdg75, +8\fdg75)$,
$(6\fdg50, +8\fdg50)$, and $(6\fdg75, +8\fdg50)$.
The direction toward V2944~Oph is in the middle of these four directions.
The other symbols/lines have the same meanings as those in Figure
\ref{distance_reddening_v1663_aql_v5116_sgr_v2575_oph_v5117_sgr}(a).
Our crossing point at $E(B-V)=0.62$ and $d=8.2$~kpc is consistent with
Marshall et al.'s (filled green squares) and Chen et al.'s (cyan-blue line)
relations.

We plot the $(B-V)_0$-$(M_V-2.5 \log f_{\rm s})$ diagram of V2944~Oph in Figure
\ref{hr_diagram_v1535_sco_v5667_sgr_v5668_sgr_v2944_oph_outburst}(d)
for $E(B-V)=0.62$ and $(m-M')_V=17.15$ in Equation 
(\ref{absolute_mag_v2944_oph}).
The track of V2944~Oph is almost on the V1500~Cyg (thick solid green line)
and PW~Vul tracks  (cyan-blue lines).
Therefore, V2944~Oph belongs to the V1500~Cyg type.
This similarity supports
our adopted values of $E(B-V)=0.62$ and $(m-M')_V=17.15$, that is,
$E(B-V)=0.62\pm0.05$ and $(m-M)_V=16.5\pm0.2$, $f_{\rm s}=1.78$,
and $d=8.2\pm1$~kpc.

We check the distance modulus of $(m-M)_V=16.5$ by comparing our model
light curve with the observation in Figure 
\ref{v2944_oph_v5666_sgr_v1369_cen_v496_sct_v_bv_ub_color_logscale}(a).
We plot a model $V$ light curve of a $0.85~M_\sun$ WD \citep[CO3, 
solid red line;][]{hac16k}, assuming that $(m-M)_V=16.5$ for V2944~Oph.
The model absolute $V$ light curve reasonably follows the observed 
apparent $V$ light curve, although the $V$ light curve has a wavy 
structure in the early phase.
This again confirms that $(m-M)_V=16.5$ is reasonable.


\begin{figure}
\plotone{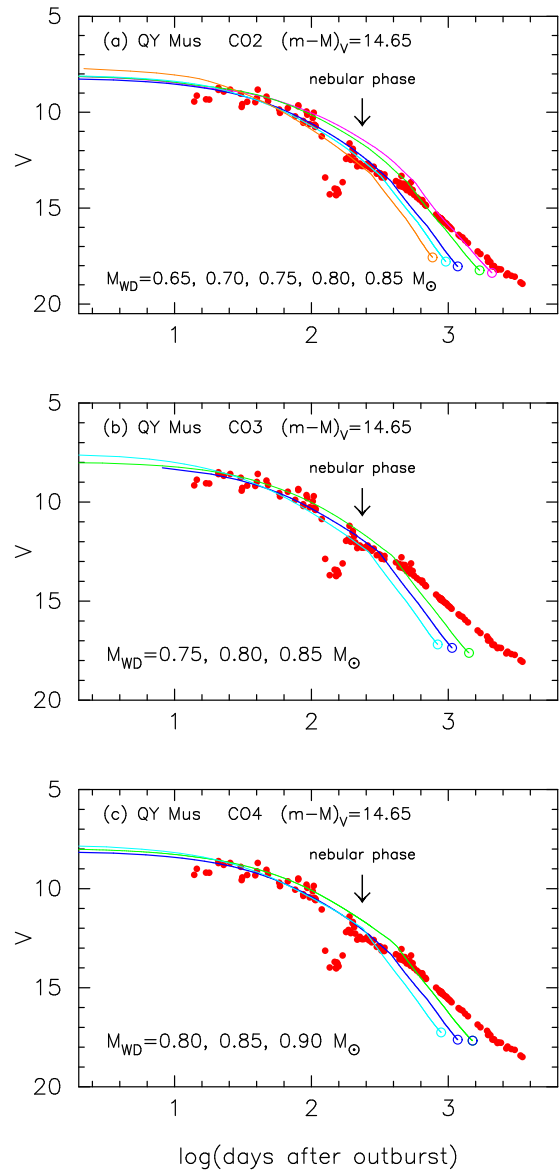}
\caption{
Model light-curve fitting with the $V$ light curve of QY~Mus for
the assumed chemical composition of (a) CO nova 2 \citep[CO2;][]{hac10k}, 
(b) CO nova 3 \citep[CO3;][]{hac16k}, 
and (c) CO nova 4 \citep[CO4;][]{hac15k}.
We assume that $(m-M)_V=14.65$ for QY~Mus.
In panel (a), we plot five WD mass models, i.e.,
$0.65$ (magenta), $0.70$ (green), $0.75$ (blue), $0.80$ (cyan),
and $0.85~M_\sun$ WD (orange line).
In panel (b), we plot three WD mass models, i.e.,
$0.75$ (green), $0.80$ (blue), and $0.85~M_\sun$ WD (cyan line).
In panel (c), we plot three WD mass models, i.e.,
$0.80$ (green), $0.85$ (blue), and $0.90~M_\sun$ WD (cyan line).
We denote the start of the nebular phase on day 235 with the downward arrow.
\label{qy_mus_x35_x45_x55_v_logscale_3fig}}
\end{figure}


\begin{figure}
\plotone{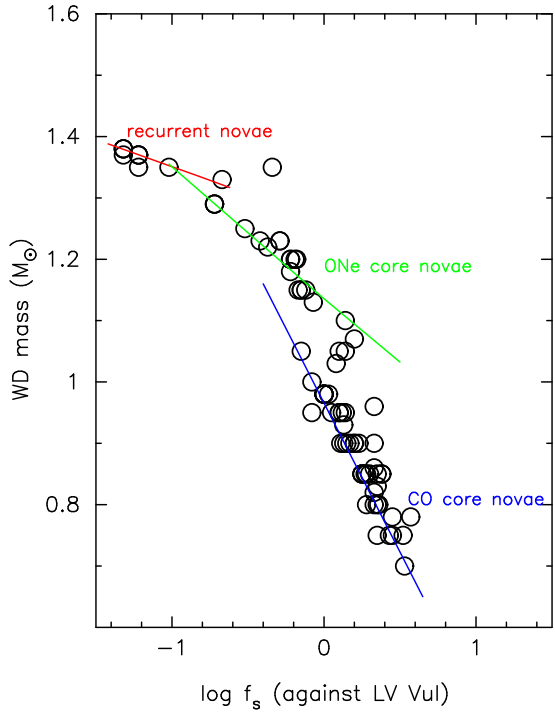}
\caption{
WD mass versus timescaling factor $\log f_{\rm s}$ for various novae.
The timescaling factor is measured against that of LV~Vul
($f_{\rm s}=1.0$ for LV~Vul).
The WD mass and $f_{\rm s}$ data are taken from Table \ref{wd_mass_novae}.
We specify three trends of $M_{\rm WD}$ versus $\log f_{\rm s}$:
recurrent novae (red line), oxygen-neon (ONe) core novae (green line),
and carbon-oxygen (CO) core novae (blue line). 
\label{timescale_wd_mass}}
\end{figure}

\section{Discussion}
\label{discussion}
\subsection{Distance determination}
\label{nova_distance}
Estimating the distance to a nova is a difficult task, and the results
are always debated \citep[e.g.,][]{schaefer18}.
Various methods have ever been proposed,
but they often have large uncertainties \citep[see, e.g.,][]{hac14k, hac15k,
hac16kb}.  A large discrepancy is reported, especially for 
the MMRD relations, which may work statistically,
but does not work well for individual novae \citep{schaefer18}.

The present work provides an alternative method for determining the
distance to a nova that has enough light/color curve data for a sufficiently
long coverage of the outburst.  
The accuracy of the distance estimate depends
mainly on the light/color curve fittings, i.e., typically 
$\Delta (m-M)_V = \pm0.1$ or $\pm0.2$, and $\Delta E(B-V)= \pm0.05$
or $\pm0.1$, which guarantee about 10\%--20\% of the distance accuracy.

\citet{schaefer18} analyzed the results of {\it Gaia} data release 2 (DR2)
and listed the trigonometric distances of 64 novae.
Among them, his seven novae dubbed as having 
``very well observed light curve ($< 30\%$ error)'' overlap
our analyzed novae in Table \ref{extinction_various_novae}. 
These distances are compared with our results 
(object, {\it Gaia} DR2/the present result) as follows: 
CI Aql, 3.14/3.3 kpc;   V705 Cas, 2.16/2.6 kpc; V1974 Cyg, 1.63/1.8 kpc;
V446 Her, 1.36/1.4 kpc; V533 Her, 1.20/1.3 kpc; V382 Vel, 1.8/1.4 kpc;
and PW Vul, 2.42/1.8 kpc.
The present results are in good agreement with Schaefer's values 
within each error box.

\subsection{Dependence of the WD mass on the chemical composition}
\label{wd_mass_determination}
The largest ambiguity in our determination of the WD mass is 
in the choice of the chemical composition of the hydrogen-rich envelope.
The chemical composition of ejecta is not well determined in many
novae. In our previous work \citep[e.g.,][]{hac16k}, we adopted
several sets of chemical composition templates
(CO1, CO2, CO3, and CO4 for CO novae and Ne1, Ne2, and Ne3
for neon novae) considering the degree of mixing between
the mass-accreting envelope and WD core material 
\citep[see][]{hac06kb, hac10k, hac15k, hac16k}.

Figure \ref{qy_mus_x35_x45_x55_v_logscale_3fig} depicts our model
light-curve fitting with QY~Mus for three different chemical compositions
(CO2, CO3, and CO4), assuming the distance
modulus in the $V$ band of $(m-M)_V=14.65$ (see 
Appendix \ref{qy_mus} for more details).  Figure
\ref{qy_mus_x35_x45_x55_v_logscale_3fig}(a) shows five WD mass
models, i.e., 0.65, 0.70, 0.75, 0.80, and $0.85~M_\sun$ WDs 
\citep[CO2,][]{hac10k}.
The $V$ light curve of $0.75~M_\sun$ WD (blue line) is in best
agreement with the observation among the five models.  
It should be noted that the $V$ magnitude of the nova, except for the
dust blackout phase until the point when the nova entered the nebular phase,
decays along with the theoretical line until day $\sim 235$.
In the nebular phase, however, strong emission lines make large
contributions to the $V$ band, the effect of which are not included
in our model light curves.  Therefore,
the observed $V$ magnitude deviates much from, and decays more 
slowly than, the model $V$ light curve in the nebular phase.
We select a $0.75~M_\sun$ WD for 
CO2 (the hydrogen content by weight is $X=0.35$).
Thus, we obtain $M_{\rm WD}=0.75\pm0.05~M_\sun$,
taking into account the fitting error.

If we increase the hydrogen content from $X=0.35$ (CO2) to $X=0.45$
\citep[CO3,][]{hac16k}, we obtain $0.80~M_\sun$ 
among 0.75, 0.80, and $0.85~M_\sun$ in Figure 
\ref{qy_mus_x35_x45_x55_v_logscale_3fig}(b).
If we further increase the hydrogen content to $X=0.55$
\citep[CO4,][]{hac15k},
we obtain $0.85~M_\sun$ among 0.80, 0.85, and $0.90~M_\sun$
in Figure \ref{qy_mus_x35_x45_x55_v_logscale_3fig}(c).
Our model light-curve fitting gives a WD mass between
$M_{\rm WD}=0.75-0.85~M_\sun$ for the chemical compositions of
$X=0.35-0.55$.  These three best-fit models reasonably reproduce
the $V$ light curve of QY~Mus.
Thus, we are able to approximate the relation between the WD mass,
$M_{\rm WD}$, and the hydrogen content, $X$, as
\begin{equation}
M_{\rm WD}/M_\sun = 0.05(X-0.55) + 0.85\pm0.05,
\end{equation}
for QY~Mus.

\subsection{Timescaling factor versus WD mass} 
\label{wd_mass_fs}
Table \ref{wd_mass_novae} provides the summary of the WD masses of the
novae in Table \ref{extinction_various_novae}.
The WD masses are estimated from direct comparison of our model
light curves with the observation.  The timescaling factor $f_{\rm s}$
of a nova is closely related to the WD mass, because a faster nova
(smaller $f_{\rm s}$) tends to host a more massive WD.
Figure \ref{timescale_wd_mass} shows the distribution of 
$\log f_{\rm s}$-$M_{\rm WD}$ for various novae in Table \ref{wd_mass_novae},
which includes both the present and the previous results 
\citep[e.g.,][]{hac16kb, hac18k, hac18kb, hac19k}.  
The three trends may represent recurrent novae (red line),
oxygen-neon (ONe) novae (green line), and carbon-oxygen (CO)
novae (blue line).  The border between the ONe and CO cores
is located around $1.0-1.1~M_\sun$.



\section{Conclusions}
\label{conclusions}
Our results are summarized as follows:\\

\noindent
{\bf 1.} Using the time-stretching method of nova light curves \citep{hac10k}
together with the time-stretched color-magnitude diagram method 
\citep{hac19k},  we estimated the distance moduli $(m-M)_V$
and color excesses $E(B-V)$ of 32 recent galactic novae.  
Tables \ref{extinction_various_novae} and
\ref{wd_mass_novae} provide a uniform data set of 73 novae obtained with
a single method, including the present and previous works. \\

\noindent
{\bf 2.} 
The present 32 nova tracks in the $(B-V)_0$-$(M_V-2.5 \log f_{\rm s})$ diagram
are divided into two types, LV~Vul and V1500~Cyg.  In general, each nova
moves from the upper right (red) to the lower left (blue) and then turns
toward the right (red) at the onset of the nebular phase.  
The LV~Vul type novae go almost straight down along the $(B-V)_0=-0.03$ line
in the middle part of the track.  The V1500~Cyg type novae are
roughly parallel to, but bluer by $\Delta (B-V)_0\sim -0.2$ mag than,
the LV~Vul type track.   This $-0.2$ mag bluer location is caused by
the higher degree of ionization state \citep{hac19k}. \\

\noindent
{\bf 3.} Our set of the distances $d$ and color excesses $E(B-V)$ 
shows a good agreement with at least one of the existing 3D dust
absorption maps.  Also, our distances show a good agreement
with {\it Gaia} data release 2 \citep{schaefer18}
in all seven overlapping novae, within the error box.\\

\noindent
{\bf 4.} The distribution of the WD masses versus the stretching factor
(i.e., decline rate) indicates three distinct tendencies 
that may represent recurrent novae, ONe novae, and CO novae. 
The border between the ONe novae and CO novae lies at $1.0-1.1~M_\sun$.\\

\noindent
{\bf 5.} Many novae, including the present 32 novae, broadly follow the
universal decline law and the present method can be applied to them.
Some exceptional novae deviate greatly from the universal decline law 
for various physical reasons and the method cannot be directly
applied to them.  \citet{hac18kb} discussed some examples 
and proposed different novae as templates depending 
on the physical reason for the deviation. Appendix \ref{m31n200812a_v1500_cyg}
explains and summarizes the results.\\

\acknowledgments
     We thank T. Iijima 
and the Astronomical Observatory of Padova (Asiago)
for the warm hospitality.
     We express our gratitude to 
the late A. Cassatella for providing us with  
UV 1455 \AA~data for {\it IUE} novae.
     We are also grateful to
the American Association of Variable Star Observers
(AAVSO) and the Variable Star Observers League of Japan (VSOLJ)
for the archival data of various novae.
  We also thank the anonymous referee for useful comments,
which improved the manuscript.
This research has been supported in part by Grants-in-Aid for
Scientific Research (15K05026, 16K05289) 
from the Japan Society for the Promotion of Science.




\appendix

\section{Light Curves and Distance Moduli of 32 Novae}
\label{light_curve_appendix}


\begin{figure}
\epsscale{0.75}
\plotone{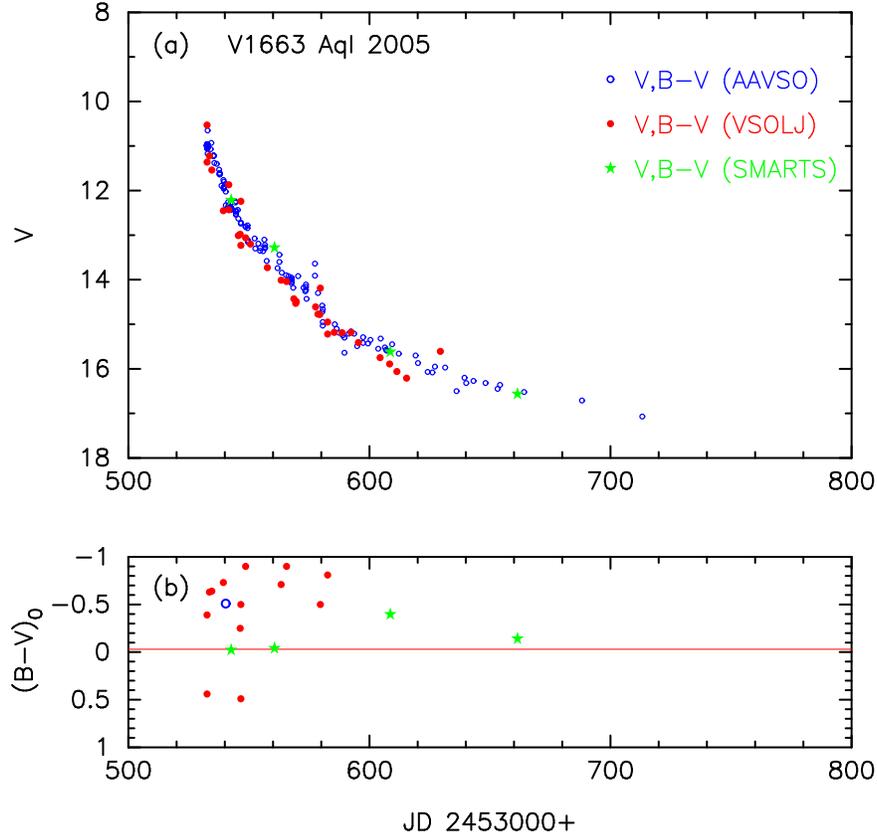}
\caption{
The light/color curves of V1663~Aql on a linear timescale.
(a) The $V$ light curve of V1663~Aql. The $BV$ data are 
taken from AAVSO (unfilled blue circles),
VSOLJ (filled red circles), and SMARTS (filled green stars).
(b) The $(B-V)_0$ color curve of V1663~Aql. 
The $(B-V)_0$ are dereddened with $E(B-V)=1.88$.
The horizontal red line shows $(B-V)_0= -0.03$, which is the
intrinsic color of optically thick free-free emission \citep{hac14k}.  
\label{v1663_aql_v_bv_ub_color_curve}}
\end{figure}


\begin{figure}
\epsscale{0.75}
\plotone{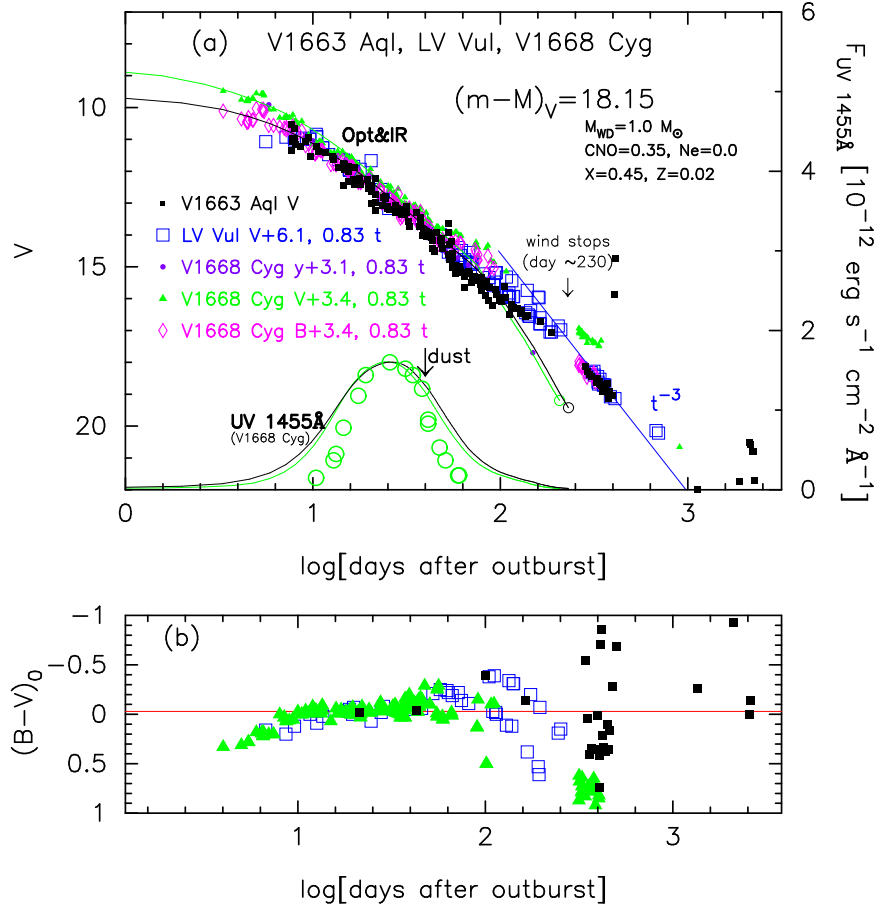}
\caption{
The light/color curves of V1663~Aql on a logarithmic timescale.
The data of V1663~Aql are the same as those in Figure
\ref{v1663_aql_v_bv_ub_color_curve}.
We add the light/color curves of LV~Vul and V1668~Cyg.
The data of LV~Vul and V1668~Cyg are the same as those in \citet{hac16k}.
Each light curve is horizontally moved by $\Delta \log t = \log f_{\rm s}$
and vertically shifted by $\Delta V$ with respect to that of V1663~Aql,
as indicated in the figure by, for example,
``LV~Vul V$+6.1$, 0.83 t,'' where $\Delta V= +6.1$ and $f_{\rm s}= 0.83$.
We also add the model light curve of a $1.0~M_\sun$ WD (CO3)
for V1663~Aql as well as a $0.98~M_\sun$ WD (CO3) model for V1668~Cyg.
The solid black and green lines
denote the model $V$ light curves of the 1.0 and $0.98~M_\sun$ WDs,
respectively.  The solid black and green lines represent the UV~1455\AA\  
flux based on the blackbody approximation of the 1.0 and $0.98~M_\sun$
WDs, respectively.  
The UV~1455\AA\  flux of V1668~Cyg (unfilled green circles) is slightly
smaller than the model light curve (solid green line) before the peak. 
This is the effect of the iron curtain \citep[e.g.,][]{hau95}. 
It slightly drops on day $\sim 60$, due to the formation of an
optically thin dust shell.
After the optically thick winds stop,
the observed $V$ magnitude decays like the straight solid blue line
of $F_\nu\propto t^{-3}$, which shows the trend of homologously
expanding ejecta.
\label{v1663_aql_lv_vul_v1668_cyg_v_bv_logscale}}
\end{figure}


\begin{figure}
\epsscale{0.55}
\plotone{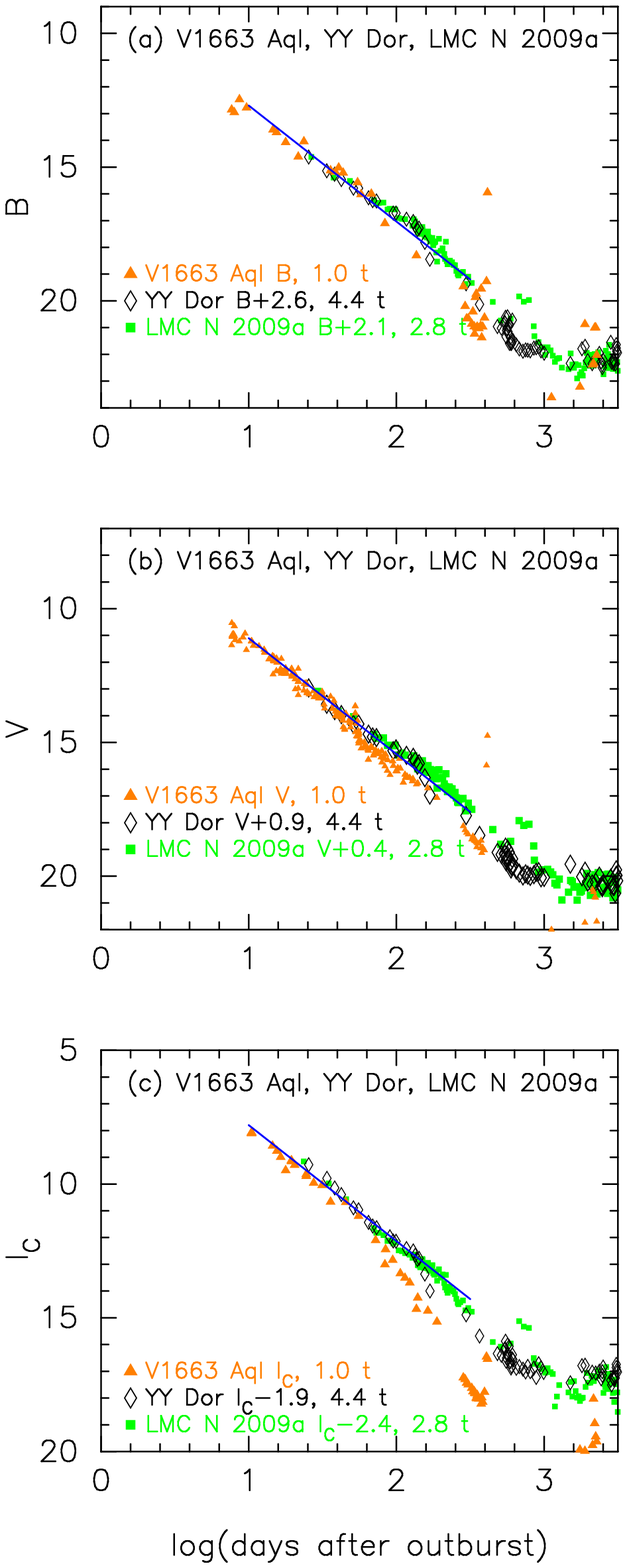}
\caption{
The (a) $B$, (b) $V$, and (c) $I_{\rm C}$ light curves of V1663~Aql,
YY~Dor, and LMC~N~2009a.
The straight solid blue lines denote the slope of $F_\nu\propto t^{-1.75}$.
The $BVI_{\rm C}$ data of V1663~Aql, YY~Dor, and LMC~N~2009a
are taken from AAVSO, VSOLJ, and SMARTS.
\label{v1663_aql_yy_dor_lmcn_2009a_b_v_i_logscale_3fig}}
\end{figure}

\subsection{V1663~Aql 2005}
\label{v1663_aql}
Figure \ref{v1663_aql_v_bv_ub_color_curve} shows (a) the $V$ light curve
and (b) $(B-V)_0$ color evolutions of V1663~Aql.  The $BV$ data are taken
from the archives of AAVSO, VSOLJ, and SMARTS.
The $(B-V)_0$ is dereddened with $E(B-V)=1.88$ as obtained in Section
\ref{v1663_aql_cmd}.
Figure \ref{v1663_aql_lv_vul_v1668_cyg_v_bv_logscale} shows the light/color
curves on a logarithmic timescale as well as those of LV~Vul and V1668~Cyg.
Each light curve is horizontally moved by $\Delta \log t = \log f_{\rm s}$
and vertically shifted by $\Delta V$ with respect to that of V1663~Aql,
as indicated in the figure.  For example, ``LV~Vul V$+6.1$, 0.83 t'' 
means that $\Delta V= +6.1$ and $f_{\rm s}= 0.83$.
These three $V$ light curves overlap each other.
Applying Equation (\ref{distance_modulus_general_temp}) to them,
we have the relation
\begin{eqnarray}
(m&-&M)_{V, \rm V1663~Aql} \cr 
&=& (m - M + \Delta V)_{V, \rm LV~Vul} - 2.5 \log 0.83 \cr
&=& 11.85 + 6.1\pm0.2 + 0.20 = 18.15\pm0.2 \cr
&=& (m - M + \Delta V)_{V, \rm V1668~Cyg} - 2.5 \log 0.83 \cr
&=& 14.6 + 3.4\pm0.2 + 0.20 = 18.2\pm0.2,
\label{distance_modulus_v1663_aql}
\end{eqnarray}
where we adopt $(m-M)_{V, \rm LV~Vul}=11.85$ and
$(m-M)_{V, \rm V1668~Cyg}=14.6$, both from \citet{hac19k}.
Thus, we obtain $(m-M)_V=18.15\pm0.1$ for V1663~Aql and
$f_{\rm s}=0.83$ against that of LV~Vul.  The distance is
estimated to be $d=2.9\pm0.3$~kpc from Equation 
(\ref{distance_modulus_rv}).
From Equations (\ref{time-stretching_general}),
(\ref{distance_modulus_general_temp}), and
(\ref{distance_modulus_v1663_aql}),
we have the relation
\begin{eqnarray}
(m- M')_{V, \rm V1663~Aql} 
& \equiv & (m_V - (M_V - 2.5\log f_{\rm s}))_{\rm V1663~Aql} \cr
&=& \left( (m-M)_V + \Delta V \right)_{\rm LV~Vul} \cr
&=& 11.85 + 6.1\pm0.2 = 17.95\pm0.2.
\label{absolute_mag_v1663_aql}
\end{eqnarray}

We further check the distance and reddening with the time-stretching method.
Figure \ref{v1663_aql_yy_dor_lmcn_2009a_b_v_i_logscale_3fig} shows
the $B$, $V$, and $I_{\rm C}$ light curves of V1663~Aql
together with those of the LMC novae YY~Dor and LMC~N~2009a.
We also plot the solid blue line with the slope of $F_\nu\propto t^{-1.75}$,
which is the typical decay slope of the universal decline law
\citep{hac06kb}. 
YY Dor and LMC N 2009a are members of the Large Magellanic Cloud (LMC).
We adopt $\mu_{0,\rm LMC}=18.493\pm0.048$ \citep{pie13} and
$E(B-V)=0.12$ \citep{imara07} for the LMC novae.
Then, we obtain $(m-M)_V= \mu_0 + A_V= 18.49 + 3.1\times 0.12 = 18.86$
toward the LMC novae.
From Equation (\ref{distance_modulus_general_temp_b})
for the $B$ band in Figure
\ref{v1663_aql_yy_dor_lmcn_2009a_b_v_i_logscale_3fig}(a),
we obtain
\begin{eqnarray}
(m&-&M)_{B, \rm V1663~Aql} \cr
&=& ((m - M)_B + \Delta B)_{\rm YY~Dor} - 2.5 \log 4.4 \cr
&=& 18.98 + 2.6\pm0.2 - 1.6 = 19.98\pm0.2 \cr
&=& ((m - M)_B + \Delta B)_{\rm LMC~N~2009a} - 2.5 \log 2.8 \cr
&=& 18.98 + 2.1\pm0.2 - 1.1 = 19.98\pm0.2.
\label{distance_modulus_b_v1663_aql_yy_dor_lmcn2009a}
\end{eqnarray}
Here, we adopt $(m-M)_B = 18.49 + 4.1\times 0.12= 18.98$ for the LMC novae
from Equation (\ref{distance_modulus_rb}).
Thus, we obtain $(m-M)_{B, \rm V1663~Aql}= 19.98\pm0.2$.

For the $V$ light curves in Figure
\ref{v1663_aql_yy_dor_lmcn_2009a_b_v_i_logscale_3fig}(b),
we similarly obtain
\begin{eqnarray}
(m&-&M)_{V, \rm V1663~Aql} \cr
&=& ((m - M)_V + \Delta V)_{\rm YY~Dor} - 2.5 \log 4.4 \cr
&=& 18.86 + 0.9\pm0.2 - 1.6 = 18.16\pm0.2 \cr
&=& ((m - M)_V + \Delta V)_{\rm LMC~N~2009a} - 2.5 \log 2.8 \cr
&=& 18.86 + 0.4\pm0.2 -1.1 = 18.16\pm0.2.
\label{distance_modulus_v_v1663_aql_yy_dor_lmcn2009a}
\end{eqnarray}
Thus, we obtain $(m-M)_{V, \rm V1663~Aql}= 18.16\pm0.2$,
which is consistent with Equation (\ref{distance_modulus_v1663_aql}).

We apply Equation (\ref{distance_modulus_general_temp_i}) for
the $I_{\rm C}$-band to Figure
\ref{v1663_aql_yy_dor_lmcn_2009a_b_v_i_logscale_3fig}(c) and obtain
\begin{eqnarray}
(m&-&M)_{I, \rm V1663~Aql} \cr
&=& ((m - M)_I + \Delta I_C)_{\rm YY~Dor} - 2.5 \log 4.4 \cr
&=& 18.67 - 1.9\pm0.3 - 1.6 = 15.17\pm 0.3 \cr
&=& ((m - M)_I + \Delta I_C)_{\rm LMC~N~2009a} - 2.5 \log 2.8 \cr
&=& 18.67 - 2.4\pm0.3 -1.1 = 15.17\pm 0.3,
\label{distance_modulus_i_v1663_aql_yy_dor_lmcn2009a}
\end{eqnarray}
where we adopt
$(m-M)_I = 18.49 + 1.5\times 0.12 = 18.67$ for the LMC novae 
from Equation (\ref{distance_modulus_ri}).
Thus, we obtain $(m-M)_{I, \rm V1663~Aql}= 15.17\pm0.2$.

We plot $(m-M)_B= 19.98$, $(m-M)_V= 18.16$, and $(m-M)_I= 15.17$ 
by the thin solid magenta, blue, and cyan lines, respectively,  
in Figure
\ref{distance_reddening_v1663_aql_v5116_sgr_v2575_oph_v5117_sgr}(a).
The three lines cross at $d=2.9$~kpc and $E(B-V)=1.88$.


\begin{figure}
\epsscale{0.75}
\plotone{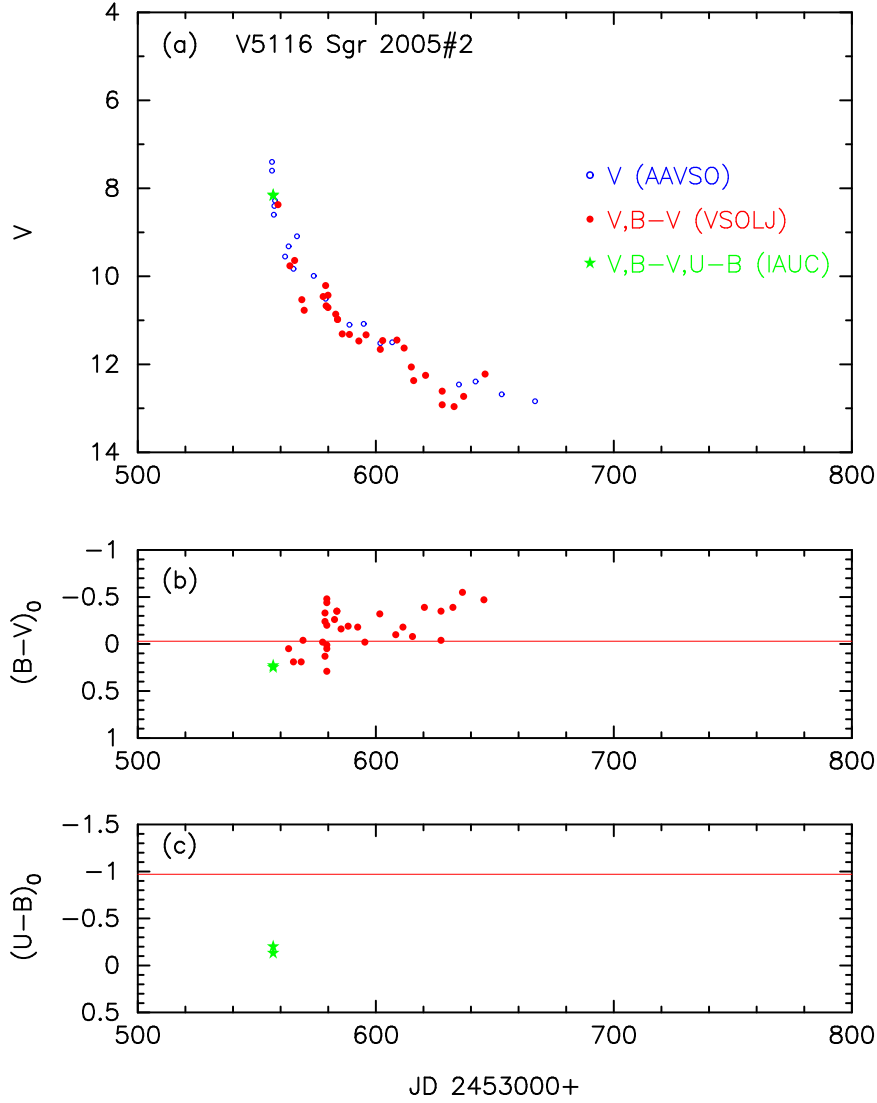}
\caption{
Same as Figure \ref{v1663_aql_v_bv_ub_color_curve}, but for V5116~Sgr.
(a) The $V$ data are taken from AAVSO (unfilled blue circles).
The $BV$ data are from VSOLJ (filled red circles) and the $UBV$ data
are from \citet{gilm05} (filled green stars).
(b) The $(B-V)_0$ and (c) $(U-B)_0$ are dereddened with $E(B-V)=0.23$.
The horizontal red lines show $(B-V)_0= -0.03$ and
$(U-B)_0= -0.97$, which are the intrinsic colors
of optically thick free-free emission \citep{hac14k}.  
\label{v5116_sgr_v_bv_ub_color_curve}}
\end{figure}


\begin{figure}
\epsscale{0.75}
\plotone{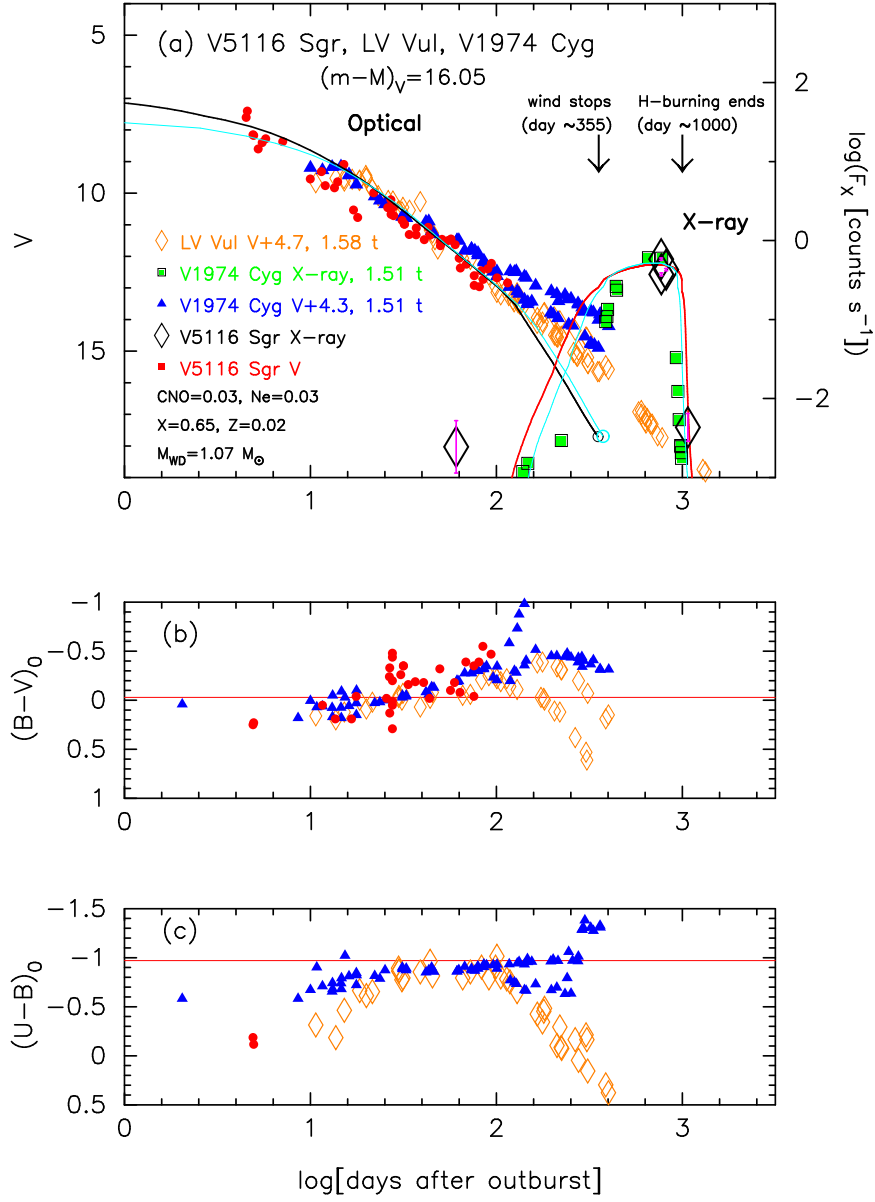}
\caption{
Same as Figure \ref{v1663_aql_lv_vul_v1668_cyg_v_bv_logscale},
but for V5116~Sgr.  We add the light/color curves of LV~Vul and V1974~Cyg.
The data of V5116~Sgr are the same as those in Figure
\ref{v5116_sgr_v_bv_ub_color_curve}.  We also add the X-ray data of
V5116~Sgr and V1974~Cyg. The data of V1974~Cyg
are all the same as those in Figures 38--42 of \citet{hac16k}.
The timescale of V5116~Sgr is longer than that of V1974~Cyg
by a factor of 1.51.
We added a model light curve of a $1.07~M_\sun$ WD (Ne3) for V5116~Sgr.
The solid black line denotes our model $V$ light curve ($1.07~M_\sun$)
of photospheric plus free-free emission.
The solid red line represents the supersoft X-ray flux ($1.07~M_\sun$)
based on the blackbody approximation \citep[see, e.g.,][]{hac16k}.
We also add the model light curve of a $0.98~M_\sun$ WD 
\citep[CO3, solid cyan lines;][]{hac16k} for V1974~Cyg.
\label{v5116_sgr_v1974_cyg_x65z02o03ne03_v_bv_ub_color_logscale}}
\end{figure}


\begin{figure}
\epsscale{0.55}
\plotone{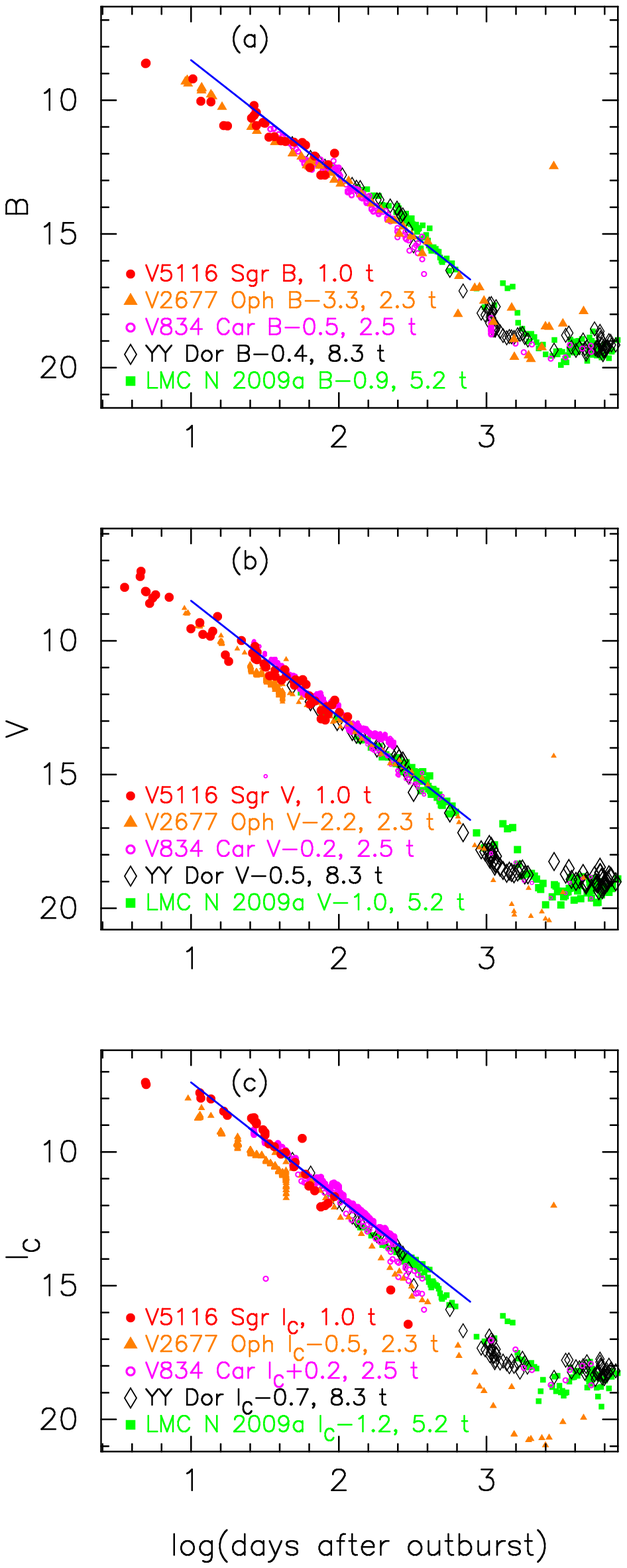}
\caption{
Same as Figure \ref{v1663_aql_yy_dor_lmcn_2009a_b_v_i_logscale_3fig},
but for V5116~Sgr.
The (a) $B$, (b) $V$, and (c) $I_{\rm C}$ light curves of V5116~Sgr
as well as those of V2677~Oph, V834~Car, YY~Dor, and LMC~N~2009a.
The $BV$ data of V5116~Sgr are the same as those in Figure
\ref{v5116_sgr_v_bv_ub_color_curve}.  The $I_{\rm C}$ data are taken
from \citet{gilm05}, AAVSO, and VSOLJ.
\label{v5116_sgr_v2677_oph_v834_car_yy_dor_lmcn_2009a_b_v_i_logscale_3fig}}
\end{figure}

\subsection{V5116~Sgr 2005\#2}
\label{v5116_sgr}
Figure \ref{v5116_sgr_v_bv_ub_color_curve} shows (a) the $V$, (b) $(B-V)_0$,
and (c) $(U-B)_0$ evolutions of V5116~Sgr.  Here, $(B-V)_0$ and $(U-B)_0$
are dereddened with $E(B-V)=0.23$ as obtained in Section \ref{v5116_sgr_cmd}.
The $V$ light curve of V5116~Sgr is similar to that of V1974~Cyg
as shown in Figure
\ref{v5116_sgr_v1974_cyg_x65z02o03ne03_v_bv_ub_color_logscale}.
Here, we plot the light/color curves of V5116~Sgr, LV~Vul, and
V1974~Cyg on logarithmic timescales.
These three $V$ light curves overlap each other.
The $V$ magnitude has slightly different values among various
observatories when strong emission lines contribute to the $V$ filter.
This is because strong [\ion{O}{3}] lines
contribute to the blue edge of $V$ filter and a small difference
in the $V$-filter response makes a large difference in the $V$ magnitude
and, as a result, in the $B-V$ color.  This effect is clearly seen in 
Figure \ref{v5116_sgr_v1974_cyg_x65z02o03ne03_v_bv_ub_color_logscale}(b),
especially in the nebular phase ($t > 100$ days).
For this reason, we do not require stringent overlapping 
in the nebular phase.
Applying Equation (\ref{distance_modulus_general_temp}) to them,
we have the relation
\begin{eqnarray}
(m&-&M)_{V, \rm V5116~Sgr} \cr 
&=& (m - M + \Delta V)_{V, \rm LV~Vul} - 2.5 \log 1.58 \cr
&=& 11.85 + 4.7\pm0.2 - 0.50 = 16.05\pm0.2 \cr
&=& (m - M + \Delta V)_{V, \rm V1974~Cyg} - 2.5 \log 1.51 \cr
&=& 12.2 + 4.3\pm0.2 - 0.45 = 16.05\pm0.2,
\label{distance_modulus_v5116_sgr}
\end{eqnarray}
where we adopt $(m-M)_{V, \rm LV~Vul}=11.85$ and
$(m-M)_{V, \rm V1974~Cyg}=12.2$, both from \citet{hac19k}.
Thus, we obtain $(m-M)_V=16.05\pm0.1$ for V5116~Sgr, consistent with
Hachisu \& Kato's (2010) results.  
From Equations (\ref{time-stretching_general}),
(\ref{distance_modulus_general_temp}), and
(\ref{distance_modulus_v5116_sgr}),
we have the relation
\begin{eqnarray}
(m- M')_{V, \rm V5116~Sgr} 
&\equiv & (m_V - (M_V - 2.5\log f_{\rm s}))_{\rm V5116~Sgr} \cr
&=& \left( (m-M)_V + \Delta V \right)_{\rm LV~Vul} \cr
&=& 11.85 + 4.7\pm0.2 = 16.55\pm0.2.
\label{absolute_mag_v5116_sgr}
\end{eqnarray}

We obtain the distance and reddening with the time-stretching method.  Figure 
\ref{v5116_sgr_v2677_oph_v834_car_yy_dor_lmcn_2009a_b_v_i_logscale_3fig}
shows the $B$, $V$, and $I_{\rm C}$ light curves of V5116~Sgr 
together with those of V2677~Oph, V834~Car, and the LMC novae YY~Dor
and LMC~N~2009a.
We apply Equation (\ref{distance_modulus_general_temp_b})
for the $B$ band to Figure
\ref{v5116_sgr_v2677_oph_v834_car_yy_dor_lmcn_2009a_b_v_i_logscale_3fig}(a)
and obtain
\begin{eqnarray}
(m&-&M)_{B, \rm V5116~Sgr} \cr
&=& ((m - M)_B + \Delta B)_{\rm V2677~Oph} - 2.5 \log 2.3 \cr
&=& 20.5 - 3.3\pm0.2 - 0.92 = 16.28\pm0.2 \cr
&=& ((m - M)_B + \Delta B)_{\rm V834~Car} - 2.5 \log 2.5 \cr
&=& 17.75 - 0.5\pm0.2 - 0.97 = 16.28\pm0.2 \cr
&=& ((m - M)_B + \Delta B)_{\rm YY~Dor} - 2.5 \log 8.3 \cr
&=& 18.98 - 0.4\pm0.2 - 2.3 = 16.28\pm0.2 \cr
&=& ((m - M)_B + \Delta B)_{\rm LMC~N~2009a} - 2.5 \log 5.2 \cr
&=& 18.98 - 0.9\pm0.2 - 1.8 = 16.28\pm0.2,
\label{distance_modulus_b_v5116_sgr_v2677_oph_v834_car_yy_dor_lmcn2009a}
\end{eqnarray}
where we adopt $(m-M)_{B, \rm V2677~Oph}= 19.2 + 1.3= 20.5$ from 
Appendix \ref{v2677_oph} and
$(m-M)_{B, \rm V834~Car}= 17.25 + 0.50= 17.75$ from 
Appendix \ref{v834_car}.
Thus, we have $(m-M)_{B, \rm V5116~Sgr}= 16.28\pm0.1$.

For the $V$ light curves in Figure
\ref{v5116_sgr_v2677_oph_v834_car_yy_dor_lmcn_2009a_b_v_i_logscale_3fig}(b),
we similarly obtain
\begin{eqnarray}
(m&-&M)_{V, \rm V5116~Sgr} \cr   
&=& ((m - M)_V + \Delta V)_{\rm V2677~Oph} - 2.5 \log 2.3 \cr
&=& 19.2 - 2.2\pm0.2 - 0.92 = 16.08\pm0.2 \cr
&=& ((m - M)_V + \Delta V)_{\rm V834~Car} - 2.5 \log 2.5 \cr
&=& 17.25 - 0.2\pm0.2 - 0.97 = 16.08\pm0.2 \cr
&=& ((m - M)_V + \Delta V)_{\rm YY~Dor} - 2.5 \log 8.3 \cr
&=& 18.86 - 0.5\pm0.2 - 2.3 = 16.06\pm0.2 \cr
&=& ((m - M)_V + \Delta V)_{\rm LMC~N~2009a} - 2.5 \log 5.2 \cr
&=& 18.86 - 1.0\pm0.2 - 1.8 = 16.06\pm0.2,
\label{distance_modulus_v_v5116_sgr_v2677_oph_v834_car_yy_dor_lmcn2009a}
\end{eqnarray}
where we adopt $(m-M)_{V, \rm V2677~Oph}= 19.2$ and 
$(m-M)_{V, \rm V834~Car}= 17.25$ from Appendixes \ref{v2677_oph} and
\ref{v834_car}, respectively.
We have $(m-M)_{V, \rm V5116~Sgr}= 16.07\pm0.1$, which is
consistent with Equation (\ref{distance_modulus_v5116_sgr}).

We apply Equation (\ref{distance_modulus_general_temp_i}) for
the $I_{\rm C}$-band to Figure
\ref{v5116_sgr_v2677_oph_v834_car_yy_dor_lmcn_2009a_b_v_i_logscale_3fig}(c)
and obtain
\begin{eqnarray}
(m&-&M)_{I, \rm V5116~Sgr} \cr
&=& ((m - M)_I + \Delta I_C)_{\rm V2677~Oph} - 2.5 \log 2.3 \cr
&=& 17.12 - 0.5\pm0.2 - 0.92 = 15.7\pm 0.2 \cr
&=& ((m - M)_I + \Delta I_C)_{\rm V834~Car} - 2.5 \log 2.5 \cr
&=& 16.45 + 0.2\pm0.2 - 0.97 = 15.68\pm 0.2 \cr
&=& ((m - M)_I + \Delta I_C)_{\rm YY~Dor} - 2.5 \log 8.3 \cr
&=& 18.67 - 0.7\pm0.2 - 2.3 = 15.67\pm 0.2 \cr
&=& ((m - M)_I + \Delta I_C)_{\rm LMC~N~2009a} - 2.5 \log 5.2 \cr
&=& 18.67 - 1.2\pm0.2 - 1.8 = 15.67\pm 0.2,
\label{distance_modulus_i_v5116_sgr_v2677_oph_v834_car_yy_dor_lmcn2009a}
\end{eqnarray}
where we adopt $(m-M)_{I, \rm V2677~Oph}= 19.2 - 1.6\times 1.3= 17.12$
from Appendix \ref{v2677_oph} and $(m-M)_{I, \rm V834~Car}= 17.25 - 
1.6\times 0.50= 16.45$ from Appendix \ref{v834_car}.
Thus, we have $(m-M)_{I, \rm V5116~Sgr}= 15.68\pm0.1$.

We plot $(m-M)_B= 16.28$, $(m-M)_V= 16.07$, and $(m-M)_I= 15.68$,
which cross at $d=12$~kpc and $E(B-V)=0.23$,
 in Figure
\ref{distance_reddening_v1663_aql_v5116_sgr_v2575_oph_v5117_sgr}(b).
Thus, we have $E(B-V)=0.23\pm0.05$ and $d=12\pm2$~kpc.


\begin{figure}
\epsscale{0.75}
\plotone{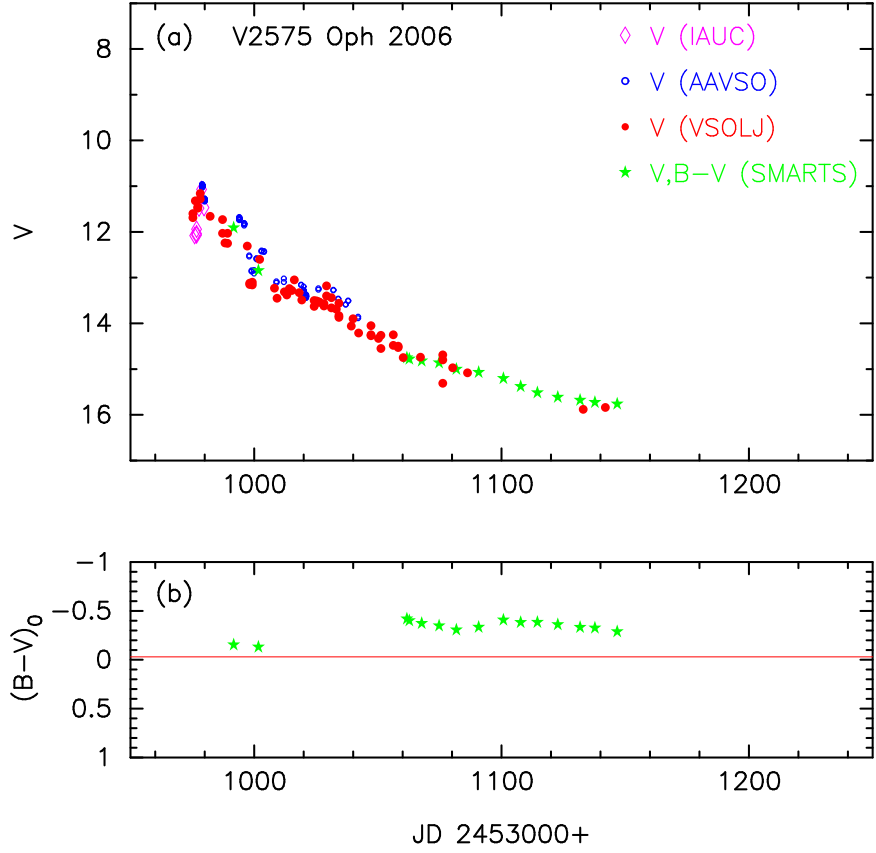}
\caption{
Same as Figure \ref{v1663_aql_v_bv_ub_color_curve}, but for V2575~Oph.
(a) The $V$ data are taken from IAU Circular No. 8671, 1
\citep[unfilled magenta diamonds;][]{poj06},
AAVSO (unfilled blue circles), and VSOLJ (filled red circles).
The $BV$ data are taken from SMARTS (filled green stars).
(b) The $(B-V)_0$ are dereddened with $E(B-V)=1.43$.
\label{v2575_oph_v_bv_ub_color_curve}}
\end{figure}


\begin{figure}
\epsscale{0.75}
\plotone{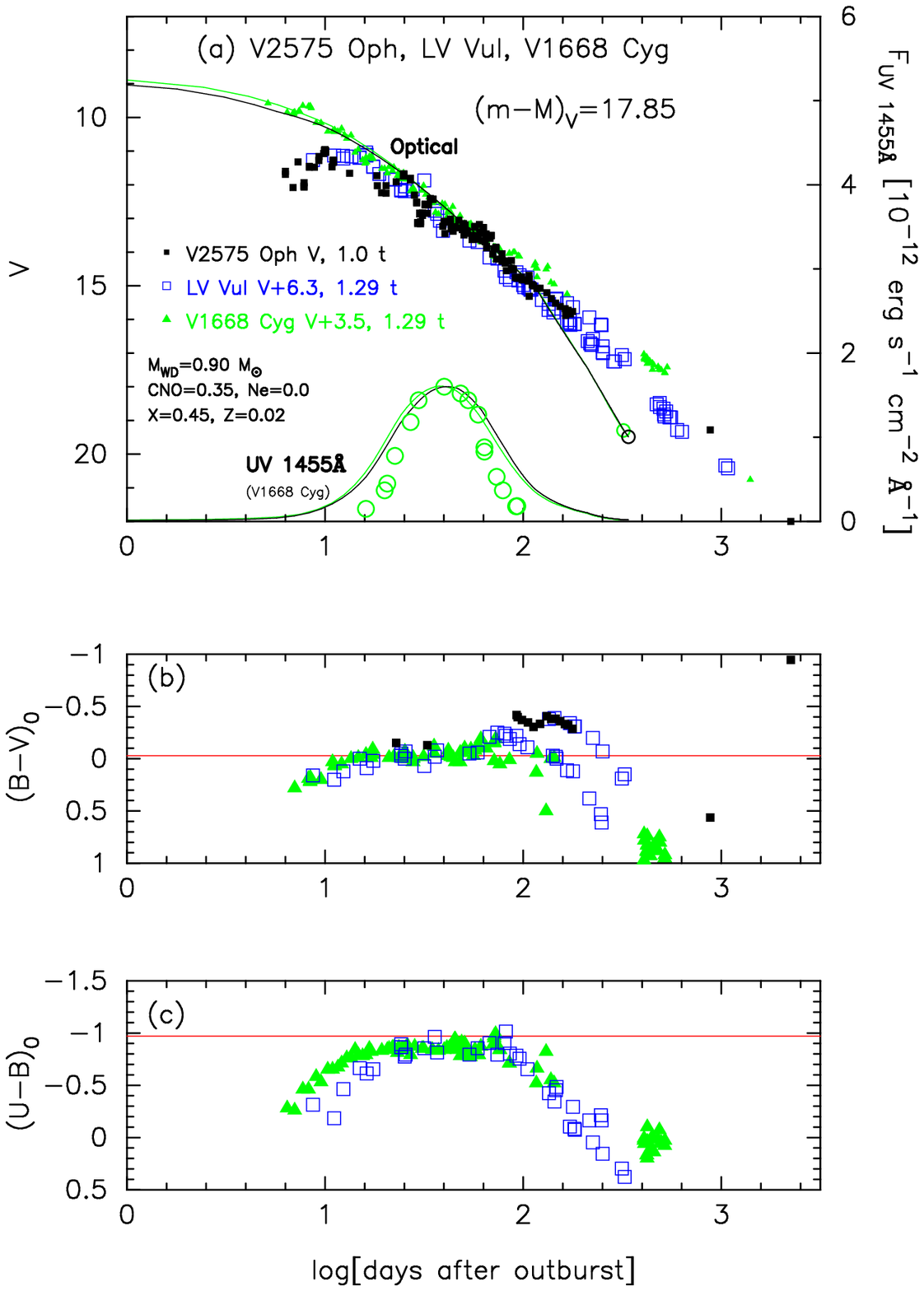}
\caption{
Same as Figure \ref{v1663_aql_lv_vul_v1668_cyg_v_bv_logscale},
but for V2575~Oph.  We add the light/color curves of LV~Vul and V1668~Cyg.
The data of V2575~Oph are the same as those in Figure
\ref{v2575_oph_v_bv_ub_color_curve}.  The data of V1668~Cyg and LV~Vul
are all the same as those in Figures 1 and 4 of \citet{hac16kb}.  We added
model light curves of a $0.90~M_\sun$ WD (solid black lines) for V2575~Oph
and a $0.98~M_\sun$ WD \citep[CO3, solid green lines;][]{hac16k}
for V1668~Cyg.  
The upper solid black and green lines denote the $V$ light curves
of photospheric plus free-free emission for the $0.90~M_\sun$ and
$0.98~M_\sun$ WDs.
The lower solid black and green lines represent
the UV~1455\AA\  flux based on the
blackbody approximation \citep[see, e.g.,][]{hac16k}.
\label{v2575_oph_v1668_cyg_lv_vul_v_bv_ub_logscale}}
\end{figure}


\begin{figure}
\epsscale{0.55}
\plotone{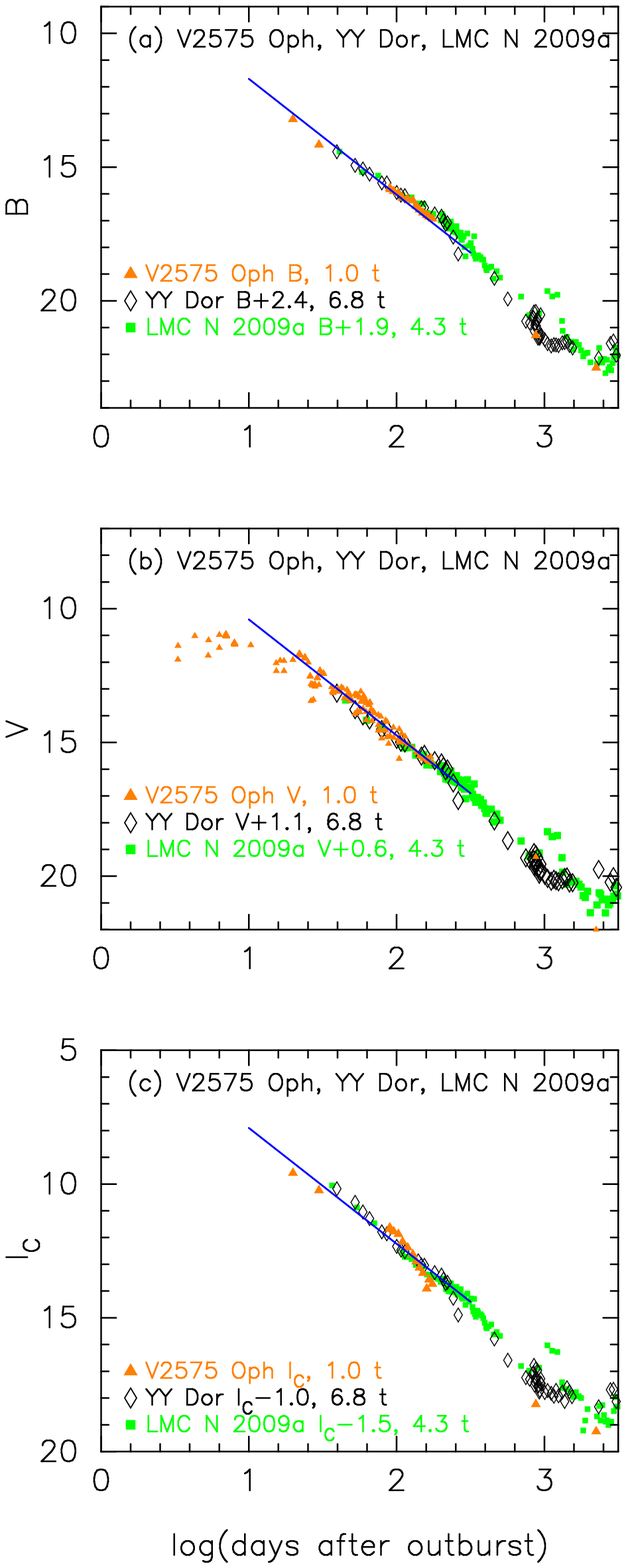}
\caption{
Same as Figure \ref{v1663_aql_yy_dor_lmcn_2009a_b_v_i_logscale_3fig},
but for V2575~Oph.
The $BVI_{\rm C}$ data of V2575~Oph are taken from SMARTS.
The $V$ data of V2575~Oph are also taken from AAVSO and VSOLJ.
\label{v2575_oph_yy_dor_lmcn_2009a_b_v_i_logscale_3fig}}
\end{figure}

\subsection{V2575~Oph 2006}
\label{v2575_oph}
Figure \ref{v2575_oph_v_bv_ub_color_curve} shows (a) the $V$ and
(b) $(B-V)_0$ evolutions of V2575~Oph.
Here, $(B-V)_0$ are dereddened with $E(B-V)=1.43$ as obtained in
Section \ref{v2575_oph_cmd}.  
The $V$ light curve of V2575~Oph is similar to that of LV~Vul
and V1668~Cyg as shown in Figure
\ref{v2575_oph_v1668_cyg_lv_vul_v_bv_ub_logscale}.
These three $V$ light curves overlap each other if we properly 
stretch/squeeze the $V$ and $(B-V)_0$ light/color curves.
In more detail, the early $V$ light curve shows a wavy structure
during $t < 30$ days.  Its upper bound follows that of LV~Vul.
After $t > 30$ days, the $V$ light curve almost follows that of LV~Vul.
There are few observational points of $(B-V)_0$ before $t < 80$ days.
The $(B-V)_0$ color curve also broadly follows the upper branch of LV~Vul
after $t > 80$ days.
Applying Equation (\ref{distance_modulus_general_temp}) to them,
we have the relation
\begin{eqnarray}
(m&-&M)_{V, \rm V2575~Oph} \cr 
&=& (m - M + \Delta V)_{V, \rm LV~Vul} - 2.5 \log 1.29 \cr
&=& 11.85 + 6.3\pm0.2 - 0.28 = 17.87\pm0.2 \cr
&=& (m - M + \Delta V)_{V, \rm V1668~Cyg} - 2.5 \log 1.29 \cr
&=& 14.6 + 3.5\pm0.2 - 0.28 = 17.82\pm0.2,
\label{distance_modulus_v2575_oph}
\end{eqnarray}
where we adopt $(m-M)_{V, \rm LV~Vul}=11.85$ and 
$(m-M)_{V, \rm V1668~Cyg}=14.6$, both from \citet{hac19k}.
Therefore, we adopt $(m-M)_V=17.85\pm0.1$ and
$f_{\rm s}=1.29$ for V2575~Oph.
From Equations (\ref{time-stretching_general}),
(\ref{distance_modulus_general_temp}), and
(\ref{distance_modulus_v2575_oph}),
we have the relation 
\begin{eqnarray}
(m- M')_{V, \rm V2575~Oph} 
&\equiv & (m_V - (M_V - 2.5\log f_{\rm s}))_{\rm V2575~Oph} \cr
&=& \left( (m-M)_V + \Delta V \right)_{\rm LV~Vul} \cr
&=& 11.85 + 6.3\pm0.2 = 18.15\pm0.2.
\label{absolute_mag_v2575_oph}
\end{eqnarray}

We further check the distance and reddening with the time-stretching method.
Figure \ref{v2575_oph_yy_dor_lmcn_2009a_b_v_i_logscale_3fig} shows
the $B$, $V$, and $I_{\rm C}$ light curves of V2575~Oph
together with those of YY~Dor and LMC~N~2009a.
We apply Equation (\ref{distance_modulus_general_temp_b})
for the $B$ band to Figure
\ref{v2575_oph_yy_dor_lmcn_2009a_b_v_i_logscale_3fig}(a)
and obtain
\begin{eqnarray}
(m&-&M)_{B, \rm V2575~Oph} \cr
&=& ((m - M)_B + \Delta B)_{\rm YY~Dor} - 2.5 \log 6.8 \cr
&=& 18.98 + 2.4\pm0.2 - 2.08 = 19.3\pm0.2 \cr
&=& ((m - M)_B + \Delta B)_{\rm LMC~N~2009a} - 2.5 \log 4.3 \cr
&=& 18.98 + 1.9\pm0.2 - 1.58 = 19.3\pm0.2.
\label{distance_modulus_b_v2575_oph_yy_dor_lmcn2009a}
\end{eqnarray}
Thus, we obtain $(m-M)_{B, \rm V2575~Oph}= 19.3\pm0.1$.

For the $V$ light curves in Figure
\ref{v2575_oph_yy_dor_lmcn_2009a_b_v_i_logscale_3fig}(b),
we similarly obtain
\begin{eqnarray}
(m&-&M)_{V, \rm V2575~Oph} \cr
&=& ((m - M)_V + \Delta V)_{\rm YY~Dor} - 2.5 \log 6.8 \cr
&=& 18.86 + 1.1\pm0.2 - 2.08 = 17.88\pm0.2 \cr
&=& ((m - M)_V + \Delta V)_{\rm LMC~N~2009a} - 2.5 \log 4.3 \cr
&=& 18.86 + 0.6\pm0.2 -1.58 = 17.88\pm0.2.
\label{distance_modulus_v_v2575_oph_yy_dor_lmcn2009a}
\end{eqnarray}
Thus, we obtain $(m-M)_{V, \rm V2575~Oph}= 17.88\pm0.1$, which is
consistent with Equation (\ref{distance_modulus_v2575_oph}).

We apply Equation (\ref{distance_modulus_general_temp_i}) for
the $I_{\rm C}$-band to Figure
\ref{v2575_oph_yy_dor_lmcn_2009a_b_v_i_logscale_3fig}(c) and obtain
\begin{eqnarray}
(m&-&M)_{I, \rm V2575~Oph} \cr
&=& ((m - M)_I + \Delta I_C)_{\rm YY~Dor} - 2.5 \log 6.8 \cr
&=& 18.67 - 1.0\pm0.3 - 2.08 = 15.59\pm 0.3 \cr
&=& ((m - M)_I + \Delta I_C)_{\rm LMC~N~2009a} - 2.5 \log 4.3 \cr
&=& 18.67 - 1.5\pm0.3 -1.58 = 15.59\pm 0.3.
\label{distance_modulus_i_v2575_oph_yy_dor_lmcn2009a}
\end{eqnarray}
Thus, we obtain $(m-M)_{I, \rm V2575~Oph}= 15.59\pm0.2$.

We plot $(m-M)_B= 19.3$, $(m-M)_V= 17.88$, and $(m-M)_I= 15.59$,
which cross at $d=4.9$~kpc and $E(B-V)=1.43$, in Figure
\ref{distance_reddening_v1663_aql_v5116_sgr_v2575_oph_v5117_sgr}(c).
Thus, we obtain $E(B-V)=1.43\pm0.05$ and $d=4.9\pm0.5$~kpc.


\begin{figure}
\epsscale{0.75}
\plotone{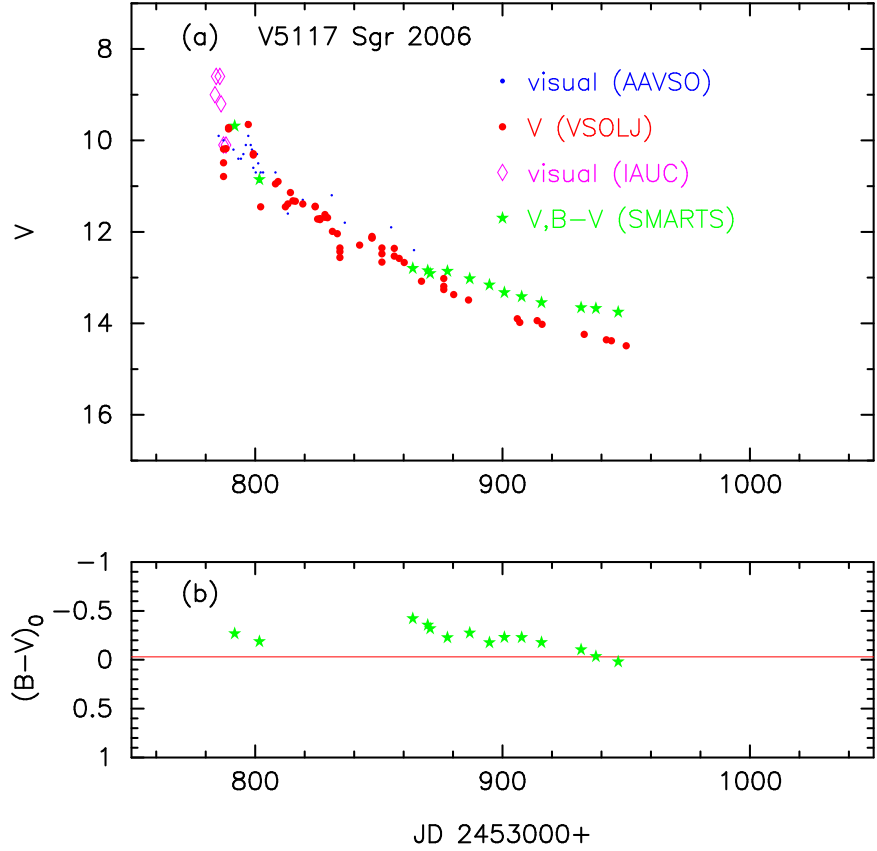}
\caption{
Same as Figure \ref{v1663_aql_v_bv_ub_color_curve}, but for V5117~Sgr.
(a) The visual data are taken from AAVSO (blue dots) and
IAU Circular 8673, 1 and 8706, 3 (unfilled magenta diamonds).
The $V$ data are from VSOLJ (filled red circles).
The $BV$ data are taken from SMARTS (filled green stars).
(b) The $(B-V)_0$ are dereddened with $E(B-V)=0.53$.
\label{v5117_sgr_v_bv_ub_color_curve}}
\end{figure}


\begin{figure}
\epsscale{0.75}
\plotone{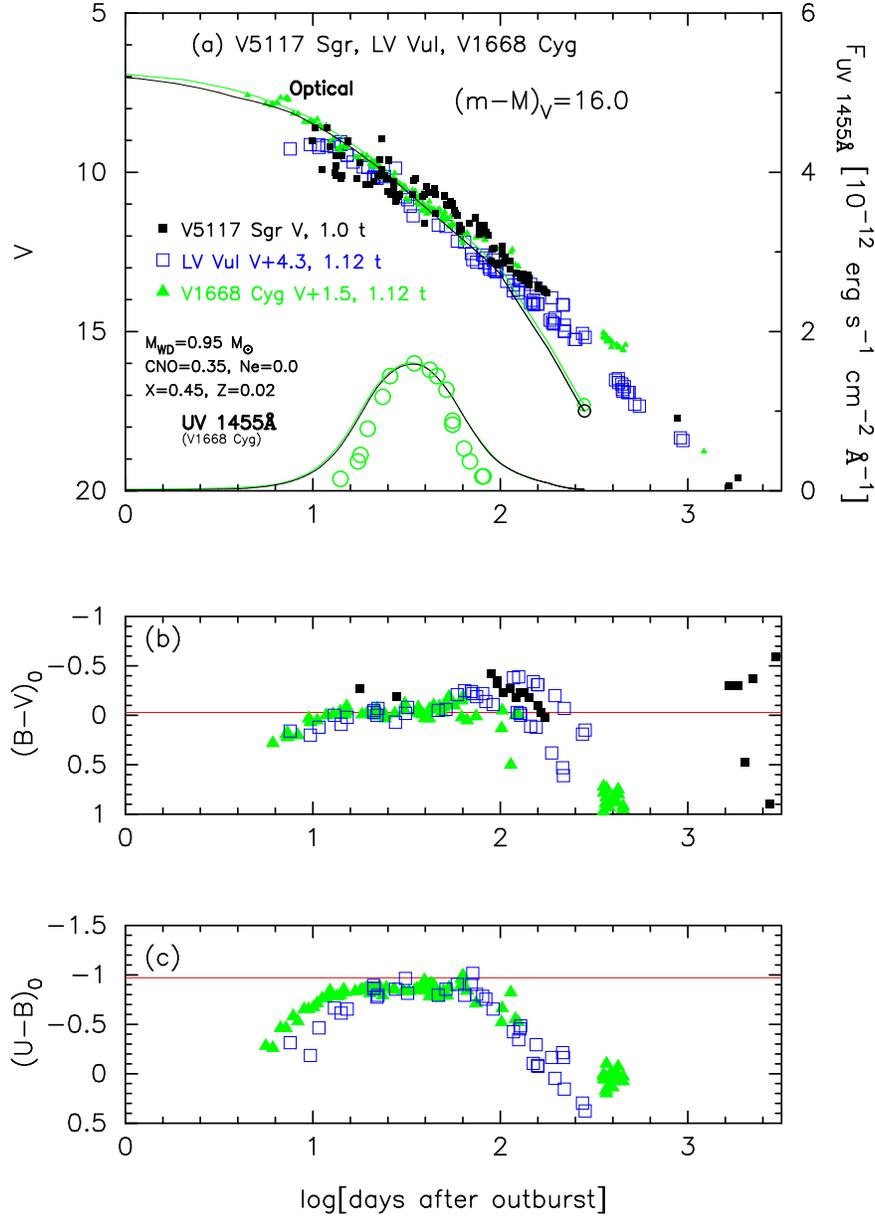}
\caption{
Same as Figures \ref{v1663_aql_lv_vul_v1668_cyg_v_bv_logscale}
and \ref{v2575_oph_v1668_cyg_lv_vul_v_bv_ub_logscale},
but for V5117~Sgr.  We add the light/color curves of LV~Vul and V1668~Cyg.
The data of V5117~Sgr are the same as those in Figure
\ref{v5117_sgr_v_bv_ub_color_curve}.  The data of LV~Vul
and V1668~Cyg are all the same as those in Figure
\ref{v2575_oph_v1668_cyg_lv_vul_v_bv_ub_logscale}.
Assuming that $(m-M)_V=16.0$, 
we added model light curves of a $0.95~M_\sun$ WD
\citep[CO3,][]{hac16k}.
The upper solid black line denotes the $V$ light curve of photospheric
plus free-free emission for a $0.95~M_\sun$ WD \citep{hac15k}.
The lower solid black line represents the UV~1455\AA\  flux based on the
blackbody approximation \citep[see, e.g.,][]{hac16k}.
The green lines denote the $0.98~M_\sun$ WD (CO3) model,
assuming that $(m-M)_V=14.6$ for V1668~Cyg.
\label{v5117_sgr_v1668_cyg_lv_vul_v_bv_ub_logscale}}
\end{figure}


\begin{figure}
\epsscale{0.55}
\plotone{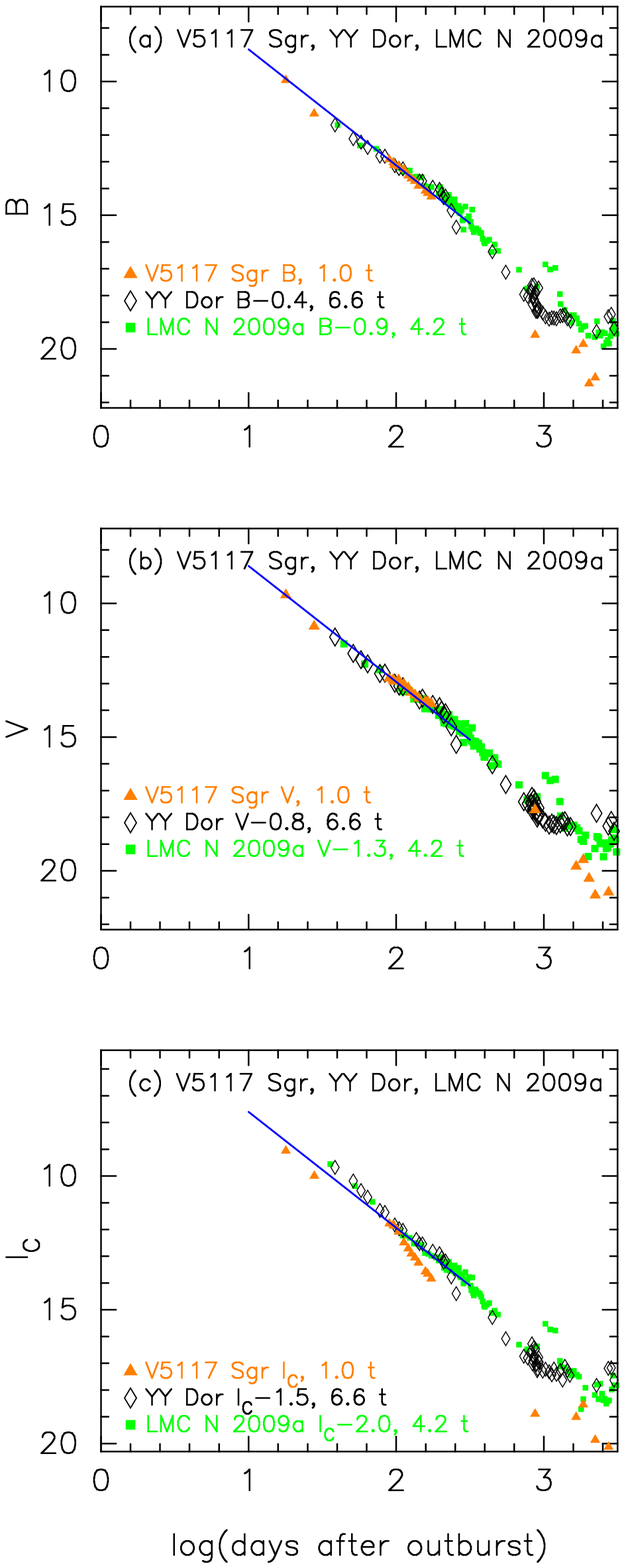}
\caption{
Same as Figure \ref{v1663_aql_yy_dor_lmcn_2009a_b_v_i_logscale_3fig},
but for V5117~Sgr.
The $BVI_{\rm C}$ data of V5117~Sgr are taken from SMARTS.
\label{v5117_sgr_yy_dor_lmcn_2009a_b_v_i_logscale_3fig}}
\end{figure}

\subsection{V5117~Sgr 2006\#2}
\label{v5117_sgr}
Figure \ref{v5117_sgr_v_bv_ub_color_curve} shows (a) the $V$ and
(b) $(B-V)_0$ evolutions of V5117~Sgr.  Here, $(B-V)_0$ are dereddened
with $E(B-V)=0.53$ as obtained in Section \ref{v5117_sgr_cmd}.
The $V$ light curve of V5117~Sgr is similar to that of LV~Vul
and V1668~Cyg as shown in Figure
\ref{v5117_sgr_v1668_cyg_lv_vul_v_bv_ub_logscale}.
The data of V5117~Sgr are the same as those in Figure
\ref{v5117_sgr_v_bv_ub_color_curve}.
The $V$ light curves of these three novae overlap each other.
Applying Equation (\ref{distance_modulus_general_temp}) to them,
we have the relation 
\begin{eqnarray}
(m&-&M)_{V, \rm V5117~Sgr} \cr 
&=& (m - M + \Delta V)_{V, \rm LV~Vul} - 2.5 \log 1.12 \cr
&=& 11.85 + 4.3\pm0.2 - 0.13 = 16.02\pm0.2 \cr
&=& (m - M + \Delta V)_{V, \rm V1668~Cyg} - 2.5 \log 1.12 \cr
&=& 14.6 + 1.5\pm0.2 - 0.13 = 15.97\pm0.2,
\label{distance_modulus_v5117_sgr}
\end{eqnarray}
where we adopt $(m-M)_{V, \rm LV~Vul}=11.85$ and
$(m-M)_{V, \rm V1668~Cyg}=14.6$, both from \citet{hac19k}.
Thus, we obtain $(m-M)_V=16.0\pm0.1$ and $f_{\rm s}=1.12$ for V5117~Sgr.
From Equations (\ref{time-stretching_general}),
(\ref{distance_modulus_general_temp}), and
(\ref{distance_modulus_v5117_sgr}),
we have the relation
\begin{eqnarray}
(m- M')_{V, \rm V5117~Sgr} 
&\equiv & (m_V - (M_V - 2.5\log f_{\rm s}))_{\rm V5117~Sgr} \cr
&=& \left( (m-M)_V + \Delta V \right)_{\rm LV~Vul} \cr
&=& 11.85 + 4.3\pm0.2 = 16.15\pm0.2.
\label{absolute_mag_v5117_sgr}
\end{eqnarray}

Figure \ref{v5117_sgr_yy_dor_lmcn_2009a_b_v_i_logscale_3fig} shows
the $B$, $V$, and $I_{\rm C}$ light curves of V5117~Sgr
together with those of YY~Dor and LMC~N~2009a.
The main part of the theoretical free-free emission light curve decays
as $F_\nu \propto t^{-1.75}$ (the universal decline law of classical novae:
straight solid blue line).  However, the $I_{\rm C}$ light-curve matching
is poor in Figure \ref{v5117_sgr_yy_dor_lmcn_2009a_b_v_i_logscale_3fig}(c).
In this case, we use only the $I_{\rm C}$ brightness near
$t\sim 100$~days, because the other $B$ and $V$ light curves follow well
the $F_\nu \propto t^{-1.75}$ law near $t\sim 100$~days, and we expect
that the $I_{\rm C}$ brightness represents the free-free flux around
this epoch.  We apply Equation (\ref{distance_modulus_general_temp_b})
for the $B$ band to Figure
\ref{v5117_sgr_yy_dor_lmcn_2009a_b_v_i_logscale_3fig}(a)
and obtain
\begin{eqnarray}
(m&-&M)_{B, \rm V5117~Sgr} \cr
&=& ((m - M)_B + \Delta B)_{\rm YY~Dor} - 2.5 \log 6.6 \cr
&=& 18.98 - 0.4\pm0.2 - 2.05 = 16.53\pm0.2 \cr
&=& ((m - M)_B + \Delta B)_{\rm LMC~N~2009a} - 2.5 \log 4.2 \cr
&=& 18.98 - 0.9\pm0.2 - 1.55 = 16.53\pm0.2.
\label{distance_modulus_b_v5117_sgr_yy_dor_lmcn2009a}
\end{eqnarray}
Thus, we obtain $(m-M)_{B, \rm V5117~Sgr}= 16.53\pm0.1$.

For the $V$ light curves in Figure
\ref{v5117_sgr_yy_dor_lmcn_2009a_b_v_i_logscale_3fig}(b),
we similarly obtain
\begin{eqnarray}
(m&-&M)_{V, \rm V5117~Sgr} \cr
&=& ((m - M)_V + \Delta V)_{\rm YY~Dor} - 2.5 \log 6.6 \cr
&=& 18.86 - 0.8\pm0.2 - 2.05 = 16.01\pm0.2 \cr
&=& ((m - M)_V + \Delta V)_{\rm LMC~N~2009a} - 2.5 \log 4.2 \cr
&=& 18.86 - 1.3\pm0.2 -1.55 = 16.01\pm0.2.
\label{distance_modulus_v_v5117_sgr_yy_dor_lmcn2009a}
\end{eqnarray}
Thus, we obtain $(m-M)_{V, \rm V5117~Sgr}= 16.01\pm0.1$, which is
consistent with Equation (\ref{distance_modulus_v5117_sgr}).

We apply Equation (\ref{distance_modulus_general_temp_i}) for
the $I_{\rm C}$-band to Figure
\ref{v5117_sgr_yy_dor_lmcn_2009a_b_v_i_logscale_3fig}(c) and obtain
\begin{eqnarray}
(m&-&M)_{I, \rm V5117~Sgr} \cr
&=& ((m - M)_I + \Delta I_C)_{\rm YY~Dor} - 2.5 \log 6.6 \cr
&=& 18.67 - 1.5\pm0.2 - 2.05 = 15.12\pm 0.2 \cr
&=& ((m - M)_I + \Delta I_C)_{\rm LMC~N~2009a} - 2.5 \log 4.2 \cr
&=& 18.67 - 2.0\pm0.2 -1.55 = 15.12\pm 0.2.
\label{distance_modulus_i_v5117_sgr_yy_dor_lmcn2009a}
\end{eqnarray}
Thus, we obtain $(m-M)_{I, \rm V5117~Sgr}= 15.12\pm0.1$.

We plot $(m-M)_B= 16.53$, $(m-M)_V= 16.01$, and $(m-M)_I= 15.12$,
which cross at $d=7.4$~kpc and $E(B-V)=0.53$, in Figure
\ref{distance_reddening_v1663_aql_v5116_sgr_v2575_oph_v5117_sgr}(d).
Thus, we obtain $E(B-V)=0.53\pm0.05$ and $d=7.4\pm0.8$~kpc.


\begin{figure}
\epsscale{0.75}
\plotone{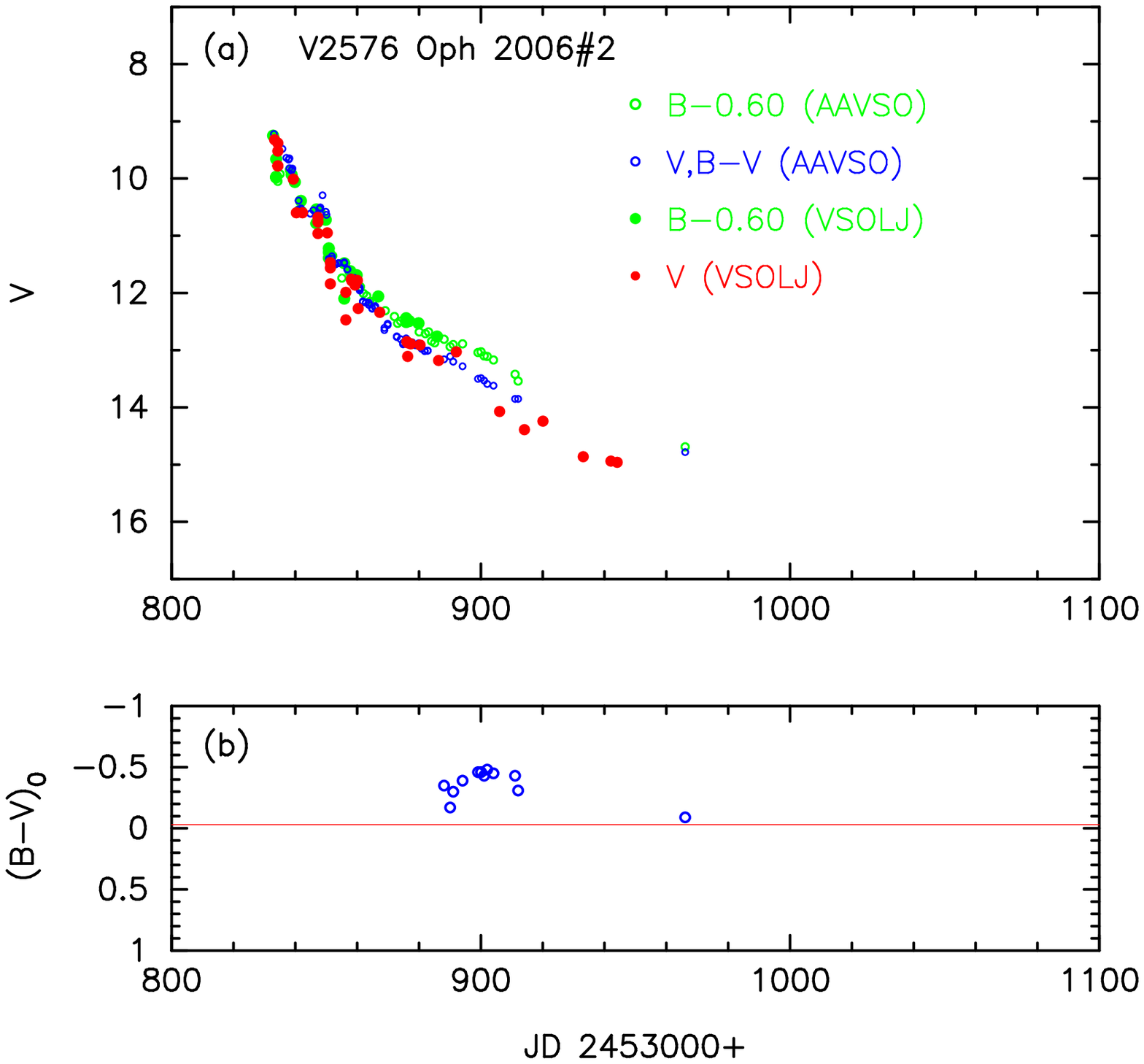}
\caption{
Same as Figure \ref{v1663_aql_v_bv_ub_color_curve}, but for V2576~Oph.
(a) The $V$ data are taken from AAVSO (unfilled blue circles) and
VSOLJ (filled red circles).  We also plot the $B$ magnitudes of
AAVSO (unfilled green circles) and VSOLJ (filled green circles),
shifted up by 0.60 mag.
(b) The $(B-V)_0$ are dereddened with $E(B-V)=0.62$.
\label{v2576_oph_v_bv_ub_color_curve}}
\end{figure}


\begin{figure}
\epsscale{0.75}
\plotone{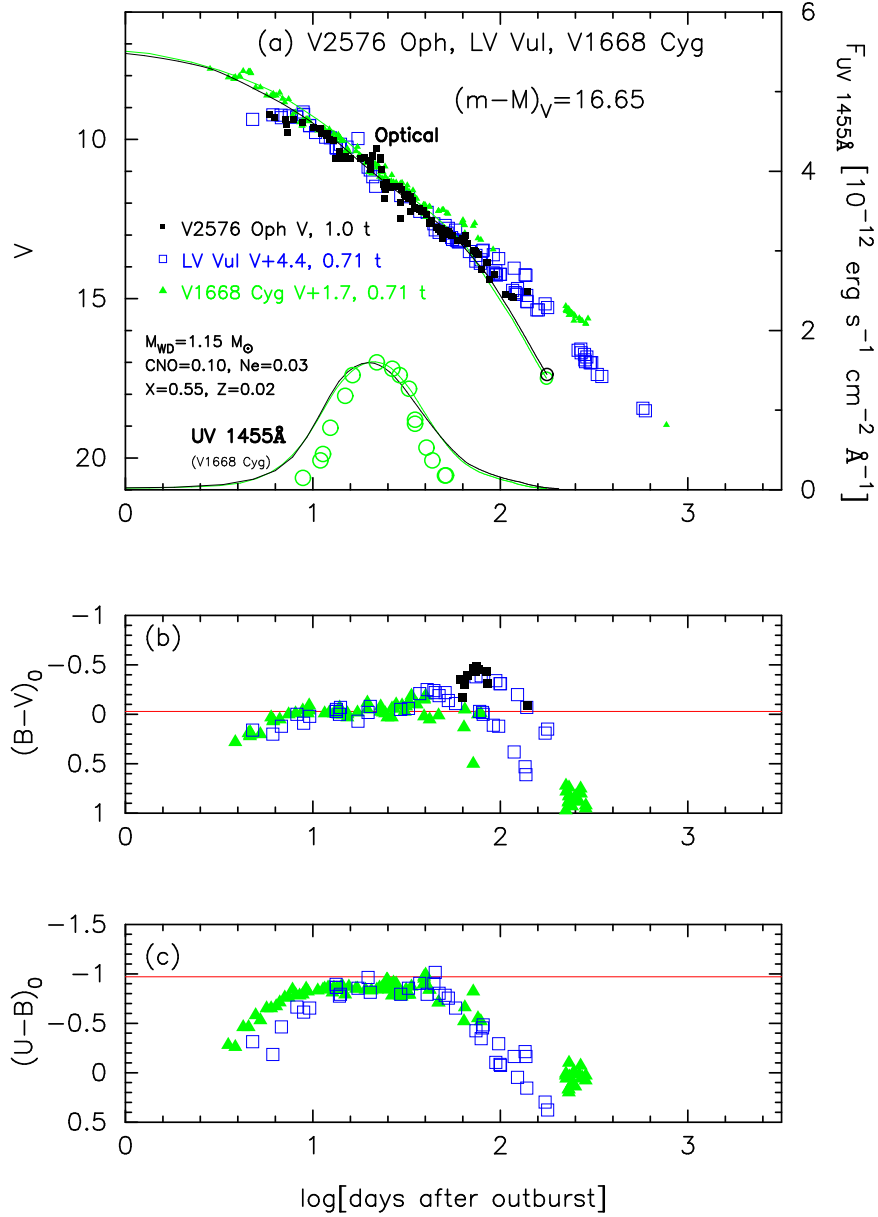}
\caption{
Same as Figure \ref{v2575_oph_v1668_cyg_lv_vul_v_bv_ub_logscale},
but for V2576~Oph.  We add the light/color curves of LV~Vul and V1668~Cyg.
The data of V2576~Oph are the same as those in Figure
\ref{v2576_oph_v_bv_ub_color_curve}.  
We add model light curves of a $1.15~M_\sun$ WD \citep[Ne2, 
solid black lines;][]{hac10k} for V2576~Oph and a $0.98~M_\sun$ WD 
\citep[CO3, solid green lines;][]{hac16k} for V1668~Cyg.
The upper solid black and green lines denote the $V$ light curves
of photospheric plus free-free emission for the $1.15~M_\sun$ WD 
for V2576~Oph and $0.98~M_\sun$ WD for V1668~Cyg, respectively.
The lower solid green lines represent
the UV~1455\AA\  flux of the $0.98~M_\sun$ WD based on the
blackbody approximation \citep[see, e.g.,][]{hac16k}.
\label{v2576_oph_v1668_cyg_lv_vul_v_bv_ub_logscale}}
\end{figure}


\begin{figure}
\epsscale{0.55}
\plotone{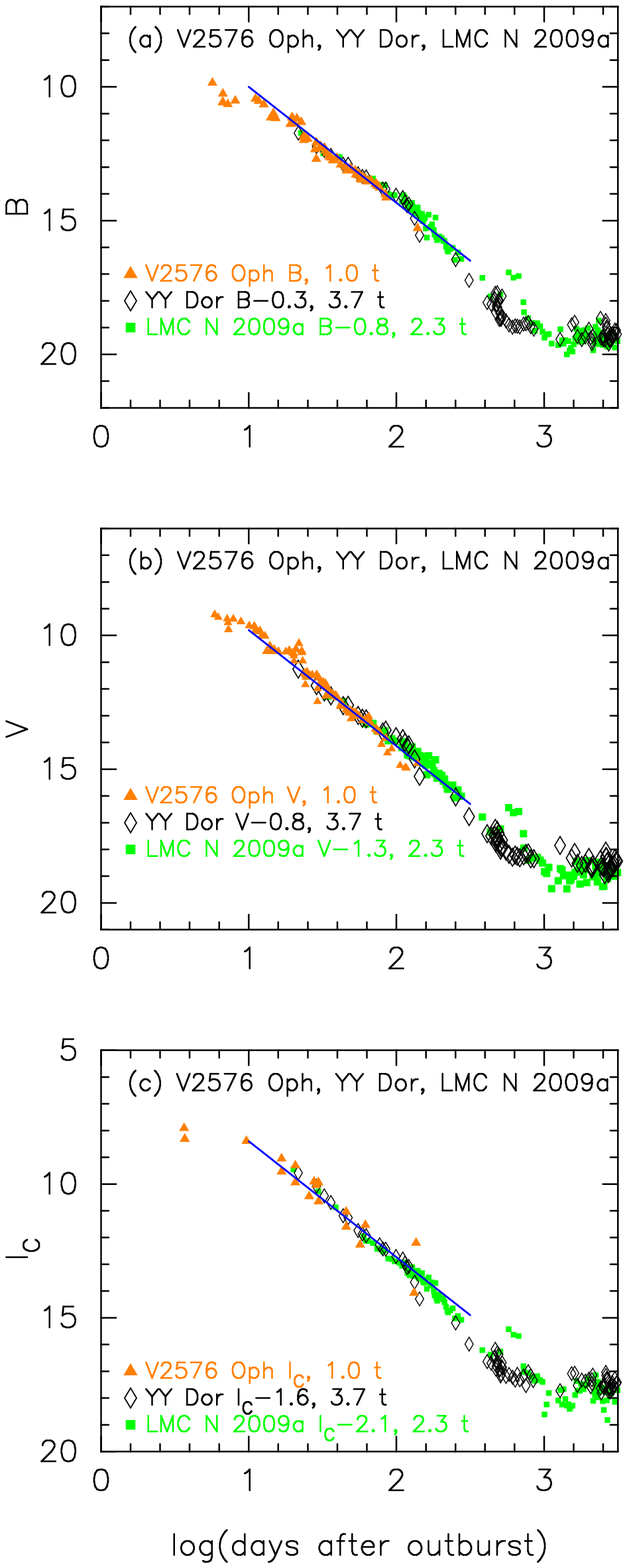}
\caption{
Same as Figure \ref{v1663_aql_yy_dor_lmcn_2009a_b_v_i_logscale_3fig},
but for V2576~Oph.
The $BV$ data of V2576~Oph are the same as those in Figure
\ref{v2576_oph_v_bv_ub_color_curve}.  The $I_{\rm C}$ data are taken
from AAVSO and VSOLJ.
\label{v2576_oph_yy_dor_lmcn_2009a_b_v_i_logscale_3fig}}
\end{figure}

\subsection{V2576~Oph 2006\#2}
\label{v2576_oph}
Figure \ref{v2576_oph_v_bv_ub_color_curve} shows (a) the $V$ and $B$,
and (b) $(B-V)_0$ evolutions of V2576~Oph.  Here, $(B-V)_0$
are dereddened with $E(B-V)=0.62$ as obtained in Section \ref{v2576_oph_cmd}.
There are plenty of $B$ and $V$ data points from the VSOLJ archive
(filled green and red circles, respectively).  However, we adopt $B-V$ data 
only when the two ($B$ and $V$) observed times are simultaneous within 
$0.15$~days.  For this constraint, we have no $B-V$ data from the VSOLJ data.
The $V$ light curve of V2576~Oph is similar to that of LV~Vul
and V1668~Cyg, as shown in Figure
\ref{v2576_oph_v1668_cyg_lv_vul_v_bv_ub_logscale}.
The data of V2576~Oph are the same as those in Figure
\ref{v2576_oph_v_bv_ub_color_curve}.
These $V$ light curves of the three novae overlap each other.
Applying Equation (\ref{distance_modulus_general_temp}) to them,
we have the relation
\begin{eqnarray}
(m&-&M)_{V, \rm V2576~Oph} \cr 
&=& (m - M + \Delta V)_{V, \rm LV~Vul} - 2.5 \log 0.71 \cr
&=& 11.85 + 4.4\pm0.2 + 0.38 = 16.63\pm0.2 \cr
&=& (m - M + \Delta V)_{V, \rm V1668~Cyg} - 2.5 \log 0.71 \cr
&=& 14.6 + 1.7\pm0.2 + 0.38 = 16.68\pm0.2.
\label{distance_modulus_v2576_oph}
\end{eqnarray}
Thus, we adopt $(m-M)_V=16.65\pm0.1$ and $f_{\rm s}=0.71$ for V2576~Oph.
From Equations (\ref{time-stretching_general}),
(\ref{distance_modulus_general_temp}), and
(\ref{distance_modulus_v2576_oph}),
we have the relation
\begin{eqnarray}
(m- M')_{V, \rm V2576~Oph} 
&\equiv & (m_V - (M_V - 2.5\log f_{\rm s}))_{\rm V2576~Oph} \cr
&=& \left( (m-M)_V + \Delta V \right)_{\rm LV~Vul} \cr
&=& 11.85 + 4.4\pm0.2 = 16.25\pm0.2.
\label{absolute_mag_v2576_oph}
\end{eqnarray}

Figure \ref{v2576_oph_yy_dor_lmcn_2009a_b_v_i_logscale_3fig} shows
the $B$, $V$, and $I_{\rm C}$ light curves of V2576~Oph
together with those of YY~Dor and LMC~N~2009a.
We apply Equation (\ref{distance_modulus_general_temp_b})
for the $B$ band to Figure
\ref{v2576_oph_yy_dor_lmcn_2009a_b_v_i_logscale_3fig}(a)
and obtain
\begin{eqnarray}
(m&-&M)_{B, \rm V2576~Oph} \cr
&=& ((m - M)_B + \Delta B)_{\rm YY~Dor} - 2.5 \log 3.7 \cr
&=& 18.98 - 0.3\pm0.2 - 1.43 = 17.25\pm0.2 \cr
&=& ((m - M)_B + \Delta B)_{\rm LMC~N~2009a} - 2.5 \log 2.3 \cr
&=& 18.98 - 0.8\pm0.2 - 0.93 = 17.25\pm0.2.
\label{distance_modulus_b_v2576_oph_yy_dor_lmcn2009a}
\end{eqnarray}
Thus, we have $(m-M)_{B, \rm V5117~Sgr}= 17.25\pm0.1$.

For the $V$ light curves in Figure
\ref{v2576_oph_yy_dor_lmcn_2009a_b_v_i_logscale_3fig}(b),
we similarly obtain
\begin{eqnarray}
(m&-&M)_{V, \rm V2576~Oph} \cr
&=& ((m - M)_V + \Delta V)_{\rm YY~Dor} - 2.5 \log 3.7 \cr
&=& 18.86 - 0.8\pm0.2 - 1.43 = 16.63\pm0.2 \cr
&=& ((m - M)_V + \Delta V)_{\rm LMC~N~2009a} - 2.5 \log 2.3 \cr
&=& 18.86 - 1.3\pm0.2 - 0.93 = 16.63\pm0.2.
\label{distance_modulus_v_v2576_oph_yy_dor_lmcn2009a}
\end{eqnarray}
We have $(m-M)_{V, \rm V5117~Sgr}= 16.63\pm0.1$, which is
consistent with Equation (\ref{distance_modulus_v2576_oph}).

We apply Equation (\ref{distance_modulus_general_temp_i}) for
the $I_{\rm C}$-band to Figure
\ref{v2576_oph_yy_dor_lmcn_2009a_b_v_i_logscale_3fig}(c) and obtain
\begin{eqnarray}
(m&-&M)_{I, \rm V2576~Oph} \cr
&=& ((m - M)_I + \Delta I_C)_{\rm YY~Dor} - 2.5 \log 3.7 \cr
&=& 18.67 - 1.6\pm0.2 - 1.43 = 15.64\pm 0.2 \cr
&=& ((m - M)_I + \Delta I_C)_{\rm LMC~N~2009a} - 2.5 \log 2.3 \cr
&=& 18.67 - 2.1\pm0.2 - 0.93 = 15.64\pm 0.2.
\label{distance_modulus_i_v2576_oph_yy_dor_lmcn2009a}
\end{eqnarray}
Thus, we have $(m-M)_{I, \rm V2576~Oph}= 15.64\pm0.1$.

We plot $(m-M)_B= 17.25$, $(m-M)_V= 16.63$, and $(m-M)_I= 15.64$,
which cross at $d=8.8$~kpc and $E(B-V)=0.62$, in Figure
\ref{distance_reddening_v2576_oph_v1281_sco_v390_nor_v597_pup}(a).
Thus, we have $E(B-V)=0.62\pm0.05$ and $d=8.8\pm1$~kpc.


\begin{figure}
\epsscale{0.75}
\plotone{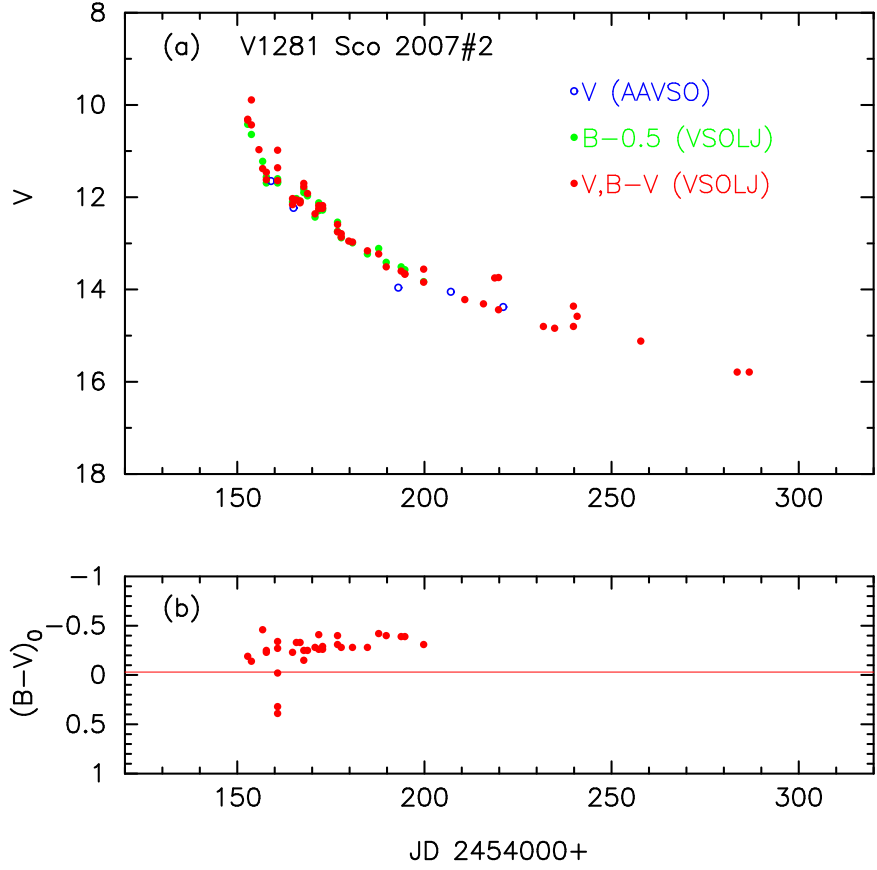}
\caption{
Same as Figure \ref{v1663_aql_v_bv_ub_color_curve}, but for V1281~Sco.
(a) The $V$ data are taken from AAVSO (unfilled blue circles) and
VSOLJ (filled red circles).  We also add the $B$ magnitude 
(filled green circles) of VSOLJ, which are shifted up by 0.5 mag.
(b) The $(B-V)_0$ are dereddened with $E(B-V)=0.82$.
\label{v1281_sco_v_bv_ub_color_curve}}
\end{figure}


\begin{figure}
\epsscale{0.75}
\plotone{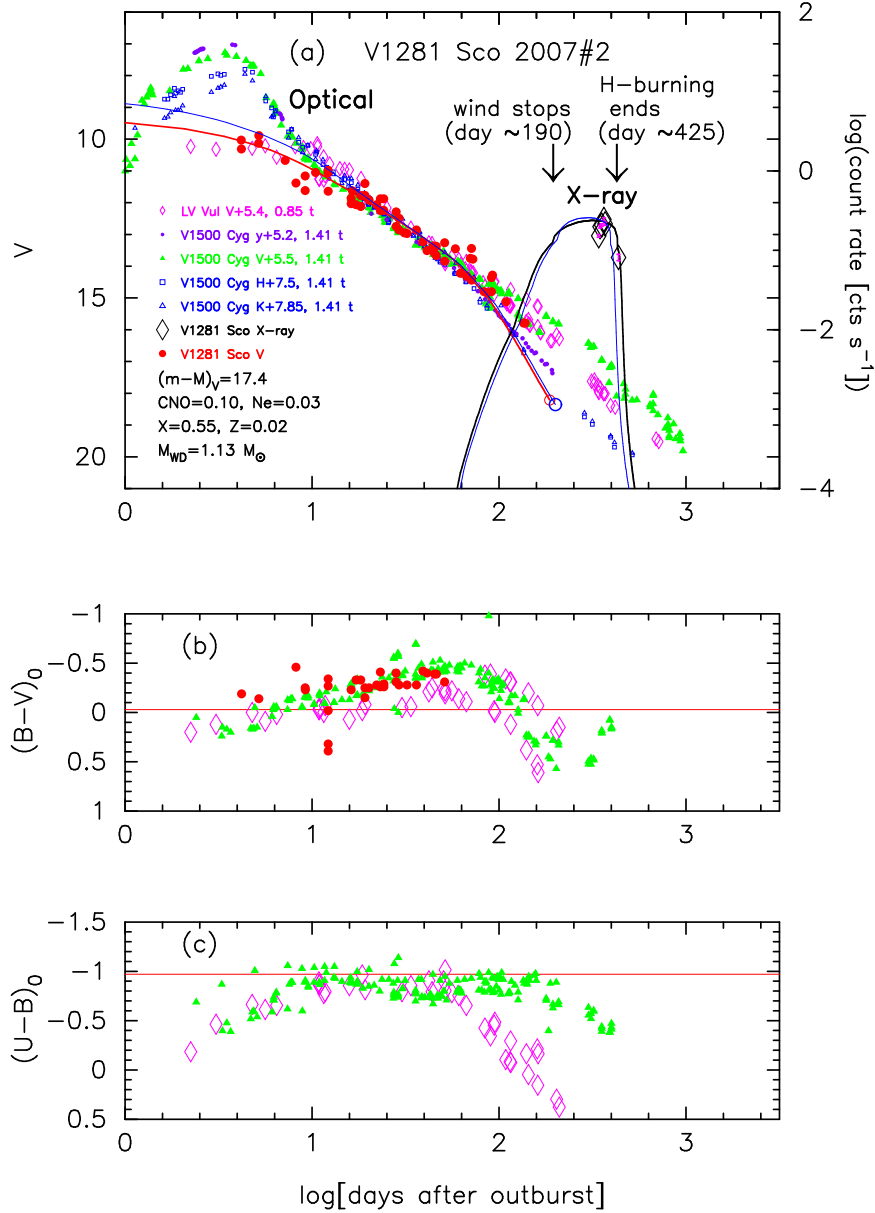}
\caption{
Same as Figure \ref{v2575_oph_v1668_cyg_lv_vul_v_bv_ub_logscale},
but for V1281~Sco.  We add the light/color curves of LV~Vul and V1500~Cyg.
The data of V1281~Sco are the same as those in Figure
\ref{v1281_sco_v_bv_ub_color_curve}.  
Assuming $(m-M)_V=17.4$, we added model light curves of a $1.13~M_\sun$ WD
\citep[Ne2;][]{hac10k}.
In panel (a), the solid red line denotes the $V$ light curve
of a photospheric plus free-free emission light curve while
the solid black line represents the blackbody supersoft X-ray
light curve, for the $1.13~M_\sun$ WD.
We also add the model light curve of a $1.20~M_\sun$ WD (Ne2, thin solid blue
lines), assuming $(m-M)_V=12.3$ for V1500~Cyg.
\label{v1281_sco_v1500_cyg_m1130_x55z02o10ne03_logscale}}
\end{figure}


\begin{figure}
\epsscale{0.55}
\plotone{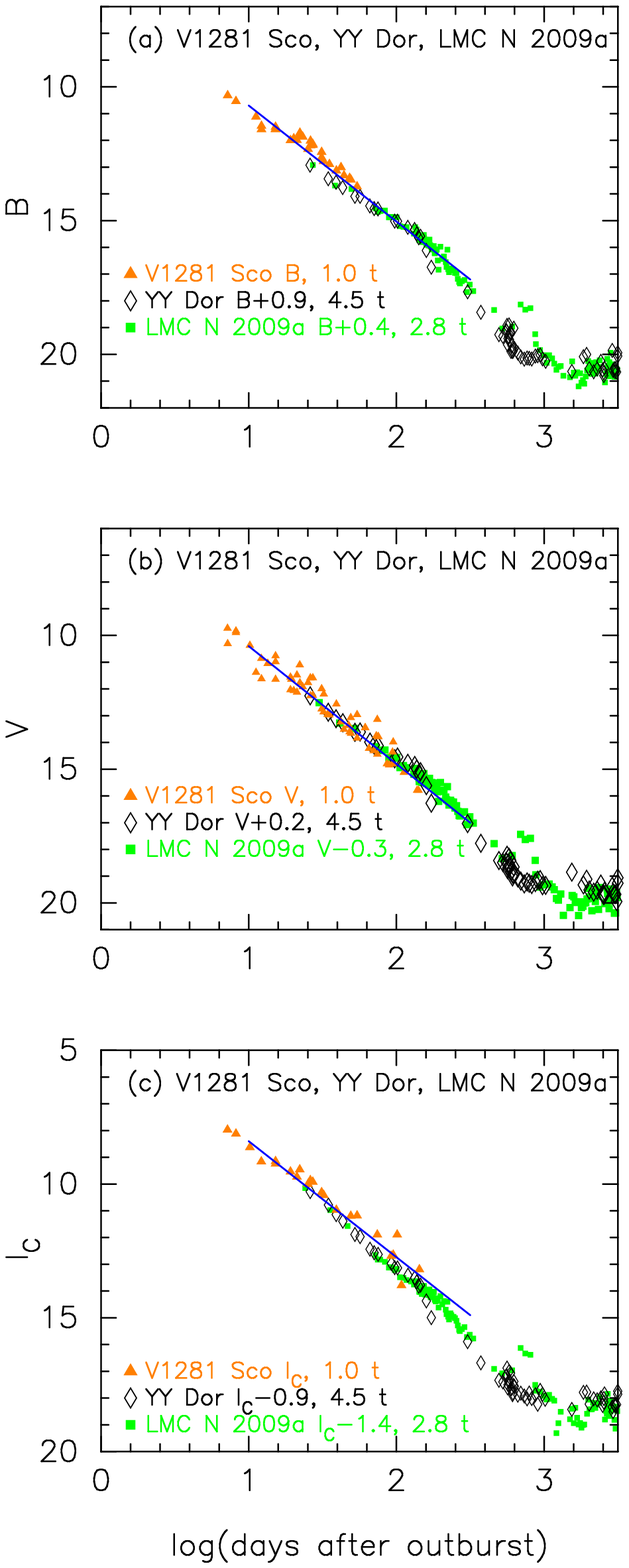}
\caption{
Same as Figure \ref{v1663_aql_yy_dor_lmcn_2009a_b_v_i_logscale_3fig},
but for V1281~Sco.
The $BV$ data of V1281~Sco are the same as those in Figure
\ref{v1281_sco_v_bv_ub_color_curve}.  The $I_{\rm C}$ data are taken
from VSOLJ.
\label{v1281_sco_yy_dor_lmcn_2009a_b_v_i_logscale_3fig}}
\end{figure}

\subsection{V1281~Sco 2007\#2}
\label{v1281_sco}
Figure \ref{v1281_sco_v_bv_ub_color_curve} shows (a) the $V$ and
(b) $(B-V)_0$ evolutions of V1281~Sco.  Here, $(B-V)_0$ are dereddened
with $E(B-V)=0.82$ as obtained in Section \ref{v1281_sco_cmd}.  
Figure \ref{v1281_sco_v1500_cyg_m1130_x55z02o10ne03_logscale} shows
the light/color evolution of V1281~Sco
as well as those of V1500~Cyg and LV~Vul.
The data of V1281~Sco are the same as those in Figure
\ref{v1281_sco_v_bv_ub_color_curve}.
The data of V1500~Cyg are the same as those in Figure 9 of \citet{hac10k}
and Figure 6 of \citet{hac14k}.
The $(B-V)_0$ color curve of V1281~Sco is similar to V1500~Cyg
rather than LV~Vul as shown in Figure
\ref{v1281_sco_v1500_cyg_m1130_x55z02o10ne03_logscale}(b).  A large part 
of these three $V$ light curves overlap each other.
Applying Equation (\ref{distance_modulus_general_temp}) to them,
we have the relation 
\begin{eqnarray}
(m&-&M)_{V, \rm V1281~Sco} \cr 
&=& (m - M + \Delta V)_{V, \rm LV~Vul} - 2.5 \log 0.85 \cr
&=& 11.85 + 5.4\pm0.2 + 0.18 = 17.43\pm0.2 \cr
&=& (m - M + \Delta V)_{V, \rm V1500~Cyg} - 2.5 \log 1.41 \cr
&=& 12.3 + 5.5\pm0.2 - 0.38 = 17.42\pm0.2,
\label{distance_modulus_v1281_sco}
\end{eqnarray}
where we adopt $(m-M)_{V, \rm LV~Vul}=11.85$ and
$(m-M)_{V, \rm V1500~Cyg}=12.3$, 
both from \citet{hac19k}.
Thus, we obtained $(m-M)_V=17.4\pm0.1$ and $f_{\rm s}=0.85$ against
LV~Vul.  The newly obtained value is slightly smaller 
than Hachisu \& Kato's (2010) value of $(m-M)_V=17.8\pm0.2$.
This is because \citet{hac10k} assumed $(m-M)_{V, \rm V1500~Cyg}=12.5$
instead of $(m-M)_{V, \rm V1500~Cyg}=12.3$, and we improved 
the vertical $V$ fit. 
From Equations (\ref{time-stretching_general}),
(\ref{distance_modulus_general_temp}), and
(\ref{distance_modulus_v1281_sco}),
we have the relation
\begin{eqnarray}
(m- M')_{V, \rm V1281~Sco} 
&\equiv & (m_V - (M_V - 2.5\log f_{\rm s}))_{\rm V1281~Sco} \cr
&=& \left( (m-M)_V + \Delta V \right)_{\rm LV~Vul} \cr
&=& 11.85 + 5.4\pm0.2 = 17.25\pm0.2.
\label{absolute_mag_v1281_sco}
\end{eqnarray}

Figure \ref{v1281_sco_yy_dor_lmcn_2009a_b_v_i_logscale_3fig} shows
the $B$, $V$, and $I_{\rm C}$ light curves of V1281~Sco
together with those of YY~Dor and LMC~N~2009a.
We apply Equation (\ref{distance_modulus_general_temp_b})
for the $B$ band to Figure
\ref{v1281_sco_yy_dor_lmcn_2009a_b_v_i_logscale_3fig}(a)
and obtain
\begin{eqnarray}
(m&-&M)_{B, \rm V1281~Sco} \cr
&=& ((m - M)_B + \Delta B)_{\rm YY~Dor} - 2.5 \log 4.5 \cr
&=& 18.98 + 0.9\pm0.2 - 1.63 = 18.25\pm0.2 \cr
&=& ((m - M)_B + \Delta B)_{\rm LMC~N~2009a} - 2.5 \log 2.8 \cr
&=& 18.98 + 0.4\pm0.2 - 1.13 = 18.25\pm0.2.
\label{distance_modulus_b_v1281_sco_yy_dor_lmcn2009a}
\end{eqnarray}
Thus, we have $(m-M)_{B, \rm V1281~Sco}= 18.25\pm0.1$.

For the $V$ light curves in Figure
\ref{v1281_sco_yy_dor_lmcn_2009a_b_v_i_logscale_3fig}(b),
we similarly obtain
\begin{eqnarray}
(m&-&M)_{V, \rm V1281~Sco} \cr
&=& ((m - M)_V + \Delta V)_{\rm YY~Dor} - 2.5 \log 4.5 \cr
&=& 18.86 + 0.2\pm0.2 - 1.63 = 17.43\pm0.2 \cr
&=& ((m - M)_V + \Delta V)_{\rm LMC~N~2009a} - 2.5 \log 2.8 \cr
&=& 18.86 - 0.3\pm0.2 -1.13 = 17.43\pm0.2.
\label{distance_modulus_v_v1281_sco_yy_dor_lmcn2009a}
\end{eqnarray}
We have $(m-M)_{V, \rm V1281~Sco}= 17.43\pm0.1$, which is
consistent with Equation (\ref{distance_modulus_v1281_sco}).

We apply Equation (\ref{distance_modulus_general_temp_i}) for
the $I_{\rm C}$-band to Figure
\ref{v1281_sco_yy_dor_lmcn_2009a_b_v_i_logscale_3fig}(c) and obtain
\begin{eqnarray}
(m&-&M)_{I, \rm V1281~Sco} \cr
&=& ((m - M)_I + \Delta I_C)_{\rm YY~Dor} - 2.5 \log 4.5 \cr
&=& 18.67 - 0.9\pm0.2 - 1.63 = 16.14\pm 0.2 \cr
&=& ((m - M)_I + \Delta I_C)_{\rm LMC~N~2009a} - 2.5 \log 2.8 \cr
&=& 18.67 - 1.4\pm0.2 -1.13 = 16.14\pm 0.2.
\label{distance_modulus_i_v1281_sco_yy_dor_lmcn2009a}
\end{eqnarray}
Thus, we have $(m-M)_{I, \rm V1281~Sco}= 16.14\pm0.1$.

We plot $(m-M)_B= 18.25$, $(m-M)_V= 17.43$, and $(m-M)_I= 16.14$,
which broadly cross at $d=9.4$~kpc and $E(B-V)=0.82$, in Figure
\ref{distance_reddening_v2576_oph_v1281_sco_v390_nor_v597_pup}(b).
Thus, we have $E(B-V)=0.82\pm0.05$ and $d=9.4\pm1$~kpc.


\begin{figure}
\epsscale{0.75}
\plotone{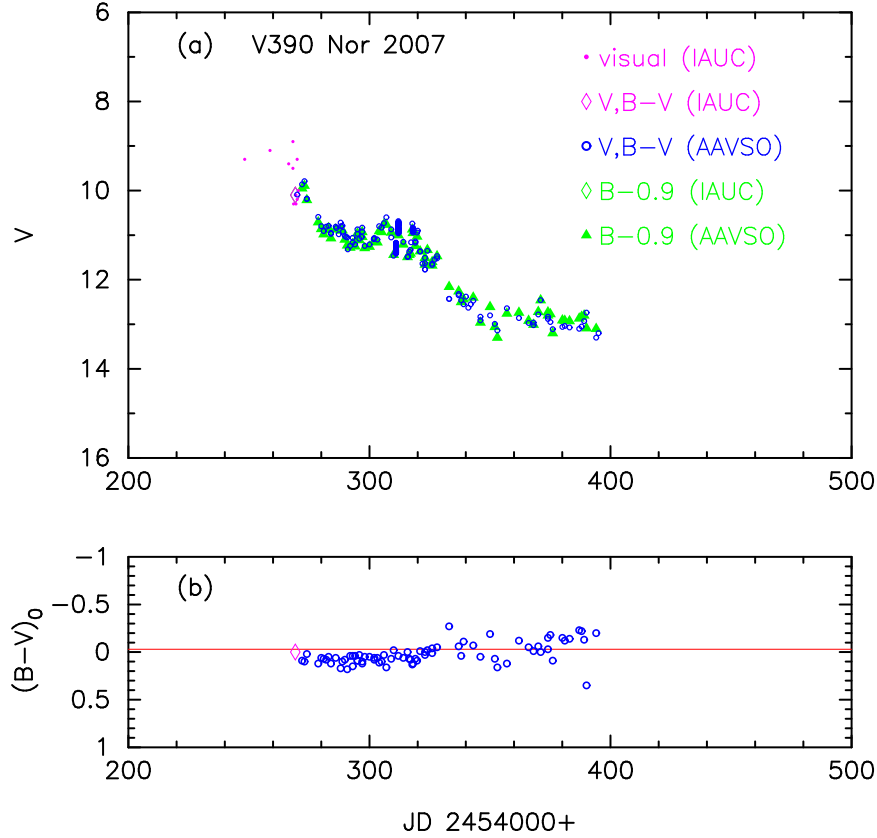}
\caption{
Same as Figure \ref{v1663_aql_v_bv_ub_color_curve}, but for V390~Nor.
(a) The $V$ (unfilled blue circles) and $B$ (filled green triangles) data
are taken from AAVSO.  The visual (magenta dots), $V$
(unfilled magenta diamonds), and $B$ (unfilled green diamond)
data are from IAU Circular 8850, 1. 
The $B$ magnitudes are shifted up by 0.9 mag.
(b) The $(B-V)_0$ are dereddened with $E(B-V)=0.89$.
\label{v390_nor_v_bv_ub_color_curve}}
\end{figure}


\begin{figure}
\epsscale{0.75}
\plotone{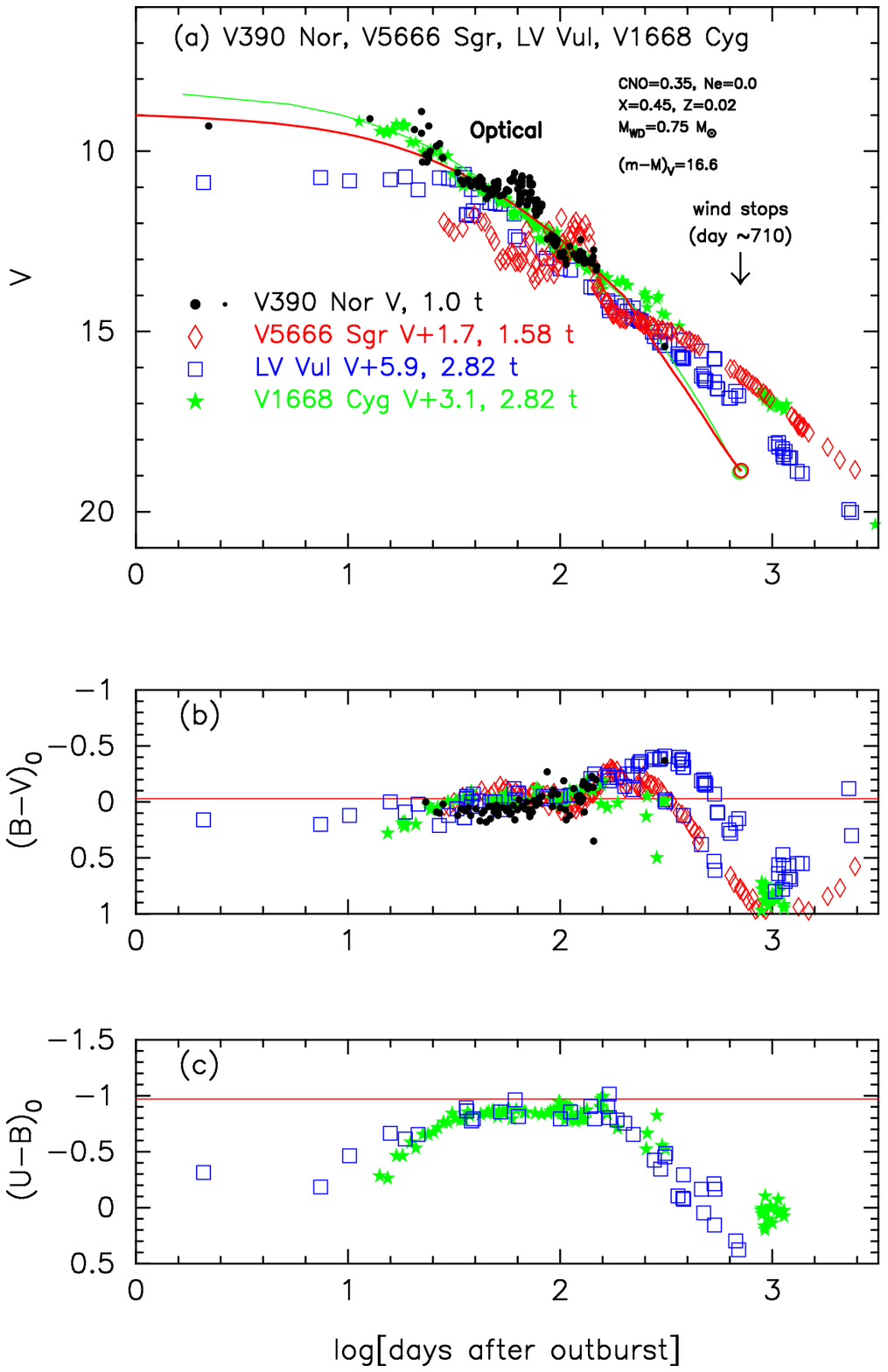}
\caption{
Same as Figure \ref{v2575_oph_v1668_cyg_lv_vul_v_bv_ub_logscale},
but for V390~Nor.  We add the light/color curves of V5666~Sgr, 
LV~Vul, and V1668~Cyg.
The data of V390~Nor are the same as those in Figure
\ref{v390_nor_v_bv_ub_color_curve}.  
The data of V5666~Sgr are the same as those in Figures 28 and 29
of \citet{hac19k}.  Assuming $(m-M)_V=16.6$, we add the model light curve
of a $0.75~M_\sun$ WD \citep[CO3, solid red line;][]{hac16k}.
We also add the model light curve of a $0.98~M_\sun$ WD 
(CO3, solid green line), assuming that $(m-M)_V= 14.6$ for V1668~Cyg.
\label{v390_nor_v5666_sgr_lv_vul_v1668_cyg_x55z02c10o10_logscale}}
\end{figure}


\begin{figure}
\epsscale{0.65}
\plotone{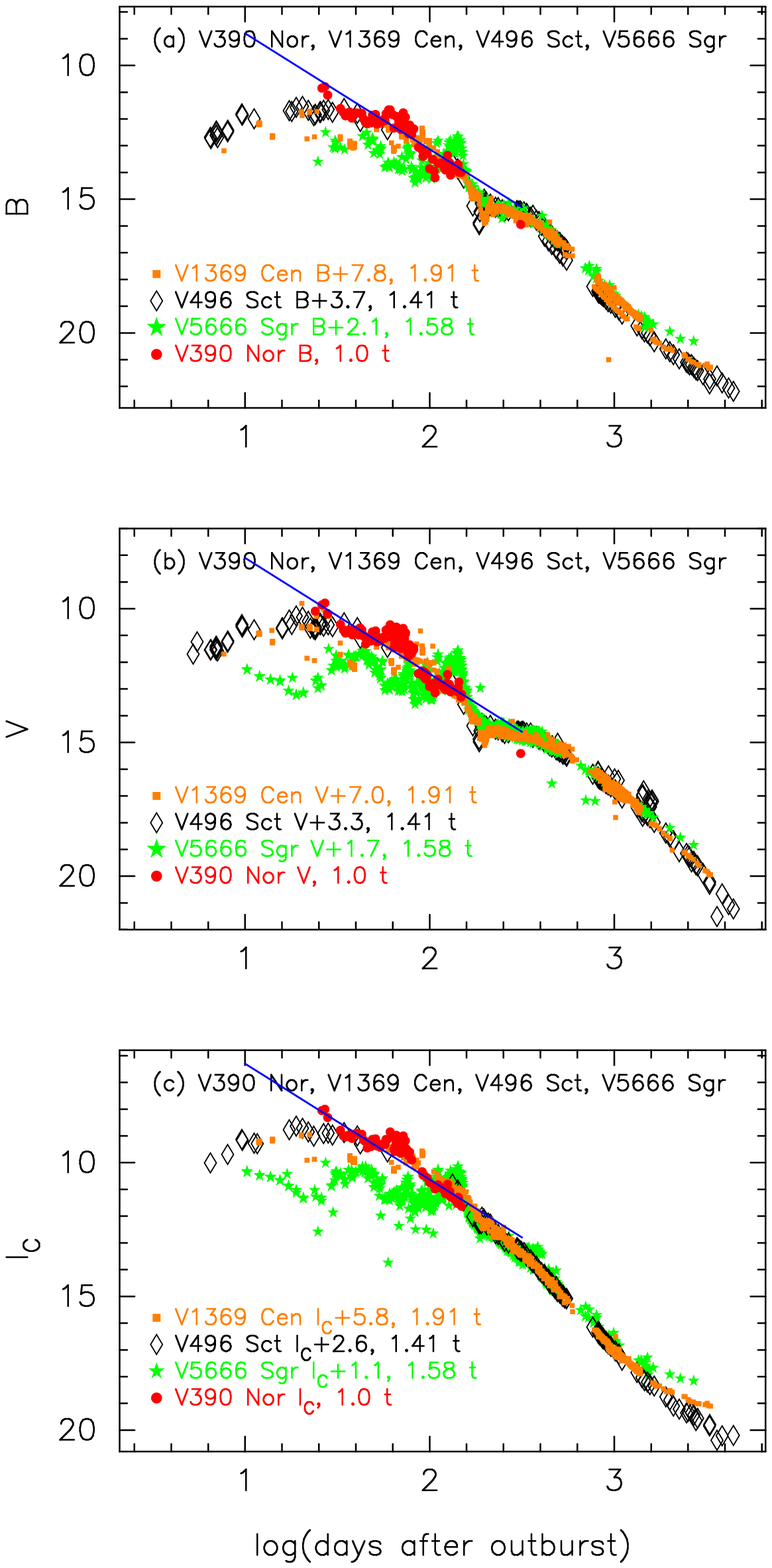}
\caption{
Same as Figure \ref{v1663_aql_yy_dor_lmcn_2009a_b_v_i_logscale_3fig},
but for V390~Nor.  We plot the 
(a) $B$, (b) $V$, and (c) $I_{\rm C}$ light curves of V390~Nor
as well as those of V1369~Cen, V496~Sct, and V5666~Sgr.
The data of V1369~Cen, V496~Sct, and V5666~Sgr are the same as
those in Figures 24, 25, 26, 28, 29, and 30 of \citet{hac19k}.
The straight solid blue line denotes the slope of $F_\nu\propto t^{-1.75}$.
The $BV$ data of V390~Nor are the same as those in Figure
\ref{v390_nor_v_bv_ub_color_curve}.  The $I_{\rm C}$ data are taken
from AAVSO.
\label{v390_nor_v1369_cen_v496_sct_v5666_sgr_b_v_i_logscale_3fig}}
\end{figure}

\subsection{V390~Nor 2007}
\label{v390_nor}
Figure \ref{v390_nor_v_bv_ub_color_curve} shows (a) the $V$, $B$, and
(b) $(B-V)_0$ evolutions of V390~Nor.  Here, $(B-V)_0$ are dereddened
with $E(B-V)=0.89$ as obtained in Section \ref{v390_nor_cmd}.
The visual magnitude (magenta dots) is taken from \citet{lil07}.
The light/color curves of V390~Nor are plotted 
in Figure \ref{v390_nor_v5666_sgr_lv_vul_v1668_cyg_x55z02c10o10_logscale}
together with the light/color curves of V5666~Sgr, LV~Vul, and V1668~Cyg.
The data of V390~Nor are the same as those in Figure
\ref{v390_nor_v_bv_ub_color_curve}.
The $BV$ data of V5666~Sgr are the same as those in Figure 28
of \citet{hac19k}.
These four $V$ light curves roughly overlap each other. 
Applying Equation (\ref{distance_modulus_general_temp}) to them,
we have the relation 
\begin{eqnarray}
(m&-&M)_{V, \rm V390~Nor} \cr 
&=& (m - M + \Delta V)_{V, \rm LV~Vul} - 2.5 \log 2.82 \cr
&=& 11.85 + 5.9\pm0.3 - 1.13 = 16.62\pm0.3 \cr
&=& (m - M + \Delta V)_{V, \rm V1668~Cyg} - 2.5 \log 2.82 \cr
&=& 14.6 + 3.1\pm0.3 - 1.13 = 16.62\pm0.3 \cr
&=& (m - M + \Delta V)_{V, \rm V5666~Sgr} - 2.5 \log 1.58 \cr
&=& 15.4 + 1.7\pm0.3 - 0.50 = 16.60\pm0.3,
\label{distance_modulus_v390_nor}
\end{eqnarray}
where we adopt 
$(m-M)_{V, \rm V5666~Sgr}=15.4$ from \citet{hac19k}.
Thus, we obtain $(m-M)_V=16.6\pm0.2$ and $f_{\rm s}=2.82$ against LV~Vul.
From Equations (\ref{time-stretching_general}),
(\ref{distance_modulus_general_temp}), and
(\ref{distance_modulus_v390_nor}),
we have the relation
\begin{eqnarray}
(m- M')_{V, \rm V390~Nor} 
&\equiv & (m_V - (M_V - 2.5\log f_{\rm s}))_{\rm V390~Nor} \cr
&=& \left( (m-M)_V + \Delta V \right)_{\rm LV~Vul} \cr
&=& 11.85 + 5.9\pm0.3 = 17.75\pm0.3.
\label{absolute_mag_v390_nor}
\end{eqnarray}

Figure \ref{v390_nor_v1369_cen_v496_sct_v5666_sgr_b_v_i_logscale_3fig}
shows the $B$, $V$, and $I_{\rm C}$ light curves of V390~Nor
together with those of V1369~Cen, V496~Sct, and V5666~Sgr.
In this case, the early-phase $BVI_{\rm C}$ light curves show wavy
(oscillatory) structures and do not seem to follow the universal decline law.
However, we can observe a moderately good overlapping of V390~Nor (filled
red circles) with V496~Sct (unfilled black diamonds) and V1369~Cen
(filled orange squares) especially near the straight solid blue lines
of $F_\nu \propto t^{-1.75}$.  We require a short period overlapping
on the solid blue line because the later observational data
show a long period overlapping with the other novae.  Applying Equation 
(\ref{distance_modulus_general_temp_b}) for the $B$ band to Figure
\ref{v390_nor_v1369_cen_v496_sct_v5666_sgr_b_v_i_logscale_3fig}(a),
we have the relation
\begin{eqnarray}
(m&-&M)_{B, \rm V390~Nor} \cr
&=& \left( (m-M)_B + \Delta B\right)_{\rm V1369~Cen} - 2.5 \log 1.91 \cr
&=& 10.36 + 7.8\pm0.3 - 0.7 = 17.46\pm0.3 \cr
&=& \left( (m-M)_B + \Delta B\right)_{\rm V496~Sct} - 2.5 \log 1.41 \cr
&=& 14.15 + 3.7\pm0.3 - 0.38 = 17.47\pm0.3 \cr
&=& \left( (m-M)_B + \Delta B\right)_{\rm V5666~Sgr} - 2.5 \log 1.58 \cr
&=& 15.9 + 2.1\pm0.3 - 0.5 = 17.5\pm0.3,
\label{distance_modulus_v396_nor_v1369_cen_v496_sct_v5666_sgr_b}
\end{eqnarray}
where we adopt $(m-M)_{B, \rm V1369~Cen}=10.25 + 1.0\times 0.11= 10.36$,
$(m-M)_{B, \rm V496~Sct}=13.7 + 1.0\times 0.45= 14.15$, and
$(m-M)_{B, \rm V5666~Sgr}=15.4 + 1.0\times 0.50=15.9$,
all from \citet{hac19k}.
Thus, we obtain $(m-M)_B=17.48\pm0.2$ for V390~Nor.

Applying Equation (\ref{distance_modulus_general_temp}) to
Figure \ref{v390_nor_v1369_cen_v496_sct_v5666_sgr_b_v_i_logscale_3fig}(b),
we have the relation
\begin{eqnarray}
(m&-&M)_{V, \rm V390~Nor} \cr
&=& \left( (m-M)_V + \Delta V\right)_{\rm V1369~Cen} - 2.5 \log 1.91 \cr
&=& 10.25 + 7.0\pm0.3 - 0.7 = 16.55\pm0.3 \cr
&=& \left( (m-M)_V + \Delta V\right)_{\rm V496~Sct} - 2.5 \log 1.41 \cr
&=& 13.7 + 3.3\pm0.3 - 0.38 = 16.62\pm0.3 \cr
&=& \left( (m-M)_V + \Delta V\right)_{\rm V5666~Sgr} - 2.5 \log 1.58 \cr
&=& 15.4 + 1.7\pm0.3 - 0.5 = 16.6\pm0.3,
\label{distance_modulus_v390_nor_v1369_cen_v496_sct_v5666_sgr_v}
\end{eqnarray}
where we adopt $(m-M)_{V, \rm V1369~Cen}=10.25$,
$(m-M)_{V, \rm V496~Sct}=13.7$, and $(m-M)_{V, \rm V5666~Sgr}=15.4$,
all from \citet{hac19k}.
Thus, we obtain $(m-M)_V=16.59\pm0.2$ for V390~Nor, which is
consistent with Equation (\ref{distance_modulus_v390_nor}).

From the $I_{\rm C}$-band data in Figure
\ref{v390_nor_v1369_cen_v496_sct_v5666_sgr_b_v_i_logscale_3fig}(c),
we obtain
\begin{eqnarray}
(m&-&M)_{I, \rm V390~Nor} \cr
&=& ((m - M)_I + \Delta I_C)_{\rm V1369~Cen} - 2.5 \log 1.91 \cr
&=& 10.07 + 5.8\pm0.3 - 0.7 = 15.17\pm0.3 \cr
&=& ((m - M)_I + \Delta I_C)_{\rm V496~Sct} - 2.5 \log 1.41 \cr
&=& 12.98 + 2.6\pm0.3 - 0.38 = 15.2\pm0.3 \cr
&=& ((m - M)_I + \Delta I_C)_{\rm V5666~Sgr} - 2.5 \log 1.58 \cr
&=& 14.6 + 1.1\pm0.3 - 0.5 = 15.2\pm0.3,
\label{distance_modulus_i_v390_nor_v1369_cen_v496_sct_v5666_sgr}
\end{eqnarray}
where we adopt $(m-M)_{I, \rm V1369~Cen}=10.25 - 1.6\times 0.11= 10.07$,
$(m-M)_{I, \rm V496~Sct}=13.7 - 1.6\times 0.45= 12.98$,
and $(m-M)_{I, \rm V5666~Sgr}=15.4 - 1.6\times 0.50= 14.6$.
Thus, we obtain $(m-M)_{I, \rm V390~Nor}= 15.19\pm0.2$.

We plot $(m-M)_B=17.48$, $(m-M)_V=16.59$, and $(m-M)_I=15.19$,
which cross at $d= 5.8$~kpc and $E(B-V)=0.89$, in Figure 
\ref{distance_reddening_v2576_oph_v1281_sco_v390_nor_v597_pup}(c).
Thus, we obtain $E(B-V)=0.89\pm0.05$ and $d= 5.8\pm0.6$~kpc.


\begin{figure}
\epsscale{0.75}
\plotone{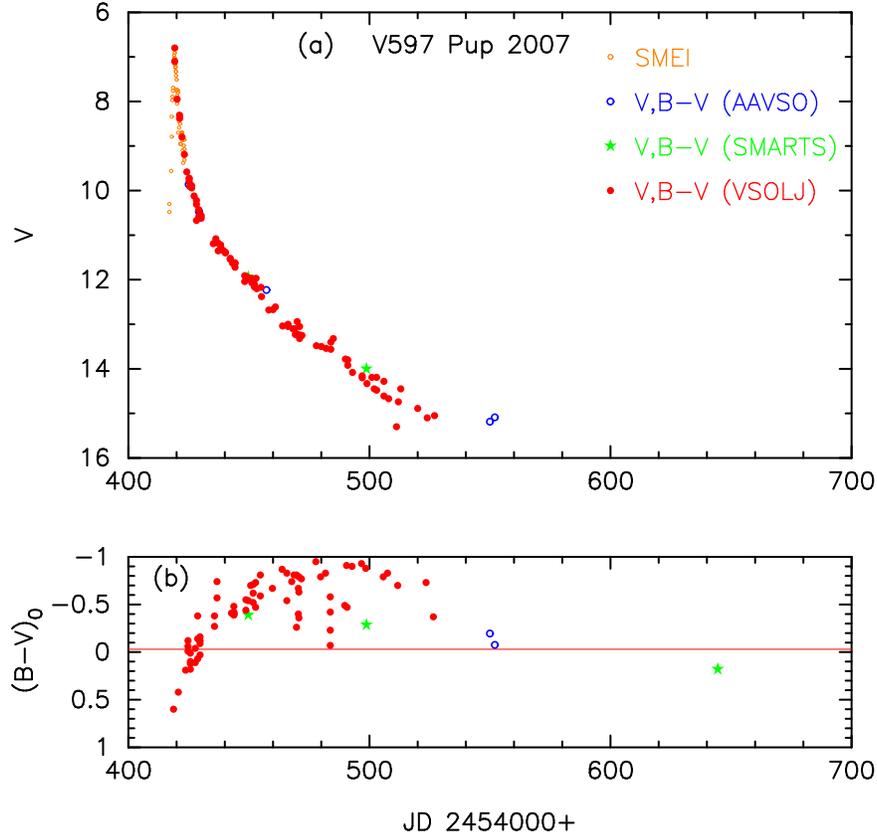}
\caption{
Same as Figure \ref{v1663_aql_v_bv_ub_color_curve}, but for V597~Pup.
(a) The $BV$ data (unfilled blue circles, filled green stars,
and filled red circles) are taken from AAVSO, SMARTS, and VSOLJ,
respectively.  We also add the data of {\it SMEI} (small unfilled
orange circles). 
(b) The $(B-V)_0$ are dereddened with $E(B-V)=0.24$.
\label{v597_pup_v_bv_ub_color_curve}}
\end{figure}


\begin{figure}
\epsscale{0.75}
\plotone{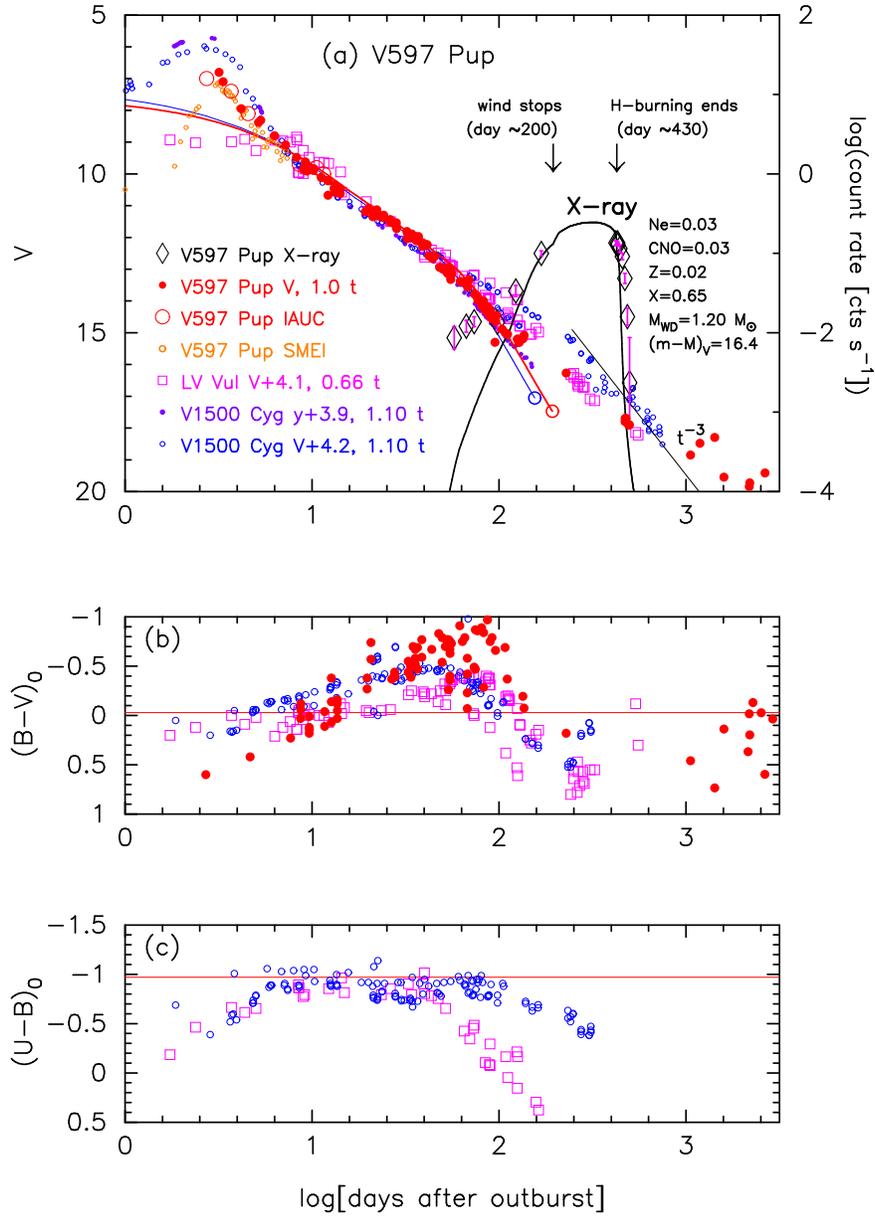}
\caption{
Same as Figure \ref{v2575_oph_v1668_cyg_lv_vul_v_bv_ub_logscale},
but for V597~Pup.  We add the light/color curves of LV~Vul and V1500~Cyg. 
The data of V597~Pup are the same as those in Figure 
\ref{v597_pup_v_bv_ub_color_curve}.  In panel (a), 
assuming that $(m-M)_V=16.4$ for V597~Pup, 
we added model light curves of a $1.2~M_\sun$ WD
\citep[Ne3;][]{hac16k}.
The solid red line denotes the $V$ light curves of blackbody plus
free-free emission while the solid black line represents the blackbody
supersoft X-ray light curve of the $1.2~M_\sun$ WD.
Assuming that $(m-M)_V=12.3$ for V1500~Cyg, we add a $1.2~M_\sun$ WD 
(Ne2) with the solid blue line.   
\label{v597_pup_v1500_cyg_lv_vul_v_bv_ub_x65z02o03ne03_logscale}}
\end{figure}


\begin{figure}
\epsscale{0.55}
\plotone{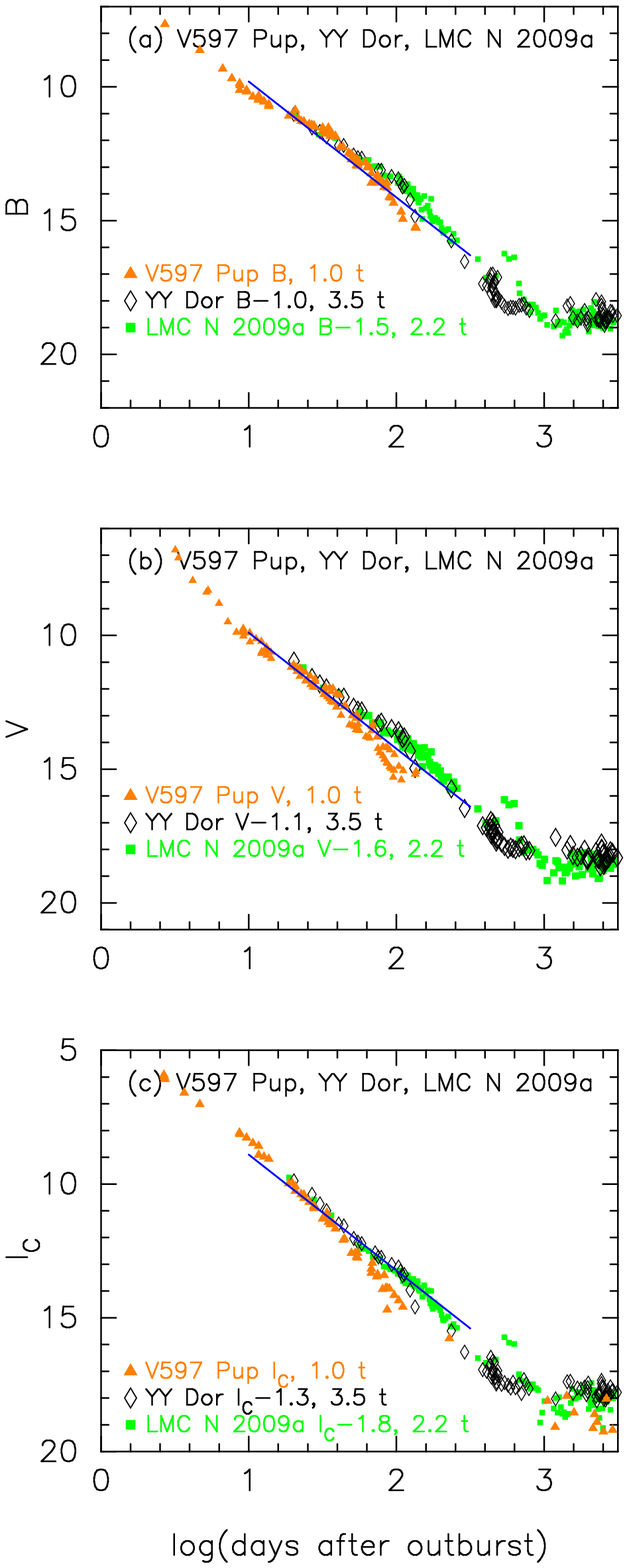}
\caption{
Same as Figure \ref{v1663_aql_yy_dor_lmcn_2009a_b_v_i_logscale_3fig},
but for V597~Pup.
The $BV$ data of V597~Pup are the same as those in Figure
\ref{v597_pup_v_bv_ub_color_curve}.  The $I_{\rm C}$ data are taken
from VSOLJ and SMARTS.
\label{v597_pup_yy_dor_lmcn_2009a_b_v_i_logscale_3fig}}
\end{figure}

\subsection{V597~Pup 2007}
\label{v597_pup}
Figure \ref{v597_pup_v_bv_ub_color_curve} shows (a) the $V$ and
{\it SMEI} magnitudes \citep{hou16},
and (b) $(B-V)_0$ color evolutions of V597~Pup.  Here, $(B-V)_0$ are
dereddened with $E(B-V)=0.24$ as obtained in Section \ref{v597_pup_cmd}.
The $V$ light curve of V597~Pup is similar to that of V1500~Cyg as shown
in Figure \ref{v597_pup_v1500_cyg_lv_vul_v_bv_ub_x65z02o03ne03_logscale},
where we add the light/color curves of LV~Vul and V1500~Cyg.
The data of V597~Pup are the same as those in Figure
\ref{v597_pup_v_bv_ub_color_curve}.    We also add two model light curves
of a $1.2~M_\sun$ WD (Ne3, solid red line) for V597~Pup
and $1.2~M_\sun$ WD (Ne2, solid blue line) for V1500~Cyg.

The optical light curve shape of V597~Pup near the peak is
largely different from our model $V$ light curve (solid red line); it is
similar to the shape of V1500~Cyg (solid blue line).
It is highly likely that V597~Pup is a superbright nova like
V1500~Cyg \citep{del91}.  The V1500~Cyg spectra in the superbright
phase show blackbody \citep{gal76, enn77} and the observed $V$ 
brightness greatly exceeds our model $V$ light curve.
After the superbright phase, the spectra changed to free-free
emission on day $\sim 5$.  After that, the $V$ light curve is
broadly approximated by our model $V$ light curve.
Thus, our time-stretching method applies to this free-free
emission-dominated phase.

These $V$ light curves overlap each other except for the superbright phase.
Applying Equation (\ref{distance_modulus_general_temp}) to them,
we have the relation 
\begin{eqnarray}
(m&-&M)_{V, \rm V597~Pup} \cr 
&=& (m - M + \Delta V)_{V, \rm LV~Vul} - 2.5 \log 0.66 \cr
&=& 11.85 + 4.1\pm0.2  + 0.45 = 16.4\pm0.2 \cr
&=& (m - M + \Delta V)_{V, \rm V1500~Cyg} - 2.5 \log 1.10 \cr
&=& 12.3 + 4.2\pm0.2  - 0.10 = 16.4\pm0.2.
\label{distance_modulus_v597_pup}
\end{eqnarray}
Thus, we obtain $(m-M)_V=16.4\pm0.1$ and $f_{\rm s}=0.66$ 
against LV~Vul.  From Equations (\ref{time-stretching_general}),
(\ref{distance_modulus_general_temp}), and
(\ref{distance_modulus_v597_pup}),
we have the relation
\begin{eqnarray}
(m- M')_{V, \rm V597~Pup} 
&\equiv & (m_V - (M_V - 2.5\log f_{\rm s}))_{\rm V597~Pup} \cr
&=& \left( (m-M)_V + \Delta V \right)_{\rm LV~Vul} \cr
&=& 11.85 + 4.1\pm0.2 = 15.95\pm0.2.
\label{absolute_mag_v597_pup}
\end{eqnarray}

Figure \ref{v597_pup_yy_dor_lmcn_2009a_b_v_i_logscale_3fig} shows
the $B$, $V$, and $I_{\rm C}$ light curves of V597~Pup
together with those of YY~Dor and LMC~N~2009a.
We apply Equation (\ref{distance_modulus_general_temp_b})
for the $B$ band to Figure
\ref{v597_pup_yy_dor_lmcn_2009a_b_v_i_logscale_3fig}(a)
and obtain
\begin{eqnarray}
(m&-&M)_{B, \rm V597~Pup} \cr
&=& ((m - M)_B + \Delta B)_{\rm YY~Dor} - 2.5 \log 3.5 \cr
&=& 18.98 - 1.0\pm0.2 - 1.35 = 16.63\pm0.2 \cr
&=& ((m - M)_B + \Delta B)_{\rm LMC~N~2009a} - 2.5 \log 2.2 \cr
&=& 18.98 - 1.5\pm0.2 - 0.85 = 16.63\pm0.2.
\label{distance_modulus_b_v597_pup_yy_dor_lmcn2009a}
\end{eqnarray}
Thus, we have $(m-M)_{B, \rm V597~Pup}= 16.63\pm0.1$.

For the $V$ light curves in Figure
\ref{v597_pup_yy_dor_lmcn_2009a_b_v_i_logscale_3fig}(b),
we similarly obtain
\begin{eqnarray}
(m&-&M)_{V, \rm V597~Pup} \cr
&=& ((m - M)_V + \Delta V)_{\rm YY~Dor} - 2.5 \log 3.5 \cr
&=& 18.86 - 1.1\pm0.2 - 1.35 = 16.41\pm0.2 \cr
&=& ((m - M)_V + \Delta V)_{\rm LMC~N~2009a} - 2.5 \log 2.2 \cr
&=& 18.86 - 1.6\pm0.2 -0.85 = 16.41\pm0.2.
\label{distance_modulus_v_v597_pup_yy_dor_lmcn2009a}
\end{eqnarray}
We have $(m-M)_{V, \rm V597~Pup}= 16.41\pm0.1$, which is
consistent with Equation (\ref{distance_modulus_v597_pup}).

We apply Equation (\ref{distance_modulus_general_temp_i}) for
the $I_{\rm C}$ band to Figure
\ref{v597_pup_yy_dor_lmcn_2009a_b_v_i_logscale_3fig}(c) and obtain
\begin{eqnarray}
(m&-&M)_{I, \rm V597~Pup} \cr
&=& ((m - M)_I + \Delta I_C)_{\rm YY~Dor} - 2.5 \log 3.5 \cr
&=& 18.67 - 1.3\pm0.2 - 1.35 = 16.02\pm 0.2 \cr
&=& ((m - M)_I + \Delta I_C)_{\rm LMC~N~2009a} - 2.5 \log 2.2 \cr
&=& 18.67 - 1.8\pm0.2 - 0.85 = 16.02\pm 0.2.
\label{distance_modulus_i_v597_pup_yy_dor_lmcn2009a}
\end{eqnarray}
Thus, we have $(m-M)_{I, \rm V597~Pup}= 16.02\pm0.1$.

We plot $(m-M)_B= 16.63$, $(m-M)_V= 16.41$, and $(m-M)_I= 16.02$,
which broadly cross at $d=13.5$~kpc and $E(B-V)=0.24$,  in Figure
\ref{distance_reddening_v2576_oph_v1281_sco_v390_nor_v597_pup}(d).
Thus, we have $E(B-V)=0.24\pm0.05$ and $d=13.5\pm2$~kpc.


\begin{figure}
\epsscale{0.75}
\plotone{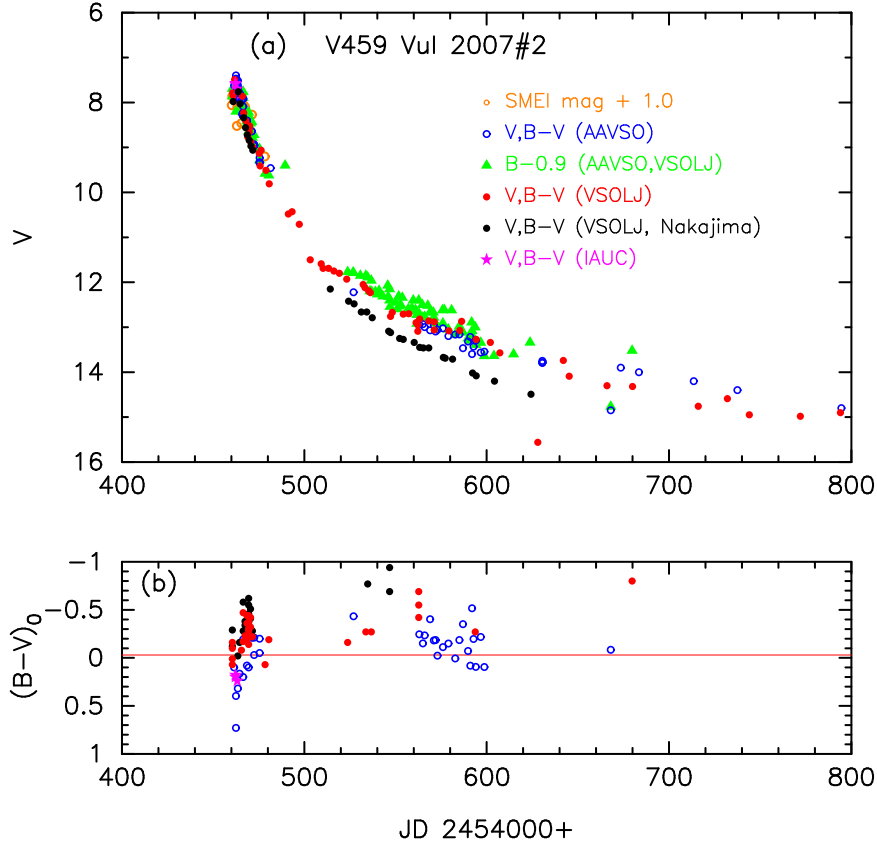}
\caption{
Same as Figure \ref{v1663_aql_v_bv_ub_color_curve}, but for V459~Vul.
(a) The $V$ (unfilled blue circles) and $B$ (filled green triangles) data
are taken from AAVSO.  The $V$ excluding Nakajima's data
(filled red circles), $V$ of Nakajima's data (filled black circles), and
$B$ (filled green triangles) data are from VSOLJ.
The {\it SMEI} data (unfilled orange circles) are also plotted.
(b) The $(B-V)_0$ are dereddened with $E(B-V)=0.90$.
\label{v459_vul_v_bv_ub_color_curve}}
\end{figure}


\begin{figure}
\epsscale{0.75}
\plotone{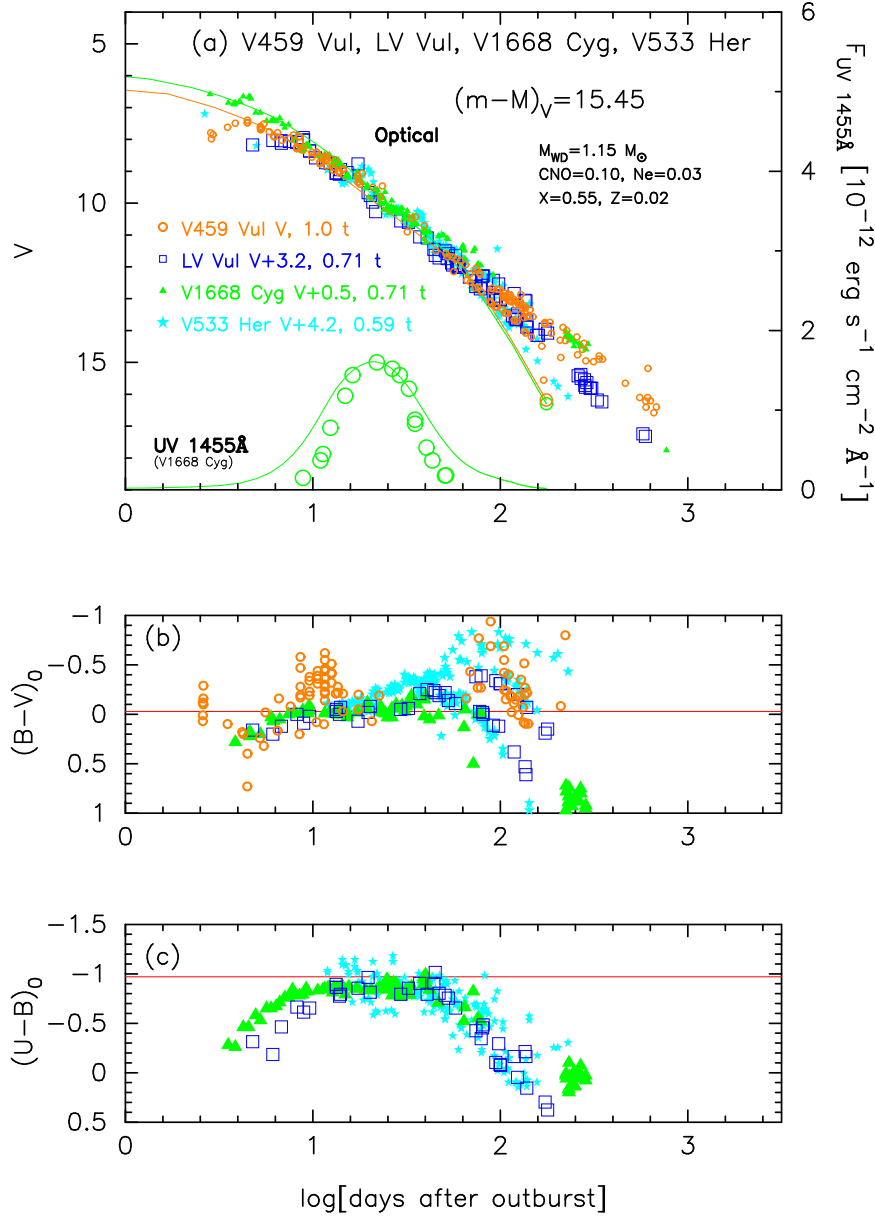}
\caption{
Same as Figure \ref{v2575_oph_v1668_cyg_lv_vul_v_bv_ub_logscale},
but for V459~Vul.  We add the light/color curves of LV~Vul, V1668~Cyg,
and V533~Her.  The data of V459~Vul are the same as those 
in Figure \ref{v459_vul_v_bv_ub_color_curve}.  
We added model light curves (solid orange lines) of a $1.15~M_\sun$ WD
\citep[Ne2;][]{hac10k},
assuming that $(m-M)_V=15.45$ for V459~Vul.   We also add 
a $0.98~M_\sun$ WD \citep[CO3, solid green lines;][]{hac16k},
assuming that $(m-M)_V=14.6$ for V1668~Cyg.
\label{v459_vul_v533_her_v1668_cyg_lv_vul_v_bv_ub_logscale}}
\end{figure}


\begin{figure}
\epsscale{0.55}
\plotone{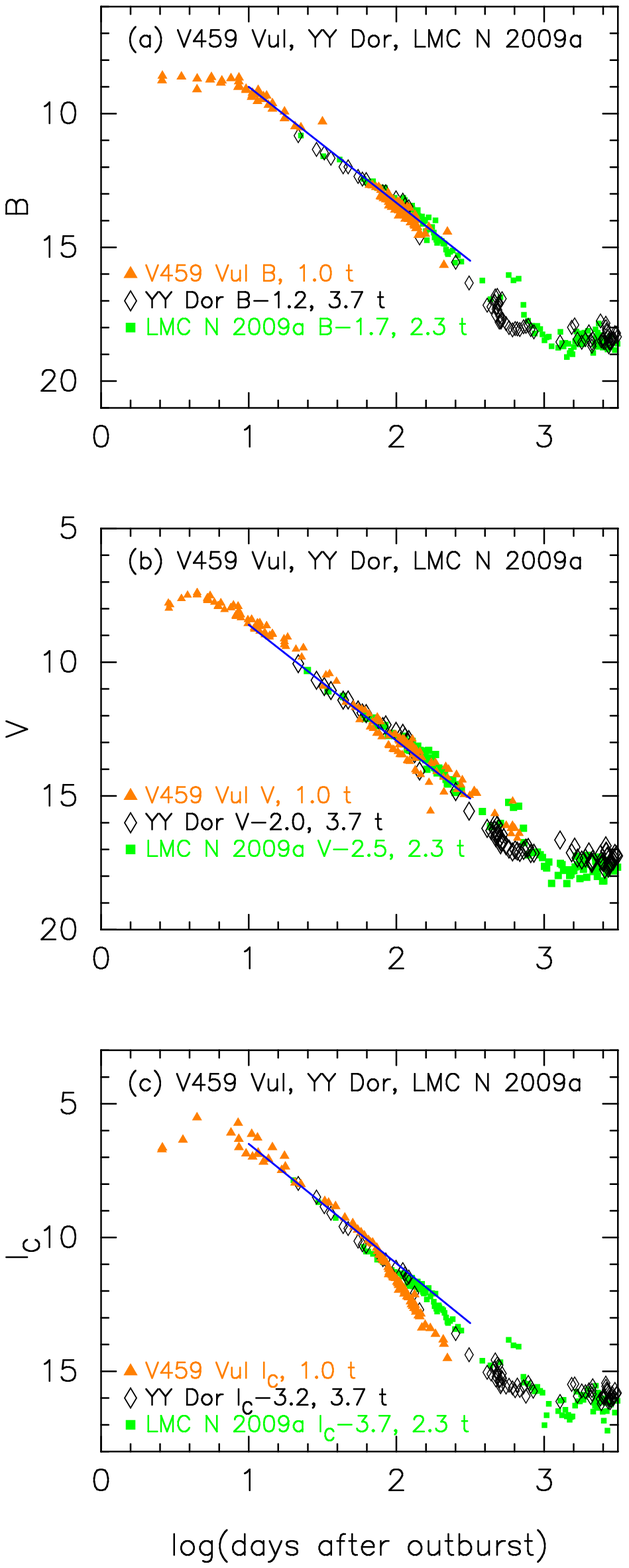}
\caption{
Same as Figure \ref{v1663_aql_yy_dor_lmcn_2009a_b_v_i_logscale_3fig},
but for V459~Vul.
The $BV$ data of V459~Vul are the same as those in Figure
\ref{v459_vul_v_bv_ub_color_curve}.  The $I_{\rm C}$ data are taken
from AAVSO and VSOLJ.
\label{v459_vul_yy_dor_lmcn_2009a_b_v_i_logscale_3fig}}
\end{figure}

\subsection{V459~Vul 2007\#2}
\label{v459_vul}
Figure \ref{v459_vul_v_bv_ub_color_curve} shows (a) the $V$, {\it SMEI},
and $B$ magnitudes, and (b) $(B-V)_0$ evolutions of V459~Vul.
Here, $(B-V)_0$ are dereddened with $E(B-V)=0.90$ as obtained
in Section \ref{v459_vul_cmd}.  The $B$ and $V$ data are taken from
CBET No.1183, AAVSO, and VSOLJ.  We also plot the {\it SMEI} magnitudes 
taken from \citet{hou16}.  Figure 
\ref{v459_vul_v533_her_v1668_cyg_lv_vul_v_bv_ub_logscale} shows  
the $V$ light and $(B-V)_0$ color curves of V459~Vul as well as 
those of LV~Vul, V1668~Cyg, and V533~Her.  The data of V533~Her are
the same as those in Figure 33 of \citet{hac19k}.
The data of V459~Vul are the same as those in Figure
\ref{v459_vul_v_bv_ub_color_curve}.
These four $V$ light curves overlap each other.
Applying Equation (\ref{distance_modulus_general_temp}) to them,
we have the relation 
\begin{eqnarray}
(m&-&M)_{V, \rm V459~Vul} \cr 
&=& (m - M + \Delta V)_{V, \rm LV~Vul} - 2.5 \log 0.71 \cr
&=& 11.85 + 3.2\pm0.2 + 0.38 = 15.43\pm0.2 \cr
&=& (m - M + \Delta V)_{V, \rm V1668~Cyg} - 2.5 \log 0.71 \cr
&=& 14.6 + 0.5\pm0.2 + 0.38 = 15.48\pm0.2 \cr
&=& (m - M + \Delta V)_{V, \rm V533~Her} - 2.5 \log 0.59 \cr
&=& 10.65 + 4.2\pm0.2 + 0.58 = 15.43\pm0.2,
\label{distance_modulus_v459_vul}
\end{eqnarray}
where we adopt 
$(m-M)_{V, \rm V533~Her}=10.65$ from \citet{hac19k}.
Thus, we obtain $(m-M)_V=15.45\pm0.1$ and $f_{\rm s}=0.71$ against LV~Vul.
From Equations (\ref{time-stretching_general}),
(\ref{distance_modulus_general_temp}), and
(\ref{distance_modulus_v459_vul}),
we have the relation
\begin{eqnarray}
(m- M')_{V, \rm V459~Vul} 
&\equiv & (m_V - (M_V - 2.5\log f_{\rm s}))_{\rm V459~Vul} \cr
&=& \left( (m-M)_V + \Delta V \right)_{\rm LV~Vul} \cr
&=& 11.85 + 3.2\pm0.2 = 15.05\pm0.2.
\label{absolute_mag_v459_vul}
\end{eqnarray}

Figure \ref{v459_vul_yy_dor_lmcn_2009a_b_v_i_logscale_3fig} shows
the $B$, $V$, and $I_{\rm C}$ light curves of V459~Vul
together with those of YY~Dor and LMC~N~2009a.
We apply Equation (\ref{distance_modulus_general_temp_b})
for the $B$ band to Figure
\ref{v459_vul_yy_dor_lmcn_2009a_b_v_i_logscale_3fig}(a)
and obtain
\begin{eqnarray}
(m&-&M)_{B, \rm V459~Vul} \cr
&=& ((m - M)_B + \Delta B)_{\rm YY~Dor} - 2.5 \log 3.7 \cr
&=& 18.98 - 1.2\pm0.2 - 1.43 = 16.35\pm0.2 \cr
&=& ((m - M)_B + \Delta B)_{\rm LMC~N~2009a} - 2.5 \log 2.3 \cr
&=& 18.98 - 1.7\pm0.2 - 0.93 = 16.35\pm0.2.
\label{distance_modulus_b_v459_vul_yy_dor_lmcn2009a}
\end{eqnarray}
Thus, we have $(m-M)_{B, \rm V459~Vul}= 16.35\pm0.1$.

For the $V$ light curves in Figure
\ref{v459_vul_yy_dor_lmcn_2009a_b_v_i_logscale_3fig}(b),
we similarly obtain
\begin{eqnarray}
(m&-&M)_{V, \rm V459~Vul} \cr
&=& ((m - M)_V + \Delta V)_{\rm YY~Dor} - 2.5 \log 3.7 \cr
&=& 18.86 - 2.0\pm0.2 - 1.43 = 15.43\pm0.2 \cr
&=& ((m - M)_V + \Delta V)_{\rm LMC~N~2009a} - 2.5 \log 2.3 \cr
&=& 18.86 - 2.5\pm0.2 - 0.93 = 15.43\pm0.2.
\label{distance_modulus_v_v459_vul_yy_dor_lmcn2009a}
\end{eqnarray}
We have $(m-M)_{V, \rm V459~Vul}= 15.43\pm0.1$, which is
consistent with Equation (\ref{distance_modulus_v459_vul}).

We apply Equation (\ref{distance_modulus_general_temp_i}) for
the $I_{\rm C}$ band to Figure
\ref{v459_vul_yy_dor_lmcn_2009a_b_v_i_logscale_3fig}(c) and obtain
\begin{eqnarray}
(m&-&M)_{I, \rm V459~Vul} \cr
&=& ((m - M)_I + \Delta I_C)_{\rm YY~Dor} - 2.5 \log 3.7 \cr
&=& 18.67 - 3.2\pm0.3 - 1.43 = 14.04\pm 0.3 \cr
&=& ((m - M)_I + \Delta I_C)_{\rm LMC~N~2009a} - 2.5 \log 2.3 \cr
&=& 18.67 - 3.7\pm0.3 -0.93 = 14.04\pm 0.3.
\label{distance_modulus_i_v459_vul_yy_dor_lmcn2009a}
\end{eqnarray}
Thus, we have $(m-M)_{I, \rm V459~Vul}= 14.04\pm0.2$.

We plot $(m-M)_B= 16.35$, $(m-M)_V= 15.43$, and $(m-M)_I= 14.04$,
which broadly cross at $d=3.4$~kpc and $E(B-V)=0.90$, in Figure
\ref{distance_reddening_v459_vul_v5579_sgr_v2670_oph_qy_mus}(a).
Thus, we obtain $E(B-V)=0.90\pm0.05$ and $d=3.4\pm0.5$~kpc.


\begin{figure}
\epsscale{0.75}
\plotone{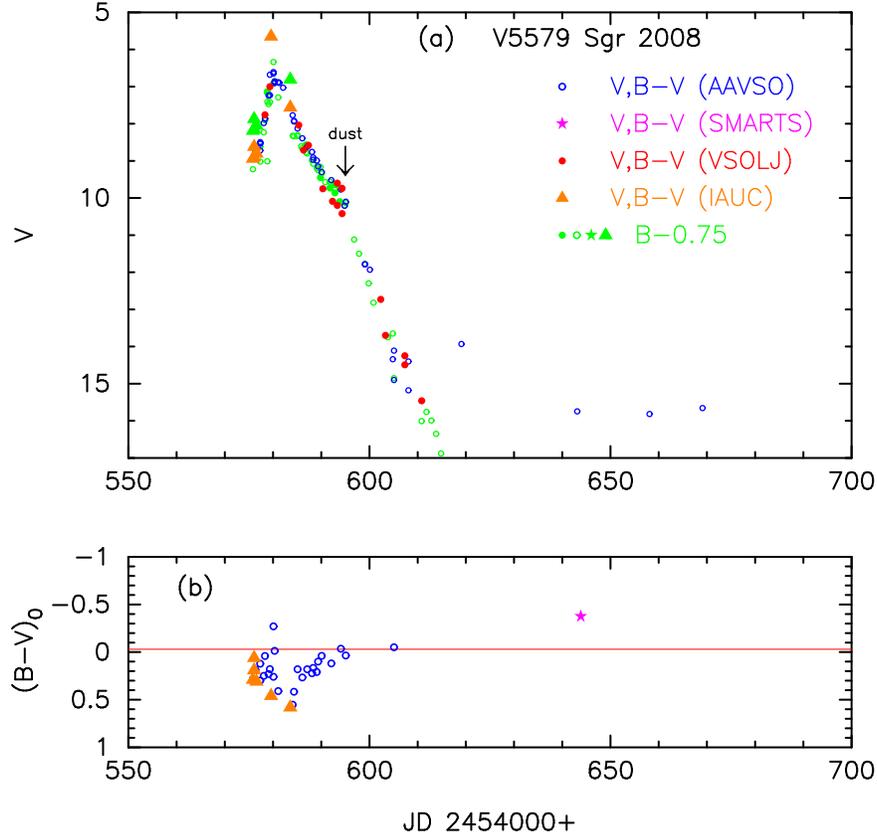}
\caption{
Same as Figure \ref{v1663_aql_v_bv_ub_color_curve}, but for V5579~Sgr.
(a) The $V$ (unfilled blue circles) and $B$ (unfilled green circles) data
are taken from AAVSO.  The $V$ (filled red circles) and
$B$ (filled green circles) data are from VSOLJ.  The $V$ 
(filled magenta stars) and $B$ (filled green stars) from SMARTS 
are also plotted.  (b) The $(B-V)_0$ are dereddened with $E(B-V)=0.82$.
\label{v5579_sgr_v_bv_ub_color_curve}}
\end{figure}


\begin{figure}
\epsscale{0.75}
\plotone{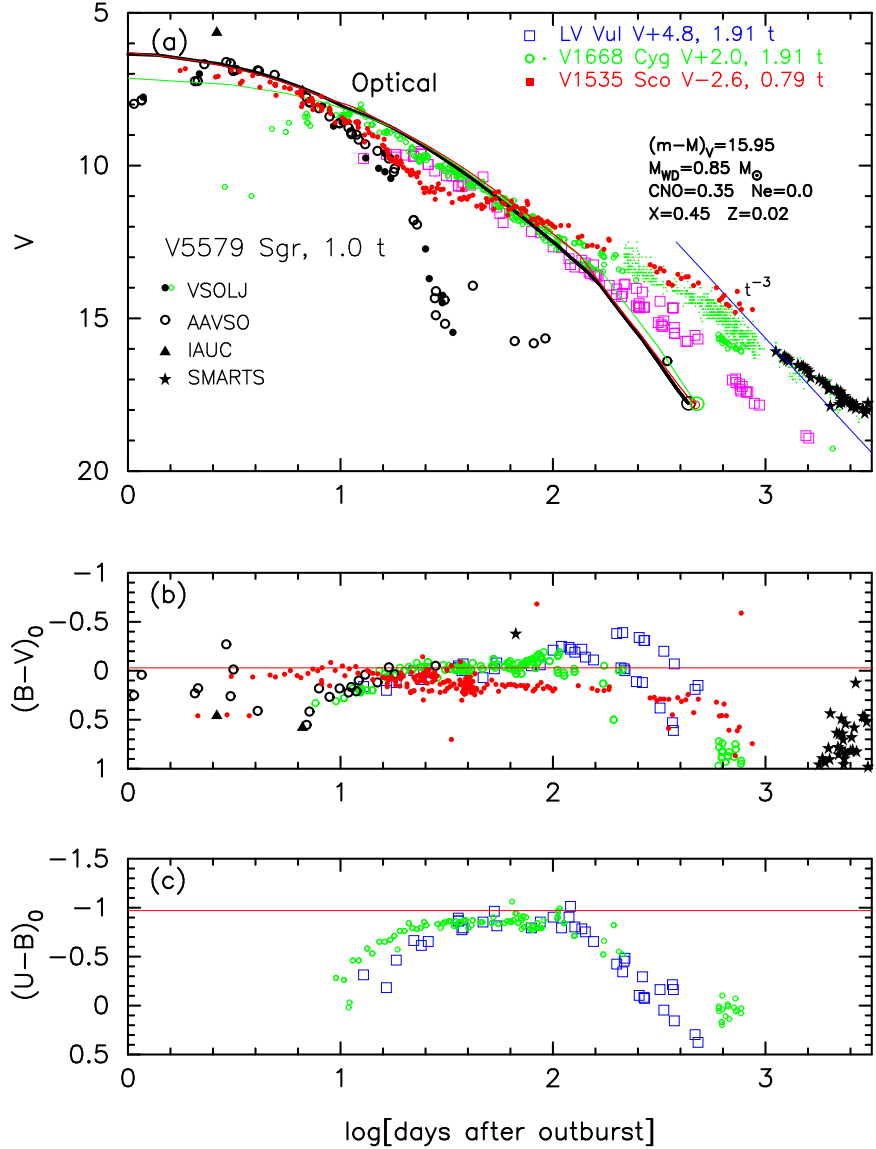}
\caption{
Same as Figure \ref{v2575_oph_v1668_cyg_lv_vul_v_bv_ub_logscale},
but for V5579~Sgr (black symbols).  We add the data of LV~Vul, V1668~Cyg,
and V1535~Sco.  The data of V5579~Sgr are the same as those in Figure
\ref{v5579_sgr_v_bv_ub_color_curve}.  The data of V1535~Sco are the
same as those in Figure \ref{v1535_sco_v_bv_ub_color_curve}.  In panel (a), 
we added a model absolute $V$ light curve (solid black line) of 
a $0.85~M_\sun$ WD \citep[CO3;][]{hac16k}, 
adopting $(m-M)_V=15.95$ for V5579~Sgr.
The solid green line denotes the $V$ light curve
of a $0.98~M_\sun$ WD (CO3) for V1668~Cyg, while the solid red line
represents the $V$ light curve of a $0.85~M_\sun$ WD
(CO4) for V1535~Sco.
Here, we assume $(m-M)_V=14.6$ for V1668~Cyg and $(m-M)_V=18.3$
for V1535~Sco.
\label{v5579_sgr_lv_vul_v1668_cyg_v1535_sco_v_bv_ub_color_logscale}}
\end{figure}


\begin{figure}
\epsscale{0.55}
\plotone{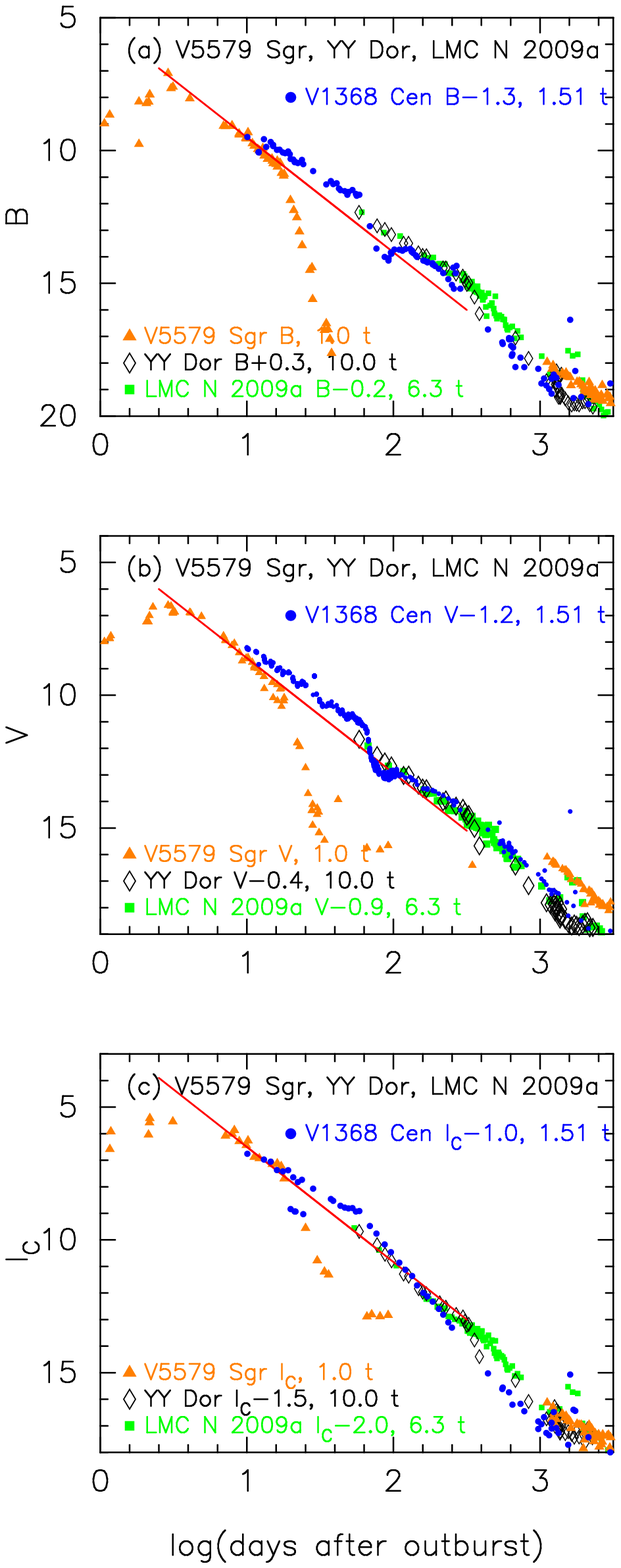}
\caption{
Same as Figure \ref{v1663_aql_yy_dor_lmcn_2009a_b_v_i_logscale_3fig},
but for V5579~Sgr.
We added the data of V1368~Cen, YY~Dor, and LMC~N~2009a.
The $BV$ data of V5579~Sgr are the same as those in Figure
\ref{v5579_sgr_v_bv_ub_color_curve}.  The $I_{\rm C}$ data are taken
from AAVSO, VSOLJ, and SMARTS.  The data of V1368~Cen are the same as
those in Figures \ref{v1368_cen_v_bv_ub_color_curve},
\ref{v1368_cen_lv_vul_v1668_cyg_os_and_v_bv_ub_logscale}, and
\ref{v1368_cen_v834_car_yy_dor_lmcn_2009a_b_v_i_logscale_3fig}.
\label{v5579_sgr_yy_dor_lmcn_2009a_b_v_i_logscale_3fig}}
\end{figure}

\subsection{V5579~Sgr 2008}
\label{v5579_sgr}
Figure \ref{v5579_sgr_v_bv_ub_color_curve} shows (a) the $V$ and $B$, and
(b) $(B-V)_0$ evolutions of V5579~Sgr.  Here, $(B-V)_0$ are dereddened
with $E(B-V)=0.82$ as obtained in Section \ref{v5579_sgr_cmd}.
Figure \ref{v5579_sgr_lv_vul_v1668_cyg_v1535_sco_v_bv_ub_color_logscale}
shows the light/color curves of V5579~Sgr
as well as those of LV~Vul, V1668~Cyg, and V1535~Sco.  
The data of V5579~Sgr are the same as those in Figure
\ref{v5579_sgr_v_bv_ub_color_curve}.
The data of V1535~Sco are the same as those in Figures 
\ref{v1535_sco_v_bv_ub_color_curve} and 
\ref{v1535_sco_lv_vul_v1668_cyg_v2468_cyg_v_bv_logscale}.
Applying Equation (\ref{distance_modulus_general_temp}) to them,
we have the relation 
\begin{eqnarray}
(m&-&M)_{V, \rm V5579~Sgr} \cr 
&=& (m - M + \Delta V)_{V, \rm LV~Vul} - 2.5 \log 1.91 \cr
&=& 11.85 + 4.8\pm0.2 - 0.7 = 15.95\pm0.2 \cr
&=& (m - M + \Delta V)_{V, \rm V1668~Cyg} - 2.5 \log 1.91 \cr
&=& 14.6 + 2.0\pm0.2 - 0.7 = 15.9\pm0.2 \cr
&=& (m - M + \Delta V)_{V, \rm V1535~Sco} - 2.5 \log 0.79 \cr
&=& 18.3 - 2.6\pm0.2 + 0.25 = 15.95\pm0.2,
\label{distance_modulus_v5579_sgr}
\end{eqnarray}
where we adopt 
$(m-M)_{V, \rm V1535~Sco}=18.3$ in Appendix \ref{v1535_sco}.
Thus, we obtain $(m-M)_V=15.95\pm0.1$ and $f_{\rm s}=1.91$ against LV~Vul.
From Equations (\ref{time-stretching_general}),
(\ref{distance_modulus_general_temp}), and
(\ref{distance_modulus_v5579_sgr}),
we have the relation
\begin{eqnarray}
(m- M')_{V, \rm V5579~Sgr} 
&\equiv & (m_V - (M_V - 2.5\log f_{\rm s}))_{\rm V5579~Sgr} \cr
&=& \left( (m-M)_V + \Delta V \right)_{\rm LV~Vul} \cr
&=& 11.85 + 4.8\pm0.2 = 16.65\pm0.2.
\label{absolute_mag_v5579_sgr}
\end{eqnarray}

Figure \ref{v5579_sgr_yy_dor_lmcn_2009a_b_v_i_logscale_3fig} shows
the $B$, $V$, and $I_{\rm C}$ light curves of V5579~Sgr
together with those of V1368~Cen, YY~Dor, and LMC~N~2009a.
We use the light curve of V1368~Cen in order to fill the gap between
the light curves of YY~Dor and LMC~N~2009a and that of V5579~Sgr.
We have already fixed the timescaling factor of V5579~Sgr, 
$f_{\rm s}= 1.91$, against LV~Vul, so we only shift the $BVI_{\rm C}$
light curves up and down, and overlap them with the other novae. 
Due to dust-shell formation, we only use the early-phase light-curve data.
In Figure \ref{v5579_sgr_yy_dor_lmcn_2009a_b_v_i_logscale_3fig},
we add straight solid red lines that represent the trend of the
universal decline law, that is, the free-free emission model 
light curve decays as $F_\nu \propto t^{-1.75}$ \citep{hac06kb}.
We do overlap the light curves on the red lines as long as
possible until the dust blackout starts.
We apply Equation (\ref{distance_modulus_general_temp_b})
for the $B$ band to Figure
\ref{v5579_sgr_yy_dor_lmcn_2009a_b_v_i_logscale_3fig}(a)
and obtain
\begin{eqnarray}
(m&-&M)_{B, \rm V5579~Sgr} \cr
&=& ((m - M)_B + \Delta B)_{\rm V1368~Cen} - 2.5 \log 1.51 \cr
&=& 18.53 - 1.3\pm0.2 - 0.45 = 16.78\pm0.2 \cr
&=& ((m - M)_B + \Delta B)_{\rm YY~Dor} - 2.5 \log 10.0 \cr
&=& 18.98 + 0.3\pm0.2 - 2.5 = 16.78\pm0.2 \cr
&=& ((m - M)_B + \Delta B)_{\rm LMC~N~2009a} - 2.5 \log 6.3 \cr
&=& 18.98 - 0.2\pm0.2 - 2.0 = 16.78\pm0.2.
\label{distance_modulus_b_v5579_sgr_yy_dor_lmcn2009a}
\end{eqnarray}
Thus, we have $(m-M)_{B, \rm V5579~Sgr}= 16.78\pm0.1$.

For the $V$ light curves in Figure
\ref{v5579_sgr_yy_dor_lmcn_2009a_b_v_i_logscale_3fig}(b),
we similarly obtain
\begin{eqnarray}
(m&-&M)_{V, \rm V5579~Sgr} \cr
&=& ((m - M)_V + \Delta V)_{\rm V1368~Cen} - 2.5 \log 1.51 \cr
&=& 17.6 - 1.2\pm0.2 - 0.45 = 15.95\pm0.2 \cr
&=& ((m - M)_V + \Delta V)_{\rm YY~Dor} - 2.5 \log 10.0 \cr
&=& 18.86 - 0.4\pm0.2 - 2.5 = 15.96\pm0.2 \cr
&=& ((m - M)_V + \Delta V)_{\rm LMC~N~2009a} - 2.5 \log 6.3 \cr
&=& 18.86 - 0.9\pm0.2 - 2.0 = 15.96\pm0.2.
\label{distance_modulus_v_v5579_sgr_yy_dor_lmcn2009a}
\end{eqnarray}
We have $(m-M)_{V, \rm V5579~Sgr}= 15.96\pm0.1$, which is
consistent with Equation (\ref{distance_modulus_v5579_sgr}).

We apply Equation (\ref{distance_modulus_general_temp_i}) for
the $I_{\rm C}$ band to Figure
\ref{v5579_sgr_yy_dor_lmcn_2009a_b_v_i_logscale_3fig}(c) and obtain
\begin{eqnarray}
(m&-&M)_{I, \rm V5579~Sgr} \cr
&=& ((m - M)_I + \Delta I_C)_{\rm V1368~Cen} - 2.5 \log 1.51 \cr
&=& 16.1 - 1.0\pm0.2 - 0.45 = 14.65\pm 0.2 \cr
&=& ((m - M)_I + \Delta I_C)_{\rm YY~Dor} - 2.5 \log 10.0 \cr
&=& 18.67 - 1.5\pm0.2 - 2.5 = 14.67\pm 0.2 \cr
&=& ((m - M)_I + \Delta I_C)_{\rm LMC~N~2009a} - 2.5 \log 6.3 \cr
&=& 18.67 - 2.0\pm0.2 - 2.0 = 14.67\pm 0.2.
\label{distance_modulus_i_v5579_sgr_yy_dor_lmcn2009a}
\end{eqnarray}
Thus, we have $(m-M)_{I, \rm V5579~Sgr}= 14.67\pm0.1$.

We plot $(m-M)_B= 16.78$, $(m-M)_V= 15.96$, and $(m-M)_I= 14.67$,
which cross at $d=4.8$~kpc and $E(B-V)=0.82$, in Figure
\ref{distance_reddening_v459_vul_v5579_sgr_v2670_oph_qy_mus}(b).
Thus, we obtain $E(B-V)=0.82\pm0.05$ and $d=4.8\pm0.5$~kpc.


\begin{figure}
\epsscale{0.75}
\plotone{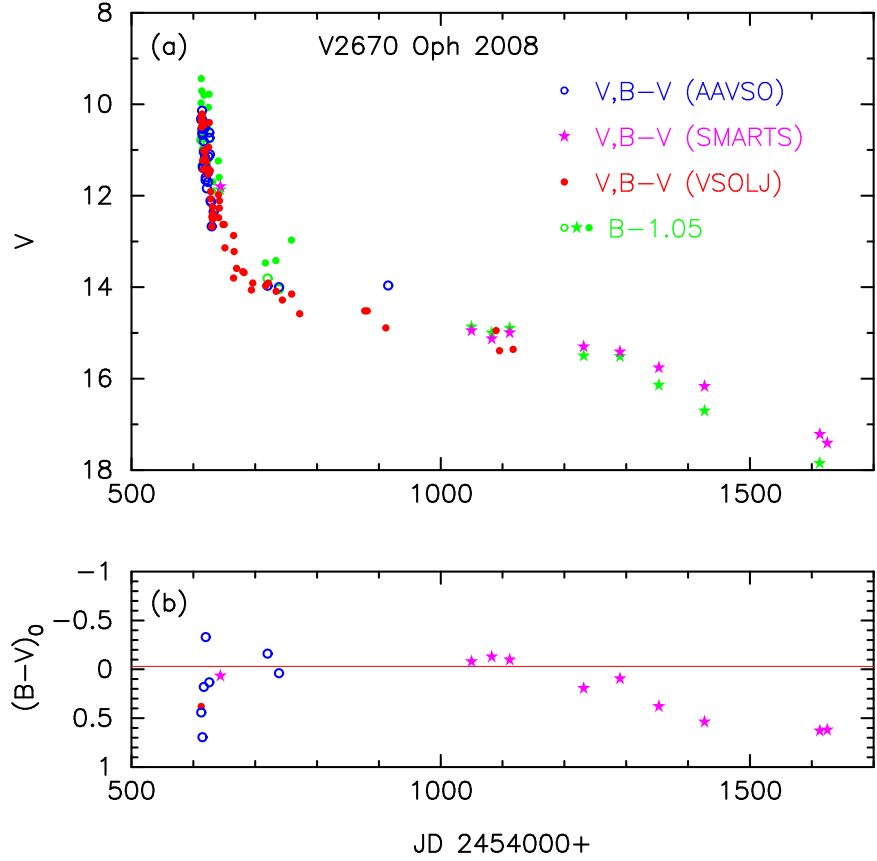}
\caption{
Same as Figure \ref{v1663_aql_v_bv_ub_color_curve}, but for V2670~Oph.
(a) The $V$ (unfilled blue circles) and $B$ (unfilled green circles) data
are taken from AAVSO.  The $V$ (filled red circles) and
$B$ (filled green circles) data are from VSOLJ.
The SMARTS data of $V$ (filled magenta stars) and $B$ (filled green stars)
are also plotted.  (b) The $(B-V)_0$ are dereddened with $E(B-V)=1.05$.
\label{v2670_oph_v_bv_ub_color_curve}}
\end{figure}


\begin{figure}
\epsscale{0.75}
\plotone{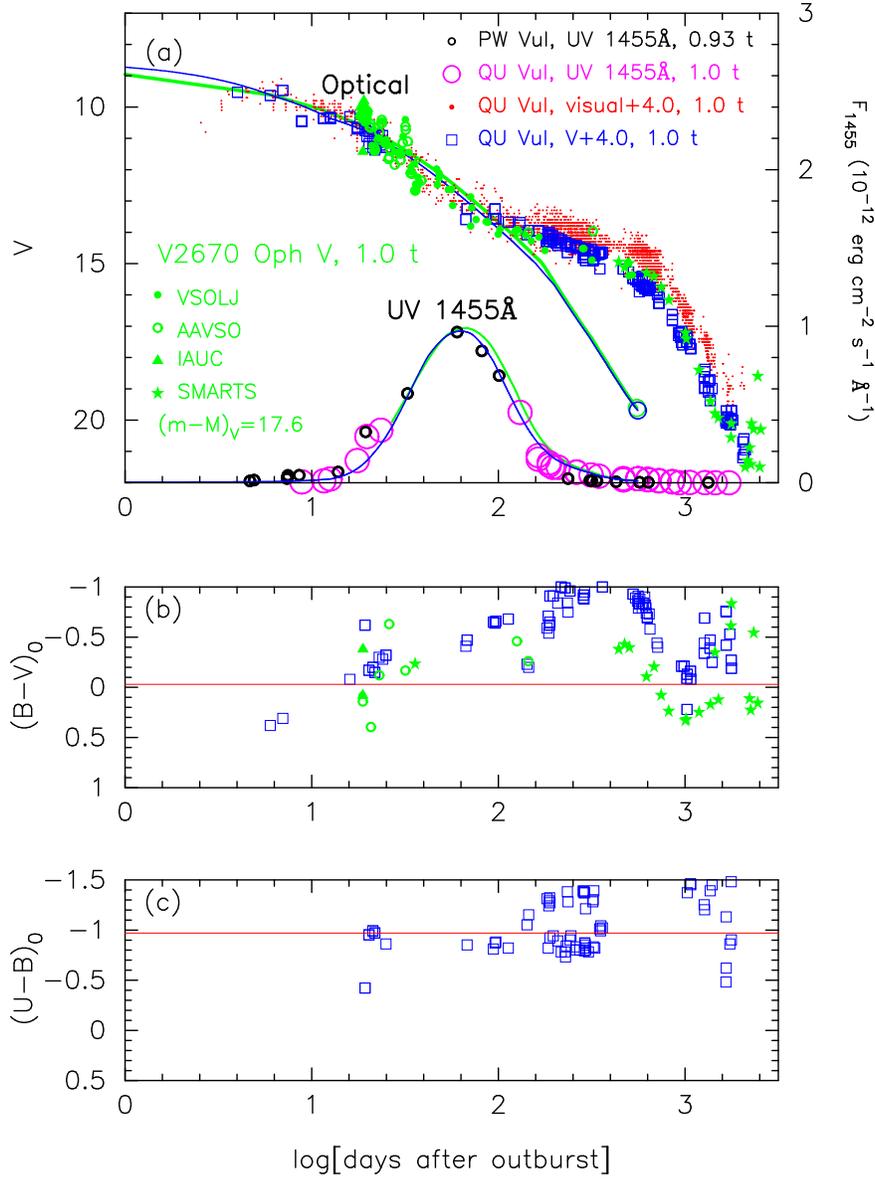}
\caption{
The light/color curves of V2670~Oph as well as those of QU~Vul.
The data of V2670~Oph are the same as those in Figure
\ref{v2670_oph_v_bv_ub_color_curve}.  The data of QU~Vul are the same as
those in Figure 10 of \citet{hac16k}.   In panel (a), we added model 
light curves (solid green lines) of a $0.80~M_\sun$ WD 
\citep[CO3;][]{hac16k},
assuming that $(m-M)_V=17.6$ for V2670~Oph.  The solid
blue lines denote the $V$ and UV~1455\AA\  light curve of
a $0.86~M_\sun$ WD \citep[Ne3;][]{hac15k},
assuming that $(m-M)_V=13.6$ for QU~Vul \citep{hac16k}.
\label{v2670_oph_qu_vul_x55z02c10o10_logscale}}
\end{figure}


\begin{figure}
\epsscale{0.75}
\plotone{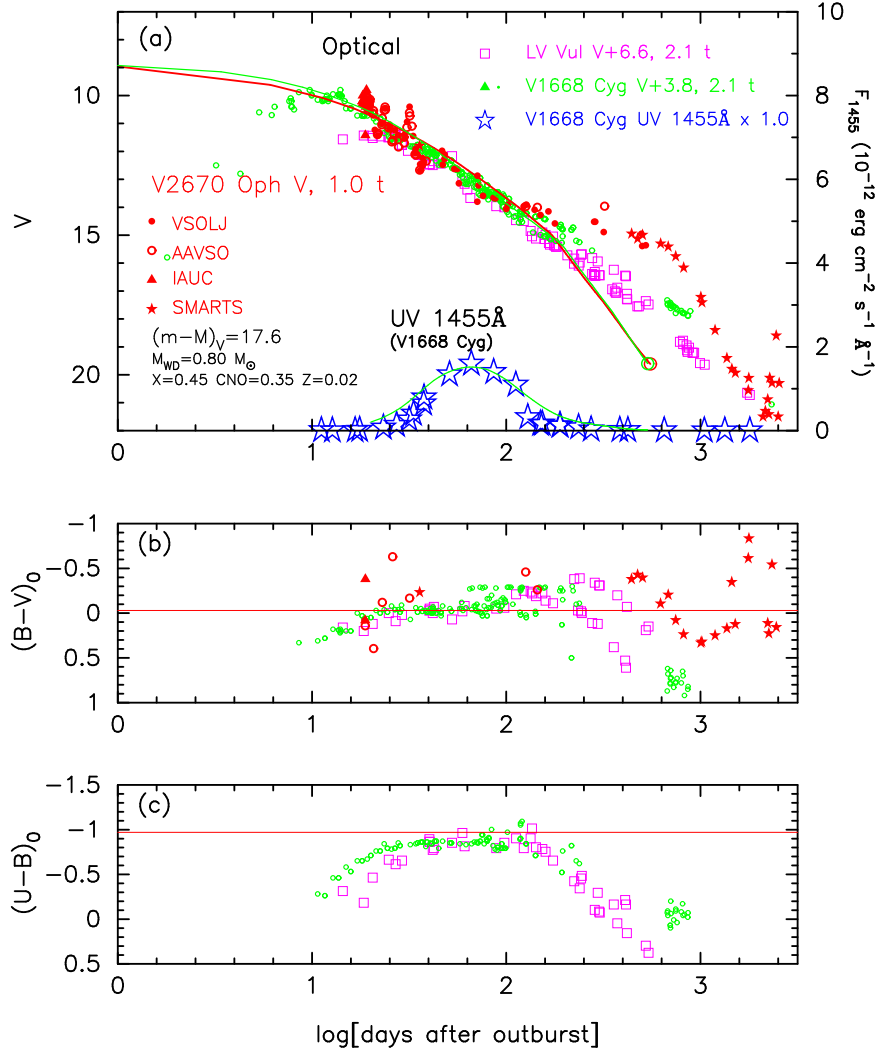}
\caption{
Same as Figure
\ref{v2575_oph_v1668_cyg_lv_vul_v_bv_ub_logscale},
but for V2670~Oph (red symbols).
The data of V2670~Oph are the same as those in Figure
\ref{v2670_oph_v_bv_ub_color_curve}.
In panel (a), we added model light curves (solid red lines) of 
a $0.80~M_\sun$ WD \citep[CO3,][]{hac16k}, 
assuming that $(m-M)_V=17.6$ for V2670~Oph.
The solid green lines denote the $V$ and UV~1455\AA\  light curve
of a $0.98~M_\sun$ WD (CO3), assuming that $(m-M)_V=14.6$ for V1668~Cyg.
\label{v2670_oph_v1668_cyg_lv_vul_v_bv_ub_color_logscale}}
\end{figure}


\begin{figure}
\epsscale{0.55}
\plotone{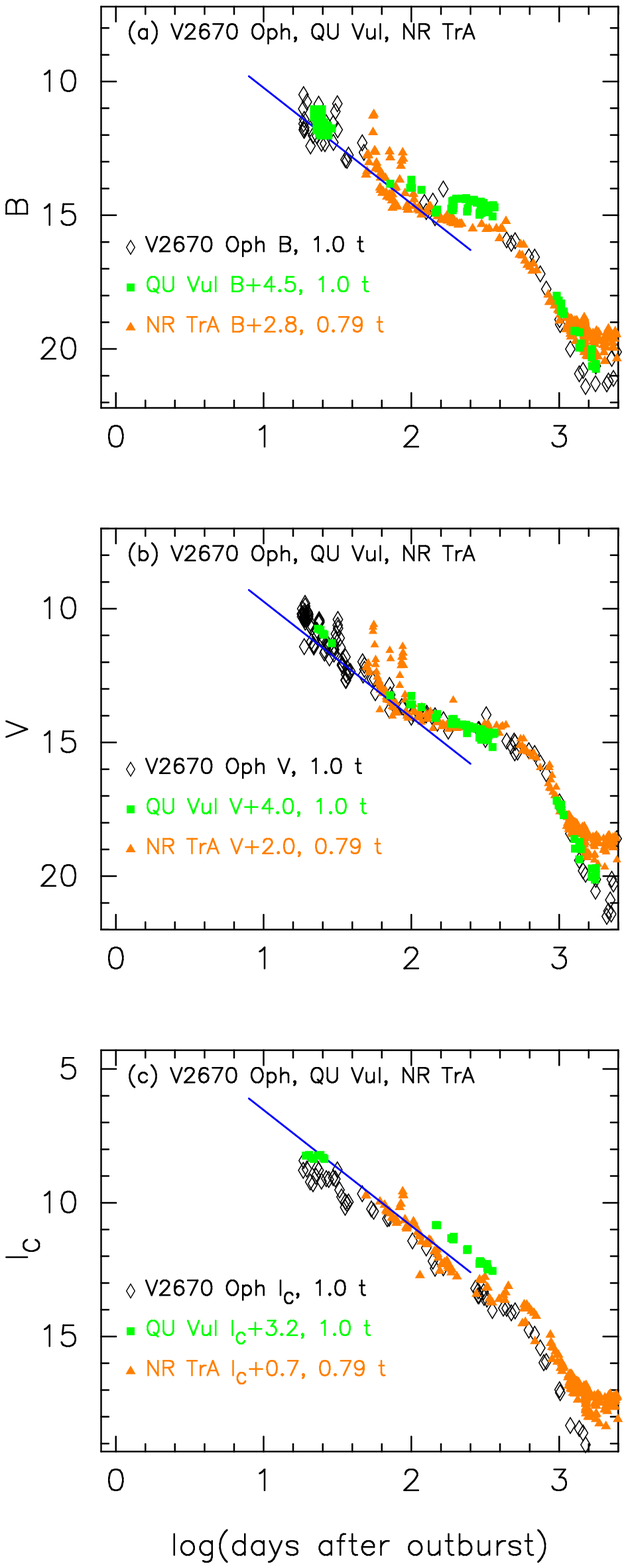}
\caption{
Same as Figure \ref{v1663_aql_yy_dor_lmcn_2009a_b_v_i_logscale_3fig},
but for V2670~Oph.
The $BV$ data of V2670~Oph are the same as those in Figure
\ref{v2670_oph_v_bv_ub_color_curve}.
The $BV$ data of QU~Vul are the same as those in Figure 10 of
\citet{hac16k}.  The $BV$ data of NR~TrA are the same as those in
Figure \ref{nr_tra_v_bv_ub_color_curve}.  The $I_{\rm C}$ data are taken
from AAVSO, VSOLJ, and SMARTS.
\label{v2670_ooph_qu_vul_nr_tra_b_v_i_logscale_3fig}}
\end{figure}

\subsection{V2670~Oph 2008\#1}
\label{v2670_oph}
Figure \ref{v2670_oph_v_bv_ub_color_curve} shows (a) the $V$ and $B$, and
(b) $(B-V)_0$ evolutions of V2670~Oph.  Here, $(B-V)_0$ are dereddened
with $E(B-V)=1.05$ as obtained in Section \ref{v2670_oph_cmd}.
Figure \ref{v2670_oph_qu_vul_x55z02c10o10_logscale} shows
the light/color curves of V2670~Oph and QU~Vul.
The timescales and $V$ light curves of these two novae are very similar.
The data of V2670~Oph are the same as those
in Figure \ref{v2670_oph_v_bv_ub_color_curve}.
The data of QU~Vul are the same as those in Figure 10 of \citet{hac16k}.
We added the UV~1455\AA\  light curves of QU~Vul and PW~Vul for comparison.
The $V$ light curves of V2670~Oph and QU~Vul overlap each other.
Applying Equation (\ref{distance_modulus_general_temp}) to them,
we have the relation 
\begin{eqnarray}
(m&-&M)_{V, \rm V2670~Oph} \cr 
&=& (m - M + \Delta V)_{V, \rm QU~Vul} - 2.5 \log 1.0 \cr
&=& 13.6 + 4.0\pm0.3 - 0.0 = 17.6\pm0.3,
\label{distance_modulus_v2670_oph_qu_vul}
\end{eqnarray}
where we adopt $(m-M)_{V, \rm QU~Vul}=13.6$ from \citet{hac16k}.
Thus, we obtain $(m-M)_V=17.6\pm0.3$ for V2670~Oph.
The timescaling factor of QU~Vul is $\log f_{\rm s}= 0.33$ against
LV~Vul, because the timescaling factor of PW~Vul is
$\log f_{\rm s}= 0.35$ against LV~Vul and the timescaling factor of
QU~Vul is smaller by $\Delta \log f_{\rm s}= -0.02$ than that of PW~Vul.

Using $(m-M)_V=17.6$ and $\log f_{\rm s}= 0.33$ for V2670~Oph,
we plot Figure \ref{v2670_oph_v1668_cyg_lv_vul_v_bv_ub_color_logscale}.
Here, we show three nova light/color curves,  V2670~Oph, LV~Vul, and 
V1668~Cyg.  
Applying Equation (\ref{distance_modulus_general_temp}) to them,
we have the relation 
\begin{eqnarray}
(m-M)_{V, \rm V2670~Oph} 
&=& (m - M + \Delta V)_{V, \rm LV~Vul} - 2.5 \log 2.1 \cr
&=& 11.85 + 6.6\pm0.3 - 0.83 = 17.62\pm0.3 \cr
&=& (m - M + \Delta V)_{V, \rm V1668~Cyg} - 2.5 \log 2.1 \cr
&=& 14.6 + 3.8\pm0.3 - 0.83 = 17.57\pm0.3,
\label{distance_modulus_v2670_oph_lv_vul}
\end{eqnarray}
where we adopt $(m-M)_{V, \rm LV~Vul}=11.85$ and
$(m-M)_{V, \rm V1668~Cyg}=14.6$, both from \citet{hac19k}.
From Equations (\ref{time-stretching_general}),
(\ref{distance_modulus_general_temp}), and
(\ref{distance_modulus_v2670_oph_lv_vul}),
we have the relation
\begin{eqnarray}
(m- M')_{V, \rm V2670~Oph} 
&\equiv & (m_V - (M_V - 2.5\log f_{\rm s}))_{\rm V2670~Oph} \cr
&=& \left( (m-M)_V + \Delta V \right)_{\rm LV~Vul} \cr 
&=& 11.85 + 6.6\pm0.3 = 18.45\pm0.3.
\label{absolute_mag_v2670_oph}
\end{eqnarray}

Figure \ref{v2670_ooph_qu_vul_nr_tra_b_v_i_logscale_3fig} shows the $B$, 
$V$, and $I_{\rm C}$ light curves of V2670~Oph together with 
those of QU~Vul and NR~TrA.  
We apply Equation (\ref{distance_modulus_general_temp_b}) for the 
$B$ band to Figure \ref{v2670_ooph_qu_vul_nr_tra_b_v_i_logscale_3fig}(a)
and obtain
\begin{eqnarray}
(m&-&M)_{B, \rm V2670~Oph} \cr
&=& ((m - M)_B + \Delta B)_{\rm QU~Vul} - 2.5 \log 1.0 \cr
&=& 14.15 + 4.5\pm0.3 - 0.0 = 18.65\pm0.3 \cr
&=& ((m - M)_B + \Delta B)_{\rm NR~TrA} - 2.5 \log 0.79 \cr
&=& 15.59 + 2.8\pm0.3 + 0.25 = 18.64\pm0.3,
\label{distance_modulus_b_v2670_oph_qu_vul_nr_tra}
\end{eqnarray}
where we adopt $(m - M)_{B, \rm QU~Vul}= 13.6 + 0.55= 14.15$
from \citet{hac16k} and $(m - M)_{B, \rm NR~TrA}= 15.35 + 0.24= 15.59$
in Appendix \ref{nr_tra}.  We obtain $(m-M)_{B, \rm V2670~Oph}= 18.65\pm0.2$.

For the $V$ light curves in Figure
\ref{nr_tra_v2670_ooph_qu_vul_b_v_i_logscale_3fig}(b),
we obtain
\begin{eqnarray}
(m&-&M)_{V, \rm V2670~Oph} \cr
&=& ((m - M)_V + \Delta V)_{\rm QU~Vul} - 2.5 \log 1.0 \cr
&=& 13.6 + 4.0\pm0.3 - 0.0 = 17.6\pm0.3 \cr
&=& ((m - M)_V + \Delta V)_{\rm NR~TrA} - 2.5 \log 0.79 \cr
&=& 15.35 + 2.0\pm0.3 + 0.25 = 17.6\pm0.3,
\label{distance_modulus_v_v2670_oph_qu_vul_nr_tra}
\end{eqnarray}
where we adopt $(m-M)_{V, \rm QU~Vul}= 13.6$ in \citet{hac16k} and
$(m-M)_{I, \rm NR~TrA}=15.35$ in Appendix \ref{nr_tra}.
We obtain $(m-M)_{V, \rm V2670~Oph}= 17.6\pm0.2$, which is
consistent with Equations (\ref{distance_modulus_v2670_oph_qu_vul})
and (\ref{distance_modulus_v2670_oph_lv_vul}).

From the $I_{\rm C}$-band data in Figure
\ref{v2670_ooph_qu_vul_nr_tra_b_v_i_logscale_3fig}(c),  we obtain
\begin{eqnarray}
(m&-&M)_{I, \rm V2670~Oph} \cr
&=& ((m - M)_I + \Delta I_C)_{\rm QU~Vul} - 2.5 \log 1.0 \cr
&=& 12.72 + 3.2\pm0.3 - 0.0 = 15.92\pm0.3 \cr
&=& ((m - M)_I + \Delta I_C)_{\rm NR~TrA} - 2.5 \log 0.79 \cr
&=& 14.97 + 0.7\pm0.3 + 0.25 = 15.92\pm0.3,
\label{distance_modulus_i_v2670_oph_qu_vul_nr_tra}
\end{eqnarray}
where we adopt $(m-M)_{I, \rm QU~Vul}=13.6 - 1.6\times 0.55= 12.72$ and
$(m-M)_{I, \rm NR~TrA}=15.35 - 1.6\times 0.24= 14.97$.
We obtain $(m-M)_{I, \rm V2670~Oph}= 15.92\pm0.2$.

We plot $(m-M)_B=18.65$, $(m-M)_V=17.6$, and $(m-M)_I=15.92$,
which broadly cross at $d=7.4$~kpc and $E(B-V)=1.05$,
in Figure \ref{distance_reddening_v459_vul_v5579_sgr_v2670_oph_qy_mus}(c).
Thus, we have $d=7.4\pm0.8$~kpc and $E(B-V)=1.05\pm0.1$.


\begin{figure}
\epsscale{0.75}
\plotone{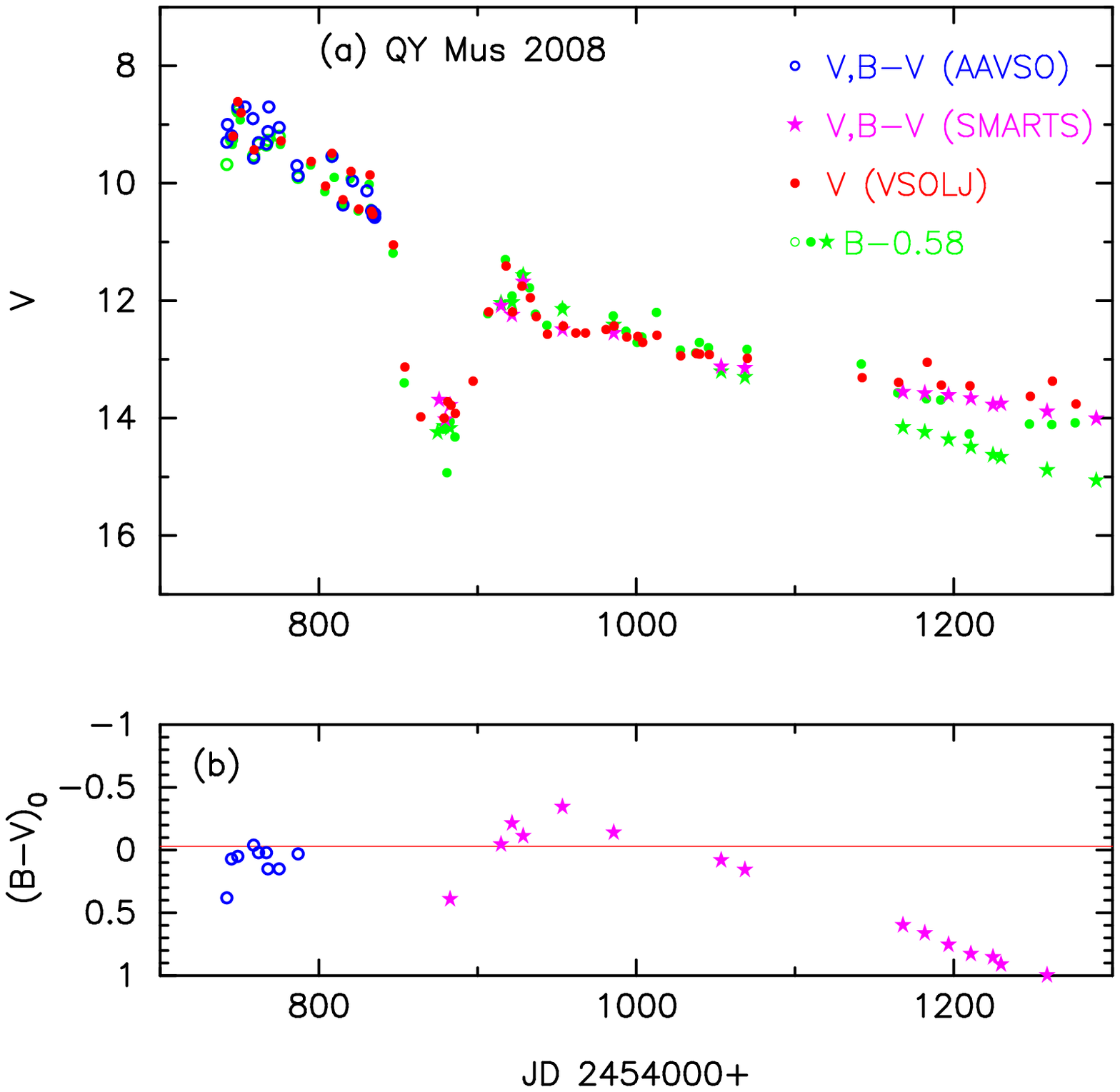}
\caption{
Same as Figure \ref{v1663_aql_v_bv_ub_color_curve}, but for QY~Mus.
(a) The $V$ (unfilled blue circles) and $B$ (unfilled green circles) data
are taken from AAVSO.  The $V$ (filled red circles) and
$B$ (filled green circles) data are from VSOLJ.
The SMARTS data of $V$ (filled magenta stars) and $B$ (filled green stars)
are also plotted.  (b) The $(B-V)_0$ are dereddened with $E(B-V)=0.58$.
\label{qy_mus_v_bv_ub_color_curve}}
\end{figure}


\begin{figure}
\epsscale{0.75}
\plotone{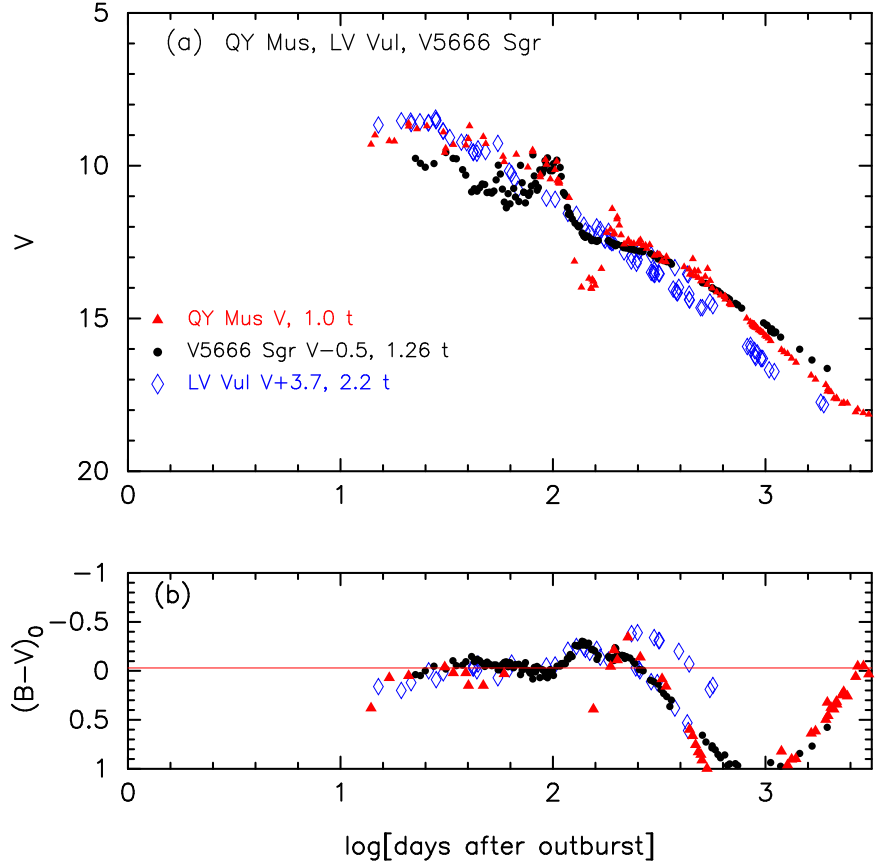}
\caption{
Same as Figure \ref{v2575_oph_v1668_cyg_lv_vul_v_bv_ub_logscale},
but for QY~Mus (filled red triangles).  We added the V5666~Sgr 
(filled black circles) and LV~Vul (unfilled blue diamonds) light/color curves.
The data of QY~Mus are the same as those in Figure
\ref{qy_mus_v_bv_ub_color_curve}.  
\label{qy_mus_v5666_sgr_lv_vul_v_bv_ub_color_logscale}}
\end{figure}


\begin{figure}
\epsscale{0.65}
\plotone{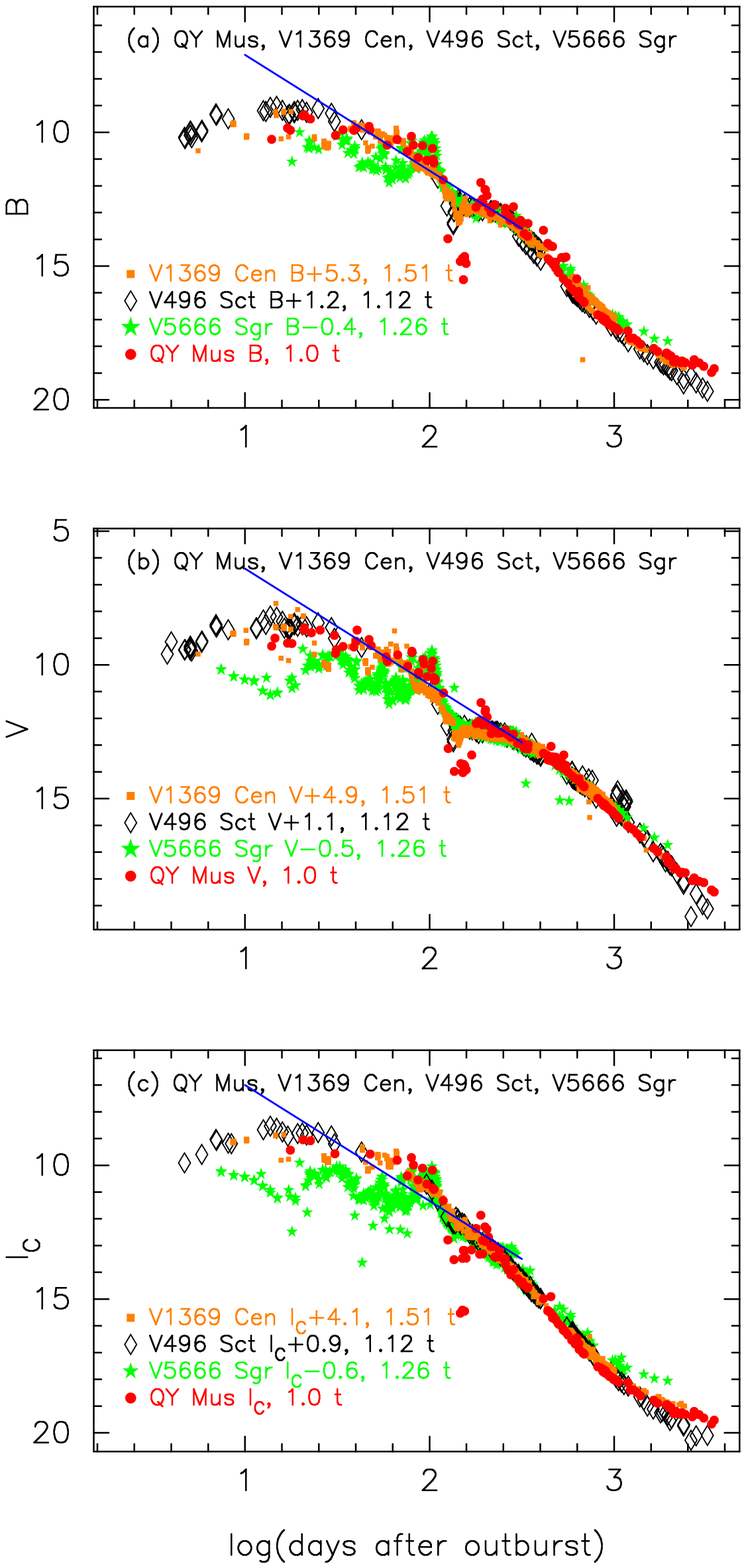}
\caption{
Same as Figure \ref{v1663_aql_yy_dor_lmcn_2009a_b_v_i_logscale_3fig},
but for QY~Mus.
The (a) $B$, (b) $V$, and (c) $I_{\rm C}$ light curves of QY~Mus
as well as those of V1369~Cen, V496~Sct, and V5666~Sgr.
The $BV$ data of QY~Mus are the same as those in Figure
\ref{qy_mus_v_bv_ub_color_curve}.  The $I_{\rm C}$ data of QY~Mus
are taken from AAVSO and SMARTS.
\label{qy_mus_v1369_cen_v496_sct_v5666_sgr_b_v_i_logscale_3fig}}
\end{figure}

\subsection{QY~Mus 2008}
\label{qy_mus}
Figure \ref{qy_mus_v_bv_ub_color_curve} shows (a) the $V$ and $B$, and
(b) $(B-V)_0$ evolutions of QY~Mus.  Here, $(B-V)_0$ are dereddened
with $E(B-V)=0.58$ as obtained in Section \ref{qy_mus_cmd}.
The $V$ light curve of QY~Mus is very similar to that of V5666~Sgr.
Therefore, we plot these two light/color curves in Figure
\ref{qy_mus_v5666_sgr_lv_vul_v_bv_ub_color_logscale}
together with LV~Vul.  We overlap these three nova light/color curves.
Applying Equation (\ref{distance_modulus_general_temp}) to them,
we have the relation 
\begin{eqnarray}
(m&-&M)_{V, \rm QY~Mus} \cr
&=& (m - M + \Delta V)_{V, \rm LV~Vul} - 2.5 \log 2.2 \cr
&=& 11.85 + 3.7\pm0.2 - 0.88 = 14.67\pm0.2 \cr
&=& (m - M + \Delta V)_{V, \rm V5666~Sgr} - 2.5 \log 1.26 \cr
&=& 15.4 - 0.5\pm0.2 - 0.25 = 14.65\pm0.2,
\label{distance_modulus_qy_mus}
\end{eqnarray}
where we adopt $(m-M)_{V, \rm LV~Vul}=11.85$ and
$(m-M)_{V, \rm V5666~Sgr}=15.4$, both from \citet{hac19k}.
Thus, we adopt $(m-M)_V=14.65\pm0.1$ and $f_{\rm s}=2.2$ against LV~Vul.
From Equations (\ref{time-stretching_general}),
(\ref{distance_modulus_general_temp}), and
(\ref{distance_modulus_qy_mus}),
we have the relation
\begin{eqnarray}
(m- M')_{V, \rm QY~Mus} 
&\equiv & (m_V - (M_V - 2.5\log f_{\rm s}))_{\rm QY~Mus} \cr
&=& \left( (m-M)_V + \Delta V \right)_{\rm LV~Vul} \cr
&=& 11.85 + 3.7\pm0.2 = 15.55\pm0.2.
\label{absolute_mag_qy_mus}
\end{eqnarray}

Figure \ref{qy_mus_v1369_cen_v496_sct_v5666_sgr_b_v_i_logscale_3fig}
shows the $B$, $V$, and $I_{\rm C}$ light curves of QY~Mus
together with those of V1369~Cen, V496~Sct, and V5666~Sgr.
The light curves overlap each other well.
Applying Equation (\ref{distance_modulus_general_temp_b})
for the $B$ band to Figure
\ref{qy_mus_v1369_cen_v496_sct_v5666_sgr_b_v_i_logscale_3fig}(a),
we have the relation
\begin{eqnarray}
(m&-&M)_{B, \rm QY~Mus} \cr
&=& \left( (m-M)_B + \Delta B\right)_{\rm V1369~Cen} - 2.5 \log 1.51 \cr
&=& 10.36 + 5.3\pm0.2 - 0.45 = 15.21\pm0.2 \cr
&=& \left( (m-M)_B + \Delta B\right)_{\rm V496~Sct} - 2.5 \log 1.12 \cr
&=& 14.15 + 1.2\pm0.2 - 0.13 = 15.22\pm0.2 \cr
&=& \left( (m-M)_B + \Delta B\right)_{\rm V5666~Sgr} - 2.5 \log 1.26 \cr
&=& 15.9 - 0.4\pm0.2 - 0.25 = 15.25\pm0.2,
\label{distance_modulus_qy_mus_v1369_cen_v496_sct_v5666_sgr_b}
\end{eqnarray}
where we adopt $(m-M)_{B, \rm V1369~Cen}=10.25 + 1.0\times 0.11= 10.36$,
$(m-M)_{B, \rm V496~Sct}=13.7 + 1.0\times 0.45= 14.15$, and
$(m-M)_{B, \rm V5666~Sgr}=15.4 + 1.0\times 0.50=15.9$,
all from \citet{hac19k}.  We obtain $(m-M)_B=15.23\pm0.1$ for QY~Mus.

Applying Equation (\ref{distance_modulus_general_temp}) to
Figure \ref{qy_mus_v1369_cen_v496_sct_v5666_sgr_b_v_i_logscale_3fig}(b),
we have the relation
\begin{eqnarray}
(m&-&M)_{V, \rm QY~Mus} \cr
&=& \left( (m-M)_V + \Delta V\right)_{\rm V1369~Cen} - 2.5 \log 1.51 \cr
&=& 10.25 + 4.9\pm0.3 - 0.45 = 14.7\pm0.2 \cr
&=& \left( (m-M)_V + \Delta V\right)_{\rm V496~Sct} - 2.5 \log 1.12 \cr
&=& 13.7 + 1.1\pm0.3 - 0.13 = 14.67\pm0.2 \cr
&=& \left( (m-M)_V + \Delta V\right)_{\rm V5666~Sgr} - 2.5 \log 1.26 \cr
&=& 15.4 - 0.5\pm0.3 - 0.25 = 14.65\pm0.2,
\label{distance_modulus_qy_mus_v1369_cen_v496_sct_v5666_sgr_v}
\end{eqnarray}
where we adopt $(m-M)_{V, \rm V1369~Cen}=10.25$,
$(m-M)_{V, \rm V496~Sct}=13.7$, and $(m-M)_{V, \rm V5666~Sgr}=15.4$,
all from \citet{hac19k}.
We obtain $(m-M)_V=14.67\pm0.1$ for QY~Mus, which is
consistent with Equation (\ref{distance_modulus_qy_mus}).

From the $I_{\rm C}$-band data in Figure
\ref{qy_mus_v1369_cen_v496_sct_v5666_sgr_b_v_i_logscale_3fig}(c),
we obtain
\begin{eqnarray}
(m&-&M)_{I, \rm QY~Mus} \cr
&=& ((m - M)_I + \Delta I_C)_{\rm V1369~Cen} - 2.5 \log 1.51 \cr
&=& 10.07 + 4.1\pm0.2 - 0.45 = 13.72\pm0.2 \cr
&=& ((m - M)_I + \Delta I_C)_{\rm V496~Sct} - 2.5 \log 1.12 \cr
&=& 12.98 + 0.9\pm0.2 - 0.13 = 13.75\pm0.2 \cr
&=& ((m - M)_I + \Delta I_C)_{\rm V5666~Sgr} - 2.5 \log 1.26 \cr
&=& 14.6 - 0.6\pm0.2 - 0.25 = 13.75\pm0.2,
\label{distance_modulus_i_qy_mus_v1369_cen_v496_sct_v5666_sgr}
\end{eqnarray}
where we adopt $(m-M)_{I, \rm V1369~Cen}=10.25 - 1.6\times 0.11= 10.07$,
$(m-M)_{I, \rm V496~Sct}=13.7 - 1.6\times 0.45= 12.98$,
and $(m-M)_{I, \rm V5666~Sgr}=15.4 - 1.6\times 0.50= 14.6$.
We obtain $(m-M)_{I, \rm QY~Mus}= 13.74\pm0.1$.

We plot $(m-M)_B=15.23$, $(m-M)_V=14.67$, and $(m-M)_I=13.74$,
which cross at $d=3.7$~kpc and $E(B-V)=0.58$,
in Figure \ref{distance_reddening_v459_vul_v5579_sgr_v2670_oph_qy_mus}(d).
Thus, we obtain $d=3.7\pm0.4$~kpc and $E(B-V)=0.58\pm0.05$.

Our model light-curve fitting gives WD masses between
$M_{\rm WD}=0.75-0.85~M_\sun$ for the hydrogen content of
$X=0.35-0.55$ by weight as shown in Figure 
\ref{qy_mus_x35_x45_x55_v_logscale_3fig}
and as discussed in Section \ref{wd_mass_determination}.
It should also be noted that our model light
curves reasonably reproduce the $V$ light curve of QY~Mus
for $(m-M)_V=14.65\pm0.1$.


\begin{figure}
\epsscale{0.75}
\plotone{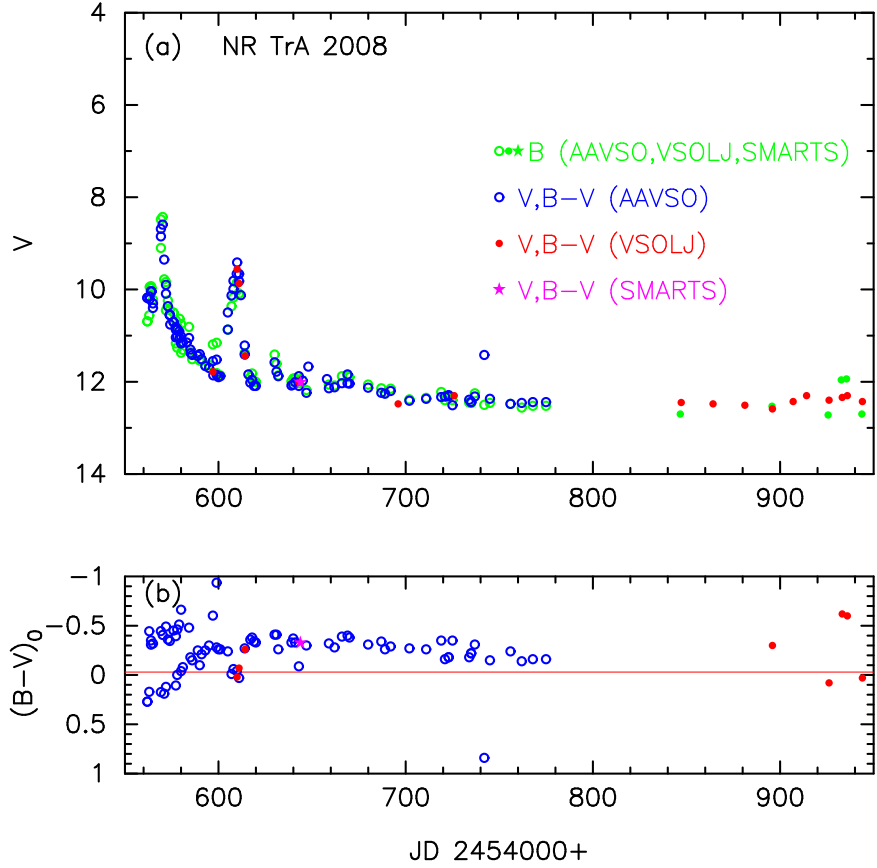}
\caption{
Same as Figure \ref{v1663_aql_v_bv_ub_color_curve}, but for NR~TrA.
(a) The $V$ (unfilled blue circles) and $B$ (unfilled green circles) data
are taken from AAVSO.  The $V$ (filled red circles) and
$B$ (filled green circles) data are from VSOLJ.
The $V$ (filled magenta stars) and
$B$ (filled green stars) data are from SMARTS.
(b) The $(B-V)_0$ are dereddened with $E(B-V)=0.24$.
\label{nr_tra_v_bv_ub_color_curve}}
\end{figure}


\begin{figure}
\epsscale{0.75}
\plotone{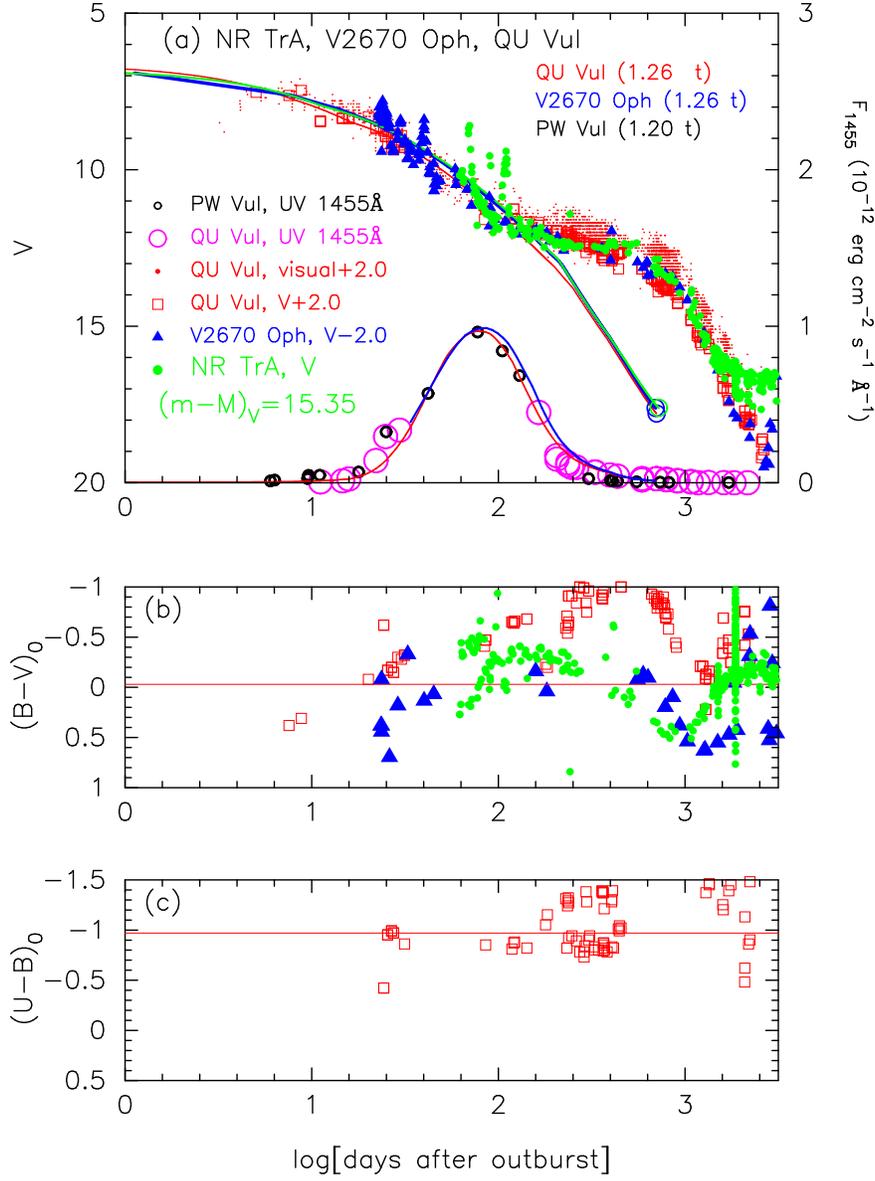}
\caption{
Same as Figure \ref{v2670_oph_qu_vul_x55z02c10o10_logscale},
but for NR~TrA (filled green circles).  
The data of NR~TrA are the same as those in Figure
\ref{nr_tra_v_bv_ub_color_curve}.  The timescales of QU~Vul
and V2670~Oph are stretched by 1.26.
The timescale of PW~Vul is also stretched by 1.20.
In panel (a), we added model light curves (solid green lines) of 
a $0.75~M_\sun$ WD \citep[CO3;][]{hac16k}, 
assuming that $(m-M)_V=15.35$ for NR~TrA.
The solid blue lines denote the $V$ and UV~1455\AA\  light curve
of a $0.80~M_\sun$ WD (CO3), assuming that $(m-M)_V=17.6$ for V2670~Oph.
The solid red lines denote the $V$ and UV~1455\AA\  light curve
of a $0.86~M_\sun$ WD (CO4), assuming that $(m-M)_V=13.6$ for QU~Vul.
\label{nr_tra_v2670_oph_qu_vul_x55z02c10o10_logscale}}
\end{figure}


\begin{figure}
\epsscale{0.75}
\plotone{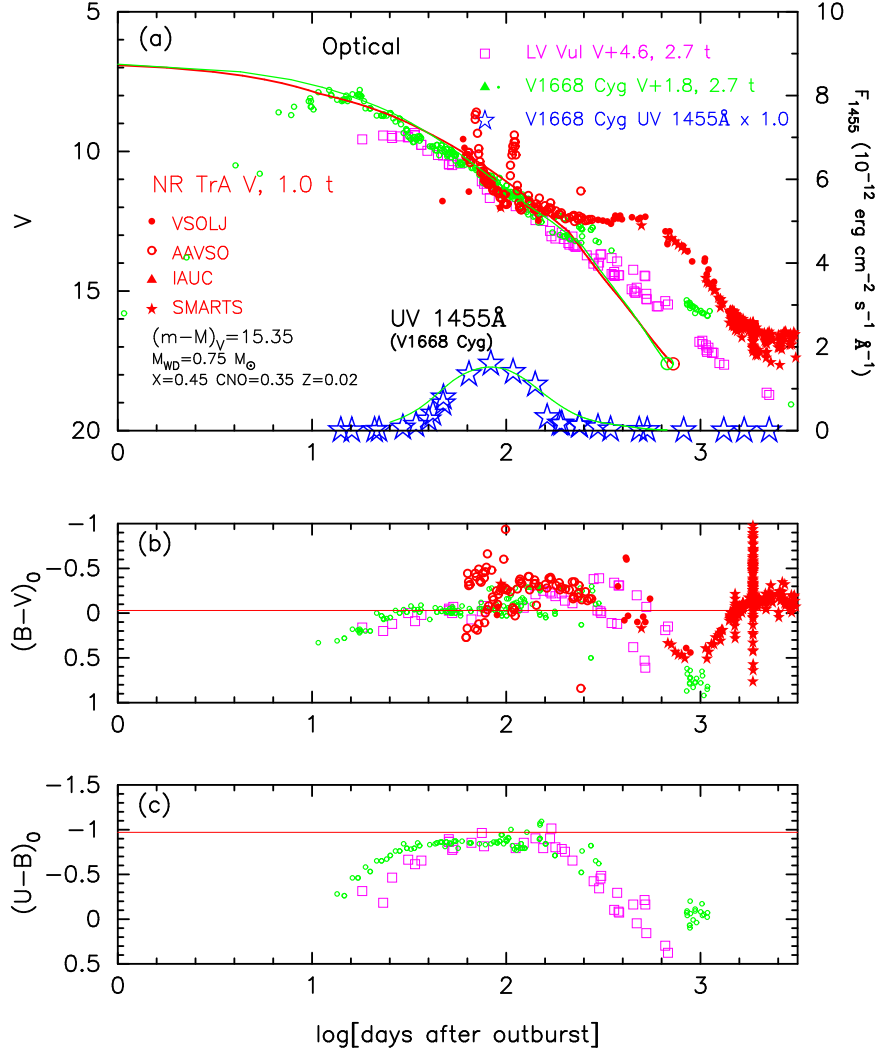}
\caption{
Same as Figure
\ref{v2670_oph_v1668_cyg_lv_vul_v_bv_ub_color_logscale},
but for NR~TrA (red symbols).
In panel (a), we added model light curves (solid red lines) of 
a $0.75~M_\sun$ WD \citep[CO3;][]{hac16k}, 
assuming that $(m-M)_V=15.35$ for NR~TrA.
The solid green lines denote the $V$ and UV~1455\AA\  light curve of 
a $0.98~M_\sun$ WD (CO3), assuming that $(m-M)_V=14.6$ for V1668~Cyg.
\label{nr_tra_v1668_cyg_lv_vul_v_bv_ub_color_logscale}}
\end{figure}


\begin{figure}
\epsscale{0.55}
\plotone{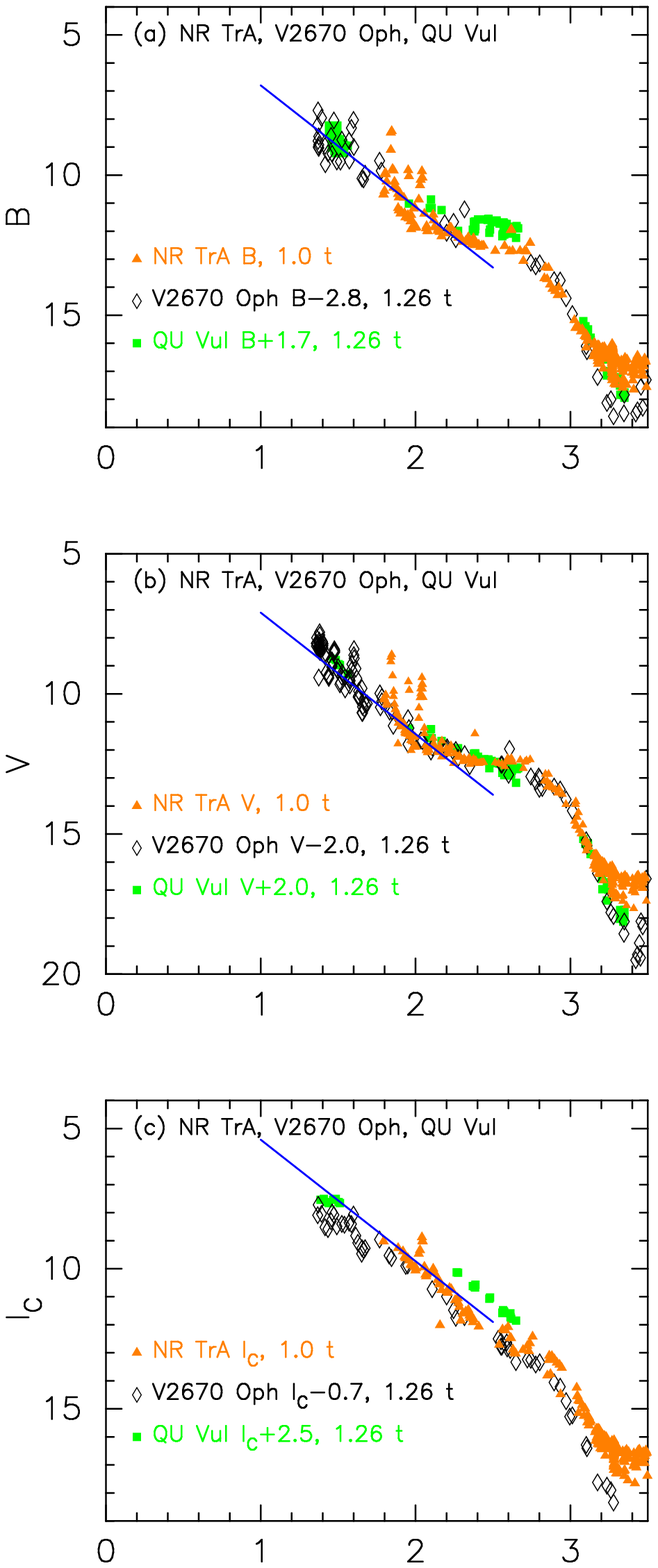}
\caption{
Same as Figure \ref{v1663_aql_yy_dor_lmcn_2009a_b_v_i_logscale_3fig},
but for NR~TrA.
The (a) $B$, (b) $V$, and (c) $I_{\rm C}$ light curves of NR~TrA
as well as those of QU~Vul and V2670~Oph.
The $BV$ data of NR~TrA are the same as those in Figure
\ref{nr_tra_v_bv_ub_color_curve}.  The $I_{\rm C}$ data of NR~TrA 
are taken from AAVSO, VSOLJ, and SMARTS.
\label{nr_tra_v2670_ooph_qu_vul_b_v_i_logscale_3fig}}
\end{figure}

\subsection{NR~TrA 2008}
\label{nr_tra}
Figure \ref{nr_tra_v_bv_ub_color_curve} shows (a) the $V$ and $B$, and
(b) $(B-V)_0$ evolutions of NR~TrA.  Here, $(B-V)_0$ are dereddened
with $E(B-V)=0.24$ as obtained in Section \ref{nr_tra_cmd}.
We plot light curves of three novae, NR~TrA, QU~Vul, and V2670~Oph
in Figure \ref{nr_tra_v2670_oph_qu_vul_x55z02c10o10_logscale}.
The decay of the $V$ light curve of NR~TrA slowed down 
in the intermediate phase, and this slow decay is
very similar to that of V2670~Oph and QU~Vul.
Applying Equation (\ref{distance_modulus_general_temp}) to them,
we have the relation 
\begin{eqnarray}
(m&-&M)_{V, \rm NR~TrA} \cr 
&=& (m - M + \Delta V)_{V, \rm QU~Vul} - 2.5 \log 1.26 \cr
&=& 13.6 + 2.0\pm0.2 - 0.25 = 15.35\pm0.2 \cr
&=& (m - M + \Delta V)_{V, \rm V2670~Oph} - 2.5 \log 1.26 \cr
&=& 17.6 - 2.0\pm0.2 - 0.25 = 15.35\pm0.2,
\label{distance_modulus_nr_tra_qu_vul}
\end{eqnarray}
where we adopt $(m-M)_{V, \rm QU~Vul}=13.6$ and 
$(m-M)_{V, \rm V2670~Oph}=17.6$, both from Appendix \ref{v2670_oph}.
Thus, we adopt $(m-M)_V=15.35\pm0.2$.
The timescale of NR~TrA is 1.26 times longer than those of QU~Vul
and V2670~Oph.  This corresponds to $\log f_{\rm s}= 0.43$ against
LV~Vul.

Using $(m-M)_V=15.35$ and $\log f_{\rm s}= 0.43$ for NR~TrA,
we plot Figure \ref{nr_tra_v1668_cyg_lv_vul_v_bv_ub_color_logscale}.
Here, we show three nova light/color curves of NR~TrA, LV~Vul, and 
V1668~Cyg.  The $V$ light curve of NR~TrA largely deviates from that of
LV~Vul and V1668~Cyg in the nebular phase during days $250-1000$.
Applying Equation (\ref{distance_modulus_general_temp}) to them,
we have the relation 
\begin{eqnarray}
(m-M)_{V, \rm NR~TrA} 
&=& (m - M + \Delta V)_{V, \rm LV~Vul} - 2.5 \log 2.7 \cr
&=& 11.85 + 4.6\pm0.3 - 1.08 = 15.37\pm0.3 \cr
&=& (m - M + \Delta V)_{V, \rm V1668~Cyg} - 2.5 \log 2.7 \cr
&=& 14.6 + 1.8\pm0.3 - 1.08 = 15.32\pm0.3,
\label{distance_modulus_nr_tra_lv_vul}
\end{eqnarray}
where we adopt $(m-M)_{V, \rm LV~Vul}=11.85$ and 
$(m-M)_{V, \rm V1668~Cyg}=14.6$, both from \citet{hac19k}.
Thus, we adopt $(m-M)_V=15.35\pm0.2$ and $f_{\rm s}=2.7$ against LV~Vul.
From Equations (\ref{time-stretching_general}),
(\ref{distance_modulus_general_temp}), and
(\ref{distance_modulus_nr_tra_lv_vul}),
we have the relation
\begin{eqnarray}
(m- M')_{V, \rm NR~TrA} 
&\equiv & (m_V - (M_V - 2.5\log f_{\rm s}))_{\rm NR~TrA} \cr
&=& \left( (m-M)_V + \Delta V \right)_{\rm LV~Vul} \cr
&=& 11.85 + 4.6\pm0.3 = 16.45\pm0.3.
\label{absolute_mag_nr_tra}
\end{eqnarray}

Figure \ref{nr_tra_v2670_ooph_qu_vul_b_v_i_logscale_3fig} shows
the $B$, $V$, and $I_{\rm C}$ light curves of NR~TrA
together with those of QU~Vul and V2670~Oph.
We apply Equation (\ref{distance_modulus_general_temp_b}) for the $B$ band to 
Figure \ref{nr_tra_v2670_ooph_qu_vul_b_v_i_logscale_3fig}(a) and obtain
\begin{eqnarray}
(m&-&M)_{B, \rm NR~TrA} \cr
&=& ((m - M)_B + \Delta B)_{\rm QU~Vul} - 2.5 \log 1.26 \cr
&=& 14.15 + 1.7\pm0.3 - 0.25 = 15.6\pm0.3 \cr
&=& ((m - M)_B + \Delta B)_{\rm V2670~Oph} - 2.5 \log 1.26 \cr
&=& 18.65 - 2.8\pm0.3 - 0.25 = 15.6\pm0.3,
\label{distance_modulus_b_nr_tra_v2670_oph_qu_vul}
\end{eqnarray}
where we adopt $(m - M)_{B, \rm QU~Vul}= 14.15$ and
$(m - M)_{B, \rm V2670~Oph}= 17.6 + 1.05= 18.65$ 
from Appendix \ref{v2670_oph}.
We have $(m-M)_{B, \rm NR~TrA}= 15.6\pm0.3$.

For the $V$ band, Figure 
\ref{nr_tra_v2670_ooph_qu_vul_b_v_i_logscale_3fig}(b) is essentially
the same as Figure \ref{nr_tra_v2670_oph_qu_vul_x55z02c10o10_logscale},
giving $(m-M)_{V, \rm NR~TrA}= 15.35\pm0.2$.

From the $I_{\rm C}$-band data in Figure
\ref{nr_tra_v2670_ooph_qu_vul_b_v_i_logscale_3fig}(c), we obtain
\begin{eqnarray}
(m&-&M)_{I, \rm NR~TrA} \cr
&=& ((m - M)_I + \Delta I_C)_{\rm QU~Vul} - 2.5 \log 1.26 \cr
&=& 12.72 + 2.5\pm0.3 - 0.25 = 14.97\pm0.3 \cr
&=& ((m - M)_I + \Delta I_C)_{\rm V2670~Oph} - 2.5 \log 1.26 \cr
&=& 15.92 - 0.7\pm0.3 - 0.25 = 14.97\pm0.3,
\label{distance_modulus_i_nr_tra_v2670_oph_qu_vul}
\end{eqnarray}
where we adopt $(m-M)_{I, \rm QU~Vul}=13.6 - 1.6\times 0.55= 12.72$ and
$(m-M)_{I, \rm V2670~Oph}=17.6 - 1.6\times 1.05= 15.92$.
We have $(m-M)_{I, \rm NR~TrA}= 14.97\pm0.2$.

We plot $(m-M)_B= 15.6$, $(m-M)_V= 15.35$, and $(m-M)_I=14.97$,
which cross at $d=8.3$~kpc and $E(B-V)=0.24$,
in Figure \ref{distance_reddening_nr_tra_v1213_cen_v5583_sgr_v5584_sgr}(a).
Thus, we obtain $d=8.3\pm1.0$~kpc and $E(B-V)=0.24\pm0.05$.


\begin{figure}
\epsscale{0.75}
\plotone{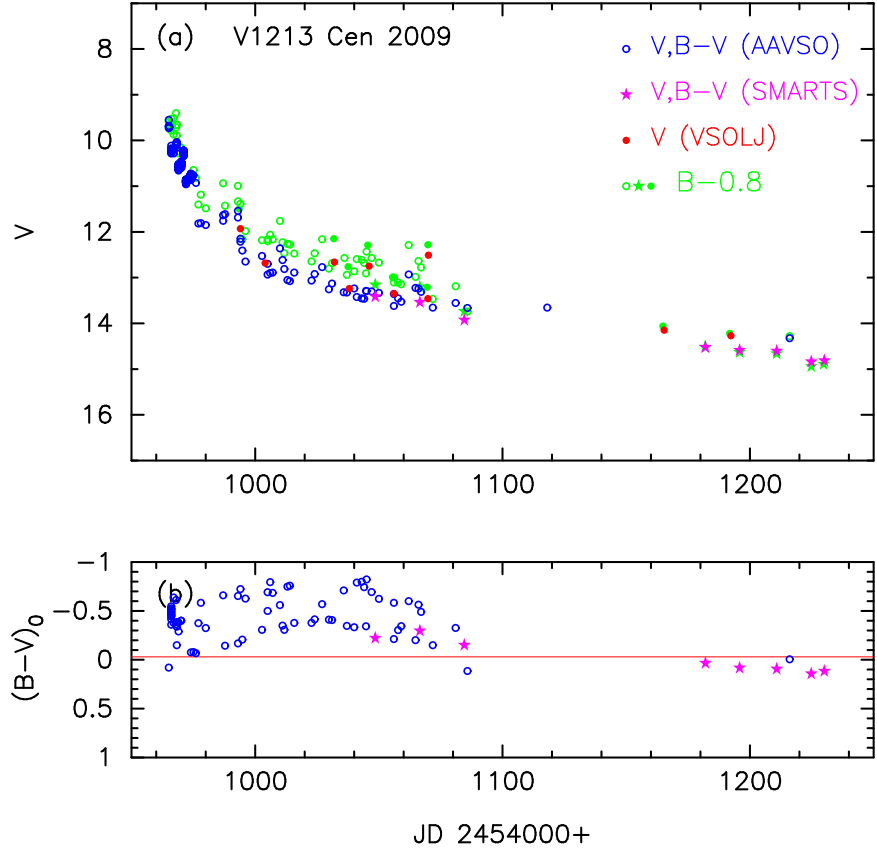}
\caption{
Same as Figure \ref{v1663_aql_v_bv_ub_color_curve}, but for V1213~Cen.
(a) The $V$ (unfilled blue circles) and $B$ (unfilled green circles) data
are taken from AAVSO.  The $V$ (filled red circles) and
$B$ (filled green circles) data are from VSOLJ.
The SMARTS data of $V$ (filled magenta stars) and $B$ (filled green stars)
are also plotted.
(b) The $(B-V)_0$ are dereddened with $E(B-V)=0.78$.
\label{v1213_cen_v_bv_ub_color_curve}}
\end{figure}


\begin{figure}
\epsscale{0.75}
\plotone{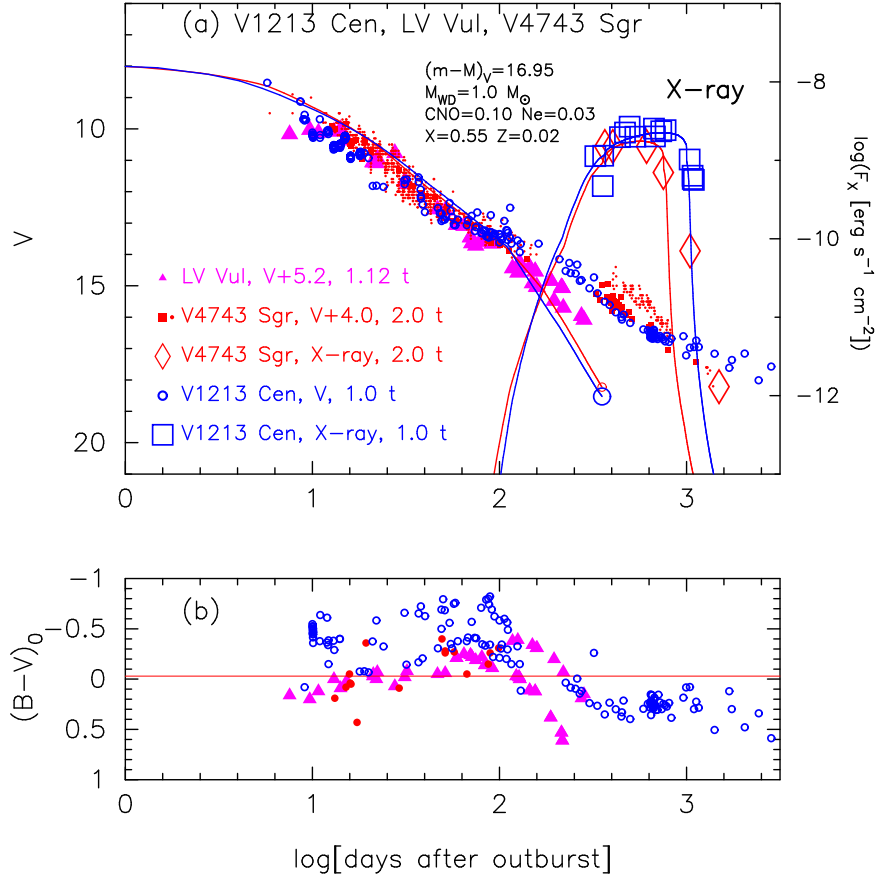}
\caption{
Same as Figure
\ref{v2575_oph_v1668_cyg_lv_vul_v_bv_ub_logscale},
but for V1213~Cen.   We add the $V$ light curves of LV~Vul and V4743~Sgr. 
The data of V1213~Cen are the same as those in Figure
\ref{v1213_cen_v_bv_ub_color_curve}.  The data of V4743~Sgr
are the same as those in Figures 18 of \citet{hac10k}.
In panel (a), we add model light curves (solid blue lines) of a
$1.0~M_\sun$ WD \citep[Ne2;][]{hac10k} both for the $V$
and X-ray (0.2--2.0~keV), assuming that $(m-M)_V=16.95$ for V1213~Cen.
The solid red lines denote the model light curves of
a $1.15~M_\sun$ WD (Ne2), 
assuming $(m-M)_V=13.7$ for V4743~Sgr \citep{hac10k}.
\label{v1213_cen_lv_vul_v4743_sgr_v_bv_ub_color_curve_logscale}}
\end{figure}


\begin{figure}
\epsscale{0.55}
\plotone{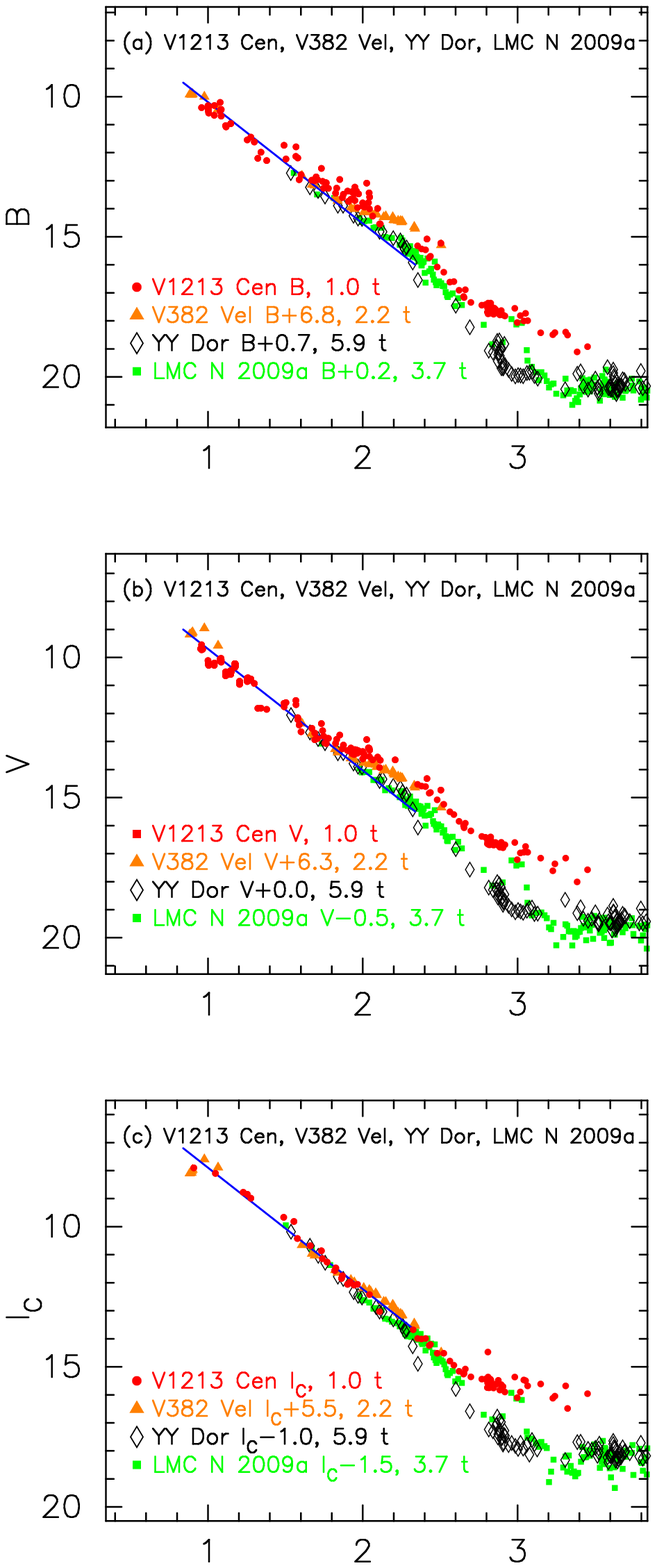}
\caption{
Same as Figure \ref{v1663_aql_yy_dor_lmcn_2009a_b_v_i_logscale_3fig},
but for V1213~Cen.
The (a) $B$, (b) $V$, and (c) $I_{\rm C}$ light curves of V1213~Cen
as well as those of V382~Vel, YY~Dor, and LMC~N~2009a.
The $BV$ data of V1213~Cen are the same as those in Figure
\ref{v1213_cen_v_bv_ub_color_curve}.  The $I_{\rm C}$ data of V1213~Cen
are taken from AAVSO, VSOLJ, and SMARTS.  The $BVI_{\rm C}$ data of
V382~Vel are the same as those in Figure 40 of \citet{hac19k}.
\label{v1213_cen_v382_vel_yy_dor_lmcn_2009a_b_v_i_logscale_3fig}}
\end{figure}

\subsection{V1213~Cen 2009}
\label{v1213_cen}
Figure \ref{v1213_cen_v_bv_ub_color_curve} shows (a) the $V$ and $B$, and
(b) $(B-V)_0$ evolutions of V1213~Cen.  Here, $(B-V)_0$ are dereddened
with $E(B-V)=0.78$ as obtained in Section \ref{v1213_cen_cmd}.
We plot light curves of three novae, V1213~Cen, LV~Vul, and V4743~Sgr
in Figure \ref{v1213_cen_lv_vul_v4743_sgr_v_bv_ub_color_curve_logscale}.
The shape of the $V$ light curve of V1213~Cen is very similar to V4743~Sgr.
Applying Equation (\ref{distance_modulus_general_temp}) to them, we have
\begin{eqnarray}
(m&-&M)_{V, \rm V1213~Cen} \cr 
&=& (m - M + \Delta V)_{V, \rm LV~Vul} - 2.5 \log 1.12 \cr
&=& 11.85 + 5.2\pm0.3 - 0.13 = 16.92\pm0.3 \cr
&=& (m - M + \Delta V)_{V, \rm V4743~Sgr} - 2.5 \log 2.0 \cr
&=& 13.7 + 4.0\pm0.3 - 0.75 = 16.95\pm0.3,
\label{distance_modulus_v1213_cen}
\end{eqnarray}
where we adopt $(m-M)_{V, \rm LV~Vul}=11.85$ from \citet{hac19k}
and $(m-M)_{V, \rm V4743~Sgr}=13.7$ from \cite{hac10k}.
Thus, we obtain $(m-M)_V=16.95\pm0.2$ and $f_{\rm s}=1.12$ against LV~Vul.
From Equations (\ref{time-stretching_general}),
(\ref{distance_modulus_general_temp}), and
(\ref{distance_modulus_v1213_cen}),
we have the relation
\begin{eqnarray}
(m- M')_{V, \rm V1213~Cen} 
&\equiv & (m_V - (M_V - 2.5\log f_{\rm s}))_{\rm V1213~Cen} \cr
&=& \left( (m-M)_V + \Delta V \right)_{\rm LV~Vul} \cr
&=& 11.85 + 5.2\pm0.3 = 17.05\pm0.3.
\label{absolute_mag_v1213_cen}
\end{eqnarray}

Figure \ref{v1213_cen_v382_vel_yy_dor_lmcn_2009a_b_v_i_logscale_3fig} shows
the $B$, $V$, and $I_{\rm C}$ light curves of V1213~Cen
together with those of V382~Vel, YY~Dor, and LMC~N~2009a.
We regard the extension of YY~Dor and LMC~N~2009a light curves
and V382~Vel light curve to overlap with that of V1213~Cen.
We apply Equation (\ref{distance_modulus_general_temp_b})
for the $B$ band to Figure
\ref{v1213_cen_v382_vel_yy_dor_lmcn_2009a_b_v_i_logscale_3fig}(a)
and obtain
\begin{eqnarray}
(m&-&M)_{B, \rm V1213~Cen} \cr
&=& ((m - M)_B + \Delta B)_{\rm V382~Vel} - 2.5 \log 2.2 \cr
&=& 11.75 + 6.8\pm0.2 - 0.85 = 17.7\pm0.2 \cr
&=& ((m - M)_B + \Delta B)_{\rm YY~Dor} - 2.5 \log 5.9 \cr
&=& 18.98 + 0.7\pm0.2 - 1.92 = 17.76\pm0.2 \cr
&=& ((m - M)_B + \Delta B)_{\rm LMC~N~2009a} - 2.5 \log 3.7 \cr
&=& 18.98 + 0.2\pm0.2 - 1.42 = 17.76\pm0.2,
\label{distance_modulus_b_v1213_cen_v382_vel_yy_dor_lmcn2009a}
\end{eqnarray}
where we adopt $(m-M)_{B, \rm V382~Vel}= 11.5 + 0.25= 11.75$ 
and $\log f_{\rm s}= -0.29$ against LV~Vul for V382~Vel from \citet{hac19k}.
Thus, we have $(m-M)_{B, \rm V1213~Cen}= 17.74\pm0.1$.

For the $V$ light curves in Figure
\ref{v1213_cen_v382_vel_yy_dor_lmcn_2009a_b_v_i_logscale_3fig}(b),
we similarly obtain
\begin{eqnarray}
(m&-&M)_{V, \rm V1213~Cen} \cr
&=& ((m - M)_V + \Delta V)_{\rm V382~Vel} - 2.5 \log 2.2 \cr
&=& 11.5 + 6.3\pm0.2 - 0.85 = 16.95\pm0.2 \cr
&=& ((m - M)_V + \Delta V)_{\rm YY~Dor} - 2.5 \log 5.9 \cr
&=& 18.86 + 0.0\pm0.2 - 1.92 = 16.94\pm0.2 \cr
&=& ((m - M)_V + \Delta V)_{\rm LMC~N~2009a} - 2.5 \log 3.7 \cr
&=& 18.86 - 0.5\pm0.2 - 1.42 = 16.94\pm0.2.
\label{distance_modulus_v_v1213_cen_v382_vel_yy_dor_lmcn2009a}
\end{eqnarray}
We have $(m-M)_{V, \rm V1213~Cen}= 16.95\pm0.1$, which is
consistent with Equation (\ref{distance_modulus_v1213_cen}).

We apply Equation (\ref{distance_modulus_general_temp_i}) for
the $I_{\rm C}$ band to Figure
\ref{v1213_cen_v382_vel_yy_dor_lmcn_2009a_b_v_i_logscale_3fig}(c) and obtain
\begin{eqnarray}
(m&-&M)_{I, \rm V1213~Cen} \cr
&=& ((m - M)_I + \Delta I_C)_{\rm V382~Vel} - 2.5 \log 2.2 \cr
&=& 11.1 + 5.5\pm0.2 - 0.85 = 15.75\pm 0.2 \cr
&=& ((m - M)_I + \Delta I_C)_{\rm YY~Dor} - 2.5 \log 5.9 \cr
&=& 18.67 - 1.0\pm0.2 - 1.92 = 15.75\pm 0.2 \cr
&=& ((m - M)_I + \Delta I_C)_{\rm LMC~N~2009a} - 2.5 \log 3.7 \cr
&=& 18.67 - 1.5\pm0.2 - 1.42 = 15.75\pm 0.2,
\label{distance_modulus_i_v1213_cen_v382_vel_yy_dor_lmcn2009a}
\end{eqnarray}
where we adopt $(m-M)_{I, \rm V382~Vel}= 11.5 - 1.6\times 0.25= 11.1$
from \citet{hac19k}.  We have $(m-M)_{I, \rm V1213~Cen}= 15.75\pm0.1$.

We plot $(m-M)_B= 17.74$, $(m-M)_V= 16.95$, and $(m-M)_I= 15.75$,
which broadly cross at $d=8.1$~kpc and $E(B-V)=0.78$,  in Figure
\ref{distance_reddening_nr_tra_v1213_cen_v5583_sgr_v5584_sgr}(b).
Thus, we have $E(B-V)=0.78\pm0.05$ and $d=8.1\pm1$~kpc.


\begin{figure}
\epsscale{0.75}
\plotone{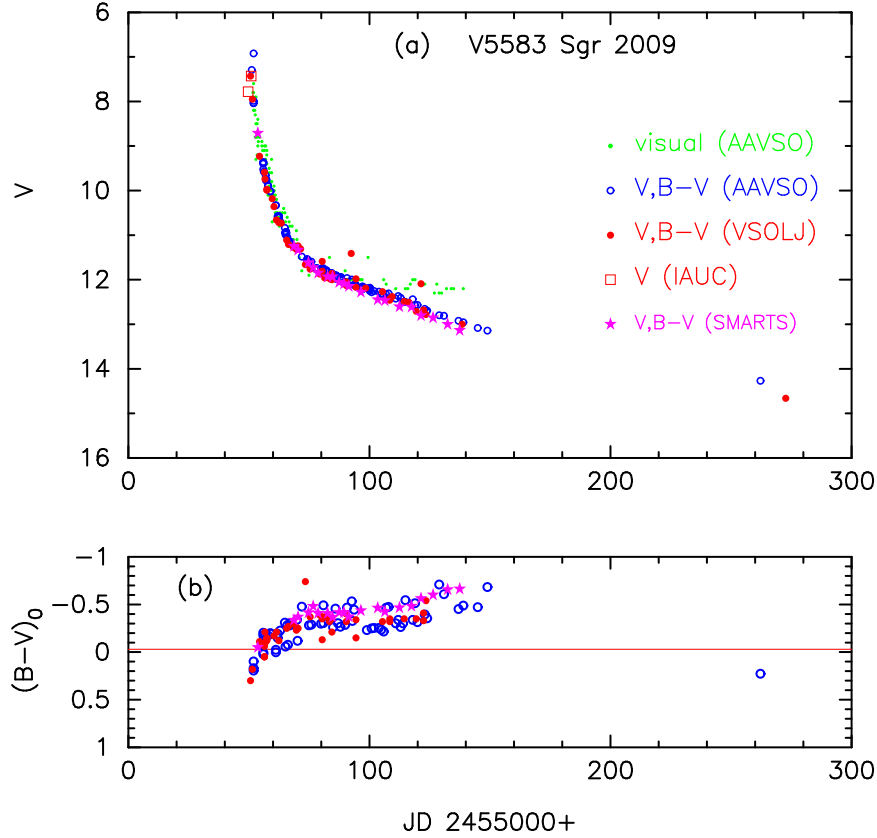}
\caption{
Same as Figure \ref{v1663_aql_v_bv_ub_color_curve}, but for V5583~Sgr.
The $V$ and $B-V$ (unfilled blue circles) and visual (green dots) data
are taken from AAVSO.  The $V$ and $B-V$ (filled red circles) data
are from VSOLJ.  The $V$ (unfilled red squares) data are from IAU Circulars.
The SMARTS data of $V$ and $B-V$ (filled magenta stars) are also plotted.
In panel (b), the $(B-V)_0$ colors are dereddened with $E(B-V)=0.30$.
\label{v5583_sgr_v_bv_ub_color_curve}}
\end{figure}


\begin{figure}
\epsscale{0.75}
\plotone{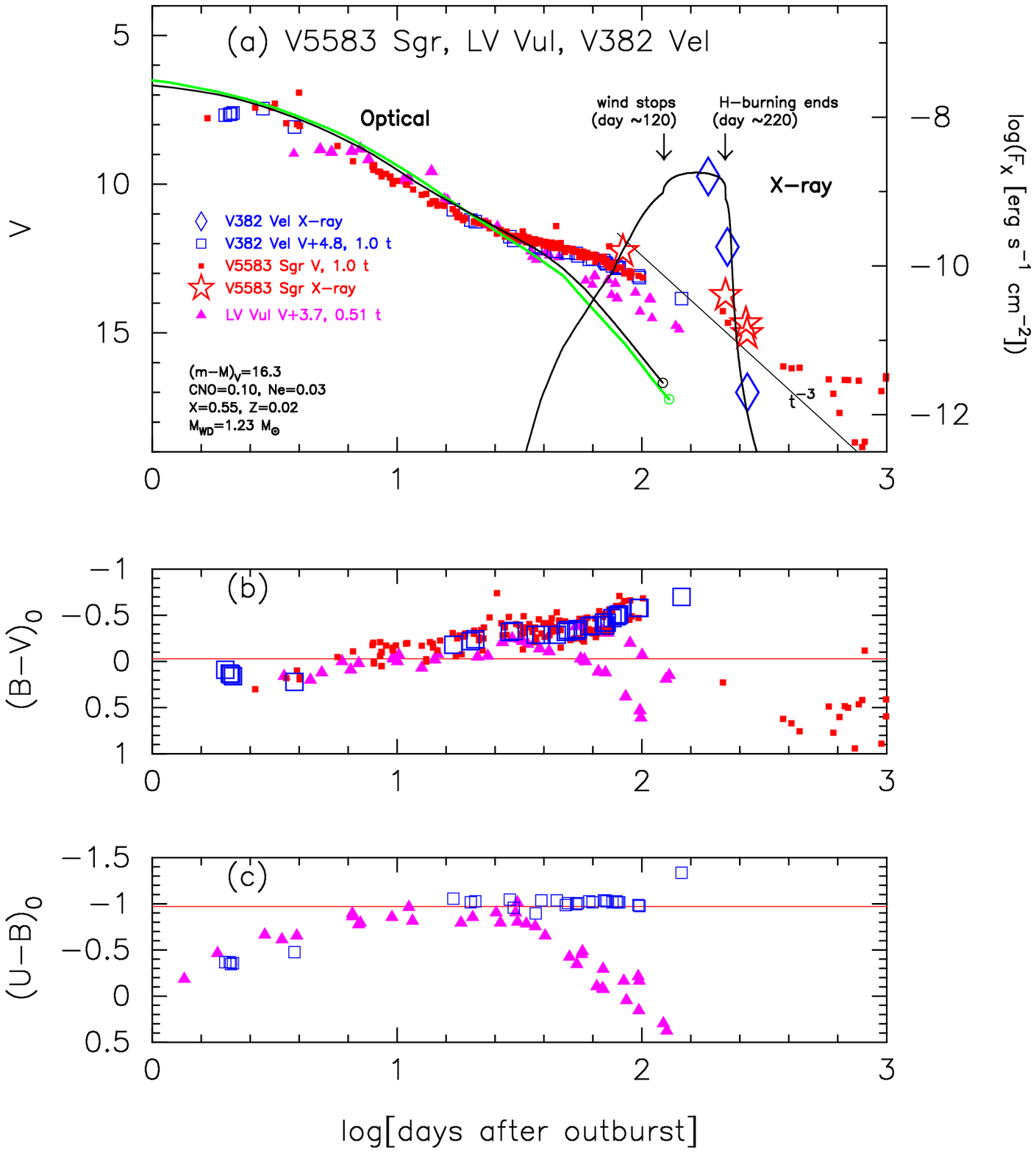}
\caption{
Same as Figure
\ref{v2575_oph_v1668_cyg_lv_vul_v_bv_ub_logscale},
but for V5583~Sgr (filled red squares).
We overplot the V382~Vel light curves by unfilled blue squares.
The data of V5583~Sgr are the same as those in Figure
\ref{v5583_sgr_v_bv_ub_color_curve}.  The data of V382~Vel
are the same as those in Figures 25--29 of \citet{hac16k}
and Figure 39 of \citet{hac16kb}.
We assume that the timescale of V382~Vel is the same as
that of V5583~Sgr, but shift down the $V$ magnitude of V382~Vel by
4.8 mag.
In panel (a), we added model light curves (solid black lines) of a
$1.23~M_\sun$ WD \citep[Ne2;][]{hac10k, hac16k} both for the $V$ and X-ray,
assuming that $(m-M)_V=16.3$ for V5583~Sgr and
$(m-M)_V=11.5$ for V382~Vel \citep{hac19k}.
For comparison, we add the light curve of a $0.98~M_\sun$ WD 
(CO3, solid green line), assuming that $(m-M)_V=11.85$ for LV~Vul.
\label{v5583_sgr_v382_vel_v_bv_ub_x55z02o10ne03_logscale}}
\end{figure}


\begin{figure}
\epsscale{0.55}
\plotone{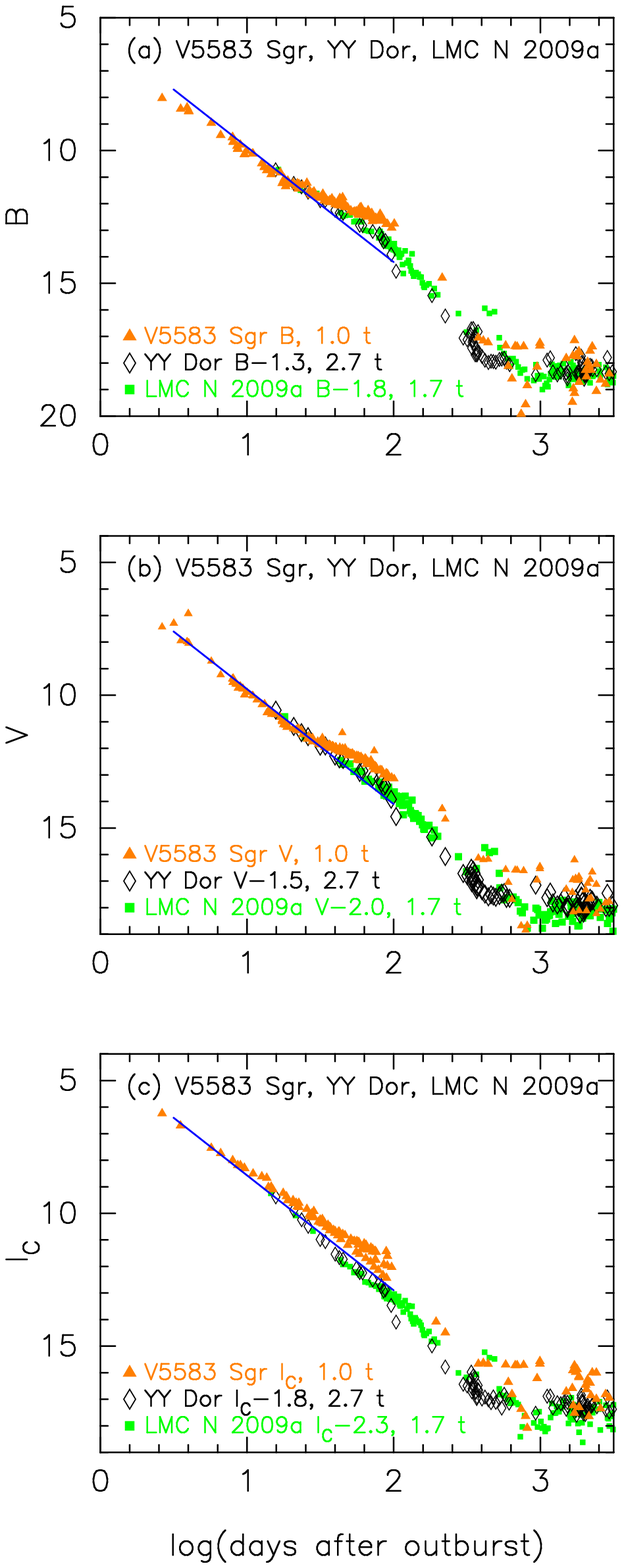}
\caption{
Same as Figure \ref{v1663_aql_yy_dor_lmcn_2009a_b_v_i_logscale_3fig},
but for V5583~Sgr.
The $BV$ data of V5583~Sgr are the same as those in Figure
\ref{v5583_sgr_v_bv_ub_color_curve}.  The $I_{\rm C}$ data of V5583~Sgr
are taken from AAVSO, VSOLJ, and SMARTS.
\label{v5583_sgr_yy_dor_lmcn_2009a_b_v_i_logscale_3fig}}
\end{figure}

\subsection{V5583~Sgr 2009\#3}
\label{v5583_sgr}
Figure \ref{v5583_sgr_v_bv_ub_color_curve} shows (a) the visual and $V$,
and (b) $(B-V)_0$ evolutions of V5583~Sgr.  Here, $(B-V)_0$ are dereddened
with $E(B-V)=0.30$ as obtained in Section \ref{v5583_sgr_cmd}.
Figure \ref{v5583_sgr_v382_vel_v_bv_ub_x55z02o10ne03_logscale} shows
the light/color curves of three novae, V5583~Sgr, LV~Vul, and V382~Vel.
The shape of the $V$ light curve of V5583~Sgr is very similar to that
of V382~Vel \citep{hac09kkm}.
The timescale of V382~Vel is almost the same as that of V5583~Sgr.
Applying Equation (\ref{distance_modulus_general_temp}) to them,
we have the relation 
\begin{eqnarray}
(m&-&M)_{V, \rm V5583~Sgr} \cr 
&=& (m - M + \Delta V)_{V, \rm LV~Vul} - 2.5 \log 0.51 \cr
&=& 11.85 + 3.7\pm0.2 + 0.73 = 16.28\pm0.2 \cr
&=& (m - M + \Delta V)_{V, \rm V382~Vel} - 2.5 \log 1.0 \cr
&=& 11.5 + 4.8\pm0.2 - 0.0 = 16.3\pm0.2,
\label{distance_modulus_v5583_sgr}
\end{eqnarray}
where we adopt $(m-M)_{V, \rm LV~Vul}=11.85$ 
and $(m-M)_{V, \rm V382~Vel}=11.5$, both from \citet{hac19k}.
Thus, we adopt $(m-M)_V=16.3\pm0.1$ and $f_{\rm s}=0.51$ against LV~Vul.
From Equations (\ref{time-stretching_general}),
(\ref{distance_modulus_general_temp}), and
(\ref{distance_modulus_v5583_sgr}),
we have the relation
\begin{eqnarray}
(m- M')_{V, \rm V5583~Sgr} 
&\equiv & (m_V - (M_V - 2.5\log f_{\rm s}))_{\rm V5583~Sgr} \cr
&=& \left( (m-M)_V + \Delta V \right)_{\rm LV~Vul} \cr
&=& 11.85 +3.7\pm0.2 = 15.55\pm0.2.
\label{absolute_mag_v5583_sgr}
\end{eqnarray}

Figure \ref{v5583_sgr_yy_dor_lmcn_2009a_b_v_i_logscale_3fig} shows
the $B$, $V$, and $I_{\rm C}$ light curves of V5583~Sgr
together with those of YY~Dor and LMC~N~2009a.
We regard the extension of the YY~Dor and LMC~N~2009a light curves
to overlap with that of V5583~Sgr.
We apply Equation (\ref{distance_modulus_general_temp_b})
for the $B$ band to Figure
\ref{v5583_sgr_yy_dor_lmcn_2009a_b_v_i_logscale_3fig}(a)
and obtain
\begin{eqnarray}
(m&-&M)_{B, \rm V5583~Sgr} \cr
&=& ((m - M)_B + \Delta B)_{\rm YY~Dor} - 2.5 \log 2.7 \cr
&=& 18.98 - 1.3\pm0.2 - 1.08 = 16.6\pm0.2 \cr
&=& ((m - M)_B + \Delta B)_{\rm LMC~N~2009a} - 2.5 \log 1.7 \cr
&=& 18.98 - 1.8\pm0.2 - 0.58 = 16.6\pm0.2.
\label{distance_modulus_b_v5583_sgr_yy_dor_lmcn2009a}
\end{eqnarray}
We have $(m-M)_{B, \rm V5583~Sgr}= 16.6\pm0.1$.

For the $V$ light curves in Figure
\ref{v5583_sgr_yy_dor_lmcn_2009a_b_v_i_logscale_3fig}(b),
we similarly obtain
\begin{eqnarray}
(m&-&M)_{V, \rm V5583~Sgr} \cr
&=& ((m - M)_V + \Delta V)_{\rm YY~Dor} - 2.5 \log 2.7 \cr
&=& 18.86 - 1.5\pm0.2 - 1.08 = 16.28\pm0.2 \cr
&=& ((m - M)_V + \Delta V)_{\rm LMC~N~2009a} - 2.5 \log 1.7 \cr
&=& 18.86 - 2.0\pm0.2 -0.58 = 16.28\pm0.2.
\label{distance_modulus_v_v5583_sgr_yy_dor_lmcn2009a}
\end{eqnarray}
We have $(m-M)_{V, \rm V5583~Sgr}= 16.28\pm0.1$, which is
consistent with Equation (\ref{distance_modulus_v5583_sgr}).

We apply Equation (\ref{distance_modulus_general_temp_i}) for
the $I_{\rm C}$ band to Figure
\ref{v5583_sgr_yy_dor_lmcn_2009a_b_v_i_logscale_3fig}(c) and obtain
\begin{eqnarray}
(m&-&M)_{I, \rm V5583~Sgr} \cr
&=& ((m - M)_I + \Delta I_C)_{\rm YY~Dor} - 2.5 \log 2.7 \cr
&=& 18.67 - 1.8\pm0.2 - 1.08 = 15.79\pm 0.2 \cr
&=& ((m - M)_I + \Delta I_C)_{\rm LMC~N~2009a} - 2.5 \log 1.7 \cr
&=& 18.67 - 2.3\pm0.2 - 0.58 = 15.79\pm 0.2.
\label{distance_modulus_i_v5583_sgr_yy_dor_lmcn2009a}
\end{eqnarray}
We have $(m-M)_{I, \rm V5583~Sgr}= 15.79\pm0.1$.

We plot $(m-M)_B= 16.6$, $(m-M)_V= 16.28$, and $(m-M)_I= 15.79$,
which broadly cross at $d=12$~kpc and $E(B-V)=0.30$, in Figure
\ref{distance_reddening_nr_tra_v1213_cen_v5583_sgr_v5584_sgr}(c).
Thus, we have $E(B-V)=0.30\pm0.05$ and $d=12\pm2$~kpc.


\begin{figure}
\epsscale{0.75}
\plotone{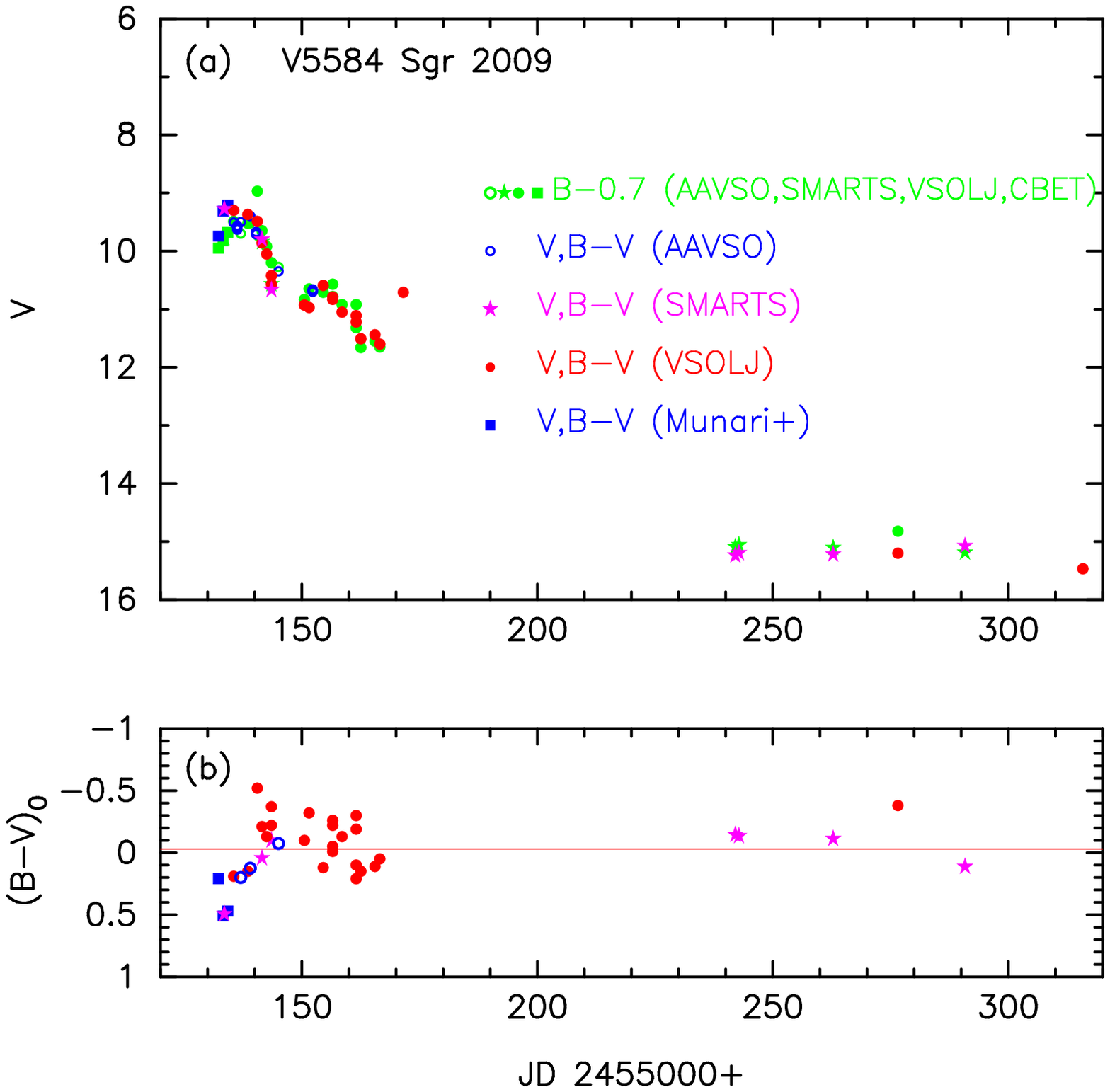}
\caption{
Same as Figure \ref{v1663_aql_v_bv_ub_color_curve}, but for V5584~Sgr.
(a) The $V$ (unfilled blue circles) and $B$ (unfilled green circles) data
are taken from AAVSO.  
The $V$ (filled magenta stars) and $B$ (filled green stars) data
are taken from SMARTS.  
The $V$ (filled red circles) and $B$ (filled green circles) data
are taken from VSOLJ.  
The $V$ (filled blue squares) and $B$ (filled green squares) data
are taken from \citet{mun09}.
(b) The $(B-V)_0$ are dereddened with $E(B-V)=0.70$.
\label{v5584_sgr_v_bv_ub_color_curve}}
\end{figure}


\begin{figure}
\epsscale{0.75}
\plotone{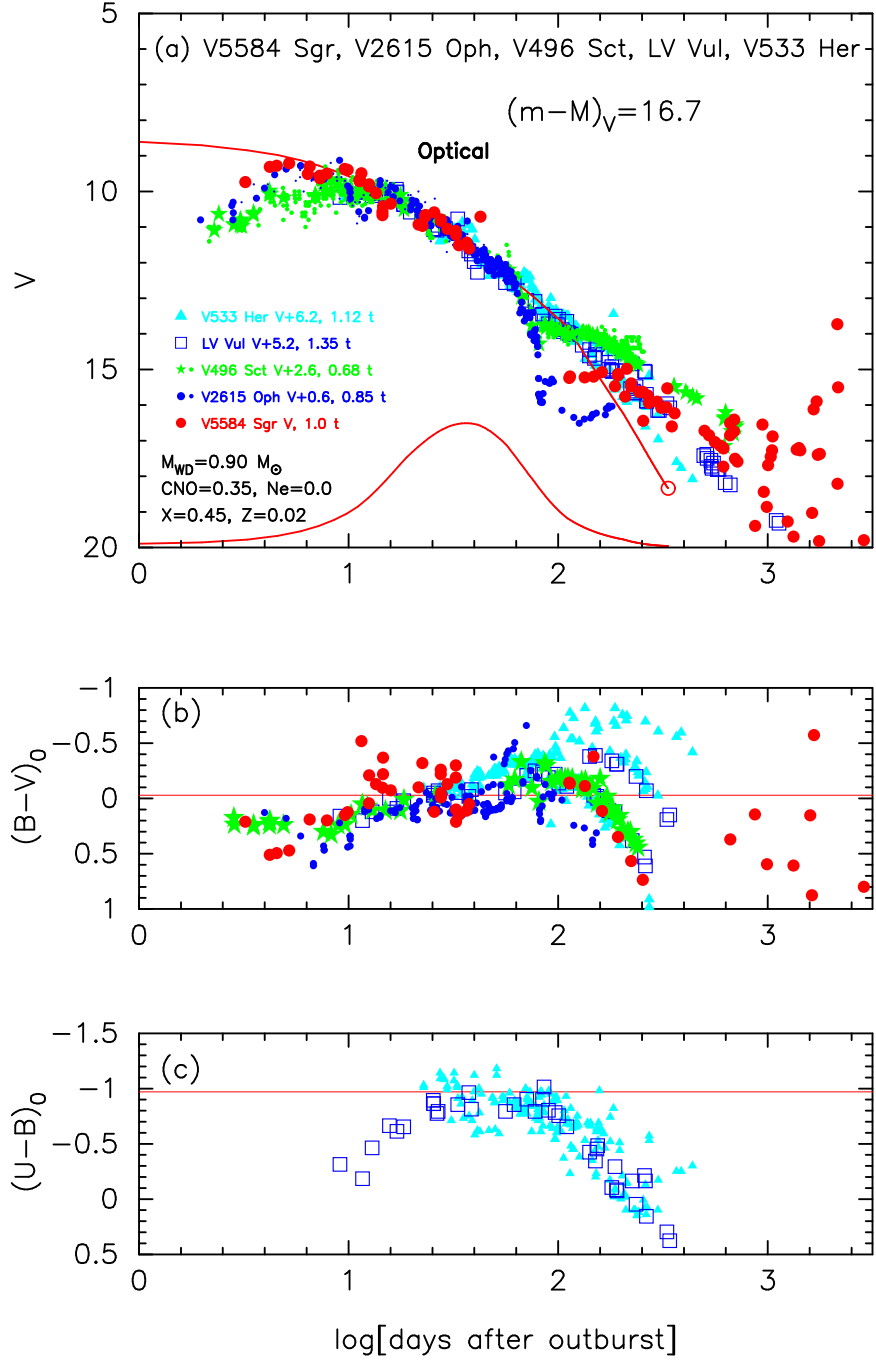}
\caption{
Same as Figure
\ref{v2575_oph_v1668_cyg_lv_vul_v_bv_ub_logscale},
but for V5584~Sgr (filled red circles).
We add the light/color curves of LV~Vul, V533~Her, V2615~Oph, and V496~Sct.
The data of V5584~Sgr are the same as those in Figure 
\ref{v5584_sgr_v_bv_ub_color_curve}.
In panel (a), we added model light curves (solid red lines) of a
$0.90~M_\sun$ WD \citep[CO3;][]{hac16k} for the $V$ (upper red line)
and UV~1455\AA\  (lower red
line), assuming that $(m-M)_V=16.7$ for V5584~Sgr.
\label{v5584_sgr_v2615_oph_v496_sct_v_bv_ub_logscale}}
\end{figure}


\begin{figure}
\epsscale{0.6}
\plotone{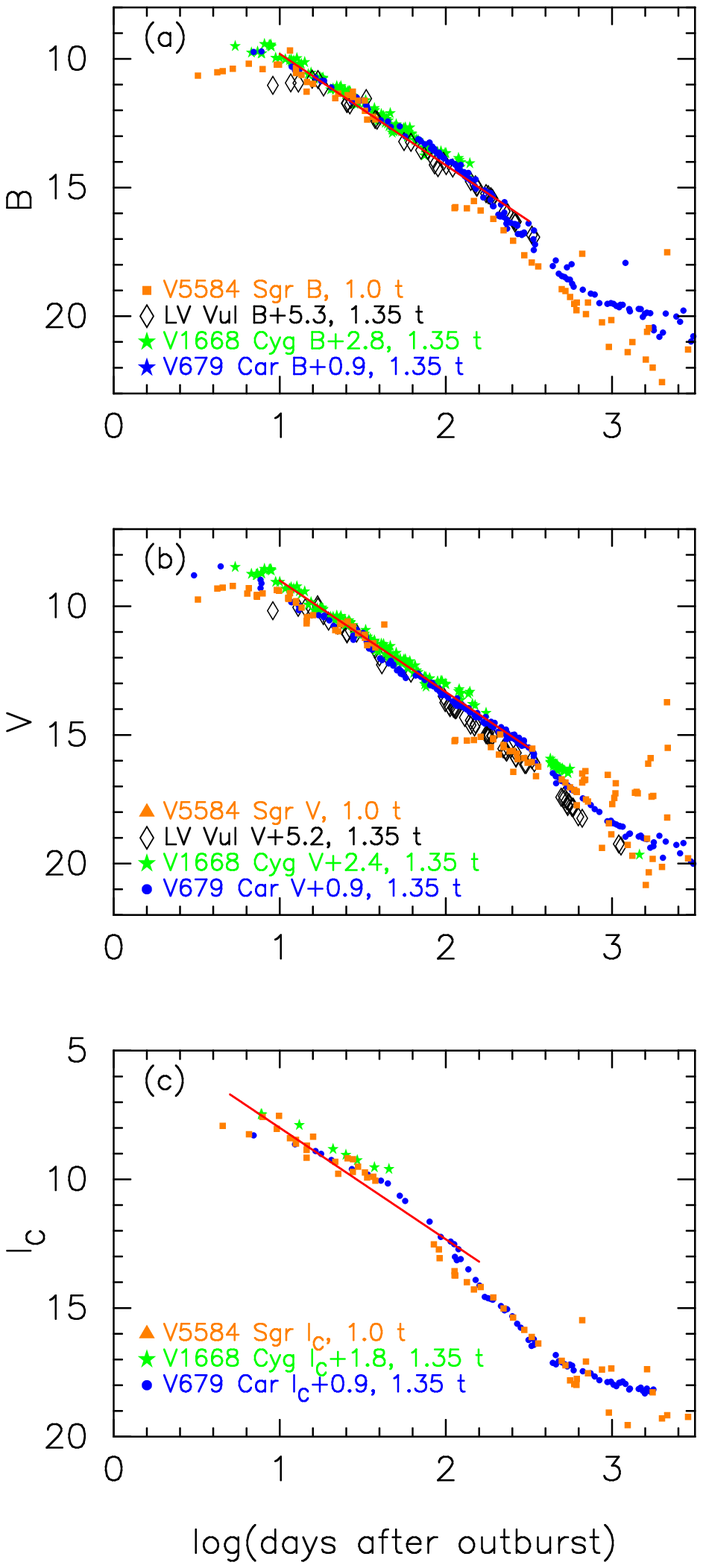}
\caption{
Same as Figure \ref{v1663_aql_yy_dor_lmcn_2009a_b_v_i_logscale_3fig},
but for V5584~Sgr.
The (a) $B$, (b) $V$, and (c) $I_{\rm C}$  light curves of V5584~Sgr
as well as those of LV~Vul, V1668~Cyg, and V679~Car.
The $BV$ data of V5584~Sgr are the same as those in Figure
\ref{v5584_sgr_v_bv_ub_color_curve}.
The $I_{\rm C}$ data of V5584~Sgr are taken from AAVSO, VSOLJ, and SMARTS.
The $BVI_{\rm C}$ data of V679~Car are the same as those in Figure 21 
of \citet{hac19k}.  
Neither $I_{\rm C}$ nor $I$ data are available for LV~Vul.
\label{v5584_sgr_lv_vul_v1668_cyg_b_v_i_logscale_3fig}}
\end{figure}

\subsection{V5584~Sgr 2009\#4}
\label{v5584_sgr}
Figure \ref{v5584_sgr_v_bv_ub_color_curve} shows (a) the $V$ and
$B$ magnitudes, and (b) $(B-V)_0$ evolutions of V5584~Sgr.  
Here, $(B-V)_0$ are dereddened with $E(B-V)=0.70$ as obtained in
Section \ref{v5584_sgr_cmd}.
Figure \ref{v5584_sgr_v2615_oph_v496_sct_v_bv_ub_logscale} shows the 
light/color curves of V5584~Sgr, LV~Vul, V533~Her, V2615~Oph, and V496~Sct.
Applying Equation (\ref{distance_modulus_general_temp}) to them,
we have the relation 
\begin{eqnarray}
(m&-&M)_{V, \rm V5584~Sgr} \cr
&=& (m - M + \Delta V)_{V, \rm LV~Vul} - 2.5 \log 1.35 \cr
&=& 11.85 + 5.2\pm0.2 - 0.33 = 16.72\pm0.2 \cr
&=& (m - M + \Delta V)_{V, \rm V533~Her} - 2.5 \log 1.12 \cr
&=& 10.65 + 6.2\pm0.2 - 0.13 = 16.72\pm0.2 \cr
&=& (m - M + \Delta V)_{V, \rm V2615~Oph} - 2.5 \log 0.85 \cr
&=& 15.95 + 0.6\pm0.2 + 0.18 = 16.73\pm0.2 \cr
&=& (m - M + \Delta V)_{V, \rm V496~Sct} - 2.5 \log 0.68 \cr
&=& 13.7 + 2.6\pm0.2 + 0.43 = 16.73\pm0.2,
\label{distance_modulus_v5584_sgr}
\end{eqnarray}
where we adopt $(m-M)_{V, \rm LV~Vul}=11.85$,
$(m-M)_{V, \rm V533~Her}=10.65$,
$(m-M)_{V, \rm V2615~Oph}=15.95$, and
$(m-M)_{V, \rm V496~Sct}=13.7$ from \cite{hac19k}.
Thus, we obtain $(m-M)_V=16.7\pm0.1$ and $f_{\rm s}=1.35$ 
against LV~Vul.
From Equations (\ref{time-stretching_general}),
(\ref{distance_modulus_general_temp}), and
(\ref{distance_modulus_v5584_sgr}),
we have the relation
\begin{eqnarray}
(m- M')_{V, \rm V5584~Sgr} 
&\equiv & (m_V - (M_V - 2.5\log f_{\rm s}))_{\rm V5584~Sgr} \cr
&=& \left( (m-M)_V + \Delta V \right)_{\rm LV~Vul} \cr
&=& 11.85 + 5.2\pm0.2 = 17.05\pm0.2.
\label{absolute_mag_v5584_sgr}
\end{eqnarray}

Figure \ref{v5584_sgr_lv_vul_v1668_cyg_b_v_i_logscale_3fig} shows
the $B$, $V$, and $I_{\rm C}$ light curves of V5584~Sgr together with
those of LV~Vul, V1668~Cyg, and V679~Car.
We apply Equation (\ref{distance_modulus_general_temp_b})
for the $B$ band to Figure
\ref{v5584_sgr_lv_vul_v1668_cyg_b_v_i_logscale_3fig}(a)
and obtain
\begin{eqnarray}
(m&-&M)_{B, \rm V5584~Sgr} \cr
&=& ((m - M)_B + \Delta B)_{\rm LV~Vul} - 2.5 \log 1.35 \cr
&=& 12.45 + 5.3\pm0.2 - 0.33 = 17.42\pm0.2 \cr
&=& ((m - M)_B + \Delta B)_{\rm V1668~Cyg} - 2.5 \log 1.35 \cr
&=& 14.9 + 2.8\pm0.2 - 0.33 = 17.37\pm0.2 \cr
&=& ((m - M)_B + \Delta B)_{\rm V679~Car} - 2.5 \log 1.35 \cr
&=& 16.79 + 0.9\pm0.2 - 0.33 = 17.36\pm0.2,
\label{distance_modulus_b_v5584_sgr_lv_vul_v1668_cyg}
\end{eqnarray}
where we adopt $(m-M)_{B, \rm LV~Vul}= 11.85 + 0.60= 12.45$,
$(m-M)_{B, \rm V1668~Cyg}= 14.6 + 0.30= 14.9$, and
$(m-M)_{B, \rm V679~Car}= 16.1 + 0.69= 16.79$,
all from \citet{hac19k}.
We have $(m-M)_{B, \rm V5584~Sgr}= 17.39\pm0.1$.

For the $V$ light curves in Figure
\ref{v5584_sgr_lv_vul_v1668_cyg_b_v_i_logscale_3fig}(b),
we similarly obtain
\begin{eqnarray}
(m&-&M)_{V, \rm V5584~Sgr} \cr
&=& ((m - M)_V + \Delta V)_{\rm LV~Vul} - 2.5 \log 1.35 \cr
&=& 11.85 + 5.2\pm0.2 - 0.33 = 16.72\pm0.2 \cr
&=& ((m - M)_V + \Delta V)_{\rm V1668~Cyg} - 2.5 \log 1.35 \cr
&=& 14.6 + 2.4\pm0.2 -0.33 = 16.67\pm0.2 \cr
&=& ((m - M)_V + \Delta V)_{\rm V679~Car} - 2.5 \log 1.35 \cr
&=& 16.1 + 0.9\pm0.2 -0.33 = 16.67\pm0.2,
\label{distance_modulus_v_v5584_sgr_lv_vul_v1668_cyg}
\end{eqnarray}
where we adopt $(m-M)_{V, \rm V679~Car}= 16.1$ from \citet{hac19k}.
We have $(m-M)_{V, \rm V5584~Sgr}= 16.69\pm0.1$, which is
consistent with Equation (\ref{distance_modulus_v5584_sgr}).

We apply Equation (\ref{distance_modulus_general_temp_i}) for
the $I_{\rm C}$ band to Figure
\ref{v5584_sgr_lv_vul_v1668_cyg_b_v_i_logscale_3fig}(c) and obtain
\begin{eqnarray}
(m&-&M)_{I, \rm V5584~Sgr} \cr
&=& ((m - M)_I + \Delta I_C)_{\rm V1668~Cyg} - 2.5 \log 1.35 \cr
&=& 14.12 + 1.8\pm0.2 - 0.33 = 15.59\pm 0.2 \cr
&=& ((m - M)_I + \Delta I_C)_{\rm V679~Car} - 2.5 \log 1.35 \cr
&=& 15.0 + 0.9\pm0.2 - 0.33 = 15.57\pm 0.2,
\label{distance_modulus_i_v5584_sgr_lv_vul_v1668_cyg}
\end{eqnarray}
where we adopt 
$(m-M)_{I, \rm V1668~Cyg}= 14.6 - 1.6\times 0.30= 14.12$ and
$(m-M)_{I, \rm V679~Car}= 16.1 - 1.6\times 0.69= 15.0$, both from 
\citet{hac19k}.  Unfortunately, no $I_{\rm C}$ or $I$ data of
LV~Vul are available.
We have $(m-M)_{I, \rm V5584~Sgr}= 15.58\pm0.1$.

We plot $(m-M)_B= 17.39$, $(m-M)_V= 16.69$, and $(m-M)_I= 15.58$,
which cross at $d=8.0$~kpc and $E(B-V)=0.70$, in Figure
\ref{distance_reddening_nr_tra_v1213_cen_v5583_sgr_v5584_sgr}(d).
Thus, we obtain $E(B-V)=0.70\pm0.05$ and $d=8.0\pm1$~kpc.


\begin{figure}
\epsscale{0.75}
\plotone{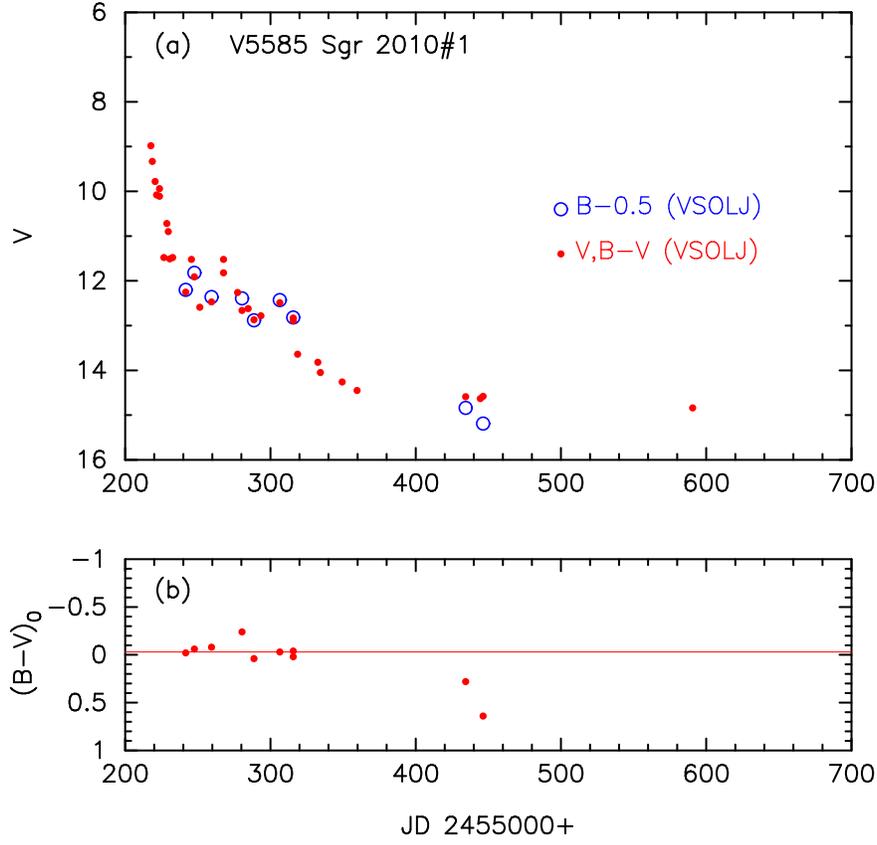}
\caption{
Same as Figure \ref{v1663_aql_v_bv_ub_color_curve}, but for V5585~Sgr.
(a) The $V$ (filled red circles) and $B$ (unfilled blue circles) data
are taken from VSOLJ.
(b) The $(B-V)_0$ are dereddened with $E(B-V)=0.47$.
\label{v5585_sgr_v_bv_ub_color_curve}}
\end{figure}


\begin{figure}
\epsscale{0.75}
\plotone{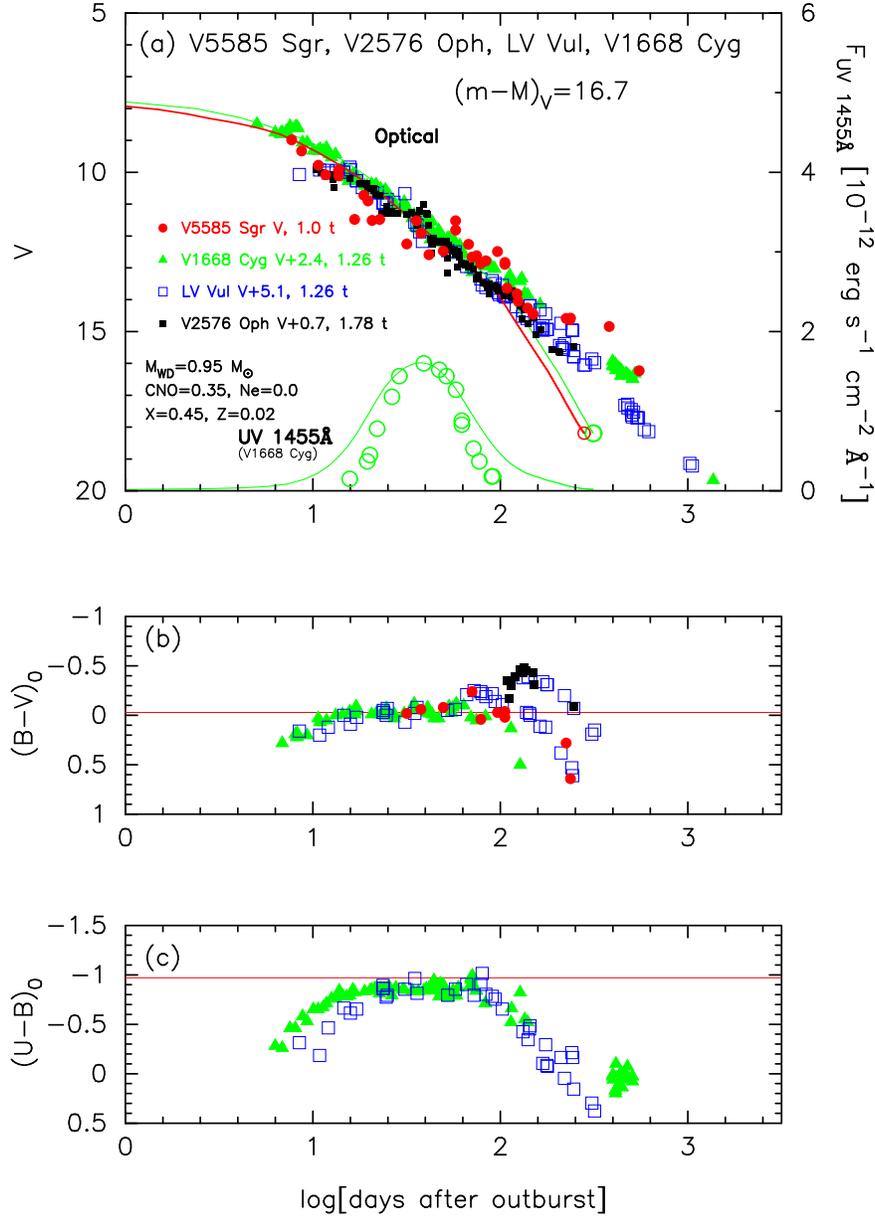}
\caption{
Same as Figure
\ref{v2575_oph_v1668_cyg_lv_vul_v_bv_ub_logscale},
but for V5585~Sgr (filled red circles).
We plot the V1668~Cyg light curves by filled green triangles,
LV~Vul by unfilled blue squares, and V2576~Oph by filled black squares.
The data of V5585~Sgr are the same as those in Figure
\ref{v5585_sgr_v_bv_ub_color_curve}.
In panel (a), we added a $V$ model light curve (solid red lines) of a
$0.95~M_\sun$ WD \citep[CO3;][]{hac16k},
assuming that $(m-M)_V=16.7$ for V5585~Sgr.
The solid green lines denote the model light curves of a $0.98~M_\sun$ WD
(CO3) for both the $V$ and UV~1455\AA\  light curves,
assuming $(m-M)_V=14.6$ for V1668~Cyg.
\label{v5585_sgr_v2576_oph_v1668_cyg_lv_vul_v_bv_ub_logscale}}
\end{figure}


\begin{figure}
\epsscale{0.55}
\plotone{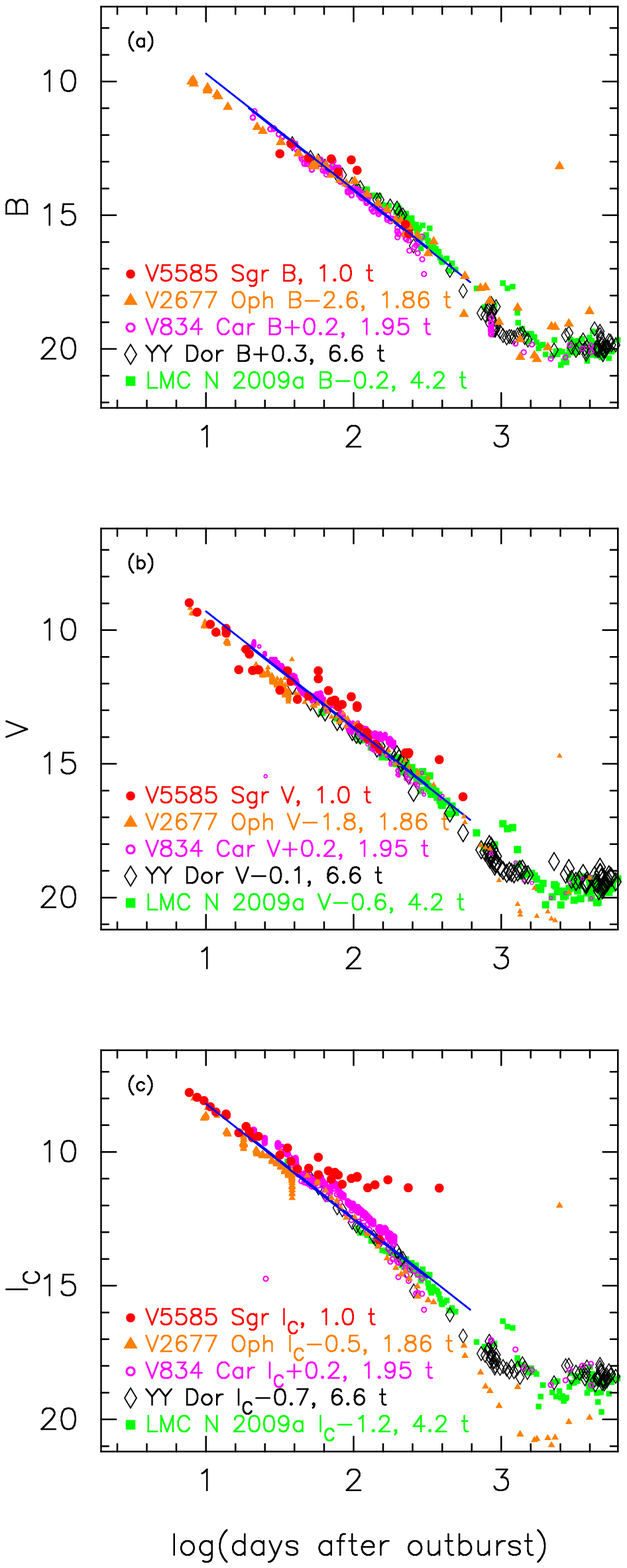}
\caption{
Same as Figure \ref{v1663_aql_yy_dor_lmcn_2009a_b_v_i_logscale_3fig},
but for V5585~Sgr.
The (a) $B$, (b) $V$, and (c) $I_{\rm C}$ light curves of V5585~Sgr
as well as those of V2677~Oph, V834~Car, YY~Dor, and LMC~N~2009a.
The $BV$ data of V5585~Sgr are the same as those in Figure
\ref{v5585_sgr_v_bv_ub_color_curve}.  The $I_{\rm C}$ data of V5585~Sgr
are taken from VSOLJ.
The $BVI_{\rm C}$ data of V2677~Oph are the same as those in Figures
\ref{v2677_oph_v_bv_ub_color_curve},
\ref{v2677_oph_v1065_cen_lv_vul_v_bv_ub_logscale}, and
\ref{v2677_oph_v834_car_yy_dor_lmcn_2009a_b_v_i_logscale_3fig}.
\label{v5585_sgr_v2677_oph_v834_car_yy_dor_lmcn_2009a_b_v_i_logscale_3fig}}
\end{figure}

\subsection{V5585~Sgr 2010}
\label{v5585_sgr}
Figure \ref{v5585_sgr_v_bv_ub_color_curve} shows the (a) $V$ and $B$,
and (b) $(B-V)_0$ evolutions of V5585~Sgr.
Here, $(B-V)_0$ are dereddened with $E(B-V)=0.47$ as obtained
in Section \ref{v5585_sgr_cmd}.
Figure \ref{v5585_sgr_v2576_oph_v1668_cyg_lv_vul_v_bv_ub_logscale} shows
the light/color curves of V5585~Sgr, LV~Vul, V1668~Cyg, and V2576~Oph.
Applying Equation (\ref{distance_modulus_general_temp}) to them,
we have the relation 
\begin{eqnarray}
(m&-&M)_{V, \rm V5585~Sgr} \cr
&=& (m - M + \Delta V)_{V, \rm LV~Vul} - 2.5 \log 1.26 \cr
&=& 11.85 + 5.1\pm0.3 - 0.25 = 16.7\pm0.3 \cr
&=& (m - M + \Delta V)_{V, \rm V1668~Cyg} - 2.5 \log 1.26 \cr
&=& 14.6 + 2.4\pm0.3 - 0.25 = 16.75\pm0.3 \cr
&=& (m - M + \Delta V)_{V, \rm V2576~Oph} - 2.5 \log 1.78 \cr
&=& 16.65 + 0.7\pm0.2 - 0.63 = 16.72\pm0.2,
\label{distance_modulus_v5585_sgr}
\end{eqnarray}
where we adopt $(m-M)_{V, \rm LV~Vul}=11.85$ and
$(m-M)_{V, \rm V1668~Cyg}=14.6$, both from \citet{hac19k},
and $(m-M)_{V, \rm V2576~Oph}=16.65$ in Appendix \ref{v2576_oph}.
Thus, we obtain $(m-M)_V=16.7\pm0.2$ and $f_{\rm s}=1.26$ against LV~Vul.
From Equations (\ref{time-stretching_general}),
(\ref{distance_modulus_general_temp}), and
(\ref{distance_modulus_v5585_sgr}),
we have the relation
\begin{eqnarray}
(m- M')_{V, \rm V5585~Sgr} 
&\equiv & (m_V - (M_V - 2.5\log f_{\rm s}))_{\rm V5585~Sgr} \cr
&=& \left( (m-M)_V + \Delta V \right)_{\rm LV~Vul} \cr
&=& 11.85 + 5.1\pm0.3 = 16.95\pm0.3.
\label{absolute_mag_v5585_sgr}
\end{eqnarray}

Figure 
\ref{v5585_sgr_v2677_oph_v834_car_yy_dor_lmcn_2009a_b_v_i_logscale_3fig}
shows the $B$, $V$, and $I_{\rm C}$ light curves of V5585~Sgr 
together with those of V2677~Oph, V834~Car, YY~Dor, and LMC~N~2009a.
We apply Equation (\ref{distance_modulus_general_temp_b})
for the $B$ band to Figure
\ref{v5585_sgr_v2677_oph_v834_car_yy_dor_lmcn_2009a_b_v_i_logscale_3fig}(a)
and obtain
\begin{eqnarray}
(m&-&M)_{B, \rm V5585~Sgr} \cr
&=& ((m - M)_B + \Delta B)_{\rm V2677~Oph} - 2.5 \log 1.86 \cr
&=& 20.5 - 2.6\pm0.2 - 0.67 = 17.23\pm0.2 \cr
&=& ((m - M)_B + \Delta B)_{\rm V834~Car} - 2.5 \log 1.95 \cr
&=& 17.75 + 0.2\pm0.2 - 0.72 = 17.23\pm0.2 \cr
&=& ((m - M)_B + \Delta B)_{\rm YY~Dor} - 2.5 \log 6.6 \cr
&=& 18.98 + 0.3\pm0.2 - 2.05 = 17.23\pm0.2 \cr
&=& ((m - M)_B + \Delta B)_{\rm LMC~N~2009a} - 2.5 \log 4.2 \cr
&=& 18.98 - 0.2\pm0.2 - 1.55 = 17.23\pm0.2,
\label{distance_modulus_b_v5585_sgr_v2677_oph_v834_car_yy_dor_lmcn2009a}
\end{eqnarray}
where we adopt $(m-M)_{B, \rm V2677~Oph}= 19.2 + 1.3= 20.5$ from 
Appendix \ref{v2677_oph} and
$(m-M)_{B, \rm V834~Car}= 17.25 + 0.50= 17.75$ from 
Appendix \ref{v834_car}.
We have $(m-M)_{B, \rm V5585~Sgr}= 17.23\pm0.1$.

For the $V$ light curves in Figure
\ref{v5585_sgr_v2677_oph_v834_car_yy_dor_lmcn_2009a_b_v_i_logscale_3fig}(b),
we similarly obtain
\begin{eqnarray}
(m&-&M)_{V, \rm V5585~Sgr} \cr   
&=& ((m - M)_V + \Delta V)_{\rm V2677~Oph} - 2.5 \log 1.86 \cr
&=& 19.2 - 1.8\pm0.2 - 0.67 = 16.73\pm0.2 \cr
&=& ((m - M)_V + \Delta V)_{\rm V834~Car} - 2.5 \log 1.95 \cr
&=& 17.25 + 0.2\pm0.2 - 0.72 = 16.73\pm0.2 \cr
&=& ((m - M)_V + \Delta V)_{\rm YY~Dor} - 2.5 \log 6.6 \cr
&=& 18.86 - 0.1\pm0.2 - 2.05 = 16.71\pm0.2 \cr
&=& ((m - M)_V + \Delta V)_{\rm LMC~N~2009a} - 2.5 \log 4.2 \cr
&=& 18.86 - 0.6\pm0.2 - 1.55 = 16.71\pm0.2.
\label{distance_modulus_v_v5585_sgr_v2677_oph_v834_car_yy_dor_lmcn2009a}
\end{eqnarray}
We have $(m-M)_{V, \rm V5585~Sgr}= 16.72\pm0.1$, which is
consistent with Equation (\ref{distance_modulus_v5585_sgr}).

We apply Equation (\ref{distance_modulus_general_temp_i}) for
the $I_{\rm C}$ band to Figure
\ref{v5585_sgr_v2677_oph_v834_car_yy_dor_lmcn_2009a_b_v_i_logscale_3fig}(c)
and obtain
\begin{eqnarray}
(m&-&M)_{I, \rm V5585~Sgr} \cr
&=& ((m - M)_I + \Delta I_C)_{\rm V2677~Oph} - 2.5 \log 1.86 \cr
&=& 17.12 - 0.5\pm0.2 - 0.67 = 15.95\pm 0.2 \cr
&=& ((m - M)_I + \Delta I_C)_{\rm V834~Car} - 2.5 \log 1.95 \cr
&=& 16.45 + 0.2\pm0.2 - 0.72 = 15.93\pm 0.2 \cr
&=& ((m - M)_I + \Delta I_C)_{\rm YY~Dor} - 2.5 \log 6.6 \cr
&=& 18.67 - 0.7\pm0.2 - 2.05 = 15.92\pm 0.2 \cr
&=& ((m - M)_I + \Delta I_C)_{\rm LMC~N~2009a} - 2.5 \log 4.2 \cr
&=& 18.67 - 1.2\pm0.2 - 1.55 = 15.92\pm 0.2,
\label{distance_modulus_i_v5585_sgr_v2677_oph_v834_car_yy_dor_lmcn2009a}
\end{eqnarray}
where we adopt $(m-M)_{I, \rm V2677~Oph}= 19.2 - 1.6\times 1.3= 17.12$
from Appendix \ref{v2677_oph} and
$(m-M)_{I, \rm V834~Car}= 17.25 - 1.6\times 0.50= 16.45$ from 
Appendix \ref{v834_car}.
We have $(m-M)_{I, \rm V5585~Sgr}= 15.93\pm0.1$.

We plot $(m-M)_B= 17.23$, $(m-M)_V= 16.72$, and $(m-M)_I= 15.93$,
which broadly cross at $d=11$~kpc and $E(B-V)=0.47$, in Figure
\ref{distance_reddening_v5585_sgr_pr_lup_v1313_sco_v834_car}(a).
Thus, we have $E(B-V)=0.47\pm0.05$ and $d=11\pm2$~kpc.


\begin{figure}
\epsscale{0.75}
\plotone{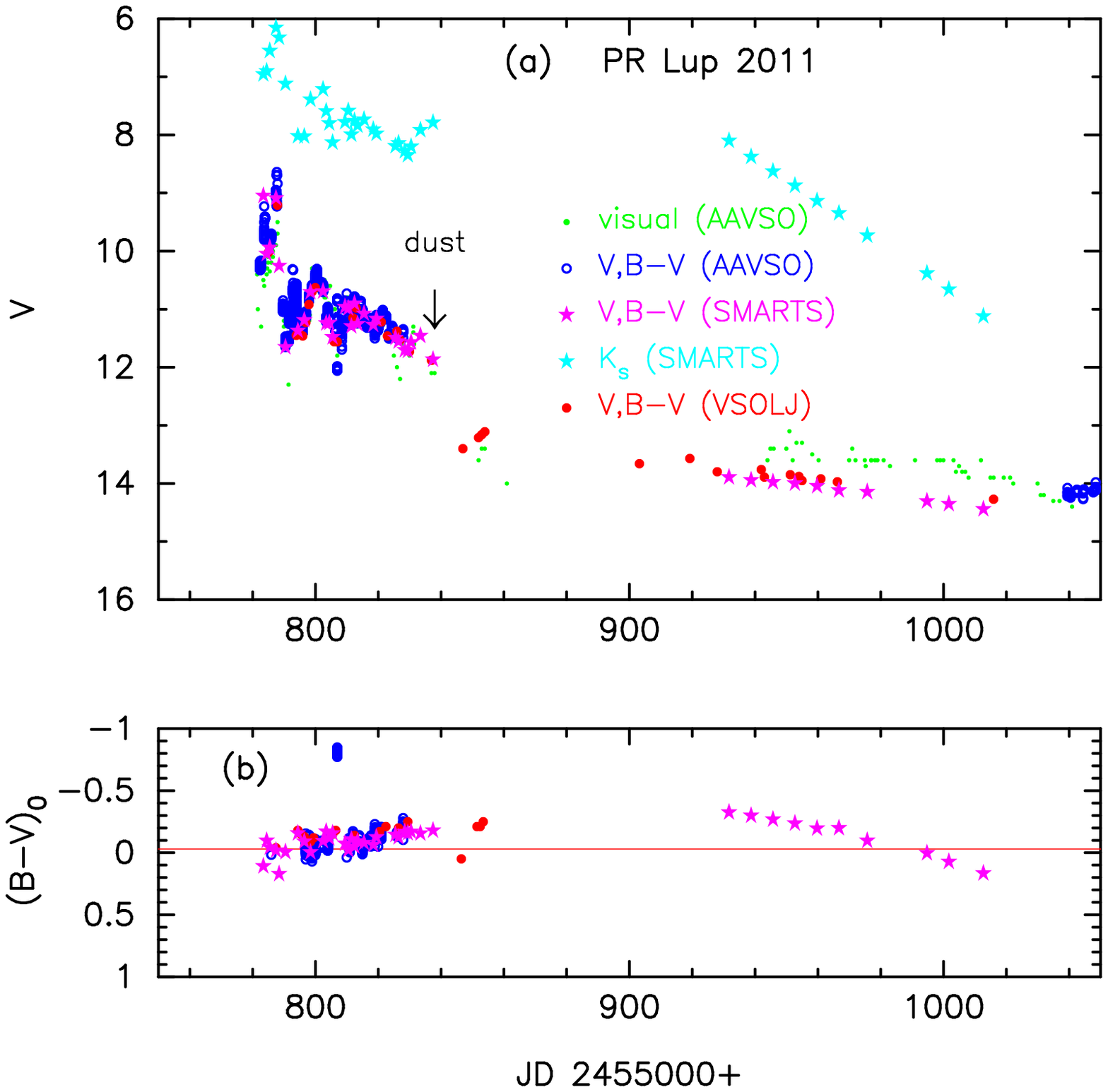}
\caption{
Same as Figure \ref{v1663_aql_v_bv_ub_color_curve}, but for PR~Lup.
(a) The visual data (green dots) are taken from AAVSO.
The $BV$ data are taken from AAVSO (unfilled blue circles),
SMARTS (filled magenta stars), and VSOLJ (filled red circles).
We add the $K_{\rm s}$ magnitudes (filled cyan stars)
which are taken from SMARTS.
A dust shell formed around the time indicated by the arrow labeled dust.
(b) The $(B-V)_0$ are dereddened with $E(B-V)=0.74$.
\label{pr_lup_v_bv_ub_color_curve}}
\end{figure}


\begin{figure}
\epsscale{0.75}
\plotone{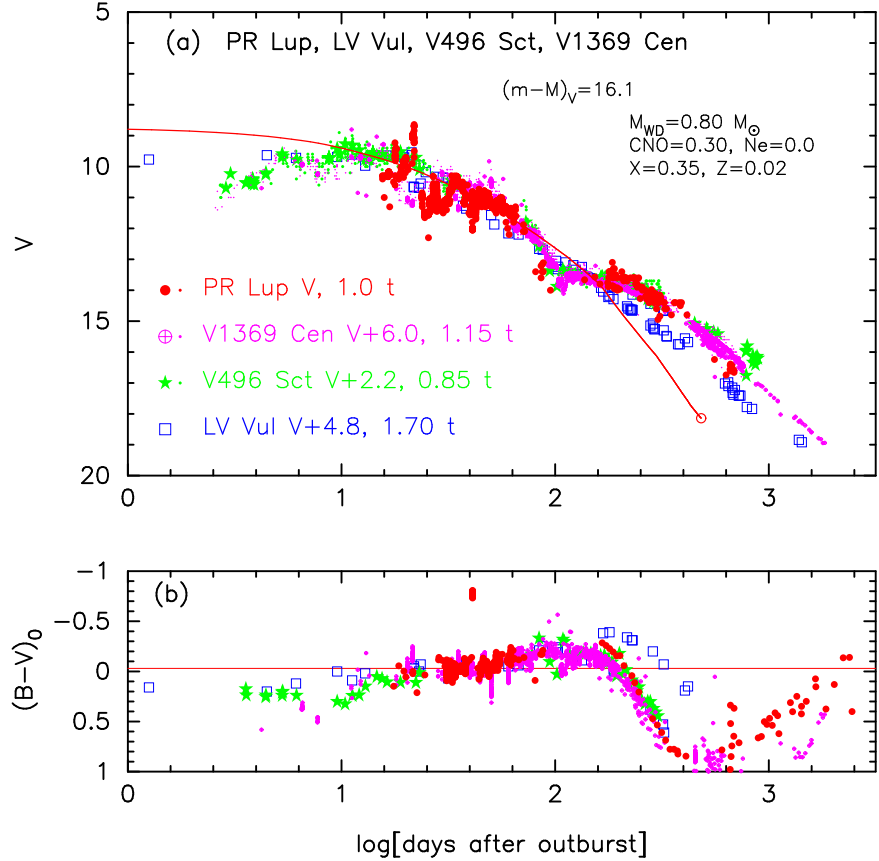}
\caption{
Same as Figure
\ref{v2575_oph_v1668_cyg_lv_vul_v_bv_ub_logscale},
but for PR~Lup (filled red circles for $V$ and red dots for visual).
We add the V1369~Cen light curves with the unfilled magenta encircled plus
($V$) and by magenta dots (visual), V496~Sct by filled green stars ($V$)
and by green dots (visual), and LV~Vul by unfilled blue squares ($V$).
The data of PR~Lup are the same as
those in Figure \ref{pr_lup_v_bv_ub_color_curve}.
The data of V1369~Cen and V496~Sct are the same as those 
in Figure 15 of \citet{hac19k}.  In panel (a),
we add the model light curve of a $0.80~M_\sun$ WD 
\citep[CO2, solid red line;][]{hac10k},
assuming that $(m-M)_V=16.1$ for PR~Lup. 
\label{pr_lup_v1369_cen_v496_sct_v_bv_ub_color_logscale}}
\end{figure}


\begin{figure}
\epsscale{0.65}
\plotone{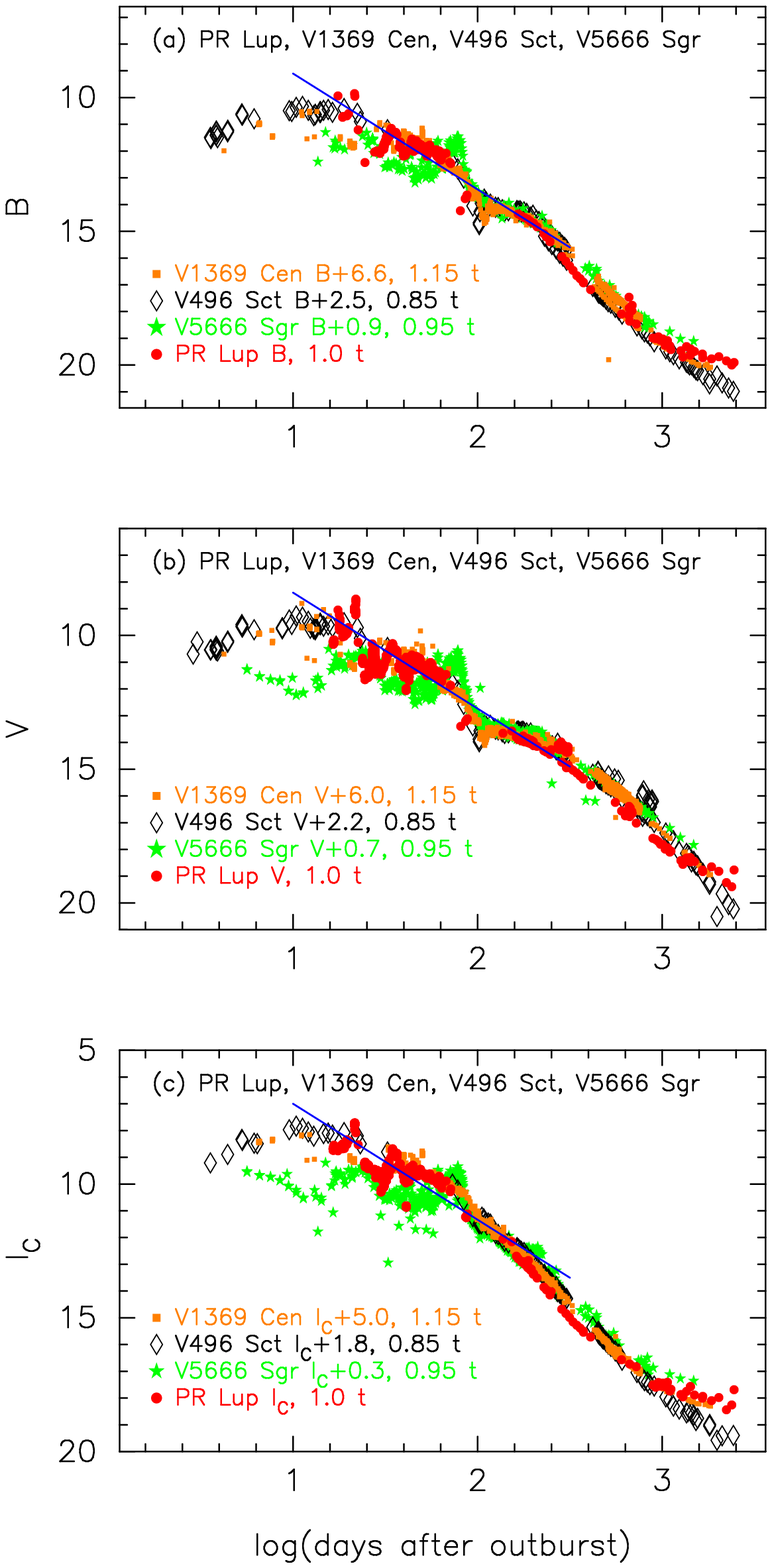}
\caption{
Same as Figure \ref{v1663_aql_yy_dor_lmcn_2009a_b_v_i_logscale_3fig},
but for PR~Lup.   We plot
the (a) $B$, (b) $V$, and (c) $I_{\rm C}$ light curves of PR~Lup
as well as those of V1369~Cen, V496~Sct, and V5666~Sgr.
The $BV$ data of PR~Lup are the same as those in Figure
\ref{pr_lup_v_bv_ub_color_curve}.  The $I_{\rm C}$ data of PR~Lup
are taken from AAVSO, VSOLJ, and SMARTS.
\label{pr_lup_v1369_cen_v496_sct_v5666_sgr_b_v_i_logscale_3fig}}
\end{figure}

\subsection{PR~Lup 2011}
\label{pr_lup}
Figure \ref{pr_lup_v_bv_ub_color_curve} shows the (a) visual, $V$, 
and $K_{\rm s}$, and (b) $(B-V)_0$ evolutions of PR~Lup.  Here, $(B-V)_0$
are dereddened with $E(B-V)=0.74$ as obtained in Section \ref{pr_lup_cmd}.
Figure \ref{pr_lup_v1369_cen_v496_sct_v_bv_ub_color_logscale} shows
the light/color curves of PR~Lup, LV~Vul, V496~Sct, and V1369~Cen.
Here we assume that PR~Lup outbursted on JD~2,455,766.0 (Day 0).
These $V$ light and $(B-V)_0$ color curves overlap each other. 
Applying Equation (\ref{distance_modulus_general_temp}) to them,
we have the relation 
\begin{eqnarray}
(m&-&M)_{V, \rm PR~Lup} \cr
&=& (m-M + \Delta V)_{V, \rm LV~Vul} - 2.5 \log 1.70 \cr
&=& 11.85 + 4.8\pm0.2 - 0.58 = 16.07\pm0.2 \cr
&=& (m-M + \Delta V)_{V, \rm V496~Sct} - 2.5 \log 0.85 \cr
&=& 13.7 + 2.2\pm0.2 + 0.18 = 16.08\pm0.2  \cr
&=& (m-M + \Delta V)_{V, \rm V1369~Cen} - 2.5 \log 1.15 \cr
&=& 10.25 + 6.0\pm0.2 - 0.15 = 16.1\pm0.2,
\label{distance_modulus_pr_lup}
\end{eqnarray}
where we adopt $(m-M)_{V, \rm LV~Vul}=11.85$,
$(m-M)_{V, \rm V496~Sct}=13.7$, and 
$(m-M)_{V, \rm V1369~Cen}=10.25$, all from \citet{hac19k}.
Thus, we obtain $(m-M)_V=16.1\pm0.1$ and $f_{\rm s}=1.70$ against LV~Vul.
From Equations (\ref{time-stretching_general}),
(\ref{distance_modulus_general_temp}), and
(\ref{distance_modulus_pr_lup}),
we have the relation
\begin{eqnarray}
(m- M')_{V, \rm PR~Lup} 
&\equiv & (m_V - (M_V - 2.5\log f_{\rm s}))_{\rm PR~Lup} \cr
&=& \left( (m-M)_V + \Delta V \right)_{\rm LV~Vul} \cr
&=& 11.85 + 4.8\pm0.2 = 16.65\pm0.2.
\label{absolute_mag_pr_lup}
\end{eqnarray}

Figure \ref{pr_lup_v1369_cen_v496_sct_v5666_sgr_b_v_i_logscale_3fig}
shows the $B$, $V$, and $I_{\rm C}$ light curves of PR~Lup
together with those of V1369~Cen, V496~Sct, and V5666~Sgr.
The light curves overlap each other well.
Applying Equation (\ref{distance_modulus_general_temp_b})
for the $B$ band to Figure
\ref{pr_lup_v1369_cen_v496_sct_v5666_sgr_b_v_i_logscale_3fig}(a),
we have the relation
\begin{eqnarray}
(m&-&M)_{B, \rm PR~Lup} \cr
&=& \left( (m-M)_B + \Delta B\right)_{\rm V1369~Cen} - 2.5 \log 1.15 \cr
&=& 10.36 + 6.6\pm0.2 - 0.15 = 16.81\pm0.2 \cr
&=& \left( (m-M)_B + \Delta B\right)_{\rm V496~Sct} - 2.5 \log 0.85 \cr
&=& 14.15 + 2.5\pm0.2 + 0.18 = 16.83\pm0.2 \cr
&=& \left( (m-M)_B + \Delta B\right)_{\rm V5666~Sgr} - 2.5 \log 0.95 \cr
&=& 15.9 + 0.9\pm0.2 + 0.05 = 16.88\pm0.2,
\label{distance_modulus_pr_lup_v1369_cen_v496_sct_v5666_sgr_b}
\end{eqnarray}
where we adopt $(m-M)_{B, \rm V1369~Cen}= 10.36$,
$(m-M)_{B, \rm V496~Sct}= 14.15$, and
$(m-M)_{B, \rm V5666~Sgr}= 15.9$ from Appendix \ref{qy_mus}.
We have $(m-M)_B=16.84\pm0.1$ for PR~Lup.

Applying Equation (\ref{distance_modulus_general_temp}) to
Figure \ref{pr_lup_v1369_cen_v496_sct_v5666_sgr_b_v_i_logscale_3fig}(b),
we have the relation
\begin{eqnarray}
(m&-&M)_{V, \rm PR~Lup} \cr
&=& \left( (m-M)_V + \Delta V\right)_{\rm V1369~Cen} - 2.5 \log 1.15 \cr
&=& 10.25 + 6.0\pm0.3 - 0.15 = 16.1\pm0.2 \cr
&=& \left( (m-M)_V + \Delta V\right)_{\rm V496~Sct} - 2.5 \log 0.85 \cr
&=& 13.7 + 2.2\pm0.3 + 0.18 = 16.08\pm0.2 \cr
&=& \left( (m-M)_V + \Delta V\right)_{\rm V5666~Sgr} - 2.5 \log 0.95 \cr
&=& 15.4 + 0.7\pm0.3 + 0.05 = 16.15\pm0.2,
\label{distance_modulus_pr_lup_v1369_cen_v496_sct_v5666_sgr_v}
\end{eqnarray}
where we adopt $(m-M)_{V, \rm V1369~Cen}=10.25$,
$(m-M)_{V, \rm V496~Sct}=13.7$, and $(m-M)_{V, \rm V5666~Sgr}=15.4$
from \citet{hac19k}.  We have $(m-M)_V=16.1\pm0.1$, which is
consistent with Equation (\ref{distance_modulus_pr_lup}).

From the $I_{\rm C}$-band data in Figure
\ref{pr_lup_v1369_cen_v496_sct_v5666_sgr_b_v_i_logscale_3fig}(c),
we obtain
\begin{eqnarray}
(m&-&M)_{I, \rm PR~Lup} \cr
&=& ((m - M)_I + \Delta I_C)_{\rm V1369~Cen} - 2.5 \log 1.15 \cr
&=& 10.07 + 5.0\pm0.2 - 0.15 = 14.92\pm0.2 \cr
&=& ((m - M)_I + \Delta I_C)_{\rm V496~Sct} - 2.5 \log 0.85 \cr
&=& 12.98 + 1.8\pm0.2 + 0.18 = 14.96\pm0.2 \cr
&=& ((m - M)_I + \Delta I_C)_{\rm V5666~Sgr} - 2.5 \log 0.95 \cr
&=& 14.6 + 0.3\pm0.2 + 0.05 = 14.95\pm0.2,
\label{distance_modulus_i_pr_lup_v1369_cen_v496_sct_v5666_sgr}
\end{eqnarray}
where we adopt $(m-M)_{I, \rm V1369~Cen}= 10.07$,
$(m-M)_{I, \rm V496~Sct}= 12.98$, and $(m-M)_{I, \rm V5666~Sgr}= 14.6$
from Appendix \ref{qy_mus}.  We have $(m-M)_I= 14.94\pm0.1$.

We plot $(m-M)_B=16.84$, $(m-M)_V=16.1$, $(m-M)_I=14.94$,
which cross at $d=5.8$~kpc and $E(B-V)=0.74$,
in Figure \ref{distance_reddening_v5585_sgr_pr_lup_v1313_sco_v834_car}(b).
Thus, we obtain $d=5.8\pm0.6$~kpc and $E(B-V)=0.74\pm0.05$.


\begin{figure}
\epsscale{0.75}
\plotone{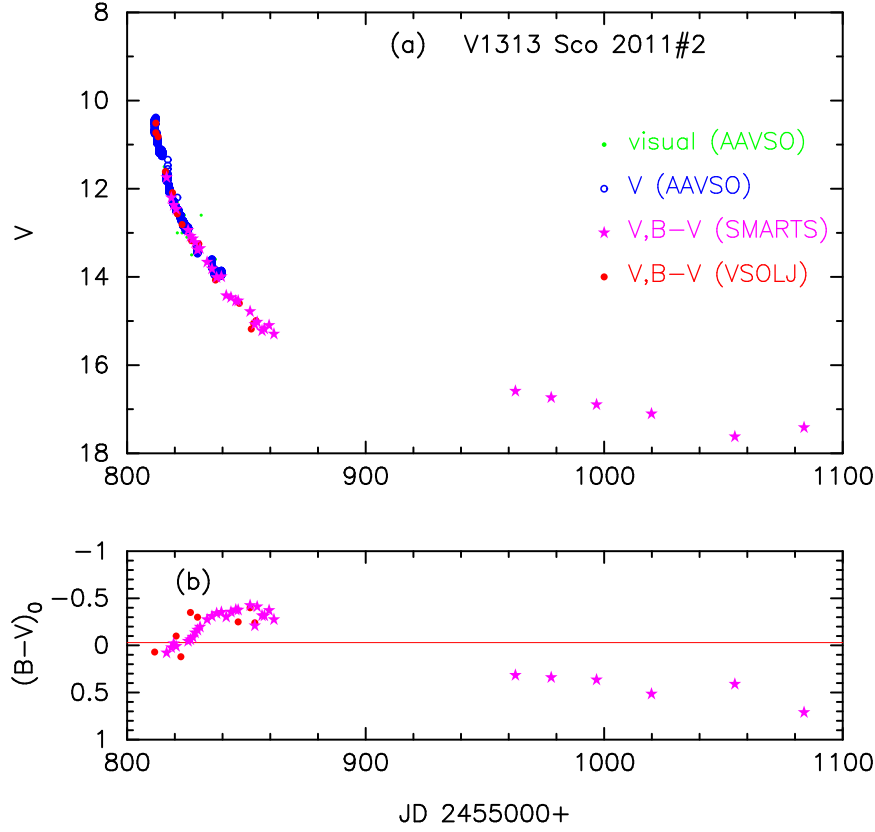}
\caption{
Same as Figure \ref{v1663_aql_v_bv_ub_color_curve}, but for V1313~Sco.
(a) The visual data (green dots) are taken from AAVSO.
The $V$ data are taken from AAVSO (unfilled blue circles),
SMARTS (filled magenta stars), and VSOLJ (filled red circles).
(b) The $(B-V)_0$ are dereddened with $E(B-V)=1.30$.
\label{v1313_sco_v_bv_ub_color_curve}}
\end{figure}


\begin{figure}
\epsscale{0.75}
\plotone{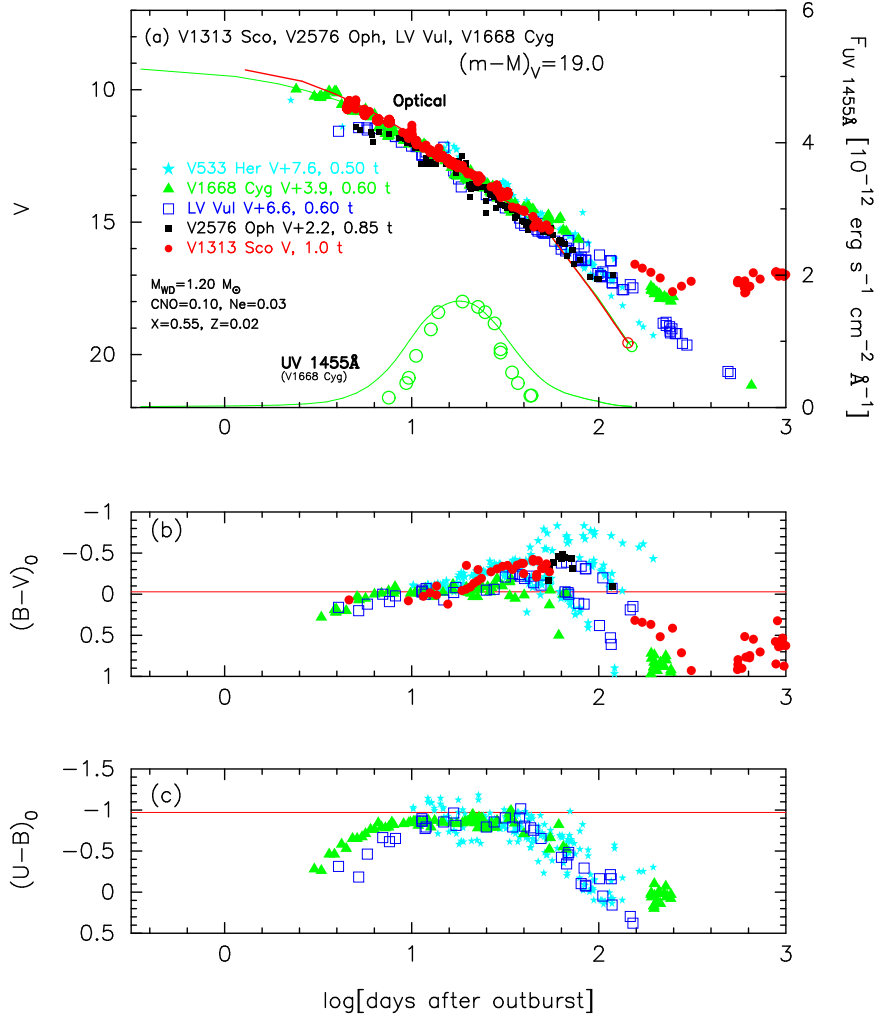}
\caption{
Same as Figure
\ref{v2575_oph_v1668_cyg_lv_vul_v_bv_ub_logscale},
but for V1313~Sco (filled red circles).
The data of V1313~Sco are the same as those in Figure
\ref{v1313_sco_v_bv_ub_color_curve}.
In panel (a), we add the model $V$ light curve (solid red line) of a
$1.20~M_\sun$ WD \citep[Ne2;][]{hac10k},
assuming that $(m-M)_V=19.0$ for V1313~Sco.
The solid green lines denote the model $V$ and UV~1455\AA\  light curves of
a $0.98~M_\sun$ WD (CO3),
assuming $(m-M)_V=14.6$ for V1668~Cyg.
\label{v1313_sco_v533_her_v2576_oph_v1668_cyg_lv_vul_v_bv_ub_logscale}}
\end{figure}


\begin{figure}
\epsscale{0.55}
\plotone{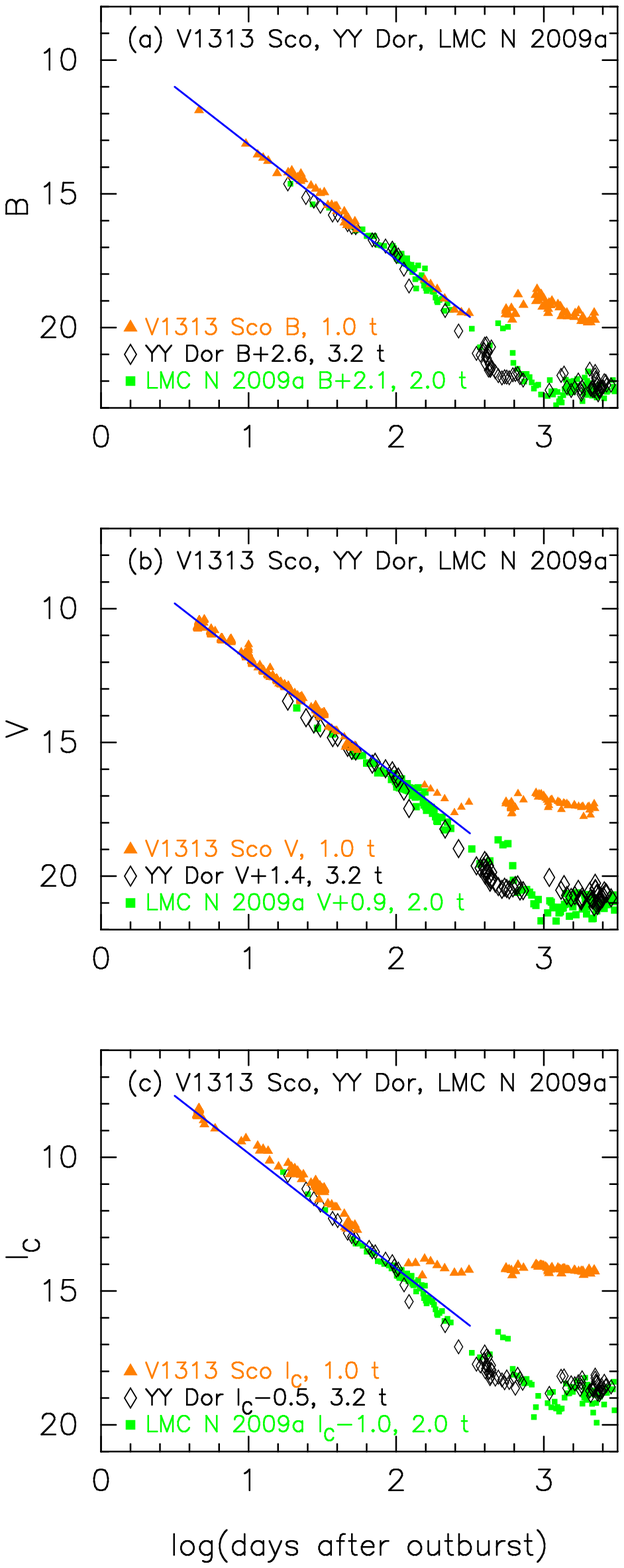}
\caption{
Same as Figure \ref{v1663_aql_yy_dor_lmcn_2009a_b_v_i_logscale_3fig},
but for V1313~Sco.
The $BV$ data of V1313~Sco are the same as those in Figure
\ref{v1313_sco_v_bv_ub_color_curve}.  The $I_{\rm C}$ data of V1313~Sco
are taken from AAVSO, VSOLJ, and SMARTS.
\label{v1313_sco_yy_dor_lmcn_2009a_b_v_i_logscale_3fig}}
\end{figure}

\subsection{V1313~Sco 2011\#2}
\label{v1313_sco}
Figure \ref{v1313_sco_v_bv_ub_color_curve} shows (a) the visual and $V$,
and (b) $(B-V)_0$ evolutions of V1313~Sco.  Here, $(B-V)_0$ are dereddened
with $E(B-V)=1.30$ as obtained in Section \ref{v1313_sco_cmd}.  
Figure \ref{v1313_sco_v533_her_v2576_oph_v1668_cyg_lv_vul_v_bv_ub_logscale}
shows the light/color curves of V1313~Sco, LV~Vul, V1668~Cyg, V533~Her, and
V2676~Oph.  
Applying Equation (\ref{distance_modulus_general_temp}) to them,
we have the relation 
\begin{eqnarray}
(m&-&M)_{V, \rm V1313~Sco} \cr
&=& (m - M + \Delta V)_{V, \rm LV~Vul} - 2.5 \log 0.60 \cr
&=& 11.85 + 6.6\pm0.2 + 0.55 = 19.0\pm0.2 \cr
&=& (m - M + \Delta V)_{V, \rm V1668~Cyg} - 2.5 \log 0.60 \cr
&=& 14.6 + 3.9\pm0.2 + 0.55 = 19.05\pm0.2 \cr
&=& (m - M + \Delta V)_{V, \rm V533~Her} - 2.5 \log 0.50 \cr
&=& 10.65 + 7.6\pm0.2 + 0.75 = 19.0\pm0.2 \cr
&=& (m - M + \Delta V)_{V, \rm V2576~Oph} - 2.5 \log 0.85 \cr
&=& 16.65 + 2.2\pm0.2 + 0.18 = 19.03\pm0.2,
\label{distance_modulus_v1313_sco}
\end{eqnarray}
where we adopt $(m-M)_{V, \rm LV~Vul}=11.85$,
$(m-M)_{V, \rm V1668~Cyg}=14.6$, and
$(m-M)_{V, \rm V533~Her}=10.65$ from \citet{hac19k},
and $(m-M)_{V, \rm V2576~Oph}=16.65$ in Appendix \ref{v2576_oph}.
Thus, we obtain $(m-M)_V=19.0\pm0.1$ and $f_{\rm s}=0.60$ against LV~Vul.
From Equations (\ref{time-stretching_general}),
(\ref{distance_modulus_general_temp}), and
(\ref{distance_modulus_v1313_sco}),
we have the relation
\begin{eqnarray}
(m- M')_{V, \rm V1313~Sco} 
&\equiv & (m_V - (M_V - 2.5\log f_{\rm s}))_{\rm V1313~Sco} \cr
&=& \left( (m-M)_V + \Delta V \right)_{\rm LV~Vul} \cr
&=& 11.85 + 6.6\pm0.2 = 18.45\pm0.2.
\label{absolute_mag_v1313_sco}
\end{eqnarray}

Figure \ref{v1313_sco_yy_dor_lmcn_2009a_b_v_i_logscale_3fig} shows
the $B$, $V$, and $I_{\rm C}$ light curves of V1313~Sco
together with those of YY~Dor and LMC~N~2009a.
We regard the extension of the YY~Dor and LMC~N~2009a light curves
to overlap with that of V1313~Sco.
We apply Equation (\ref{distance_modulus_general_temp_b})
for the $B$ band to Figure
\ref{v1313_sco_yy_dor_lmcn_2009a_b_v_i_logscale_3fig}(a)
and obtain
\begin{eqnarray}
(m&-&M)_{B, \rm V1313~Sco} \cr
&=& ((m - M)_B + \Delta B)_{\rm YY~Dor} - 2.5 \log 3.2 \cr
&=& 18.98 + 2.6\pm0.2 - 1.25 = 20.33\pm0.2 \cr
&=& ((m - M)_B + \Delta B)_{\rm LMC~N~2009a} - 2.5 \log 2.0 \cr
&=& 18.98 + 2.1\pm0.2 - 0.75 = 20.33\pm0.2.
\label{distance_modulus_b_v1313_sco_yy_dor_lmcn2009a}
\end{eqnarray}
We have $(m-M)_{B, \rm V1313~Sco}= 20.33\pm0.1$.

For the $V$ light curves in Figure
\ref{v1313_sco_yy_dor_lmcn_2009a_b_v_i_logscale_3fig}(b),
we similarly obtain
\begin{eqnarray}
(m&-&M)_{V, \rm V1313~Sco} \cr
&=& ((m - M)_V + \Delta V)_{\rm YY~Dor} - 2.5 \log 3.2 \cr
&=& 18.86 + 1.4\pm0.2 - 1.25 = 19.01\pm0.2 \cr
&=& ((m - M)_V + \Delta V)_{\rm LMC~N~2009a} - 2.5 \log 2.0 \cr
&=& 18.86 + 0.9\pm0.2 - 0.75 = 19.01\pm0.2.
\label{distance_modulus_v_v1313_sco_yy_dor_lmcn2009a}
\end{eqnarray}
We have $(m-M)_{V, \rm V1313~Sco}= 19.01\pm0.1$, which is
consistent with Equation (\ref{distance_modulus_v1313_sco}).

We apply Equation (\ref{distance_modulus_general_temp_i}) for
the $I_{\rm C}$ band to Figure
\ref{v1313_sco_yy_dor_lmcn_2009a_b_v_i_logscale_3fig}(c) and obtain
\begin{eqnarray}
(m&-&M)_{I, \rm V1313~Sco} \cr
&=& ((m - M)_I + \Delta I_C)_{\rm YY~Dor} - 2.5 \log 3.2 \cr
&=& 18.67 - 0.5\pm0.2 - 1.25 = 16.92\pm 0.2 \cr
&=& ((m - M)_I + \Delta I_C)_{\rm LMC~N~2009a} - 2.5 \log 2.0 \cr
&=& 18.67 - 1.0\pm0.2 - 0.75 = 16.92\pm 0.2.
\label{distance_modulus_i_v1313_sco_yy_dor_lmcn2009a}
\end{eqnarray}
We have $(m-M)_{I, \rm V1313~Sco}= 16.92\pm0.1$.

We plot $(m-M)_B= 20.33$, $(m-M)_V= 19.01$, and $(m-M)_I= 16.92$,
which broadly cross at $d=9.9$~kpc and $E(B-V)=1.30$, in Figure
\ref{distance_reddening_v5585_sgr_pr_lup_v1313_sco_v834_car}(c).
Thus, we obtain $E(B-V)=1.30\pm0.1$ and $d=9.9\pm2$~kpc.


\begin{figure}
\epsscale{0.75}
\plotone{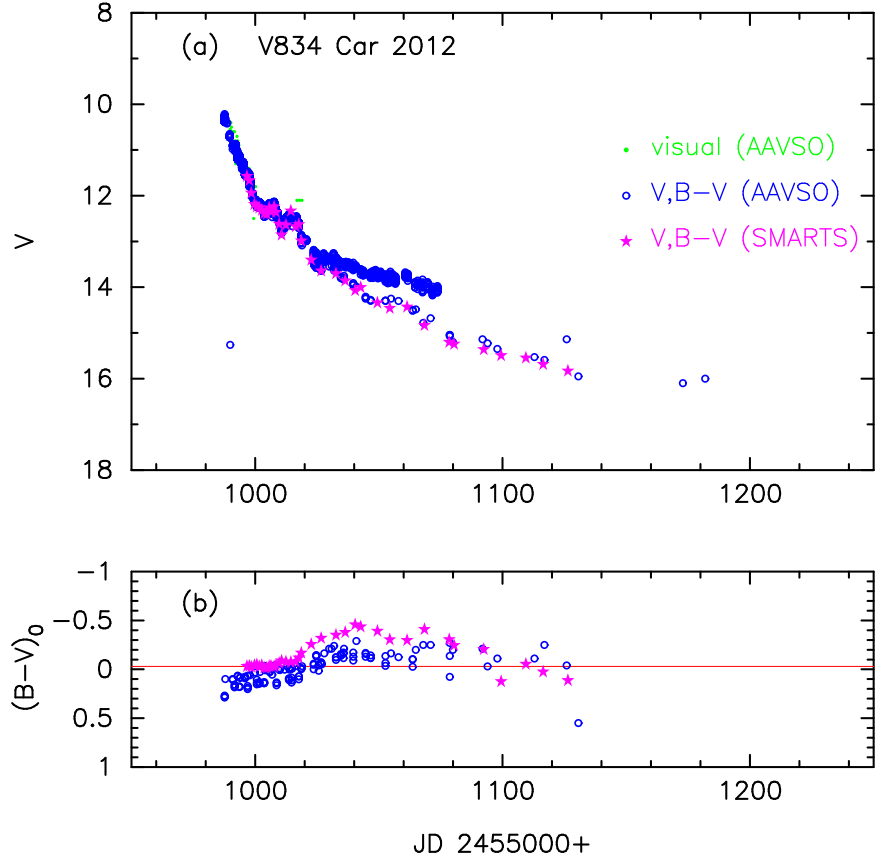}
\caption{
Same as Figure \ref{v1663_aql_v_bv_ub_color_curve}, but for V834~Car.
(a) The visual data (green dots) are taken from AAVSO.
The $V$ data are taken from AAVSO (unfilled blue circles) and
SMARTS (filled magenta stars).
(b) The $(B-V)_0$ are dereddened with $E(B-V)=0.50$.
\label{v834_car_v_bv_ub_color_curve}}
\end{figure}


\begin{figure}
\epsscale{0.75}
\plotone{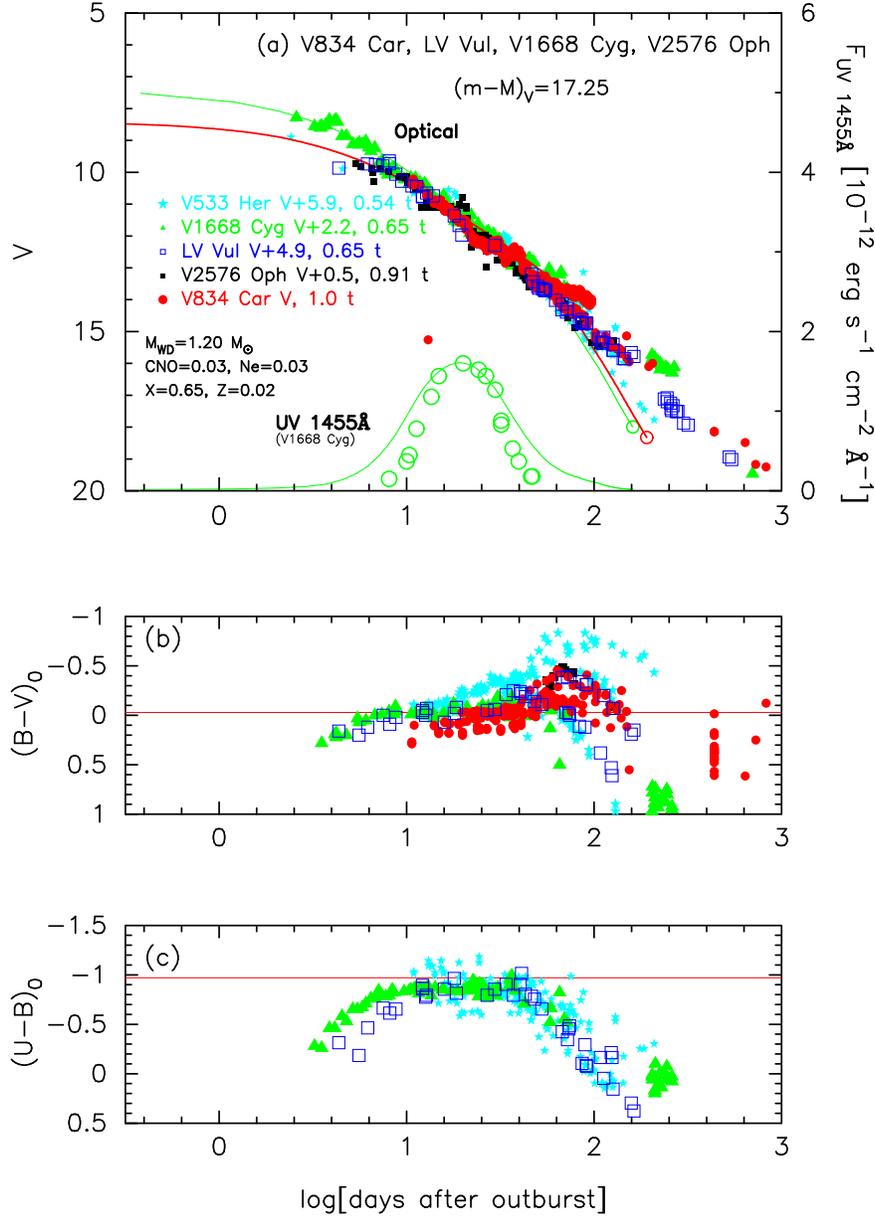}
\caption{
Same as Figure
\ref{v2575_oph_v1668_cyg_lv_vul_v_bv_ub_logscale},
but for V834~Car (filled red circles).
The data of V834~Car are the same as those in Figure
\ref{v834_car_v_bv_ub_color_curve}.
In panel (a), we added the model $V$ light
curve (solid red line) of a $1.20~M_\sun$ WD \citep[Ne3;][]{hac16k},
assuming that $(m-M)_V=17.25$ for V834~Car.
The solid green lines denote the model light curves of
a $0.98~M_\sun$ WD (CO3)
both for the $V$ and UV~1455\AA\   light curves,
assuming $(m-M)_V=14.6$ for V1668~Cyg.
\label{v834_car_v2576_oph_v1668_cyg_lv_vul_v_bv_ub_logscale}}
\end{figure}


\begin{figure}
\epsscale{0.55}
\plotone{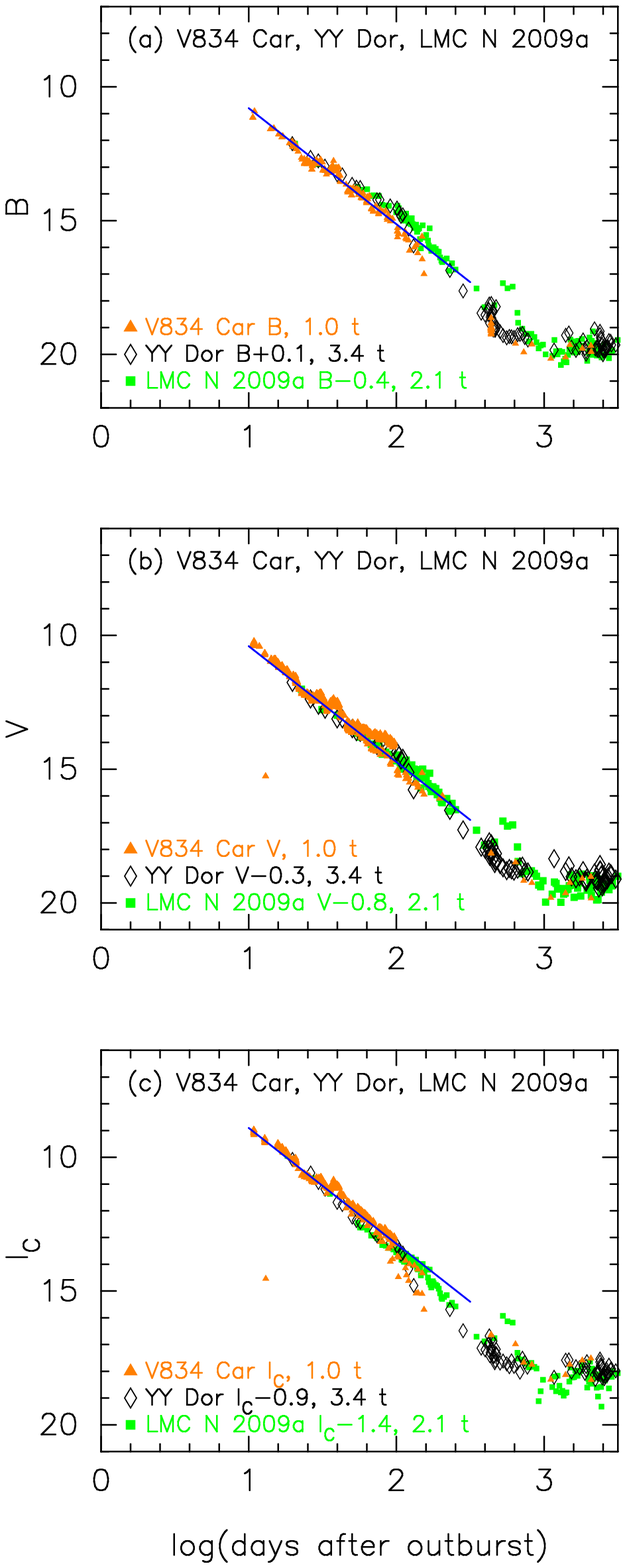}
\caption{
Same as Figure \ref{v1663_aql_yy_dor_lmcn_2009a_b_v_i_logscale_3fig},
but for V834~Car.
The $BV$ data of V834~Car are the same as those in Figure
\ref{v834_car_v_bv_ub_color_curve}.  The $I_{\rm C}$ data of V834~Car
are taken from AAVSO and SMARTS.
\label{v834_car_yy_dor_lmcn_2009a_b_v_i_logscale_3fig}}
\end{figure}

\subsection{V834~Car 2012}
\label{v834_car}
Figure \ref{v834_car_v_bv_ub_color_curve} shows the (a) $V$ and visual, and
(b) $(B-V)_0$ evolutions of V834~Car.  Here, $(B-V)_0$ are dereddened
with $E(B-V)=0.50$ as obtained in Section \ref{v834_car_cmd}.  
Figure \ref{v834_car_v2576_oph_v1668_cyg_lv_vul_v_bv_ub_logscale} shows
the light/color curves of V834~Car, LV~Vul, V1668~Cyg, V533~Her, and
V2576~Oph.  
Applying Equation (\ref{distance_modulus_general_temp}) to them,
we have the relation 
\begin{eqnarray}
(m&-&M)_{V, \rm V834~Car} \cr
&=& (m - M + \Delta V)_{V, \rm LV~Vul} - 2.5 \log 0.65 \cr
&=& 11.85 + 4.9\pm0.2 + 0.48 = 17.23\pm0.2 \cr
&=& (m - M + \Delta V)_{V, \rm V1668~Cyg} - 2.5 \log 0.65 \cr
&=& 14.6 + 2.2\pm0.2 + 0.48 = 17.28\pm0.2 \cr
&=& (m - M + \Delta V)_{V, \rm V533~Her} - 2.5 \log 0.54 \cr
&=& 10.65 + 5.9\pm0.2 + 0.68 = 17.23\pm0.2 \cr
&=& (m - M + \Delta V)_{V, \rm V2576~Oph} - 2.5 \log 0.91 \cr
&=& 16.65 + 0.5\pm0.2 + 0.10 = 17.25\pm0.2,
\label{distance_modulus_v834_car}
\end{eqnarray}
where we adopt $(m-M)_{V, \rm LV~Vul}=11.85$,
$(m-M)_{V, \rm V1668~Cyg}=14.6$, and
$(m-M)_{V, \rm V533~Her}=10.65$ from \citet{hac19k},
and $(m-M)_{V, \rm V2576~Oph}=16.65$ in Appendix \ref{v2576_oph},.
Thus, we obtain $(m-M)_V=17.25\pm0.1$ and $f_{\rm s}=0.65$ against LV~Vul.
From Equations (\ref{time-stretching_general}),
(\ref{distance_modulus_general_temp}), and
(\ref{distance_modulus_v834_car}),
we have the relation
\begin{eqnarray}
(m- M')_{V, \rm V834~Car} 
&\equiv & (m_V - (M_V - 2.5\log f_{\rm s}))_{\rm V834~Car} \cr
&=& \left( (m-M)_V + \Delta V \right)_{\rm LV~Vul} \cr
&=& 11.85 + 4.9\pm0.2 = 16.75\pm0.2.
\label{absolute_mag_v834_car}
\end{eqnarray}

Figure \ref{v834_car_yy_dor_lmcn_2009a_b_v_i_logscale_3fig} shows
the $B$, $V$, and $I_{\rm C}$ light curves of V834~Car
together with those of YY~Dor and LMC~N~2009a.
We regard the extension of the YY~Dor and LMC~N~2009a light curves
to overlap with that of V834~Car.
We apply Equation (\ref{distance_modulus_general_temp_b})
for the $B$ band to Figure
\ref{v834_car_yy_dor_lmcn_2009a_b_v_i_logscale_3fig}(a)
and obtain
\begin{eqnarray}
(m&-&M)_{B, \rm V834~Car} \cr
&=& ((m - M)_B + \Delta B)_{\rm YY~Dor} - 2.5 \log 3.4 \cr
&=& 18.98 + 0.1\pm0.2 - 1.33 = 17.75\pm0.2 \cr
&=& ((m - M)_B + \Delta B)_{\rm LMC~N~2009a} - 2.5 \log 2.1 \cr
&=& 18.98 - 0.4\pm0.2 - 0.83 = 17.75\pm0.2.
\label{distance_modulus_b_v834_car_yy_dor_lmcn2009a}
\end{eqnarray}
We have $(m-M)_{B, \rm V834~Car}= 17.75\pm0.1$.

For the $V$ light curves in Figure
\ref{v834_car_yy_dor_lmcn_2009a_b_v_i_logscale_3fig}(b),
we similarly obtain
\begin{eqnarray}
(m&-&M)_{V, \rm V834~Car} \cr   
&=& ((m - M)_V + \Delta V)_{\rm YY~Dor} - 2.5 \log 3.4 \cr
&=& 18.86 - 0.3\pm0.2 - 1.33 = 17.23\pm0.2 \cr
&=& ((m - M)_V + \Delta V)_{\rm LMC~N~2009a} - 2.5 \log 2.1 \cr
&=& 18.86 - 0.8\pm0.2 -0.83 = 17.23\pm0.2.
\label{distance_modulus_v_v834_car_yy_dor_lmcn2009a}
\end{eqnarray}
We have $(m-M)_{V, \rm V834~Car}= 17.23\pm0.1$, which is
consistent with Equation (\ref{distance_modulus_v834_car}).

We apply Equation (\ref{distance_modulus_general_temp_i}) for
the $I_{\rm C}$ band to Figure
\ref{v834_car_yy_dor_lmcn_2009a_b_v_i_logscale_3fig}(c) and obtain
\begin{eqnarray}
(m&-&M)_{I, \rm V834~Car} \cr
&=& ((m - M)_I + \Delta I_C)_{\rm YY~Dor} - 2.5 \log 3.4 \cr
&=& 18.67 - 0.9\pm0.2 - 1.33 = 16.44\pm 0.2 \cr
&=& ((m - M)_I + \Delta I_C)_{\rm LMC~N~2009a} - 2.5 \log 2.1 \cr
&=& 18.67 - 1.4\pm0.2 -0.83 = 16.44\pm 0.2.
\label{distance_modulus_i_v834_car_yy_dor_lmcn2009a}
\end{eqnarray}
We have $(m-M)_{I, \rm V834~Car}= 16.44\pm0.1$.

We plot $(m-M)_B= 17.75$, $(m-M)_V= 17.23$, and $(m-M)_I= 16.44$,
which broadly cross at $d=14$~kpc and $E(B-V)=0.50$, in Figure
\ref{distance_reddening_v5585_sgr_pr_lup_v1313_sco_v834_car}(d).
Thus, we obtain $E(B-V)=0.50\pm0.05$ and $d=14\pm2$~kpc.


\begin{figure}
\epsscale{0.75}
\plotone{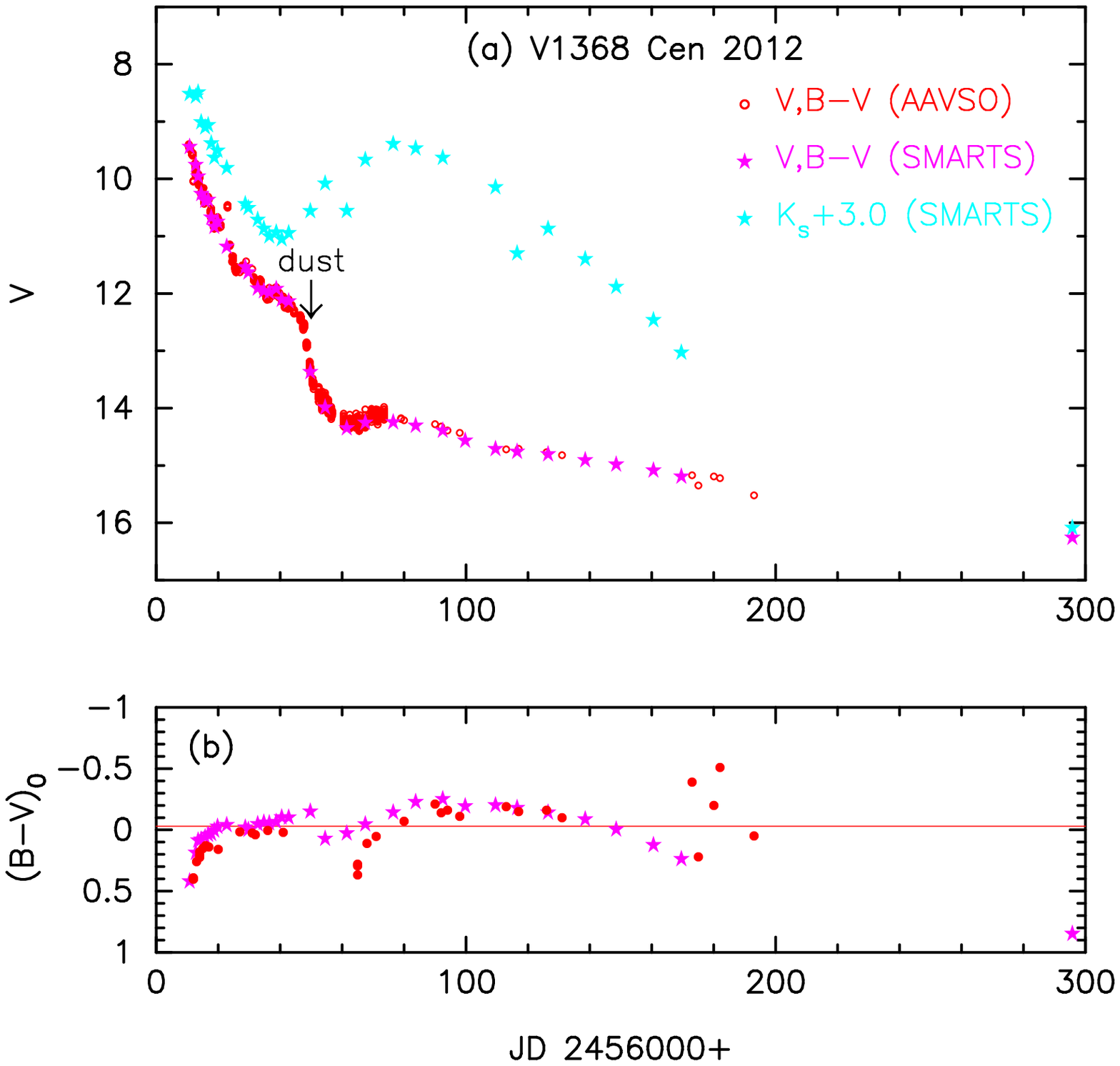}
\caption{
Same as Figure \ref{v1663_aql_v_bv_ub_color_curve}, but for V1368~Cen.
(a) The $V$ data are taken from AAVSO (unfilled red circles) and
SMARTS (filled magenta stars).  The $K_{\rm s}$ data are taken from
SMARTS (filled cyan stars).  A dust shell formed around the time
indicated by the arrow labeled dust.
(b) The $(B-V)_0$ are dereddened with $E(B-V)=0.93$.
\label{v1368_cen_v_bv_ub_color_curve}}
\end{figure}


\begin{figure}
\epsscale{0.75}
\plotone{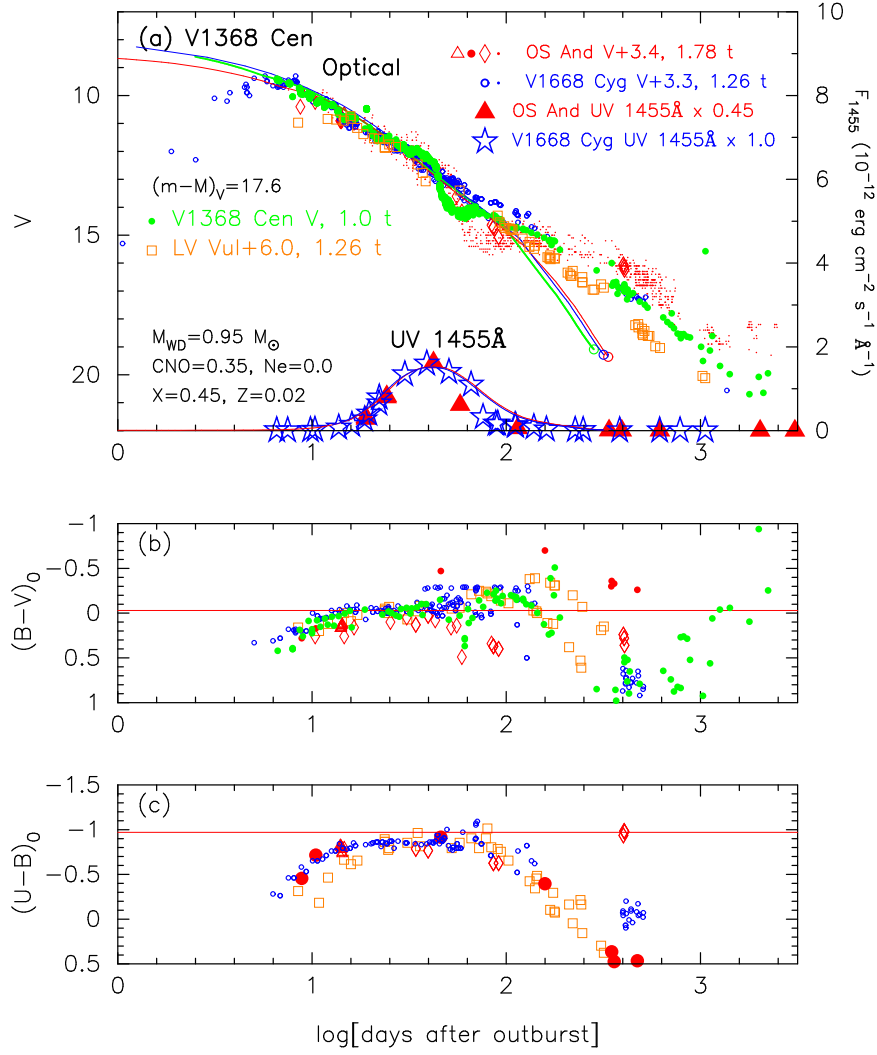}
\caption{
Same as Figure
\ref{v2575_oph_v1668_cyg_lv_vul_v_bv_ub_logscale},
but for V1368~Cen (filled green circles).
The data of V1368~Cen are the same as those in Figure
\ref{v1368_cen_v_bv_ub_color_curve}.  
In panel (a), we added a model $V$ light curve of a $0.95~M_\sun$ WD 
\citep[CO3, solid green lines;][]{hac16k},
assuming that $(m-M)_V=17.6$ for V1368~Cen.
The solid blue lines denotes the $V$ and UV~1455\AA\  light curve
of a $0.98~M_\sun$ WD (CO3),
assuming $(m-M)_V=14.6$ for V1668~Cyg, while the solid red lines
represent the $V$ and UV~1455\AA\  light curves of a $1.05~M_\sun$ WD
(CO3), assuming $(m-M)_V=14.8$ for OS~And.
\label{v1368_cen_lv_vul_v1668_cyg_os_and_v_bv_ub_logscale}}
\end{figure}


\begin{figure}
\epsscale{0.55}
\plotone{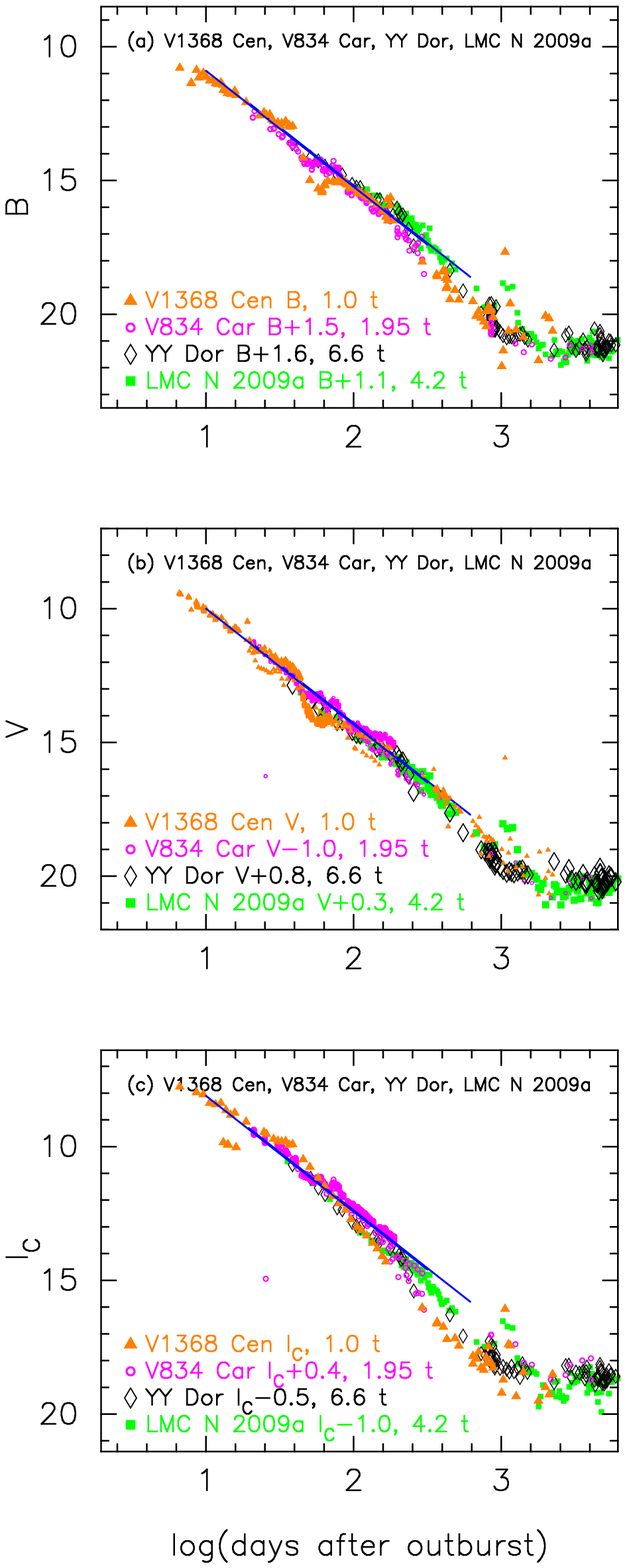}
\caption{
Same as Figure \ref{v1663_aql_yy_dor_lmcn_2009a_b_v_i_logscale_3fig},
but for V1368~Cen.
The $BV$ data of V1368~Cen are the same as those in Figure
\ref{v1368_cen_v_bv_ub_color_curve}.  The $I_{\rm C}$ data of V1368~Cen
are taken from AAVSO and SMARTS.
\label{v1368_cen_v834_car_yy_dor_lmcn_2009a_b_v_i_logscale_3fig}}
\end{figure}

\subsection{V1368~Cen 2012}
\label{v1368_cen}
Figure \ref{v1368_cen_v_bv_ub_color_curve} shows the (a) $V$ and 
$K_{\rm s}$, and (b) $(B-V)_0$ evolutions of V1368~Cen.  
Here, $(B-V)_0$ are dereddened with $E(B-V)=0.93$ as obtained in
Section \ref{v1368_cen_cmd}.
Figure \ref{v1368_cen_lv_vul_v1668_cyg_os_and_v_bv_ub_logscale} shows
the light/color curves of V1368~Cen, LV~Vul, V1668~Cyg, and OS~And.
They overlap each other.
Applying Equation (\ref{distance_modulus_general_temp}) to them,
we have the relation 
\begin{eqnarray}
(m&-&M)_{V, \rm V1368~Cen} \cr
&=& (m - M + \Delta V)_{V, \rm LV~Vul} - 2.5 \log 1.26 \cr
&=& 11.85 + 6.0\pm0.2 - 0.25 = 17.6\pm0.2 \cr
&=& (m - M + \Delta V)_{V, \rm V1668~Cyg} - 2.5 \log 1.26 \cr
&=& 14.6 + 3.3\pm0.2 - 0.25 = 17.65\pm0.2 \cr
&=& (m - M + \Delta V)_{V, \rm OS~And} - 2.5 \log 1.78 \cr
&=& 14.8 +3.4\pm0.2 - 0.63 = 17.57\pm0.2,
\label{distance_modulus_v1368_cen}
\end{eqnarray}
where we adopt $(m-M)_{V, \rm LV~Vul}=11.85$ and 
$(m-M)_{V, \rm V1668~Cyg}=14.6$ from \citet{hac19k}, and
$(m-M)_{V, \rm OS~And}=14.8$ from \citet{hac16kb}.
Thus, we obtain $(m-M)_V=17.6\pm0.1$ and $f_{\rm s}=1.26$ against LV~Vul.
From Equations (\ref{time-stretching_general}),
(\ref{distance_modulus_general_temp}), and
(\ref{distance_modulus_v1368_cen}),
we have the relation
\begin{eqnarray}
(m- M')_{V, \rm V1368~Cen} 
&\equiv & (m_V - (M_V - 2.5\log f_{\rm s}))_{\rm V1368~Cen} \cr
&=& \left( (m-M)_V + \Delta V \right)_{\rm LV~Vul} \cr
&=& 11.85 + 6.0\pm0.2 = 17.85\pm0.2.
\label{absolute_mag_v1368_cen}
\end{eqnarray}

Figure \ref{v1368_cen_v834_car_yy_dor_lmcn_2009a_b_v_i_logscale_3fig}
shows the $B$, $V$, and $I_{\rm C}$ light curves of V1368~Cen 
together with those of V834~Car, YY~Dor, and LMC~N~2009a.
We apply Equation (\ref{distance_modulus_general_temp_b})
for the $B$ band to Figure
\ref{v1368_cen_v834_car_yy_dor_lmcn_2009a_b_v_i_logscale_3fig}(a)
and obtain
\begin{eqnarray}
(m&-&M)_{B, \rm V1368~Cen} \cr
&=& ((m - M)_B + \Delta B)_{\rm V834~Car} - 2.5 \log 1.95 \cr
&=& 17.75 + 1.5\pm0.2 - 0.73 = 18.52\pm0.2 \cr
&=& ((m - M)_B + \Delta B)_{\rm YY~Dor} - 2.5 \log 6.6 \cr
&=& 18.98 + 1.6\pm0.2 - 2.05 = 18.53\pm0.2 \cr
&=& ((m - M)_B + \Delta B)_{\rm LMC~N~2009a} - 2.5 \log 4.2 \cr
&=& 18.98 + 1.1\pm0.2 - 1.55 = 18.53\pm0.2.
\label{distance_modulus_b_v1368_cen_v834_car_yy_dor_lmcn2009a}
\end{eqnarray}
We have $(m-M)_{B, \rm V1368~Cen}= 18.53\pm0.1$.

For the $V$ light curves in Figure
\ref{v1368_cen_v834_car_yy_dor_lmcn_2009a_b_v_i_logscale_3fig}(b),
we similarly obtain
\begin{eqnarray}
(m&-&M)_{V, \rm V1368~Cen} \cr   
&=& ((m - M)_V + \Delta V)_{\rm V834~Car} - 2.5 \log 1.95 \cr
&=& 17.3 - 1.0\pm0.2 - 0.73 = 17.57\pm0.2 \cr
&=& ((m - M)_V + \Delta V)_{\rm YY~Dor} - 2.5 \log 6.6 \cr
&=& 18.86 + 0.8\pm0.2 - 2.05 = 17.61\pm0.2 \cr
&=& ((m - M)_V + \Delta V)_{\rm LMC~N~2009a} - 2.5 \log 4.2 \cr
&=& 18.86 + 0.3\pm0.2 - 1.55 = 17.61\pm0.2.
\label{distance_modulus_v_v1368_cen_v834_car_yy_dor_lmcn2009a}
\end{eqnarray}
We have $(m-M)_{V, \rm V1368~Cen}= 17.6\pm0.1$, which is
consistent with Equation (\ref{distance_modulus_v1368_cen}).

We apply Equation (\ref{distance_modulus_general_temp_i}) for
the $I_{\rm C}$ band to Figure
\ref{v1368_cen_v834_car_yy_dor_lmcn_2009a_b_v_i_logscale_3fig}(c) and obtain
\begin{eqnarray}
(m&-&M)_{I, \rm V1368~Cen} \cr
&=& ((m - M)_I + \Delta I_C)_{\rm V834~Car} - 2.5 \log 1.95 \cr
&=& 16.45 + 0.4\pm0.2 - 0.73 = 16.12\pm 0.2 \cr
&=& ((m - M)_I + \Delta I_C)_{\rm YY~Dor} - 2.5 \log 6.6 \cr
&=& 18.67 - 0.5\pm0.2 - 2.05 = 16.12\pm 0.2 \cr
&=& ((m - M)_I + \Delta I_C)_{\rm LMC~N~2009a} - 2.5 \log 4.2 \cr
&=& 18.67 - 1.0\pm0.2 - 1.55 = 16.12\pm 0.2.
\label{distance_modulus_i_v1368_cen_v834_car_yy_dor_lmcn2009a}
\end{eqnarray}
We have $(m-M)_{I, \rm V1368~Cen}= 16.12\pm0.1$.

We plot $(m-M)_B= 18.53$, $(m-M)_V= 17.6$, and $(m-M)_I= 16.12$,
which broadly cross at $d=8.8$~kpc and $E(B-V)=0.93$, in Figure
\ref{distance_reddening_v1368_cen_v2676_oph_v5589_sgr_v2677_oph}(a). 
Thus, we obtain $E(B-V)=0.93\pm0.05$ and $d=8.8\pm1.0$~kpc.


\begin{figure}
\epsscale{0.75}
\plotone{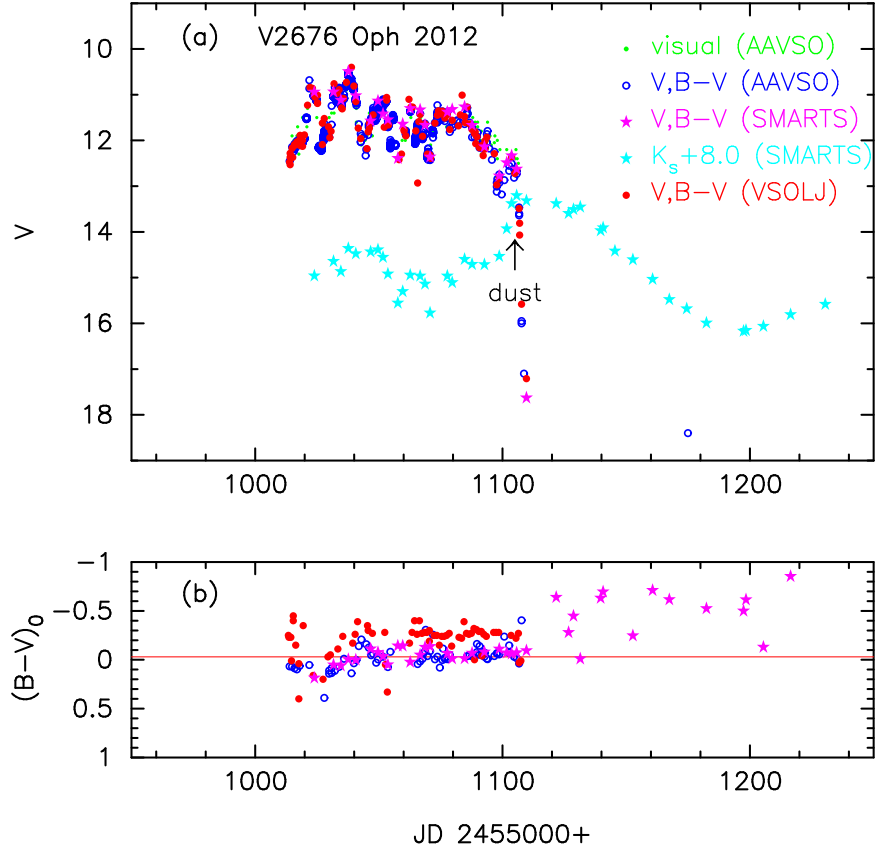}
\caption{
Same as Figure \ref{v1663_aql_v_bv_ub_color_curve}, but for V2676~Oph.
(a) The visual data (green dots) are taken from AAVSO.
The $V$ data are taken from AAVSO (unfilled blue circles),
SMARTS (filled magenta stars), and VSOLJ (filled red circles).
We add the $K_{\rm s}$ data (filled cyan stars) from SMARTS.
The dust formation epoch is denoted by the arrow labeled dust.
(b) The $(B-V)_0$ are dereddened with $E(B-V)=0.90$.
\label{v2676_oph_v_bv_ub_color_curve}}
\end{figure}


\begin{figure}
\epsscale{0.75}
\plotone{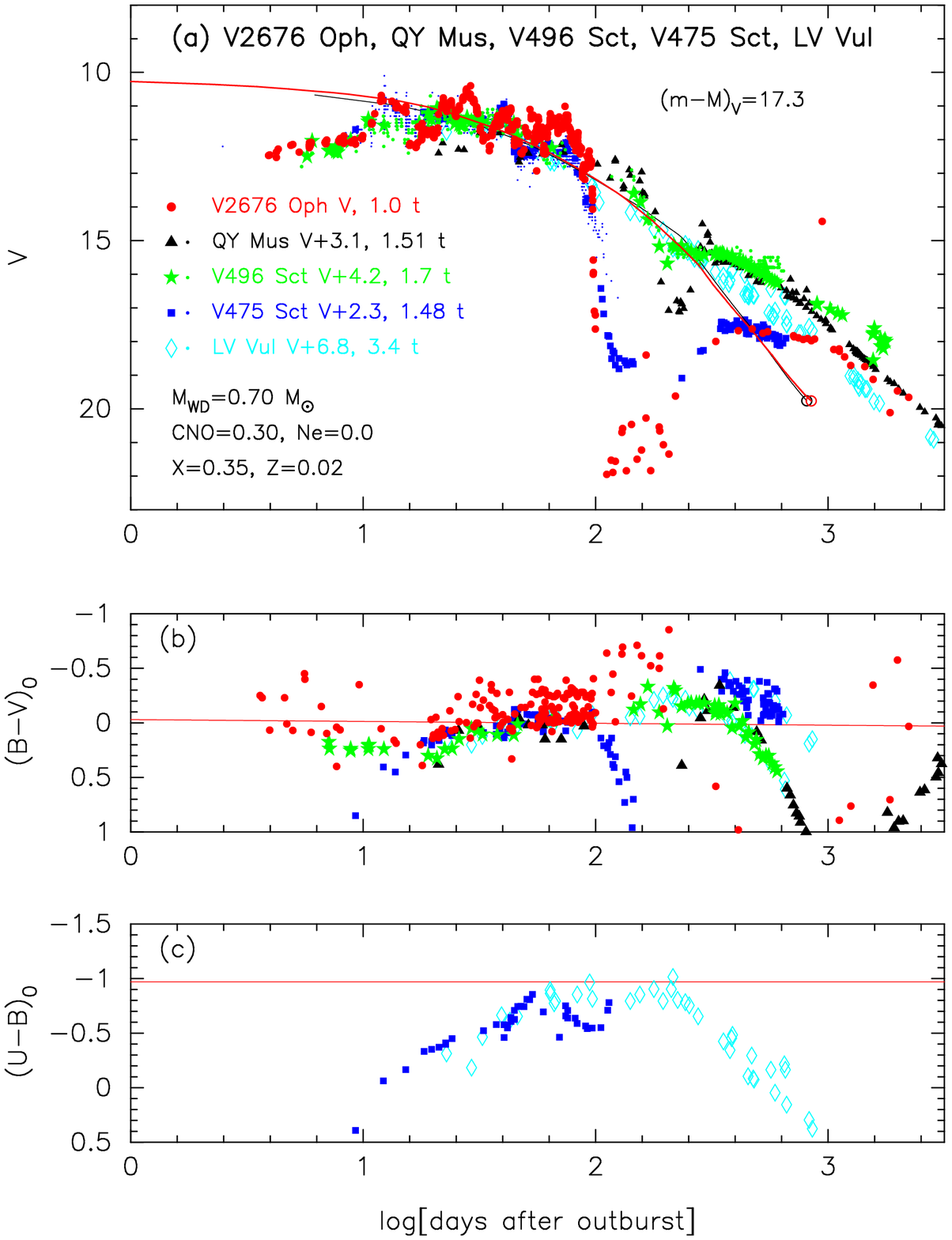}
\caption{
Same as Figure
\ref{v2575_oph_v1668_cyg_lv_vul_v_bv_ub_logscale},
but for V2676~Oph (filled red circles).  The data of V2676~Oph are
the same as those in Figure \ref{v2676_oph_v_bv_ub_color_curve}.
The data of V475~Sct are the same as those in Figures 49 and 50 of 
\citet{hac16kb}. 
In panel (a), we added the model $V$ light curve of a $0.70~M_\sun$ WD 
\citep[CO2, solid red line;][]{hac10k}, 
assuming that $(m-M)_V=17.3$ for V2676~Oph.
The solid black line denotes the $V$ light curve of a $0.80~M_\sun$ WD
\citep[CO3;][]{hac16k}, assuming $(m-M)_V=14.65$ for QY~Mus.
\label{v2676_oph_v496_sct_v475_sct_lv_vul_v_bv_ub_color_logscale}}
\end{figure}


\begin{figure}
\epsscale{0.6}
\plotone{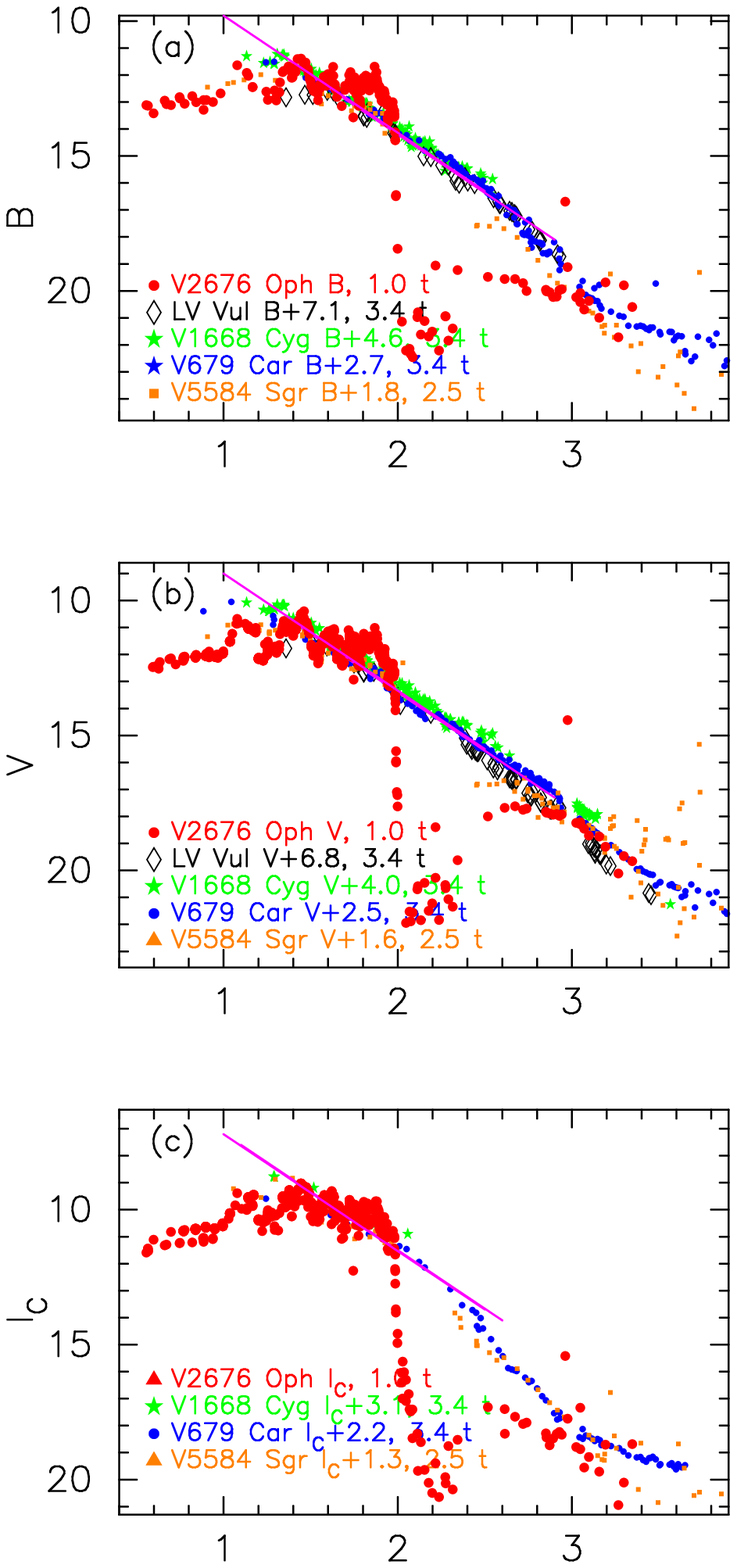}
\caption{
Same as Figure \ref{v1663_aql_yy_dor_lmcn_2009a_b_v_i_logscale_3fig},
but for V2676~Oph.
The (a) $B$, (b) $V$, and (c) $I_{\rm C}$  light curves of V2676~Oph
as well as those of V5584~Sgr, 679~Car, LV~Vul, and V1668~Cyg.
The $BV$ data of V2676~Oph are the same as those in Figure
\ref{v2676_oph_v_bv_ub_color_curve}.
The $I_{\rm C}$ data of V2676~Oph are taken from AAVSO, VSOLJ, and SMARTS.
\label{v2676_oph_v5584_sgr_v679_car_lv_vul_v1668_cyg_b_v_i_logscale_3fig}}
\end{figure}

\subsection{V2676~Oph 2012}
\label{v2676_oph}
Figure \ref{v2676_oph_v_bv_ub_color_curve} shows the (a) visual, $V$, and
$K_{\rm s}$, and (b) $(B-V)_0$ evolutions of V2676~Oph.  Here, $(B-V)_0$ are
dereddened with $E(B-V)=0.90$ as obtained in Section \ref{v2676_oph_cmd}.
Figure \ref{v2676_oph_v496_sct_v475_sct_lv_vul_v_bv_ub_color_logscale}
shows the light/color curves of V2676~Oph as well as those of QY~Mus, 
V496~Sct, V475~Sct, and LV~Vul.  The $V$ light curve of V2676~Oph
overlaps well with that of V475~Sct.
Applying Equation (\ref{distance_modulus_general_temp}) to them,
we have the relation 
\begin{eqnarray}
(m&-&M)_{V, \rm V2676~Oph} \cr
&=& (m - M + \Delta V)_{V, \rm LV~Vul} - 2.5 \log 3.4 \cr
&=& 11.85 + 6.8\pm0.3 - 1.33 = 17.32\pm0.3 \cr
&=& (m - M + \Delta V)_{V, \rm V496~Sct} - 2.5 \log 1.7 \cr
&=& 13.7 +4.2\pm0.3 - 0.58 = 17.32\pm0.3 \cr
&=& (m - M + \Delta V)_{V, \rm QY~Mus} - 2.5 \log 1.51 \cr
&=& 14.65 + 3.1\pm0.3 - 0.45 = 17.3\pm0.3 \cr
&=& (m - M + \Delta V)_{V, \rm V475~Sct} - 2.5 \log 1.48 \cr
&=& 15.4 + 2.3\pm0.3 - 0.43 = 17.27\pm0.3,
\label{distance_modulus_v2676_oph}
\end{eqnarray}
where we adopt $(m-M)_{V, \rm LV~Vul}=11.85$ and
$(m-M)_{V, \rm V496~Sct}=13.7$ from \citet{hac19k}, 
$(m-M)_{V, \rm QY~Mus}=14.65$ from Section \ref{qy_mus_cmd}, and
$(m-M)_{V, \rm V475~Sct}=15.4$ from \citet{hac16kb}.
Thus, we obtain $(m-M)_V=17.3\pm0.2$ and $f_{\rm s}=3.4$ against LV~Vul.
From Equations (\ref{time-stretching_general}),
(\ref{distance_modulus_general_temp}), and
(\ref{distance_modulus_v2676_oph}),
we have the relation
\begin{eqnarray}
(m- M')_{V, \rm V2676~Oph} 
&\equiv & (m_V - (M_V - 2.5\log f_{\rm s}))_{\rm V2676~Oph} \cr
&=& \left( (m-M)_V + \Delta V \right)_{\rm LV~Vul} \cr
&=& 11.85 + 6.8\pm0.3 = 18.65\pm0.3.
\label{absolute_mag_v2676_oph}
\end{eqnarray}

Figure 
\ref{v2676_oph_v5584_sgr_v679_car_lv_vul_v1668_cyg_b_v_i_logscale_3fig}
shows the $B$, $V$, and $I_{\rm C}$ light curves of V2676~Oph together with
those of V5584~Sgr, V679~Car, LV~Vul, and V1668~Cyg.
We apply Equation (\ref{distance_modulus_general_temp_b})
for the $B$ band to Figure
\ref{v2676_oph_v5584_sgr_v679_car_lv_vul_v1668_cyg_b_v_i_logscale_3fig}(a)
and obtain
\begin{eqnarray}
(m&-&M)_{B, \rm V2676~Oph} \cr
&=& ((m - M)_B + \Delta B)_{\rm LV~Vul} - 2.5 \log 3.4 \cr
&=& 12.45 + 7.1\pm0.2 - 1.32 = 18.23\pm0.2 \cr
&=& ((m - M)_B + \Delta B)_{\rm V1668~Cyg} - 2.5 \log 3.4 \cr
&=& 14.9 + 4.6\pm0.2 - 1.32 = 18.18\pm0.2  \cr
&=& ((m - M)_B + \Delta B)_{\rm V679~Car} - 2.5 \log 3.4 \cr
&=& 16.79 + 2.7\pm0.2 - 1.32 = 18.17\pm0.2  \cr
&=& ((m - M)_B + \Delta B)_{\rm V5584~Sgr} - 2.5 \log 2.5 \cr
&=& 17.4 + 1.8\pm0.2 - 1.0 = 18.2\pm0.2,
\label{distance_modulus_b_v2676_oph_v5584_sgr_lv_vul_v1668_cyg}
\end{eqnarray}
where we adopt $(m-M)_{B, \rm LV~Vul}= 11.85 + 0.60= 12.45$,
$(m-M)_{B, \rm V1668~Cyg}= 14.6 + 0.30= 14.9$, and 
$(m-M)_{B, \rm V679~Car}= 16.1 + 0.69= 16.79$ from \citet{hac19k},
and $(m-M)_{B, \rm V5584~Sgr}= 16.7 + 0.70= 17.4$ from Appendix
\ref{v5584_sgr}.
We have $(m-M)_{B, \rm V2676~Oph}= 18.2\pm0.1$.

For the $V$ light curves in Figure
\ref{v2676_oph_v5584_sgr_v679_car_lv_vul_v1668_cyg_b_v_i_logscale_3fig}(b),
we similarly obtain
\begin{eqnarray}
(m&-&M)_{V, \rm V2676~Oph} \cr
&=& ((m - M)_V + \Delta V)_{\rm LV~Vul} - 2.5 \log 3.4 \cr
&=& 11.85 + 6.8\pm0.2 - 1.32 = 17.33\pm0.2 \cr
&=& ((m - M)_V + \Delta V)_{\rm V1668~Cyg} - 2.5 \log 3.4 \cr
&=& 14.6 + 4.0\pm0.2 - 1.32 = 17.28\pm0.2  \cr
&=& ((m - M)_V + \Delta V)_{\rm V679~Car} - 2.5 \log 3.4 \cr
&=& 16.1 + 2.5\pm0.2 - 1.32 = 17.28\pm0.2 \cr
&=& ((m - M)_V + \Delta V)_{\rm V5584~Sgr} - 2.5 \log 2.5 \cr
&=& 16.7 + 1.6\pm0.2 - 1.0 = 17.3\pm0.2,
\label{distance_modulus_v_v2676_oph_v5584_sgr_lv_vul_v1668_cyg}
\end{eqnarray}
where we adopt $(m-M)_{V, \rm LV~Vul}= 11.85$ and
$(m-M)_{V, \rm V1668~Cyg}= 14.6$, $(m-M)_{V, \rm V679~Car}= 16.1$
from \citet{hac19k},
and $(m-M)_{V, \rm V5584~Sgr}= 16.7$ from Section \ref{v5584_sgr}.
We have $(m-M)_{V, \rm V2676~Oph}= 17.3\pm0.1$, which is
consistent with Equation (\ref{distance_modulus_v2676_oph}).

We apply Equation (\ref{distance_modulus_general_temp_i}) for
the $I_{\rm C}$ band to Figure
\ref{v2676_oph_v5584_sgr_v679_car_lv_vul_v1668_cyg_b_v_i_logscale_3fig}(c)
and obtain
\begin{eqnarray}
(m&-&M)_{I, \rm V2676~Oph} \cr
&=& ((m - M)_I + \Delta I_C)_{\rm V1668~Cyg} - 2.5 \log 3.4 \cr
&=& 14.12 + 3.1\pm0.2 - 1.32 = 15.9\pm 0.2 \cr
&=& ((m - M)_I + \Delta I_C)_{\rm V679~Car} - 2.5 \log 3.4 \cr
&=& 15.0 + 2.2\pm0.2 - 1.32 = 15.88\pm 0.2 \cr
&=& ((m - M)_I + \Delta I_C)_{\rm V5584~Sgr} - 2.5 \log 2.5 \cr
&=& 15.58 + 1.3\pm0.2 - 1.0 = 15.88\pm 0.2,
\label{distance_modulus_i_v2676_oph_v5584_sgr_lv_vul_v1668_cyg}
\end{eqnarray}
where we adopt 
$(m-M)_{I, \rm V1668~Cyg}= 14.6 - 1.6\times 0.30= 14.12$, and
$(m-M)_{I, \rm V679~Car}= 16.1 - 1.6\times 0.69= 15.0$
from \citet{hac19k}, and  
$(m-M)_{I, \rm V5584~Sgr}= 16.7 - 1.6\times 0.70= 15.58$ from
Appendix \ref{v5584_sgr}.
Unfortunately, neither $I_{\rm C}$ nor $I$ data of
LV~Vul are available.
We have $(m-M)_{I, \rm V2676~Oph}= 15.88\pm0.1$.

We plot $(m-M)_B= 18.2$, $(m-M)_V= 17.3$, and $(m-M)_I= 15.88$,
which cross at $d=8.0$~kpc and $E(B-V)=0.90$, in Figure
\ref{distance_reddening_v1368_cen_v2676_oph_v5589_sgr_v2677_oph}(b).
Thus, we obtain $E(B-V)=0.90\pm0.05$ and $d=8.0\pm1$~kpc.


\begin{figure}
\epsscale{0.75}
\plotone{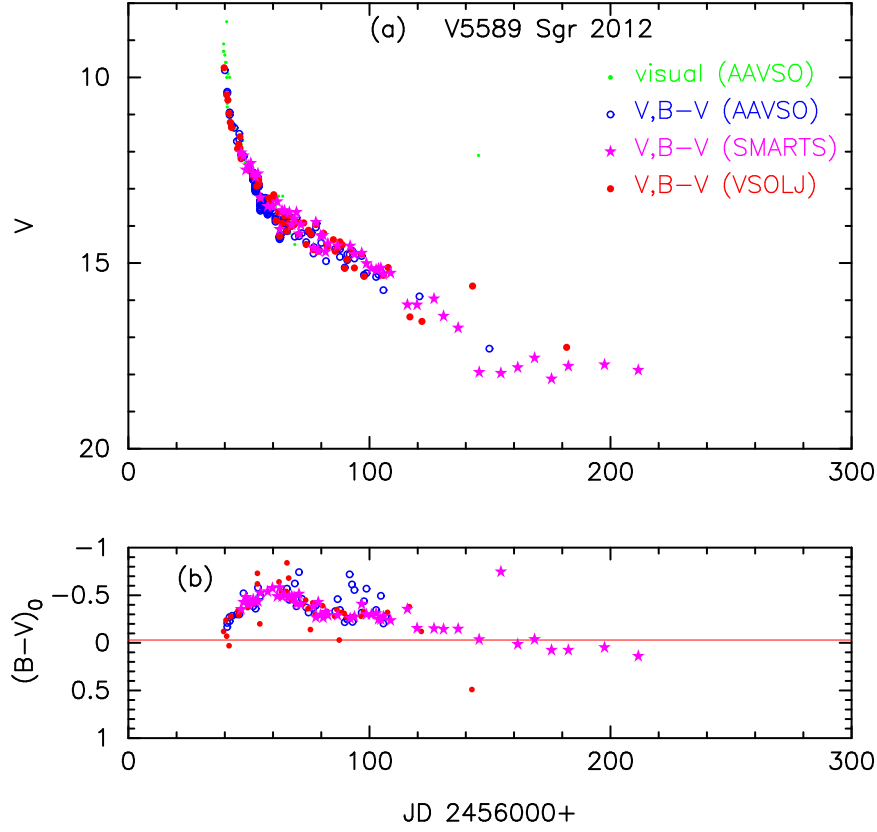}
\caption{
Same as Figure \ref{v1663_aql_v_bv_ub_color_curve}, but for V5589~Sgr.
(a) The visual data (green dots) are taken from AAVSO.
The $V$ data are taken from AAVSO (unfilled blue circles),
SMARTS (filled magenta stars), and VSOLJ (filled red circles).
(b) The $(B-V)_0$ are dereddened with $E(B-V)=0.84$.
\label{v5589_sgr_v_bv_ub_color_curve}}
\end{figure}


\begin{figure}
\epsscale{0.75}
\plotone{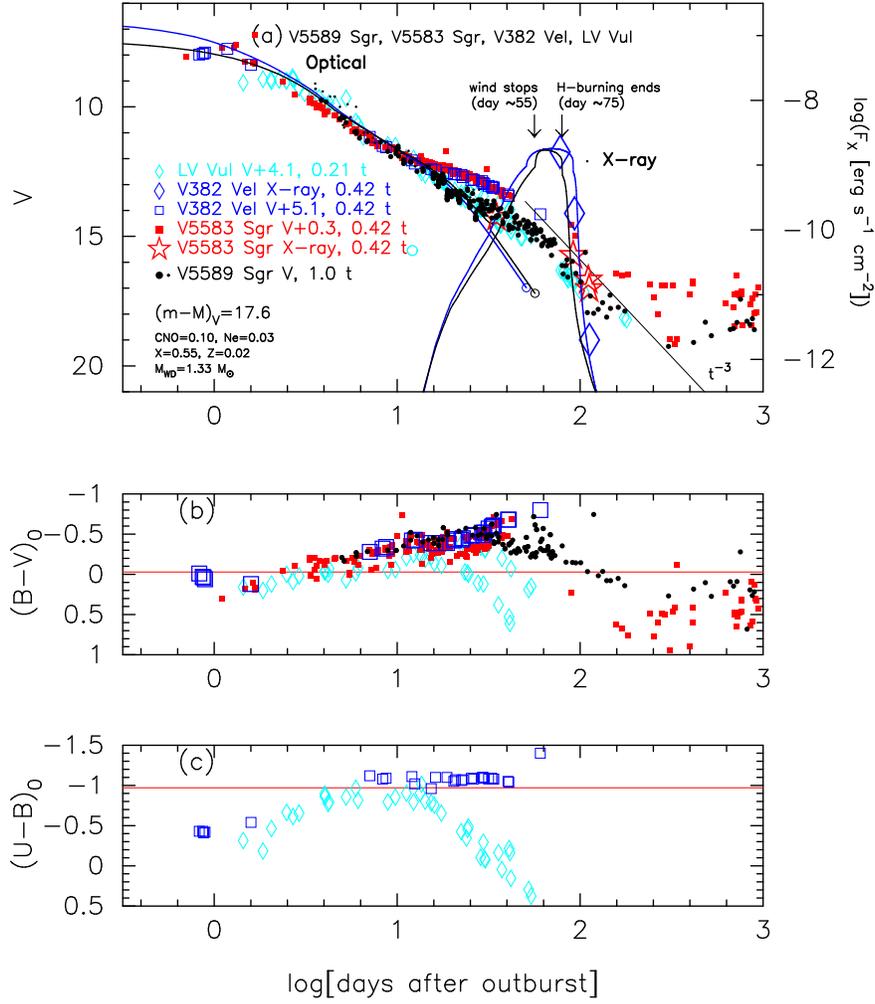}
\caption{
Same as Figure
\ref{v2575_oph_v1668_cyg_lv_vul_v_bv_ub_logscale},
but for V5589~Sgr (black dots for visual and filled black circles for $V$
and $B-V$).  We add the three novae, V5583~Sgr, V382~Vel, and LV~Vul,
the $V$ light curves of which are similar to those of V5589~Sgr.
The data of V5589~Sgr are the same as those in Figure
\ref{v5589_sgr_v_bv_ub_color_curve}.
In panel (a), we add the model light curve (solid black lines) of 
a $1.33~M_\sun$ WD \citep[Ne2;][]{hac10k}, assuming that 
$(m-M)_V=17.6$ for V5589~Sgr.  We also add the model light curves
(solid blue lines) of a $1.23~M_\sun$ WD (Ne2), 
assuming that $(m-M)_V=11.5$ for V382~Vel \citep{hac16k}.
\label{v5589_sgr_v5583_sgr_v382_vel_v_bv_ub_logscale}}
\end{figure}


\begin{figure}
\epsscale{0.55}
\plotone{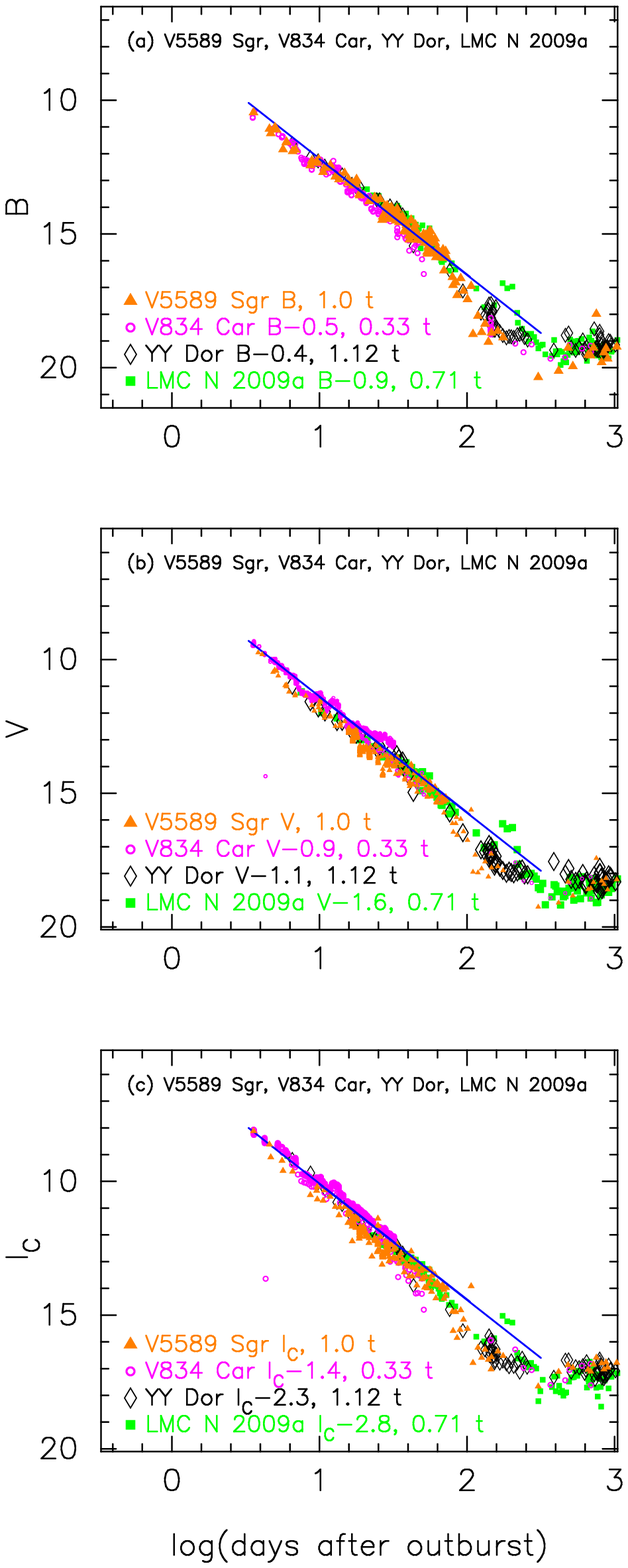}
\caption{
Same as Figure \ref{v1663_aql_yy_dor_lmcn_2009a_b_v_i_logscale_3fig},
but for V5589~Sgr.
The $BV$ data of V5589~Sgr are the same as those in Figure
\ref{v5589_sgr_v_bv_ub_color_curve}.  The $I_{\rm C}$ data of V5589~Sgr
are taken from AAVSO, VSOLJ, and SMARTS.
\label{v5589_sgr_v834_car_yy_dor_lmcn_2009a_b_v_i_logscale_3fig}}
\end{figure}


\begin{figure*}
\plotone{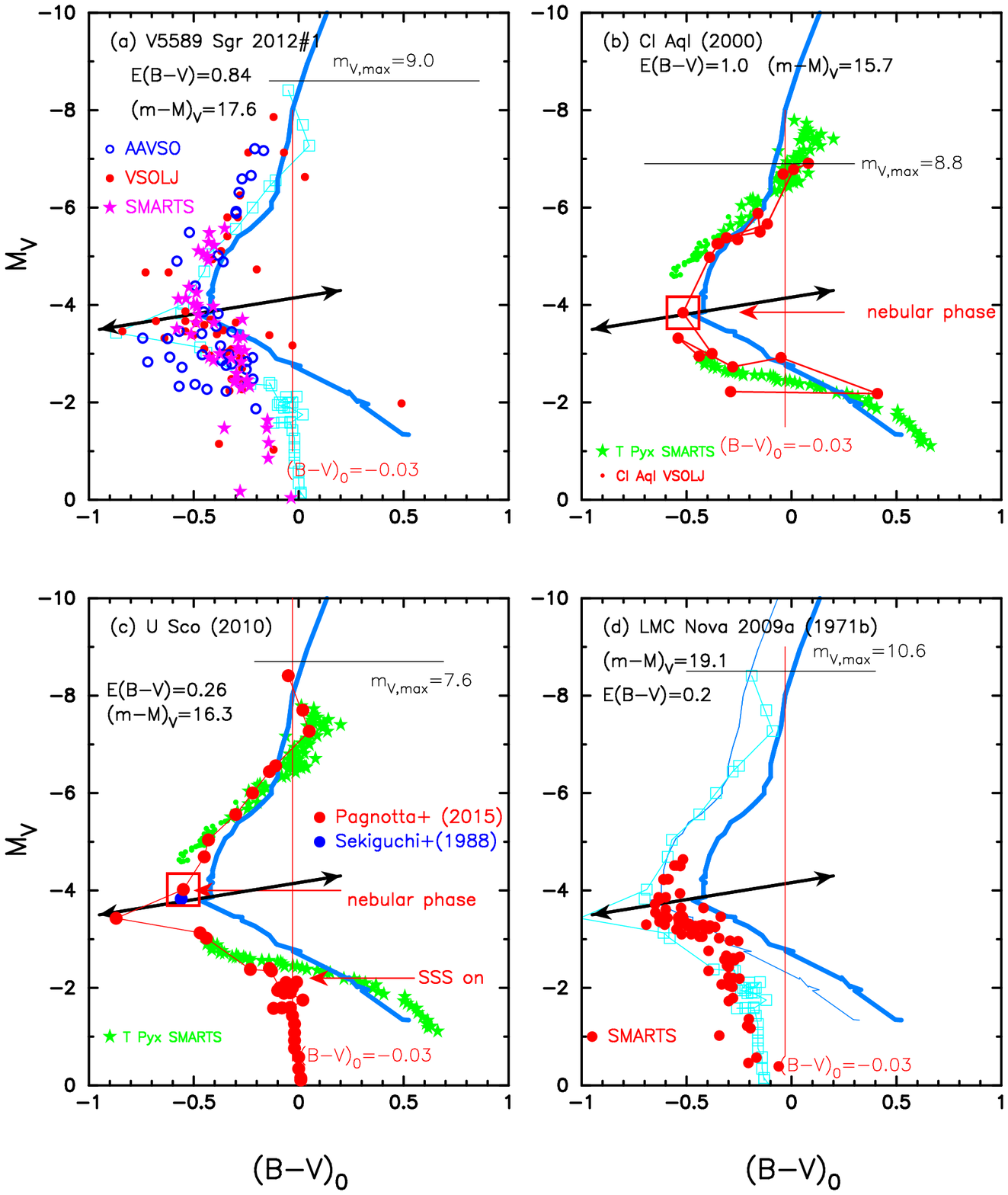}
\caption{
Color-magnitude diagram for
(a) V5589~Sgr, (b) CI~Aql, (c) U~Sco, and (d) LMC~N~2009a.
The thick solid cyan-blue line is the V1500~Cyg track.
The two-headed black arrow represents the locations of the turning point
of the tracks \citep{hac16kb}.
In panel (a), we add the track of U~Sco with the unfilled cyan squares
connected by the solid cyan line.  In panels (b) and (c), we also plot the 
T~Pyx (2011) outburst (filled green stars, taken from SMARTS).
In panel (d), the very thin solid cyan-blue line corresponds to
the V1500~Cyg track, which is blue-shifted by $\Delta (B-V)_0=-0.2$.
The unfilled cyan squares connected by a thin solid cyan line represent
the U~Sco track, which is blue-shifted by $\Delta (B-V)_0=-0.14$.
The blue-shifted nature of LMC~N~2009a is due to the low (subsolar)
metallicity of the LMC \citep[e.g.,][]{hac18kb}.
\label{hr_diagram_v5589_sgr_u_sco_ci_aql_lmcn2009a_outburst_mv}}
\end{figure*}

\subsection{V5589~Sgr 2012\#1}
\label{v5589_sgr}
Figure \ref{v5589_sgr_v_bv_ub_color_curve} shows the (a) visual and $V$,
and (b) $(B-V)_0$ evolutions of V5589~Sgr.  Here, $(B-V)_0$ are
dereddened with $E(B-V)=0.84$ as obtained in Section \ref{v5589_sgr_cmd}.
Figure \ref{v5589_sgr_v5583_sgr_v382_vel_v_bv_ub_logscale} shows
the light/color curves of V5589~Sgr, V5583~Sgr, V382~Vel, and  LV~Vul.
Applying Equation (\ref{distance_modulus_general_temp}) to them,
we have the relation 
\begin{eqnarray}
(m&-&M)_{V, \rm V5589~Sgr} \cr
&=& (m - M + \Delta V)_{V, \rm LV~Vul} - 2.5 \log 0.21 \cr
&=& 11.85 + 4.1\pm0.2 + 1.68 = 17.63\pm0.2 \cr
&=& (m - M + \Delta V)_{V, \rm V382~Vel} - 2.5 \log 0.42 \cr
&=& 11.5 + 5.1\pm0.2 + 0.95 = 17.55\pm0.2 \cr
&=& (m - M + \Delta V)_{V, \rm V5583~Sgr} - 2.5 \log 0.42 \cr
&=& 16.3 + 0.3\pm0.2 + 0.95 = 17.55\pm0.2,
\label{distance_modulus_v5589_sgr}
\end{eqnarray}
where we adopt $(m-M)_{V, \rm LV~Vul}=11.85$ and 
$(m-M)_{V, \rm V382~Vel}=11.5$ from \citet{hac19k},
and $(m-M)_{V, \rm V5583~Sgr}=16.3$ in Section \ref{v5583_sgr_cmd}.
Thus, we adopt $(m-M)_V=17.6\pm0.1$ and $f_{\rm s}=0.21$ against LV~Vul.
From Equations (\ref{time-stretching_general}),
(\ref{distance_modulus_general_temp}), and
(\ref{distance_modulus_v5589_sgr}),
we have the relation
\begin{eqnarray}
(m- M')_{V, \rm V5589~Sgr} 
&\equiv & (m_V - (M_V - 2.5\log f_{\rm s}))_{\rm V5589~Sgr} \cr
&=& \left( (m-M)_V + \Delta V \right)_{\rm LV~Vul} \cr
&=& 11.85 + 4.1\pm0.2 = 15.95\pm0.2.
\label{absolute_mag_v5589_sgr}
\end{eqnarray}

Figure \ref{v5589_sgr_v834_car_yy_dor_lmcn_2009a_b_v_i_logscale_3fig}
shows the $B$, $V$, and $I_{\rm C}$ light curves of V5589~Sgr 
together with those of V834~Car, YY~Dor, and LMC~N~2009a.
We apply Equation (\ref{distance_modulus_general_temp_b})
for the $B$ band to Figure
\ref{v5589_sgr_v834_car_yy_dor_lmcn_2009a_b_v_i_logscale_3fig}(a)
and obtain
\begin{eqnarray}
(m&-&M)_{B, \rm V5589~Sgr} \cr
&=& ((m - M)_B + \Delta B)_{\rm V834~Car} - 2.5 \log 0.33 \cr
&=& 17.75 - 0.5\pm0.2 + 1.2 = 18.45\pm0.2 \cr
&=& ((m - M)_B + \Delta B)_{\rm YY~Dor} - 2.5 \log 1.12 \cr
&=& 18.98 - 0.4\pm0.2 - 0.13 = 18.45\pm0.2 \cr
&=& ((m - M)_B + \Delta B)_{\rm LMC~N~2009a} - 2.5 \log 0.71 \cr
&=& 18.98 - 0.9\pm0.2 + 0.37 = 18.45\pm0.2.
\label{distance_modulus_b_v5589_sgr_v834_car_yy_dor_lmcn2009a}
\end{eqnarray}
We have $(m-M)_{B, \rm V5589~Sgr}= 18.45\pm0.1$.

For the $V$ light curves in Figure
\ref{v5589_sgr_v834_car_yy_dor_lmcn_2009a_b_v_i_logscale_3fig}(b),
we similarly obtain
\begin{eqnarray}
(m&-&M)_{V, \rm V5589~Sgr} \cr   
&=& ((m - M)_V + \Delta V)_{\rm V834~Car} - 2.5 \log 0.33 \cr
&=& 17.3 - 0.9\pm0.2 + 1.2 = 17.6\pm0.2 \cr
&=& ((m - M)_V + \Delta V)_{\rm YY~Dor} - 2.5 \log 1.12 \cr
&=& 18.86 - 1.1\pm0.2 - 0.13 = 17.63\pm0.2 \cr
&=& ((m - M)_V + \Delta V)_{\rm LMC~N~2009a} - 2.5 \log 0.71 \cr
&=& 18.86 - 1.6\pm0.2 + 0.37 = 17.63\pm0.2.
\label{distance_modulus_v_v5589_sgr_v834_car_yy_dor_lmcn2009a}
\end{eqnarray}
We have $(m-M)_{V, \rm V5589~Sgr}= 17.62\pm0.1$, which is
consistent with Equation (\ref{distance_modulus_v1368_cen}).

We apply Equation (\ref{distance_modulus_general_temp_i}) for
the $I_{\rm C}$ band to Figure
\ref{v5589_sgr_v834_car_yy_dor_lmcn_2009a_b_v_i_logscale_3fig}(c) and obtain
\begin{eqnarray}
(m&-&M)_{I, \rm V5589~Sgr} \cr
&=& ((m - M)_I + \Delta I_C)_{\rm V834~Car} - 2.5 \log 0.33 \cr
&=& 16.45 - 1.4\pm0.2 + 1.2 = 16.25\pm 0.2 \cr
&=& ((m - M)_I + \Delta I_C)_{\rm YY~Dor} - 2.5 \log 1.12 \cr
&=& 18.67 - 2.3\pm0.2 - 0.13 = 16.24\pm 0.2 \cr
&=& ((m - M)_I + \Delta I_C)_{\rm LMC~N~2009a} - 2.5 \log 0.71 \cr
&=& 18.67 - 2.8\pm0.2 + 0.37 = 16.24\pm 0.2.
\label{distance_modulus_i_v5589_sgr_v834_car_yy_dor_lmcn2009a}
\end{eqnarray}
We have $(m-M)_{I, \rm V5589~Sgr}= 16.25\pm0.1$.

We plot $(m-M)_B= 18.45$, $(m-M)_V= 17.62$, and $(m-M)_I= 16.25$,
which cross at $d=10$~kpc and $E(B-V)=0.84$, in Figure
\ref{distance_reddening_v1368_cen_v2676_oph_v5589_sgr_v2677_oph}(c).
Thus, we obtain $E(B-V)=0.84\pm0.05$ and $d=10\pm1$~kpc.

The orbital period of V5589~Sgr is $P_{\rm orb}=1.5923$~days, and
its companion is a subgiant \citep{mro15}.
This binary parameter reminds us of U~Sco, the orbital
period of which is $P_{\rm orb}=1.23$~days \citep[e.g.,][]{schaefer11}.
We plot the color-magnitude diagram of (a) V5589~Sgr as well as that of
the recurrent novae (b) CI~Aql, (c) U~Sco, and (d) LMC~N~2009a in Figure
\ref{hr_diagram_v5589_sgr_u_sco_ci_aql_lmcn2009a_outburst_mv}.
The orbital periods of CI~Aql and LMC~N~2009a are $P_{\rm orb}=0.618$~days
\citep[e.g.,][]{schaefer11} and $P_{\rm orb}=1.19$~days 
\citep{bod16}, respectively.
The track of V5589~Sgr is very close to the U~Sco track in the
color-magnitude diagram.  In these binary systems, the accretion disk is so
bright in the SSS phase, and it makes the $B-V$ color blue, 
$(B-V)_0\sim -0.2$ to $0.0$ in the later phase ($M_V\gtrsim-2$).

\citet{hac16kb} empirically found that the turning point (or inflection
point) of nova tracks in the color-magnitude diagram can be clearly
identified in many of the V1500~Cyg-type novae, and that the start of 
the nebular phase almost coincides with the turning point, 
and that they are located on the two-headed black arrow in Figure
\ref{hr_diagram_v5589_sgr_u_sco_ci_aql_lmcn2009a_outburst_mv}.
The turning point of V5589~Sgr is not clearly identified, but
its location is almost on the two-headed arrow.  This may support
that our distance modulus of $(m-M)_V=17.6$ is reasonable.


\begin{figure}
\epsscale{0.75}
\plotone{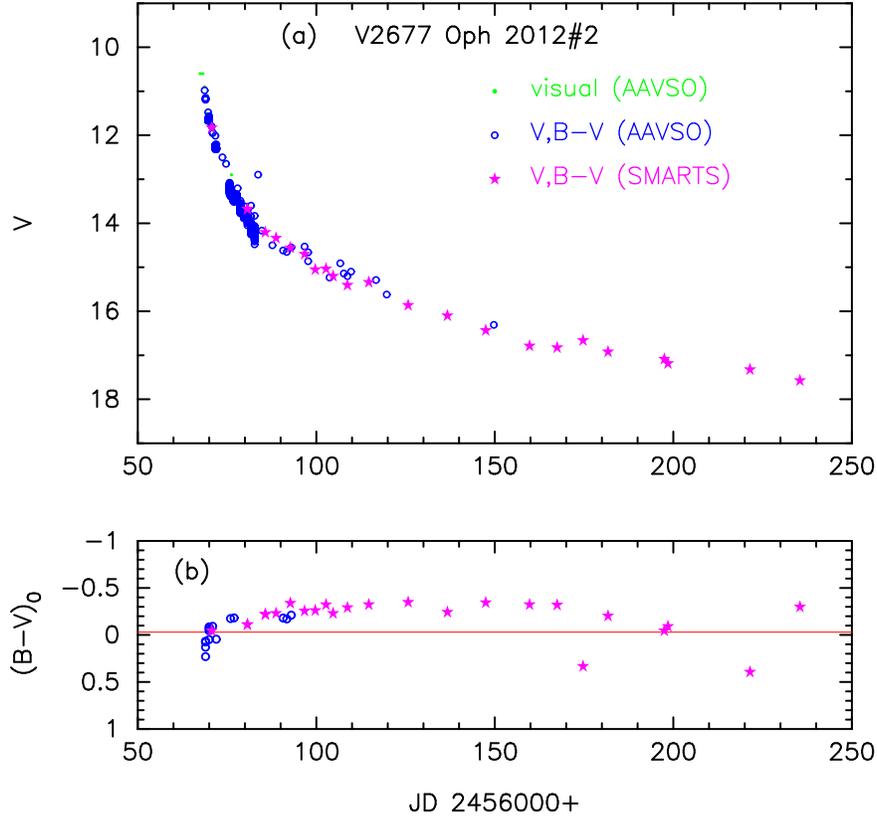}
\caption{
Same as Figure \ref{v1663_aql_v_bv_ub_color_curve}, but for V2677~Oph.
(a) The visual data (green dots) are taken from AAVSO.
The $BV$ data are taken from AAVSO (unfilled blue circles) and
SMARTS (filled magenta stars).
(b) The $(B-V)_0$ are dereddened with $E(B-V)=1.30$.
\label{v2677_oph_v_bv_ub_color_curve}}
\end{figure}


\begin{figure}
\epsscale{0.75}
\plotone{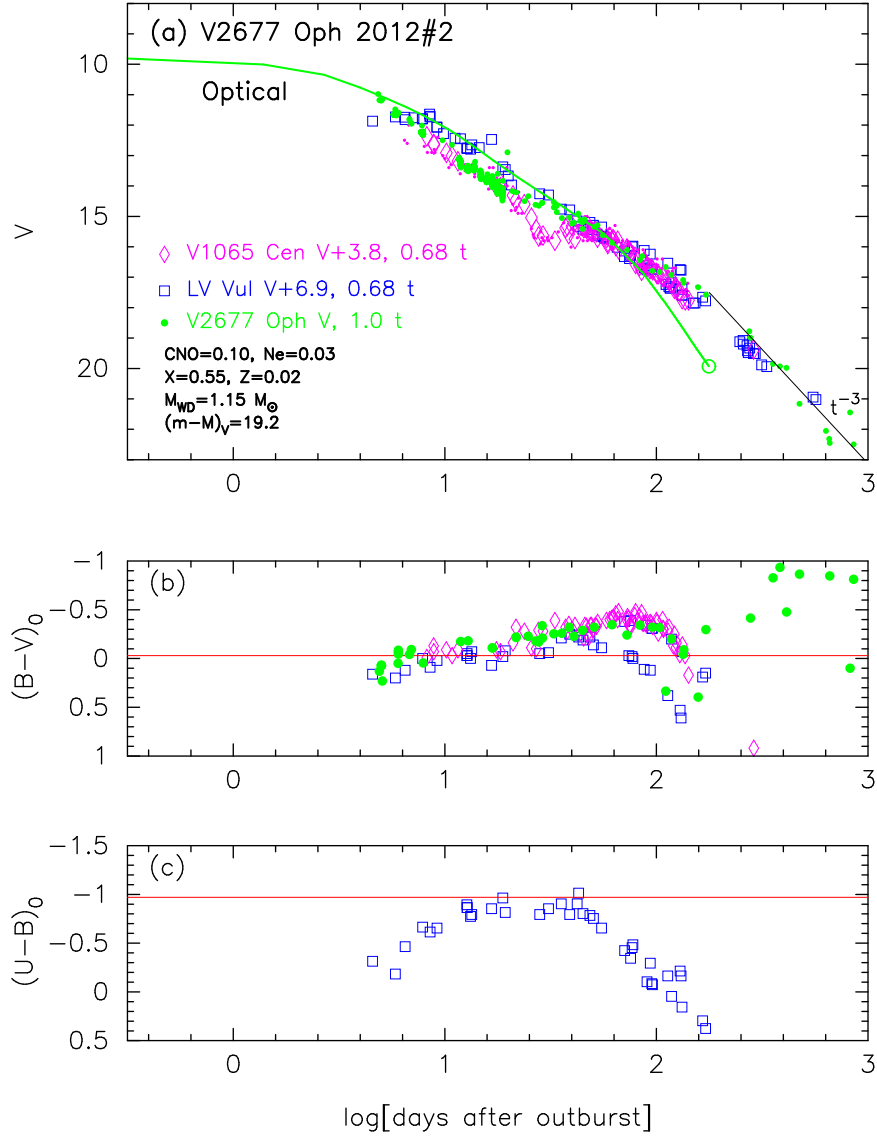}
\caption{
Same as Figure
\ref{v2575_oph_v1668_cyg_lv_vul_v_bv_ub_logscale},
but for V2677~Oph (filled green circles).  The data of V2677~Oph are
the same as those in Figure \ref{v2677_oph_v_bv_ub_color_curve}.  
The other data are the same as those in Figure 4 of \citet{hac18k}.
Assuming that $(m-M)_V=19.2$, we plot a model light curve
of a $1.15~M_\sun$ WD (solid green line) \citep[Ne2;][]{hac10k}.
\label{v2677_oph_v1065_cen_lv_vul_v_bv_ub_logscale}}
\end{figure}


\begin{figure}
\epsscale{0.6}
\plotone{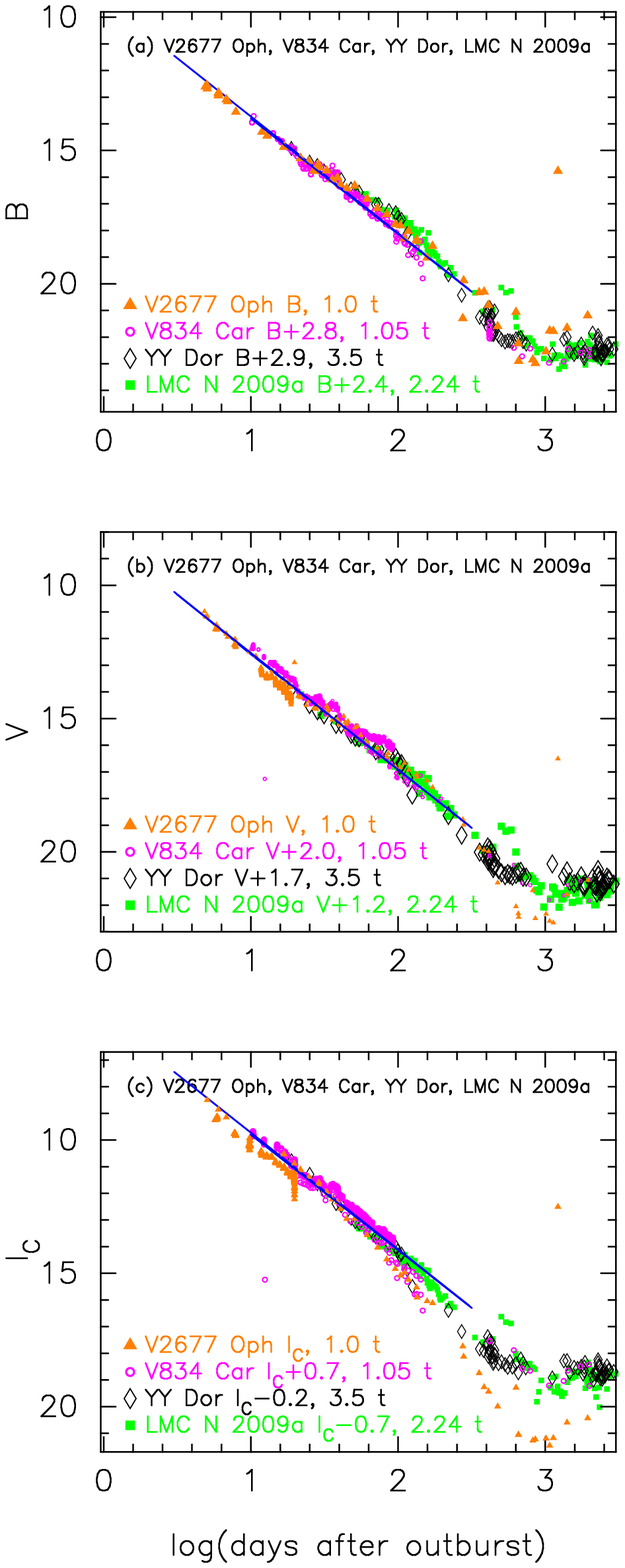}
\caption{
Same as Figure \ref{v1663_aql_yy_dor_lmcn_2009a_b_v_i_logscale_3fig},
but for V2677~Oph.
The $BV$ data of V2677~Oph are the same as those in Figure
\ref{v2677_oph_v_bv_ub_color_curve}.  The $I_{\rm C}$ data of V2677~Oph
are taken from AAVSO and SMARTS.
\label{v2677_oph_v834_car_yy_dor_lmcn_2009a_b_v_i_logscale_3fig}}
\end{figure}

\subsection{V2677~Oph 2012\#2}
\label{v2677_oph}
Figure \ref{v2677_oph_v_bv_ub_color_curve} shows the (a) visual, $V$, and
(b) $(B-V)_0$ evolutions of V2677~Oph.  Here, $(B-V)_0$ are dereddened
with $E(B-V)=1.30$ as explained in Section \ref{v2677_oph_cmd}.
Figure \ref{v2677_oph_v1065_cen_lv_vul_v_bv_ub_logscale} shows
the light/color curves of V2677~Oph, LV~Vul, and V1065~Cen.  
Applying Equation (\ref{distance_modulus_general_temp}) to them,
we have the relation 
\begin{eqnarray}
(m&-&M)_{V, \rm V2677~Oph} \cr
&=& (m - M + \Delta V)_{V, \rm LV~Vul} - 2.5 \log 0.68 \cr
&=& 11.85 + 6.9\pm0.2 + 0.43 = 19.18\pm0.2 \cr
&=& (m - M + \Delta V)_{V, \rm V1065~Cen} - 2.5 \log 0.68 \cr
&=& 15.0 + 3.8\pm0.2 + 0.43 = 19.23\pm0.2,
\label{distance_modulus_v2677_oph}
\end{eqnarray}
where we adopt 
$(m-M)_{V, \rm LV~Vul}=11.85$ from \citet{hac19k} and
$(m-M)_{V, \rm V1065~Cen}=15.0$ from \citet{hac18k}.
Thus, we obtain $(m-M)_V=19.2\pm0.1$ and $f_{\rm s}=0.68$ against LV~Vul.
From Equations (\ref{time-stretching_general}),
(\ref{distance_modulus_general_temp}), and
(\ref{distance_modulus_v2677_oph}),
we have the relation
\begin{eqnarray}
(m- M')_{V, \rm V2677~Oph} 
&\equiv & (m_V - (M_V - 2.5\log f_{\rm s}))_{\rm V2677~Oph} \cr
&=& \left( (m-M)_V + \Delta V \right)_{\rm LV~Vul} \cr
&=& 11.85 + 6.9\pm0.2 = 18.75\pm0.2.
\label{absolute_mag_v2677_oph}
\end{eqnarray}

Figure \ref{v2677_oph_v834_car_yy_dor_lmcn_2009a_b_v_i_logscale_3fig}
shows the $B$, $V$, and $I_{\rm C}$ light curves of V2677~Oph 
together with those of V834~Car, YY~Dor, and LMC~N~2009a.
We apply Equation (\ref{distance_modulus_general_temp_b})
for the $B$ band to Figure
\ref{v2677_oph_v834_car_yy_dor_lmcn_2009a_b_v_i_logscale_3fig}(a)
and obtain
\begin{eqnarray}
(m&-&M)_{B, \rm V2677~Oph} \cr
&=& ((m - M)_B + \Delta B)_{\rm V834~Car} - 2.5 \log 1.05 \cr
&=& 17.75 + 2.8\pm0.2 - 0.05 = 20.5\pm0.2 \cr
&=& ((m - M)_B + \Delta B)_{\rm YY~Dor} - 2.5 \log 3.5 \cr
&=& 18.98 + 2.9\pm0.2 - 1.38 = 20.5\pm0.2 \cr
&=& ((m - M)_B + \Delta B)_{\rm LMC~N~2009a} - 2.5 \log 2.24 \cr
&=& 18.98 + 2.4\pm0.2 - 0.88 = 20.5\pm0.2.
\label{distance_modulus_b_v2677_oph_v834_car_yy_dor_lmcn2009a}
\end{eqnarray}
We have $(m-M)_{B, \rm V2677~Oph}= 20.5\pm0.1$.

For the $V$ light curves in Figure
\ref{v2677_oph_v834_car_yy_dor_lmcn_2009a_b_v_i_logscale_3fig}(b),
we similarly obtain
\begin{eqnarray}
(m&-&M)_{V, \rm V2677~Oph} \cr   
&=& ((m - M)_V + \Delta V)_{\rm V834~Car} - 2.5 \log 1.05 \cr
&=& 17.25 + 2.0\pm0.2 - 0.05 = 19.2\pm0.2 \cr
&=& ((m - M)_V + \Delta V)_{\rm YY~Dor} - 2.5 \log 3.5 \cr
&=& 18.86 + 1.7\pm0.2 - 1.38 = 19.18\pm0.2 \cr
&=& ((m - M)_V + \Delta V)_{\rm LMC~N~2009a} - 2.5 \log 2.24 \cr
&=& 18.86 + 1.2\pm0.2 - 0.88 = 19.18\pm0.2.
\label{distance_modulus_v_v2677_oph_v834_car_yy_dor_lmcn2009a}
\end{eqnarray}
We have $(m-M)_{V, \rm V2677~Oph}= 19.18\pm0.1$, which is
consistent with Equation (\ref{distance_modulus_v2677_oph}).

We apply Equation (\ref{distance_modulus_general_temp_i}) for
the $I_{\rm C}$ band to Figure
\ref{v2677_oph_v834_car_yy_dor_lmcn_2009a_b_v_i_logscale_3fig}(c) and obtain
\begin{eqnarray}
(m&-&M)_{I, \rm V2677~Oph} \cr
&=& ((m - M)_I + \Delta I_C)_{\rm V834~Car} - 2.5 \log 1.05 \cr
&=& 16.45 + 0.7\pm0.2 - 0.05 = 17.1\pm 0.2 \cr
&=& ((m - M)_I + \Delta I_C)_{\rm YY~Dor} - 2.5 \log 3.5 \cr
&=& 18.67 - 0.2\pm0.2 - 1.38 = 17.09\pm 0.2 \cr
&=& ((m - M)_I + \Delta I_C)_{\rm LMC~N~2009a} - 2.5 \log 2.24 \cr
&=& 18.67 - 0.7\pm0.2 - 0.88 = 17.09\pm 0.2.
\label{distance_modulus_i_v2677_oph_v834_car_yy_dor_lmcn2009a}
\end{eqnarray}
We have $(m-M)_{I, \rm V2677~Oph}= 17.1\pm0.1$.

We plot $(m-M)_B= 20.5$, $(m-M)_V= 19.18$, and $(m-M)_I= 17.1$,
which cross at $d=10.7$~kpc and $E(B-V)=1.30$, in Figure
\ref{distance_reddening_v1368_cen_v2676_oph_v5589_sgr_v2677_oph}(d).
Thus, we obtain $E(B-V)=1.30\pm0.10$ and $d=10.7\pm2$~kpc.


\begin{figure}
\epsscale{0.75}
\plotone{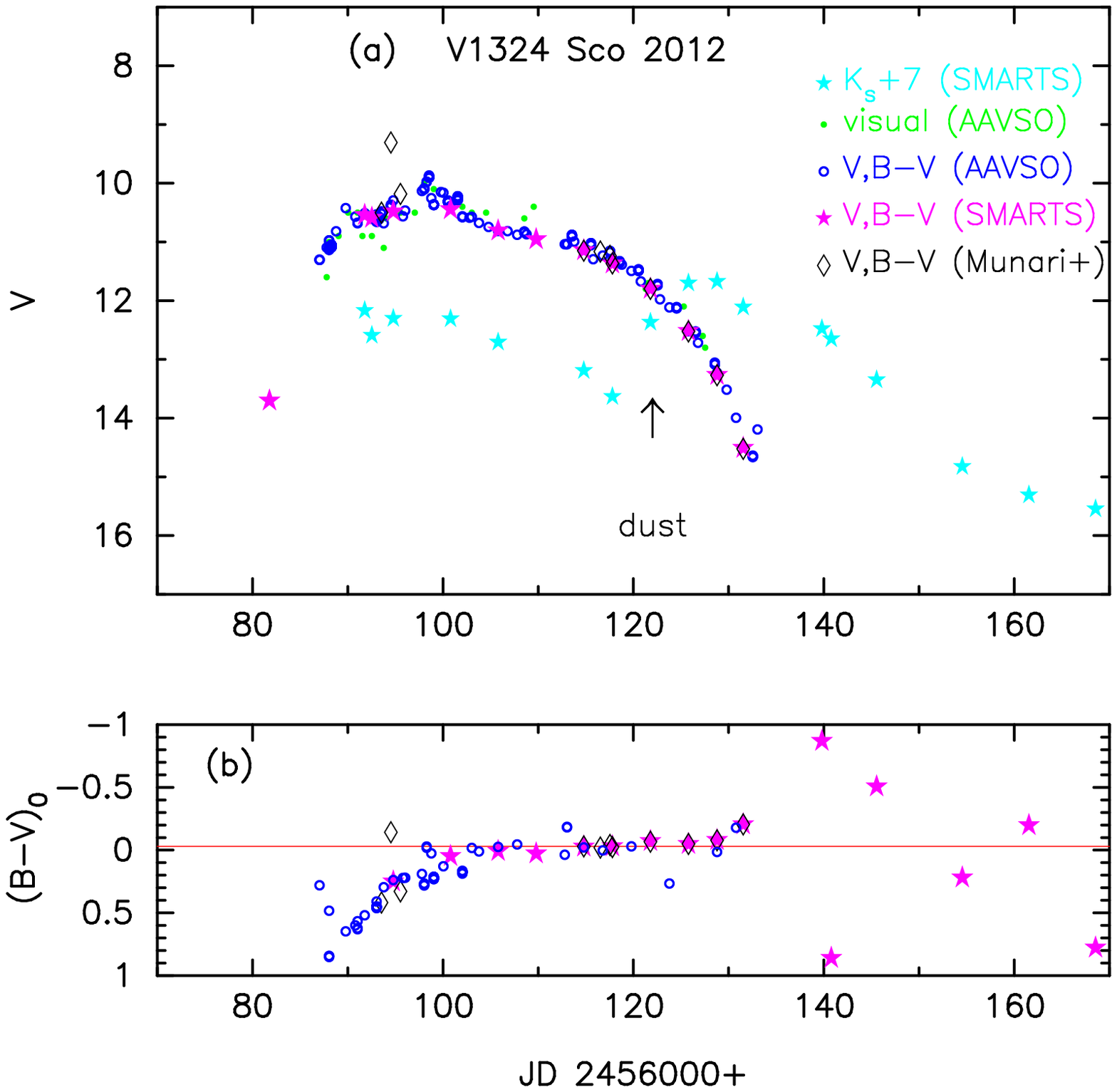}
\caption{
Same as Figure \ref{v1663_aql_v_bv_ub_color_curve}, but for V1324~Sco.
(a) The visual data (green dots) are taken from AAVSO.
The $BV$ data are taken from AAVSO (unfilled blue circles), 
SMARTS (filled magenta stars), and \citet[][unfilled black diamonds]{mun15wh}.
The $K_{\rm s}$ data are taken from SMARTS (filled cyan stars).
The rise of the $K_{\rm s}$ magnitude indicates the formation of a dust shell
as shown by the arrow labeled ``dust.''
(b) The $(B-V)_0$ are dereddened with $E(B-V)=1.32$.
\label{v1324_sco_v_bv_ub_color_curve}}
\end{figure}


\begin{figure}
\epsscale{0.75}
\plotone{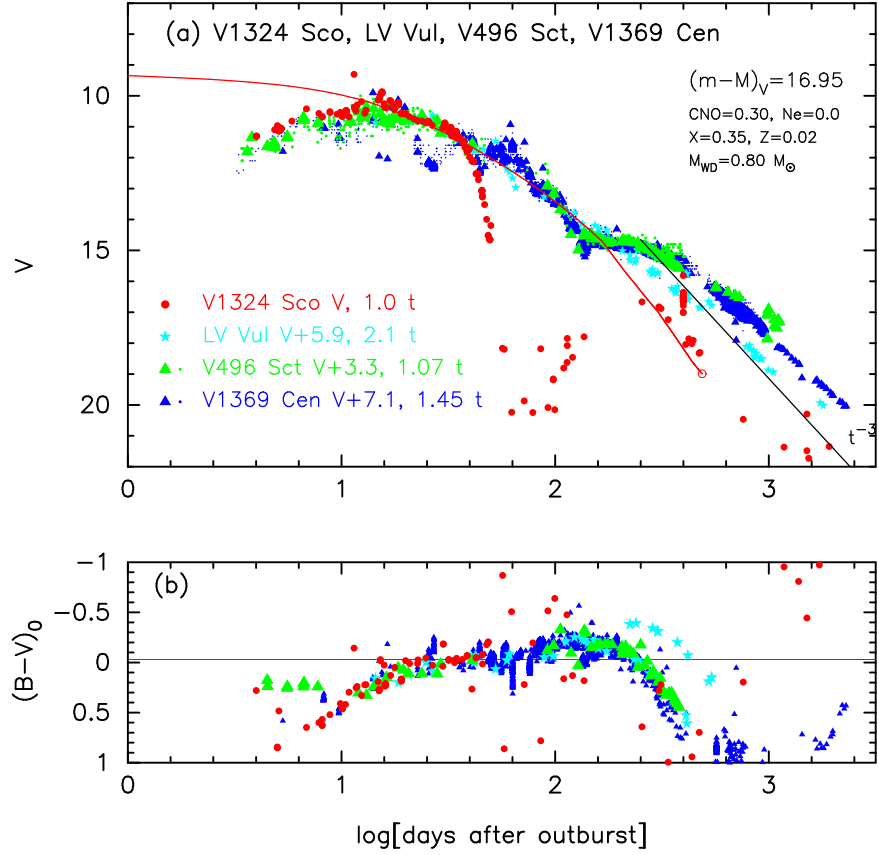}
\caption{
Same as Figure
\ref{v2575_oph_v1668_cyg_lv_vul_v_bv_ub_logscale},
but for V1324~Sco (filled red circles).  The data of V1324~Sco are
the same as those in Figure \ref{v1324_sco_v_bv_ub_color_curve}.
We add the model $V$ light curve of a $0.80~M_\sun$ WD 
\citep[CO2, solid red line;][]{hac10k}, 
assuming $(m-M)_V= 16.95$ for V1324~Sco.
\label{v1324_sco_lv_vul_v496_sct_v1369_cen_v_bv_ub_color_logscale}}
\end{figure}


\begin{figure}
\epsscale{0.65}
\plotone{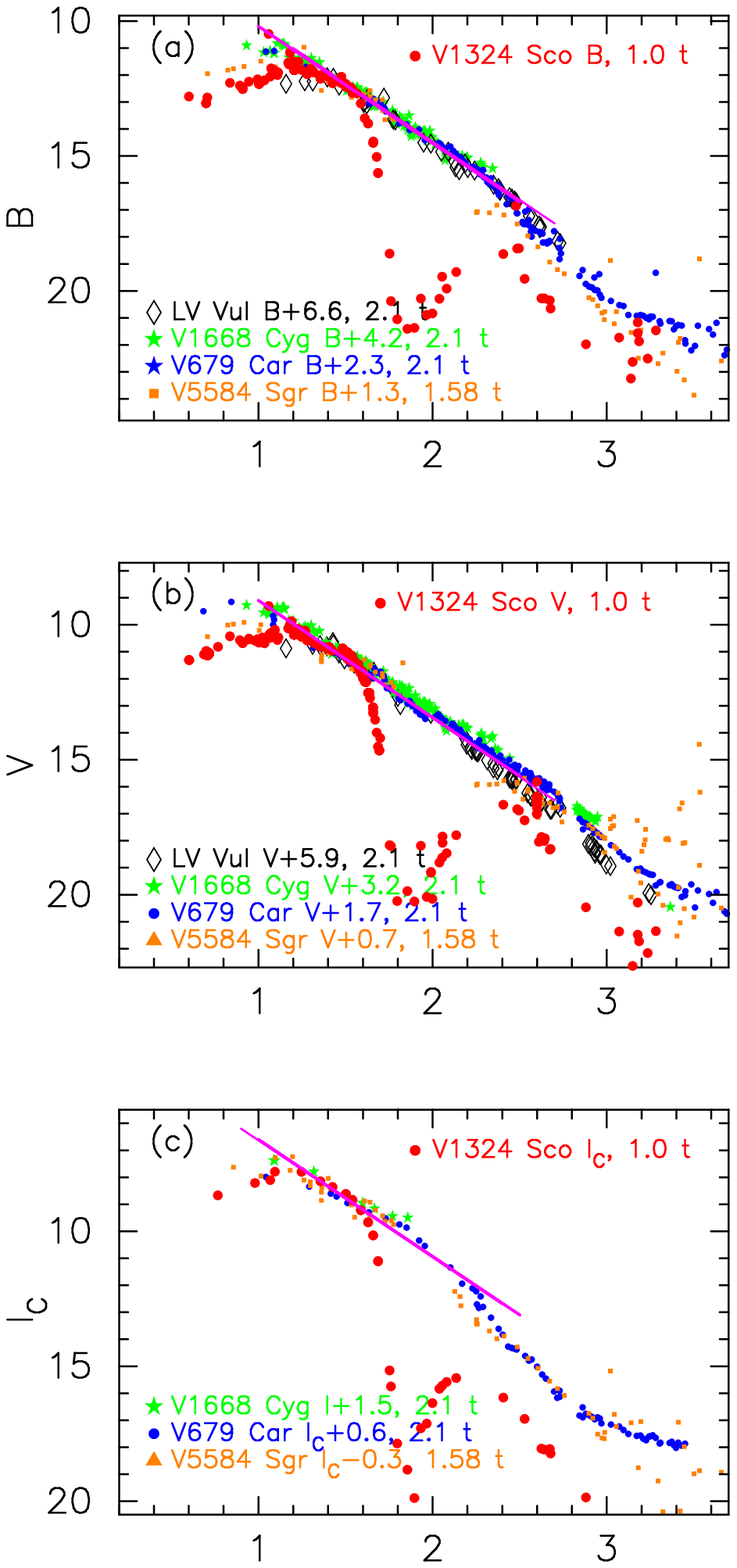}
\caption{
Same as Figure \ref{v1663_aql_yy_dor_lmcn_2009a_b_v_i_logscale_3fig},
but for V1324~Sco.
The (a) $B$, (b) $V$, and (c) $I_{\rm C}$  light curves of V1324~Sco
as well as those of V5584~Sgr, V679~Car, LV~Vul, and V1668~Cyg.
The $BV$ data of V1324~Sco are the same as those in Figure
\ref{v1324_sco_v_bv_ub_color_curve}.
The $I_{\rm C}$ data of V1324~Sco are taken from \citet{mun15wh},
AAVSO, and SMARTS.
No $I_{\rm C}$ data are available for LV~Vul.
\label{v1324_sco_v5584_sgr_v679_car_lv_vul_v1668_cyg_b_v_i_logscale_3fig}}
\end{figure}

\subsection{V1324~Sco 2012}
\label{v1324_sco}
Figure \ref{v1324_sco_v_bv_ub_color_curve} shows the (a) visual, $V$,
$K_{\rm s}$, and (b) $(B-V)_0$ evolutions of V1324~Sco.  Here, $(B-V)_0$ are
dereddened with $E(B-V)=1.32$ as obtained in Section \ref{v1324_sco_cmd}.
Figure \ref{v1324_sco_lv_vul_v496_sct_v1369_cen_v_bv_ub_color_logscale} 
shows the light/color curves of V1324~Sco 
as well as those of LV~Vul, V496~Sct, and V1369~Cen.
The $V$ light and $(B-V)_0$ color curves of these four novae overlap
each other except for the dust blackout phase.
Applying Equation (\ref{distance_modulus_general_temp}) to them,
we have the relation 
\begin{eqnarray}
(m&-&M)_{V, \rm V1324~Sco} \cr
&=& (m - M + \Delta V)_{V, \rm LV~Vul} - 2.5 \log 2.1 \cr
&=& 11.85 + 5.9\pm0.3 - 0.82 = 16.93\pm0.3 \cr
&=& (m - M + \Delta V)_{V, \rm V496~Sct} - 2.5 \log 1.07 \cr
&=& 13.7 + 3.3\pm0.3 - 0.07 = 16.93\pm0.3 \cr
&=& (m - M + \Delta V)_{V, \rm V1369~Cen} - 2.5 \log 1.45 \cr
&=& 10.25 + 7.1\pm0.3 - 0.4 = 16.95\pm0.3,
\label{distance_modulus_v1324_sco}
\end{eqnarray}
where we adopt $(m-M)_{V, \rm LV~Vul}=11.85$,
$(m-M)_{V, \rm V496~Sct}=13.7$, and
$(m-M)_{V, \rm V1369~Cen}=10.25$ from \citet{hac19k}.
Thus, we obtain $(m-M)_V=16.95\pm0.2$ for V1324~Sco.
From Equations (\ref{time-stretching_general}),
(\ref{distance_modulus_general_temp}), and
(\ref{distance_modulus_v1324_sco}),
we have the relation
\begin{eqnarray}
(m- M')_{V, \rm V1324~Sco} 
&\equiv & (m_V - (M_V - 2.5\log f_{\rm s}))_{\rm V1324~Sco} \cr
&=& \left( (m-M)_V + \Delta V \right)_{\rm LV~Vul} \cr
&=& 11.85 + 5.9\pm0.3 = 17.75\pm0.3.
\label{absolute_mag_v1324_sco}
\end{eqnarray}

Figure 
\ref{v1324_sco_v5584_sgr_v679_car_lv_vul_v1668_cyg_b_v_i_logscale_3fig} 
shows the $B$, $V$, and $I_{\rm C}$ light curves of V1324~Sco 
together with those of V679~Car, V5584~Sgr, LV~Vul, and V1668~Cyg.
We apply Equation (\ref{distance_modulus_general_temp_b})
for the $B$ band to Figure
\ref{v1324_sco_v5584_sgr_v679_car_lv_vul_v1668_cyg_b_v_i_logscale_3fig}(a)
and obtain
\begin{eqnarray}
(m&-&M)_{B, \rm V1324~Sco} \cr
&=& ((m - M)_B + \Delta B)_{\rm LV~Vul} - 2.5 \log 2.1 \cr
&=& 12.45 + 6.6\pm0.2 - 0.82 = 18.23\pm0.2 \cr
&=& ((m - M)_B + \Delta B)_{\rm V1668~Cyg} - 2.5 \log 2.1 \cr
&=& 14.9 + 4.2\pm0.2 - 0.82 = 18.28\pm0.2  \cr
&=& ((m - M)_B + \Delta B)_{\rm V679~Car} - 2.5 \log 2.1 \cr
&=& 16.79 + 2.3\pm0.2 - 0.82 = 18.27\pm0.2  \cr
&=& ((m - M)_B + \Delta B)_{\rm V5584~Sgr} - 2.5 \log 1.58 \cr
&=& 17.4 + 1.3\pm0.2 - 0.5 = 18.2\pm0.2,
\label{distance_modulus_b_v1324_sco_v2676_oph_v5584_sgr_lv_vul_v1668_cyg}
\end{eqnarray}
where we adopt $(m-M)_{B, \rm LV~Vul}= 11.85 + 0.60= 12.45$,
$(m-M)_{B, \rm V1668~Cyg}= 14.6 + 0.30= 14.9$, and 
$(m-M)_{B, \rm V679~Car}= 16.1 + 0.69= 16.79$ from \citet{hac19k},
and $(m-M)_{B, \rm V5584~Sgr}= 16.7 + 0.70= 17.4$ from Appendix
\ref{v5584_sgr}.
We have $(m-M)_{B, \rm V1324~Sco}= 18.24\pm0.1$.

For the $V$ light curves in Figure
\ref{v1324_sco_v5584_sgr_v679_car_lv_vul_v1668_cyg_b_v_i_logscale_3fig}(b),
we similarly obtain
\begin{eqnarray}
(m&-&M)_{V, \rm V1324~Sco} \cr
&=& ((m - M)_V + \Delta V)_{\rm LV~Vul} - 2.5 \log 2.1 \cr
&=& 11.85 + 5.9\pm0.2 - 0.82 = 16.93\pm0.2 \cr
&=& ((m - M)_V + \Delta V)_{\rm V1668~Cyg} - 2.5 \log 2.1 \cr
&=& 14.6 + 3.2\pm0.2 - 0.82 = 16.98\pm0.2  \cr
&=& ((m - M)_V + \Delta V)_{\rm V679~Car} - 2.5 \log 2.1 \cr
&=& 16.1 + 1.7\pm0.2 - 0.82 = 16.98\pm0.2 \cr
&=& ((m - M)_V + \Delta V)_{\rm V5584~Sgr} - 2.5 \log 1.58 \cr
&=& 16.7 + 0.7\pm0.2 - 0.5 = 16.9\pm0.2,
\label{distance_modulus_v_v1324_sco_v2676_oph_v5584_sgr_lv_vul_v1668_cyg}
\end{eqnarray}
where we adopt $(m-M)_{V, \rm LV~Vul}= 11.85$,
$(m-M)_{V, \rm V1668~Cyg}= 14.6$, and $(m-M)_{V, \rm V679~Car}= 16.1$
from \citet{hac19k},
$(m-M)_{V, \rm V5584~Sgr}= 16.7$ from Appendix \ref{v5584_sgr}.
We have $(m-M)_{V, \rm V1324~Sco}= 16.95\pm0.1$, which is
consistent with Equation (\ref{distance_modulus_v1324_sco}).

We apply Equation (\ref{distance_modulus_general_temp_i}) for
the $I_{\rm C}$ band to Figure
\ref{v1324_sco_v5584_sgr_v679_car_lv_vul_v1668_cyg_b_v_i_logscale_3fig}(c)
and obtain
\begin{eqnarray}
(m&-&M)_{I, \rm V1324~Sco} \cr
&=& ((m - M)_I + \Delta I_C)_{\rm V1668~Cyg} - 2.5 \log 2.1 \cr
&=& 14.12 + 1.5\pm0.2 - 0.82 = 14.8\pm 0.2 \cr
&=& ((m - M)_I + \Delta I_C)_{\rm V679~Car} - 2.5 \log 2.1 \cr
&=& 15.0 + 0.6\pm0.2 - 0.82 = 14.78\pm 0.2 \cr
&=& ((m - M)_I + \Delta I_C)_{\rm V5584~Sgr} - 2.5 \log 1.58 \cr
&=& 15.58 - 0.3\pm0.2 - 0.5 = 14.78\pm 0.2,
\label{distance_modulus_i_v1324_sco_v2676_oph_v5584_sgr_lv_vul_v1668_cyg}
\end{eqnarray}
where we adopt 
$(m-M)_{I, \rm V1668~Cyg}= 14.6 - 1.6\times 0.30= 14.12$ and
$(m-M)_{I, \rm V679~Car}= 16.1 - 1.6\times 0.69= 15.0$
from \citet{hac19k}, and
$(m-M)_{I, \rm V5584~Sgr}= 16.7 - 1.6\times 0.70= 15.58$ from
Appendix \ref{v5584_sgr}.
Unfortunately, neither $I_{\rm C}$ nor $I$ data of
LV~Vul are available.
We have $(m-M)_{I, \rm V1324~Sco}= 14.78\pm0.2$.

We plot $(m-M)_B= 18.24$, $(m-M)_V= 16.95$, and $(m-M)_I= 14.78$,
which cross at $d=3.7$~kpc and $E(B-V)=1.32$, in Figure
\ref{distance_reddening_v1324_sco_v5592_sgr_v962_cep_v2659_cyg}(a).
Thus, we obtain $E(B-V)=1.32\pm0.1$ and $d=3.7\pm0.6$~kpc.


\begin{figure}
\epsscale{0.75}
\plotone{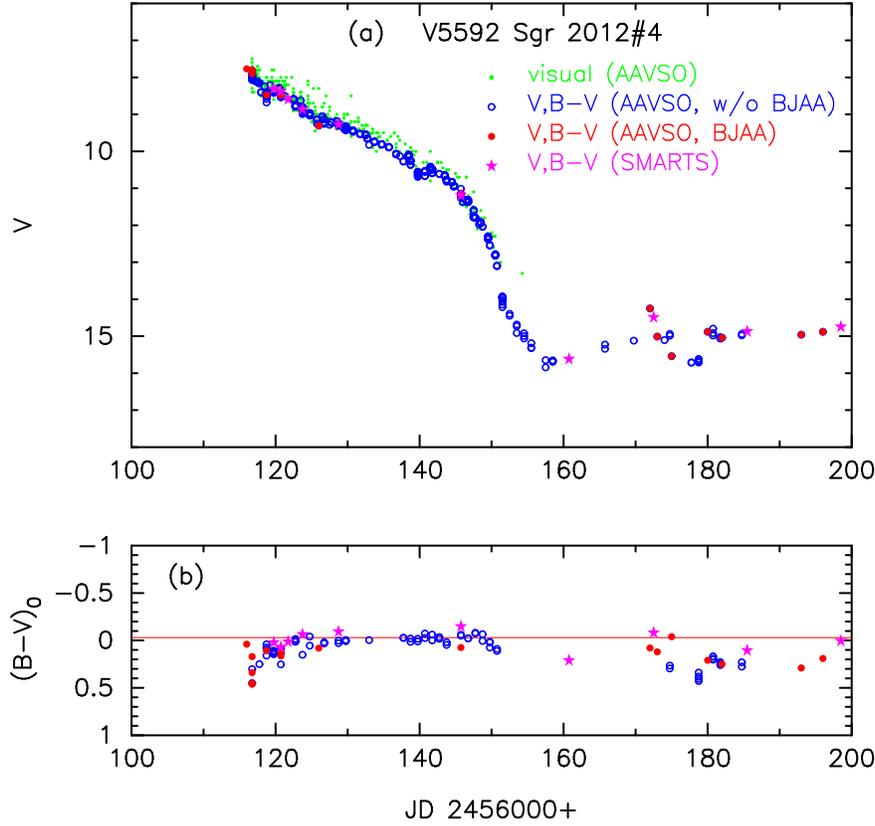}
\caption{
Same as Figure 
\ref{v1663_aql_v_bv_ub_color_curve}, but for V5592~Sgr.
(a) The visual data (green dots) are taken from AAVSO.
The $BV$ data are taken from AAVSO (unfilled blue circles and filled
red circles) and SMARTS (filled magenta stars).
(b) The $(B-V)_0$ are dereddened with $E(B-V)=0.33$.
The $B-V$ of BJAA's data (filled red circles) taken from
AAVSO are shifted redward by 0.20 mag.   See the text for details.
\label{v5592_sgr_v_bv_ub_color_curve}}
\end{figure}


\begin{figure}
\epsscale{0.75}
\plotone{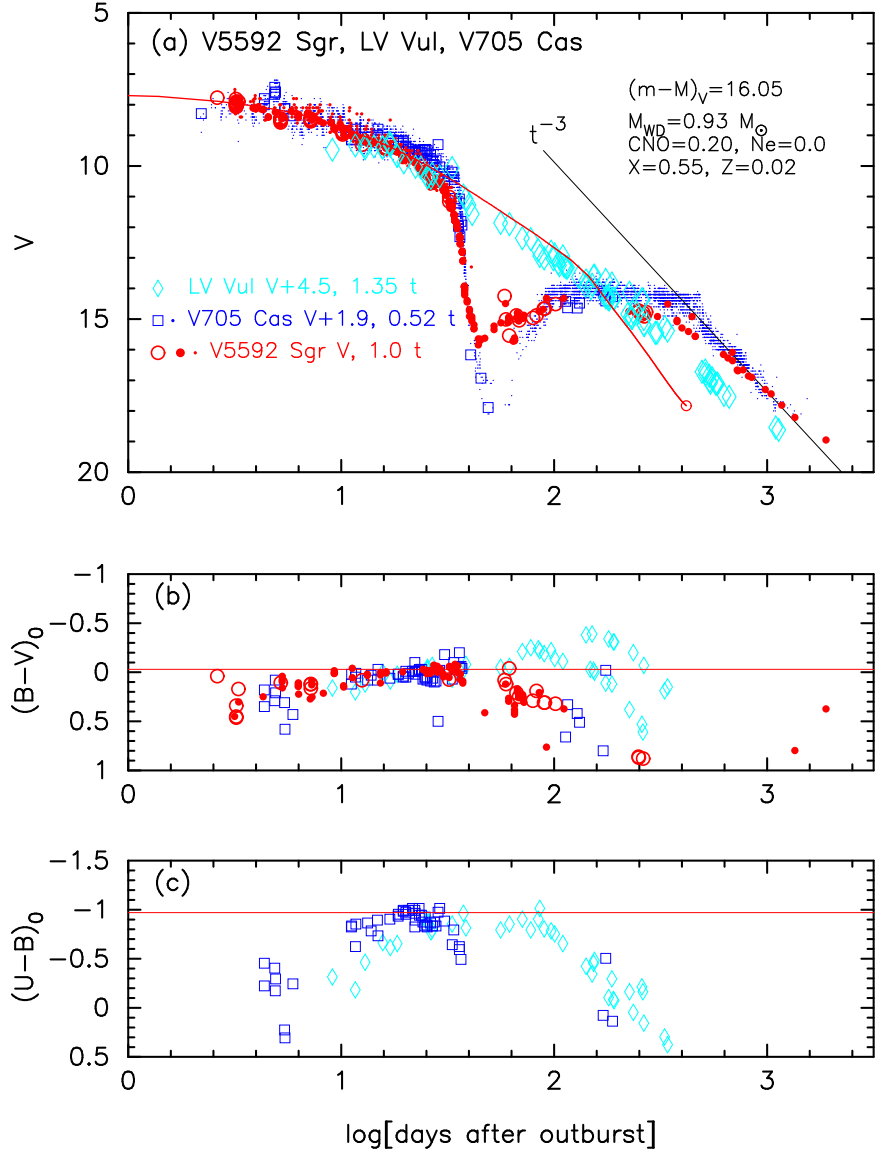}
\caption{
Same as Figure 
\ref{v2575_oph_v1668_cyg_lv_vul_v_bv_ub_logscale},
but for V5592~Sgr (red dots for visual, unfilled red circles and filled red
squares for $V$).  The data of V5592~Sgr are the same as those in Figure 
\ref{v5592_sgr_v_bv_ub_color_curve}.
We add the model $V$ light curve of a $0.93~M_\sun$ WD 
\citep[CO4, solid red line, ][]{hac15k},
assuming $(m-M)_V= 16.05$ for V5592~Sgr.
\label{v5592_sgr_v705_cas_lv_vul_v_bv_ub_color_logscale}}
\end{figure}


\begin{figure}
\epsscale{0.65}
\plotone{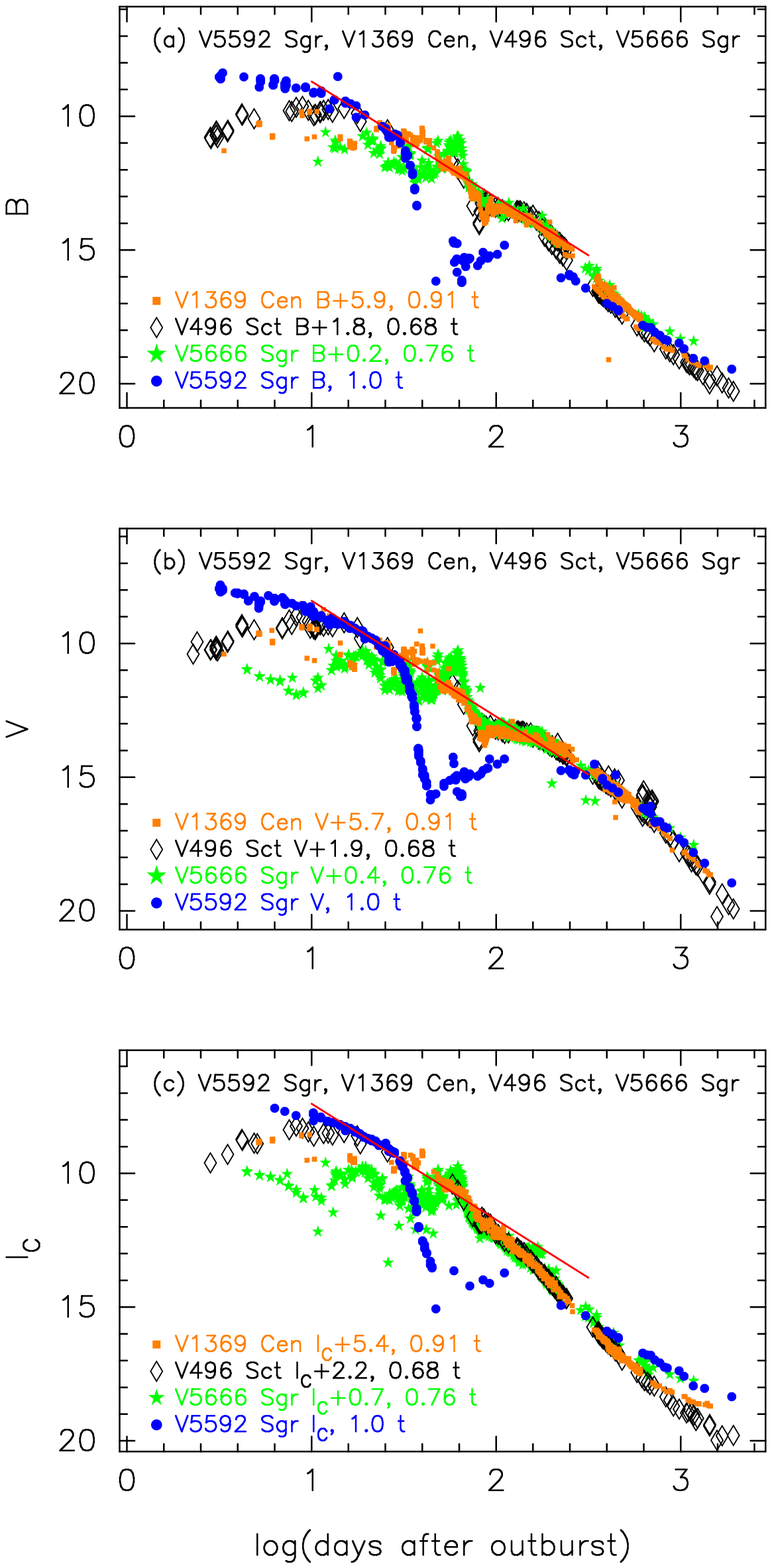}
\caption{
Same as Figure \ref{v1663_aql_yy_dor_lmcn_2009a_b_v_i_logscale_3fig},
but for V5592~Sgr.   We plot
the (a) $B$, (b) $V$, and (c) $I_{\rm C}$ light curves of V5592~Sgr
as well as those of V1369~Cen, V496~Sct, and V5666~Sgr.
The $BV$ data of V5592~Sgr are the same as those in Figure
\ref{v5592_sgr_v_bv_ub_color_curve}.  The $I_{\rm C}$ data of V5592~Sgr
are taken from AAVSO and SMARTS.
\label{v5592_sgr_v1369_cen_v496_sct_v5666_sgr_b_v_i_logscale_3fig}}
\end{figure}

\subsection{V5592~Sgr 2012\#4}
\label{v5592_sgr}
Figure \ref{v5592_sgr_v_bv_ub_color_curve} shows the (a) visual, $V$,
and (b) $(B-V)_0$ evolutions of V5592~Sgr.  Here, $(B-V)_0$ are dereddened
with $E(B-V)=0.33$ as obtained in Section \ref{v5592_sgr_cmd}.
The BJAA's (observer's ID) $B-V$ data of AAVSO are systematically bluer
by 0.20 mag than the other $B-V$ data in AAVSO, so we shifted BJAA's 
$B-V$ data by 0.20 mag redder in the figure.  As a result, these 
two $B-V$ data in the AAVSO archive overlap each other in Figure 
\ref{v5592_sgr_v_bv_ub_color_curve}(b).

Figure \ref{v5592_sgr_v705_cas_lv_vul_v_bv_ub_color_logscale} 
shows the light/color curves of V5592~Sgr as well as those of
LV~Vul and V705~Cas.
We overlap these $V$ light and $(B-V)_0$ color curves.
Applying Equation (\ref{distance_modulus_general_temp}) to them,
we have the relation 
\begin{eqnarray}
(m&-&M)_{V, \rm V5592~Sgr} \cr
&=& (m - M + \Delta V)_{V, \rm LV~Vul} - 2.5 \log 1.35 \cr
&=& 11.85 + 4.5\pm0.3 - 0.33 = 16.02\pm0.3 \cr
&=& (m - M + \Delta V)_{V, \rm V705~Cas} - 2.5 \log 0.52 \cr
&=& 13.45 + 1.9\pm0.3 + 0.7 = 16.05\pm0.3,
\label{distance_modulus_v5592_sgr}
\end{eqnarray}
where we adopt $(m-M)_{V, \rm LV~Vul}=11.85$
and $(m-M)_{V, \rm V705~Cas}=13.45$ from \citet{hac19k}.
Thus, we obtain $(m-M)_V=16.05\pm0.2$ and $f_{\rm s}=1.35$ against LV~Vul.
From Equations (\ref{time-stretching_general}),
(\ref{distance_modulus_general_temp}), and
(\ref{distance_modulus_v5592_sgr}), we have the relation
\begin{eqnarray}
(m- M')_{V, \rm V5592~Sgr} 
&\equiv & (m_V - (M_V - 2.5\log f_{\rm s}))_{\rm V5592~Sgr} \cr
&=& \left( (m-M)_V + \Delta V \right)_{\rm LV~Vul} \cr
&=& 11.85 + 4.5\pm0.3 = 16.35\pm0.3.
\label{absolute_mag_v5592_sgr}
\end{eqnarray}

Figure \ref{v5592_sgr_v1369_cen_v496_sct_v5666_sgr_b_v_i_logscale_3fig}
shows the $B$, $V$, and $I_{\rm C}$ light curves of V5592~Sgr
together with those of V1369~Cen, V496~Sct, and V5666~Sgr.
The light curves overlap each other.
Applying Equation (\ref{distance_modulus_general_temp_b})
for the $B$ band to Figure
\ref{v5592_sgr_v1369_cen_v496_sct_v5666_sgr_b_v_i_logscale_3fig}(a),
we have the relation
\begin{eqnarray}
(m&-&M)_{B, \rm V5592~Sgr} \cr
&=& \left( (m-M)_B + \Delta B\right)_{\rm V1369~Cen} - 2.5 \log 0.91 \cr
&=& 10.36 + 5.9\pm0.2 + 0.1 = 16.36\pm0.2 \cr
&=& \left( (m-M)_B + \Delta B\right)_{\rm V496~Sct} - 2.5 \log 0.68 \cr
&=& 14.15 + 1.8\pm0.2 + 0.43 = 16.38\pm0.2 \cr
&=& \left( (m-M)_B + \Delta B\right)_{\rm V5666~Sgr} - 2.5 \log 0.76 \cr
&=& 15.9 + 0.2\pm0.2 + 0.3 = 16.4\pm0.2,
\label{distance_modulus_v5592_sgr_v1369_cen_v496_sct_v5666_sgr_b}
\end{eqnarray}
where we adopt $(m-M)_{B, \rm V1369~Cen}= 10.36$,
$(m-M)_{B, \rm V496~Sct}= 14.15$, and
$(m-M)_{B, \rm V5666~Sgr}= 15.9$ from Appendix \ref{qy_mus}.
We have $(m-M)_B=16.38\pm0.1$ for V5592~Sgr.

Applying Equation (\ref{distance_modulus_general_temp}) to
Figure \ref{v5592_sgr_v1369_cen_v496_sct_v5666_sgr_b_v_i_logscale_3fig}(b),
we have the relation
\begin{eqnarray}
(m&-&M)_{V, \rm V5592~Sgr} \cr
&=& \left( (m-M)_V + \Delta V\right)_{\rm V1369~Cen} - 2.5 \log 0.91 \cr
&=& 10.25 + 5.7\pm0.3 + 0.1 = 16.05\pm0.2 \cr
&=& \left( (m-M)_V + \Delta V\right)_{\rm V496~Sct} - 2.5 \log 0.68 \cr
&=& 13.7 + 1.9\pm0.3 + 0.43 = 16.03\pm0.2 \cr
&=& \left( (m-M)_V + \Delta V\right)_{\rm V5666~Sgr} - 2.5 \log 0.76 \cr
&=& 15.4 + 0.4\pm0.3 + 0.3  = 16.1\pm0.2,
\label{distance_modulus_v5592_sgr_v1369_cen_v496_sct_v5666_sgr_v}
\end{eqnarray}
where we adopt $(m-M)_{V, \rm V1369~Cen}=10.25$,
$(m-M)_{V, \rm V496~Sct}=13.7$, and $(m-M)_{V, \rm V5666~Sgr}=15.4$
from \citet{hac19k}.  We have $(m-M)_V=16.06\pm0.1$, which is
consistent with Equation (\ref{distance_modulus_v5592_sgr}).

From the $I_{\rm C}$-band data in Figure
\ref{v5592_sgr_v1369_cen_v496_sct_v5666_sgr_b_v_i_logscale_3fig}(c),
we obtain
\begin{eqnarray}
(m&-&M)_{I, \rm V5592~Sgr} \cr
&=& ((m - M)_I + \Delta I_C)_{\rm V1369~Cen} - 2.5 \log 0.91 \cr
&=& 10.07 + 5.4\pm0.2 + 0.1 = 15.57\pm0.2 \cr
&=& ((m - M)_I + \Delta I_C)_{\rm V496~Sct} - 2.5 \log 0.68 \cr
&=& 12.98 + 2.2\pm0.2 + 0.43 = 15.61\pm0.2 \cr
&=& ((m - M)_I + \Delta I_C)_{\rm V5666~Sgr} - 2.5 \log 0.76 \cr
&=& 14.6 + 0.7\pm0.2 + 0.3 = 15.6\pm0.2,
\label{distance_modulus_i_v5592_sgr_v1369_cen_v496_sct_v5666_sgr}
\end{eqnarray}
where we adopt $(m-M)_{I, \rm V1369~Cen}= 10.07$,
$(m-M)_{I, \rm V496~Sct}= 12.98$, and $(m-M)_{I, \rm V5666~Sgr}= 14.6$
from Appendix \ref{qy_mus}.
We obtain $(m-M)_{I, \rm V5592~Sgr}= 15.59\pm0.1$.

We plot $(m-M)_B=16.38$, $(m-M)_V=16.06$, and $(m-M)_I=15.59$,
which cross at $d=10$~kpc and $E(B-V)=0.33$, in Figure
\ref{distance_reddening_v1324_sco_v5592_sgr_v962_cep_v2659_cyg}(b).
Thus, we obtain $d=10\pm1$~kpc and $E(B-V)=0.33\pm0.05$.


\begin{figure}
\epsscale{0.75}
\plotone{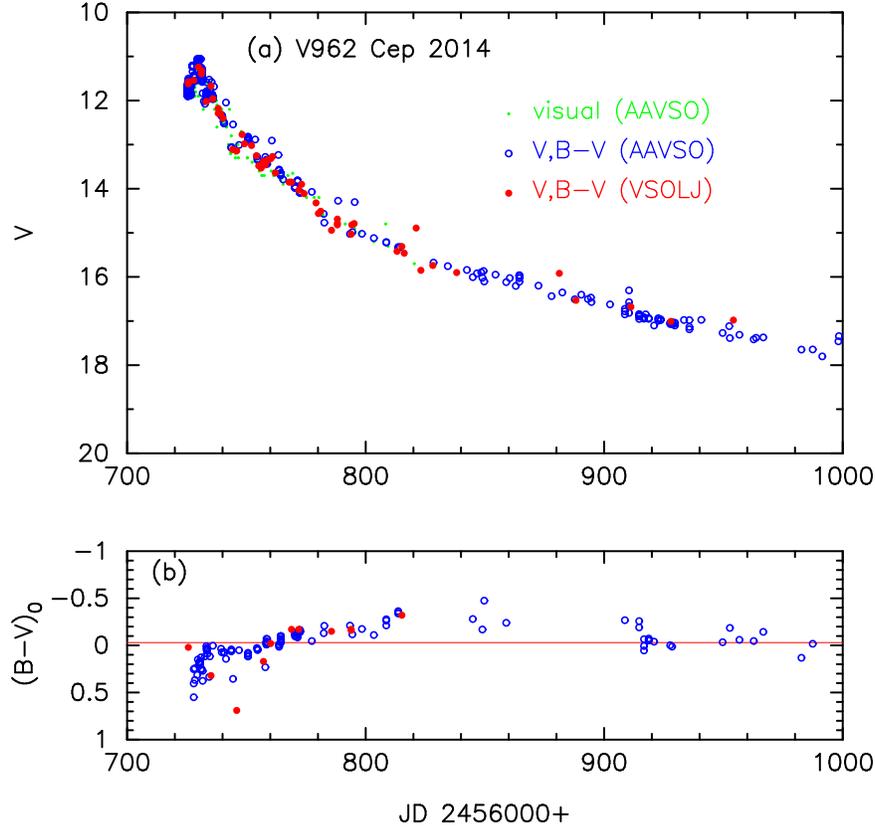}
\caption{
Same as Figure \ref{v1663_aql_v_bv_ub_color_curve}, but for V962~Cep.
(a) The visual data (green dots) are taken from AAVSO.
The $BV$ data are taken from AAVSO (unfilled blue circles)
and VSOLJ (filled red circles).
(b) The $(B-V)_0$ are dereddened with $E(B-V)=1.10$.
\label{v962_cep_v_bv_ub_color_curve}}
\end{figure}


\begin{figure}
\epsscale{0.75}
\plotone{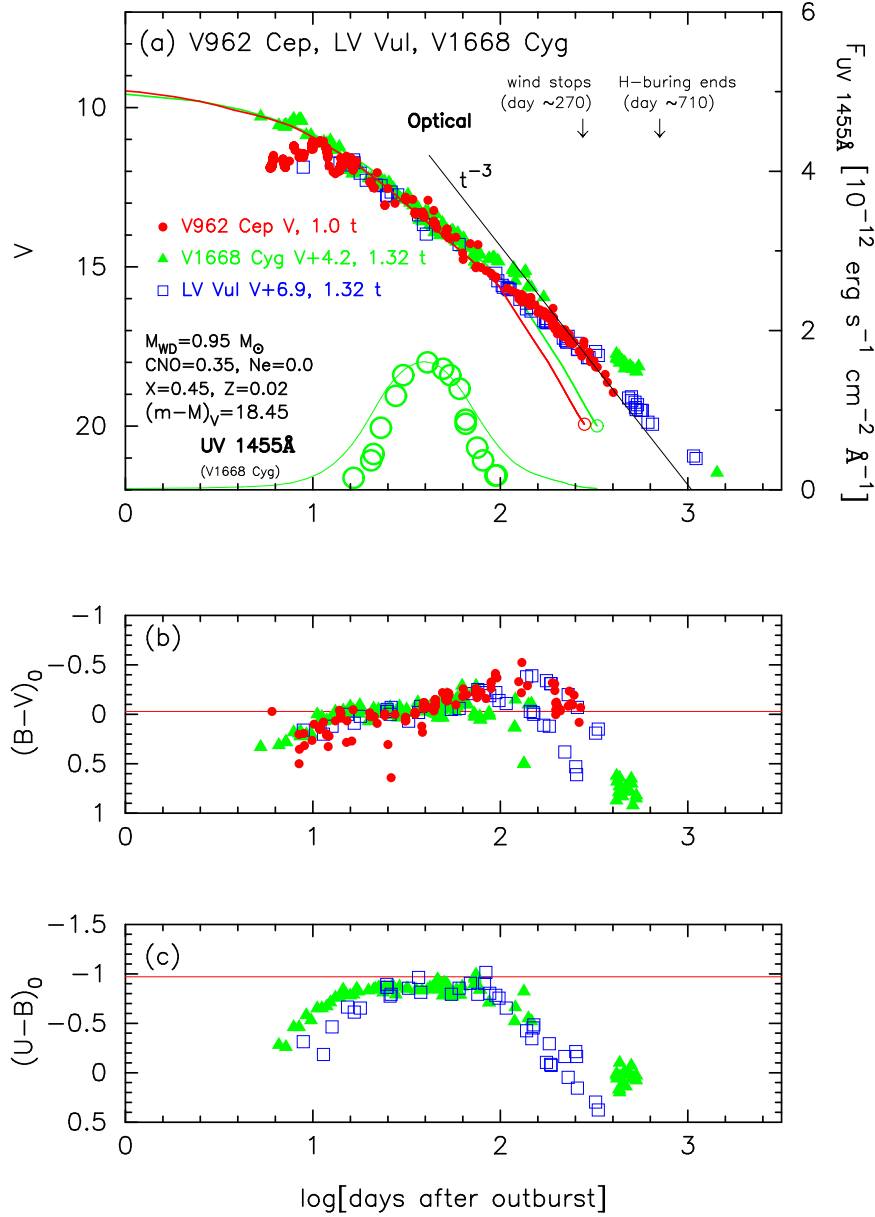}
\caption{
Same as Figure 
\ref{v2575_oph_v1668_cyg_lv_vul_v_bv_ub_logscale},
but for V962~Cep (filled red circles).  The data of V962~Cep are
the same as those in Figure \ref{v962_cep_v_bv_ub_color_curve}.
We add the model $V$ light curve of a $0.95~M_\sun$ WD 
\citep[CO3, solid red line;][]{hac16k}, 
assuming $(m-M)_V= 18.45$ for V962~Cep.  
The solid green lines denote the $V$ and UV~1455\AA\  light curves
of a $0.98~M_\sun$ WD (CO3), assuming $(m-M)_V=14.6$ for V1668~Cyg.
\label{v962_cep_v1668_cyg_lv_vul_v_bv_ub_logscale}}
\end{figure}


\begin{figure}
\epsscale{0.55}
\plotone{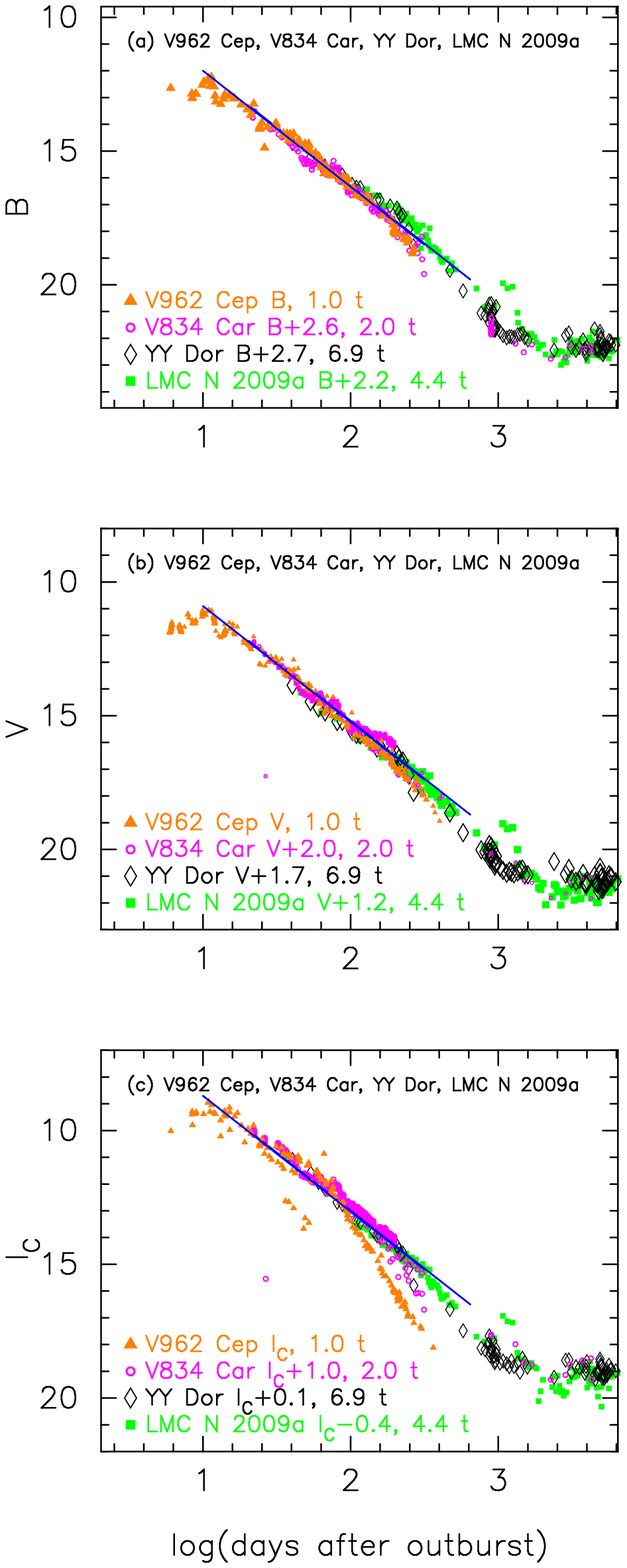}
\caption{
Same as Figure \ref{v1663_aql_yy_dor_lmcn_2009a_b_v_i_logscale_3fig},
but for V962~Cep.
The $BV$ data of V962~Cep are the same as those in Figure
\ref{v962_cep_v_bv_ub_color_curve}.  The $I_{\rm C}$ data of V962~Cep
are taken from AAVSO and VSOLJ.
\label{v962_cep_v834_car_yy_dor_lmcn_2009a_b_v_i_logscale_3fig}}
\end{figure}

\subsection{V962~Cep 2014}
\label{v962_cep}
Figure \ref{v962_cep_v_bv_ub_color_curve} shows the (a) visual, $V$,
and (b) $(B-V)_0$ evolutions of V962~Cep.  Here, $(B-V)_0$ are dereddened
with $E(B-V)=1.10$ as obtained in Section \ref{v962_cep_cmd}.  
Figure \ref{v962_cep_v1668_cyg_lv_vul_v_bv_ub_logscale} shows
the light/color curves of V962~Cep as well as those of LV~Vul and V1668~Cyg.
Applying Equation (\ref{distance_modulus_general_temp}) to them,
we have the relation 
\begin{eqnarray}
(m&-&M)_{V, \rm V962~Cep} \cr 
&=& (m - M + \Delta V)_{V, \rm LV~Vul} - 2.5 \log 1.32 \cr
&=& 11.85 + 6.9\pm0.2 - 0.30 = 18.45\pm0.2 \cr
&=& (m - M + \Delta V)_{V, \rm V1668~Cyg} - 2.5 \log 1.32 \cr
&=& 14.6 + 4.2\pm0.2 - 0.30 = 18.50\pm0.2,
\label{distance_modulus_v962_cep}
\end{eqnarray}
where we adopt $(m-M)_{V, \rm LV~Vul}=11.85$ and 
$(m-M)_{V, \rm V1668~Cyg}=14.6$ from \citet{hac19k}.
Thus, we obtain $(m-M)_V=18.45\pm0.1$ and $f_{\rm s}=1.32$ against LV~Vul.
From Equations (\ref{time-stretching_general}),
(\ref{distance_modulus_general_temp}), and
(\ref{distance_modulus_v962_cep}),
we have the relation
\begin{eqnarray}
(m- M')_{V, \rm V962~Cep} 
&\equiv & (m_V - (M_V - 2.5\log f_{\rm s}))_{\rm V962~Cep} \cr
&=& \left( (m-M)_V + \Delta V \right)_{\rm LV~Vul} \cr
&=& 11.85 + 6.9\pm0.2 = 18.75\pm0.2.
\label{absolute_mag_v962_cep}
\end{eqnarray}

Figure \ref{v962_cep_v834_car_yy_dor_lmcn_2009a_b_v_i_logscale_3fig}
shows the $B$, $V$, and $I_{\rm C}$ light curves of V962~Cep 
together with those of V834~Car, YY~Dor, and LMC~N~2009a.
We apply Equation (\ref{distance_modulus_general_temp_b})
for the $B$ band to Figure
\ref{v962_cep_v834_car_yy_dor_lmcn_2009a_b_v_i_logscale_3fig}(a)
and obtain
\begin{eqnarray}
(m&-&M)_{B, \rm V962~Cep} \cr
&=& ((m - M)_B + \Delta B)_{\rm V834~Car} - 2.5 \log 2.0 \cr
&=& 17.75 + 2.6\pm0.2 - 0.78 = 19.57\pm0.2 \cr
&=& ((m - M)_B + \Delta B)_{\rm YY~Dor} - 2.5 \log 6.9 \cr
&=& 18.98 + 2.7\pm0.2 - 2.1 = 19.58\pm0.2 \cr
&=& ((m - M)_B + \Delta B)_{\rm LMC~N~2009a} - 2.5 \log 4.4 \cr
&=& 18.98 + 2.2\pm0.2 - 1.6 = 19.58\pm0.2,
\label{distance_modulus_b_v962_cep_v834_car_yy_dor_lmcn2009a}
\end{eqnarray}
where we adopt $(m-M)_{B, \rm V834~Car}= 17.25 + 0.50 = 17.75$.
We have $(m-M)_{B, \rm V962~Cep}= 19.58\pm0.1$.

For the $V$ light curves in Figure
\ref{v962_cep_v834_car_yy_dor_lmcn_2009a_b_v_i_logscale_3fig}(b),
we similarly obtain
\begin{eqnarray}
(m&-&M)_{V, \rm V962~Cep} \cr   
&=& ((m - M)_V + \Delta V)_{\rm V834~Car} - 2.5 \log 2.0 \cr
&=& 17.25 + 2.0\pm0.2 - 0.78 = 18.47\pm0.2 \cr
&=& ((m - M)_V + \Delta V)_{\rm YY~Dor} - 2.5 \log 6.9 \cr
&=& 18.86 + 1.7\pm0.2 - 2.1 = 18.46\pm0.2 \cr
&=& ((m - M)_V + \Delta V)_{\rm LMC~N~2009a} - 2.5 \log 4.4 \cr
&=& 18.86 + 1.2\pm0.2 - 1.6 = 18.46\pm0.2,
\label{distance_modulus_v_v962_cep_v834_car_yy_dor_lmcn2009a}
\end{eqnarray}
where we adopt $(m-M)_{V, \rm V834~Car}= 17.25$.
We have $(m-M)_{V, \rm V962~Cep}= 18.46\pm0.1$, which is
consistent with Equation (\ref{distance_modulus_v962_cep}).

We apply Equation (\ref{distance_modulus_general_temp_i}) for
the $I_{\rm C}$ band to Figure
\ref{v962_cep_v834_car_yy_dor_lmcn_2009a_b_v_i_logscale_3fig}(c) and obtain
\begin{eqnarray}
(m&-&M)_{I, \rm V962~Cep} \cr
&=& ((m - M)_I + \Delta I_C)_{\rm V834~Car} - 2.5 \log 2.0 \cr
&=& 16.45 + 1.0\pm0.2 - 0.78 = 16.67\pm 0.2 \cr
&=& ((m - M)_I + \Delta I_C)_{\rm YY~Dor} - 2.5 \log 6.9 \cr
&=& 18.67 + 0.1\pm0.2 - 2.1 = 16.67\pm 0.2 \cr
&=& ((m - M)_I + \Delta I_C)_{\rm LMC~N~2009a} - 2.5 \log 4.4 \cr
&=& 18.67 - 0.4\pm0.2 - 1.6 = 16.67\pm 0.2,
\label{distance_modulus_i_v962_cep_v834_car_yy_dor_lmcn2009a}
\end{eqnarray}
where we adopt $(m-M)_{I, \rm V834~Car}= 17.25 - 1.6\times 0.50 = 16.45$.
We have $(m-M)_{I, \rm V962~Cep}= 16.67\pm0.1$.

We plot $(m-M)_B= 19.58$, $(m-M)_V= 18.46$, and $(m-M)_I= 16.67$,
which cross at $d=10.2$~kpc and $E(B-V)=1.10$, in Figure
\ref{distance_reddening_v1324_sco_v5592_sgr_v962_cep_v2659_cyg}(c).
Thus, we obtain $E(B-V)=1.10\pm0.10$ and $d=10.2\pm2$~kpc.


\begin{figure}
\epsscale{0.75}
\plotone{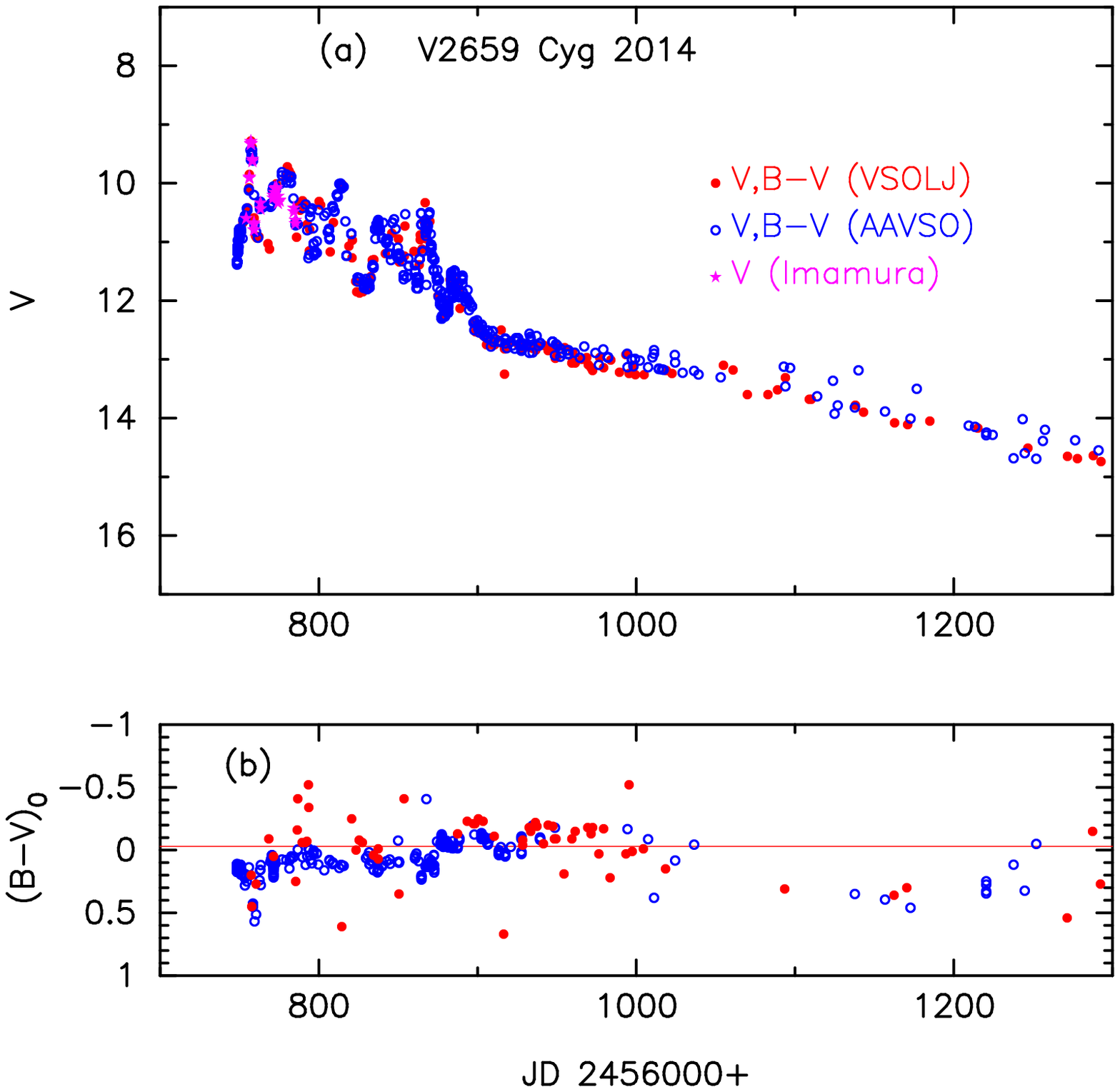}
\caption{
Same as Figure \ref{v1663_aql_v_bv_ub_color_curve}, but for V2659~Cyg.
(a) The $V$ data (filled magenta stars) are taken from \citet{ima14}.
The $BV$ data are from AAVSO (unfilled blue circles) and 
VSOLJ (filled red circles).
(b) The $(B-V)_0$ are dereddened with $E(B-V)=0.80$.
\label{v2659_cyg_v_bv_ub_color_curve}}
\end{figure}


\begin{figure}
\epsscale{0.75}
\plotone{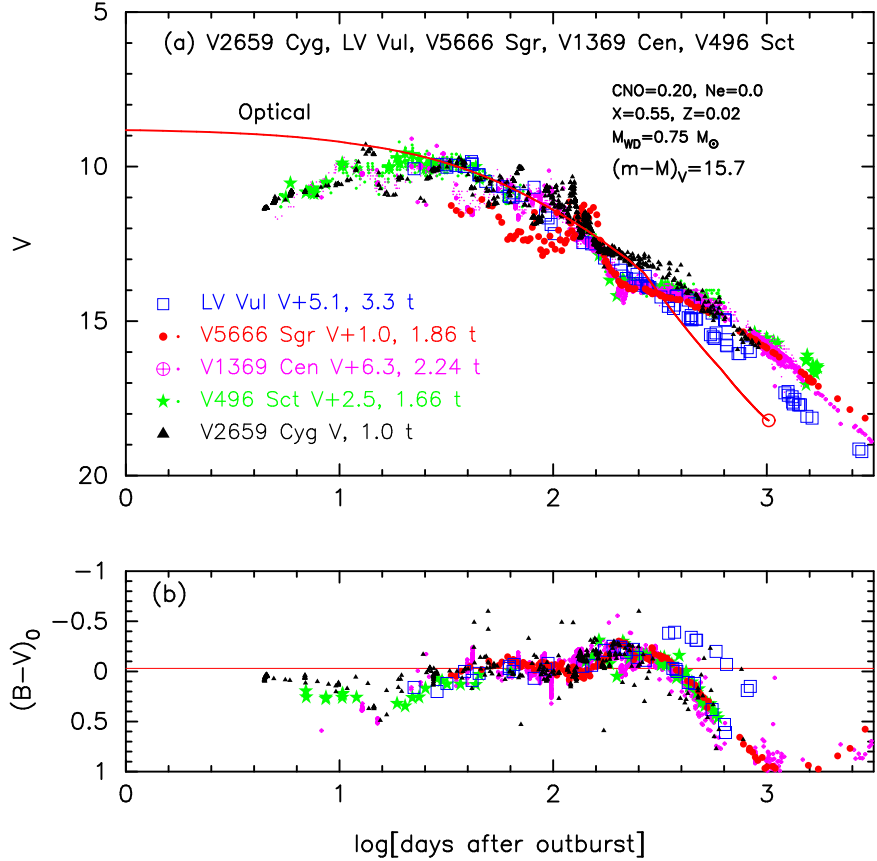}
\caption{
Same as Figure 
\ref{v2575_oph_v1668_cyg_lv_vul_v_bv_ub_logscale},
but for V2659~Cyg (filled black triangles).  We add the light/color curves
of LV~Vul, V5666~Sgr, V1369~Cen, and V496~Sct.  The data of V2659~Cyg are
the same as those in Figure \ref{v2659_cyg_v_bv_ub_color_curve}.
In panel (a), we add the model $V$ light curve of a $0.75~M_\sun$ WD
\citep[CO4, solid red line;][]{hac15k}, 
assuming $(m-M)_V= 15.7$ for V2659~Cyg.  
\label{v2659_cyg_v5666_sgr_v1369_cen_v496_sct_lv_vul_v_bv_ub_color_logscale}}
\end{figure}


\begin{figure}
\epsscale{0.65}
\plotone{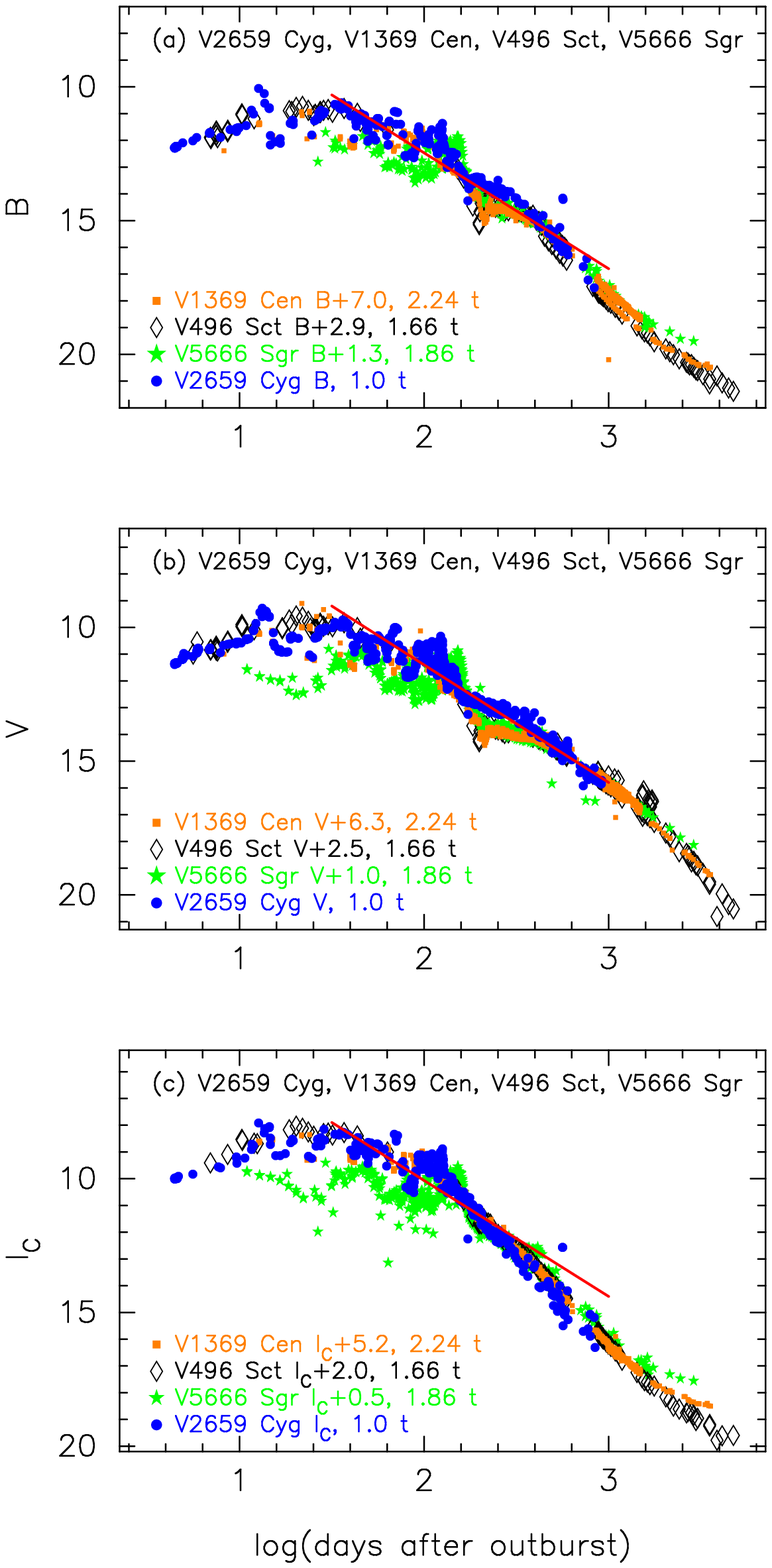}
\caption{
Same as Figure \ref{v1663_aql_yy_dor_lmcn_2009a_b_v_i_logscale_3fig},
but for V2659~Cyg.   We plot
the (a) $B$, (b) $V$, and (c) $I_{\rm C}$ light curves of V2659~Cyg
as well as those of V1369~Cen, V496~Sct, and V5666~Sgr.
The $BV$ data of V2659~Cyg are the same as those in Figure
\ref{v2659_cyg_v_bv_ub_color_curve}.  The $I_{\rm C}$ data of V2659~Cyg
are taken from AAVSO and VSOLJ.
\label{v2659_cyg_v1369_cen_v496_sct_v5666_sgr_b_v_i_logscale_3fig}}
\end{figure}

\subsection{V2659~Cyg 2014}
\label{v2659_cyg}
Figure \ref{v2659_cyg_v_bv_ub_color_curve} shows the (a) $V$ light and
(b) $(B-V)_0$ color curves of V2659~Cyg on a linear timescale.  Here, 
$(B-V)_0$ are dereddened with $E(B-V)=0.80$ as explained in 
Section \ref{v2659_cyg_cmd}.  Figure
\ref{v2659_cyg_v5666_sgr_v1369_cen_v496_sct_lv_vul_v_bv_ub_color_logscale}
shows the $V$ light and $(B-V)_0$ color curves of V2659~Cyg, 
LV~Vul, V5666~Sgr, V1369~Cen, and V496~Sct.
The light curves of V2659~Cyg, V5666~Sgr, V1369~Cen, and V496~Sct
are so similar to each other.  
Applying Equation (\ref{distance_modulus_general_temp}) to them,
we have the relation 
\begin{eqnarray}
(m&-&M)_{V, \rm V2659~Cyg} \cr
&=& (m-M + \Delta V)_{V, \rm LV~Vul} - 2.5 \log 3.3 \cr
&=& 11.85 + 5.1\pm0.3 - 1.30 = 15.65\pm0.3 \cr
&=& (m-M + \Delta V)_{V, \rm V5666~Sgr} - 2.5 \log 1.86 \cr
&=& 15.4 + 1.0\pm0.3 - 0.68 = 15.72\pm0.3 \cr
&=& (m-M + \Delta V)_{V, \rm V1369~Cen} - 2.5 \log 2.24 \cr
&=& 10.25 + 6.3\pm0.3 - 0.88  = 15.67\pm0.3 \cr
&=& (m-M + \Delta V)_{V, \rm V496~Sct} - 2.5 \log 1.66 \cr
&=& 13.7 + 2.5\pm0.3 - 0.55 = 15.65\pm0.3,
\label{distance_modulus_v2659_cyg}
\end{eqnarray}
where we adopt $(m-M)_{V, \rm LV~Vul}=11.85$,
$(m-M)_{V, \rm V5666~Sgr}=15.4$, 
$(m-M)_{V, \rm V1369~Cen}=10.25$, and
$(m-M)_{V, \rm V496~Sct}=13.7$ in \citet{hac19k}.
Thus, we obtained $(m-M)_V=15.65\pm0.1$ and $f_{\rm s}=3.3$ against LV~Vul.
From Equations (\ref{time-stretching_general}),
(\ref{distance_modulus_general_temp}), and
(\ref{distance_modulus_v2659_cyg}),
we have the relation
\begin{eqnarray}
(m- M')_{V, \rm V2659~Cyg} 
&\equiv & (m_V - (M_V - 2.5\log f_{\rm s}))_{\rm V2659~Cyg} \cr
&=& \left( (m-M)_V + \Delta V \right)_{\rm LV~Vul} \cr
&=& 11.85 + 5.1\pm0.3 = 16.95\pm0.3.
\label{absolute_mag_v2659_cyg}
\end{eqnarray}

Figure \ref{v2659_cyg_v1369_cen_v496_sct_v5666_sgr_b_v_i_logscale_3fig}
shows the $B$, $V$, and $I_{\rm C}$ light curves of V2659~Cyg
together with those of V1369~Cen, V496~Sct, and V5666~Sgr.
The light curves overlap each other.  Applying Equation 
(\ref{distance_modulus_general_temp_b}) for the $B$ band to Figure
\ref{v2659_cyg_v1369_cen_v496_sct_v5666_sgr_b_v_i_logscale_3fig}(a),
we have the relation
\begin{eqnarray}
(m&-&M)_{B, \rm V2659~Cyg} \cr
&=& \left( (m-M)_B + \Delta B\right)_{\rm V1369~Cen} - 2.5 \log 2.24 \cr
&=& 10.36 + 7.0\pm0.2 - 0.88 = 16.48\pm0.2 \cr
&=& \left( (m-M)_B + \Delta B\right)_{\rm V496~Sct} - 2.5 \log 1.66 \cr
&=& 14.15 + 2.9\pm0.2 - 0.55 = 16.5\pm0.2 \cr
&=& \left( (m-M)_B + \Delta B\right)_{\rm V5666~Sgr} - 2.5 \log 1.86 \cr
&=& 15.9 + 1.3\pm0.2 - 0.68 = 16.52\pm0.2,
\label{distance_modulus_v2659_cyg_v1369_cen_v496_sct_v5666_sgr_b}
\end{eqnarray}
where we adopt $(m-M)_{B, \rm V1369~Cen}= 10.25 + 0.11 = 10.36$,
$(m-M)_{B, \rm V496~Sct}= 13.7 + 0.45 = 14.15$, and
$(m-M)_{B, \rm V5666~Sgr}= 15.4 + 0.50 = 15.9$ from Appendix \ref{qy_mus}.
We have $(m-M)_B=16.5\pm0.1$ for V2659~Cyg.

Applying Equation (\ref{distance_modulus_general_temp}) to
Figure \ref{v2659_cyg_v1369_cen_v496_sct_v5666_sgr_b_v_i_logscale_3fig}(b),
we have the relation
\begin{eqnarray}
(m&-&M)_{V, \rm V2659~Cyg} \cr
&=& \left( (m-M)_V + \Delta V\right)_{\rm V1369~Cen} - 2.5 \log 2.24 \cr
&=& 10.25 + 6.3\pm0.2 - 0.88 = 15.67\pm0.2 \cr
&=& \left( (m-M)_V + \Delta V\right)_{\rm V496~Sct} - 2.5 \log 1.66 \cr
&=& 13.7 + 2.5\pm0.2 - 0.55 = 15.65\pm0.2 \cr
&=& \left( (m-M)_V + \Delta V\right)_{\rm V5666~Sgr} - 2.5 \log 1.86 \cr
&=& 15.4 + 1.0\pm0.2 - 0.68 = 15.72\pm0.2,
\label{distance_modulus_v2659_cyg_v1369_cen_v496_sct_v5666_sgr_v}
\end{eqnarray}
where we adopt $(m-M)_{V, \rm V1369~Cen}=10.25$,
$(m-M)_{V, \rm V496~Sct}=13.7$, and $(m-M)_{V, \rm V5666~Sgr}=15.4$
from \citet{hac19k}.  We have $(m-M)_V=15.68\pm0.1$, which is
consistent with Equation (\ref{distance_modulus_v2659_cyg}).

From the $I_{\rm C}$-band data in Figure
\ref{v2659_cyg_v1369_cen_v496_sct_v5666_sgr_b_v_i_logscale_3fig}(c),
we obtain
\begin{eqnarray}
(m&-&M)_{I, \rm V2659~Cyg} \cr
&=& ((m - M)_I + \Delta I_C)_{\rm V1369~Cen} - 2.5 \log 2.24 \cr
&=& 10.07 + 5.2\pm0.2 - 0.88 = 14.39\pm0.2 \cr
&=& ((m - M)_I + \Delta I_C)_{\rm V496~Sct} - 2.5 \log 1.66 \cr
&=& 12.98 + 2.0\pm0.2 - 0.55 = 14.43\pm0.2 \cr
&=& ((m - M)_I + \Delta I_C)_{\rm V5666~Sgr} - 2.5 \log 1.86 \cr
&=& 14.6 + 0.5\pm0.2 - 0.68 = 14.42\pm0.2,
\label{distance_modulus_i_v2659_cyg_v1369_cen_v496_sct_v5666_sgr}
\end{eqnarray}
where we adopt $(m-M)_{I, \rm V1369~Cen}= 10.07$,
$(m-M)_{I, \rm V496~Sct}= 12.98$, and $(m-M)_{I, \rm V5666~Sgr}= 14.6$
from Appendix \ref{qy_mus}.
We have $(m-M)_{I, \rm V2659~Cyg}= 14.41\pm0.1$.

We plot $(m-M)_B=16.5$, $(m-M)_V=15.68$, and $(m-M)_I=14.41$,
which cross at $d=4.4$~kpc and $E(B-V)=0.80$, in Figure
\ref{distance_reddening_v1324_sco_v5592_sgr_v962_cep_v2659_cyg}(d).
Thus, we obtain $d=4.4\pm0.5$~kpc, $E(B-V)=0.80\pm0.05$.


\begin{figure}
\epsscale{0.75}
\plotone{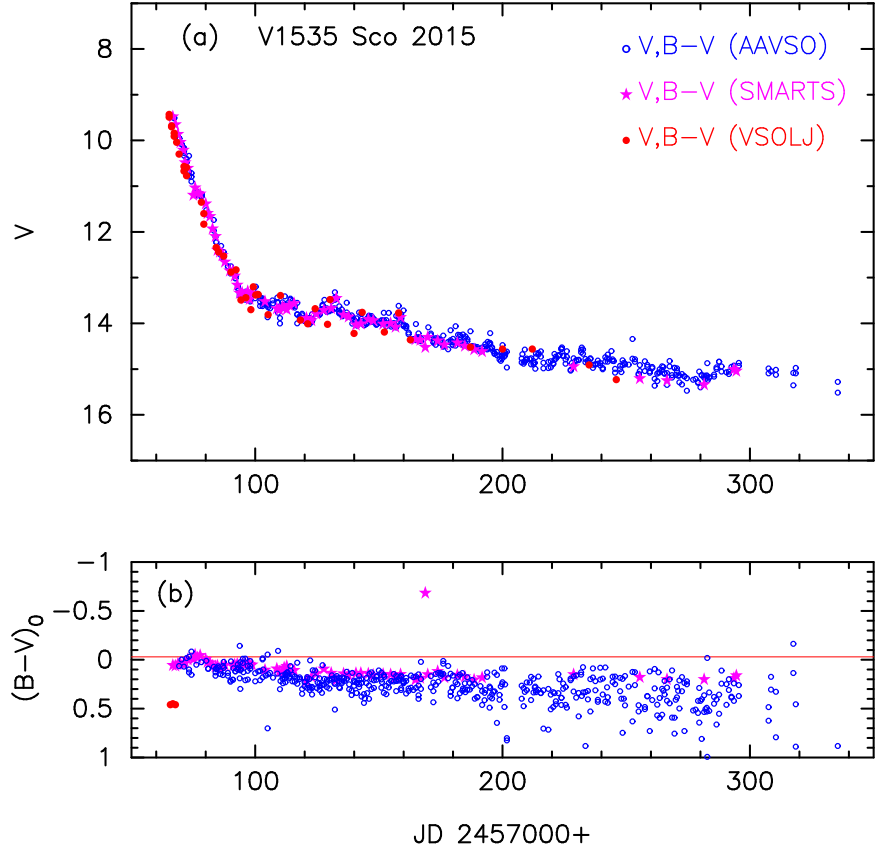}
\caption{
Same as Figure \ref{v1663_aql_v_bv_ub_color_curve}, 
but for V1535~Sco.  
(a) The $BV$ data are taken from AAVSO (unfilled blue circles),
SMARTS (filled magenta stars), and VSOLJ (filled red circles).
(b) The $(B-V)_0$ are dereddened with $E(B-V)=0.78$.
\label{v1535_sco_v_bv_ub_color_curve}}
\end{figure}


\begin{figure}
\epsscale{0.75}
\plotone{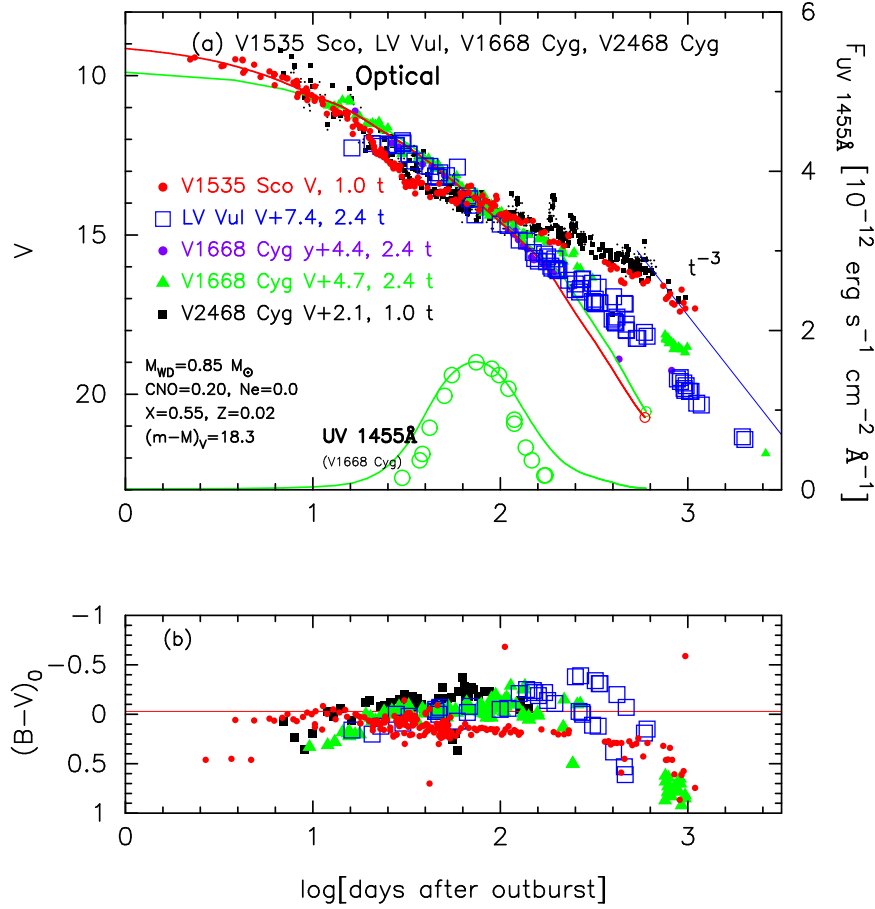}
\caption{
Same as Figure 
\ref{v2575_oph_v1668_cyg_lv_vul_v_bv_ub_logscale},
but for V1535~Sco (filled red circles).  
We plot the (a) $V$ light and (b) $(B-V)_0$ color curves of V1535~Sco
as well as those of LV~Vul, V1668~Cyg, and V2468~Cyg.  In panel (a),
we add the model $V$ light curve (solid red line) of a $0.85~M_\sun$ WD
\citep[CO4;][]{hac15k}, assuming
that $(m-M)_V=18.3$ for V1535~Sco.  We also add model $V$ and 
UV~1455\AA\  light curves of a $0.98~M_\sun$ WD 
\citep[CO3, solid green lines;][]{hac16k}, assuming
that $(m-M)_V=14.6$ for V1668~Cyg. 
\label{v1535_sco_lv_vul_v1668_cyg_v2468_cyg_v_bv_logscale}}
\end{figure}


\begin{figure}
\plottwo{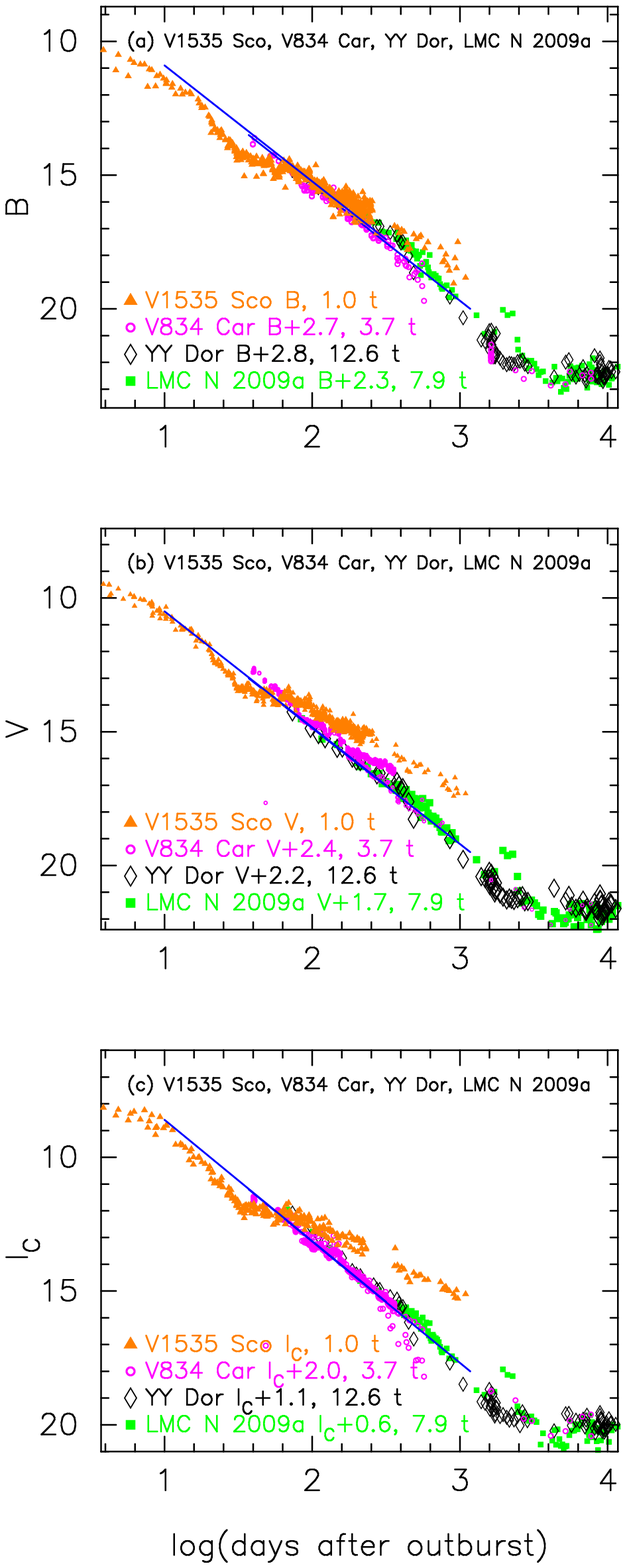}{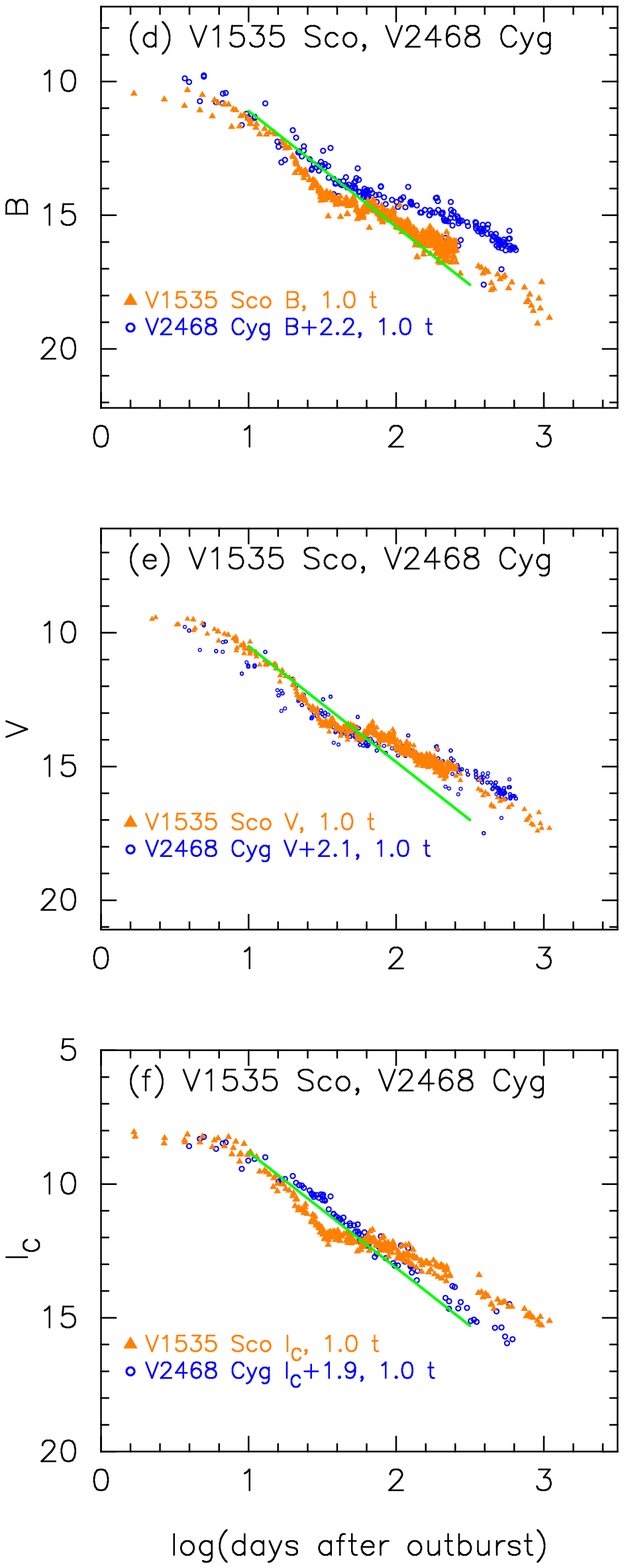}
\caption{
Same as Figure \ref{v1663_aql_yy_dor_lmcn_2009a_b_v_i_logscale_3fig},
but for V1535~Sco.
The $BV$ data of V1535~Sco are the same as those in Figure
\ref{v1535_sco_v_bv_ub_color_curve}.  The $I_{\rm C}$ data of V1535~Sco
are taken from AAVSO, VSOLJ, and SMARTS.
The $BVI_{\rm C}$ data of V2468~Cyg are the same as those in Figure
47 of \citet{hac19k}.
\label{v1535_sco_v834_car_yy_dor_lmcn_2009a_b_v_i_logscale_3fig}}
\end{figure}

\subsection{V1535~Sco 2015}
\label{v1535_sco}
Figure \ref{v1535_sco_v_bv_ub_color_curve} shows the (a) $V$ light and
(b) $(B-V)_0$ color curves of V1535~Sco.  Here, $(B-V)_0$ are dereddened
with $E(B-V)=0.78$ as obtained in Section \ref{v1535_sco_cmd}.  
Figure \ref{v1535_sco_lv_vul_v1668_cyg_v2468_cyg_v_bv_logscale} shows
the $V$ light and $(B-V)_0$ color curves of V1535~Sco
as well as those of LV~Vul, V1668~Cyg, and V2468~Cyg.
Applying Equation (\ref{distance_modulus_general_temp}) to them,
we have the relation 
\begin{eqnarray}
(m&-&M)_{V, \rm V1535~Sco} \cr 
&=& (m - M + \Delta V)_{V, \rm LV~Vul} - 2.5 \log 2.4 \cr
&=& 11.85 + 7.4\pm0.2 - 0.95 = 18.3\pm0.2 \cr
&=& (m - M + \Delta V)_{V, \rm V1668~Cyg} - 2.5 \log 2.4 \cr
&=& 14.6 + 4.7\pm0.2 - 0.95 = 18.35\pm0.2 \cr
&=& (m - M + \Delta V)_{V, \rm V2468~Cyg} - 2.5 \log 1.0 \cr
&=& 16.2 + 2.1\pm0.2 - 0.0 = 18.3\pm0.2,
\label{distance_modulus_v1535_sco}
\end{eqnarray}
where we adopt $(m-M)_{V, \rm LV~Vul}=11.85$, 
$(m-M)_{V, \rm V1668~Cyg}=14.6$, and 
$(m-M)_{V, \rm V2468~Cyg}=16.2$ in \citet{hac19k}.
Thus, we obtain $(m-M)_V=18.3\pm0.1$ and $f_{\rm s}= 2.4$ against LV~Vul. 
From Equations (\ref{time-stretching_general}),
(\ref{distance_modulus_general_temp}), and
(\ref{distance_modulus_v1535_sco}),
we have the relation
\begin{eqnarray}
(m- M')_{V, \rm V1535~Sco} 
&\equiv & (m_V - (M_V - 2.5\log f_{\rm s}))_{\rm V1535~Sco} \cr
&=& \left( (m-M)_V + \Delta V \right)_{\rm LV~Vul} \cr
&=& 11.85 + 7.4\pm0.2 = 19.25\pm0.2.
\label{absolute_mag_v1535_sco}
\end{eqnarray}

Figure \ref{v1535_sco_v834_car_yy_dor_lmcn_2009a_b_v_i_logscale_3fig}
shows the $B$, $V$, and $I_{\rm C}$ light curves of V1535~Sco together 
with those of V2468~Cyg, V834~Car, YY~Dor, and LMC~N~2009a.
We apply Equation (\ref{distance_modulus_general_temp_b})
for the $B$ band to Figures
\ref{v1535_sco_v834_car_yy_dor_lmcn_2009a_b_v_i_logscale_3fig}(a)
and (d) and obtain
\begin{eqnarray}
(m&-&M)_{B, \rm V1535~Sco} \cr
&=& ((m - M)_B + \Delta B)_{\rm V2468~Cyg} - 2.5 \log 1.0 \cr
&=& 16.85 + 2.2\pm0.2 - 0.0 = 19.05\pm0.2 \cr
&=& ((m - M)_B + \Delta B)_{\rm V834~Car} - 2.5 \log 3.7 \cr
&=& 17.75 + 2.7\pm0.2 - 1.43 = 19.02\pm0.2 \cr
&=& ((m - M)_B + \Delta B)_{\rm YY~Dor} - 2.5 \log 12.6 \cr
&=& 18.98 + 2.8\pm0.2 - 2.75 = 19.03\pm0.2 \cr
&=& ((m - M)_B + \Delta B)_{\rm LMC~N~2009a} - 2.5 \log 7.9 \cr
&=& 18.98 + 2.3\pm0.2 - 2.25 = 19.03\pm0.2.
\label{distance_modulus_b_v1535_sco_v834_car_yy_dor_lmcn2009a}
\end{eqnarray}
We have $(m-M)_{B, \rm V1535~Sco}= 19.04\pm0.1$.

For the $V$ light curves in Figures
\ref{v1535_sco_v834_car_yy_dor_lmcn_2009a_b_v_i_logscale_3fig}(b)
and (e), we similarly obtain
\begin{eqnarray}
(m&-&M)_{V, \rm V1535~Sco} \cr   
&=& ((m - M)_V + \Delta V)_{\rm V2468~Cyg} - 2.5 \log 1.0 \cr
&=& 16.2 + 2.1\pm0.2 - 0.0 = 18.3\pm0.2 \cr
&=& ((m - M)_V + \Delta V)_{\rm V834~Car} - 2.5 \log 3.7 \cr
&=& 17.3 + 2.4\pm0.2 - 1.43 = 18.27\pm0.2 \cr
&=& ((m - M)_V + \Delta V)_{\rm YY~Dor} - 2.5 \log 12.6 \cr
&=& 18.86 + 2.2\pm0.2 - 2.75 = 18.31\pm0.2 \cr
&=& ((m - M)_V + \Delta V)_{\rm LMC~N~2009a} - 2.5 \log 7.9 \cr
&=& 18.86 + 1.7\pm0.2 - 2.25 = 18.31\pm0.2.
\label{distance_modulus_v_v1535_sco_v834_car_yy_dor_lmcn2009a}
\end{eqnarray}
We have $(m-M)_{V, \rm V1535~Sco}= 18.3\pm0.1$, which is
consistent with Equation (\ref{distance_modulus_v1535_sco}).

We apply Equation (\ref{distance_modulus_general_temp_i}) for
the $I_{\rm C}$ band to Figures
\ref{v1535_sco_v834_car_yy_dor_lmcn_2009a_b_v_i_logscale_3fig}(c) and 
(f) and obtain
\begin{eqnarray}
(m&-&M)_{I, \rm V1535~Sco} \cr
&=& ((m - M)_I + \Delta I_C)_{\rm V2468~Cyg} - 2.5 \log 1.0 \cr
&=& 15.16 + 1.9\pm0.2 - 0.0 = 17.06\pm 0.2 \cr
&=& ((m - M)_I + \Delta I_C)_{\rm V834~Car} - 2.5 \log 3.7 \cr
&=& 16.45 + 2.0\pm0.2 - 1.43 = 17.02\pm 0.2 \cr
&=& ((m - M)_I + \Delta I_C)_{\rm YY~Dor} - 2.5 \log 12.6 \cr
&=& 18.67 + 1.1\pm0.2 - 2.75 = 17.02\pm 0.2 \cr
&=& ((m - M)_I + \Delta I_C)_{\rm LMC~N~2009a} - 2.5 \log 7.9 \cr
&=& 18.67 + 0.6\pm0.2 - 2.25 = 17.02\pm 0.2.
\label{distance_modulus_i_v1535_sco_v834_car_yy_dor_lmcn2009a}
\end{eqnarray}
We have $(m-M)_{I, \rm V1535~Sco}= 17.03\pm0.1$.

We plot $(m-M)_B= 19.04$, $(m-M)_V= 18.3$, and $(m-M)_I= 17.03$,
which broadly cross at $d=15$~kpc and $E(B-V)=0.78$, in Figure
\ref{distance_reddening_v1535_sco_v5667_sgr_v5668_sgr_v2944_oph}(a).
Thus, we obtain $E(B-V)=0.78\pm0.05$ and $d=15\pm2$~kpc.


\begin{figure}
\epsscale{0.75}
\plotone{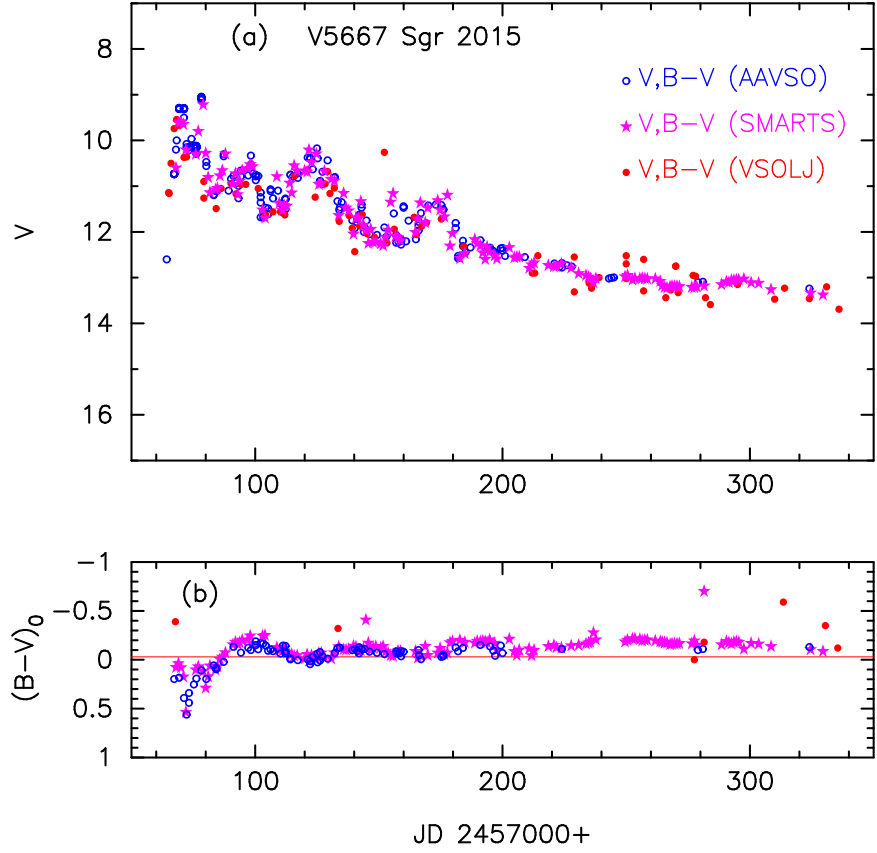}
\caption{
Same as Figure \ref{v1663_aql_v_bv_ub_color_curve}, 
but for V5667~Sgr.
(a) The $BV$ data are taken from AAVSO (unfilled blue circles),
SMARTS (filled magenta stars), and VSOLJ (filled red circles).
(b) The $(B-V)_0$ are dereddened with $E(B-V)=0.63$.
\label{v5667_sgr_v_bv_ub_color_curve}}
\end{figure}


\begin{figure}
\epsscale{0.75}
\plotone{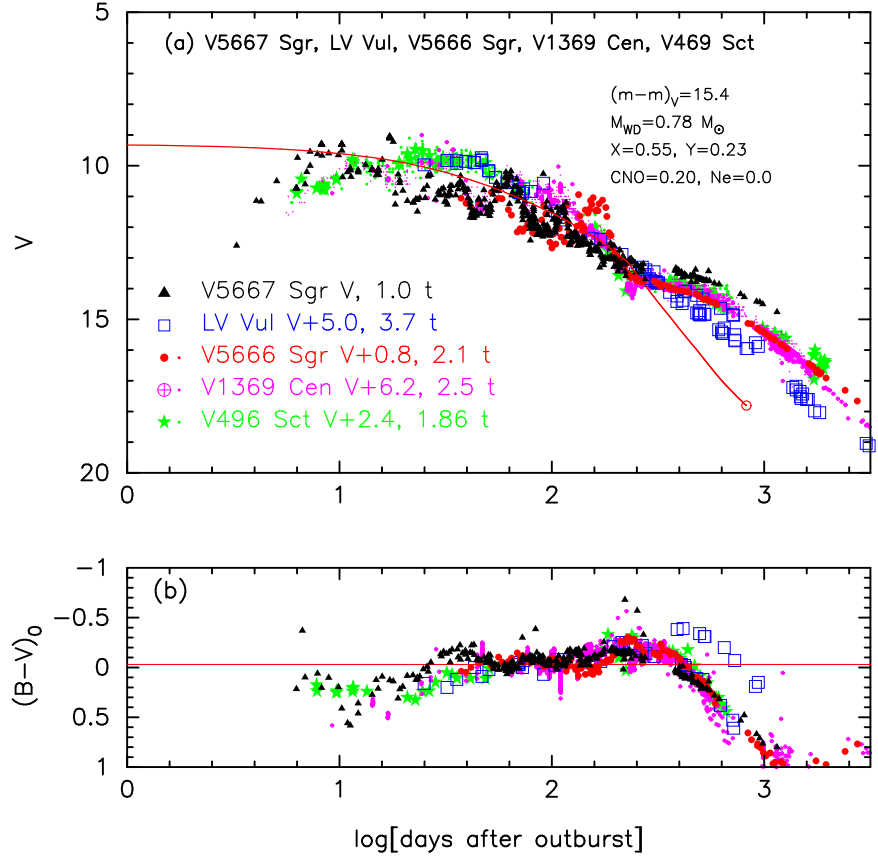}
\caption{
Same as Figure 
\ref{v2575_oph_v1668_cyg_lv_vul_v_bv_ub_logscale},
but for V5667~Sgr (filled black triangles).  We plot the (a) $V$ light
and (b) $(B-V)_0$ color curves of V5667~Sgr as well as those of LV~Vul, 
V5666~Sgr, V1369~Cen, and V496~Sct.  In panel (a), 
we add the model $V$ light curve (solid red line) of a $0.78~M_\sun$ WD
\citep[CO4;][]{hac15k}, assuming that $(m-M)_V= 15.4$ for V5667~Sgr. 
\label{v5667_sgr_lv_vul_v5666_sgr_v1369_cen_v496_sct_v_bv_ub_color_logscale}}
\end{figure}


\begin{figure}
\epsscale{0.65}
\plotone{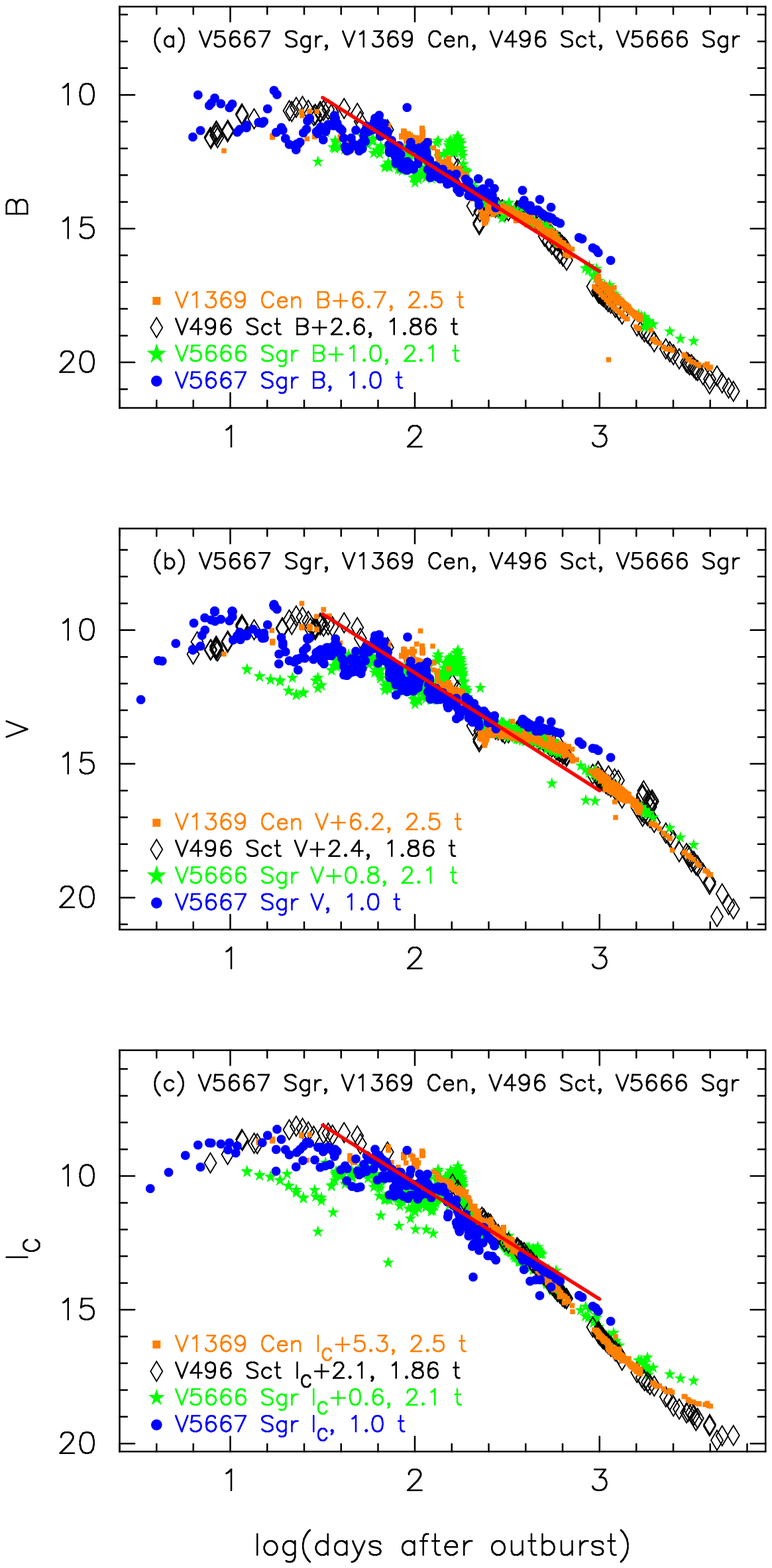}
\caption{
Same as Figure \ref{v1663_aql_yy_dor_lmcn_2009a_b_v_i_logscale_3fig},
but for V5667~Sgr.
The (a) $B$, (b) $V$, and (c) $I_{\rm C}$ light curves of V5667~Sgr
as well as those of V1369~Cen, V496~Sct, and V5666~Sgr.
The $BV$ data of V5667~Sgr are the same as those in Figure
\ref{v5667_sgr_v_bv_ub_color_curve}.  The $I_{\rm C}$ data of V5667~Sgr
are taken from VSOLJ and SMARTS.
\label{v5667_sgr_v1369_cen_v496_sct_v5666_sgr_b_v_i_logscale_3fig}}
\end{figure}

\subsection{V5667~Sgr 2015\#1}
\label{v5667_sgr}
Figure \ref{v5667_sgr_v_bv_ub_color_curve} shows the (a) $V$ and
(b) $(B-V)_0$ evolutions of V5667~Sgr.
Here, $(B-V)_0$ are dereddened with $E(B-V)=0.63$ as obtained in
Section \ref{v5667_sgr_cmd}.   Figure
\ref{v5667_sgr_lv_vul_v5666_sgr_v1369_cen_v496_sct_v_bv_ub_color_logscale}
shows the $V$ light and $(B-V)_0$ color curves of V5667~Sgr
as well as those of LV~Vul, V5666~Sgr, V1369~Cen, and V496~Sct.
Applying Equation (\ref{distance_modulus_general_temp}) to them,
we have the relation 
\begin{eqnarray}
(m&-&M)_{V, \rm V5667~Sgr} \cr 
&=& (m-M + \Delta V)_{V, \rm LV~Vul} - 2.5 \log 3.7 \cr
&=& 11.85 + 5.0\pm0.3 - 1.43 = 15.37\pm0.3 \cr
&=& (m-M + \Delta V)_{V, \rm V496~Sct} - 2.5 \log 1.86 \cr
&=& 13.7 + 2.4\pm0.3 - 0.68 = 15.43\pm0.3 \cr
&=& (m-M + \Delta V)_{V, \rm V1369~Cen} - 2.5 \log 2.5 \cr
&=& 10.25 + 6.2\pm0.3 - 1.0  = 15.45\pm0.3 \cr
&=& (m-M + \Delta V)_{V, \rm V5666~Sgr} - 2.5 \log 2.1 \cr
&=& 15.4 + 0.8\pm0.3 - 0.8 = 15.4\pm0.3,
\label{distance_modulus_v5667_sgr}
\end{eqnarray}
where we adopt $(m-M)_{V, \rm LV~Vul}=11.85$,
$(m-M)_{V, \rm V496~Sct}=13.7$,
$(m-M)_{V, \rm V1369~Cen}=10.25$, and
$(m-M)_{V, \rm V5666~Sgr}=15.4$ from \citet{hac19k}.
Thus, we obtained $f_{\rm s}=3.7$ against LV~Vul and
$(m-M)_{V, \rm V5667~Sgr}=15.4\pm0.2$.
From Equations (\ref{time-stretching_general}),
(\ref{distance_modulus_general_temp}), and
(\ref{distance_modulus_v5667_sgr}),
we have the relation
\begin{eqnarray}
(m- M')_{V, \rm V5667~Sgr} 
&\equiv & (m_V - (M_V - 2.5\log f_{\rm s}))_{\rm V5667~Sgr} \cr
&=& \left( (m-M)_V + \Delta V \right)_{\rm LV~Vul} \cr
&=& 11.85 + 5.0\pm0.3 = 16.85\pm0.3.
\label{absolute_mag_v5667_sgr}
\end{eqnarray}

Figure \ref{v5667_sgr_v1369_cen_v496_sct_v5666_sgr_b_v_i_logscale_3fig}
shows the $B$, $V$, and $I_{\rm C}$ light curves of V5667~Sgr
together with those of V1369~Cen, V496~Sct, and V5666~Sgr.
Applying Equation (\ref{distance_modulus_general_temp_b})
for the $B$ band to Figure
\ref{v5667_sgr_v1369_cen_v496_sct_v5666_sgr_b_v_i_logscale_3fig}(a),
we have the relation
\begin{eqnarray}
(m&-&M)_{B, \rm V5667~Sgr} \cr
&=& \left( (m-M)_B + \Delta B\right)_{\rm V1369~Cen} - 2.5 \log 2.5 \cr
&=& 10.36 + 6.7\pm0.3 - 1.0 = 16.06\pm0.3 \cr
&=& \left( (m-M)_B + \Delta B\right)_{\rm V496~Sct} - 2.5 \log 1.86 \cr
&=& 14.15 + 2.6\pm0.3 - 0.68 = 16.07\pm0.3 \cr
&=& \left( (m-M)_B + \Delta B\right)_{\rm V5666~Sgr} - 2.5 \log 2.1 \cr
&=& 15.9 + 1.0\pm0.3 - 0.8 = 16.1\pm0.3,
\label{distance_modulus_v5667_sgr_v1369_cen_v496_sct_v5666_sgr_b}
\end{eqnarray}
where we adopt $(m-M)_{B, \rm V1369~Cen}= 10.36$,
$(m-M)_{B, \rm V496~Sct}= 14.15$, and
$(m-M)_{B, \rm V5666~Sgr}= 15.9$ from Appendix \ref{qy_mus}.
We have $(m-M)_B=16.08\pm0.2$ for V5667~Sgr.

Applying Equation (\ref{distance_modulus_general_temp}) to
Figure \ref{v5667_sgr_v1369_cen_v496_sct_v5666_sgr_b_v_i_logscale_3fig}(b),
we have the relation
\begin{eqnarray}
(m&-&M)_{V, \rm V5667~Sgr} \cr
&=& \left( (m-M)_V + \Delta V\right)_{\rm V1369~Cen} - 2.5 \log 2.5 \cr
&=& 10.25 + 6.2\pm0.3 - 1.0 = 15.45\pm0.3 \cr
&=& \left( (m-M)_V + \Delta V\right)_{\rm V496~Sct} - 2.5 \log 1.86 \cr
&=& 13.7 + 2.4\pm0.3 - 0.68 = 15.42\pm0.3 \cr
&=& \left( (m-M)_V + \Delta V\right)_{\rm V5666~Sgr} - 2.5 \log 2.1 \cr
&=& 15.4 + 0.8\pm0.3 - 0.8 = 15.4\pm0.3,
\label{distance_modulus_v5667_sgr_v1369_cen_v496_sct_v5666_sgr_v}
\end{eqnarray}
where we adopt $(m-M)_{V, \rm V1369~Cen}=10.25$,
$(m-M)_{V, \rm V496~Sct}=13.7$, and $(m-M)_{V, \rm V5666~Sgr}=15.4$
from \citet{hac19k}.  We have $(m-M)_V=15.42\pm0.2$, which is
essentially the same as Equation (\ref{distance_modulus_v5667_sgr}).

From the $I_{\rm C}$-band data in Figure
\ref{v5667_sgr_v1369_cen_v496_sct_v5666_sgr_b_v_i_logscale_3fig}(c),
we obtain
\begin{eqnarray}
(m&-&M)_{I, \rm V5667~Sgr} \cr
&=& ((m - M)_I + \Delta I_C)_{\rm V1369~Cen} - 2.5 \log 2.5 \cr
&=& 10.07 + 5.3\pm0.3 - 1.0 = 14.37\pm0.3 \cr
&=& ((m - M)_I + \Delta I_C)_{\rm V496~Sct} - 2.5 \log 1.86 \cr
&=& 12.98 + 2.1\pm0.3 - 0.68 = 14.4\pm0.3 \cr
&=& ((m - M)_I + \Delta I_C)_{\rm V5666~Sgr} - 2.5 \log 2.1 \cr
&=& 14.6 + 0.6\pm0.3 - 0.8 = 14.4\pm0.3,
\label{distance_modulus_i_v5667_sgr_v1369_cen_v496_sct_v5666_sgr}
\end{eqnarray}
where we adopt $(m-M)_{I, \rm V1369~Cen}= 10.07$,
$(m-M)_{I, \rm V496~Sct}= 12.98$, and $(m-M)_{I, \rm V5666~Sgr}= 14.6$
from Appendix \ref{qy_mus}.
We have $(m-M)_{I, \rm V5667~Sgr}= 14.39\pm0.2$.

We plot $(m-M)_B=16.08$, $(m-M)_V=15.42$, and $(m-M)_I=14.39$,
which cross at $d=4.9$~kpc and $E(B-V)=0.63$, in Figure
\ref{distance_reddening_v1535_sco_v5667_sgr_v5668_sgr_v2944_oph}(b).
Thus, we obtain $d=4.9\pm0.5$~kpc and $E(B-V)=0.63\pm0.05$.


\begin{figure}
\epsscale{0.75}
\plotone{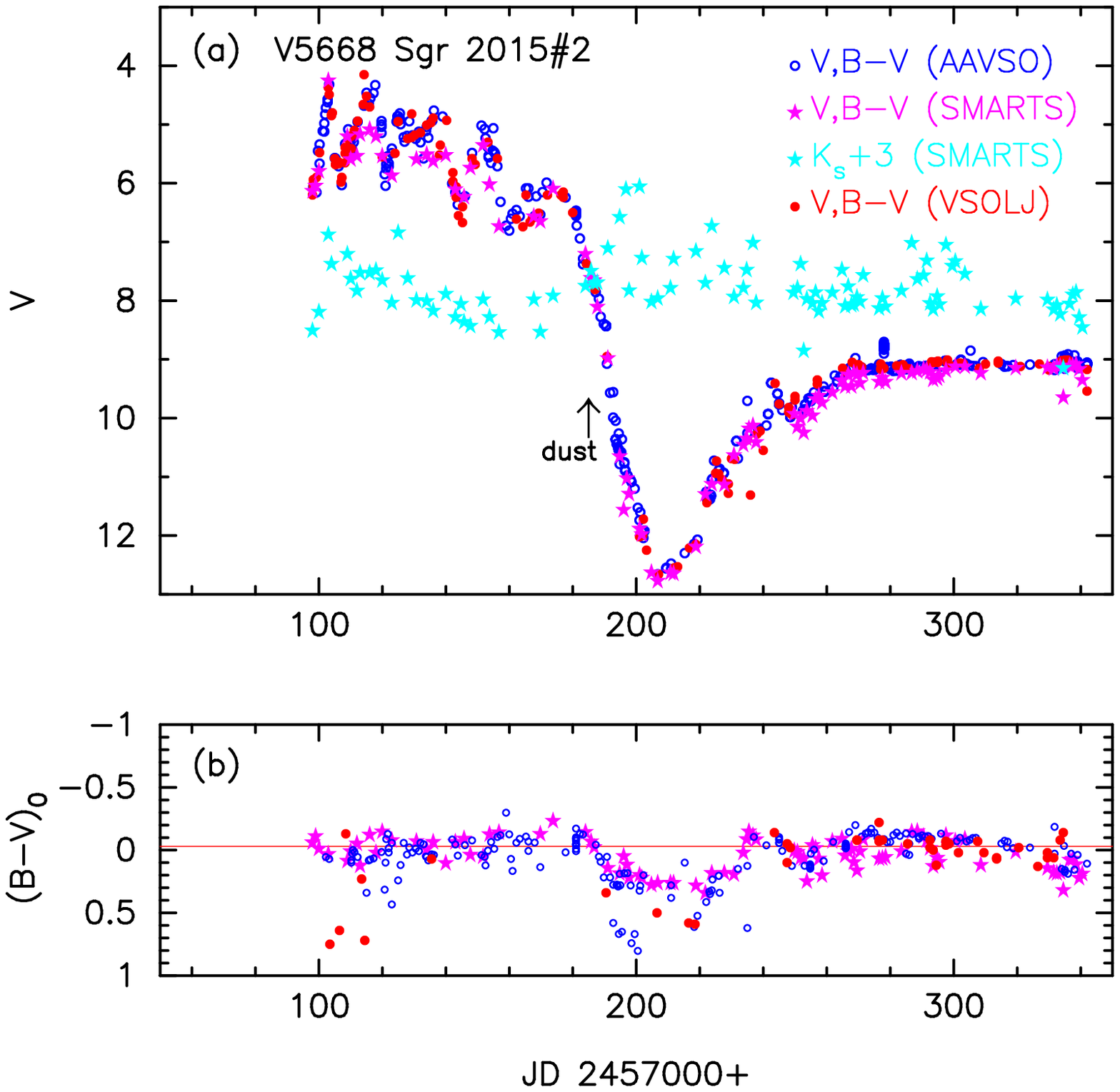}
\caption{
Same as Figure \ref{v1663_aql_v_bv_ub_color_curve}, but for V5668~Sgr.
(a) The $BV$ data are taken from AAVSO (unfilled blue circles),
SMARTS (filled magenta stars), and VSOLJ (filled red circles).
The $K_{\rm s}$ data are taken from SMARTS (filled cyan stars).
We denote the epoch when a dust shell formed by the arrow labeled dust.
(b) The $(B-V)_0$ are dereddened with $E(B-V)=0.20$.
\label{v5668_sgr_v_bv_ub_color_curve}}
\end{figure}


\begin{figure}
\epsscale{0.75}
\plotone{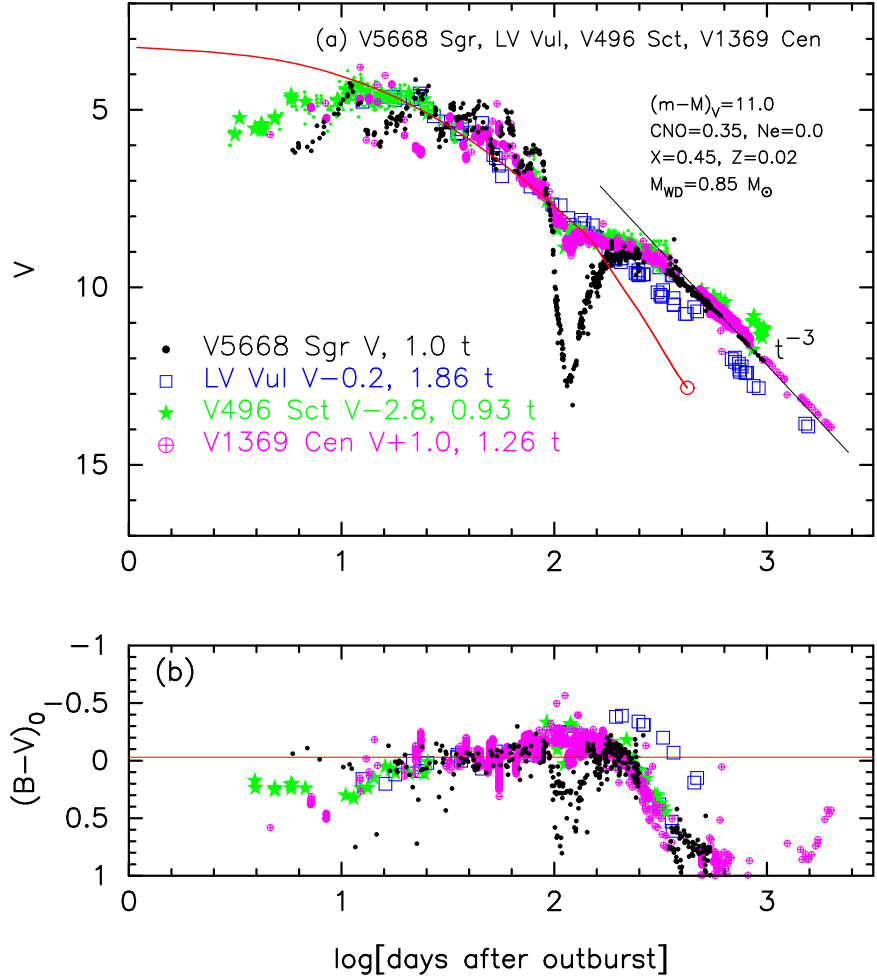}
\caption{
Same as Figure \ref{v2575_oph_v1668_cyg_lv_vul_v_bv_ub_logscale},
but for V5668~Sgr (filled black circles).  We plot the (a) $V$ light
and (b) $(B-V)_0$ color curves of V5668~Sgr as well as those of
LV~Vul, V496~Sct, and V1369~Cen.  The data of V5668~Sgr are
the same as those in Figure \ref{v5668_sgr_v_bv_ub_color_curve}.
In panel (a), we add a model $V$ light curve of a $0.85~M_\sun$ WD
\citep[CO3, solid red line;][]{hac16k}, 
assuming that $(m-M)_V=11.0$ for V5668~Sgr.
\label{v5668_sgr_lv_vul_v496_sct_v1369_cen_v_bv_ub_color_logscale}}
\end{figure}


\begin{figure}
\epsscale{0.65}
\plotone{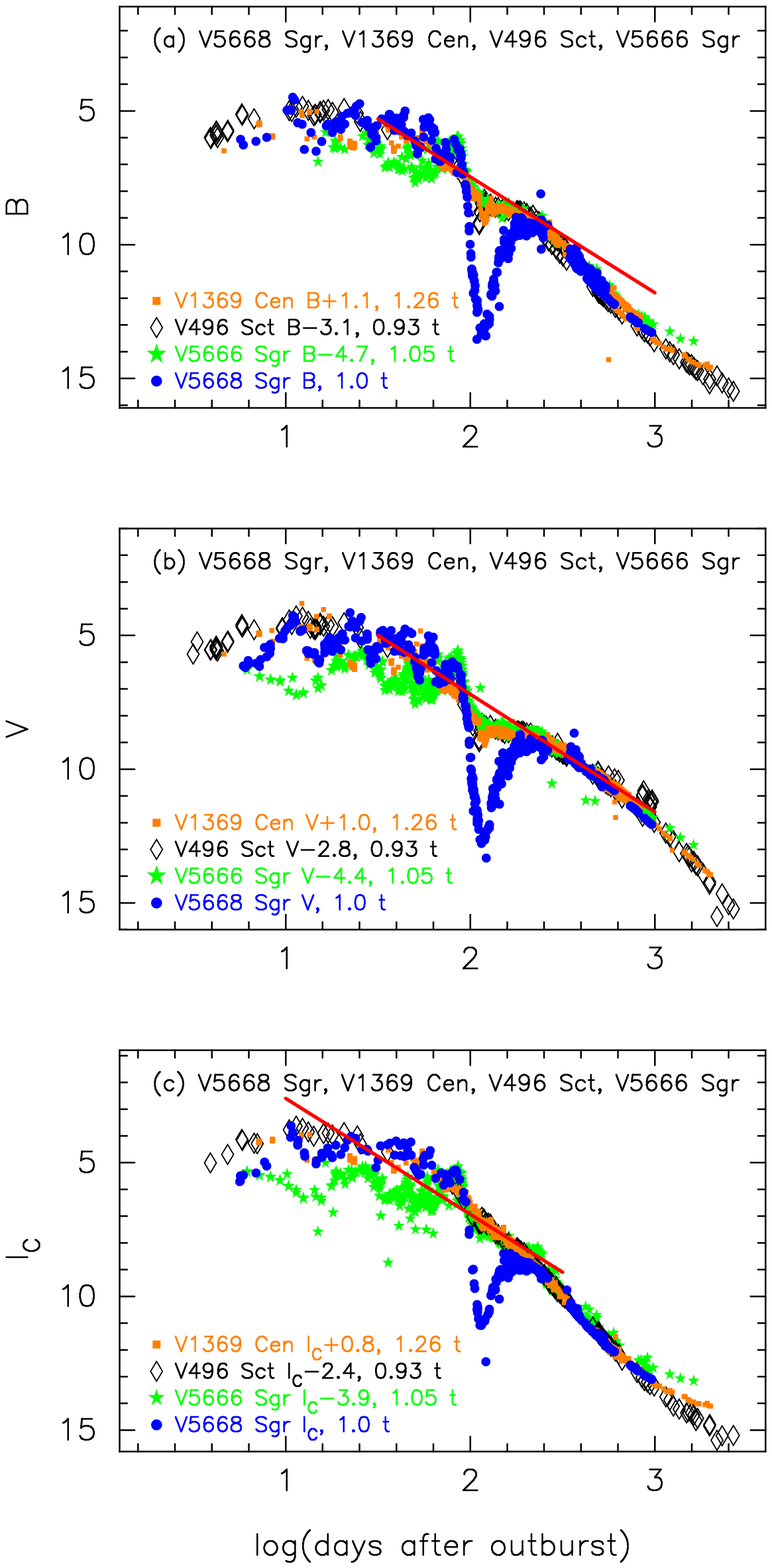}
\caption{
Same as Figure \ref{v1663_aql_yy_dor_lmcn_2009a_b_v_i_logscale_3fig},
but for V5668~Sgr.
The (a) $B$, (b) $V$, and (c) $I_{\rm C}$ light curves of V5668~Sgr
as well as those of V1369~Cen, V496~Sct, and V5666~Sgr.
The $BV$ data of V5668~Sgr are the same as those in Figure
\ref{v5668_sgr_v_bv_ub_color_curve}.  The $I_{\rm C}$ data of V5668~Sgr
are taken from AAVSO, VSOLJ, and SMARTS.
\label{v5668_sgr_v1369_cen_v496_sct_v5666_sgr_b_v_i_logscale_3fig}}
\end{figure}

\subsection{V5668~Sgr 2015\#2}
\label{v5668_sgr}
Figure \ref{v5668_sgr_v_bv_ub_color_curve} shows the (a) $V$, $K_{\rm s}$,
and (b) $(B-V)_0$ evolutions of V5668~Sgr.
Here, $(B-V)_0$ are dereddened with $E(B-V)=0.20$ as obtained in
Section \ref{v5668_sgr_cmd}.
Figure \ref{v5668_sgr_lv_vul_v496_sct_v1369_cen_v_bv_ub_color_logscale}
shows the light/color curves of V5668~Sgr
as well as those of LV~Vul, V496~Sct, and V1369~Cen.
Applying Equation (\ref{distance_modulus_general_temp}) to them,
we have the relation 
\begin{eqnarray}
(m&-&M)_{V, \rm V5668~Sgr} \cr
&=& (m - M + \Delta V)_{V, \rm LV~Vul} - 2.5 \log 1.86 \cr
&=& 11.85 - 0.2\pm0.2 - 0.68 = 10.97\pm0.2 \cr
&=& (m - M + \Delta V)_{V, \rm V496~Sct} - 2.5 \log 0.93 \cr
&=& 13.7 - 2.8\pm0.2 + 0.08 = 10.98\pm0.2 \cr
&=& (m - M + \Delta V)_{V, \rm V1369~Cen} - 2.5 \log 1.26 \cr
&=& 10.25 + 1.0\pm0.2 - 0.25 = 11.0\pm0.2,
\label{distance_modulus_v5568_sgr_lv_vul}
\end{eqnarray}
where we adopt $(m-M)_{V, \rm LV~Vul}=11.85$ and
$(m-M)_{V, \rm V496~Sct}=13.7$, and
$(m-M)_{V, \rm V1369~Cen}=10.25$ from \citet{hac19k}.
Thus, we obtain $(m-M)_V=11.0\pm0.1$ and $f_{\rm s}=1.86$ against LV~Vul.
From Equations (\ref{time-stretching_general}),
(\ref{distance_modulus_general_temp}), and
(\ref{distance_modulus_v5568_sgr_lv_vul}),
we have the relation
\begin{eqnarray}
(m- M')_{V, \rm V5668~Sgr} 
&\equiv & (m_V - (M_V - 2.5\log f_{\rm s}))_{\rm V5668~Sgr} \cr
&=& \left( (m-M)_V + \Delta V \right)_{\rm LV~Vul} \cr
&=& 11.85 - 0.2\pm0.2 = 11.65\pm0.2.
\label{absolute_mag_v5668_sgr}
\end{eqnarray}

Figure \ref{v5668_sgr_v1369_cen_v496_sct_v5666_sgr_b_v_i_logscale_3fig}
shows the $B$, $V$, and $I_{\rm C}$ light curves of V5668~Sgr
together with those of V1369~Cen, V496~Sct, and V5666~Sgr.
Applying Equation (\ref{distance_modulus_general_temp_b})
for the $B$ band to Figure
\ref{v5668_sgr_v1369_cen_v496_sct_v5666_sgr_b_v_i_logscale_3fig}(a),
we have the relation
\begin{eqnarray}
(m&-&M)_{B, \rm V5668~Sgr} \cr
&=& \left( (m-M)_B + \Delta B\right)_{\rm V1369~Cen} - 2.5 \log 1.26 \cr
&=& 10.36 + 1.1\pm0.3 - 0.25 = 11.21\pm0.3 \cr
&=& \left( (m-M)_B + \Delta B\right)_{\rm V496~Sct} - 2.5 \log 0.93 \cr
&=& 14.15 - 3.1\pm0.3 + 0.08 = 11.13\pm0.3 \cr
&=& \left( (m-M)_B + \Delta B\right)_{\rm V5666~Sgr} - 2.5 \log 1.05 \cr
&=& 15.9 - 4.7\pm0.3 - 0.05 = 11.15\pm0.3,
\label{distance_modulus_v5668_sgr_v1369_cen_v496_sct_v5666_sgr_b}
\end{eqnarray}
where we adopt $(m-M)_{B, \rm V1369~Cen}= 10.36$,
$(m-M)_{B, \rm V496~Sct}= 14.15$, and
$(m-M)_{B, \rm V5666~Sgr}= 15.9$ from Appendix \ref{qy_mus}.
We have $(m-M)_B=11.16\pm0.2$ for V5668~Sgr.

Applying Equation (\ref{distance_modulus_general_temp}) to
Figure \ref{v5668_sgr_v1369_cen_v496_sct_v5666_sgr_b_v_i_logscale_3fig}(b),
we have the relation
\begin{eqnarray}
(m&-&M)_{V, \rm V5668~Sgr} \cr
&=& \left( (m-M)_V + \Delta V\right)_{\rm V1369~Cen} - 2.5 \log 1.26 \cr
&=& 10.25 + 1.0\pm0.3 - 0.25 = 11.0\pm0.3 \cr
&=& \left( (m-M)_V + \Delta V\right)_{\rm V496~Sct} - 2.5 \log 0.93 \cr
&=& 13.7 - 2.8\pm0.3 + 0.08 = 10.98\pm0.3 \cr
&=& \left( (m-M)_V + \Delta V\right)_{\rm V5666~Sgr} - 2.5 \log 1.05 \cr
&=& 15.4 - 4.4\pm0.3 - 0.05 = 10.95\pm0.3,
\label{distance_modulus_v5668_sgr_v1369_cen_v496_sct_v5666_sgr_v}
\end{eqnarray}
where we adopt $(m-M)_{V, \rm V1369~Cen}=10.25$,
$(m-M)_{V, \rm V496~Sct}=13.7$, and $(m-M)_{V, \rm V5666~Sgr}=15.4$
from \citet{hac19k}.
We have $(m-M)_V=10.98\pm0.2$, which is essentially
the same as Equation (\ref{distance_modulus_v5568_sgr_lv_vul}).

From the $I_{\rm C}$-band data in Figure
\ref{v5668_sgr_v1369_cen_v496_sct_v5666_sgr_b_v_i_logscale_3fig}(c),
we obtain
\begin{eqnarray}
(m&-&M)_{I, \rm V5668~Sgr} \cr
&=& ((m - M)_I + \Delta I_C)_{\rm V1369~Cen} - 2.5 \log 1.26 \cr
&=& 10.07 + 0.8\pm0.3 - 0.25 = 10.62\pm0.3 \cr
&=& ((m - M)_I + \Delta I_C)_{\rm V496~Sct} - 2.5 \log 0.93 \cr
&=& 12.98 - 2.4\pm0.3 + 0.08 = 10.66\pm0.3 \cr
&=& ((m - M)_I + \Delta I_C)_{\rm V5666~Sgr} - 2.5 \log 1.05 \cr
&=& 14.6 - 3.9\pm0.3 - 0.05 = 10.65\pm0.3,
\label{distance_modulus_i_v5668_sgr_v1369_cen_v496_sct_v5666_sgr}
\end{eqnarray}
where we adopt $(m-M)_{I, \rm V1369~Cen}= 10.07$,
$(m-M)_{I, \rm V496~Sct}= 12.98$, and $(m-M)_{I, \rm V5666~Sgr}= 14.6$
from Appendix \ref{qy_mus}.
We have $(m-M)_{I, \rm V5668~Sgr}= 10.64\pm0.2$.

We plot $(m-M)_B=11.16$, $(m-M)_V=10.98$, and $(m-M)_I=10.64$,
which cross at $d=1.2$~kpc and $E(B-V)=0.20$, in Figure
\ref{distance_reddening_v1535_sco_v5667_sgr_v5668_sgr_v2944_oph}(c).
Thus, we obtain $d=1.2\pm0.2$~kpc and $E(B-V)=0.20\pm0.03$.


\begin{figure}
\epsscale{0.75}
\plotone{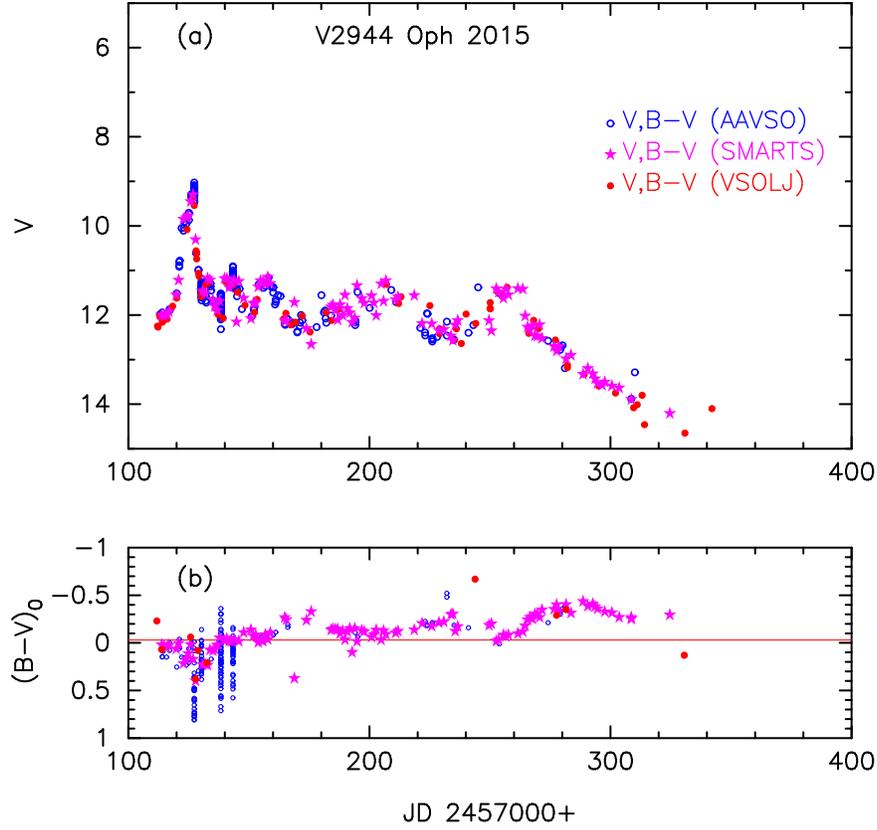}
\caption{
Same as Figure \ref{v1663_aql_v_bv_ub_color_curve}, but for V2944~Oph.
(a) The $V$ data are taken from AAVSO (unfilled blue circles),
SMARTS (filled magenta stars), and VSOLJ (filled red circles).
(b) The $(B-V)_0$ are dereddened with $E(B-V)=0.62$.
\label{v2944_oph_v_bv_ub_color_curve}}
\end{figure}


\begin{figure}
\epsscale{0.75}
\plotone{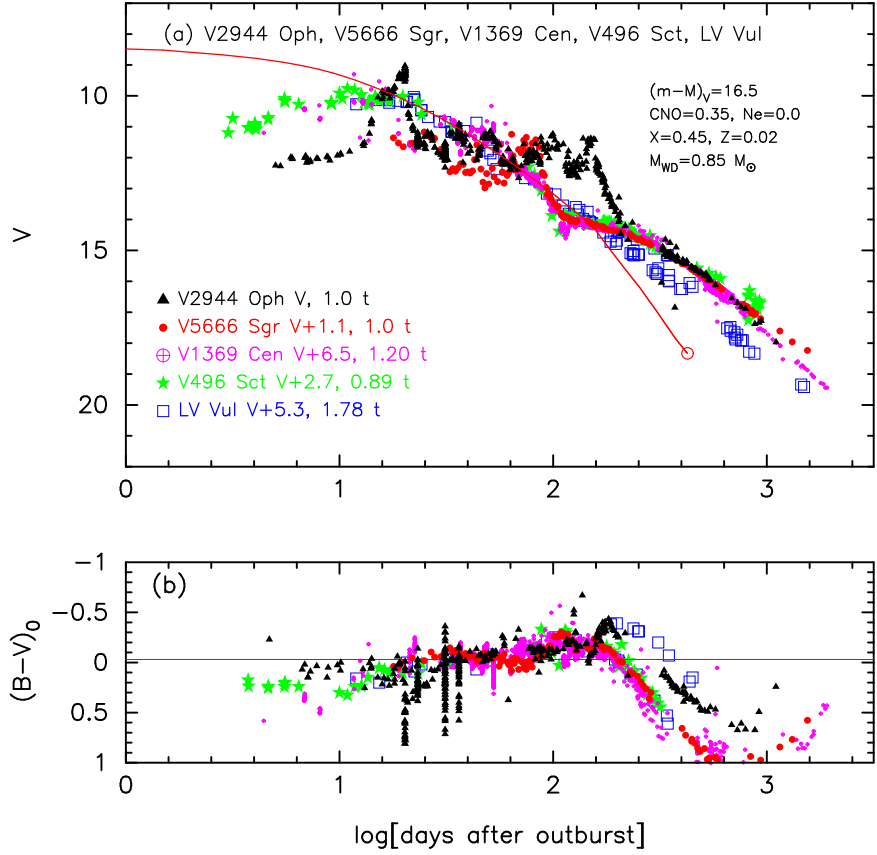}
\caption{
Same as Figure 
\ref{v2575_oph_v1668_cyg_lv_vul_v_bv_ub_logscale},
but for V2944~Oph (filled black triangles).  
We plot the (a) $V$ light
and (b) $(B-V)_0$ color curves of V2944~Oph as well as those of
V5666~Sgr, V1369~Cen, V496~Sct, and LV~Vul.  The data of V2944~Oph are
the same as those in Figure \ref{v2944_oph_v_bv_ub_color_curve}.
In panel (a), we add the model $V$ light curve of a $0.85~M_\sun$ WD
\citep[CO3, solid red line;][]{hac16k},
assuming that $(m-M)_V=16.5$ for V2944~Oph.
\label{v2944_oph_v5666_sgr_v1369_cen_v496_sct_v_bv_ub_color_logscale}}
\end{figure}


\begin{figure}
\epsscale{0.65}
\plotone{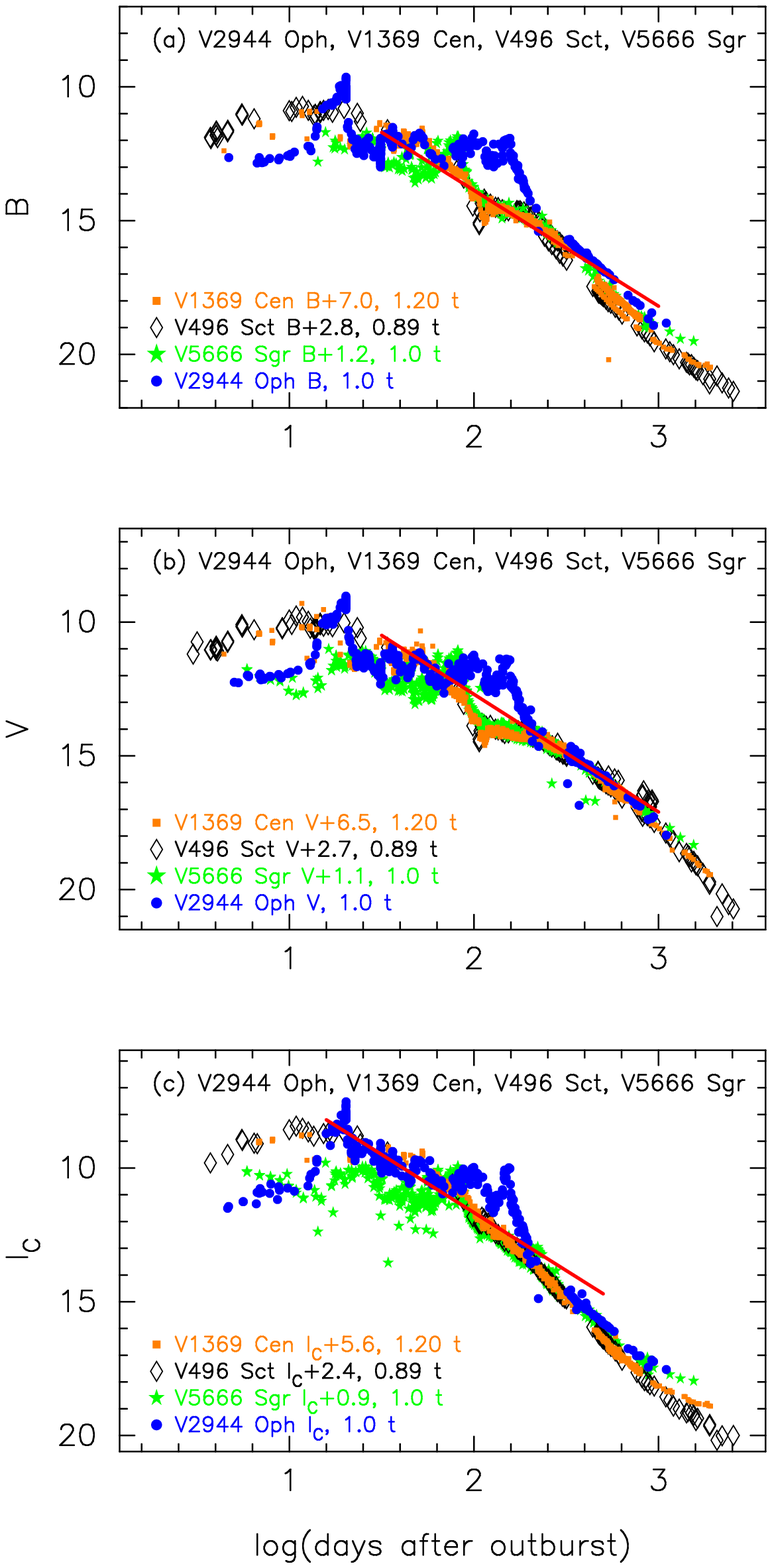}
\caption{
Same as Figure \ref{v1663_aql_yy_dor_lmcn_2009a_b_v_i_logscale_3fig},
but for V2944~Oph.
The (a) $B$, (b) $V$, and (c) $I_{\rm C}$ light curves of V2944~Oph
as well as those of V1369~Cen, V496~Sct, and V5666~Sgr.
The $BV$ data of V2944~Oph are the same as those in Figure
\ref{v2944_oph_v_bv_ub_color_curve}.  The $I_{\rm C}$ data of V2944~Oph
are taken from AAVSO, VSOLJ, and SMARTS.
\label{v2944_oph_v1369_cen_v496_sct_v5666_sgr_b_v_i_logscale_3fig}}
\end{figure}

\subsection{V2944~Oph 2015}
\label{v2944_oph}
Figure \ref{v2944_oph_v_bv_ub_color_curve} shows the (a) $V$ and
(b) $(B-V)_0$ evolutions of V2944~Oph.  Here, $(B-V)_0$ are dereddened
with $E(B-V)=0.62$ as obtained in Section \ref{v2944_oph_cmd}.
Figure \ref{v2944_oph_v5666_sgr_v1369_cen_v496_sct_v_bv_ub_color_logscale}
shows the $V$ light and $(B-V)_0$ color
curves of V2944~Oph, V5666~Sgr, V1369~Cen, V496~Sct, and LV~Vul.
Applying Equation (\ref{distance_modulus_general_temp}) to them,
we have the relation 
\begin{eqnarray}
(m&-&M)_{V, \rm V2944~Oph} \cr
&=& (m-M + \Delta V)_{V, \rm LV~Vul} - 2.5 \log 1.78 \cr
&=& 11.85 + 5.3\pm0.3 - 0.63 = 16.52\pm0.3 \cr
&=& (m-M + \Delta V)_{V, \rm V496~Sct} - 2.5 \log 0.89 \cr
&=& 13.7 + 2.7\pm0.3 + 0.13 = 16.53\pm0.3 \cr
&=& (m-M + \Delta V)_{V, \rm V1369~Cen} - 2.5 \log 1.20 \cr
&=& 10.25 + 6.5\pm0.3 - 0.2  = 16.55\pm0.3 \cr
&=& (m-M + \Delta V)_{V, \rm V5666~Sgr} - 2.5 \log 1.0 \cr
&=& 15.4 + 1.1\pm0.3 + 0.0 = 16.5\pm0.3,
\label{distance_modulus_v2944_oph}
\end{eqnarray}
where we adopt $(m-M)_{V, \rm LV~Vul}=11.85$,
$(m-M)_{V, \rm V496~Sct}=13.7$,
$(m-M)_{V, \rm V1369~Cen}=10.25$, and
$(m-M)_{V, \rm V5666~Sgr}=15.4$ from \citet{hac19k},
Thus, we obtained $f_{\rm s}=1.78$ against LV~Vul and
$(m-M)_{V, \rm V2944~Oph}=16.5\pm0.2$.
From Equations (\ref{time-stretching_general}),
(\ref{distance_modulus_general_temp}), and
(\ref{distance_modulus_v2944_oph}),
we have the relation
\begin{eqnarray}
(m- M')_{V, \rm V2944~Oph} 
&\equiv & (m_V - (M_V - 2.5\log f_{\rm s}))_{\rm V2944~Oph} \cr
&=& \left( (m-M)_V + \Delta V \right)_{\rm LV~Vul} \cr
&=& 11.85 + 5.3\pm0.3 = 17.15\pm0.3.
\label{absolute_mag_v2944_oph}
\end{eqnarray}

Figure \ref{v2944_oph_v1369_cen_v496_sct_v5666_sgr_b_v_i_logscale_3fig}
shows the $B$, $V$, and $I_{\rm C}$ light curves of V2944~Oph
together with those of V1369~Cen, V496~Sct, and V5666~Sgr.
Applying Equation (\ref{distance_modulus_general_temp_b})
for the $B$ band to Figure
\ref{v2944_oph_v1369_cen_v496_sct_v5666_sgr_b_v_i_logscale_3fig}(a),
we have the relation
\begin{eqnarray}
(m&-&M)_{B, \rm V2944~Oph} \cr
&=& \left( (m-M)_B + \Delta B\right)_{\rm V1369~Cen} - 2.5 \log 1.20 \cr
&=& 10.36 + 7.0\pm0.3 - 0.2 = 17.16\pm0.3 \cr
&=& \left( (m-M)_B + \Delta B\right)_{\rm V496~Sct} - 2.5 \log 0.89 \cr
&=& 14.15 + 2.8\pm0.3 + 0.13 = 17.08\pm0.3 \cr
&=& \left( (m-M)_B + \Delta B\right)_{\rm V5666~Sgr} - 2.5 \log 1.0 \cr
&=& 15.9 + 1.2\pm0.3 - 0.0 = 17.1\pm0.3,
\label{distance_modulus_v2944_oph_v1369_cen_v496_sct_v5666_sgr_b}
\end{eqnarray}
where we adopt $(m-M)_{B, \rm V1369~Cen}= 10.36$,
$(m-M)_{B, \rm V496~Sct}= 14.15$, and
$(m-M)_{B, \rm V5666~Sgr}= 15.9$ from Appendix \ref{qy_mus}.
We have $(m-M)_B=17.11\pm0.2$ for V2944~Oph.

Applying Equation (\ref{distance_modulus_general_temp}) to
Figure \ref{v2944_oph_v1369_cen_v496_sct_v5666_sgr_b_v_i_logscale_3fig}(b),
we have the relation
\begin{eqnarray}
(m&-&M)_{V, \rm V2944~Oph} \cr
&=& \left( (m-M)_V + \Delta V\right)_{\rm V1369~Cen} - 2.5 \log 1.20 \cr
&=& 10.25 + 6.5\pm0.3 - 0.2 = 16.55\pm0.3 \cr
&=& \left( (m-M)_V + \Delta V\right)_{\rm V496~Sct} - 2.5 \log 0.89 \cr
&=& 13.7 + 2.7\pm0.3 + 0.13 = 16.53\pm0.3 \cr
&=& \left( (m-M)_V + \Delta V\right)_{\rm V5666~Sgr} - 2.5 \log 1.0 \cr
&=& 15.4 + 1.1\pm0.3 - 0.0 = 16.5\pm0.3,
\label{distance_modulus_v2944_oph_v1369_cen_v496_sct_v5666_sgr_v}
\end{eqnarray}
where we adopt $(m-M)_{V, \rm V1369~Cen}=10.25$,
$(m-M)_{V, \rm V496~Sct}=13.7$, and $(m-M)_{V, \rm V5666~Sgr}=15.4$
from \citet{hac19k}.
We have $(m-M)_{V, \rm V2944~Oph}=16.5\pm0.2$, which is essentially
the same as Equation (\ref{distance_modulus_v2944_oph}).

From the $I_{\rm C}$-band data in Figure
\ref{v2944_oph_v1369_cen_v496_sct_v5666_sgr_b_v_i_logscale_3fig}(c),
we obtain
\begin{eqnarray}
(m&-&M)_{I, \rm V2944~Oph} \cr
&=& ((m - M)_I + \Delta I_C)_{\rm V1369~Cen} - 2.5 \log 1.20 \cr
&=& 10.07 + 5.6\pm0.3 - 0.2 = 15.47\pm0.3 \cr
&=& ((m - M)_I + \Delta I_C)_{\rm V496~Sct} - 2.5 \log 0.89 \cr
&=& 12.98 + 2.4\pm0.3 + 0.13 = 15.51\pm0.3 \cr
&=& ((m - M)_I + \Delta I_C)_{\rm V5666~Sgr} - 2.5 \log 1.0 \cr
&=& 14.6 + 0.9\pm0.3 - 0.0 = 15.5\pm0.3,
\label{distance_modulus_i_v2944_oph_v1369_cen_v496_sct_v5666_sgr}
\end{eqnarray}
where we adopt $(m-M)_{I, \rm V1369~Cen}= 10.07$,
$(m-M)_{I, \rm V496~Sct}= 12.98$, and $(m-M)_{I, \rm V5666~Sgr}= 14.6$
from Appendix \ref{qy_mus}.
We have $(m-M)_{I, \rm V2944~Oph}= 15.5\pm0.2$.

We plot $(m-M)_B=17.11$, $(m-M)_V=16.5$, and $(m-M)_I=15.5$,
which cross at $d=8.2$~kpc and $E(B-V)=0.62$, in Figure
\ref{distance_reddening_v1535_sco_v5667_sgr_v5668_sgr_v2944_oph}(d).
Thus, we obtain $d=8.2\pm1$~kpc and $E(B-V)=0.62\pm0.05$.


\begin{figure}
\plotone{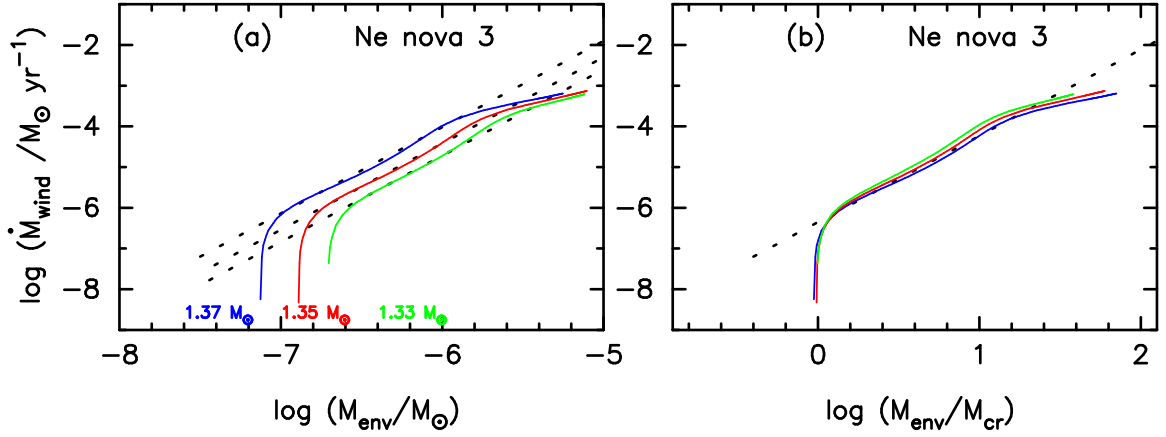}
\caption{
(a) Wind mass-loss rate versus hydrogen-rich envelope mass of
optically thick wind solutions for three WD masses with the chemical
composition of Ne nova 3 (Ne3).
The solid blue, red, and green lines denote $1.37~M_\sun$, $1.35~M_\sun$,
and $1.33~M_\sun$  WD models, respectively.
(b) The horizontal axis is scaled by the critical mass of each
hydrogen-rich envelope.  The critical mass $M_{\rm cr}$
is the minimum mass for wind solutions.  The three (blue, red, and green)
lines almost overlap.
\label{dmdt_env_mass_scaling_relation_x65z02o03ne03}}
\end{figure}


\begin{figure}
\plotone{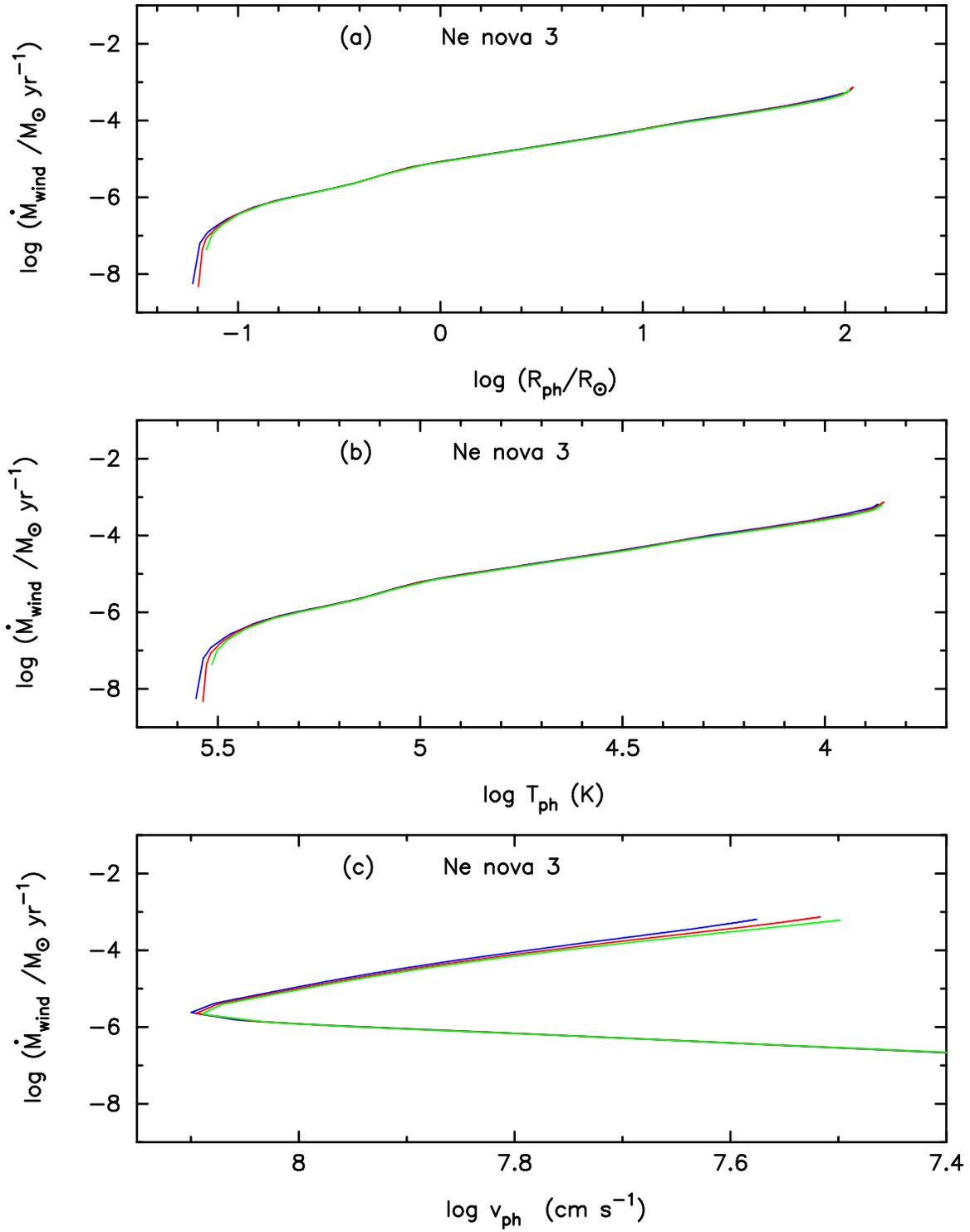}
\caption{
Wind mass-loss rate (a) versus photospheric radius,
(b) versus photospheric temperature, and (c) versus photospheric velocity
for the three ($1.37$, $1.35$, and $1.33~M_\sun$) WD mass models in Figure
\ref{dmdt_env_mass_scaling_relation_x65z02o03ne03}.
The three (blue, red, and green) lines almost overlap each other. 
\label{dmdt_ph_rad_temp_vel_scaling_relation_x65z02o03ne03}}
\end{figure}

\section{Timescaling Law of Free-Free Emission Light Curves}
\label{timescaling_law_free-free_emission}

The timescaling law of nova light curves can be explained with
optically thick wind solutions on WDs.  
\citet{kat94h} calculated the nova evolution for various WD masses 
from $0.5~M_\sun$ to $1.38~M_\sun$ based on the optically thick
wind theory.  They obtained the wind mass-loss rate $\dot M_{\rm wind}$,
photospheric temperature $T_{\rm ph}$, velocity $v_{\rm ph}$,
and radius $R_{\rm ph}$ for a specific envelope mass $M_{\rm env}$
and WD mass $M_{\rm WD}$.  We plot such results in Figures
\ref{dmdt_env_mass_scaling_relation_x65z02o03ne03} and
\ref{dmdt_ph_rad_temp_vel_scaling_relation_x65z02o03ne03}
for three WD masses of 1.33, 1.35, and $1.37~M_\sun$ 
\citep[Ne3;][]{hac16k}.
See Figure 6 of \citet{kat94h} for less massive WDs. 

Figure \ref{dmdt_env_mass_scaling_relation_x65z02o03ne03}(a) shows
the wind mass-loss rate versus hydrogen-rich envelope mass of
optically thick wind solutions for three WD masses of
$1.37~M_\sun$ (blue), $1.35~M_\sun$ (red), and $1.33~M_\sun$ (green).
The wind mass-loss rate decreases with the decreasing envelope mass.
The wind stops at the critical envelope mass $M_{\rm cr}$.
If we normalize the envelope mass by each critical mass,
these three lines almost overlap each other as shown in Figure
\ref{dmdt_env_mass_scaling_relation_x65z02o03ne03}(b).
We show only the three WD masses in this figure, but 
obtained the similar tendency of envelope solutions for other
WD masses from $0.5~M_\sun$ to $1.33~M_\sun$ with the same or
different chemical compositions \citep[see, e.g., Figure 6 of][]{kat94h}.

Figure \ref{dmdt_ph_rad_temp_vel_scaling_relation_x65z02o03ne03}
shows (a) the wind mass-loss rate versus the photospheric radius,
(b) the wind mass-loss rate versus the photospheric temperature,
and (c) the wind mass-loss rate versus the photospheric velocity,
for the three WD masses of $1.37~M_\sun$ (Ne3, blue),
$1.35~M_\sun$ (Ne3, red), and $1.33~M_\sun$ (Ne3, green),
which are the same model sequences
in Figure \ref{dmdt_env_mass_scaling_relation_x65z02o03ne03}.
It is clearly shown that these three lines almost overlap each other.
We obtained a similar tendency of the envelope solutions for other
WD masses with the same or different chemical compositions
\citep[see, e.g., Figure 7 of][]{kat94h}.

Our free-free emission flux is approximately calculated from
$F_\nu \propto \dot M^2_{\rm wind} / v^2_{\rm ph} R_{\rm ph}$ 
\citep{hac06kb}.  The above overlapping solutions directly means that
\begin{equation}
\left[ {{\dot M^2_{\rm wind}} \over{v^2_{\rm ph} R_{\rm ph}}} \right]^{ 
\{M_{\rm WD}\}}_{(t)}
=
\left[ {{\dot M^2_{\rm wind}} \over{v^2_{\rm ph} R_{\rm ph}}} \right]^{ 
\{1.37~M_\sun\}}_{(t')}
=
\left[ {{\dot M^2_{\rm wind}} \over{v^2_{\rm ph} R_{\rm ph}}} \right]^{ 
\{1.35~M_\sun\}}_{(t'')}
=
\left[ {{\dot M^2_{\rm wind}} \over{v^2_{\rm ph} R_{\rm ph}}} \right]^{ 
\{1.33~M_\sun\}}_{(t''')},
\label{free-free_flux_scale}
\end{equation}
for the same wind mass-loss rate of $\dot M_{\rm wind}$.
The only difference is the timescale.  The time is calculated from
the decreasing rate of the envelope mass, that is,
\begin {equation}
t \approx \int {{d M_{\rm env}} \over {\dot M_{\rm wind}}}
= M_{\rm cr} \int {{d (M_{\rm env}/M_{\rm cr})} \over {\dot M_{\rm wind}}}
= M_{\rm cr}(M_{\rm WD}) \tau,
\end{equation}
where
\begin {equation}
\tau \equiv \int {{d (M_{\rm env}/M_{\rm cr})} \over {\dot M_{\rm wind}}}.
\end{equation}
The overlapping of lines in Figure 
\ref{dmdt_env_mass_scaling_relation_x65z02o03ne03}(b) guarantees
that
\begin {equation}
t' \approx M_{\rm cr}(1.37~M_\sun) \tau,
\end{equation}
\begin {equation}
t'' \approx M_{\rm cr}(1.35~M_\sun) \tau,
\end{equation}
\begin {equation}
t''' \approx M_{\rm cr}(1.33~M_\sun) \tau.
\end{equation}
These relations read
\begin {equation}
{t \over {M_{\rm cr}(M_{\rm WD})}} 
\approx {t' \over {M_{\rm cr}(1.37~M_\sun)}}
\approx {t'' \over {M_{\rm cr}(1.35~M_\sun)}}
\approx {t''' \over {M_{\rm cr}(1.33~M_\sun)}}.
\label{time-scaling_law_wd_masses}
\end{equation}
Thus, the free-free emission light curves obey a timescaling law,
and its timescaling factor is approximated by the ratio of the critical
envelope masses.  From Equation (\ref{time-scaling_law_wd_masses}),
the fluxes in Equation (\ref{free-free_flux_scale}) are the
time-stretched free-free fluxes.  We define these fluxes as
\begin{equation}
F^{ \{M_{\rm WD}\}} (t)\equiv \left[ {{\dot M^2_{\rm wind}} 
\over{v^2_{\rm ph} R_{\rm ph}}} \right]^{ \{M_{\rm WD}\}}_{(t)},
\label{ff_flux_wdwd}
\end{equation}
\begin{equation}
F'^{ \{1.37~M_\sun \}} (t')\equiv \left[ {{\dot M^2_{\rm wind}} 
\over{v^2_{\rm ph} R_{\rm ph}}} \right]^{ \{1.37~M_\sun \}}_{(t')},
\label{ff_flux_wd137}
\end{equation}
\begin{equation}
F''^{ \{1.35~M_\sun \}} (t'')\equiv \left[ {{\dot M^2_{\rm wind}} 
\over{v^2_{\rm ph} R_{\rm ph}}} \right]^{ \{1.35~M_\sun \}}_{(t'')},
\label{ff_flux_wd135}
\end{equation}
and
\begin{equation}
F'''^{ \{1.33~M_\sun \}} (t''')\equiv \left[ {{\dot M^2_{\rm wind}} 
\over{v^2_{\rm ph} R_{\rm ph}}} \right]^{ \{1.33~M_\sun \}}_{(t''')}.
\label{ff_flux_wd133}
\end{equation}
We formulate the conversion from $(t, F(t))$ to $(t', F'(t'))$ by
the time stretch of $t'= t / f_{\rm s}$.
Applying $t'= t / f_{\rm s}$ to the $V$ fluxes of
Equation (\ref{ff_flux_wd137}), 
we take the time derivative of energy $E$ as
\begin {equation}
F'(t') \equiv {{d E} \over {d t'}} = f_{\rm s} {{d E} \over {d t}}\equiv
f_{\rm s} F(t/f_{\rm s}),
\end{equation}
and obtain
\begin{equation}
F^{ \{M_{\rm WD}\}} (t)\equiv \left[ {{\dot M^2_{\rm wind}} 
\over{v^2_{\rm ph} R_{\rm ph}}} \right]^{ \{M_{\rm WD}\}}_{(t)}
= \left[ {{\dot M^2_{\rm wind}} \over{v^2_{\rm ph} R_{\rm ph}}} \right]^{ 
\{1.37~M_\sun\}}_{(t')}
\equiv F'^{ \{1.37~M_\sun \}} (t') 
= f_{\rm s} F^{ \{1.37~M_\sun \}}(t/f_{\rm s}).
\label{free-free_flux_conversion}
\end{equation}
Then, we take the absolute magnitude of the $V$ flux and obtain
\begin{eqnarray}
M^{\{M_{\rm WD}\}}_V(t) &=& M'^{\{1.37 ~M_\sun\}}_V(t') \cr
&=& M^{\{1.37 ~M_\sun\}}_V(t/f_{\rm s}) - 2.5 \log f_{\rm s},
\end{eqnarray}
where $M_V$ is the absolute $V$ magnitude of the free-free emission
light curve.
Here, the conversion from $(t, F(t))$ to $(t', F'(t'))$ corresponds
to the conversion from $(t, M_V(t))$ to $(t', M'_V(t'))$.
Thus, we derive Equation (\ref{time-stretching_general}).

For the UV~1455\AA\  light curves, we calculate the flux based on blackbody
approximation, that is,
$F_\nu =  4\pi R^2_{\rm ph} B_\nu(T_{\rm ph})$, where 
$B_\nu(T_{\rm ph})$ is Planckian.  Using the above overlapping solutions
in Figure \ref{dmdt_ph_rad_temp_vel_scaling_relation_x65z02o03ne03},
we obtain
\begin{equation}
\left[ R^2_{\rm ph} B_\nu(T_{\rm ph}) \right]^{\{M_{\rm WD}\}}_{(t)}
=
\left[ R^2_{\rm ph} B_\nu(T_{\rm ph}) \right]^{\{1.37~M_\sun\}}_{(t')}
=
\left[ R^2_{\rm ph} B_\nu(T_{\rm ph}) \right]^{\{1.35~M_\sun\}}_{(t'')}
=
\left[ R^2_{\rm ph} B_\nu(T_{\rm ph}) \right]^{\{1.33~M_\sun\}}_{(t''')},
\label{uv1455aa_flux_scale}
\end{equation}
for the same wind mass-loss rate of $\dot M_{\rm wind}$.
Therefore, the UV~1455\AA\  fluxes also follow the same timescaling law
as that of the free-free emission light curves from Equation 
(\ref{time-scaling_law_wd_masses}).  Similarly, we have
\begin{equation}
F'_\nu (t') \equiv {{d E_\nu} \over {d t'}} 
= f_{\rm s} {{d E_\nu} \over {d t}}\equiv f_{\rm s} F_\nu (t/f_{\rm s}).
\end{equation}
This reads
\begin{equation}
F_\nu^{\{M_{\rm WD}\}}(t) \equiv
\left[ 4\pi R^2_{\rm ph} B_\nu(T_{\rm ph}) \right]^{\{M_{\rm WD}\}}_{(t)}
= \left[ 4\pi R^2_{\rm ph} B_\nu(T_{\rm ph}) \right]^{\{1.37~M_\sun\}}_{(t')}
\equiv F{'}^{\{1.37~M_\sun\}}_\nu (t')
= f_{\rm s} F_\nu^{\{1.37~M_\sun\}}(t/f_{\rm s}),
\label{scaling_law_uv1455aa_flux}
\end{equation}
for the UV~1455\AA\  blackbody flux.  The UV~1455\AA\   flux obeys
the same scaling law as the free-free emission flux, i.e.,
Equations (\ref{free-free_flux_conversion}) 
and (\ref{scaling_law_uv1455aa_flux}).
Thus, our free-free and UV~1455\AA\  fluxes obey the same timescaling law,
and their fluxes overlap independently of the WD mass
if we properly stretch/squeeze the timescale.
Such results are shown in Figure 8 of \citet{hac06kb} for 
various WD masses and chemical compositions.   

It should be noted that the soft X-ray light curves do not obey
the same timescaling law.  
In the supersoft X-ray source (SSS) phase, the optically thick winds
had already stopped and the hydrogen-rich envelope mass decreases
by hydrogen burning, not by wind mass-loss.  Thus, the SSS phase does not
follow the same timescaling law of the universal decline law.
The model soft X-ray light curves are calculated from a blackbody
approximation like the UV~1455\AA\  light curves
\citep[e.g.,][]{hac06kb, hac10k}, but its flux increases
after optically thick winds stop.  This means that we do not use 
the above overlapping solutions in Figure
\ref{dmdt_ph_rad_temp_vel_scaling_relation_x65z02o03ne03}
and the relation of Equation (\ref{time-scaling_law_wd_masses})
because $\dot M_{\rm wind}=0$ after the optically thick winds stop. 
Therefore, the soft X-ray light curves do not follow the same
timescaling law as those of free-free emission and UV~1455\AA\  
light curves.  A detailed explanation is given by \citet{hac10k},
and their detailed results are shown in their Figure 6.


\begin{figure}
\plotone{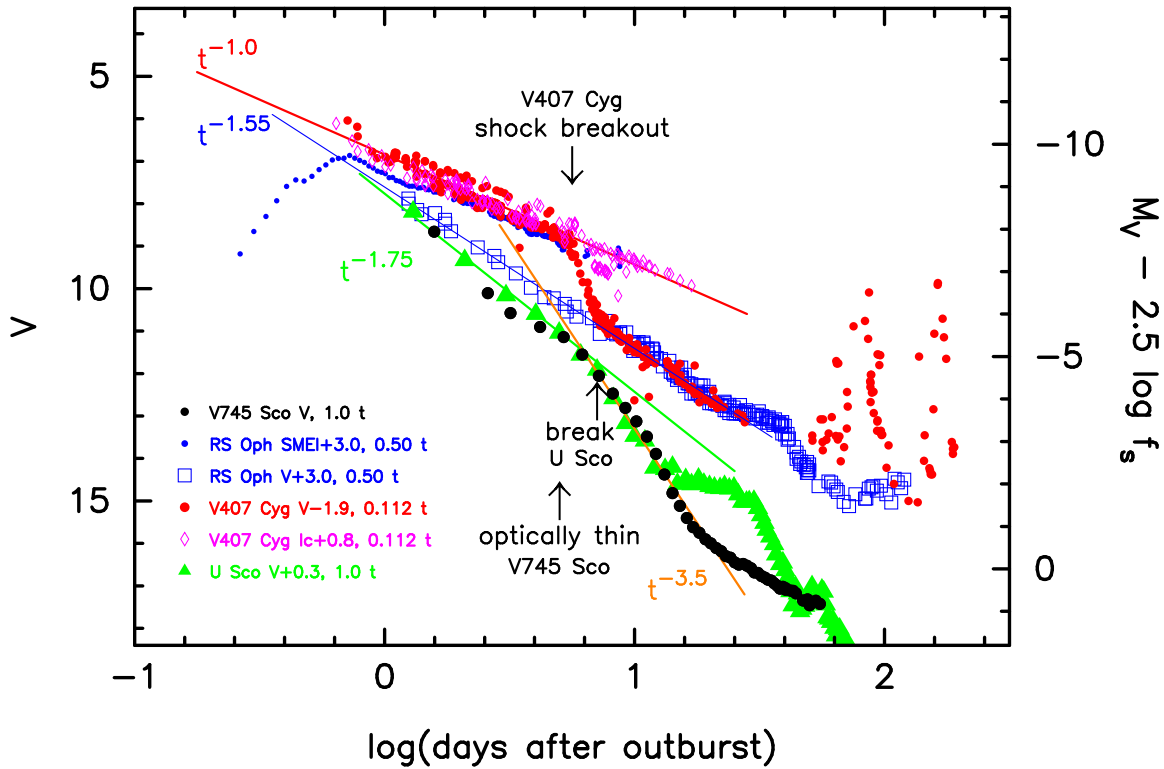}
\caption{
Nova $V$ light curves for V745~Sco, U~Sco, RS~Oph, and V407~Cyg on 
logarithmic timescales.  We add the $I_{\rm C}$ light curves of V407~Cyg and
the {\it SMEI} magnitudes \citep{hou10} of RS~Oph for comparison.
We also add the time-stretched absolute magnitude, $M_V - 2.5 \log f_{\rm s}$,
against V745~Sco ($f_{\rm s}=1$ for V745~Sco), in the right ordinate.
The $V$ light curves of the V745~Sco (2014) and U~Sco (2010) outbursts are
plotted with the filled black circles and green triangles, respectively. 
The $V$ and $I_{\rm C}$ light curves of V407~Cyg are represented by
the filled red circles and unfilled magenta diamonds, respectively. 
The {\it SMEI} and $V$ magnitudes of the RS~Oph (2006) outburst are
depicted by the blue dots and unfilled blue squares, respectively. 
The sources of the data are all the same as those in \citet{hac18kb}.
The solid red, blue, green, and orange lines represent
the decline trends of $F_\nu \propto t^{-1.0}$, $t^{-1.55}$,
$t^{-1.75}$, and $t^{-3.5}$, respectively. 
\label{v745_sco_u_sco_v407_cyg_rs_oph_v_template_no2}}
\end{figure}

\section{Comparison of M31N~2008-12\lowercase{a} and V745~Sco with 
V1500~Cyg and LV~Vul}
\label{m31n200812a_v1500_cyg}
The universal decline law can be applied to many novae as already shown in
the main text.  \citet{hac18kb} analyzed fast and recurrent novae and 
showed that some of them deviate from the universal decline law.
As shown in Figure \ref{v745_sco_u_sco_v407_cyg_rs_oph_v_template_no2},
V407~Cyg shows a decay trend of $F_\nu \propto t^{-1.0}$ in the early phase,
which is different from the universal decline law of $F_\nu \propto t^{-1.75}$.
V745~Sco does not have a substantial slope of $F_\nu \propto t^{-1.75}$,
but rather a steeper decline of $F_\nu \propto t^{-3.5}$.
Here, we explain the reason why these light curves deviate 
from the universal decline law
and discuss the applicability of the time-stretching method to such novae.

We first explain the reason for $F_\nu \propto t^{-1.0}$, which is
much more gradual than the universal decline of $F_\nu \propto t^{-1.75}$,
and then the reason for $F_\nu \propto t^{-3.5}$, which is much steeper
than the universal decline.  The companion star to the WD of V407~Cyg 
is a Mira \citep{mun90, kol98, kol03}.
The Mira companion emits massive cool winds and, as a result,
CSM accumulates around the binary \citep[e.g.,][]{moh12}.  
V407~Cyg outbursted in 2010 as a classical nova \citep{nis10}.
Just after the nova explosion, the nova ejecta
plunge into the CSM and create a strong shock 
\citep[e.g.,][]{orl12, pan15}.  \citet{sho11} and \citet{iij15} showed
clear evidence of a strong shock in their spectra.
Such a CSM shock has been observed in many Type IIn supernovae.
\citet{mor13} presented a decay trend of $F_\nu \propto t^{-1}$ in SN~2005ip.
Thus, we regard the decay trend of $F_\nu \propto t^{-1}$ to be an effect of
strong CSM shock.  Both the $V$ and $I_{\rm C}$ light curves show
a similar trend as shown in Figure
\ref{v745_sco_u_sco_v407_cyg_rs_oph_v_template_no2}.
\citet{hac18kb} explained the decline trend as follows: 
just after the nova outburst, the nova ejecta created a strong CSM shock.
This shock-heating made a significant contribution to the continuum 
flux (both in the $V$ and $I_{\rm C}$ bands).
The strong shock has broken out of the CSM on day $\sim 45$.
Because the shock-heating disappears, the continuum flux drops and decays as 
$F_\nu \propto t^{-1.55}$ like RS~Oph (Figure 
\ref{v745_sco_u_sco_v407_cyg_rs_oph_v_template_no2}).

RS~Oph decays as $F_\nu \propto t^{-1.55}$ in the $V$ band 
(Figure \ref{v745_sco_u_sco_v407_cyg_rs_oph_v_template_no2}).
It should be noted that the radio flux of RS~Oph also follows 
$F_\nu \propto t^{-1.55}$ \citep[see Figure 18 of][]{hac18kb}.  
The companion star to the WD is not a Mira but a red giant in RS~Oph.
Cool winds in RS~Oph are much less massive than in V407~Cyg.
The ejecta created a CSM shock, but the CSM shock is much weaker than
in V407~Cyg.  We suppose that the CSM shock is not strong enough
to increase the continuum flux to $F_\nu \propto t^{-1.0}$ in RS~Oph.
However, the {\it SMEI} light curve of RS~Oph follows 
$L_{\rm SMEI}\propto t^{-1.0}$, where $L_{\rm SMEI}$ 
is the {\it SMEI} band luminosity (peak quantum efficiency at 7000\AA\   
with an FWHM of 3000\AA).  This is because the CSM shock 
increases the H$\alpha$ flux, which is included in the {\it SMEI} band.

V745~Sco, T~CrB, and V1534~Sco also have a red-giant companion,
but do not show evidence of a CSM shock in their $V$ light curves.
Their $V$ light curves almost overlap the V838~Her $V$ light curve 
\citep[see, e.g.,][]{hac18kb}.  This implies that their CSM is extremely
less massive because V838~Her has a main-sequence companion \citep[e.g.,
][]{ing92, lei92}, and we expect no CSM shock in V838~Her.
The X-ray fluxes were observed with {\it Swift} in V407~Cyg, RS~Oph,
V745~Sco, and V1534~Sco. The origin of these early phase fluxes is
shock-heating.  On the other hand, \citet{mun18} showed
no deceleration of the V1534~Sco ejecta.
We suppose that the decay trend of the $V$ light curve changes
from an extremely strong CSM shock to a relatively weak shock,
that is, from the trend of V407~Cyg ($F_\nu \propto t^{-1.0}$) to RS~Oph
($F_\nu \propto t^{-1.55}$), and finally to V745~Sco and V1534~Sco 
\citep{hac18kb} as shown in Figure 
\ref{v745_sco_u_sco_v407_cyg_rs_oph_v_template_no2}.

Next, we explain the steep decline of $F_\nu \propto t^{-3.5}$.
The slope of the nova decay trend changes from
$F_\nu \propto t^{-1.75}$ (universal decline) 
to $F_\nu \propto t^{-3.5}$ (rapid decline) on day $\sim 8$
(upward black arrow labeled ``break U~Sco''
in Figure \ref{v745_sco_u_sco_v407_cyg_rs_oph_v_template_no2}).
This change in the slope is universal among the model light curves (see 
Figure \ref{all_mass_v1668_cyg_x45z02c15o20_real_scale_universal_no2}).
\citet{hac06kb} called this change the break.
It accompanies the quick drop in the wind mass-loss rate and the rapid
shrinking of the photosphere.  In the case of V1668~Cyg in Figure
\ref{all_mass_v1668_cyg_x45z02c15o20_real_scale_universal_no2},
this change is almost coincident with the start of the nebular phase
on day $\sim 100$.  The break in the V745~Sco $V$
light curve almost coincides with that of U~Sco
in Figure \ref{v745_sco_u_sco_v407_cyg_rs_oph_v_template_no2}.
This epoch is close to the epoch when the ejecta of V745~Sco became
optically thin (upward black arrow labeled ``optically thin V745~Sco'').
We regard the change from $F_\nu \propto t^{-1.75}$ to
$F_\nu \propto t^{-3.5}$ to be the transition of the ejecta
from being optically thick to thin.

A nova reaches point A or a brighter point on the model light curves
in Figure \ref{all_mass_v1668_cyg_x45z02c15o20_calib_linear_m098} if
the ignition mass is large.
Then, the nova declines along the model light curve with
decreasing envelope mass, that is, it passes through points B and C,
and eventually goes down through the break point.
If the initial envelope mass is very small, a nova cannot reach 
the universal decline segment ($F_\nu \propto t^{-1.75}$).
It starts from some place on the rapid decline segment 
($F_\nu \propto t^{-3.5}$).  Thus, its maximum brightness is much
fainter than that of normal decline type, novae which start 
from some place on the universal decline segment.
Such novae, having very small ignition masses, like M31N~2008-12a,
are located much below the empirical MMRD relations 
(Figure \ref{max_t3_scale_no8}).

\citet{hac18kb} categorized these novae into three types:\\
{\bf 1.} Novae that follow the trend of the universal decline
law ($F_\nu \propto t^{-1.75}$) in the early phase, as shown in Figure 
\ref{v745_sco_u_sco_v407_cyg_rs_oph_v_template_no2},  
are called normal-decline (or U~Sco) types.\\
{\bf 2.} Novae that start from some place on the rapid decline
($F_\nu \propto t^{-3.5}$) segment after the break, as shown in Figure 
\ref{v745_sco_u_sco_v407_cyg_rs_oph_v_template_no2}, are called
rapid-decline (or V745~Sco) types.  The change in the slope from 
$F_\nu \propto t^{-1.75}$ to $F_\nu \propto t^{-3.5}$
corresponds to the break in the model $V$ light curves in Figure
\ref{all_mass_v1668_cyg_x45z02c15o20_real_scale_universal_no2}. \\
{\bf 3.} Novae that follow a slow decline trend of $F_\nu \propto t^{-1.0}$
in the early phase, 
as shown in Figure \ref{v745_sco_u_sco_v407_cyg_rs_oph_v_template_no2},  
are called CSM-shock (or RS~Oph) types.
Novae of this type show a strong CSM shock interaction,
which contributes significantly to the $F_\nu$ flux.

These three kinds of very fast (or recurrent) novae sometimes do not
follow the universal decline law, and we cannot directly apply
Equation (\ref{distance_modulus_general_temp}) to them because it is
based on the universal decline law.  
The physical reasons for the deviations are well understood as 
already explained above.  Therefore, \citet{hac18kb} proposed
different template novae instead of V1500 Cyg and LV Vul.
For example, CSM-shock-type novae broadly follow the $V$ or $I_{\rm C}$
light curve of V407 Cyg (or RS Oph) in the early phase.
If we properly stretch the timescales of such novae 
and can overlap their $V$ light curves with
one of the template novae V407~Cyg and RS~Oph,
we apply Equation (\ref{distance_modulus_general_temp}) to them.
Such an example is LMC N 2013 as discussed by \citet{hac18kb}. 
\citet{hac18kb} also examined the case of rapid-decline-type novae,
that is, V745 Sco, T CrB, V838 Her, and V1534 Sco in our Galaxy, 
LMC N 2012a in the LMC, and M31N 2008-12a in M31.  They overlapped
the $V$ light curve of T~CrB, V838~Her, and V1534~Sco with that of V745~Sco
by stretching their timescales and obtained reasonable 
$(m-M)_V$ for T~CrB, V838~Her, and V1534~Sco with
Equation (\ref{distance_modulus_general_temp}).


\begin{figure}
\epsscale{0.75}
\plotone{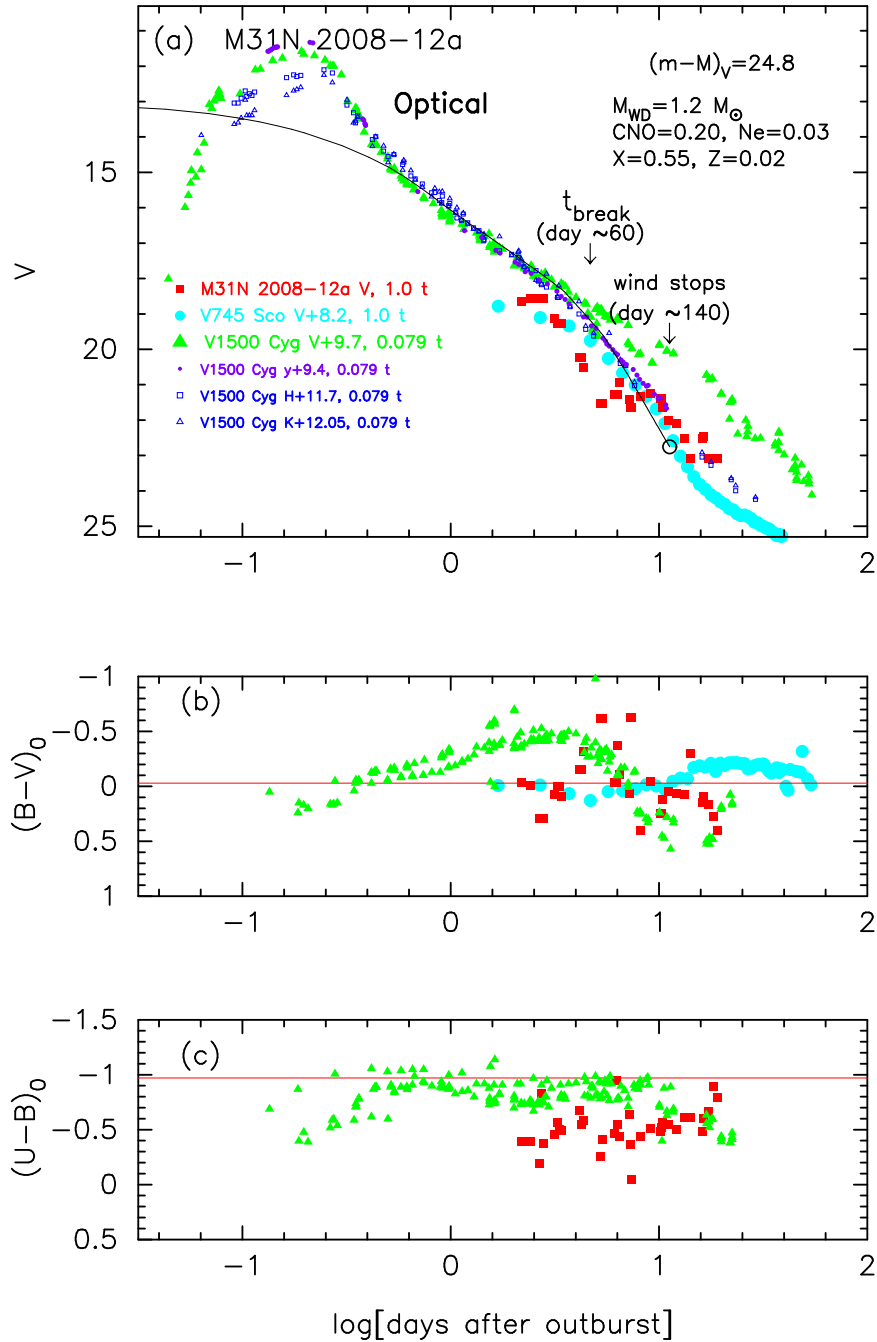}
\caption{
Same as Figure \ref{v597_pup_v1500_cyg_lv_vul_v_bv_ub_x65z02o03ne03_logscale},
but for M31N~2008-12a.  
We add the light/color curves of V745~Sco and V1500~Cyg. 
In panel (a), assuming that $(m-M)_V=12.3$ for V1500~Cyg, 
we add the model $V$ light curve of a $1.2~M_\sun$ WD 
\citep[Ne2, solid black line;][]{hac10k}.
We also add two epochs of the V1500~Cyg light curves:
one is the break (about day $\sim 60$) in the light curves 
\citep{hac14k} and the other is the epoch when the wind stops
(about day $\sim 140$) for our $1.2~M_\sun$ WD model \citep{hac14k}.
\label{m31_12a_v745_sco_v1500_cyg_v_bv_ub_logscale}}
\end{figure}


\begin{figure}
\epsscale{0.75}
\plotone{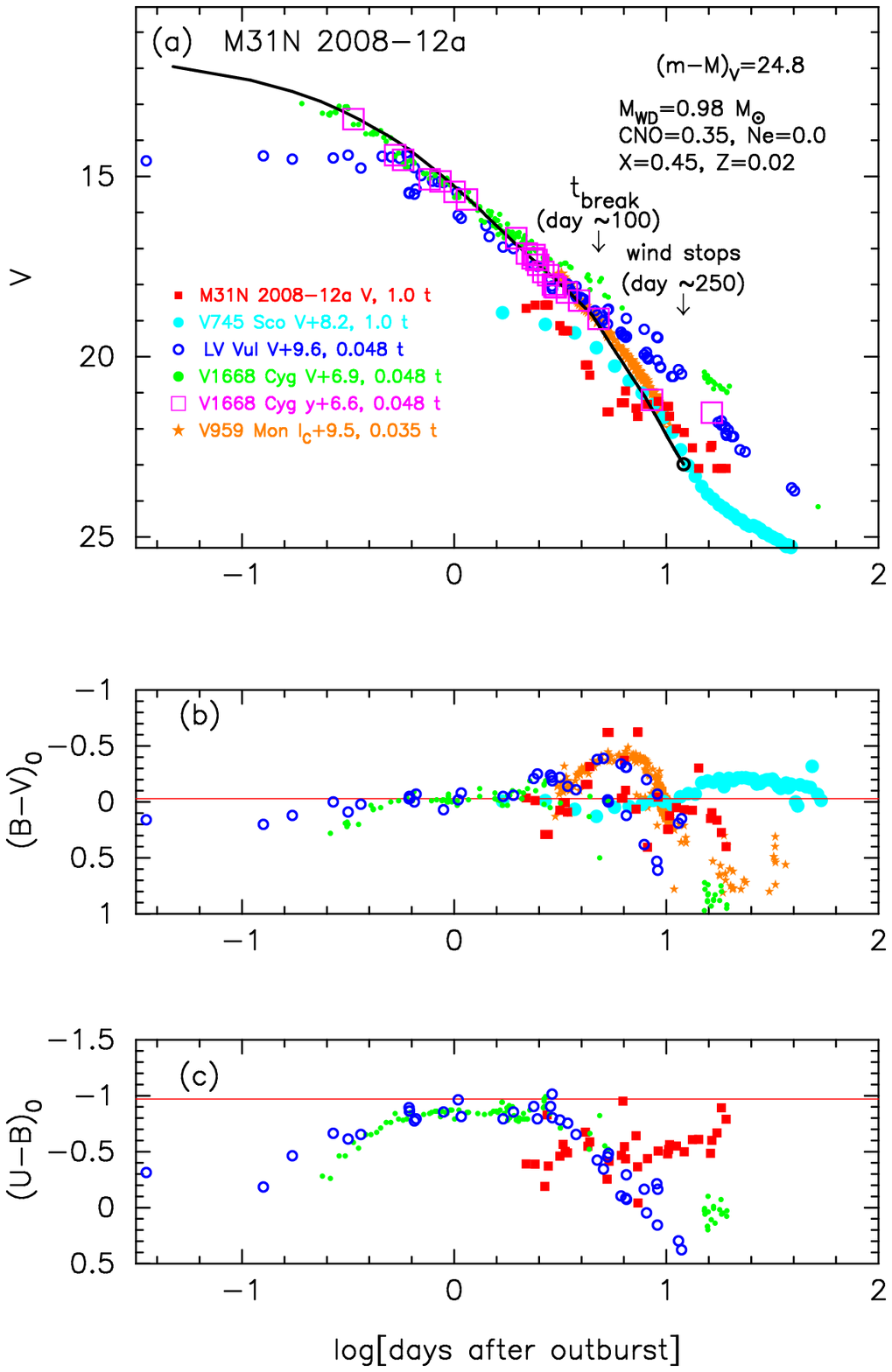}
\caption{
Same as Figure \ref{v1663_aql_lv_vul_v1668_cyg_v_bv_logscale},
but for M31N~2008-12a.  
We add the light/color curves of V745~Sco, LV~Vul, V1668~Cyg,
and V959~Mon.  The data of V959~Mon are taken from AAVSO, VSOLJ,
SMARTS, and \citet{mun13b}.
In panel (a), assuming that $(m-M)_V=14.6$ for V1668~Cyg, 
we add the model $V$ light curve of a $0.98~M_\sun$ WD 
\citep[CO3, solid black line;][]{hac16k}.
We also add two epochs of the V1668~Cyg light curves:
one is the break (about day $\sim 100$) in the $y$ light curve 
and the other is the epoch when the optically thick wind stops
(about day $\sim 250$) for our $0.98~M_\sun$ WD model \citep{hac18k}.
\label{m31_12a_v745_sco_lv_vul_v1668_cyg_v_bv_ub_logscale}}
\end{figure}

Finally, we compare the light curve of M31N~2008-12a with the template
novae V1500~Cyg and LV~Vul and show that the $V$ light curve of M31N
2008-12a does not overlap with the template nova V1500~Cyg or LV~Vul,
that is, Equation (\ref{distance_modulus_general_temp}) cannot directly
apply to the pair of M31N~2008-12a and V1500~Cyg (or LV~Vul).
We first depict M31N~2008-12a, V745 Sco, and V1500~Cyg in Figure 
\ref{m31_12a_v745_sco_v1500_cyg_v_bv_ub_logscale}.  We next plot
M31N 2008-12a, V745~Sco, LV~Vul, V1668~Cyg, and V959~Mon
in Figure \ref{m31_12a_v745_sco_lv_vul_v1668_cyg_v_bv_ub_logscale}.
We always include the light/color curves of V745 Sco because V745 Sco is a
template nova for the rapid-decline-type novae like M31N 2008-12a.
Then, we discuss the relation between M31N~2008-12a and V1500~Cyg (or LV~Vul),
and describe how to apply Equation (\ref{distance_modulus_general_temp})
to the pair of M31N~2008-12a and V1500~Cyg (or LV~Vul).

\citet{hac18kb} discussed that M31N~2008-12a and V745~Sco belong to
the rapid-decline type.  They determined the timescaling factor $f_{\rm s}$
mainly from the rapid decline segment of $F_\nu \propto t^{-3.5}$
together with X-ray phase (see also Figures
\ref{v745_sco_u_sco_v407_cyg_rs_oph_v_template_no2},
\ref{m31_12a_v745_sco_v1500_cyg_v_bv_ub_logscale},
and \ref{m31_12a_v745_sco_lv_vul_v1668_cyg_v_bv_ub_logscale}).
It should be noted that our light-curve fitting method
is not only based on the shape of the $V$ light curve 
but also includes global evolutionary features that represent the
physical properties of novae.  In this fitting, we first
overlap the $V$ light curve of M31N 2008-12a with that of V745 Sco
because V745 Sco is a template nova of the rapid-decline type.  This
process was described in detail in \citet{hac18kb}.  We consider
the pair of M31N 2008-12a and V745~Sco as a whole.  We match the starts
of the nebular phases both for M31N 2008-12a and V1500~Cyg
in Figure \ref{m31_12a_v745_sco_v1500_cyg_v_bv_ub_logscale}.
The nebular phase of V1500~Cyg started almost 60 days after the outburst.
The start of the nebular phase of M31N 2008-12a is not clear, but the
SSS phase started at least six days after the peak.  This means that
we set the phase of day 5 of M31N 2008-12a at the phase of day 60 of
V1500 Cyg. 

We try to overlap the rapid-decline ($F_\nu\propto t^{-3.5}$)
segment of V1500~Cyg with that of M31N 2008-12a and V745 Sco.
V1500 Cyg shows no break in the $V$ light curve, so
we cannot find the rapid-decline segment.
Instead, we find a clear break on the $y$- and NIR-band
light curves as shown in Figure 
\ref{m31_12a_v745_sco_v1500_cyg_v_bv_ub_logscale}.
This is because strong [\ion{O}{3}] lines cloud the break on the $V$ light
curve.  Fortunately, $y$ and NIR $H$- and $K$-band data are available
for V1500 Cyg.  Therefore, we use the $y$-band light curve to detect
the break in the light curve.
These $y$ band and NIR $H$ and $K$ bands have no strong effect from the
emission lines.  The $V$ light curves of V745 Sco and M31N 2008-12a also have
no strong effect from emission lines because their ejecta are much less
massive than the ejecta of V1500 Cyg.  Therefore, the $V$ light curves of
V745 Sco and M31N 2008-12a broadly follow the rapid-decline part of
V1500 Cyg on the $y$ light curve.
We add the model $V$ light curve of a $1.2~M_\sun$ WD
\citep[Ne2;][]{hac10k}.  
The model $V$ light curve broadly follows the break of the $y$ light curve
of V1500~Cyg.  
Applying Equation (\ref{distance_modulus_general_temp}) to them,
we have the relation 
\begin{eqnarray}
(m&-&M)_{V, \rm M31N~2008-12a} \cr 
&=& (m - M + \Delta V)_{V, \rm V1500~Cyg} - 2.5 \log 0.079 \cr
&=& 12.3 + 9.7\pm0.4 + 2.75 = 24.75\pm0.4 \cr
&=& (m - M + \Delta V)_{V, \rm V745~Sco} - 2.5 \log 1.0 \cr
&=& 16.6 + 8.2\pm0.4 + 0.0 = 24.8\pm0.4,
\label{distance_modulus_m31n200812a_v1500_cyg}
\end{eqnarray}
where we adopt $(m-M)_{V, \rm V1500~Cyg}=12.3$ from \citet{hac19k}
and $(m-M)_{V, \rm V745~Sco}=16.6$ from \citet{hac18kb}.
Thus, we obtain $(m-M)_V=24.8\pm0.3$ and $f_{\rm s}= 0.079$ against
V1500~Cyg.   

We also try to overlap LV~Vul (and V1668~Cyg) with M31N 2008-12a
and V745 Sco.  LV Vul shows no break in the $V$ light curve as shown
in Figure \ref{m31_12a_v745_sco_lv_vul_v1668_cyg_v_bv_ub_logscale}.
V1668 Cyg has no break on the $V$ light curve, too, but does show
a break on the $y$ light curve. This is because strong [\ion{O}{3}] lines
contribute to the broadband $V$ in the nebular
phase, but not to the intermediate band $y$ 
\citep[see, e.g., Figure 1 of][]{mun13b}. 
The $y$ band is designed to avoid strong emission lines such as 
[\ion{O}{3}], so it is useful to detect the break.
The number of the V1668~Cyg $y$ data points after the break is very few,
so we add the $I_{\rm C}$-band light curve of V959~Mon, the data of
which are taken from SMARTS and \citet{mun13b}.  The timescaling
factor, reddening, and distance modulus in the $V$ band of V959~Mon
are taken from \citet{hac18k}.   
The $I_{\rm C}$ band is also not affected by strong emission lines
in the nebular phase so that the $I_{\rm C}$ light curve of V959~Mon
does show a similar shape to the $y$-band light curve of V1668~Cyg.    
The $V$ light curves of V745 Sco and M31N 2008-12a have no strong effects
of emission lines because their ejecta are much less massive than 
the ejecta of LV~Vul and V1668~Cyg.  Therefore, the $V$ light curves of
V745 Sco and M31N 2008-12a broadly follow the rapid-decline part of
V1668~Cyg on the $y$-band light curve after the break.
We also add a model $V$ light curve of a $0.98~M_\sun$ WD
\citep[CO3;][]{hac16k}.  
The model $V$ light curve broadly follows the break of the $y$ light
curve of V1668~Cyg.  

We place the phase of day 5 of M31N 2008-12a (start of the nebular phase)
at the phase of day 100 (time of the break $=$ start of the nebular phase)
of V1668~Cyg to adjust both timescales.  This suggests that
$f_{\rm s}\approx 0.05$ against the timescales of LV~Vul and V1668~Cyg,
because those of LV~Vul and V1668~Cyg are the same 
\citep[see, e.g.,][]{hac19k}.  We finally obtain the fitting locations
of M31N 2008-12a, V745~Sco, LV~Vul, V1668~Cyg, and V959~Mon as shown
in Figure \ref{m31_12a_v745_sco_lv_vul_v1668_cyg_v_bv_ub_logscale}.
Applying Equation (\ref{distance_modulus_general_temp}) to them,
we have the relation 
\begin{eqnarray}
(m&-&M)_{V, \rm M31N~2008-12a} \cr 
&=& (m - M + \Delta V)_{V, \rm LV~Vul} - 2.5 \log 0.048 \cr
&=& 11.85 + 9.6\pm0.4 + 3.3 = 24.75\pm0.4 \cr
&=& (m - M + \Delta V)_{V, \rm V1668~Cyg} - 2.5 \log 0.048 \cr
&=& 14.6 + 6.9\pm0.4 + 3.3 = 24.8\pm0.4 \cr
&=& (m - M + \Delta V)_{V, \rm V745~Sco} - 2.5 \log 1.0 \cr
&=& 16.6 + 8.2\pm0.4 + 0.0 = 24.8\pm0.4,
\label{distance_modulus_m31n200812a_lv_vul}
\end{eqnarray}
where we adopt $(m-M)_{V, \rm LV~Vul}=11.85$ and
$(m-M)_{V, \rm V1668~Cyg}=14.6$ from \citet{hac19k}.
We obtain $(m-M)_V=24.8\pm0.3$ and $f_{\rm s}= 0.048$ 
($\log f_{\rm s}= -1.32$) against LV~Vul. 

We summarize this section as follows:\\
1. There are exceptions, the $V$ light curves of which deviate largely
from the universal decline law.\\
2. We cannot directly apply the template nova V1500 Cyg or LV Vul 
(or V1668~Cyg) to these exceptional novae in the context of Equation 
(\ref{distance_modulus_general_temp}), because their $V$ light curves 
never overlap with the $V$ light curve of V1500 Cyg or LV Vul (or V1668~Cyg).\\
3. \citet{hac18kb} examined very fast (or recurrent) novae and proposed
different templates depending on the physical reason of deviation.
Adopting different template novae, they applied Equation 
(\ref{distance_modulus_general_temp}) to the exceptions.\\
4. For the special case of M31N 2008-12a, its $V$ light curve does not
overlap that of V1500~Cyg or LV~Vul (or V1668~Cyg) in the context of
Equation (\ref{distance_modulus_general_temp}).  M31N 2008-12a belongs to
the rapid-decline type, and the template is V745~Sco \citep{hac18kb}.
Therefore, we consider the pair of M31N 2008-12a and V745~Sco as a whole.
If we use the V1500 Cyg $y$ light curve,
we can find an overlapping part on the rapid-decline segment and apply
Equation (\ref{distance_modulus_general_temp}) to the pair of 
M31N 2008-12a and V745~Sco.  Similarly, if we use the V1668~Cyg $y$ light
curve, we are able to find an overlapping part on the rapid-decline segment
of V1668~Cyg and apply Equation (\ref{distance_modulus_general_temp}). \\

\clearpage







\end{document}